%% file: Thesis.tex
\documentclass[11pt, a4paper, oneside]{Thesis} % Paper size, default font size and one-sided paper
\usepackage[sorting=none]{biblatex}
\graphicspath{{./Figures/}} % Specifies the directory where pictures are stored
\hypersetup{urlcolor=blue, colorlinks=true} % Colors hyperlinks in blue - change to black if annoying
\title{\ttitle} % Defines the thesis title - don't touch this
\begin{document}

\frontmatter % Use roman page numbering style (i, ii, iii, iv...) for the pre-content pages

\setstretch{1.3} % Line spacing of 1.3

% Define the page headers using the FancyHdr package and set up for one-sided printing
\fancyhead{} % Clears all page headers and footers
\rhead{\thepage} % Sets the right side header to show the page number
\lhead{} % Clears the left side page header

\pagestyle{fancy} % Finally, use the "fancy" page style to implement the FancyHdr headers

\newcommand{\HRule}{\rule{\linewidth}{0.5mm}} % New command to make the lines in the title page

%----------------------------------------------------------------------------------------
%	TITLE PAGE
%----------------------------------------------------------------------------------------
\input{./TitlePage/TitlePage}

\clearpage % Start a new page

%----------------------------------------------------------------------------------------
%	ACKNOWLEDGEMENTS
%----------------------------------------------------------------------------------------

\setstretch{1.3} % Reset the line-spacing to 1.3 for body text (if it has changed)

\acknowledgements{\addtocontents{toc}{\vspace{1em}} % Add a gap in the Contents, for aesthetics
\small{
The hard work on the thesis is left behind and I realize that my research would not be as good as it is now without support and assistance of many wonderful people.

First of all, I would like to express my sincere gratitude to my thesis advisor Dr. Alexander Semenov. I am indebted to him for his supervision, encouragement, patience, and rigorous training during all that time. His help has directly contributed to my understanding of difficult issues of polymer physics, influenced my way of thinking about and performing scientific research. It has definitely lifted my skills to a new level of professionalism.  

I would like also to thank Professor Jörg Baschnagel. He enraptures everyone who knows him. His personal attitude and scientific discussions have always commanded a great respect. Moreover, Prof. Baschnagel together with Prof. Günter Reiter organized the great \href{http://www.softmattergraduate.uni-freiburg.de/}{IRTG} project devoted to the collaboration between Strasbourg and Freiburg Universities. I am personally grateful for the opportunity to be a part of this project and work with amazing people. In particular, gorgeous coordinators of the project Christelle Vergnat, Amandine Henckel and Birgitta Zovko who were always open for any questions and help, and friendly and kind collaborators Prof. Eckhard Bartsch and PhD student Jochen Schneider from the University of Freiburg for their hospitality during my 3-month internship in their laboratory. 

My deepest appreciation also goes to the \href{http://en.wikipedia.org/wiki/Institut_Charles_Sadron}{Institute Charles Sadron} for its friendly atmosphere and to all the people who made this atmosphere feasible. Furthermore, I would like to express my gratitude to all people in our theory and simulation group, whose company I have enjoyed within these years. I would like to especially mention Dr. Olivier Benzerara who has helped me a lot during my PhD, Dr. Patrycja Polińska, Dr. Gi-moon Nam, Dr. Stephan Frey, Dr. Fabian Weysser, Dr. Julia Zabel, Dr. Nava Schulman, Dr. Diddo Diddens, Dr. Falko Ziebert, Julien Fierling and Céline Ruscher for their amazing company in our office and wonderful talks. I acknowledge many pleasant discussions with Prof. Sergei Obukhov during his stay in Strasbourg every summer.

I express my heartfelt gratitude to my friends Artyom Kovalenko, Filipp Napolskiy, Sergey Pronkin, Alexander Shakhov, my numerous \href{http://en.wikipedia.org/wiki/MEPhI}{MEPhI} friends, my English teachers Michael Leonard and John Abrams who helped and supported me within these years.

Last but not least, I am thankful to my family for their love and unconditional support during my study. In particular, the patience and understanding of my mum, dad and sister during the years are greatly appreciated.
}}
% Resume of the thesis in French
\input{./resume/resume_fr}

\input{./resume/resume_eng}

\clearpage % Start a new page
%----------------------------------------------------------------------------------------
%	LIST OF CONTENTS/FIGURES/TABLES PAGES
%----------------------------------------------------------------------------------------

\pagestyle{fancy} % The page style headers have been "empty" all this time, now use the "fancy" headers as defined before to bring them back
\lhead{\emph{Contents}} % Set the left side page header to "Contents"
\tableofcontents % Write out the Table of Contents

%\lhead{\emph{List of Figures}} % Set the left side page header to "List of Figures"
%\listoffigures % Write out the List of Figures

%\lhead{\emph{List of Tables}} % Set the left side page header to "List of Tables"
%\listoftables % Write out the List of Tables

%----------------------------------------------------------------------------------------
%	ABBREVIATIONS
%----------------------------------------------------------------------------------------

\clearpage % Start a new page

\setstretch{1.5} % Set the line spacing to 1.5, this makes the following tables easier to read

\lhead{\emph{Abbreviations}} % Set the left side page header to "Abbreviations"
\listofsymbols{ll} % Include a list of Abbreviations (a table of two columns)
{
\textbf{ODE} & \textbf{O}rdinary \textbf{D}ifferential \textbf{E}quation \\
\textbf{PDE} & \textbf{P}artial \textbf{D}ifferential \textbf{E}quation \\
\textbf{SCFT} & \textbf{S}elf \textbf{C}onsistent \textbf{F}ield \textbf{T}eory \\
\textbf{GSD} & \textbf{G}round \textbf{S}tate \textbf{D}ominance \\
\textbf{RPA} & \textbf{R}andom \textbf{P}hase \textbf{A}pproximation	\\
\textbf{GSDE} & \textbf{G}round \textbf{S}tate \textbf{D}ominance \textbf{E}xtended\\
\textbf{WAR} & \textbf{W}eak \textbf{A}dsorption \textbf{R}egime\\
\textbf{SAR} & \textbf{S}trong \textbf{A}dsorption \textbf{R}egime\\
\textbf{PI} & \textbf{P}olymer \textbf{I}nduced\\
\textbf{FPI} & \textbf{F}ree \textbf{P}olymer \textbf{I}nduced\\
\textbf{hr} & \textbf{h}igh \textbf{r}esolution\\
\textbf{sh} & \textbf{sh}ort-range\\
\textbf{lr} & \textbf{l}ong-\textbf{r}ange\\
\textbf{pr} & \textbf{p}urely \textbf{r}epulsive\\
\textbf{PS} & \textbf{P}olystyrene \\
\textbf{PEG} & \textbf{P}oly\textbf{e}thylene \textbf{g}lycol\\
%\textbf{Acronym} & \textbf{W}hat (it) \textbf{S}tands \textbf{F}or \\
}

%----------------------------------------------------------------------------------------
%	PHYSICAL CONSTANTS/OTHER DEFINITIONS
%----------------------------------------------------------------------------------------

\clearpage % Start a new page

\lhead{\emph{Physical Constants}} % Set the left side page header to "Physical Constants"

\listofconstants{lrcl} % Include a list of Physical Constants (a four column table)
{
Planck constant & $h$ & $=$ & $6.626\ 069\ 57\times10^{-34}\ \mbox{J}\cdot\mbox{s}$ \\
Boltzmann constant & $k_B$ & $=$ & $1.380\ 648\ 8\times10^{-23}\ \mbox{J}/\mbox{K}$ \\  
Boltz. const. at room temperature ($T=298^{\circ}K$) & $k_BT$ & $=$ & $4.11\times10^{-21}\ \mbox{J}$\\
Avogadro constant & $N_{A}$ & $=$ & $6.022\ 141\ 29 \times10^{23}$\\
% Constant Name & Symbol & = & Constant Value (with units) \\
}

%----------------------------------------------------------------------------------------
%	SYMBOLS
%----------------------------------------------------------------------------------------

\clearpage % Start a new page

\lhead{\emph{Symbols}} % Set the left side page header to "Symbols"

\listofnomenclature{lll} % Include a list of Symbols (a three column table)
{
$a_s$ & polymer statistical segment & $\mbox{\textup{\AA}}$ \\
$M_0$ & molar mass of a monomer unit &  $\mbox{g}/\mbox{mol}$ \\
$R_g$ & radius of gyration &  $\mbox{nm}$ \\
$R_c$ & radius of a colloid &  $\mbox{nm}$ \\
$\text{v}, \text{w}$ & second and third virial coefficients &  $\mbox{\textup{\AA}}^3$, $\mbox{\textup{\AA}}^6$ \\
$v_N, w_N$ & reduced second and third virial coefficients &   \\
$c_b\ (\phi_b)$ & bulk monomer concentration (volume fraction) &  $1/\mbox{\textup{\AA}}^3$\\
$N$ & polymerization index &  \\
$\Pi_b$ & osmotic pressure &  $1/\mbox{\textup{\AA}}^3$\\
$A_H$ & Hamaker constant &  $\mbox{J}$\\
$ q(x, s)$ & distribution function of a chain of length, $s$ with one end fixed at $x$&  \\
$ w(x)$ & self-consistent field&  \\
$\Omega (W)$ & thermodynamic potenial &  $\mbox{J}$\\
$ \hat{Q}$ & dimensionless partion function of a chain &  \\
$\hat{\Omega} (\hat{W})$ & dimensionless thermodynamic potenial &  \\
$h\ (\bar{h})$ & distance (dimensionless) between colloidal surfaces &  $\mbox{nm}$ \\
$h_m$ & mid-plate separation &  $\mbox{nm}$ \\
$\xi\ (\bar{\xi})$ & correlation (dimensionless) length &  $\mbox{nm}$ \\
$ U^*$ & barrier maximum for spherical geometry & $\mbox{J}$ \\
% Symbol & Name & Unit \\
}

%----------------------------------------------------------------------------------------
%	DEDICATION
%----------------------------------------------------------------------------------------

%\setstretch{1.3} % Return the line spacing back to 1.3

%\pagestyle{empty} % Page style needs to be empty for this page

%\dedicatory{To France for its elegance \ldots

%To Strasbourg for its loveliness \ldots

%To Institut Charles Sadron for its friendliness \ldots
%}% % Dedication text

\addtocontents{toc}{\vspace{2em}} % Add a gap in the Contents, for aesthetics

%----------------------------------------------------------------------------------------
%	THESIS CONTENT - CHAPTERS
%----------------------------------------------------------------------------------------

\mainmatter % Begin numeric (1,2,3...) page numbering

\pagestyle{fancy} % Return the page headers back to the "fancy" style

% Include the chapters of the thesis as separate files from the Chapters folder
% Uncomment the lines as you write the chapters

\input{./Chapters/colloids}

\input{./Chapters/polymers} 
\input{./Chapters/repulsion}
\input{./Chapters/adsorption} 
\input{./Chapters/brushes}

\input{./Chapters/conclusion}

%----------------------------------------------------------------------------------------
%	THESIS CONTENT - APPENDICES
%----------------------------------------------------------------------------------------

\addtocontents{toc}{\vspace{2em}} % Add a gap in the Contents, for aesthetics

\appendix % Cue to tell LaTeX that the following 'chapters' are Appendices

% Include the appendices of the thesis as separate files from the Appendices folder
% Uncomment the lines as you write the Appendices

\input{./Appendices/AppendixA}
\input{./Appendices/AppendixB}
\input{./Appendices/AppendixC}

\clearpage % Start a new page

\lhead{\emph{List of Figures}} % Set the left side page header to "List of Figures"
\listoffigures % Write out the List of Figures

\lhead{\emph{List of Tables}} % Set the left side page header to "List of Tables"
\listoftables % Write out the List of Tables

\addtocontents{toc}{\vspace{2em}} % Add a gap in the Contents, for aesthetics

\backmatter

%\clearpage % Start a new page
%----------------------------------------------------------------------------------------
%	BIBLIOGRAPHY
%----------------------------------------------------------------------------------------
\label{Bibliography}
\lhead{\emph{Bibliography}} % Change the page header to say "Bibliography"
\printbibliography

%\bibliographystyle{unsrtnat} % Use the "unsrtnat" BibTeX style for formatting the Bibliography

%\bibliography{Bibliography} % The references (bibliography) information are stored in the file named "Bibliography.bib"

\end{document}

%% file: TitlePage/TitlePage.tex
\begin{titlepage}
\begin{center}

\textsc{\LARGE \href{https://www.unistra.fr} % Your university's URL
                {Université de Strasbourg}}\\[0.5cm] % University name
\textsc{\Large Doctoral Thesis}\\[0.5cm] % Thesis type

\large \textit{Thèse présentée pour obtenir le titre de Docteur de l’Université de Strasbourg\\ 
			   Discipline : Physique}\\[0.5cm] % University requirement text

\HRule \\[0.4cm] % Horizontal line
{\huge \bfseries Theory of colloidal stabilization by unattached polymers}\\[0.4cm] % Thesis title
\HRule \\[1.5cm] % Horizontal line
 
\begin{minipage}{0.4\textwidth}
\begin{flushleft} \large
\emph{Author:}\\
\href{http://www.linkedin.com/pub/alexey-shvets/63/152/b02}{Alexey Shvets} % Author name - remove the \href bracket to remove the link
\end{flushleft}
\end{minipage}
\begin{minipage}{0.4\textwidth}
\begin{flushright} \large
\emph{Directeur de thèse:} \\
\href{http://www-ics.u-strasbg.fr/spip.php?article290}{Dr. Alexander Semenov} % Supervisor name - remove the \href bracket to remove the link  
\end{flushright}
\end{minipage}\\[1cm]

\begin{minipage}{0.7\textwidth}
\begin{flushleft} 
\textit{Membres du Jury:}\\[0.3cm] 
\emph{Rapporteur externe:} 
\href{https://www.google.com}{Dr. A. Subbotin}\\
Institute of Petrochemical Synthesis, Russian Academy of Sciences, Moscow, Russia \\[0.3cm]
\emph{Rapporteur externe:} 
\href{http://www.personal.reading.ac.uk/~sms06al2/}{Dr. A. Likhtman}\\
Professor,  University of Reading, Reading, UK \\[0.3cm]
\emph{Examinateur interne:} 
\href{http://www-ics.u-strasbg.fr/spip.php?article128}{Dr. J. Baschnagel}\\
Professeur, Université de Strasbourg, Strasbourg, France \\[0.3cm]
\emph{Examinateur externe:} 
\href{http://www.phys.ufl.edu/~sergei/}{Dr. S. Obukhov}\\
Professor, University of Florida, Gainesville, USA\\[0.3cm]
\emph{Examinateur externe:} 
\href{http://www.physics.nyu.edu/Grosberg/}{Dr. A. Grosberg}\\
Professor, New York University, New York, USA
\end{flushleft}
\end{minipage}
\begin{minipage}{0.2\textwidth}
\begin{flushleft} 
\end{flushleft}
\end{minipage}\\[0.8cm]
\includegraphics[scale=0.1]{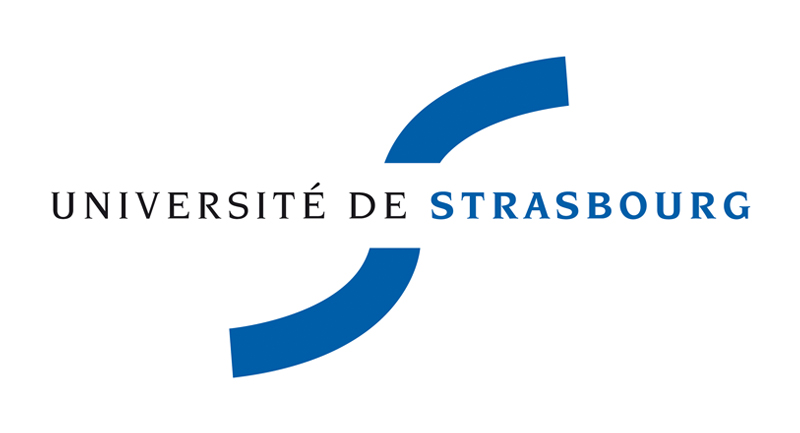} 
\includegraphics[scale=0.1]{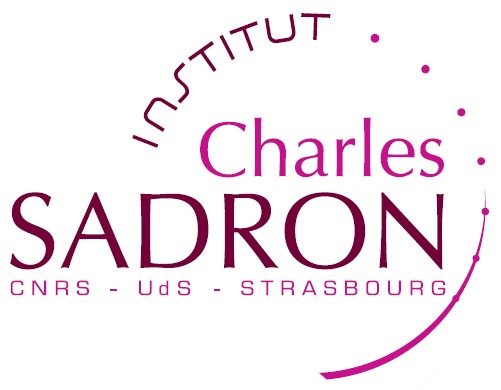} 
\includegraphics[scale=0.4]{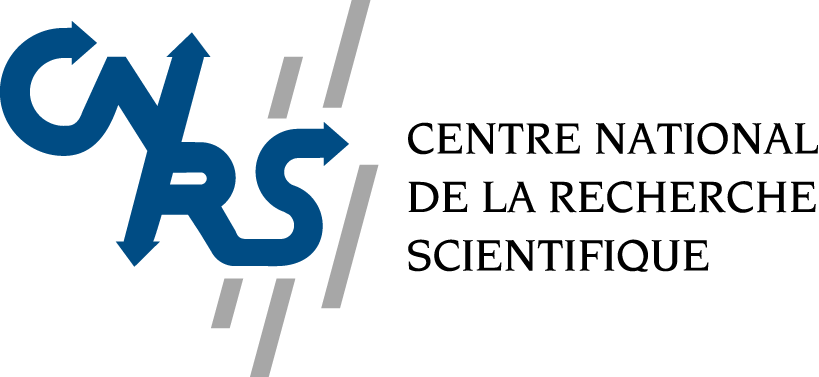} 
\includegraphics[scale=0.2]{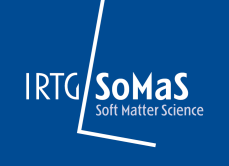} \\[0.5cm]
%\groupname\\\deptname\\[2cm] % Research group name and department name

{\large Strasbourg, France}\\
{\large May 2014}\\[2cm] % Date
%{\large \today}\\[2cm] % Date
 
\vfill
\end{center}

\end{titlepage}

%% file: resume/resume_fr.tex
\chapter{Résumé} % Main chapter title
\label{chap:resume} % For referencing the chapter elsewhere, use \ref{Chapter1} 
\lhead{\emph{Résumé}} % This is for the header on each page - perhaps a shortened title
\textbf{Motivation}. Les dispersions colloïdales ont beaucoup d’applications technologiques importantes [1]. 
Le terme "solution colloïdale" est appliqué pour le cas de particules ayant un diamètre entre 1 $nm$ et 1 $\mu m$ dispersées dans un milieu. 
A cause du mouvement brownien, les particules ont des collisions fréquentes entre elles. Les forces d’attraction de van der Waals, dérivant de potentiels à longue 
portés, conduisent à l’agrégation et à la précipitation des particules. Plusieurs méthodes ont été proposées pour diminuer ou contrebalancer l’effet d’attraction 
de van der Waals et augmenter la stabilité colloïdale. Par exemple, le choix du solvant possédant l’indice de réfraction le plus proche possible de celui des particules peut 
diminuer les forces de van der Waals [2]. D'autres facteurs influencent la stabilité comme les interactions électrostatiques et les interactions spécifiques 
liées aux chaînes de polymères. Dans le cas des polymères, les chaînes peuvent être greffées à la surface des particules ou être dissoutes 
dans le solvant (chaînes libres).

\begin{figure}[ht!]
\begin{minipage}[h]{0.45\linewidth}
\center{\includegraphics[width=1\linewidth]{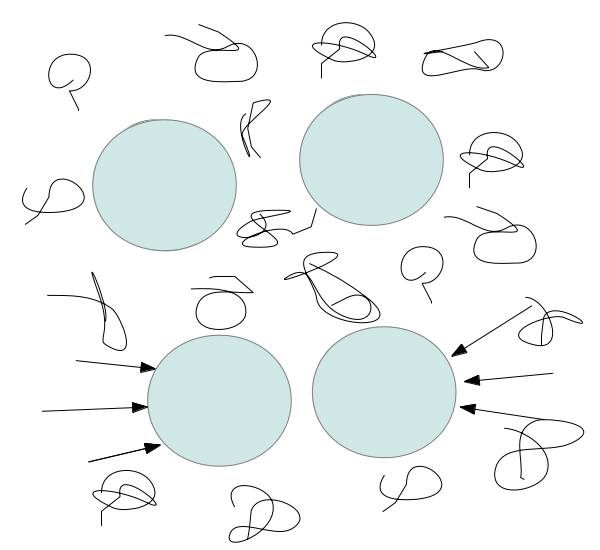}}
\caption{\small{Schéma des forces de déplétion attractives entre les particules induites par les chaînes de polymères dans une solution diluée.}}
\label{depletion_attraction_fig}
\end{minipage}
\hfill
\begin{minipage}[h]{0.45\linewidth}
\center{\includegraphics[width=1\linewidth]{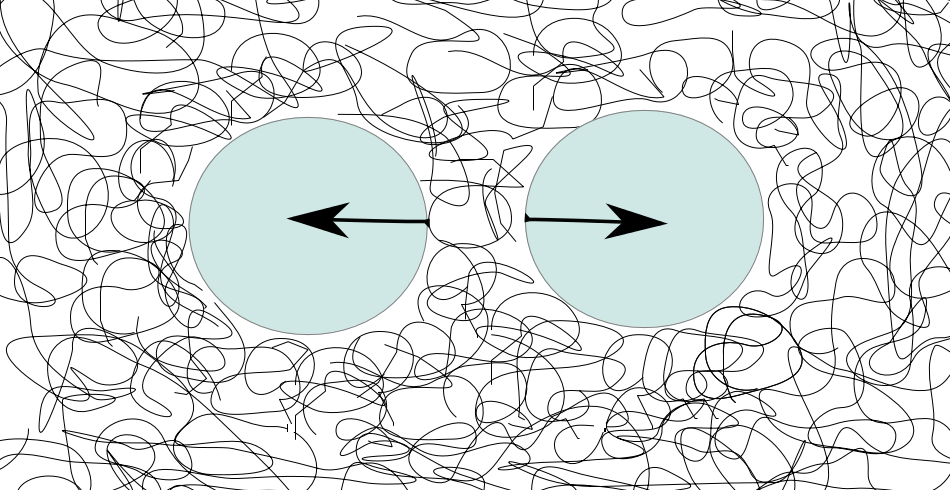}}
\caption{\small{Schéma des forces de déplétion répulsives entre les particules qui peuvent apparaître dans une solution concentrée de polymères.}}
\label{depletion_stabilization_fig}
\end{minipage}
\end{figure}

Les forces de déplétion attractives induites par des polymères (PI, "polymer-induced") sont bien étudiées dans la littérature [7-10]. 
Ces interactions ont lieu entre des particules dans une solution de polymères dilués, Fig.$\ref{depletion_attraction_fig}$. 
Quand des chaînes de polymères sont exclues de la région entre de particules, la pression osmotique à l’extérieur est supérieure à celle entre les particules, 
ce qui conduit à une force d’attraction. Cependant, dans certains cas l’augmentation de la concentration de polymères dans le système peut rendre la dispersion 
plus stable [1]. Ce phénomène est appelé la stabilisation par déplétion (Fig.$\ref{depletion_stabilization_fig}$). 
     
%%%%%%%%%%%%%%%%%%%%%%%%%%%%%%%%%%%%%%%%%%%%%%%%%%%%%%%%%%%%%%%%%%%%%%%%%%%%%%%%%%%%%%%%%%%%%%%%%%%%%%%%%%%%%%%%%%%%%%%%%%%%%%%
%        depletion attraction and depletion stabilization
%%%%%%%%%%%%%%%%%%%%%%%%%%%%%%%%%%%%%%%%%%%%%%%%%%%%%%%%%%%%%%%%%%%%%%%%%%%%%%%%%%%%%%%%%%%%%%%%%%%%%%%%%%%%%%%%%%%%%%%%%%%%%%%
Il possède aussi une nature entropique mais il est lié au comportement spécifique des extrémités des chaînes [3, 4]. L’influence de cet effet est souvent sous estimée
par la communauté de la matière molle, cependant ce type de stabilisation peut ouvrir de nouvelles voies d’élaboration de colloïdes stables.

Dans ce travail de thèse, nous avons étudié l’effet de la stabilisation par déplétion dans le cas des chaînes de polymères libres (FPI, "free polymer induced interaction"). Des modèles théoriques précédents portent un caractère trop simplifié et utilisent des approximations sans vérification. De plus, l’influence des paramètres de la solution, c’est-à-dire, de la structure de polymères et de son interaction avec la surface de particule, n’a pas été étudiée. Nous avons choisi les objectifs suivants:\\
1) élucider la nature de l’interaction FPI et caractériser de façon quantitative le rayon d'action et la magnitude et aussi le potentiel d’interaction;\\
2) établir l’influence des paramètres macroscopiques, mésoscopiques et moléculaires;\\
3) généraliser des approches numériques et analytiques contemporaines pour une gamme plus large de masses moléculaires et de concentrations ainsi que pour différents types d’interaction entre le polymère et la surface;\\
4) identifier les conditions optimales pour observer la stabilisation FPI;\\
5) identifier et sélectionner les paramètres du modèle tel qu'ils soient compatibles avec les paramètres du modèle experimentale utilisés par nos collaborateurs expérimentateur de l'université Freiburg de manière à pouvoir le plus qualitiativement possible comparer les résultats expérimentaux et théoriques.

Pour atteindre ces objectifs nous avons analysé les interactions complexes colloïde-solvent, colloïde-polymère, polymère-solvent et la conformation des polymères.
Dans ce but nous avons dévellopé un modèle SCFT (self-consistent field theory) pour les systèmes colloïde/polymère. Dans cette approche les chaînes de polymères sont traitées comme des chemins aléatoires propagés dans un champ généré par l'ensembles des segments de polymères.
La SCFT sert comme base pour nos calcules numériques relatif à la distribution des configurations de chaînes, aux divers paramètres thermodynamiques, ainsi qu'aux profiles d'énergies.   

Il est à noter que nous avons aussi traité le problème de manière analytique.
L'approche analytique classique est la thèorie des interactions FPI devellopée par De Gennes et al., celle-ci est basé sur l'approximation GSD (ground state dominance) qui est valide pour des chaînes infiniment longues.
Nous avons amélioré cette approche en prenant en compte dans la théorie l'effet de taille fini des chaînes [3, 4] et en montrant que ceci peut modifier qualitativement le potentiel d'interaction entre particules colloïdales.
Cette avancé théorique que nous appelons GSDE [5] (où E correspond à end-effects) a montré être en bon agrément avec les résultats numériques obtenues grâce à la SCFT.
De plus, nous avons trouvé que ces deux approches (analytique avec la GSDE et numérique avec la SCFT) se complémentes l'une et l'autre et permettent de quantifier pleinement les interactions FPI dans une large gamme de conditions experimentales correspondant aux régimes les plus importants.       

\textbf{Résultats obtenus}. 

\begin{figure}[ht!]
\begin{minipage}[h]{0.45\linewidth}
\center{\includegraphics[width=1\linewidth]{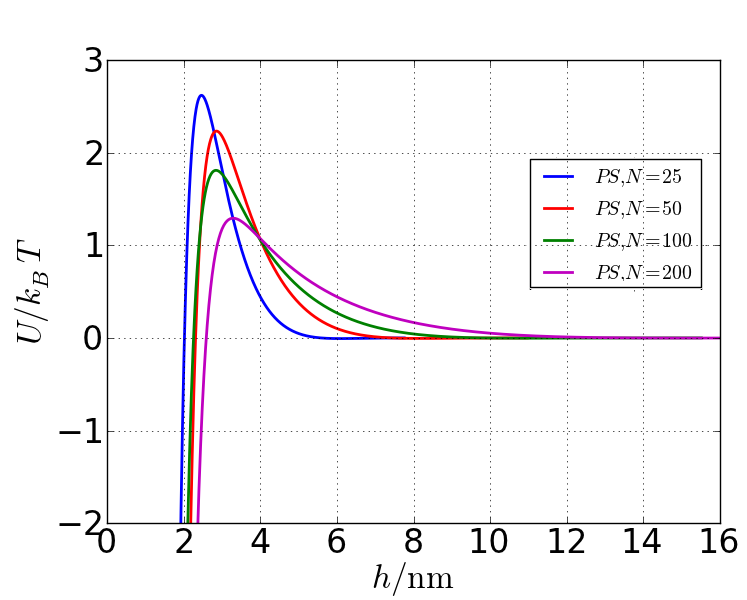}}
\caption{\small{Thermodynamic potentials calculated using the analytical de Gennes theory and numerical  SCFT results.}}
\label{therm_pot_vs_h_ps_fig}
\end{minipage}
\hfill
\begin{minipage}[h]{0.45\linewidth}
\center{\includegraphics[width=1\linewidth]{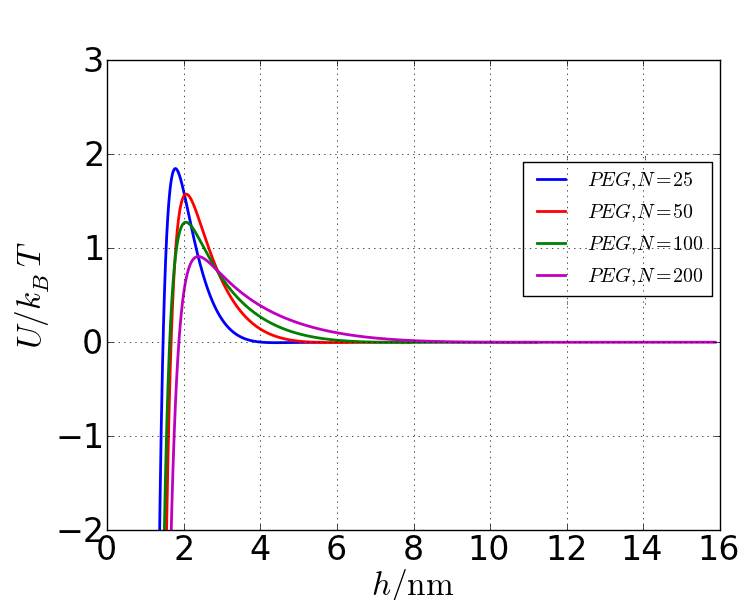}}
\caption{\small{Comparison of thermodynamic potentials calculated using the analytical GSDE theory and numerical SCFT results.}}
\label{therm_pot_vs_h_peg_fig}
\end{minipage}
\end{figure}

\textbf{1) Interactions entre particules purement répulsives immergées dans une solution de polymères}.
Dans ce cas le potentiel d'interaction FPI entre deux particules colloïdales dans une solution de polymères semi-dilués généralement présente un pic d'énergie associé aux effets de bouts de chaînes. La barrière d'énergie $U_{m}$ augmente avec la concentration de polymères $\phi$  dans le régime semi-dilué; $U_{m}$ sature ou décroit légerement dans le régime concentré, voir figures \ref{therm_pot_vs_h_ps_fig}--\ref{therm_pot_vs_h_peg_fig} (correspond à un système de Poly-styrène (PS) dilués dans du toluène et de Poly-éthylène Glycol (PEG) dilués en solvant aqueux).
Dans tout les cas considérés l'énergie de répulsion due aux polymères libres dans le régime concentré est de l'ordre de $U_m=1-3~k_BT$  pour des particules colloïdales de taille $R_c=100nm$. Pour des chaînes de polymère de tailles comprisent entre $N = 25-200$ monomères, la barrière de potentiel est effective à une séparation entre les deux surfaces solides de l'ordre de $h\sim 2nm$.
Ces résultats publiés [5] montrent que la stabilité colloïdale peut être amélioré de manière significative dans une solution concentrée de chaînes libres.
Par contre  dans la plupart des cas une pré-stabilisation par d'autre moyen sera nécessaire. 
De plus il est important de considerer les interactions FPI en les combinant avec d'autres méthode de stabilisation des solutions de colloïdes (adsorption réversible ou irréversible sur une surface de polymères, couches de polymères attachés de manière covalente à une surface, surface permettant une pénetration mesurée).

%%%%%%%%%%%%%%%%%%%%%%%%%%%%%%%%%%%%%%%%%%%%%%%%%%%%%%%%%%%%%%%%%%%%%%%%%%%%%%%%%%%%%%%%%%%%%%%%%%%%%%%%%%%%%%%%%%%%%%%%%%%%%%%
%        Comparison between SCFT and GSD and GSDE
%%%%%%%%%%%%%%%%%%%%%%%%%%%%%%%%%%%%%%%%%%%%%%%%%%%%%%%%%%%%%%%%%%%%%%%%%%%%%%%%%%%%%%%%%%%%%%%%%%%%%%%%%%%%%%%%%%%%%%%%%%%%%%%
\textbf{2)Interaction entre particules dans le cas d’adsorption réversible des chaînes de polymère à leur surface.}
De plus nous avons considéré le cas de surfaces de colloïde interagissant de manière faible ou modéré avec les segments des polymères monodisperses dilués, ceci conduit à la formation de couches de polymères adsorbées de manière réversible. 
Comme précédement, en combinant les approches GSDE et SCFT qui ont été généralisées pour le cas d'une adsorption, nous obtenons un potentiel effectif d'interaction avec la surface d'un colloïde qui est fonction de la largeur de la séparation et de paramètres externes comme la concentration en monomère du milieu et la masse moléculaire. 
En génerale nos résultats pour le potentiel d'interaction sont consistant avec les calcules [6] réalisés pour un régime de faible adsorption. 
Nous avons aussi découvert que la hauteur de la barrière de potentiel est une fonction qui croît avec la force d'adsorption, mais ceci est à relativiser du fait l'acroissement de concentration en monomère à la surface est limité par les effets de volume exclu ce qui limitent une trop forte adsorption.
En accord avec cette restriction la valeur de la barrière est fixé à $U_m=1-3k_BT$ pour tout les systèmes colloïde/polymère (solutions de PS et PEG).   

\textbf{3)Interaction entre particules dans le cas d’adsorption irréversible des chaînes de polymère à leur surface.} 
%%%%%%%%%%%%%%%%%%%%%%%%%%%%%%%%%%%%%%%%%%%%%%%%%%%%%%%%%%%%%%%%%%%%%%%%%%%%%%%%%%%%%%%%%%%%%%%%%%%%%%%%%%%%%%%%%%%%%%%%%%%%%%%
%        U_m vs phi real
%%%%%%%%%%%%%%%%%%%%%%%%%%%%%%%%%%%%%%%%%%%%%%%%%%%%%%%%%%%%%%%%%%%%%%%%%%%%%%%%%%%%%%%%%%%%%%%%%%%%%%%%%%%%%%%%%%%%%%%%%%%%%%%
\begin{figure}[ht!]
\begin{minipage}[h]{0.45\linewidth}
\center{\includegraphics[width=1\linewidth]{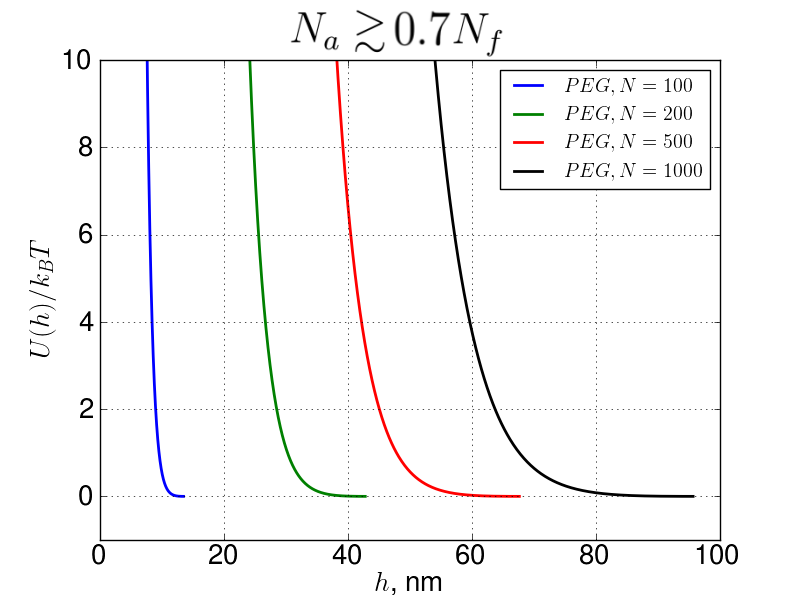}}
\caption{\small{Particles, $R_c=200\,nm$ dans PEG.}}
\label{therm_pot_real_na1_fig}
\end{minipage}
\hfill
\begin{minipage}[h]{0.45\linewidth}
\center{\includegraphics[width=1\linewidth]{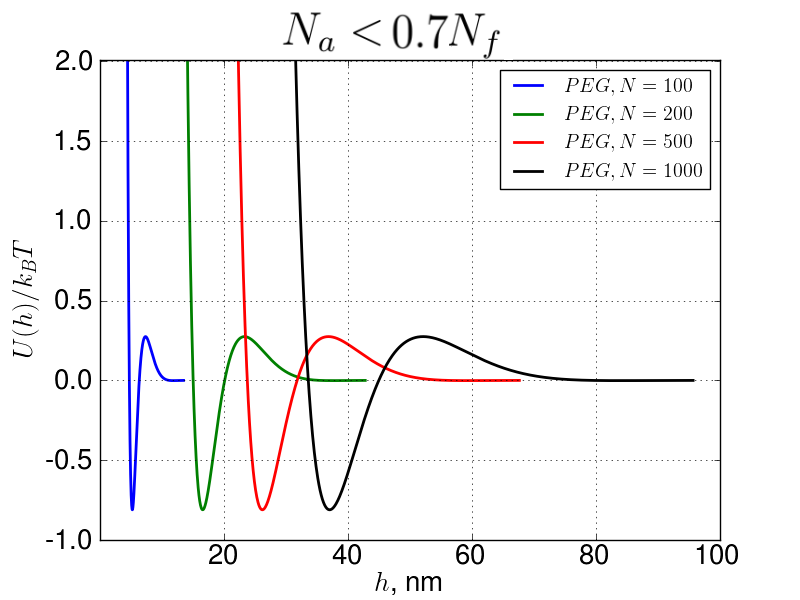}}
\caption{\small{Particles, $R_c=200\,nm$, dans PEG.}}
\label{therm_pot_real_na2d3_fig}
\end{minipage}
\end{figure} 
Dans le but d'amélioré la répulsion induite par les polymères libre, nous avons prolongé nos études en considérant la possibilité d'une adsorption irréversible à la surface des colloïdes d'une partie des polymères libre en solution.
Dans ce système modèle les colloïdes couverts par une couche mince de polymères interagissent les uns avec les autres et avec les polymères libres en solution.   
L'épaisseur et la densité de la couche adsorbée ont été modifiés dans le but de trouver les meilleurs conditions pour la stabilité.
Nous avons observé que les profiles de potentiel d'interaction  sont significativement affectés par la présence de ces couches adsorbées aussi dans le cas où les chaînes adsorbées sont de même taille de les chaînes de polymère libres que dans le cas où les chaînes adsorbées sont plus courtes (voir figures \ref{therm_pot_real_na1_fig}, \ref{therm_pot_real_na2d3_fig} ). 
Le point le plus important que nous avons pu découvrir sont des effets cumulées de stabilisation des polymères libres en solution en tandem avec les couches faiblement adsorbées.
Pour les tailles typiques de particules colloïdales que nous considérons la barrière d'énergie potentiel resultante est suffisament importante pour empecher la floculation ou la coagulation des colloïdes.
Cet effet peut être ramené à une modulation (par les chaînes adsorbées) de la distribution des polymères non attachés qui acroit l'énergie libre et la pression des bouts de chaînes.

{\raggedright \textbf{Référence:}}\\
	1) Napper D. H. \textit{Stabilization of Colloidal Dispersions}, Academic Press, \textbf{1984}.\\
	2) Israelachvili J.N. \textit{Intermolecular and Surface Forces}, 3 ed., Academic Press, \textbf{2011}. \\			
	3) Semenov A.N. \textit{J. Phys II} \textbf{1996}, 6, 1759.\\
	4) Semenov A.N. \textit{Macromolecules} \textbf{2008}, 41, 2243 \\
	5) Shvets A. A., Semenov, A. N., \textit{J. Chem. Phys.}, \textbf{2013} 139, 054905  \\
	6) Skau K.I.,  Blokhuis E.M., Avalos J.B., \textit{J. Chem. Phys.} \textbf{2003}, 119, 3483\\
	7) De Gennes P.G. \textit{Macromolecules} \textbf{1981}, 14, 1637.\\
	8) De Gennes P.G. \textit{Macromolecules} \textbf{1982}, 15, 492. \\	
	9) Scheijtens J.M.H.M., Fleer G. J. \textit{Advances in Colloidal and Surface science}, \textbf{1982}, 16, 361-380.\\
	10) Asakura S., Oosawa F. J. \textit{Chem. Phys}, \textbf{1954} 22, 1255.

%% file: resume/resume_eng.tex
\chapter{Resume}
\textbf{Motivation}. Stable colloidal dispersions with evenly distributed particles are important for many technological applications [1]. 
Due to Brownian motion colloidal particles have constant collisions with each other which often lead to their aggregation driven by the long range 
van der Waals attraction. As a result the colloidal systems often tend to precipitate. A number of methods have been devised to minimize the effect 
of long-range van der Waals attraction between colloidal particles or to override the influence of the attraction in order to provide the colloidal stability [2]. 

\textbf{Aims}. In the PhD project we investigated the colloidal stabilization in solutions of free polymers which is commonly referred to as depletion stabilization. 
Previous theoretical studies of free-polymer induced (FPI) stabilization were based on oversimplified models involving uncontrolled approximations [1]. 
Even the most basic features of the depletion stabilization phenomenon were unknown. It was unclear how the PI repulsion depends on the solution parameters, 
polymer structure and monomer/surface interactions. Thus, the following problems arose naturally as the main goals of this project: 

1) To elucidate the nature of the free polymer induced interaction and to quantitatively assess their range and magnitude as well as their potential for colloidal stabilization. \\
2) To establish how the stabilization effect of FPI interaction depends on the main macroscopic, mesoscopic and molecular parameters of the system. \\
3) To generalize both the numerical and analytical SCFT approaches to the depletion stabilization phenomenon to account for wider concentration and molecular weight regimes and for surface effects (surface repulsion vs. attraction, reversible vs. irreversible adsorption). \\
4) To identify the optimum regimes for exploiting the FPI interaction stabilization effect.\\
5) To chose or optimize model parameters such that the calculation can be applied as closely as possible the parameters of the experimental model systems studied by the experimentally in the University of Freiburg (Prof. Eckhard Bartsch's group) in order to allow for a comparison of experimental and theoretical results.

To reach these goals we had to analyse the intricate interplay of colloid-solvent, colloid-polymer, polymer-solvent interactions and  polymer chains conformations. 
To this end we developped the self-consistent field theory (SCFT) for the colloid/polymer systems. In this approach polymer chains are treated as random walks 
propagating in the molecular field created by all the polymer segments. The SCFT then served as a basis for our numerical calculations of the chain conformational 
distributions, and of  various thermodynamic parameters and interaction energy profiles. 

Noteworthy, we also treated the problem analytically. The classical analytical theory of FPI interactions was developped by De Gennes et al. [8]-[11] 
based on the ground state dominance (GSD) approximation which is strictly valid for very long (infinite) chains. We significantly improved their approach by 
taking into account the finite chain length effects [4, 5] and showing that they can qualitatively change the interaction potential between colloidal particles. 
This advanced theory which we called the GSDE theory [6] (where "E" stands for the end-effects) is shown to be in good agreement with computational SCFT results.   
Moreover, we find that the two approaches (analytical GSDE and numerical SCFT) consistently complement each other allowing to fully quantify the FPI interactions 
in a wide range of conditions in the most important regimes.

\textbf{Results and outlook:}. 

\begin{figure}[ht!]|
\begin{minipage}[h]{0.45\linewidth}
\center{\includegraphics[width=1\linewidth]{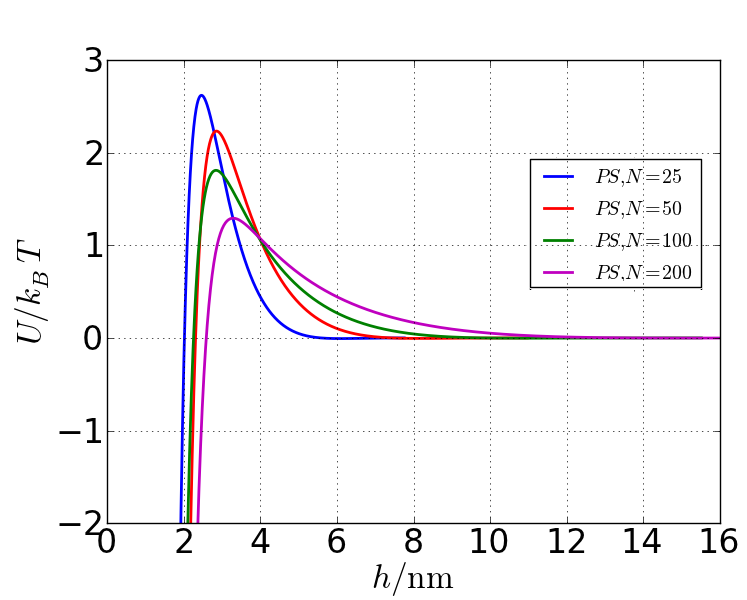}}
\caption{\small{Thermodynamic potentials calculated using the analytical de Gennes theory and numerical  SCFT results.}}
\label{therm_pot_vs_h_ps_eng_fig}
\end{minipage}
\hfill
\begin{minipage}[h]{0.45\linewidth}
\center{\includegraphics[width=1\linewidth]{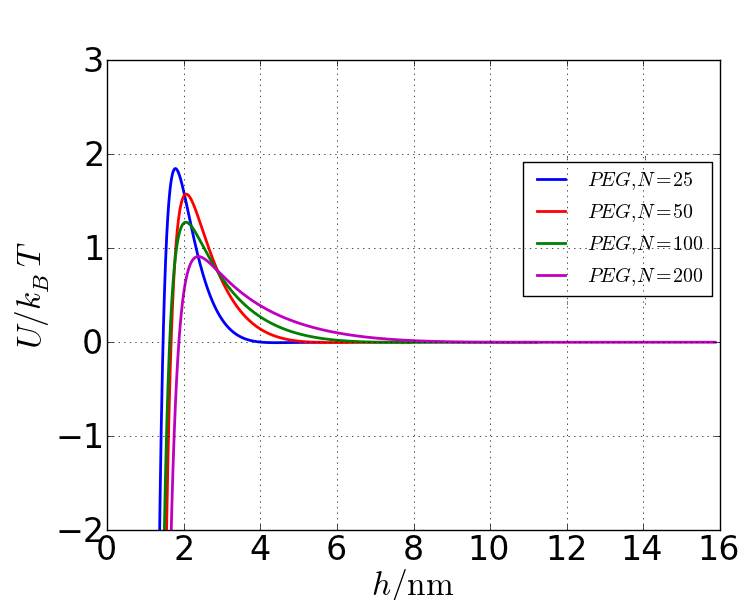}}
\caption{\small{Comparison of thermodynamic potentials calculated using the analytical GSDE theory and numerical SCFT results.}}
\label{therm_pot_vs_h_peg_eng_fig}
\end{minipage}
\end{figure}

\textbf{1) Purely repulsive colloidal surfaces.}
In this case the potential of the FPI interaction between two colloidal particles in a semidilute polymer solution generically shows a repulsion energy peak 
associated with the polymer chain end effects. The barrier  height $U_m$ increases with bulk polymer concentration $\phi$ in  the semidilute regime; $U_m$ 
saturates or slightly decreases in  the concentrated regime, see Figs.\ref{therm_pot_vs_h_ps_eng_fig}, \ref{therm_pot_vs_h_peg_eng_fig} (prepared for real systems of 
polystyrene (PS) in toluene and polyethylene glycol (PEG) in water).  In all the cases considered the repulsion energy due to free polymers in the concentrated 
solution regime is about $U_m=1-3 k_BT$ for colloidal particle size $R_c=100 nm$. For the typical range of free chain polymerization degree, $N=25-200$, 
the potential barrier is developed at separations $h \sim 2nm$ between the solid surfaces. These results published in [6] show that colloidal stability can be 
significantly improved  in a concentrated solution of free chains. However, in most cases a pre-stabilization by other means would be required. 
Therefore, it is important to consider FPI interaction in combination with other means enhancing the colloid stabilization effect (reversible or irreversible 
polymer surface adsorption, grafted polymer layers, soft penetrable rather than solid interacting surfaces). 

%%%%%%%%%%%%%%%%%%%%%%%%%%%%%%%%%%%%%%%%%%%%%%%%%%%%%%%%%%%%%%%%%%%%%%%%%%%%%%%%%%%%%%%%%%%%%%%%%%%%%%%%%%%%%%%%%%%%%%%%%%%%%%%
%        Comparison between SCFT and GSD and GSDE
%%%%%%%%%%%%%%%%%%%%%%%%%%%%%%%%%%%%%%%%%%%%%%%%%%%%%%%%%%%%%%%%%%%%%%%%%%%%%%%%%%%%%%%%%%%%%%%%%%%%%%%%%%%%%%%%%%%%%%%%%%%%%%%
\textbf{2)Reversibly adsorbed colloidal surfaces.}
Further, we considered colloidal solid surfaces with weak or moderate affinity for segments of dissolved homopolymers leading to formation
of reversible polymer adsorption layers. As before, combining GSDE and SCFT approaches,
which have been generalized for the adsorption case, we obtained the colloidal interaction
potential as a function of separation and external parameters like bulk monomer concentration
and chain length. Our general results for the interaction potential turn out to be nicely
consistent with calculations [7] done in the weak adsorption regime. We found that the
barrier height is an increasing function of adsorption strength, but due to the elevated
monomer concentration at the surface this increase is limited for strong adsorption. In
accordance with this restriction, the value of the barrier is $U_m=1-3k_BT$ for the same
colloid/polymer systems (PS and PEG solutions).

\textbf{3)Colloidal surfaces with soft shell layer.} 
%%%%%%%%%%%%%%%%%%%%%%%%%%%%%%%%%%%%%%%%%%%%%%%%%%%%%%%%%%%%%%%%%%%%%%%%%%%%%%%%%%%%%%%%%%%%%%%%%%%%%%%%%%%%%%%%%%%%%%%%%%%%%%%
%        U_m vs phi real
%%%%%%%%%%%%%%%%%%%%%%%%%%%%%%%%%%%%%%%%%%%%%%%%%%%%%%%%%%%%%%%%%%%%%%%%%%%%%%%%%%%%%%%%%%%%%%%%%%%%%%%%%%%%%%%%%%%%%%%%%%%%%%%
\begin{figure}[ht!]|
\begin{minipage}[h]{0.45\linewidth}
\center{\includegraphics[width=1\linewidth]{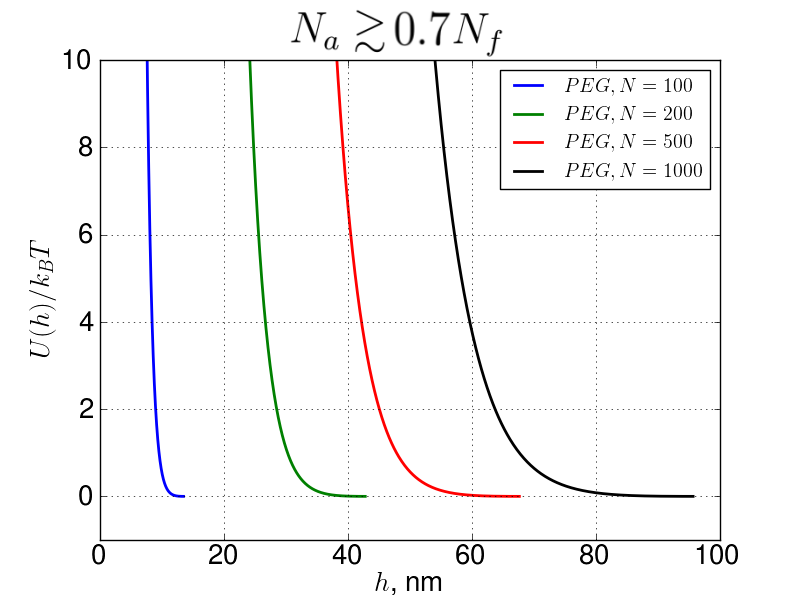}}
\caption{\small{Particles, $R_c=200\,nm$ dans PEG.}}
\label{therm_pot_real_na1_eng_fig}
\end{minipage}
\hfill
\begin{minipage}[h]{0.45\linewidth}
\center{\includegraphics[width=1\linewidth]{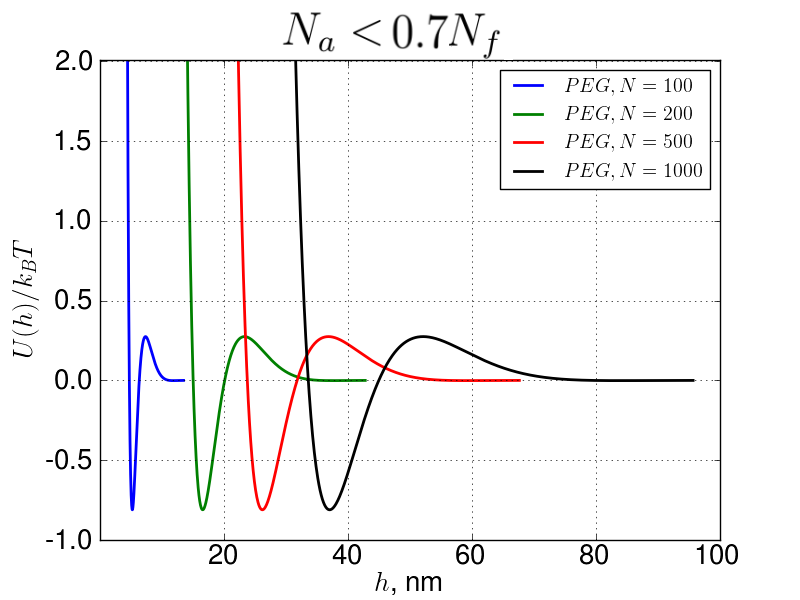}}
\caption{\small{Particles, $R_c=200\,nm$, dans PEG.}}
\label{therm_pot_real_na2d3_eng_fig}
\end{minipage}
\end{figure}        
Seeking to enhance the free polymer mediated repulsion, we advanced our studies to consider the possibility of irreversible adsorption of some free polymers on the colloidal surfaces. In these model systems colloidal particles
covered by the adsorbed soft layers interact with each other and with free polymers in solution. The thickness and density of the layers were varied in the effort to find the best
conditions for the stabilization. We find that the interaction potential profiles are significantly affected by the presence of soft layers both in the case when adsorbed and free chains are
identical and when adsorbed chains are shorter (see Figs. \ref{therm_pot_real_na1_eng_fig}, \ref{therm_pot_real_na2d3_eng_fig}). Most importantly, we
revealed a strongly enhanced cumulative stabilization effect of free polymers in tandem with soft adsorbed layers. For typical sizes of colloidal particles, the resultant potential energy
barrier is high enough to prevent colloid coagulation/flocculation. This effect can be traced back to a favorable modulation (by the adsorbed chains) of the distribution of unattached
polymers increasing the free energy and the pressure of their free ends.\\\\

{\raggedright \textbf{References:}}\\
	1) Napper D. H. \textit{Stabilization of Colloidal Dispersions}, Academic Press, \textbf{1984}.\\
	2) Israelachvili J.N. \textit{Intermolecular and Surface Forces}, 3 ed., Academic Press, \textbf{2011}. \\		
	3) Derjaguin B.V., \textit{Kolloid Z.} (in German) \textbf{1934}, 69 (2), 155–164.\\
	4) Semenov A.N. \textit{J. Phys II} \textbf{1996}, 6, 1759.\\
	5) Semenov A.N. \textit{Macromolecules} \textbf{2008}, 41, 2243 \\
	6) Shvets A. A., Semenov, A. N., \textit{J. Chem. Phys.}, \textbf{2013} 139, 054905  \\
	7) Skau K.I.,  Blokhuis E.M., Avalos J.B., \textit{J. Chem. Phys.} \textbf{2003}, 119, 3483\\
	8) De Gennes P.G. \textit{Macromolecules} \textbf{1981}, 14, 1637.\\
	9) De Gennes P.G. \textit{Macromolecules} \textbf{1982}, 15, 492. \\	
	10) Scheijtens J.M.H.M., Fleer G. J. \textit{Advances in Colloidal and Surface science}, \textbf{1982}, 16, 361-380.\\
	11) Asakura S., Oosawa F. J. \textit{Chem. Phys}, \textbf{1954} 22, 1255.\\

%% file: Chapters/colloids.tex
% Chapter 1
\chapter{Introduction. Colloids and their interactions} % Main chapter title
\label{chap:Chapter1} % For referencing the chapter elsewhere, use \ref{Chapter1} 
\lhead{Chapter 1. \emph{Introduction}} % This is for the header on each page - perhaps a shortened title

\section{Colloidal dispersions. Colloids} 
Colloidal dispersions are indispensable in today's daily life. When we speak about colloids, we imply a solution that has particles ranging between $1$ $nm$ and $1$ $\mu m$ in diameter, that are able to remain evenly distributed throughout the solution. This means that colloidal particles are sub-micron sized substances dispersed in medium of low-molecular weight solvent. Usually, the dispersed substance is referred to as being in the dispersed phase, while the substance in which it is dispersed is in the continuous phase.                          
             
From a physical point of view colloidal particles are characterized by Brownian motion, originating from thermal energy of order $k_BT$ for each particle. On the Earth, all particles are subject to gravity. The probability to find a particle at a height $h$ above the surface of the earth is given by the barometric distribution:
$$
	    P(h) = \exp\left(-\frac{m^*gh}{k_BT}\right) = \exp\left(-\frac{h}{l_{sed}}\right)
$$
where $m^*$ is the effective mass of the colloidal particle(mass of the particle minus mass of the displaced solvent), $T$ is temperature and $k_B$ is Boltzmann's constant. We also introduced here the sedimentation length $l_{sed}=k_BT/m^*g$ characterizing the spatially inhomogeneous distribution due to the gravitational impact. Therefore, to have evenly distributed colloids throughout the solution, we should consider a system that satisfies the condition $h<l_{sed}$. We now turn to the justification of the numerical values that restrict the size of the colloidal particles. To this end, we should compare colloidal particles with their surroundings: low molecular weight solvent molecules on one side of the scale and macroscopic bodies on the other side. Since we consider low molecular weight solvent as a continuous medium, we should ignore the detailed atomic structure of colloidal particles. Thus, we obtain the lower limit for the size of the colloidal particles. On the other hand, the influence of the gravitational field imposes its own limitations. When it becomes comparable
with the thermal motion of colloidal particles or, in other words, the size of colloidal particle becomes comparable with the sedimentation length, i.e
$$
            d \sim l_{sed} = \frac{k_BT}{m^*g} = \frac{6k_BT}{\rho^*\pi d^3 g}
$$
which leads to the following upper limit on the size of the colloidal particles
$$
            d \sim \left(\frac{6k_BT}{\pi\rho^* g}\right)^{1/4}
$$            
For a particle with an excess density $\rho^*=1g/cm^3$, this relation leads to the aforementioned value $d\sim 1 \mu m$. From the last relationship one can notice that if we make density of a particle equal to the density of solvent we do not get the upper bound for the size of the particle. 
	    
Colloidal particles are able to move due to diffusion. In accordance with random nature of the thermal motion, in a time $t$, a particle usually diffuses the distance $l\simeq\sqrt{2Dt}$. For a spherical particle, the diffusion constant is given by Stokes-Einstein relation
$D=k_BT/(3\pi\eta d)$, where $\eta$ is the viscosity of the solvent, $d$ is diameter of the spherical particle. Therefore, a particle diffuses a distance comparable to its own diameter in time
$$
	    t_d \simeq \frac{3\pi \eta d^3}{2k_BT}
$$
For a particle of size $d=1\mu m$ in water with viscosity $\eta = 0.0089[g/cm/s]$ \cite{CRC_handbook_90ed}, this time is approximately $100$ seconds. For the bigger particles the diffusion time is much longer. 
	    
Colloidal science has long been in focus owing to impressive array of applications it provided in technologically and biologically relevant areas. The importance of colloidal systems is confirmed by the following industrial examples: paints, inks, glues, emulsions, gels, lubricants, foams, different food ingredients, drinks such as beer and wine, cosmetic and agricultural products. In pharmaceutical industry many drug components have colloidal characteristics. In biology, colloids manifest in the form of biopolymers (folded nucleic acids, proteins and polysaccharides), different organelles inside a cell, prokaryotes (in particular bacteria) and viruses.
In this light, a simplified view of blood treats it as a colloidal dispersion of red corpuscles in a liquid \cite{Napper_book}. Probably, the most famous example of colloidal solution in everyday life is milk. Milk contains a series of globular (folded) proteins along with 
lipids assembled in micelles and vesicles. All of these entities represent colloids dispersed in water \cite{Lekkerkerker_book}. 
                        
The main problems in colloidal science are typically related to stability of colloidal solution. The term stability is used in quite special 
meaning and may refer to \textbf{thermodynamic} as well as to \textbf{kinetic} stability. Thermodynamically stable or metastable means that the system is in a equilibrium state corresponding to a local minimum of the appropriate thermodynamic potential for given constraints on the system like Gibbs energy at constant temperature and pressure. The thermodynamic stability can be defined in the absolute sense. For example, if several states are accessible to the system under certain conditions, the state with the lowest value of the potential is called the stable state, while the other states are referred to as metastable. Transitions between metastable and stable states occur at rates which depend on the magnitude of the appropriate activation energy barriers which separate them. Sometimes, the inverse rate may  greatly exceed the possible monitoring time. Most colloidal systems are metastable or unstable with respect to the separate bulk phases. On the other hand, kinetic stability means that the particles do not aggregate at a significant rate. In this case, the aggregation rate is completely determined by the interaction between colloidal particles due to the fact that the range of interaction between colloidal particles is significantly greater than the range of the molecular interactions. In contrast to the thermodynamic stability, the kinetic stability can not be defined in an absolute sense. For any particular system, the kinetic stability is characterized by certain time scale during which the system is stable. For example, for some practical applications, fresh milk under normal condition is kinetically stable for few hours. For the ink in a pen this time can reach several years. Otherwise, for some experimental reasons the characteristic time scale for the aforementioned systems could be just a fraction of a second. Therefore, even for the same system, this time can be different. The kinetic stability is most typical for the colloidal systems and usually it is referred to as just the colloidal stability \cite{Derjaguin_book, Hiemenz_book}.
            
In an unstable system the colloidal particles may adhere to each other and form aggregates of increasing size that under the action of gravity may precipitate out.	 An initially formed aggregate is called a \textbf{floc} and the process of its formation is called \textbf{flocculation}. The floc may or may not separate out. If the aggregate changes to a denser form, it is said to undergo \textbf{coagulation}. An aggregate usually separates out either by sedimentation (if it is more dense than the medium) or by creaming (if it less dense than the medium). The terms flocculation and coagulation have often been used interchangeably. Usually coagulation is irreversible whereas flocculation can be reversed by the process of deflocculation. Next, based on the character of interaction between colloids and liquid, we distinguish \textbf{lyophilic} colloids that are strongly interacting with surrounding liquid from \textbf{lyophobic} colloids which are weakly interacting with liquid molecules or do not interact at all. In most cases, lyophilic colloids include sols of organic substances like gelatin, gum, starch, proteins etc, which are stable in liquid due to the strong interaction with liquid molecules. In contrast with that,lyophobic colloids include sols of inorganic substances like Arsenic ($As_2S_3$), Iron ($Fe(OH)_3$), Platinum etc, which are less stable due to weak interaction between colloidal particles and liquid molecules \cite{VO_DLVO, www_lyophobic}. 
Therefore, the study of colloidal stabilization implies the study of the kinetic stabilization of lyophobic colloids. 
                                            
Most of the concepts that will be developed in the thesis are valid for different geometries of colloidal particles. Moreover, we focus our attention only on solutions of spherical mono-disperse particles. The specific properties of elongated, rod-like, disk-like or flexible colloidal systems are not considered.
%%%%%%%%%%%%%%%%%%%%%%%%%%%%%%%%%%%%%%%%%%%%%%%%%%%%%%%%%%%%%%%%%%%%%%%%%%%%%%%%%%%%%%%%%%%%%%%%%%%%%%%%%%%%%%%%%%%%%%%%%%%%%%%%%%%%%%%%%%%%%%%%%%%%%%%%%%%%%
%           The condition for the kinetic stability
%%%%%%%%%%%%%%%%%%%%%%%%%%%%%%%%%%%%%%%%%%%%%%%%%%%%%%%%%%%%%%%%%%%%%%%%%%%%%%%%%%%%%%%%%%%%%%%%%%%%%%%%%%%%%%%%%%%%%%%%%%%%%%%%%%%%%%%%%%%%%%%%%%%%%%%%%%%%%
\section{The condition for the kinetic stability}
In order to make quantitative assessment of the condition to the kinetic stabilization, we have to consider the movement of spherical colloidal particles in a viscous medium which undergo Brownian motion. Denote the colloidal radius as $R_c$. In a dilute colloidal solution the probability of simultaneous collisions of three or more particles is very low, thus we can restrict our consideration only by pairwise collisions and describe coagulation in terms of consecutive doublet formation involving collisions between two single particles. 
	    
Therefore, by virtue of the coagulation kinetic \cite{Derjaguin_book, Hiemenz_book, Russel_book}, the formation of a dimer from two identical colloidal particles can be described by the following differential equation
$$
	    \frac{\mathrm{d}N}{\mathrm{d}t} = -kN^2
$$
where $N$ is the number of free colloid particles per $cm^3$, $k$ is a rate constant. After integration we obtain
$$
	    \frac{1}{N_t} - \frac{1}{N_0} = kt
$$
Let us rewrite the expression in more appropriate form introducing new variable $n=N_t/N_0 < 1$ that represent the fraction of the remaining particles in comparison with their original amount at the time, $t$. Then we can write the rate equation as
$$
	    \frac{1}{N_0}\left\{\frac{1-n}{n}\right\} = kt
$$
The rate constant, $k$ depends on the repulsive interaction between the particles and corresponds to slow (Fuchs) kinetics (due to the potential barrier only a small part of the collisions leads to aggregation). The constant can be represented via rate constant, $k_0$ of the fast (Smoulokowski) kinetics (when there are no repulsive barriers and each collision leads to aggregation) as
$$
	    k = \frac{k_0}{W} 
$$
where $W\geqslant 1$ is the \textbf{stability ratio} that takes into account the loss of energy of the incident particles due to potential barrier between the particles. It is known \cite{Derjaguin_book, Hiemenz_book, Russel_book} that the fast kinetic rate constant is $k_0=4k_BT/3\eta$, where $\eta$ is the viscosity of the corresponding medium. Thus, for the stability ratio we can write: 
\begin{equation}
\label{intr_colloids_stab_ratio_via_k0}
	    W = N_0k_0\left(\frac{n}{1-n}\right)t
\end{equation}
On the other hand, using the result obtained in \cite{Derjaguin_book, Hiemenz_book, Russel_book} the expression for the stability ratio is 
$$
	    W = 2R_c\int\limits_{2R_c}^{\infty} e^{U(r)}\frac{\mathrm{d}r}{r^2} 
$$
here the integration variable, $r$ corresponds to the distance between centers of colloidal particles and, as before, $R_c$ is the colloidal radius. To prevent some of the collisions from aggregation, suppose that the potential $U(r)$ has always range with $U(r)>0$ and generally it is quite close to the surface of the colloidal particle. When a pair of particles overcome this barrier, they fall into the strong aggregation range and stick to each other. One can notice, from the above equation, that the corresponding stability ratio grows when we increase the potential barrier between particles and $W \gtrapprox 1$ for quite low barriers with $U(r) \simeq 0$. For simplicity, we consider the rectangular potential barrier:
$$
	    U(r) = \left\{
	    \begin{array}{rl}
	      %-\infty, & r < 2R_c\\\\
	      U_m,     & 2R_c< r < 2R_c + r_0\\\\
	      0,       & r > 2R_c + r_0\\
	    \end{array} \right.
$$	
where $r_0$ is the range of the potential. Thus, the stability ratio can be written as
$$
	    W = 2R_c\int\limits_{2R_c}^{\infty} e^{U(r)}\frac{\mathrm{d}r}{r^2} = 2R_ce^{U_m}\int\limits_{2R_c}^{2R_c+r_0}\frac{\mathrm{d}r}{r^2} =
	    e^{U_m}\left\{\frac{r_0/R_c}{2 + r_0/R_c}\right\} 
$$
Usually for colloid systems $R_c\gg r_0$ and finally we can write: 
\begin{equation}
\label{intr_colloids_stab_ratio_via_pot}
	    W = \frac{1}{2}\left(\frac{r_0}{R_c}\right)e^{U_m}
\end{equation}
Combining Eq.(\ref{intr_colloids_stab_ratio_via_k0}) and Eq.(\ref{intr_colloids_stab_ratio_via_pot}) for the stability ratio, we can extract the dependence of the barrier height as a function of other parameters, namely:
\begin{equation}
\label{intr_colloids_pot_bar_max_general}
	    U_m = \ln\left\{2N_0k_0\left(\frac{R_c}{r_0}\right)\left(\frac{n}{1-n}\right)t\right\}
\end{equation}
Instead of initial concentration of colloids  $N_0$ it is more convenient to use the initial volume fraction, $\phi_0$ occupied by the colloids. For this, we know that all of the colloids have ideal spherical form. The volume of one particle is $V_1 = 4\pi R_c^3/3$ and the corresponding initial number of the particles can be expressed as 
$$
	    N_0 = \frac{\phi_0}{V_1} = \frac{3\phi_0}{4\pi R_c^3}
$$
In order to simplify Eq.(\ref{intr_colloids_pot_bar_max_general}), we exclude all the numerical multipliers which produce the net numerical constant almost equal to $1$. Thus, we have:
\begin{equation}
\label{intr_colloids_pot_bar_max}
	    U_m = \ln\left\{\frac{\phi_0 k_BT t}{R_c^2r\eta}\right\}
\end{equation}
where we also supposed that $50\%$ of initial colloids are coagulated at the time $t$, i.e. we set $n=0.5$. Calculate now the value of the 
barrier height for the some realistic systems.

\begin{figure}[ht!]
\begin{minipage}[h]{0.5\linewidth}
\center{\includegraphics[width=1\linewidth]{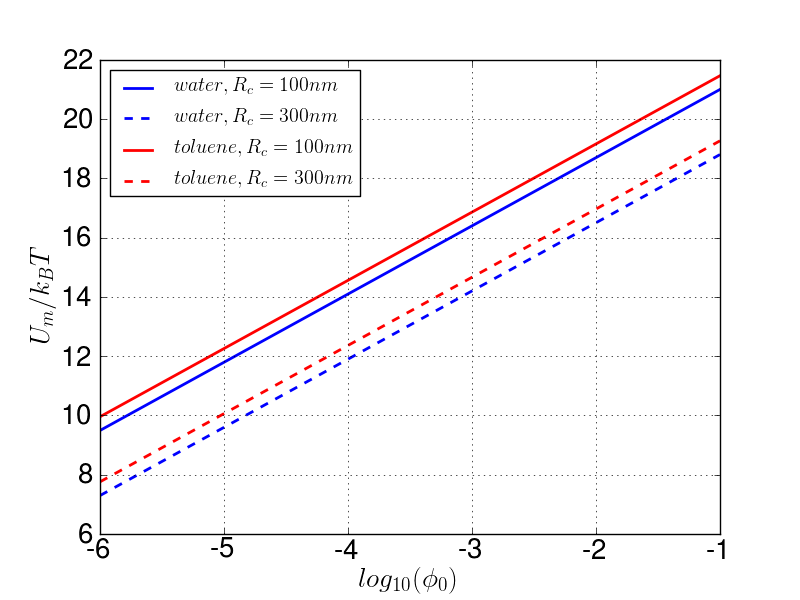}}
\caption{\small{The dependence of the barrier maximum on the initial fraction of colloids described by Eq.(\ref{intr_colloids_pot_bar_max}) for simple liquids.}}
\label{colloids_um_vs_log10_phi0_pure_fig}
\end{minipage}
\hfill
\begin{minipage}[h]{0.5\linewidth}
\center{\includegraphics[width=1\linewidth]{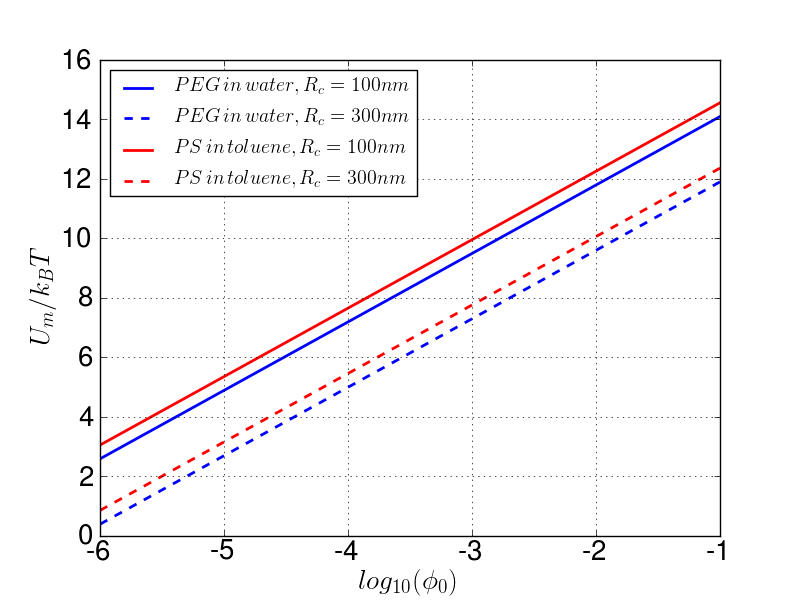}}
\caption{\small{The dependence of the barrier maximum on the initial fraction of colloids described by eq.(\ref{intr_colloids_pot_bar_max}) for the polymer solutions of polystyrene and polyethylene glycol dissolved in toluene and water correspondingly. The polymers with chain lengths $N \simeq 10^5$ dissolved in solvent with the volume fraction $\phi_p \sim 0.1$.}}
\label{colloids_um_vs_log10_phi0_polymers_fig}
\end{minipage}
\end{figure}

\textbf{Real examples}. \\
Let us use 1 day i.e $t=24h\times 60min\times 60s = 86400[s]$ for the reference time. The typical range of the interaction potential barrier is $r_0=2\div3nm$ and we set $r_0=3nm$. For the colloids we consider two sizes: $R_{c1}=100nm=10^{-5}cm$ and $R_{c2}=300nm=3\times 10^{-5}cm$. 
Then, as a solvent we consider the following liquids:\\
a)\emph{Simpe solutions}: \cite{CRC_handbook_90ed}: \textbf{water} with viscosity $\eta^w = 0.0089[g/cm/s]$ and \textbf{toluene} with $\eta^t = 0.0056[g/cm/s]$. The typical values of the potential barrier height for those system as a function of initial colloidal volume fraction are presented in Fig.\ref{colloids_um_vs_log10_phi0_pure_fig}.\\
b)\emph{Polymer solution}. Consider certain polymer solutions like polystyrene (PS) dissolved in toluene and polyethylene glycol (PEG) dissolved in water. We suppose that these polymer solutions are in marginal solvent regimes with polymer volume fraction, $\phi_p = 0.1$. 
We consider polymer chains with length $N=10^5$ and for simplicity assume that they are unentangled. In this case, the viscosity of the solution will be different in comparison with the previous example and in accordance with \cite{Rubinstein_book} for the semidilute polymer solution viscosity we can write
$$
	    \eta \simeq \eta_s(1 + N \phi_p^2)
$$
where $\eta_s$ is the viscosity of the corresponding solvent. The last expression is oversimplified, but nevertheless we limit ourselves, for the assessment of the barrier height, by this expression. For polymer solutions that we listed above the product is $N \phi_p^2 \simeq 10^3$. Correspondingly, we can simplify the above expression for viscosity as $\eta \simeq \eta_s N \phi_p^2$. This leads to the additional term $-\log(N\phi_p^2)$ in comparison with Eq.(\ref{intr_colloids_pot_bar_max}) which decreases the barrier height as shown in Fig.\ref{colloids_um_vs_log10_phi0_polymers_fig}.	    
%%%%%%%%%%%%%%%%%%%%%%%%%%%%%%%%%%%%%%%%%%%%%%%%%%%%%%%%%%%%%%%%%%%%%%%%%%%%%%%%%%%%%%%%%%%%%%%%%%%%%%%%%%%%%%%%%%%%%%%%%%%%%%%
%        Umax vs polymer volume fraction
%%%%%%%%%%%%%%%%%%%%%%%%%%%%%%%%%%%%%%%%%%%%%%%%%%%%%%%%%%%%%%%%%%%%%%%%%%%%%%%%%%%%%%%%%%%%%%%%%%%%%%%%%%%%%%%%%%%%%%%%%%%%%%%
            
%%%%%%%%%%%%%%%%%%%%%%%%%%%%%%%%%%%%%%%%%%%%%%%%%%%%%%%%%%%%%%%%%%%%%%%%%%%%%%%%%%%%%%%%%%%%%%%%%%%%%%%%%%%%%%%%%%%%%%%%%%%%%%%%%%%%%%%%%%%%%%%%%%%%%%%%%%%%%
%           Van der Waals forces for colloidal system.
%%%%%%%%%%%%%%%%%%%%%%%%%%%%%%%%%%%%%%%%%%%%%%%%%%%%%%%%%%%%%%%%%%%%%%%%%%%%%%%%%%%%%%%%%%%%%%%%%%%%%%%%%%%%%%%%%%%%%%%%%%%%%%%%%%%%%%%%%%%%%%%%%%%%%%%%%%%%%
\section{Forces between colloids} 
Usually, colloidal suspension is a solution of relatively large particles in a simple low-molecular weight solvent. Normally, the degrees of freedom of solvent molecules can be eliminated \cite{Poon_book, Onsager_1949}. This simplification leads to the description of the system of colloidal particles interacting via some effective mean field potential that takes into account the most significant solvent properties. 
When a spherical colloidal particle is immersed in the bulk, the solvent structure is locally perturbed around it, up to a certain distance from its surface. When two such particles approach each other, their solvent shells begin to overlap and the local fluid structure becomes highly directional. The nature, the strength and even the sign of the resultant interaction are not obvious and depend on the details of the solvent-solvent and solvent-colloid direct interactions. For a concentrated colloidal suspension, the character of this interaction becomes incredibly complex since three-body effects start to play their role. Therefore, for simplicity, we will take into consideration only dilute colloidal solutions and keep in mind only two-body interactions acting between colloidal particles. We consider some typical potentials acting between colloidal particles below. 
\subsection{Derjaguin approximation} 
\label{sec:colloids_derjaguin}
\begin{figure}[ht!]
\center{\includegraphics[width=0.5\linewidth]{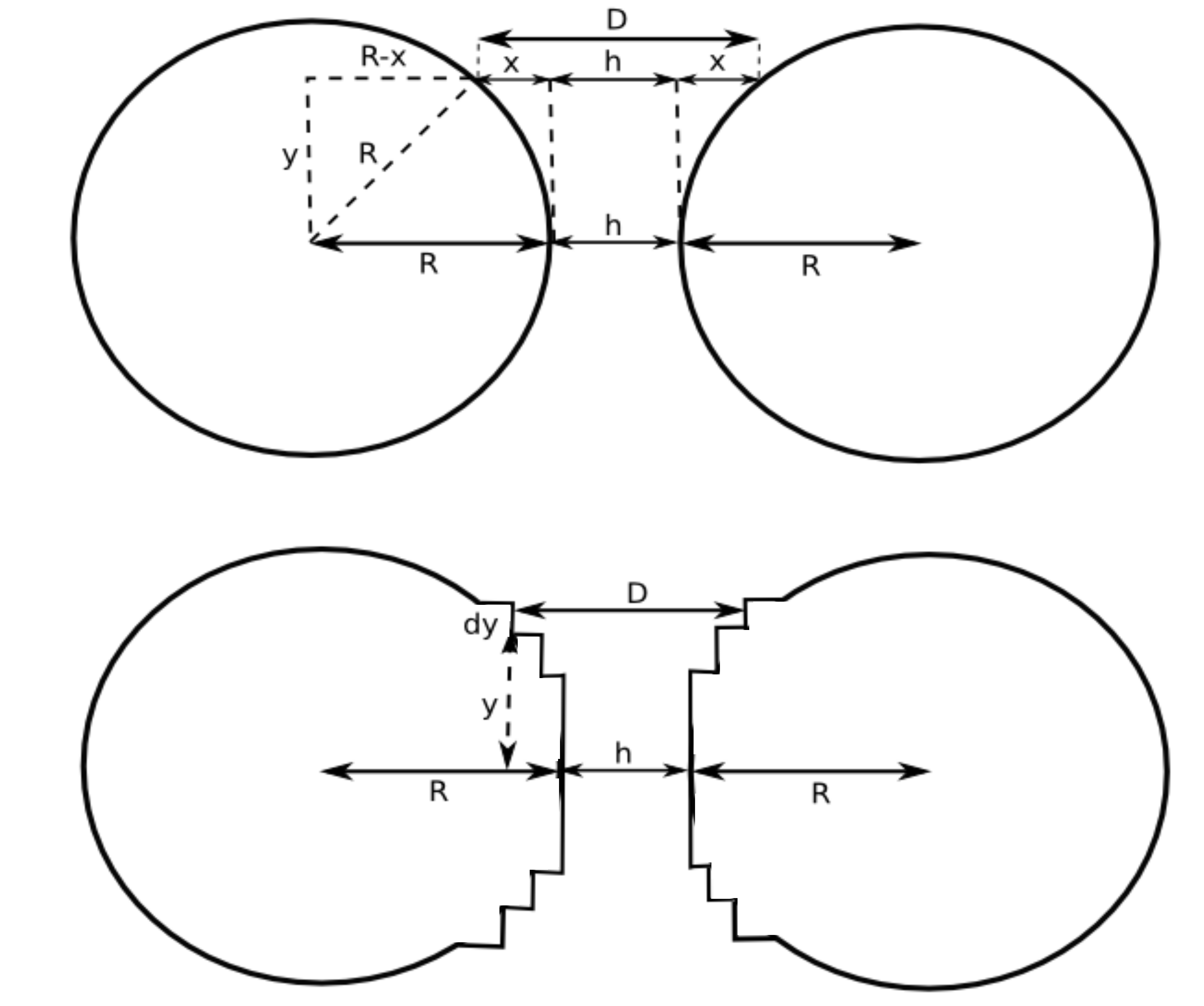}}
\caption{\small{The Derjaguin approximation, which relates the interaction potential between two spherical particles to the energy per unit area of two flat surfaces by Eq.(\ref{intr_colloids_derjaguin_approx}).}}
\label{colloids_derjaguin_fig}
\end{figure}

Before we start a reviewing the interactions between colloidal particles, let us consider very important approximation originally proposed by Derjaguin in \cite{Derjaguin_approx}. The Derjaguin approximation expresses the force acting between curved bodies of finite size in terms of the force acting between two planar infinite walls. This approximation is widely used to estimate the forces between colloidal particles, since forces between two planar bodies are often much easier to calculate. We restrict out consideration by only spherical particles with the same radii, $R$. In addition, instead of force acting between the particles we emphasize on the interaction potential. In this case, the Derjaguin approximation corresponds to replacement of the spherical surface by a collection of flat rings. Consider two spherical particles separated by center-to-center distance $r=2R+h$ as shown in Fig.\ref{colloids_derjaguin_fig}. 
%%%%%%%%%%%%%%%%%%%%%%%%%%%%%%%%%%%%%%%%%%%%%%%%%%%%%%%%%%%%%%%%%%%%%%%%%%%%%%%%%%%%%%%%%%%%%%%%%%%%%%%%%%%%%%%%%%%%%%%%%%%%%%%
%        The Derjaguin approximation
%%%%%%%%%%%%%%%%%%%%%%%%%%%%%%%%%%%%%%%%%%%%%%%%%%%%%%%%%%%%%%%%%%%%%%%%%%%%%%%%%%%%%%%%%%%%%%%%%%%%%%%%%%%%%%%%%%%%%%%%%%%%%%%
	    
The distance between parallel disks can be expressed as $D = 2x + h$. Then, applying Pythagorean theorem, we have $(R-x)^2 + y^2 = R^2$. This expression leads to $x/R - x^2/(2R^2) = y^2/(2R^2)$. When the range of interaction is short in comparison to the colloidal size, it is sufficient to consider only small values of $h/R$ or $y/R$. Therefore, for $y\ll R$ the last expression can be represented as $x \simeq y^2/(2R)$. In this case, the distance between parallel disks can be written as $D \simeq h + y^2/R$ and accordingly increment: $\mathrm{d}D = 2y\mathrm{d}y/R$. Thereby, the interaction between two spherical particles can be written as the sum of the interactions of flat rings with radius $y$ and surface areas: $2\pi y \mathrm{d}y$ separated by distance $D$ from each other. If we denote the interaction potential in the case of flat plates as $W(h)$ and correspondingly for the spherical particles as $W_s(h)$, then: 
$$
	    W_s(h) = \sum\limits_{y=0}^{\infty}2\pi y \mathrm{d}y W(D) \simeq \int\limits_0^{\infty}\mathrm{d}y 2\pi y W(D)
$$
where we assumed that the interaction is sufficiently short-ranged and also neglected the contribution of disks with high values of $D$.  	     Finally, using that $\mathrm{d}D = 2y\mathrm{d}y/R$, we can write:
\begin{equation}
\label{intr_colloids_derjaguin_approx}
	    W_s(h) = \pi R \int\limits_h^{\infty}\mathrm{d}D\,W(D)
\end{equation}
This is very important expression that will be used throughout this thesis where we often will be synonymously called it as the Derjaguin transformation.
\subsection{Hard-core repulsion}	    	    
In a colloidal system, particles have well-defined size and shape preventing interpenetration of particles to each other. This interaction is called \textbf{hard-core} interaction and is always present in colloidal systems. The hard-core repulsion is a characteristic of inherent colloid-colloid interaction that is independent of the surrounding solvent.
\subsection{Van der Waals interaction}	    
\label{sec:colloids_VdW}
The Van der Waals interaction probably is the most important interaction in the surface and colloidal chemistry that acts between
atoms, molecules and particles. The interaction has its origin in different kinds of dipole-dipole interactions such as interactions between
two permanent dipoles (\textbf{Keesom}), a permanent dipole and an induced dipole (\textbf{Debye}), and two instantaneously induced dipoles (\textbf{London}) arising from the cooperative oscillations of electron clouds when molecules are closeby. All of these interactions obey the law $W_{dd} = -C/r^6$, but the last one, known as dispersion interaction, is always present and contribute to the long-range attraction between colloidal particles \cite{Napper_book}. This attraction follows from summation of all atomic van der Waals interactions. Dielectric properties of colloidal particles and intervening medium determine the strength of the interaction. For two spherical colloidal particles with radius $R_c$ the nonretarded van der Waals attraction\footnote{All random dipole-dipole interactions follow the inverse sixth-power law except the so-called retarded van der Waals attraction, which varies with the inverse seventh power of the separation. The result of the retarded regime is well know Casimir effect \cite{Hiemenz_book}. In the future, we will use only nonretarded interaction and therefore will omit the term nonretarded.} can be written as \cite{Israelachvili_2011}
$$
	    W_{vdw}(h) = -\frac{A_H}{6}\left\{\frac{2R_c^2}{(4R_c+h)h} + \frac{2R_c^2}{(2R_c + h)^2} + \ln \frac{(4R_c+h)h}{(2R_c+h)^2}\right\}
$$
where $A_H$ is the \emph{Hamaker constant} and $h$ is the closest distance between spherical surfaces. In most cases the typical range of the noticeable interaction is much less than the particle size, so that it is reasonable to consider the asymptotic of the expression at $h\ll R_c$ i.e.:
\begin{equation}
\label{intr_colloids_vdw}
	    W_{vdw}(h) = -\frac{A_HR_c}{12h}
\end{equation}       
This expression is widely used in the colloid physics community. One may notice, that the Hamaker constant has energetic dimension or, in other words, can be expressed in $k_BT$. The Hamaker constant can be analytically represented using the Lifshitz general theory of van der Waals forces \cite{Lifshitz_vdw, Israelachvili_2011}. In the simplest form the Hamaker constant for two macroscopic phases with the same refractive index $n_1$ (dielectric constant $\varepsilon_1$) interacting across a medium with refractive index $n_3$ (dielectric constant $\varepsilon_3$) can be written as: 
\begin{equation}
\label{intr_colloids_hamaker}
	     A_H \simeq \frac{3}{4}k_BT\left(\frac{\varepsilon_1-\varepsilon_3}{\varepsilon_1+\varepsilon_3}\right)^2 + 
	               \frac{3h\nu_e}{16\sqrt{2}}\frac{(n_1^2-n_3^2)^2}{(n_1^2+n_3^2)^{3/2}}
\end{equation}	    
where $h=6.626\times 10^{-34} \,[J\cdot s]$ is the Planck's constant, $\nu_e$ is the main electronic absorption frequency lying in $UV$ range and usually around $3\times 10^{15}\,[Hz]$ \cite{Israelachvili_2011}. The first term in the above expression gives the zero frequency energy of the van der Waals interaction. It includes the Keesom and Debye contributions. The second term gives the dispersion energy and includes the London contribution. This expression is not exact, but the total error normally does not exceed 5$\%$ \cite{Israelachvili_2011}. Next, the expression for the Hamaker constant is always positive, thereby the interaction between two colloidal 
particles in a medium is always attractive\footnote{The interaction between two particle composed from different substance in some cases can be repulsive, for more details see \cite{Israelachvili_2011}.}. The contribution of dispersion energy can be very high when the refractive 
index of one media is much higher than of the other, i.e for $n_1\gg n_3$. Meanwhile, the purely entropic zero-frequency contribution of the Hamaker constant never exceeds $3k_BT/4$. 
	    
For the computation of the Hamaker constant the \emph{combining relations} are often used. These relations obtain approximate values for unknown Hamaker constants in terms of the known ones. Let $A_{11}$ be a Hamaker constant for media $1$ (colloids) interacting across a vacuum and $A_{33}$ be a Hamaker constant for media $3$ (solvent) interacting across a vacuum. Then, the Hamaker constant for media $1$ interacting across medium $3$ is 
$$
            A_{131} \simeq \left(\sqrt{A_{11}} - \sqrt{A_{33}}\right)^2
$$
The last expression is known as the combining relation \cite{Israelachvili_2011}. The values for the Hamaker constants, $A_{ii}$ are usually tabulated and can be found in many handbooks. The relation is applicable only when the dispersion interaction dominates over zero-frequency interactions and breaks down when applied to the media with high dielectric constants such as water or, when the zero-frequency contribution is large \cite{Israelachvili_2011}.
  	   	    
The long range van der Waals attraction strongly influences the colloidal stability. Consider several examples of it. The Hamaker constant of gold particles interacting across water is $A_H \simeq 28 k_BT_{room}$ \cite{Parsegian_2005}\footnote{In \cite{Israelachvili_2011} the author gives the value $A_H \simeq 40\times 10^{-20}[J]$ which for $T = 298 K$ is $A_H \simeq 100k_BT$, where $k_BT = 4.11\times10^{-21}[J]$. We will use the value given in the main text as it is just an illustrative example.}. For particles with the size, say $R_c = 100 nm$, located at a distance $h = 10nm$, it leads to $W_{vdw} \simeq -24\, k_BT$. It is much more than the energy of ordinary thermal motion. In the absence of any other repulsive interactions overriding the attraction it will inevitably lead to a very rapid aggregation between the particles in the solution and precipitation of the gold aggregates. 
	    
In \cite{Noskov_ps_vdw} the Hamaker constant was defined for polystyrene latex-particles: $A_H\simeq 3.2\times 10^{-21} [J] \simeq 4 k_BT$. 	For the same size of the particles separated by the same distance it produces $W_{vdw} \simeq -3\, k_BT$. This value is much less than for the gold particles in water, but still more than enough to provoke aggregation between particles. There are also methods capable to minimize the long-range van der Waals attraction between colloidal particles or to override the influence of the attraction in order to provide the colloidal stability. To reduce van der Waals attraction between colloidal particles immersed in a solvent, we must pick up the refractive index of the solvent as close as possible to the refractive index of colloids (see Eq.(\ref{intr_colloids_hamaker})). In addition, interactions of electrostatic origin can be present, and additives to the dispersion, like polymers, may influence the interaction between the colloidal particles. In the case of polymers, they can be grafted on the particle surfaces or be free in the solution. We consider several other types of interactions which, together with the van der Waals interaction can prevent or accelerate the aggregation of colloidal particles.	    	    
\subsection{Electrostatic interaction}
\begin{figure}[ht!]
\center{\includegraphics[width=0.5\linewidth]{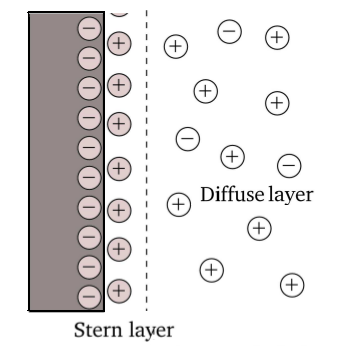}}
\caption{\small{Charged colloidal surface immersed in electrolyte solution.}}
\label{colloids_in_electrolyte_fig}
\end{figure}

Stabilization of colloids due to electrostatic forces plays the central role in modern colloidal science and its applications. Even originally neutral surface of colloids in a liquid may be charged by dissociation of surface groups or by adsorption of charged molecules from the surrounding solution \cite{Israelachvili_2011}.	In electrolyte solvent, this leads to formation of a surface potential which will attract counter-ions from the surrounding solution and repel co-ions. In equilibrium, the charged colloidal surface is balanced by oppositely charged counter-ions in solution making the combination of surface and surrounding solution electrically neutral. The region in the vicinity of the surface with high concentration of ions is called the \emph{electrical double layer}. The electrical double layer can be approximately divided into two spatial domains. Ions whose charge is opposite to the sign of the surface charge (counter-ions) are in the region closest to the surface and strongly bound to it. This immovable layer is called the Stern or Helmholtz layer. The region adjacent to the Stern layer is called the diffuse layer and mostly contains loosely associated counter-ions  that are comparatively mobile (see Fig.\ref{colloids_in_electrolyte_fig}). The ion concentration in the bulk solution defines the width of the electrical double layer, so that at high enough concentration of the ions the charges on the colloidal surface are screened. \cite{Russel_book}. 
%%%%%%%%%%%%%%%%%%%%%%%%%%%%%%%%%%%%%%%%%%%%%%%%%%%%%%%%%%%%%%%%%%%%%%%%%%%%%%%%%%%%%%%%%%%%%%%%%%%%%%%%%%%%%%%%%%%%%%%%%%%%%%%
%        Charged surface in electrolyte
%%%%%%%%%%%%%%%%%%%%%%%%%%%%%%%%%%%%%%%%%%%%%%%%%%%%%%%%%%%%%%%%%%%%%%%%%%%%%%%%%%%%%%%%%%%%%%%%%%%%%%%%%%%%%%%%%%%%%%%%%%%%%%%
	    
Colloidal particles are in constant motion. Due to it the particles randomly collide with each other resulting in the overlap between the double layers. Therefore, dispersed like-charged colloid repel each other upon approach. The length scale, over which the force is significant, is defined by the Debye length	    
$$
	    \lambda_D = \sqrt{\frac{1}{8\pi\lambda_B I}}
$$
where $I = \frac{1}{2}\sum_i\rho_{\infty, i}z_i^2$  is the molar ionic strength of electrolyte consistent from different types of ions $i$ with valency $z_i$ and bulk concentration $\rho_{\infty, i}$. For convenience, we have identified separately the Bjerrum length:
$$
	    \lambda_B = \frac{e^2}{4\pi \varepsilon_0\varepsilon k_BT}
$$
where $e$ is elementary charge and $\varepsilon$ the relative dielectric constant. The Bjerrum length is the distance between two unit charges, when their Coulomb energy is equal to the thermal energy $k_BT$ \cite{Israelachvili_2011}. Further, for simplicity, we will consider only one-component salt with the valency $z$. In this case the ionic strength is: $I = \rho_{\infty}z^2$, where $\rho_{\infty}$ is molar concentration of the salt in the bulk.	Therefore,
$$
	    \lambda_D = \sqrt{\frac{1}{8\pi z^2\lambda_B \rho_{\infty}}}
$$
The interaction potential between two charged flat plates due to repulsion between double layers is given \cite{Israelachvili_2011} by the following expression:
$$
	    W_{dl}(h) = \frac{Z}{\pi \lambda_D}\exp\left(-\frac{h}{\lambda_D}\right) 
$$
This expression is valid when the flat plates are sufficiently far apart, since the potential profiles originating from each individual surface will not be much perturbed by the presence of the other surface. This approximation suggests that one can simply superpose the potentials profiles originating from each surface and quantitatively we can write for it $h\gtrsim\lambda_D$. Then, applying the Derjaguin transformation Eq.(\ref{intr_colloids_derjaguin_approx}) the interaction potential for two spherical particles, with radii $R_c$, can be written as
\begin{equation}
\label{intr_colloids_pot_dl}
	    W_{dl}(h) = ZR_c\exp\left(-\frac{h}{\lambda_D}\right)
\end{equation}
In this expression the multiplier $Z$ is\footnote{It should be noted that the multiplier is independent of concentration of ions.}
$$
	    Z = 64\pi k_BT\rho_{\infty}\gamma^2\lambda_D^2 = 8 \frac{k_BT\gamma^2}{z^2\lambda_B} = 32\pi \varepsilon_0\varepsilon \left(k_BT/ze\right)^2\gamma^2
$$
where $\gamma=\tanh(ze\psi_0/4k_BT)$ is determined through the surface potential $\psi_0$ and never exceeds $1$. Instead of surface potential, the expression for $\gamma$  can be written in terms of surface charge density $\sigma$. To this end, we can use the Grahame equation that for symmetric electrolytes with valency $z$ has the following form:
$$
	    \sigma = \sqrt{8\rho_{\infty}\varepsilon_0\varepsilon k_BT}\sinh\left(\frac{ze\psi_0}{2k_BT}\right) 
$$
More general cases of the Grahame equation are considered in \cite{Israelachvili_2011}. The Grahame equation can be simplified in a case of weak surface potential when $\sinh(x)$ can be expanded to the Taylor series. Further, we restrict ourselves only by first order approximation. In such a way, we obtain quite simple dependence between surface charge density and surface potential, namely, $\sigma \simeq \varepsilon_0\varepsilon\psi_0/\lambda_D$ that lead to: 
$$
	    \gamma \simeq \frac{ze\psi_0}{4k_BT} \simeq \frac{ze\lambda_D\sigma}{4\varepsilon_0\varepsilon k_BT}
$$
This expression, in turn, simplifies the Eq.(\ref{intr_colloids_pot_dl}) in the case of weak surface potential and monovalent salt:
$$
	    W_{dl}(h) \simeq 2\pi R_c \varepsilon_0\varepsilon \psi_0^2\exp\left(-\frac{h}{\lambda_D}\right) \simeq \frac{2\pi R_c\lambda_D^2\sigma^2}{\varepsilon_0\varepsilon}\exp\left(-\frac{h}{\lambda_D}\right)
$$
Despite that this relationship is obtained with restriction of monovalent salt, it is valid for any electrolytes \cite{Israelachvili_2011} in approximation of weak surface potential. On the other hand, in the case of the strong surface potential when $ze\psi_0 \gg 4k_BT$ we can write $\gamma \simeq 1$ and accordingly,
$$
	    W_{dl}(h) \simeq 32\pi R_c\varepsilon_0\varepsilon \left(k_BT/ze\right)^2\exp\left(-\frac{h}{\lambda_D}\right) 
$$
Thus, the interaction energy does not depend on the surface potential caused by the substantial screening of the colloid surface.	
\subsection{DLVO interaction}	    
\begin{figure}[ht!]
\begin{minipage}[h]{0.5\linewidth}
\center{\includegraphics[width=1\linewidth]{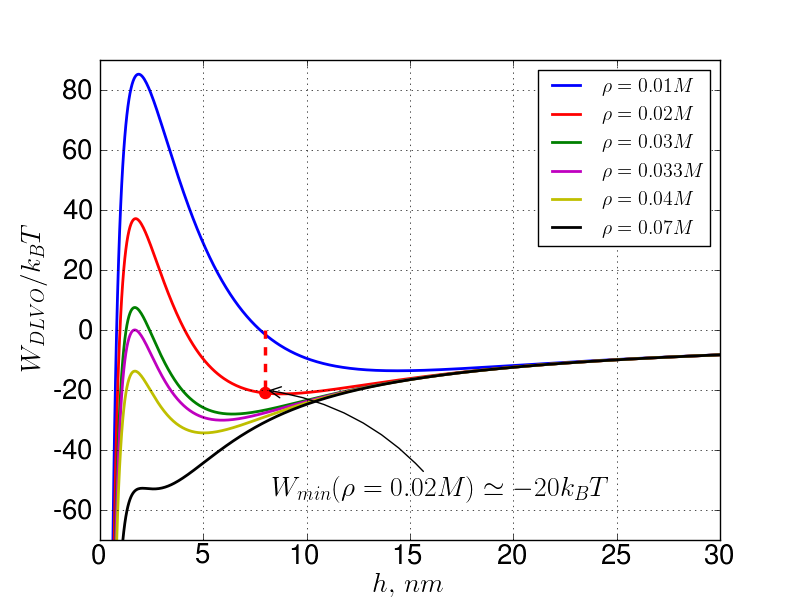}}
\caption{\small{The DLVO interaction potential (in units $k_BT$) as a function of distance between two identical gold spherical particles with $R_c = 100nm$ in aqueous solution, containing different concentration of monovalent salt ($NaCl$). The Hamaker constant is $A_H = 28k_BT$ and the surface potential we set to $\psi_0 = -70mV$.}}
\label{colloids_dlvo_potentials_fig}
\end{minipage}
\hfill
\begin{minipage}[h]{0.5\linewidth}
\center{\includegraphics[width=1\linewidth]{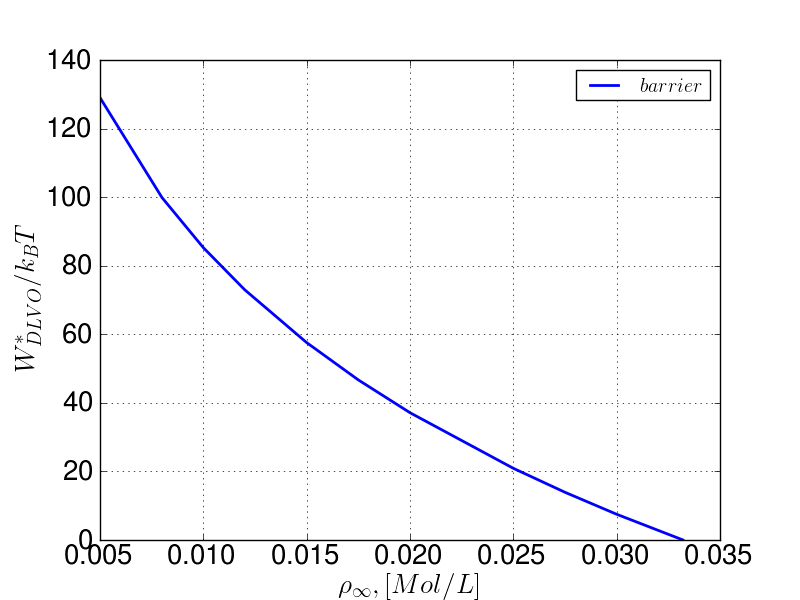}}
\caption{\small{The dependence of the barrier height on ion concentration (monovalent salt ($NaCl$)) due to Eq.(\ref{intr_colloids_dlvo_barrier}). The calculations was done for two identical gold particles with $R_c = 100nm$ in aqueous solution.
The Hamaker constant is $A_H = 28k_BT$ and the surface potential we set to $\psi_0 = -70mV$.}}
\label{colloids_dlvo_barriers_vs_conc_fig}
\end{minipage}
\end{figure}

The abbreviation DLVO comes from the first letters of the names of the author who first developed the theory to explain the aggregation of aqueous dispersions, namely Derjaguin, Landau, Verwey and Overbeek (see \cite{Derjaguin_DLVO, VO_DLVO}). Assuming the additive nature of the interaction of the double layer Eq.(\ref{intr_colloids_pot_dl}) and van der Waals Eq.(\ref{intr_colloids_vdw}) the total DLVO potential is given by	    	    	    
\begin{equation}
\label{intr_colloids_pot_dlvo}
	    W_{DLVO}(h) \simeq W_{dl}(h) + W_{VdW}(h) = 8 k_BT\gamma^2\left(\frac{R_c}{z^2\lambda_B}\right)\exp\left(-\frac{h}{\lambda_D}\right) - \frac{A_HR_c}{12h}
\end{equation}
Note that at short distances, the dispersion interaction always wins since the first term is always limited. The dispersion interaction always leads to aggregation between colloids. However, the electrostatic repulsive interaction usually prevents the colloidal particles to get close enough to each other or in other words to fall into the primary minimum of the DLVO potential. Thus, to prevent the irreversible coagulation, we should pick up the conditions imposed on this potential. These conditions should ensure the kinetic stability. It is obvious that the conditions are associated with the DLVO potential barrier height. Let us consider a specific example: colloidal gold particles in $NaCl$ aqueous solution (see Fig.\ref{colloids_dlvo_potentials_fig}). When dissolved in water, the sodium chloride dissociate to the monovalent $Na^{+}$ and $Cl^{-}$ ions become surrounded by the polar water molecules. In this case, we fixed the Hamaker constant $A_H=28k_BT$, dielectric permittivity of water $\varepsilon\simeq 80$, colloidal size $R_c = 100nm$. For the surface potential we assumed that it is equal to $\psi_0 \simeq -70$ $mV$ and also suppose that the quantity is unchanged for considered ion concentrations \cite{VO_DLVO}.
%%%%%%%%%%%%%%%%%%%%%%%%%%%%%%%%%%%%%%%%%%%%%%%%%%%%%%%%%%%%%%%%%%%%%%%%%%%%%%%%%%%%%%%%%%%%%%%%%%%%%%%%%%%%%%%%%%%%%%%%%%%%%%%
%        DLVO vs h
%%%%%%%%%%%%%%%%%%%%%%%%%%%%%%%%%%%%%%%%%%%%%%%%%%%%%%%%%%%%%%%%%%%%%%%%%%%%%%%%%%%%%%%%%%%%%%%%%%%%%%%%%%%%%%%%%%%%%%%%%%%%%%%
Therefore, we only have one variable parameter, namely, the concentration of ions in solution. The gold particles are negatively charged and repel each other. At low and intermediate salt concentration, the electrostatic barrier prevents the colloidal particles from aggregation.
When we increase the salt concentration, the Debye radius becomes smaller leading to a decrease of the electrostatic repulsion. The gold particles moving due to thermal motion have a greater chance to get closer to each other at a range of a few angstroms. At such scale, the dominant role of the van der Waals attraction, leads to the aggregation of the colloidal particles and as a final result to their precipitation. The condition for the aggregation, due to insufficient barrier, is $\rho_{\infty} > \rho_{\infty, c}$, where $\rho_{\infty, c}$ is the critical ion concentration at which the barrier disappears. One may notice in Fig.\ref{colloids_dlvo_potentials_fig} a secondary minimum which manifests itself at large inter-particle separations. This secondary minimum is much weaker in comparison to the primary minimum. In addition, the secondary attraction well can give rise to a reversible adhesion if the depth of the well exceeds $W_{DLVO}(h_2) \simeq 1.5 k_BT$. These flocculated aggregates are sufficiently stable. Moreover, the aggregates can not be broken up by Brownian motion, but may dissociate under an externally applied force such as shaking or upon lowering salt concentration. The depth of the secondary minimum decreases upon increasing the ion concentration until it reaches a certain value when the system has only one global minimum (see Fig.\ref{colloids_dlvo_potentials_fig}). 
	  	  
The critical condition for the complete loss of stability is identified with the disappearance of the repulsive barrier. This means that the following two conditions are take place:
\begin{equation}
\label{intr_colloids_dlvo_max_cond}
	  0 = \frac{\partial W_{DLVO}(h)}{\partial h} \rightarrow \frac{8k_BT\gamma^2}{z^2\lambda_B\lambda_{D,c}}\exp\left({-\frac{h_c}{\lambda_{D,c}}}\right) = \frac{A_H}{12h^2_c}
\end{equation}
This is the condition for a maximum and    
$$
	  0 = W_{DLVO}(h) \rightarrow \frac{8k_BT\gamma^2}{z^2\lambda_B}\exp\left({-\frac{h_c}{\lambda_{D,c}}}\right) = \frac{A_H}{12h_c}
$$
which along with the above condition ensures that this maximum is equal to zero. Further, dividing the first expression by the second one, we get the following expression $h_c = \lambda_{D,c}$, where subscription $c$ refers to critical condition. Finally, substituting this expression into one of the above conditions, we obtain an expression for the critical Debye length, namely,
$$
	  \lambda_{D, c} = \frac{2.72z^2}{96\gamma^2}\left(\frac{A_h}{k_BT}\right)\lambda_B  
$$
where we used that $\exp(1) \simeq 2.72$ . Correspondingly, for the critical ion concentration we can write
\begin{equation}
\label{intr_colloids_dlvo_conc_crit}
	  \rho_{\infty, c} = \frac{8(12)^2\left(k_BT\right)^2}{(2.72)^2\pi \lambda_B^3A_H^2}\frac{\gamma^4}{z^6} \sim \frac{1}{z^6}
\end{equation}	    
It is interesting to look at this expression in cases of strong and weak surface potential. As we have already seen in the case of strong surface potential $\gamma \simeq 1$. This expression virtually does not change its form. But in the case of weak surface potential $\gamma \simeq ze\psi_0/(4k_BT)$ and we can write 
$$
	  \rho_{\infty, c} = \frac{9}{2(2.72)^2\pi \lambda_B^3A_H^2}\frac{\left(e\psi_0\right)^4}{z^2\left(k_BT\right)^2} \sim \frac{1}{z^2}
$$	  
The critical concentration in the case of gold colloidal particles placed in aqueous $NaCl$ solution is equal to $\rho_{\infty} = 0.0332 M$ (the parameters for the system can be found from the subscription in Fig.\ref{colloids_dlvo_potentials_fig}).  

Let us find how the barrier hight depends on ion concentration. For that, based on Eq.(\ref{intr_colloids_dlvo_max_cond}), we can write the following equation
$$
	  \Gamma \bar{h}_m^2e^{-(\bar{h}_m-1)} = 1    
$$
This is definition the positions of the barrier peak. For convenience, we introduced dimensionless parameter 
\begin{equation}
\label{intr_colloids_dlvo_gamma}
	  \Gamma = \frac{96\gamma^2}{z^2}\left(\frac{k_BT}{A_H}\right)\left(\frac{\lambda_D}{\lambda_B}\right) = \left(\frac{\lambda_{D}}{\lambda_{D, c}}\right) = \sqrt{\frac{\rho_{\infty, c}}{\rho_{\infty}}}
\end{equation}
and dimensionless spatial root $\bar{h}_m = h/\lambda_D$ that corresponds to peak position. It is obvious that the condition for the existence of a repulsive barrier is $\Gamma > 1$ (or in terms of concentration $\rho_{\infty} < \rho_{\infty, c}$). We can rewrite the expression for the DLVO potential barrier height, Eq.(\ref{intr_colloids_pot_dlvo}), using Eq(\ref{intr_colloids_dlvo_gamma}):	  
\begin{equation}
\label{intr_colloids_dlvo_barrier}
	  W_{DLVO}^*(\bar{h}_m) \simeq \frac{A_H}{12\Gamma}\left(\frac{R_c}{\lambda_{D, c}}\right)\left\{\Gamma e^{-(\bar{h}_m-1)} - \frac{1}{\bar{h}_m}\right\} = 
	                               \frac{A_H}{12\Gamma \bar{h}_m^2}\left(\frac{R_c}{\lambda_{D, c}}\right)\left(1 - \bar{h}_m\right)
\end{equation}
Recall that $\bar{h}_m = 1$ corresponds to the case when the barrier is equal to zero. For other cases with non negative barrier, we have\footnote{Solutions with $\bar{h} > 1$ correspond to the secondary minimum.} $\bar{h}_m < 1$. Correspondingly, the dependence of the barrier height as a function of bulk ion concentration is present in Fig.\ref{colloids_dlvo_barriers_vs_conc_fig}. The height of the barrier indicates how stable the system is. The discussion on limitations of DLVO theory and its improvements is given, for example, in \cite{Ninham_dlvo_limitations}.  
\subsection{Steric interaction due to soluble polymers} 
Apart from the electrostatic stabilization, the colloidal particles can be stabilized by polymers attached to a colloidal surface. When two surfaces covered by polymers approach each other, the local osmotic pressure strongly increases due to steric hindrance of the polymers on the faced colloids. This competition between the polymers for the same volume leads to repulsive interaction and is usually referred to steric or overlap repulsion. Correspondingly, stabilization involving this interaction is known as steric stabilization\cite{Napper_book, Israelachvili_2011}. This interaction plays the important role in many natural and practical systems.  
	  
\begin{figure}[ht!]
\center{\includegraphics[width=1.0\linewidth]{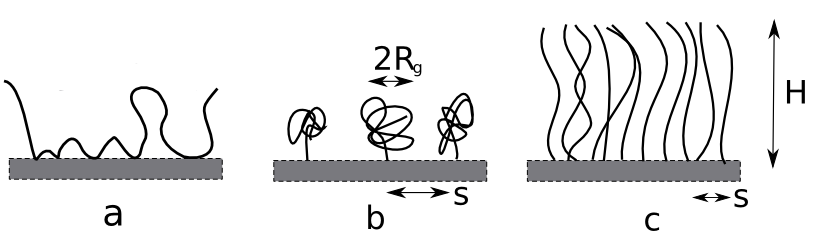}}
\caption{\small{Different cases of attached polymers to a surface: a) a layer of adsorbed polymers, b) mushrooms, c) a brush.}}
\label{colloids_attached_polymers_fig}
\end{figure}

The steric interaction is quite complex and depends on many parameters such as the polymer coverage of the colloidal surface, the quality of the solvent and the mechanism how polymers are bonded to the surface. Polymers can be attached to the surface by different ways. The most widely used examples of attached chains are adsorbed chains, mushrooms and brushes (see fig.\ref{colloids_attached_polymers_fig}).	 The polymers adsorbed on the surface contain many segments sticked to the surface with certain strength. Meanwhile, mushrooms and brushes are chemically bound to the surface by one end. The number of attached chains per unit area of the surface $\sigma \simeq 1/s^2$, where $s$ is a mean distance between two grafting site, is constant and the bonds are not allowed to shift laterally. For a low density of chains, the surface is covered with a number of separated polymers, each of height and size given by $R_g$. The polymers do not overlap with each other. This is referred to as the mushroom regime. When the density of polymer chains on the surface is high, $\sigma \gg 1/R_g^2$, we talk about a polymer brush. The thickness of a brush $H$ is substantially larger than the radius of gyration. 	  
%%%%%%%%%%%%%%%%%%%%%%%%%%%%%%%%%%%%%%%%%%%%%%%%%%%%%%%%%%%%%%%%%%%%%%%%%%%%%%%%%%%%%%%%%%%%%%%%%%%%%%%%%%%%%%%%%%%%%%%%%%%%%%%
%        Charged surface in electrolyte
%%%%%%%%%%%%%%%%%%%%%%%%%%%%%%%%%%%%%%%%%%%%%%%%%%%%%%%%%%%%%%%%%%%%%%%%%%%%%%%%%%%%%%%%%%%%%%%%%%%%%%%%%%%%%%%%%%%%%%%%%%%%%%%

The attached polymers are responsible for the formation of soft layer that should prevent colloids from aggregation caused by the van der Waals attraction. It is obvious that, to maximize the repulsive interaction between colloids, the particles should satisfy the certain conditions, namely the soft layer must be sufficiently dense and the thickness of the layer should be enough to override the van der Waals attraction. With this in mind, let us consider in more details a polymer brush in good solvent. Several models have been proposed to describe the steric repulsion \cite{Napper_book} caused by the polymer brushes. We will follow arguments \cite{Alexander_1977, deGennes_brushes} forming the basis of Alexander-deGennes model of polymer brush and its subsequent improvements made in \cite{Milner_Witten_Cates_theory}. Approximations in Alexander-deGennes model in particularly assume that the grafted polymers are uniformly stretched. Therefore, the free end of each chain is located at the edge of the brush at a distance $H$ from the surface. In our subsequent analysis, we shall mainly follow \cite{Butt_2006, Milner_Witten_Cates_theory}
	  
First of all, we need to find the thickness $H$ of the unperturbed brush. It is determined by a balance between two effects. On the one hand, the excluded volume of polymer segments causes repulsive interaction which tends to swell the brush. In opposition to this, the chain elasticity caused by the configurational entropy resists stretching. In the next chapter we find the expression for the conformational entropy of an ideal chain, Eq.(\ref{intr_polymer_free_en_spring}) which for the stretched chain corresponds to elastic energy contribution. Setting the brush thickness $H$ equal to the extension of the chain, the force is $3k_BTH/(Na_s^2)$. In order to obtain the repulsive energy contribution in the brush, we use the expression for the energy density of polymer segments in a solvent $k_BT\text{v}c^2/2$, where $\text{v}$ is the second virial coefficient (see Eq.(\ref{intr_polymer_flory_huggins_en_virial})). Thereby, we need to know the segment concentration. Due to step-like character of the concentration profile the monomers are distributed uniformly and the concentration of segments is $c\simeq N/Hs^2$. The free energy per chain is obtained by multiplying the density with the volume of one chain $Hs^2$. Therefore, this energy is equal to $k_BT\text{v}N^2/2Hs^2$, which is independent of grafted density. To change the stretching distance $H$, a force $-k_BT\text{v}N^2/2H^2s^2$ must be applied. The minus sign indicates that this force  is directed oppositely to the elastic force and tends to expand the brush. The total force per chain is
$$
	  \Pi_1 \simeq \frac{3k_BTH}{Na_s^2} - \frac{k_BT\text{v}N^2\sigma}{2H^2}
$$
In equilibrium the net force must be equal to zero. This leads to a brush thickness
\begin{equation}
\label{intr_colloids_MWC_step_brush_h}
	  H \simeq N\left(\frac{\text{v}a_s^2\sigma}{6}\right)^{1/3}
\end{equation}
As one can notice, the brush thickness increases linearly with the number of segments and also weakly increases with the grafting density. 
Moreover, the brush thickness continuously increases with the solvent quality, as reflected in the excluded volume parameter.
	  
Now, to get an expression for interaction energy, let two polymer-coated surfaces approach each other. At a distance $h =2H$ between flat surfaces, the polymer chains start to overlap or the polymer layers undergo compressions. This leads to increase in the local segment density of polymer chains which will result in a repulsion. The interaction energy per unit area is obtained from interaction energy per polymer chain by multiplication on $\sigma$ is given by
\begin{equation}
\label{intr_colloids_MWC_step_brush}
\begin{array}{l}
	  W(h) \simeq  2\sigma\int\limits_{H}^{h/2}\mathrm{d}H\, \Pi_1 = \frac{3H^2k_BT\sigma}{Na_s^2}\left\{\left(\frac{h}{2H}\right)^2 + 2\left(\frac{2H}{h}\right) - 3\right\} = \\
	          = k_BTH\left(\frac{9\text{v}\sigma^4}{2a_s^4}\right)^{1/3}\left\{\left(\frac{h}{2H}\right)^2 + 2\left(\frac{2H}{h}\right) - 3\right\}, \quad h < 2H   
\end{array}
\end{equation}
where we used symmetry with respect to mid-plane, which leads to the factor 2. The last expression can be applied only for $h<2H$.
	  
In order to derive Eq.(\ref{intr_colloids_MWC_step_brush}), we used that the segments are homogeneously distributed in the volume $Hs^2$ . 
Outside, the segment concentration is assumed to be zero.  Semenov \cite{Semenov_JETP_1985} and later Milner, Witten, Cates \cite{Milner_Witten_Cates_parabolic_dens} improved this step profile for the segment density of the unperturbed brush. They proposed a more realistic parabolic profile 
$$
	 c(x) \simeq \frac{\pi^2H^2}{8\text{v}N}\left\{1 - \left(\frac{x}{H}\right)^2\right\}
$$
which vanishes smoothly at the outer edge of the brush. The thickness of uncompressed brush in this case is defined as
\begin{equation}
\label{intr_colloids_MWC_parab_brush_h}
	 H = N\left(\frac{12\text{v}a_s^2\sigma}{\pi^2}\right)^{1/3}
\end{equation}
The interaction energy per unit area \cite{Milner_Witten_Cates_theory, Milner_Witten_Cates_parabolic_dens, McLean_mwc_parab_Brush} is given by
\begin{equation}
\label{intr_colloids_MWC_parab_brush}
	  W(h) \simeq k_BTH\left(\frac{\pi^4\text{v}\sigma^4}{144a_s^4}\right)^{1/3}\left\{\left(\frac{h}{2H}\right)^2 + \left(\frac{2H}{h}\right) - \frac{1}{5}\left(\frac{h}{2H}\right)^{5} - \frac{9}{5}\right\}, \quad h < 2H   
\end{equation}         
         
In addition to aforementioned examples, a commonly used expressions for describing the steric force per unit area between two parallel planar plates were given by de Gennes \cite{deGennes_brushes} as the following
\begin{equation}
\label{intr_colloids_AdG_step_brush_force}
	  \Pi(h) \simeq k_BT\sigma^{3/2}\left\{\left(\frac{2H}{h}\right)^{9/4} - \left(\frac{h}{2H}\right)^{3/4} \right\}, \quad h < 2H
\end{equation}
This expression is obtained for homogeneous segment concentration of a brush (step profile for the segment concentration) using scaling theory description. The interaction potential between two flat surfaces can be obtained by integrating $\Pi(h)$ over the plate separation distance (from $h$ to $2H$) which leads to
\begin{equation}
\label{intr_colloids_AdG_step_brush}
	  W(h) \simeq \frac{8k_BTH\sigma^{3/2}}{35}\left\{7\left(\frac{2H}{h}\right)^{5/4} + 5\left(\frac{h}{2H}\right)^{7/4} -12\right\}, \quad h < 2H
\end{equation}         
For the thickness of the uncompressed brush deGennes proposed the following expression
\begin{equation}
\label{intr_colloids_AdG_parab_brush_h}
	  H \simeq N\left(a_s^{5}\sigma\right)^{1/3}
\end{equation}	  
This is roughly the same expression for brush thickness as we saw in Eqs.(\ref{intr_colloids_MWC_parab_brush_h},\ref{intr_colloids_MWC_step_brush_h}), assuming that $\text{v} \simeq a_s^3$. It was shown \cite{Israelachvili_2011} that the Eq.(\ref{intr_colloids_AdG_step_brush}) is roughly exponential when the parameter $h/2H$ is varied in the range $0.2$ to $0.9$ and is given by the following formula
$$
	  \Pi(h) \simeq 100k_BT \sigma^{3/2} e^{-\pi h/H}
$$
This is just an approximation of Eq.(\ref{intr_colloids_AdG_step_brush}) and does not have a physical origin. Correspondingly, the interaction potential between two flat surfaces can be obtained by integrating $\Pi(h)$ over the plate separation distance:
\begin{equation}
\label{intr_colloids_AdG_exp}
	  W(h) \simeq \frac{100k_BT}{\pi} H\sigma^{3/2} e^{-\pi h/H}
\end{equation}
	  
\begin{figure}[ht!]
\begin{minipage}[h]{0.5\linewidth}
\center{\includegraphics[width=1\linewidth]{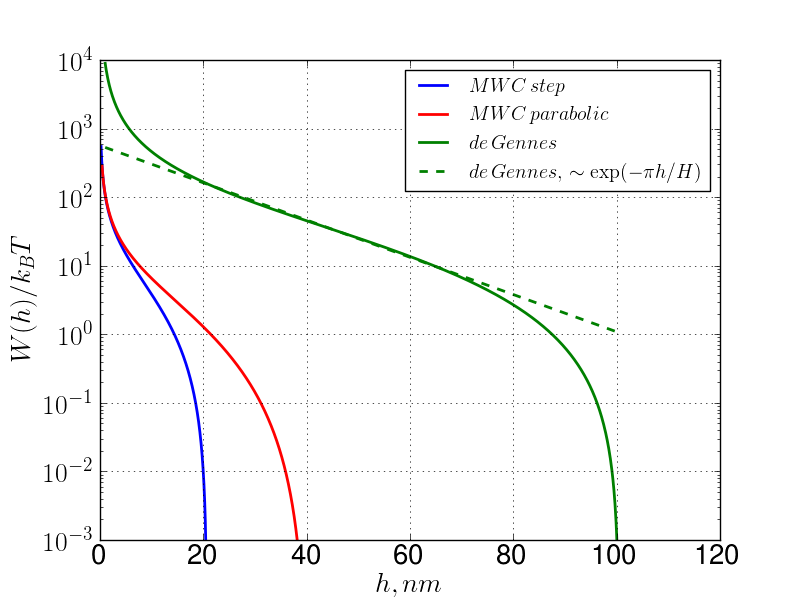}}
\caption{\small{The interaction potential per unit area calculated for polystyrene brush immersed in toluene 
		for the Milner, Witten, Cates model with a step Eq.(\ref{intr_colloids_MWC_step_brush}) and parabolic segment 
		profile Eq.(\ref{intr_colloids_MWC_parab_brush}), and the de Gennes model Eq.(\ref{intr_colloids_AdG_step_brush}).
		The following parameters have been used: $N=100$, $a_s = 7.6 \textup{\AA}$, $\text{v}=23 \textup{\AA}^3$, $\sigma = 5\times 10^{17}m^{-2}$.}}
\label{colloids_brush_comparison_fig}
\end{minipage}
\hfill
\begin{minipage}[h]{0.5\linewidth}
\center{\includegraphics[width=1\linewidth]{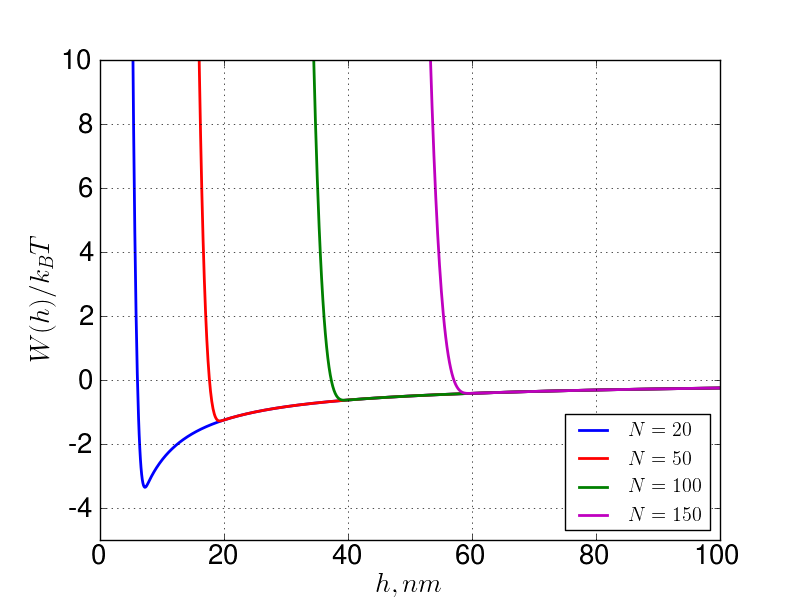}}
\caption{\small{The total interaction potential per unit area calculated between two mica colloids with $R_c=100nm$ coated by polyethylene 	brush immersed in toluene. We used the Milner, Witten, Cates model with a parabolic segment profile Eq.(\ref{intr_colloids_MWC_parab_brush_sperical}). The following parameters have been used: $N=100$, $a_s = 7.6 \textup{\AA}$, $\text{v}=23 \textup{\AA}^3$, $\sigma = 5\times 10^{17}m^{-2}$ and $A_H=3k_BT$.}}
\label{colloids_brush_therm_pot_rc100nm_mwc_parab_fig}
\end{minipage}
\end{figure}

In Fig.\ref{colloids_brush_comparison_fig} we have shown in the real variables all of the above expressions for the interaction energy. 
We chose polystyrene brush in toluene ($a_s = 7.6 \textup{\AA}$, $\text{v}=23 \textup{\AA}^3$) as a reference system. Each chain of length $N=100$ attached to a plate with the anchored density $\sigma = 5\times 10^{17}m^{-2}$. As one can notice the difference is huge and related in particularly with that we used $\text{v} \simeq 0.05a_s^3$. Despite that, all of the expression for interaction energy Eqs.(\ref{intr_colloids_MWC_step_brush}, \ref{intr_colloids_AdG_step_brush}, \ref{intr_colloids_MWC_parab_brush}) can be used to fit force curves quite well. Eq.(\ref{intr_colloids_AdG_step_brush_force}) is widely used due to its advantage of having only two parameters, $\sigma$ and $H$, which both have a physical meaning. In order to improve and generalize the interaction potential between brushes attached to different geometries many studies have been done since that time,both analytical \cite{Zhulina_1991a, Zhulina_1991b, Zhulina_1991c, Wiesner_langmuir} and computational studies \cite{Baschnagel_2004, Matsen_brush_2007, Matsen_brush_2009}. In addition good reviews about brushes and interactions between brush layers were given in \cite{Milner_science, Curriea_brush_review}.	  	  	  	  	  
%%%%%%%%%%%%%%%%%%%%%%%%%%%%%%%%%%%%%%%%%%%%%%%%%%%%%%%%%%%%%%%%%%%%%%%%%%%%%%%%%%%%%%%%%%%%%%%%%%%%%%%%%%%%%%%%%%%%%%%%%%%%%%%
%        steric vs h
%%%%%%%%%%%%%%%%%%%%%%%%%%%%%%%%%%%%%%%%%%%%%%%%%%%%%%%%%%%%%%%%%%%%%%%%%%%%%%%%%%%%%%%%%%%%%%%%%%%%%%%%%%%%%%%%%%%%%%%%%%%%%%%
	  
For further analysis of the steric interaction, we should compare it with van der Waals attraction Eq.(\ref{intr_colloids_vdw}). In order to implement that we will work only with Eq.(\ref{intr_colloids_MWC_parab_brush}) which we should recalculate for the spherical geometry.
In addition, the Derjaguin approximation is quite reasonable as long as the range of the interaction potential is much smaller than the sphere radius. Thereby, we obtain
\begin{equation}
\label{intr_colloids_MWC_parab_brush_sperical}
	  W_s(h) \simeq k_BTR_cH^2\left(\frac{\pi^7\text{v}\sigma^4}{18a_s^4}\right)^{1/3}f\left(\frac{h}{2H}\right), \quad h < 2H   
\end{equation}
where $H$ is defined in Eq.(\ref{intr_colloids_MWC_parab_brush}) and 
\begin{equation}
\label{intr_colloids_MWC_parab_brush_sperical_f}	  
	  f(x) = \frac{1}{30}\left(x^6 -10x^3 +54x - 30\log(x) -45\right)
\end{equation}
For the van der Waals interaction energy we will use Eq.(\ref{intr_colloids_vdw}) as before. We have chosen colloids made from mica with\footnote{We also ignore the influence of the polystyrene brush on the Hamaker constant since its refractive index is quite close to refractive index of toluene.} $A_H = 3k_BT$ \cite{Parsegian_2005} and $R_c = 100nm$ and for polymer brush we used the same parameters as we considered in Fig.\ref{colloids_brush_comparison_fig}. We depicted in Fig.\ref{colloids_brush_therm_pot_rc100nm_mwc_parab_fig}, the total interaction energy as a function of distance between the closest points of the spheres for different brush sizes. As one can observe from the picture, the steric part of the potential grows very fast in comparison with van der Waals. This potential contains only one minimum caused by van der Waals attraction. The depth of this minimum depends on the size of the brush and the strength of van der Waals interaction at distances $h\simeq 2H$. Thus, in order to estimate the size of the chain which leads to just a weak flocculation $W_{tot} \lesssim 0.2 k_BT$,
we need to consider only van der Waals contribution at distances $h\simeq 2H$ which for $H$ defined in Eq.(\ref{intr_colloids_MWC_parab_brush_h}) leads to condition
$$
	   N > \frac{A_HR_c}{5}\left(\frac{\pi^2}{12\text{v}a_s^2\sigma}\right)^{1/3} 
$$
providing just a weak flocculation that is colloid stability.
\subsection{Depletion interaction} 
\begin{figure}[ht!]
\center{\includegraphics[width=0.5\linewidth]{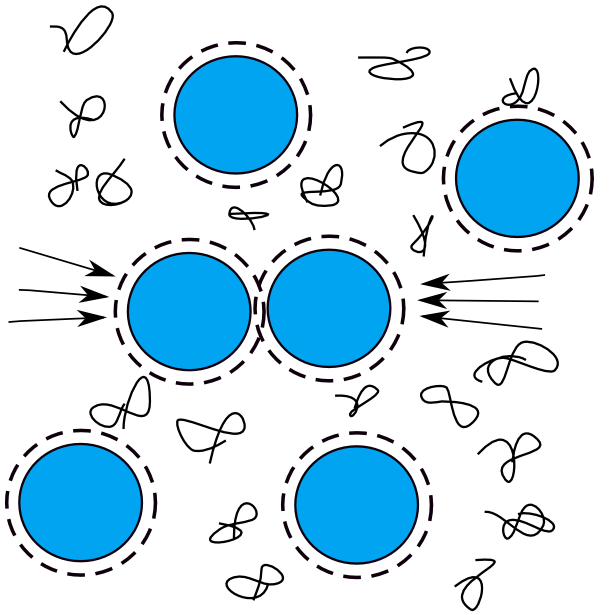}}
\caption{\small{Schematic representation of the depletion attraction mechanism for colloidal particles in a solution of unattached polymers.}}
\label{colloids_depletion_attraction_fig}
\end{figure}

Now we move from the interactions which help to prevent aggregation between colloidal particles to the interaction, which provokes aggregation between the particles. Suppose, for example, we have colloidal particles immersed in a solution of non-adsorbing polymers. 
Due to the steric repulsion between colloidal surface with monomers a polymer chain loses conformational entropy in the region and as a consequence the effective depletion layer is emerged near the surface. The mechanism responsible for the attraction between colloidal particles originates from the presence of the depletion layer. We depicted several colloidal particles immersed in solution of unattached polymers (see Fig.\ref{colloids_depletion_attraction_fig}). The depletion layers are indicated by the dashed circles around the colloids. When the depletion layers overlap the volume available for free chains increases. Thus, the free energy of the polymers is getting lower when colloids are close to each other. This effect suggests the existence of some attractive force between the particles. The force manifests itself even in the lack of direct attraction between colloids or between colloid and polymer \cite{Vrij_1976, Lekkerkerker_book}. 
	  
Let us consider two colloidal particles with diameter $2R_c$, each surrounded by a depletion layer with thickness $\delta$. In case of small polymer concentration, when the chains are not overlapped, the attraction energy (the depletion potential) can be calculated as a product of 
ideal osmotic pressure, $\Pi = n_bk_BT$, where $n_b$ is a bulk number density of the polymer coils times the overlap volume of the depletion layers, $V_{ov}$. Therefore, the depletion potential between two hard spheres equals
\begin{equation}
\label{intr_colloids_depletion_AO}	  
	    W(h) = \left\{
	    \begin{array}{rl}
	      %-\infty, & r < 2R_c\\\\
	      -PV_{ov}(h),     & 0< h < 2\delta\\\\
	                0,     &    h > 2\delta\\
	    \end{array} \right.
\end{equation}
where 
$$
	  V_{ov} = \frac{\pi}{6}(2\delta - h)^2(3R_c + 2\delta +h/2)
$$
is the overlap volume. This result was firstly derived by Asakura and Oosawa \cite{AO_a, AO_b} regarding polymers as pure hard spheres. Later Vrij \cite{Vrij_1976} obtained the same result by describing polymer chains as penetrable hard spheres. One can notice from Eq.(\ref{intr_colloids_depletion_AO}) that the depletion attraction is determined by the size $2\delta$ which is for dilute polymer solution is quite close to $2R_g$. In addition, the attraction increases with increase of osmotic pressure, thus with polymer concentration. Therefore, the depletion mechanism offers the possibility to modify the range and the strength of the attraction between colloids independently.
%%%%%%%%%%%%%%%%%%%%%%%%%%%%%%%%%%%%%%%%%%%%%%%%%%%%%%%%%%%%%%%%%%%%%%%%%%%%%%%%%%%%%%%%%%%%%%%%%%%%%%%%%%%%%%%%%%%%%%%%%%%%%%%
%        Depletion attraction
%%%%%%%%%%%%%%%%%%%%%%%%%%%%%%%%%%%%%%%%%%%%%%%%%%%%%%%%%%%%%%%%%%%%%%%%%%%%%%%%%%%%%%%%%%%%%%%%%%%%%%%%%%%%%%%%%%%%%%%%%%%%%%%

For the semidilute dilute polymer solution De Gennes et al \cite{deGennes_1981, deGennes_1982, Joanny_1979} developed classical analytical theory of FPI interactions.
The interaction potential between two flat plates with pure repulsive surfaces is\footnote{This expression will be derived and investigated in Sec.\ref{sec:repulsion_gsd_free_en}} 
\begin{equation}
\label{intr_colloids_depletion_gsd}	  
W_{gs}(h) = -\frac{16c_ba^2}{\xi}\exp\left(-\frac{h}{\xi}\right), \quad h\gg \xi
\end{equation}
where $c_b$ is the bulk monomer concentration, $\xi$ is the correlation length in the polymer solution and $a_s^2=6a$ is the polymer statistical segment.
Applying the Derjaguin approximation, Eq.(\ref{intr_colloids_derjaguin_approx}), we obtain
\begin{equation}
\label{intr_colloids_depletion_gsd_Rc}	  
W_{gs}(h) = -16\pi R_c c_ba^2\exp\left(-\frac{h}{\xi}\right), \quad h\gg \xi
\end{equation}	  
%\subsection{Mixing Interaction.}  Small review of mixing interactions. Often in practical systems there are multiple types of interactions,

%% file: Chapters/polymers.tex
% Chapter 2
\chapter{Basic statistical properties of polymers} % Main chapter title
\label{chap:Chapter2} % For referencing the chapter elsewhere, use \ref{Chapter1} 
\lhead{Chapter 2. \emph{Introduction. Polymers}} % This is for the header on each page - perhaps a shortened title

\section{Polymer chain. Structure and architectures}
\begin{figure}[ht!]
\center{\includegraphics[width=0.6\linewidth]{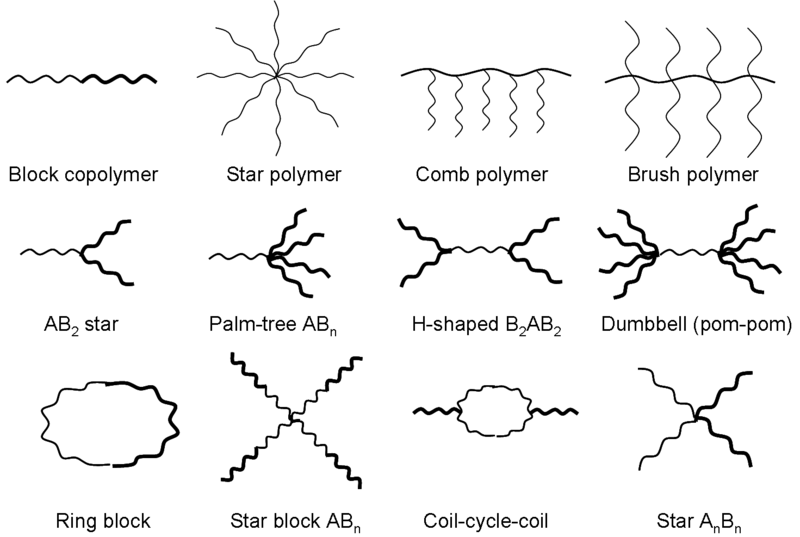}}
\caption{\small{Different polymer architectures. The picture is adopted from \cite{www_wiki_polymer_architecture}.}}
\label{intr_polymer_architecture_fig}
\end{figure}

\textbf{Polymers}, also known as \textbf{macromolecules}, usually represent long molecules (with high molecular weight $M$) composed of repeated units (monomers) of relatively small and simple molecules which are linked together by covalent bonds \cite{Strobl_book}. The number of monomer units that constitute polymer chain is called the \textbf{polymerisation index} $N = M/M_0$, where $M_0$ is the molecular weight of one unit. Polymer molecules consisting of identical monomers are usually called \textbf{homopolymers}, while polymers comprising monomers of different kinds are referred to as \textbf{heteropolymers} or \textbf{copolymers}. The most interesting and important among them are biological macromolecules such as DNA consisting of four different types of monomers. Another example is provided by proteins consisting of 20 different types of monomeric units. Some heteropolymers are not biological, but are artificially synthesized. Often the following types of copolymers are distinguished:
\textbf{statistical copolymer} represents a totally random sequence of different monomers; sequence of \textbf{block-copolymer} has a number of blocks of repeating identical monomers, such as $AAAAAAABBBBBCCCC$;
\textbf{alternating copolymer} is just a simple periodic sequence, such as $ABABABABABABABAB$ that, in turn, can be treated as a homopolymer whose repeating units are $AB$ each. Linear homopolymer chains are the most basic macromolecular structures that serve as building blocks of 
more complex molecular structures. In particular, polymers can have different kinds of branched architectures \cite{Rubinstein_book}.
A branched polymer molecule is composed of a main chain with one or more substituent side chains or branches. 
Different types of branched polymers include star polymers, comb polymers, brush polymers, ladders, dendrimers etc can be seen in
Fig.\ref{intr_polymer_architecture_fig}.
%%%%%%%%%%%%%%%%%%%%%%%%%%%%%%%%%%%%%%%%%%%%%%%%%%%%%%%%%%%%%%%%%%%%%%%%%%%%%%%%%%%%%%%%%%%%%%%%%%%%%%%%%%%%%%%%%%%%%%%%%%%%%%%
%        polymer architecture
%%%%%%%%%%%%%%%%%%%%%%%%%%%%%%%%%%%%%%%%%%%%%%%%%%%%%%%%%%%%%%%%%%%%%%%%%%%%%%%%%%%%%%%%%%%%%%%%%%%%%%%%%%%%%%%%%%%%%%%%%%%%%%%
          
Synthetic polymers such as polystyrene (PS) and polyethylene (PE) are composed of flexible chains which are soluble in a variety of organic solvents like toluene, cyclohexane, etc, but insoluble in water. Another class of polymers are water soluble polymers. They either have strong dipolar groups which are compatible with the strong polarizability of the aqueous media (e.g., polyethylene oxide) or they carry charged groups. Charged polymers are also known as polyelectrolytes. They are extensively studied not only because of their numerous industrial applications, but also from a pure scientific interest. The physics of polyelectrolytes includes the physics of charged systems as well as the physics of polymers \cite{Oosawa_pe, Forster_Schmidt, Joanny_1996}. One important example related to polyelectrolytes and water soluble polymers is associating polymers. In cases when the copolymers have both hydrophobic and hydrophilic groups, they will self-assemble in solution to form meso-structures such as lamellae, cylinders and spheres dispersed in solution \cite{Matsen_1994, Israelachvili_2011}.                     
          
Materials composed of polymer molecules are usually called polymeric materials and have completely different properties from materials 
composed from low molecular weight molecules. Due to the specific properties, the polymeric materials are widely used in applications.          
As with most other substances, physical properties of polymers can be understood via statistical mechanics. To apply statistical mechanics 
we should consider a large assembly of molecules. However, in the case of polymers, macromolecules themselves consisted of many monomeric units and we need to investigate the properties of an isolated macromolecule. To achieve the conditions for the study of the isolated macromolecule it is enough to place the polymer in very dilute solution thereby eliminating any interaction with other polymers.
                    
\section{Polymer models of an ideal chain}	    	  
The statistical thermodynamics of flexible polymers is well developed \cite{Flory_book_1953, Grosberg_1994, Rubinstein_book}. In contrast to other molecules or particles, polymer chains contain not only translational and rotational degrees of freedom, but also a vast number of conformational degrees of freedom where we understand conformation as a spatial arrangement of the atoms constituting the macromolecule.     This fact plays a crucial role in determining their behavior in solution and at surfaces \cite{Netz_report_2003}. The main parameters used to characterize a polymer chain are the polymerization index $N$ , which counts the number of repeat units or monomers along the chain, and the monomer size $b$, is the size of one monomer or the distance between two neighboring monomers. The monomer size ranges from few Angstroms for synthetic polymers to a few nanometers for biopolymers \cite{Flory_book_1953}.
	  
In this section, we review some common theoretical models for linear polymers. The most important fact in the theoretical description of the generic properties of macromolecules is that the polymer chains consist of a very large number $N$ of repeating units. Thus, one can suppose that when studying some averages related to polymer conformations, the most probable conformations dominate the averages. This fact considerably simplifies the analysis.
%%%%%%%%%%%%%%%%%%%%%%%%%%%%%%%%%%%%%%%%%%%%%%%%%%%%%%%%%%%%%%%%%%%%%%%%%%%%%%%%%%%%%%%%%%%%%%%%%%%%%%%%%%%%%%%%%%%%%%%%%%%%%%%
%        The simplification of chains
%%%%%%%%%%%%%%%%%%%%%%%%%%%%%%%%%%%%%%%%%%%%%%%%%%%%%%%%%%%%%%%%%%%%%%%%%%%%%%%%%%%%%%%%%%%%%%%%%%%%%%%%%%%%%%%%%%%%%%%%%%%%%%%
\begin{figure}[ht!]
\center{\includegraphics[width=1.0\linewidth]{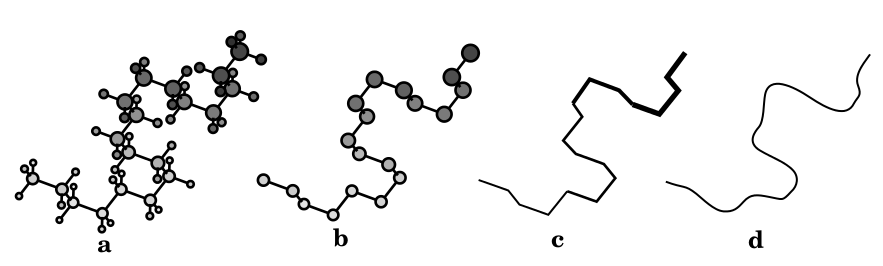}}
\caption{\small{Simplification of a polymer chain conformation from an atomistic model (a) to main-chain atoms only (b), and then to bonds 					on the main chain only (c), and finally to a flexible thread model (d). The picture is adopted from \cite{Teraoka_book}.}}
\label{polymers_chain_simpflification_fig}
\end{figure}	  	  
Moreover, we represent the polymer chain as a sequence of spherical particles connected via rigid rods or springs.
Even in such molecular models, we must adopt the idea of coarse-graining in polymer systems, which involves getting rid of as many degrees of freedom as possible neglecting chemical details whilst preserving the important universal features of their behavior 
(see Fig.\ref{polymers_chain_simpflification_fig}). For this reason, we usually assume that the size of the repeating unit of the molecular model is larger than the atomic scale. Therefore, the essential features of chains will be absorbed in a few characteristics such as: connectivity of repeating units, flexibility of chains and excluded volume interactions. Such simplified models of polymers are very helpful, because they allow to form a relationship between the conformation of the chain and physical properties of a polymer substance.
	  
The simplest theoretical description of flexible chain conformations is achieved with the so-called \textbf{freely jointed chain} model \cite{Grosberg_1994}, where a polymer consists of $N + 1$ beads correspondingly numbered as $0, 1, \ldots N$ and is connected by $N$ rods. The positions of the beads are defined by $\Gamma = \{\boldsymbol{r}_0, \boldsymbol{r}_1, \ldots, \boldsymbol{r}_N\}$. Alternatively, we can use another representation of the conformation $\Gamma$ using position of the chain end, $\boldsymbol{r}_0$ and bond vectors, $\boldsymbol{b}_i = \boldsymbol{r}_i - \boldsymbol{r}_{i-1}$ as an internal degrees of freedom. In this case, any positions of the chain can be represented as 	    
$$
          \boldsymbol{r}_i = \boldsymbol{r}_0 + \sum\limits_{m=1}^{i}\boldsymbol{b}_m, \quad i=0, 1, \ldots, N
$$
Each bond vector has a fixed length $|\boldsymbol{b}_i|$ = b corresponding to the Kuhn length\footnote{which is related, but not identical to the monomer size} and it is allowed to rotate freely. The main advantage of this model is that due to the simplicity, all interesting observables (such as chain size or distribution functions) can be calculated with relative ease.
	  
In this simple model, two arbitrary bond vectors are uncorrelated. The thermal average over the scalar product of two different bond vectors vanishes, $\langle\boldsymbol{b}_i\boldsymbol{b}_k\rangle$ = 0 for $i\ne k$, while the mean squared bond vector length is simply given by $\langle\boldsymbol{b}_i^2\rangle = b^2$. It follows that the mean squared end-to-end radius $R^2$ is proportional to the number of beads,
\begin{equation}
\label{intr_polymer_freely_joint}
	  R^2 = Nb^2
\end{equation}
The same result is obtained for the mean square displacement of a freely diffusing particle where the same underlying physical principle is used, namely the statistics of Markov processes. The chain conformation, $\Gamma$ can be interpreted as a \textit{random walk} path where the bond vectors between beads, $\{\boldsymbol{b}_i\}$ corresponds to sequential random steps (see Fig.\ref{polymers_random_walk_fig}).
	 	           	 
In order to improve the freely jointed chain model consider now \textbf{freely rotating chain} model which is closer to real synthetic polymers. In this model, different chain conformations are produced by torsional rotations of the polymer backbone bonds of length $b$ at fixed bond angle $\theta$. The correlation between two bond vectors does not vanish and is given by \cite{Grosberg_1994}
$$
	  \langle\boldsymbol{b}_i\boldsymbol{b}_{k}\rangle = b^2 (\cos\theta)^{|i-k|}
$$
The mean-squared end-to-end radius for this model in the limit of long chains ($N \rightarrow \infty$) is 
\begin{equation}
\label{intr_polymer_freely_rotate}
	  R^2 = Na_s^2, \quad a_s^2 = b^2\frac{1+\cos\theta}{1-\cos\theta}	
\end{equation}
where $a_s$ is the statistical segment of the chain. In order to make the connection between the two models, we observe that the contour length for freely rotating chain model is $L = Nb \cos(\theta/2)$ 
while for freely jointed chain model $L = Nb$. Using the scaling relation $R^2 = l_K L$, which is true for freely jointed chain as a definition for the Kuhn length $l_K$, we obtain for the freely rotating chain model 
$$
l_K = b\frac{1+\cos\theta}{\cos(\theta/2)(1-\cos\theta)}  
$$
where the Kuhn length $l_K$ is now interpreted as an effective monomer size. For a typical carbon backbone a bond angle is $\theta\simeq 70^{\circ}$ \cite{Flory_book_1969} and thus the relation between the Kuhn length and the monomer size is $l_K \simeq 2.5b$. For a typical bond length of $b \simeq 0.15 nm$ it leads to a Kuhn length of $l_K \simeq 0.38 nm$. It is obvious that, the Kuhn length $l_K$ is always larger than the monomer size $b$. Thus, we have just shown that it is possible to use the simple freely joined chain model also to describe more detailed chain structures if the Kuhn length is considered as an effective length which takes into account correlations between chemical bonds. 
Using the definition of the Kuhn segment $R= l_KL$ and the statistical segment $R=a_sN$, we can relate them with each other:
$$
a_s^2 = L_1l_K
$$
where $L_1=L/N$ is the contour length per repeat unit.
%%%%%%%%%%%%%%%%%%%%%%%%%%%%%%%%%%%%%%%%%%%%%%%%%%%%%%%%%%%%%%%%%%%%%%%%%%%%%%%%%%%%%%%%%%%%%%%%%%%%%%%%%%%%%%%%%%%%%%%%%%%%%%%%%%%%%%%%%%%%%%%%%%%%%%%%%%%%%
%           Universal behavior
%%%%%%%%%%%%%%%%%%%%%%%%%%%%%%%%%%%%%%%%%%%%%%%%%%%%%%%%%%%%%%%%%%%%%%%%%%%%%%%%%%%%%%%%%%%%%%%%%%%%%%%%%%%%%%%%%%%%%%%%%%%%%%%%%%%%%%%%%%%%%%%%%%%%%%%%%%%%%
\section{Universal behavior}
The models discussed so far  do not account for interactions between monomers which are not necessarily close neighbours along the backbone. 
Including these interactions will give a different scaling behavior for long polymer chains. The root mean square end-to-end radius, can be written for $N\rightarrow\infty$ in general form as 
\begin{equation}
\label{intr_polymer_universal_behavior}
R \simeq a_sN^{\nu}
\end{equation}
For an ideal polymer chain (no interactions between monomers), Eq.(\ref{intr_polymer_freely_joint}) implies $\nu=1/2$. This holds only for polymers where the attraction between monomers (in comparison with the monomer–solvent interaction) cancels the steric repulsion 
(since the monomers cannot penetrate each other). It can be achieved in the condition of “theta” solvents. More generally, polymers in solution can experience three types of solvent conditions, with theta solvent condition being intermediate between "good" and "bad" (sometimes called "poor") solvent regimes. In good solvents the monomer-solvent interaction is more favorable than the monomer-monomer one. 
Single polymer chains in good solvents have “swollen” spatial configurations dominated by the steric repulsion, characterized by an exponent $\nu=3/5$. This spatial size of a polymer coil is much smaller than the extended contour length $L \sim a_sN$, but larger than the size of an ideal chain  $R = a_sN^{1/2}$. In the opposite case of "bad"  solvent regime, the effective interaction between monomers is attractive, leading to collapse of the chains and to their precipitation from solution (phase separation between the polymer and the solvent). In this case, the polymer size is characterized by an exponent $\nu=1/3$. The solvent quality depends mainly on the specific chemistry determining the interaction between the solvent molecules and monomers. It can be changed, for example, by varying the temperature.  	  	  		 
%%%%%%%%%%%%%%%%%%%%%%%%%%%%%%%%%%%%%%%%%%%%%%%%%%%%%%%%%%%%%%%%%%%%%%%%%%%%%%%%%%%%%%%%%%%%%%%%%%%%%%%%%%%%%%%%%%%%%%%%%%%%%%%%%%%%%%%%%%%%%%%%%%%%%%%%%%%%%
%           The Gaussian chain model
%%%%%%%%%%%%%%%%%%%%%%%%%%%%%%%%%%%%%%%%%%%%%%%%%%%%%%%%%%%%%%%%%%%%%%%%%%%%%%%%%%%%%%%%%%%%%%%%%%%%%%%%%%%%%%%%%%%%%%%%%%%%%%%%%%%%%%%%%%%%%%%%%%%%%%%%%%%%%
\section{The Gaussian chain model}
In many theoretical calculations aimed at elucidating large-scale properties, the simplification is fulfiled even a step further and a 
continuous model is used as schematically shown in Fig.\ref{polymers_chain_simpflification_fig}d. In such models the polymer backbone is replaced by a continuous line and all microscopic details are neglected. The basis of this model constitutes the \textbf{Gaussian spring-bead} model. Since this is a foundation of all further work, we will focus in more detail on it.
          
\begin{figure}[ht!]
\center{\includegraphics[width=1\linewidth]{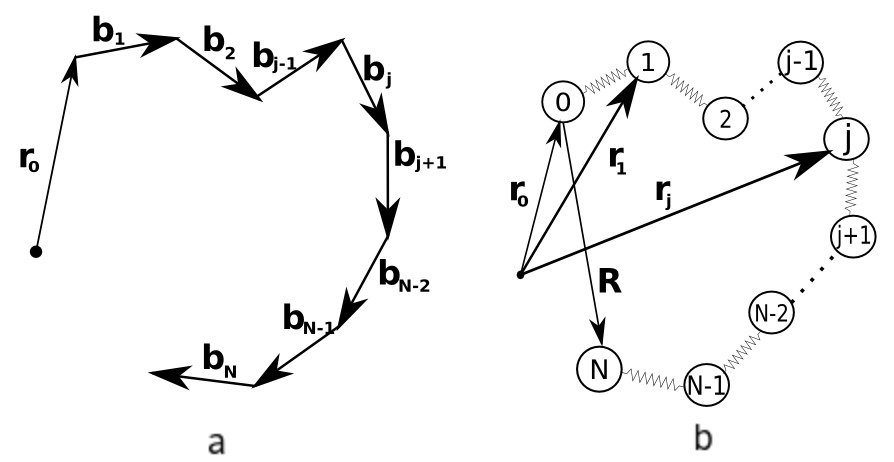}}
\caption{\small{Freely jointed chain model (left) and bead-spring model(right).}}
\label{polymers_random_walk_fig}
\end{figure}

\begin{figure}[ht!]
\center{\includegraphics[width=1\linewidth]{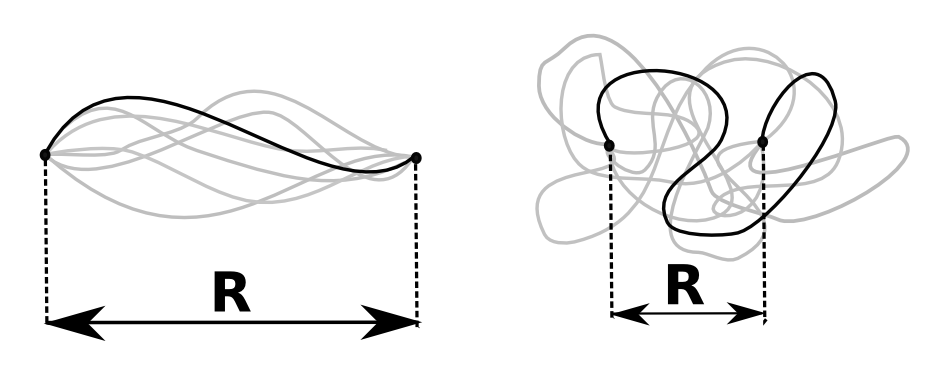}}
\caption{\small{The possible conformations for the stretched chains(left) and free chains(right).}}
\label{entropy_loss_fig}
\end{figure}

The standard Gaussian spring-bead model represents a chain of interacting spherically symmetric beads strung on a thread with Gaussian correlations of adjacent beads positions (see Fig.\ref{polymers_random_walk_fig}). We take as a basis the canonical distribution for an ideal chain(interaction only between adjacent monomers). The statistical weight of the conformation is 
$$
	    \rho(\Gamma) = e^{-U(\boldsymbol{r}_1-\boldsymbol{r}_0)}e^{-U(\boldsymbol{r}_2-\boldsymbol{r}_1)}\ldots  
		           e^{-U(\boldsymbol{r}_N-\boldsymbol{r}_{N-1})}
$$
where potential is already in reduced variables i.e in $k_BT$ units. Let us introduce the now function
$$
	    g(\boldsymbol{b}_i) = e^{-U(\boldsymbol{b}_i)}
$$
Then, the expression for the statistical weight can be rewritten via the function as 
$$
          \rho(\Gamma) = g(\boldsymbol{b}_1) g(\boldsymbol{b}_2)\ldots g(\boldsymbol{b}_N) = \prod\limits_{i=1}^{N}g(\boldsymbol{b}_i)
$$
For the standard Gaussian model the correlations between adjacent beads can be formally represented by a harmonic bond potential, 
$$
	    U(\boldsymbol{r}_i - \boldsymbol{r}_{i-1}) = U(\boldsymbol{b}_i) = \kappa \frac{|\boldsymbol{b}_i|^2}{2}
$$
with the spring constant, $\kappa = 3/b^2$. Thereby, the function $g(\boldsymbol{b})$, for the standard Gaussian model, is 
$$
            g(\boldsymbol{b}_i) = e^{-\frac{3|\boldsymbol{b}_i|^2}{2b^2}}
$$
%%%%%%%%%%%%%%%%%%%%%%%%%%%%%%%%%%%%%%%%%%%%%%%%%%%%%%%%%%%%%%%%%%%%%%%%%%%%%%%%%%%%%%%%%%%%%%%%%%%%%%%%%%%%%%%%%%%%%%%%%%%%%%%
%        bead spring model
%%%%%%%%%%%%%%%%%%%%%%%%%%%%%%%%%%%%%%%%%%%%%%%%%%%%%%%%%%%%%%%%%%%%%%%%%%%%%%%%%%%%%%%%%%%%%%%%%%%%%%%%%%%%%%%%%%%%%%%%%%%%%%%
The probability density of the corresponding conformation $\Gamma$ is 
$$
	    P(\Gamma) = \frac{1}{Z_{0}}\prod\limits_{i=1}^{N}e^{-\frac{3|\boldsymbol{r}_i-\boldsymbol{r}_{i-1}|^2}{2b^2}} = 
			\frac{1}{Z_{0}}\prod\limits_{i=1}^{N}e^{-\frac{3|\boldsymbol{b}_i|^2}{2b^2}}
$$
with
$$
	    Z_{0} = \int\mathrm{d}\boldsymbol{r}_0\int\mathrm{d}\boldsymbol{r}_1\ldots\int\mathrm{d}\boldsymbol{r}_N 
		     e^{-\frac{3}{2b^2}\sum\limits_{i=1}^{N}|\boldsymbol{r}_i-\boldsymbol{r}_{i-1}|^2} = 
	            V\int\mathrm{d}\boldsymbol{b}_1\int\mathrm{d}\boldsymbol{b}_1\ldots\int\mathrm{d}\boldsymbol{b}_N e^{-\frac{3}{2b^2}\sum\limits_{i=1}^{N}|\boldsymbol{b}_i|^2}
$$
where $V$ is the volume of the system. The Jacobian of the transformation $\{\boldsymbol{r}_i\}\rightarrow \{\boldsymbol{b}_i\}$ is just equals to unity. Thus, every integral can be calculated and
$$
	    Z_0 = V\left(\frac{2\pi b}{3}\right)^{3N/2}
$$
This quantity corresponds to the total number of possible conformations of the ideal chain multiplied by volume of the system which measures the accessible positions of the center of mass of the chain. Now normalize the function, $g(\boldsymbol{b}_i)$ so that its integral is just unity
$$
            g(\boldsymbol{b}_i) \rightarrow p(\boldsymbol{b}_i) = \left(\frac{3}{2\pi b^2}\right)^{3/2}e^{-\frac{3|\boldsymbol{b}_i|^2}{2b^2}}
$$
So, for probability distribution we can write
$$
	    P(\Gamma) = \prod\limits_{i=1}^{N}p(\boldsymbol{b}_i)
$$
Using the last expression, we can derive useful expression with certain constraints. For example, the probability density to find a chain with specified distance, $\boldsymbol{R} = \boldsymbol{r}_N-\boldsymbol{r}_0$ between its ends is given by
$$
	    P(\boldsymbol{R}) = \int\mathrm{d}\boldsymbol{r}_0\int\mathrm{d}\boldsymbol{r}_1\ldots\int\mathrm{d}\boldsymbol{r}_N 
				 \delta(\boldsymbol{r}_N-\boldsymbol{r}_0 -\boldsymbol{R})P(\Gamma) = 
				 \left(\frac{3}{2\pi Nb^2}\right)^{3/2}\exp\left(-\frac{3|\boldsymbol{R}|^2}{2Nb^2}\right)
$$
Thereby, the probability distribution function for end to end vector of a chain is a Gaussian distribution. We can use this result to calculate the free energy (or conformational entropy) of an ideal chain with fixed end to end vector, $\boldsymbol{R}$
\begin{equation}
\label{intr_polymer_free_en_spring}
	    F_0(\boldsymbol{R}) = - TS_0(\boldsymbol{R}) = k_BT\ln\left(\frac{Z_0}{V}\times P(\boldsymbol{R})\right) = \frac{3k_BT|\boldsymbol{R}|^2}{2Nb^2} + C
\end{equation}
we used that the number of all possible conformations of a chain with fixed vector between its ends is given as $Z_0$ corresponding to all possible conformation of an ideal Gaussian chain without constraints multiplied by the probability of the chain with fixed end to end vector. One can notice in Fig.\ref{entropy_loss_fig} and notice that the number of possible conformations for the stretched chain is less than for the free chain. The stretched chain has a lower conformational entropy and higher free energy.
%%%%%%%%%%%%%%%%%%%%%%%%%%%%%%%%%%%%%%%%%%%%%%%%%%%%%%%%%%%%%%%%%%%%%%%%%%%%%%%%%%%%%%%%%%%%%%%%%%%%%%%%%%%%%%%%%%%%%%%%%%%%%%%
%        entropy loss
%%%%%%%%%%%%%%%%%%%%%%%%%%%%%%%%%%%%%%%%%%%%%%%%%%%%%%%%%%%%%%%%%%%%%%%%%%%%%%%%%%%%%%%%%%%%%%%%%%%%%%%%%%%%%%%%%%%%%%%%%%%%%%%
	    
Now, one can notice that the parameter $b$ is just root-mean-squre length of the bond, namely,
$$
	    \langle\boldsymbol{b}_i\boldsymbol{b}_i\rangle = \int\mathrm{d}\boldsymbol{b}_i(\boldsymbol{b}_i\boldsymbol{b}_i)p(\boldsymbol{b}_i) = 
	    \left(\frac{3}{2\pi b^2}\right)^{3/2}\int\mathrm{d}\boldsymbol{b}_i\boldsymbol{b}_i^2e^{-\frac{3\boldsymbol{b}_i^2}{2b^2}} = 
	    4\pi\left(\frac{3}{2\pi b^2}\right)^{3/2}\int\limits_0^{\infty}\mathrm{d}b_ib_i^4e^{-\frac{3b_i^2}{2b^2}} = b^2
$$	    
The random walk model that we used to derive all of the above expressions has some drawbacks. The most important of them is that beads do not have a physical size that could prevent the walk to go back at the same position. It corresponds to the overlap between monomers. 	    
In order to improve the model we can introduce the excluded volume interaction between monomers which prevents two monomers from occupying the same place (see Fig.\ref{excluded_volume_fig}). Now we are not interested in the specific form of the interaction and consider just arbitrary external field acting on each monomer.
	    	    
\subsection{External field}	
\begin{figure}[ht!]
\begin{minipage}[ht]{0.5\linewidth}
\center{\includegraphics[width=1\linewidth]{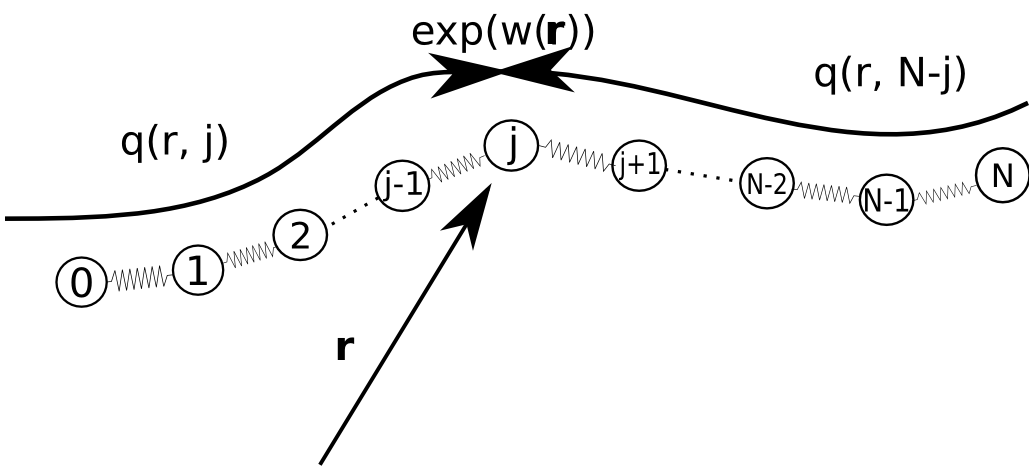}}
\caption{\small{The statistical independence of sub-chains with $j+1$ and $N-j+1$ beads.}}
\label{bead_spring_factor_fig}
\end{minipage}
\hfill
\begin{minipage}[ht]{0.5\linewidth}
\center{\includegraphics[width=1\linewidth]{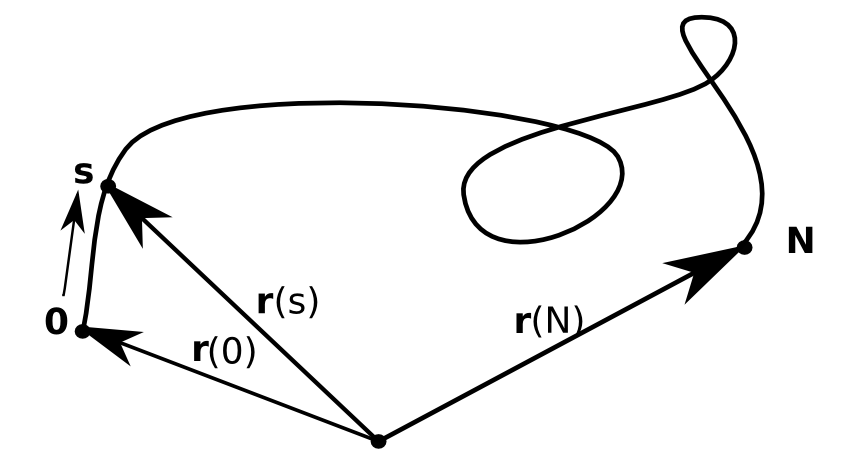}}
\caption{\small{Parametrization of the continuous chain.}}
\label{continuous_chain_fig}
\end{minipage}
\end{figure}

Now we generalize the previous model to take into account external field acting on each monomer of the chain at position, $\boldsymbol{r}_i$,
$$
	    U(\{\boldsymbol{r}_i\}) = U_0(\{\boldsymbol{r}_i\}) + U_{ext}(\{\boldsymbol{r}_i\}) = \sum\limits_{i=0}^{N}\frac{3|\boldsymbol{r}_i - \boldsymbol{r}_{i-1}|^2}{2b^2} 
	    + \sum\limits_{i=0}^{N}w(\boldsymbol{r}_i) 
$$
where the first term, $U_0(\{\boldsymbol{r}_i\})$ is the harmonic stretching energy for the discrete Gaussian chain. The second term, $U_{ext}(\{\boldsymbol{r}_i\})$ represents the interaction energy of each beads with the external field. We can introduce the microscopic monomer (bead) density 
\begin{equation}
\label{intr_polymer_density_discrete}
          \hat{\rho}(\boldsymbol{r}) = \sum\limits_{i=0}^{N}\delta(\boldsymbol{r}-\boldsymbol{r}_{i}) 
\end{equation}
Hence,
$$
	   \int\mathrm{d}\boldsymbol{r}\hat{\rho}(\boldsymbol{r}) = \sum\limits_{i=0}^{N}\int\mathrm{d}\boldsymbol{r}\delta(\boldsymbol{r}-\boldsymbol{r}_{i}) = 
								      \sum_{i=0}^{N}1 = N + 1 
$$
It corresponds to the total number of monomers in the chain. The expression for the interaction energy of the monomers with the external potential we can rewrite as 
$$
	   U_{ext}(\{\boldsymbol{r}_i\}; [w]) = \int\mathrm{d}\boldsymbol{r}w(\boldsymbol{r})\hat{\rho}(\boldsymbol{r}) 
$$
In the case of a non constant external field the probability density is 
$$
	   P(\Gamma) = \frac{1}{Z[w]}e^{-\frac{3}{2b^2}\sum\limits_{i=1}^{N}|\boldsymbol{r}_i-\boldsymbol{r}_{i-1}|^2 - \sum\limits_{i=0}^{N}w(\boldsymbol{r}_i)} 
$$
where 
$$
	   Z[w] = \int\mathrm{d}\boldsymbol{r}_0\int\mathrm{d}\boldsymbol{r}_1\ldots\int\mathrm{d}\boldsymbol{r}_N 
		     e^{-\frac{3}{2b^2}\sum\limits_{i=1}^{N}|\boldsymbol{r}_i-\boldsymbol{r}_{i-1}|^2 - \sum\limits_{i=0}^{N}w(\boldsymbol{r}_i)} 
$$
is the partition function of the Gaussian chain in external field. Let us find another represent of the partition function. For this end consider now normalized partition function
$$
	   Q[w] = \frac{Z[w]}{Z_0} = 
	   \frac{\int\mathrm{d}\Gamma\, \exp(-U(\{\boldsymbol{r}_i\}))}{V\left(\int\mathrm{d}\boldsymbol{b}\exp\left(\frac{3\boldsymbol{b}^2}{2b^2}\right)\right)^N} 
$$
where we denoted $\mathrm{d}\Gamma = \mathrm{d}\boldsymbol{r}_0\mathrm{d}\boldsymbol{r}_1\ldots\mathrm{d}\boldsymbol{r}_N$. Then, denote also that
$$
	   \Phi(\boldsymbol{r}) = \frac{e^{-U_0(\boldsymbol{r})}}{\int\mathrm{d}\boldsymbol{r} e^{-U_0(\boldsymbol{r})}} = 
		                  \left(\frac{3}{2\pi b^2}\right)^{3/2} \exp{\left(-\frac{3|\boldsymbol{r}|^2}{2b^2}\right)}
$$
We can rewrite the expression for the normalized partition function, $Q[w]$ as
\begin{equation}
\label{intr_polymer_part_fun_beadspring}
           Q[w] = \frac{1}{V}\int\mathrm{d}\boldsymbol{r}_N \,e^{-w(\boldsymbol{r}_N)} \int\mathrm{d}\boldsymbol{r}_{N-1}\,\Phi(\boldsymbol{r}_N-\boldsymbol{r}_{N-1})e^{-w(\boldsymbol{r}_{N-1})}
		  \ldots
                 \int\mathrm{d}\boldsymbol{r}_{0}\Phi(\boldsymbol{r}_{1}- \boldsymbol{r}_{0})\,e^{-w(\boldsymbol{r}_{0})} %e^{-w(\boldsymbol{r}_{N-1})}\, \Phi(\boldsymbol{r}_{N-1}- \boldsymbol{r}_{N-2})
\end{equation}
Let us introduce the function $q(\boldsymbol{r}, j; [w])$, so 
$$
	  q(\boldsymbol{r}, 0; [w]) = e^{-w(\boldsymbol{r})}
$$
For other $j=1,\ldots, N$ we can write the recursive relationship as
$$
	  q(\boldsymbol{r}, j; [w]) = e^{-w(\boldsymbol{r})}\int\mathrm{d}\boldsymbol{r}'\, \Phi(\boldsymbol{r} - \boldsymbol{r}')\,q(\boldsymbol{r}', j-1; [w])
$$
Therefore, the expression for the normalized partition function can be written as
$$
	  Q[w] = \frac{1}{V}\int\mathrm{d}\boldsymbol{r}\, q(\boldsymbol{r}, N, [w])
$$
the function, $q(\boldsymbol{r}, j; [w])$ is a statistical weight for a chain with $j+1$ beads and position of the end at $\boldsymbol{r}$ that is the functional of the external field, $w(\boldsymbol{r})$ and usually it is called as a \textit{chain propagator}. We can rewrite the above expression via subchain propagators.Suppose we have 2 sub-chains composed from $j+1$ and $N-j+1$ beads connected at bead, $j$ with position vector $\boldsymbol{r} = \boldsymbol{r}_j$ (see Fig.\ref{bead_spring_factor_fig}). Using the expression for the normalized partition function Eq.(\ref{intr_polymer_part_fun_beadspring}) and definition for the propagator we can write
\begin{equation}
\label{intr_polymer_part_fun_beadspring_factor}
           Q[w] = \frac{1}{V}\int\mathrm{d}\boldsymbol{r}\,q(\boldsymbol{r}, N-j; [w])\,e^{w(\boldsymbol{r})}\,q(\boldsymbol{r}, j, [w]) 
\end{equation}
where the factor $e^{w(\boldsymbol{r})}$ introduced to cancel the excess contribution, $e^{-w(\boldsymbol{r})}$ originated from two joined ends. The physical meaning of the expression is that the conformations of sub-chains with $N+1$ and $N-j+1$ beads are statistically independent (see Fig.\ref{bead_spring_factor_fig}).
%%%%%%%%%%%%%%%%%%%%%%%%%%%%%%%%%%%%%%%%%%%%%%%%%%%%%%%%%%%%%%%%%%%%%%%%%%%%%%%%%%%%%%%%%%%%%%%%%%%%%%%%%%%%%%%%%%%%%%%%%%%%%%%
%        factorization
%%%%%%%%%%%%%%%%%%%%%%%%%%%%%%%%%%%%%%%%%%%%%%%%%%%%%%%%%%%%%%%%%%%%%%%%%%%%%%%%%%%%%%%%%%%%%%%%%%%%%%%%%%%%%%%%%%%%%%%%%%%%%%%
%%%%%%%%%%%%%%%%%%%%%%%%%%%%%%%%%%%%%%%%%%%%%%%%%%%%%%%%%%%%%%%%%%%%%%%%%%%%%%%%%%%%%%%%%%%%%%%%%%%%%%%%%%%%%%%%%%%%%%%%%%%%%%%
%        therm pot A = 200, A = 500
%%%%%%%%%%%%%%%%%%%%%%%%%%%%%%%%%%%%%%%%%%%%%%%%%%%%%%%%%%%%%%%%%%%%%%%%%%%%%%%%%%%%%%%%%%%%%%%%%%%%%%%%%%%%%%%%%%%%%%%%%%%%%%%

\subsection{Continuous chain}	    
Now we parametrize a chain by a variable $s$ that increases continuously along its length. At the first-monomer end, $s = 0$,  and at the 
other end, $s = N$. Using this parametrization, define functions, $\boldsymbol{r}_{i}(s)$, that specify the space curve occupied by homopolymer (see Fig.\ref{continuous_chain_fig}). The microscopic monomer density is defined as 
\begin{equation}
\label{intr_polymer_density_continuous}
          \hat{\rho}(\boldsymbol{r}) = \int\limits_{0}^{N}\mathrm{d}s\,\delta(\boldsymbol{r}-\boldsymbol{r}(s)) 
\end{equation}
The potential energy related with the external field is
$$
	   U_{ext}(\boldsymbol{r}; [w]) = \int\mathrm{d}\boldsymbol{r}'w(\boldsymbol{r}')\hat{\rho}(\boldsymbol{r}') =  
	                                 \int\limits_{0}^{N}\mathrm{d}s\int\mathrm{d}\boldsymbol{r}'w(\boldsymbol{r}')\delta(\boldsymbol{r}'-\boldsymbol{r}(s)) = 
	                                 \int\limits_{0}^{N}\mathrm{d}s\,w(\boldsymbol{r}(s)) 
$$
For entropic contribution in the continuous case, we can write
$$
	   \frac{3}{2b^2\Delta s}\sum\limits_{i=0}^n|\boldsymbol{r}_i - \boldsymbol{r}_{i-1}|^2 = 
	   \frac{3}{2b^2}\sum\limits_{i=0}^n\left|\frac{\boldsymbol{r}_i - \boldsymbol{r}_{i-1}}{\Delta s}\right|^2\Delta s \simeq 
	   \frac{3}{2b^2}\int\limits_{0}^N\mathrm{d}s\,|\boldsymbol{r}'(s)|^2
$$
Then, in order to make the transition from the discrete case to the continuous limit let us split the integration curve $s\in[0..N]$ by the discrete variable, $s_i=i\Delta s_n$ with mesh size $\Delta s_n = N/n$ so that we can use the approximation $\{\boldsymbol{r}(s_i)\}_{i=0}^{n}$ as the continuous line, $\boldsymbol{r}(s)$. Therefore, the expression for the canonical partition function in the continuous limit can be written as
$$
\begin{array}{l}
	   Q[w] = \int\mathcal{D}\boldsymbol{r} \exp\left(-\frac{3}{2b^2}\int\limits_{0}^N\mathrm{d}s\,|\boldsymbol{r}'(s)|^2 - \int\limits_{0}^{N}\mathrm{d}s\,w(\boldsymbol{r}(s)) \right) \equiv   
	    \\\\
	   \lim_{n\to\infty} \frac{1}{V}\int\mathrm{d}\boldsymbol{r}_0\ldots \int\mathrm{d}\boldsymbol{r}_n\left(\frac{3}{2\pi b^2\Delta s_n}\right)^{3n/2}
	   \exp\left(-\frac{3}{2\pi b^2\Delta s_n}\sum_{i=0}^{n}|\boldsymbol{r}_i - \boldsymbol{r}_{i-1}|^2 - \sum_{i=0}^{n}(w(\boldsymbol{r}(s_i))\Delta s_n)\right)
\end{array}
$$
where the integral measure 
\begin{equation}
\label{intr_polymer_wiener_measure}
	  \tilde{\mathcal{D}}\boldsymbol{r} \equiv \mathcal{D}\boldsymbol{r} \exp\left(-\frac{3}{2b^2}\int\limits_{0}^N\mathrm{d}s\,|\boldsymbol{r}'(s)|^2\right) 
\end{equation}
is usually used in the probability theory and called as \textit{Wiener measure}. Now, as for the discrete case we can obtain another representation of the partition function via propagator, $q(\boldsymbol{r}, s; [w])$. For that we consider $n$th approximation for the partition function
$$
          Q_n[w] = \frac{1}{V}\int\mathrm{d}\boldsymbol{r}_n e^{-\Delta s_n w(\boldsymbol{r}_n)}
          \int\mathrm{d}\boldsymbol{r}_{n-1} \Phi_n(\boldsymbol{r}_n - \boldsymbol{r}_{n-1})\,e^{-\Delta s_n w(\boldsymbol{r}_{n-1})}   
          \ldots \int\mathrm{d}\boldsymbol{r}_0 \Phi_n(\boldsymbol{r}_1 - \boldsymbol{r}_{0})\,e^{-\Delta s_n w(\boldsymbol{r}_{0})}      
$$
where the Gaussian function is
$$
	   \Phi_n(\boldsymbol{r}) = \left(\frac{3}{2\pi b^2\Delta s_n}\right)^{3/2} \exp{\left(-\frac{3|\boldsymbol{r}|^2}{2b^2\Delta s_n}\right)}
$$
One can notice that for quite small $\Delta s_n$ the Gaussian function is localized only around $\boldsymbol{r} = 0$.
As before, we can write the expression for the normalized partition function as
\begin{equation}
\label{intr_polymer_part_continuous}
	  Q[w] = \frac{1}{V}\int\mathrm{d}\boldsymbol{r}\, q(\boldsymbol{r}, N, [w])  
\end{equation}
with 
\begin{equation}
\label{intr_polymer_prop_q0_continuous}
	  q(\boldsymbol{r}, 0; [w]) = e^{-\Delta s_n w(\boldsymbol{r})}
\end{equation}
For other $j=1,\ldots, N$ the recursive relationship is
$$
	  q(\boldsymbol{r}, s+\Delta s_n; [w]) = e^{-\Delta s_n w(\boldsymbol{r})}\int\mathrm{d}\boldsymbol{r}'\, 
					          \Phi_n(\boldsymbol{r} - \boldsymbol{r}')\,q(\boldsymbol{r}', s; [w])
$$
This integral equation is equivalent to the diffusion equation. 
\subsection{Edwards equation}	  
Below, for brevity, we omit writing the field, $w(\boldsymbol{r})$ in the function $q(\boldsymbol{r}, s)$ keep in mind that 
it is still functional of the field, $w(\boldsymbol{r})$. For small enough $\Delta s_n$ the function, 
$\Phi_n(\boldsymbol{r} - \boldsymbol{r}')$ is localized only around $\boldsymbol{r}' = \boldsymbol{r}$. Let us expand the function, $q(\boldsymbol{r}, s)$  in the Taylor series around that point:
$$
	  q(\boldsymbol{r}, s+\Delta s_n) = e^{-\Delta s_n w(\boldsymbol{r})}\int\mathrm{d}\boldsymbol{r}'\, 
	  \Phi_n(\boldsymbol{r} - \boldsymbol{r}')\,\left(q(\boldsymbol{r}, s) + \frac{\partial q(\boldsymbol{r}, s)}{\partial x_i}\Delta x_i + 
	  \frac{1}{2} \frac{\partial^2 q(\boldsymbol{r}, s)}{\partial x_i\partial x_j}\Delta x_i\Delta x_j + \ldots\right)
$$	
where $\Delta x_i = (\boldsymbol{r} - \boldsymbol{r}')_i$ is the $i$th projection of the vector and we used the Einstein convention for the index summation. Using that 
$$
	  \sqrt{\frac{3}{2\pi b^2}}\int\limits_{-\infty}^{\infty}\mathrm{d}x\,x^n\,e^{-\frac{3x^2}{2b^2}} = \left\{
	    \begin{array}{rl}	      
	      \left(\frac{b}{\sqrt{3}}\right)^{n}(n-1)!,     & \text{n are even}\\\\
	      0,                 & \text{n are odd }\\
	    \end{array} \right.
$$	
If we take into account the normalized condition $\int\mathrm{d}\boldsymbol{r}\Phi_n(\boldsymbol{r}) = 1$, we can write
$$
          q(\boldsymbol{r}, s+\Delta s_n) \simeq e^{-\Delta s_n w(\boldsymbol{r})}\left(q(\boldsymbol{r}, s) + 
		                                 \frac{\Delta s_n b^2}{6}\Delta q(\boldsymbol{r}, s) + \mathcal{O}(\Delta s_n^2)\right)
$$
where the symbol, $\Delta$ before the function, $q(\boldsymbol{r}, s)$ denotes the Laplace operator acting on the space variable of the corresponding function. Expand in the Taylor series the first term of the r.h.s in the above expression
$$
	  q(\boldsymbol{r}, s+\Delta s_n) \simeq \left(1 - \Delta s_n w(\boldsymbol{r}) + \mathcal{O}(\Delta s_n^2)\right) 
	  \left(q(\boldsymbol{r}, s) + \frac{\Delta s_n b^2}{6}\Delta q(\boldsymbol{r}, s) + \mathcal{O}(\Delta s_n^2)\right)
$$
Leaving only first order terms for the $\Delta s_n$, we can write
$$
	  \frac{q(\boldsymbol{r}, s+\Delta s_n) - \boldsymbol{r}, s)}{\Delta s_n} \simeq \frac{b^2}{6}\Delta q(\boldsymbol{r}, s) - 
						    w(\boldsymbol{r})q(\boldsymbol{r}, s) + \mathcal{O}(\Delta s_n)
$$
In the continuous limit at $n\to\infty$ we get a diffusion-like equation
\begin{equation}
\label{intr_polymer_edwards_eq_field}
	  \frac{\partial q(\boldsymbol{r}, s)}{\partial s} = \frac{b^2}{6}\Delta q(\boldsymbol{r}, s) - w(\boldsymbol{r})q(\boldsymbol{r}, s) 
\end{equation}
This equation was firstly derived in \cite{Edwards_1965} and in polymer physics society is called \textit{Edwards equation}. 
In this way to find a propagator, $q(\boldsymbol{r}, s)$ we should solve the diffusion equation. The initial condition to 
Eq.(\ref{intr_polymer_edwards_eq_field}) follows from Eq.(\ref{intr_polymer_prop_q0_continuous}); at $\Delta s_n \to 0$ it is given by  $q(\boldsymbol{r}, 0; [w]) = 1$. The boundary conditions depend on the specific problem and for example for the system squeezed inside impenetrable walls the boundary condition is $q(\boldsymbol{r}, s)=0$, where $\boldsymbol{r}\in S_{\boldsymbol{r}}$, the surface boundary.
In the limit of continuous chain, $\Delta s_n\to 0$, the expression for the normalized partition function written as two subchain propagator, Eq.(\ref{intr_polymer_part_fun_beadspring_factor}), can be presented as
\begin{equation}
\label{intr_polymer_part_fun_continuous_factor}
           Q[w] = \frac{1}{V}\int\mathrm{d}\boldsymbol{r}\,q(\boldsymbol{r}, N-s)q(\boldsymbol{r}, s) 
\end{equation}
This expression in the continuous case shows the statistical independence of the sub-chains.
	  
\textbf{Edwards Hamiltonian.}
%%%%%%%%%%%%%%%%%%%%%%%%%%%%%%%%%%%%%%%%%%%%%%%%%%%%%%%%%%%%%%%%%%%%%%%%%%%%%%%%%%%%%%%%%%%%%%%%%%%%%%%%%%%%%%%%%%%%%%%%%%%%%%%
%        excluded volume illustration and density from microscopic to macroscopic
%%%%%%%%%%%%%%%%%%%%%%%%%%%%%%%%%%%%%%%%%%%%%%%%%%%%%%%%%%%%%%%%%%%%%%%%%%%%%%%%%%%%%%%%%%%%%%%%%%%%%%%%%%%%%%%%%%%%%%%%%%%%%%%
\begin{figure}[ht!]
\begin{minipage}{0.5\linewidth}
\center{\includegraphics[width=1.0\linewidth]{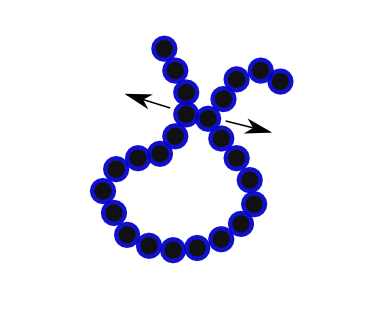}}
\caption{\small{The illustration of excluded volume effect. The interaction between non adjacent units when they approach each other is always repulsive.}}
\label{excluded_volume_fig}
\end{minipage}
\hfill
\begin{minipage}{0.5\linewidth}
\center{\includegraphics[width=1.0\linewidth]{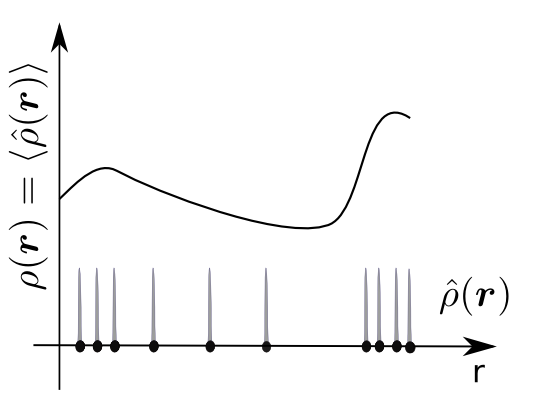}}
\caption{\small{An averaged density profile that corresponds to microscopic density. 
		The idea for the picture is adopted from \cite{Kawakatsu_book}.}}
\label{density_micro_macro_fig}
\end{minipage}
\end{figure}	  
Usually, the Hamiltonian of the continuous Gaussian chain is introduced. The Hamiltonian consists of entropic and external field parts
$$
	  \mathcal{H}[\boldsymbol{r}, w, l] = \mathcal{H}_0 + \mathcal{H}_w = \frac{3}{2b^2}\int\limits_{0}^l\mathrm{d}s\,|\boldsymbol{r}'(s)|^2 + \int\limits_{0}^{l}\mathrm{d}s\,w(\boldsymbol{r}(s))
$$
where the square brackets are used to denote that it is a functional of the corresponding functions. In particular, for two body interactions
$$
	  \mathcal{H}_w[\boldsymbol{r}, w, l] = \frac{1}{2}\int\limits_{0}^{l}\mathrm{d}s\mathrm{d}s'\,U\left(\boldsymbol{r}(s)-\boldsymbol{r}(s')\right)
$$
For very short ranged repulsive potential, it can be rewritten as
$$
	  \mathcal{H}_w[\boldsymbol{r}, w, l] =  \frac{\text{v}}{2}\int\limits_{0}^{l}\mathrm{d}s\mathrm{d}s'\,\delta\left(\boldsymbol{r}(s)-\boldsymbol{r}(s')\right)
$$	  
where $\text{v}$ is an excluded volume parameter (see Fig.\ref{excluded_volume_fig}).	  	  

Using that notation we can write the expression for the normalized partition function in the continuous case as 
\begin{equation}
\label{intr_polymer_part_fun_continuous_canonocal}
	  Q[w] = \int\mathcal{D}\boldsymbol{r}\,e^{-\mathcal{H}[\boldsymbol{r}, w, N]} = \int\tilde{\mathcal{D}}\boldsymbol{r}\,e^{-\mathcal{H}_w[\boldsymbol{r}, w, N]}    
\end{equation}
The corresponding statistical weight of the chain with one end fixed at $\boldsymbol{r}$ and length of the chain, $s$, is
$$
	  q(\boldsymbol{r}, s) = \int\mathcal{D}\boldsymbol{r}\,\delta(\boldsymbol{r} - \boldsymbol{r}(s))\,e^{-\mathcal{H}[\boldsymbol{r}, w, s]}  
$$
Since we know the partition function, we can evaluate the average segment distribution. 
\subsection{Monomer concentration}
Following standard statistical mechanics, the average is computed by weighting each configuration using the Boltzmann factor
\begin{equation}
\begin{array}{c}
\label{intr_polymer_part_fun_continuous_canonocal_conc}
	  \rho(\boldsymbol{r}) = \langle\hat{\rho}(\boldsymbol{r})\rangle = \frac{1}{Q[w]}\int\mathcal{D}\boldsymbol{r}\,\hat{\rho}\,e^{-\mathcal{H}[\boldsymbol{r}, w, N]}
	                       = \frac{1}{Q[w]}\int\limits_0^N\mathrm{d}s\int\mathcal{D}\boldsymbol{r}\,\delta(\boldsymbol{r} - \boldsymbol{r}(s))\,e^{-\mathcal{H}[\boldsymbol{r}, w, N]} =\\
	                       = \frac{1}{Q[w]}\int\limits_0^N\mathrm{d}s\,q(\boldsymbol{r}, s)q(\boldsymbol{r}, N-s)
\end{array}
\end{equation}
where we used $\mathcal{H}[\boldsymbol{r}, w, N] = \mathcal{H}[\boldsymbol{r}, w, s] + \mathcal{H}[\boldsymbol{r}, w, N - s]$
and expression for microscopic density is given in Eq.(\ref{intr_polymer_density_continuous}). This equation provides the practical means of calculating the average segment concentration.  
%%%%%%%%%%%%%%%%%%%%%%%%%%%%%%%%%%%%%%%%%%%%%%%%%%%%%%%%%%%%%%%%%%%%%%%%%%%%%%%%%%%%%%%%%%%%%%%%%%%%%%%%%%%%%%%%%%%%%%%%%%%%%%%%%%%%%%%%%%%%%%%%%%%%%%%%%%%%%
%         The radius of Gyration
%%%%%%%%%%%%%%%%%%%%%%%%%%%%%%%%%%%%%%%%%%%%%%%%%%%%%%%%%%%%%%%%%%%%%%%%%%%%%%%%%%%%%%%%%%%%%%%%%%%%%%%%%%%%%%%%%%%%%%%%%%%%%%%%%%%%%%%%%%%%%%%%%%%%%%%%%%%%%
\section{The radius of Gyration}
The root mean square end-to-end distance is not the only one length characterizing the size of a polymer coil. For example, a polymer
can have many ends or no ends at all. A different way to characterize the chain size is called Gyration radius \cite{Semenov_review_2012}. 
It is defined as the mean-square distance from a monomer unit to the center of mass $\boldsymbol{r}_{cm} = \frac{1}{N}\sum_i\boldsymbol{r}_i$ i.e
\begin{equation}
\label{intr_polymer_freely_rotate_Rg}
	  R_g^2 = \langle\frac{1}{N}\sum\limits_{i}(\boldsymbol{r}_i - \boldsymbol{r}_{cm})^2\rangle = \frac{1}{2N^2}\sum\limits_{i,j}\langle|\boldsymbol{r}_i - \boldsymbol{r}_{j}|^2\rangle
\end{equation}
where $N$ is the total number of monomer units and $\langle\ldots\rangle$ is an ensemble average. For an ideal linear chain it can be easily shown \cite{Rubinstein_book} that
$$
	  R_g^2 = \frac{Na_s^2}{6} = \frac{\langle R^2\rangle}{6}, \quad N\gg 1
$$
where $\langle R^2\rangle$ is the mean square end-to-end distance of the ideal linear chain, $a_s$ is the statistical segment.
%%%%%%%%%%%%%%%%%%%%%%%%%%%%%%%%%%%%%%%%%%%%%%%%%%%%%%%%%%%%%%%%%%%%%%%%%%%%%%%%%%%%%%%%%%%%%%%%%%%%%%%%%%%%%%%%%%%%%%%%%%%%%%%%%%%%%%%%%%%%%%%%%%%%%%%%%%%%%
%           Polymer solution
%%%%%%%%%%%%%%%%%%%%%%%%%%%%%%%%%%%%%%%%%%%%%%%%%%%%%%%%%%%%%%%%%%%%%%%%%%%%%%%%%%%%%%%%%%%%%%%%%%%%%%%%%%%%%%%%%%%%%%%%%%%%%%%%%%%%%%%%%%%%%%%%%%%%%%%%%%%%%
\section{Polymer solution}	    
In the previous sections we discussed the statistical properties of a single polymer chain. We turn now to a discussion of the essential properties of the many-chain systems. Depending on composition of the system, we usually distinguish the following many-chain systems: \textbf{polymer melt} that consists only of one type of polymer chains, \textbf{polymer blend} that consists of different kinds of polymer chains and \textbf{polymer solution} is a mixture of the polymer chains with low molecular mass solvent. 	    
%%%%%%%%%%%%%%%%%%%%%%%%%%%%%%%%%%%%%%%%%%%%%%%%%%%%%%%%%%%%%%%%%%%%%%%%%%%%%%%%%%%%%%%%%%%%%%%%%%%%%%%%%%%%%%%%%%%%%%%%%%%%%%%
%        solution regimes
%%%%%%%%%%%%%%%%%%%%%%%%%%%%%%%%%%%%%%%%%%%%%%%%%%%%%%%%%%%%%%%%%%%%%%%%%%%%%%%%%%%%%%%%%%%%%%%%%%%%%%%%%%%%%%%%%%%%%%%%%%%%%%%
\begin{figure}[ht!]
\center{\includegraphics[width=1\linewidth]{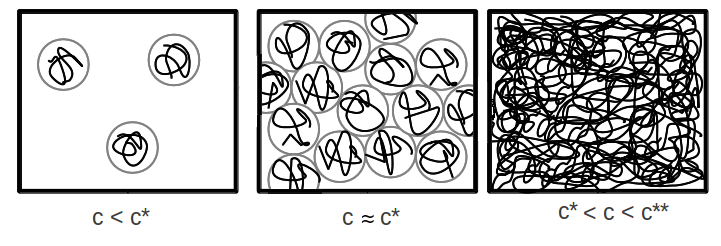}}
\caption{\small{Several concentration regimes of polymer solutions(from left to right):1) a dilute solution ($c<c^*$), 
	       2) a solution at overlap concentration ($c\simeq c^*$), 3) a solution above overlap concentration ($c>c^*$).}}
\label{diff_solutions_regimes_fig}
\end{figure}	    

For our reasons, the most interesting case is the polymer solution which can be categorized by different concentration regimes (see Fig.\ref{diff_solutions_regimes_fig}). The \textit{dilute regime} is defined by $c < c^*$ , for which $c$ denotes the monomer concentration (per unit volume) and $c^*$ is the concentration where individual chains start to overlap. Clearly, the overlap concentration is reached when the average bulk monomer concentration exceeds the monomer concentration inside a polymer coil. In order to estimate the overlap concentration $c^*$ , we simply note that the average monomer concentration inside a coil with radius $R\sim a_sN^{\nu}$ is given by
\begin{equation}
\label{intr_polymer_star}
c^* \simeq \frac{N}{R^3} \sim N^{1-3\nu}a_s^{-3}
\end{equation}
For ideal chains with $\nu=1/2$ the overlap concentration scales as a $a_s^3c^*\sim N^{-1/2}$ and thus decreases slowly as the polymerization index $N$ increases. On the other hand, for swollen chains with $\nu=3/5$, the overlap concentration scales as a $a_s^3c^*\sim N^{-4/5}$ and thus decreases more rapidly with increasing chain length. The crossover to the concentrated or melt-like regime occurs when the monomer concentration in the solution reaches the local monomer concentration inside a Gaussian blob \cite{deGennes_book, Netz_report_2003}, which for good solvent conditions is given by
\begin{equation}
\label{intr_polymer_star_star}
	    c^{**} \simeq \frac{g}{\xi^3} \sim \frac{\text{v}}{a_s^{6}}
\end{equation}
where  $g\simeq (a_s^3/\text{v})^2$ is the minimal number of monomers in the chain, below which the chain statistics is unperturbed by the interactions, $\xi = a_sg^{\nu}$ is a blob size and \text{v} is the second-virial coefficient which describe repulsive interactions between monomers. 
	    
It was shown that the semi-dilute regime is obtained for concentrations $c^* < c < c^{**}$ . This is quite wide concentration range, that 
spans for long chains under good solvent conditions. Thus, it is very important for typical applications.	    
%%%%%%%%%%%%%%%%%%%%%%%%%%%%%%%%%%%%%%%%%%%%%%%%%%%%%%%%%%%%%%%%%%%%%%%%%%%%%%%%%%%%%%%%%%%%%%%%%%%%%%%%%%%%%%%%%%%%%%%%%%%%%%%%%%%%%%%%%%%%%%%%%%%%%%%%%%%%%
%           Flory-Huggins model
%%%%%%%%%%%%%%%%%%%%%%%%%%%%%%%%%%%%%%%%%%%%%%%%%%%%%%%%%%%%%%%%%%%%%%%%%%%%%%%%%%%%%%%%%%%%%%%%%%%%%%%%%%%%%%%%%%%%%%%%%%%%%%%%%%%%%%%%%%%%%%%%%%%%%%%%%%%%%
\subsection{Flory-Huggins model}	     
\begin{figure}[ht!]
\center{\includegraphics[width=0.5\linewidth]{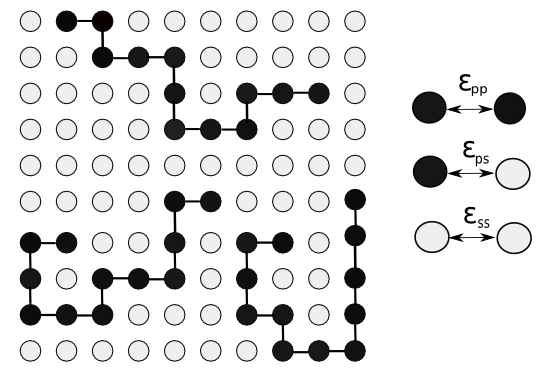}}
\caption{\small{Lattice model of polymer solution and interaction between sites.}}
\label{lattice_model_fig}
\end{figure}

Flory-Huggins theory of polymer solutions is a mean field theory based on a lattice model. The two dimensional lattice model is represented in Fig.\ref{lattice_model_fig}. The lattice has $n$ sites and each site occupied either by one monomer of the polymer or by the solvent molecule. For simplicity, we assume that all monomers have the same size as solvent molecules and equal to the volume of the one site of the cubic lattice, i.e $b^3$. 
%%%%%%%%%%%%%%%%%%%%%%%%%%%%%%%%%%%%%%%%%%%%%%%%%%%%%%%%%%%%%%%%%%%%%%%%%%%%%%%%%%%%%%%%%%%%%%%%%%%%%%%%%%%%%%%%%%%%%%%%%%%%%%%
%        lattice model
%%%%%%%%%%%%%%%%%%%%%%%%%%%%%%%%%%%%%%%%%%%%%%%%%%%%%%%%%%%%%%%%%%%%%%%%%%%%%%%%%%%%%%%%%%%%%%%%%%%%%%%%%%%%%%%%%%%%%%%%%%%%%%%
Let us put randomly $n_p$ polymers on the lattice placing successively $N$ monomers for each polymer chain(random walk on the lattice), so the rest $n_s = n - Nn_p$ sites are filled by the solvent molecules. As a parameter of the system that controls the composition of the solution is used the volume fraction of polymers in solution $\phi = n_pN/n$ and correspondingly the volume fraction of the solvent is $1 - \phi = n_s/n$, obtained from the incompressibility condition. In order to study the thermodynamic properties of this system, we must construct an expression for the free energy of the lattice polymer solution. The Helmholtz free energy, $F = U - TS$ includes two components: 1) the entropic contribution, which describes the number of possible arrangements of the chain on the lattice and 2) internal energy contribution, which takes into account the pairwise interactions between adjacent sites. In the Flory-Huggins theory
the entropic term is calculated using the combinatorial analysis for the $N$ link self-avoiding random walk with volume fraction, $\phi$ 
it is equals to	    
$$
	    -\frac{S}{k_BTn} = \frac{\phi}{N}\ln\left(\frac{\phi}{N}\right) + (1-\phi)\ln(1-\phi)
$$
where the first term is related to the translational entropy of the chain, the second one can be associated with the translational entropy of solvent molecules. Extracting from the above expression the weighted average of the entropy of pure polymer and entropy of pure solvent, we obtain the expression for the entropy of mixing 
$$
	    -\frac{S_{mix}}{k_BTn} = -\frac{S}{k_BTn} - \frac{\phi}{N}\ln\left(\frac{1}{N}\right) - (1-\phi)\ln(1)
$$
or, finally 
$$
	    -\frac{S_{mix}}{k_BTn} = \frac{\phi}{N}\ln\left(\phi\right) + (1-\phi)\ln(1-\phi)
$$
In order to construct an expression for the internal energy we ignore any correlations between monomers along the chain(the chain connectivity). In general, the internal energy includes three kinds of pairwise interactions between adjacent lattice sites (in dimensionless units): 1) polymer-polymer interaction energy $\varepsilon_{pp}$, 2) polymer-solvent interaction energy $\varepsilon_{ps}$ and 3) solvent-solvent interaction energy $\varepsilon_{ss}$. Using the corresponding volume fractions as a weighted coefficients, we can write the expression for the internal energy per one site:
$$
	    \frac{U}{k_BTn} = \varepsilon_{pp}\frac{\phi^2}{2} + \varepsilon_{ps}\phi(1 - \phi) + \varepsilon_{ss}\frac{(1 - \phi)^2}{2}
$$
where the terms for the polymer-polymer and the solvent-solvent interaction divided by two since for the pairwise interaction they are counted twice. Extracting from the expression the contributions from internal energy of pure components with corresponding weighted averages, we get the expression for the internal energy of mixing per lattice site:
$$
	    \frac{U_{mix}}{k_BTn} = \frac{U}{k_BTn} - \varepsilon_{pp}\frac{\phi}{2} - \varepsilon_{ss}\frac{(1-\phi)}{2} = 
	    \frac{1}{2}\phi(1 - \phi)\left(2\varepsilon_{ps} - \varepsilon_{pp} - \varepsilon_{ss}\right) = \chi\phi(1 - \phi)
$$
Instead of three parameters that characterize interactions between different sites, we introduced one, 
$\chi = (2\varepsilon_{ps} - \varepsilon_{pp} - \varepsilon_{ss})/2$ that is called as the \textit{Flory-Huggins interaction parameter}.
The parameter plays the role of a measure of the chemical dissimilarity between the solvent and the polymer.
Finally, we can write the expression for the free energy of mixing $F_{mix} = U_{mix} - TS_{mix}$ in the Flory-Huggins model
\begin{equation}
\label{intr_polymer_flory_huggins_en}
	    f \equiv  \frac{F_{mix}}{nk_BT}  = \frac{\phi}{N}\ln\left(\phi\right) + (1-\phi)\ln(1-\phi) + \chi\phi(1 - \phi)
\end{equation}
where $f$ is the dimensionless free energy density. Note that $F_{mix}\rightarrow 0$ at $\phi\rightarrow 0$ and $\phi\rightarrow 1$. 
In the future, we omit $k_BT$ units. 
%%%%%%%%%%%%%%%%%%%%%%%%%%%%%%%%%%%%%%%%%%%%%%%%%%%%%%%%%%%%%%%%%%%%%%%%%%%%%%%%%%%%%%%%%%%%%%%%%%%%%%%%%%%%%%%%%%%%%%%%%%%%%%%%%%%%%%%%%%%%%%%%%%%%%%%%%%%%%
%           Osmotic pressure.
%%%%%%%%%%%%%%%%%%%%%%%%%%%%%%%%%%%%%%%%%%%%%%%%%%%%%%%%%%%%%%%%%%%%%%%%%%%%%%%%%%%%%%%%%%%%%%%%%%%%%%%%%%%%%%%%%%%%%%%%%%%%%%%%%%%%%%%%%%%%%%%%%%%%%%%%%%%%%            
\subsection{Osmotic pressure} 
\label{sec:polymers_osmotic_pressure}
	    The Flory-Huggins theory should be considered in its prediction for the measurable quantities such as the osmotic pressure of a polymer solution. 
	    The osmotic pressure is the pressure difference required to maintain equilibrium across a semi-permeable
	    membrane that separates the solution from a reservoir with solvent \cite{Rubinstein_book}. 
	    It can be shown that the osmotic pressure $\Pi$ is defined as the change rate 
	    of the total free energy of the system $F_{mix} = nf$ with respect to the volume at constant number of polymers:	    
$$
	    \Pi = - \frac{\partial F_{mix}}{\partial V}\Big\vert_{n_p}
$$	    
	    The volume fraction $\phi$ of polymers each of length $N$ is the ratio of their volume to the volume of the system, $\phi = b^3nN/V$.
	    It helps us to rewrite the derivative with respect to volume $V$ in terms of the derivative with respect to volume fraction at constant number of polymers
$$
	    \partial V = -\frac{(b^3nN)}{\phi}\partial\phi   
$$
	    Note also that the total number of sites $n$ can be expressed in terms of the number of polymers, namely $n = n_pN/\phi$. Therefore, for the osmotic pressure
	    we can write
$$
	    \Pi = \frac{\phi^2}{b^3}\frac{\partial f/\phi}{\partial\phi}\Big\vert_{n_p} = \frac{k_BT}{b^3}\left(\frac{\phi}{N} - \ln(1-\phi) - \phi -\chi\phi^2\right)
$$
	    For small polymer volume fraction this expression can be written in the form of the virial expansion
\begin{equation}
\label{intr_flory_huggins_presure}
	    \Pi = k_BT \left\{\frac{c_b}{N} + \left(1 - 2\chi\right)b^3\frac{c_b^2}{2} + \frac{b^6}{3}c_b^3 + \ldots\right\} = k_BT \left\{\frac{c_b}{N} + \frac{\text{v}c_b^2}{2} + \frac{\text{w}c_b^3}{3} + \ldots\right\}
\end{equation}
	    where we introduced the bulk monomer concentration $c_b=\phi/b^3$. The first term in the above expression gives the osmotic pressure $\Pi_{ideal}$ of the 
	    ideal solution at the low concentration limit. This relation is a general statement of the van't Hoff law. It claims that each solute contributes $k_BT$ to the osmotic pressure. 
	    When polymer concentration increases, $\Pi$ starts to deviate from the dilute limit mainly due to the second term unless $\text{v}=0$. 
	    In Eq.(\ref{intr_flory_huggins_presure}) we denoted 
$$
	    \text{v} = b^3\left(1 - 2\chi\right)  \quad \text{and} \quad \text{w} = b^6
$$
	    Here $\text{v}, \text{w}$ are the second and third virial coefficients respectively. These coefficients are responsible for binary, ternary interactions
	    of monomeric units since, for example, the term $\phi^2/2$ corresponds to the probability of two monomers to occupy the adjacent 
	    lattice sites \cite{Lifshitz_1978, Grosberg_1994}. 
	    The coefficients depend only on the interaction potential between the units, e.g. (see \cite{Gould_Tobochnik_2010, Semenov_review_2012})
$$
	    \text{v} = -\int\mathrm{d}\boldsymbol{r}f_M(\boldsymbol{r}) \quad \text{and}\quad 
	    \text{w} = -\int\mathrm{d}\boldsymbol{r}\mathrm{d}\boldsymbol{r'}f_M(\boldsymbol{r})f_M(\boldsymbol{r'})f_M(\boldsymbol{r}-\boldsymbol{r'})
$$
	    where 
$$
	    f_M(\boldsymbol{r}) \equiv \exp\left(-\frac{U(\boldsymbol{r})}{k_BT}\right) - 1 
$$
	    is the Mayer function which is defined as the difference between Boltzmann factor at distance $\boldsymbol{r}$ and that at infinite separation. 
	    Depending on the temperature, the excluded volume parameter can be positive or negative. For the case of the effective attraction
	    between monomers, when the direct monomer-solvent interaction is less favorable (the interaction potential has lower minimum) than
	    the direct monomer-monomer interaction, the excluded volume could be negative.	    
	    However, if the monomer-solvent interaction is more favorable than monomer-monomer interaction, the excluded volume is positive.
	    Close to the $\theta$ temperature (or above it), $\text{v}(T)$ can be approximated as
$$
	    \text{v} = b^3\left(1 - \frac{\theta}{T}\right)
$$
	    The third virial coefficient $\text{w}$ is normally positive and slightly depends on temperature \cite{Grosberg_1994}.
	    
	    Along with Eq.(\ref{intr_flory_huggins_presure}), we will actively use the expression for the free energy density written in 
	    the third virial approximation \cite{Grosberg_1994} (here we use $k_BT$ as the energy unit)
\begin{equation}
\label{intr_polymer_flory_huggins_en_virial}
	    f_b(c_b) \simeq f^0_b(c_b) + f_{int}(c_b) =  \frac{c_b}{N}\ln\frac{c_b}{Ne} + \frac{\text{v}}{2}c_b^2 + \frac{\text{w}}{6}c_b^3
\end{equation}
	    where the first term $f^0_b$ is the ideal-gas constribution to the free energy density. The bulk chemical potential is defined as 	    	    
\begin{equation}
\label{intr_polymer_flory_huggins_chem_pot_virial}
	    \mu_b(c_b) = \frac{\partial f_b}{\partial c_b} = \mu^0_{b} + \mu_{int} = \frac{1}{N}\ln\frac{c_b}{Ne} + \frac{1}{N} + \text{v}c_b + \frac{\text{w}}{2}c_b^2 
\end{equation}
	    correspondingly, we again can write the expression for the osmotic pressure
\begin{equation}
\label{intr_polymer_flory_huggins_pressure_virial}
	    \Pi_b = \mu_b(c_b) c_b - f_b(c_b) = \Pi_0 + \Pi_{int} = \frac{c_b}{N} + \frac{\text{v}}{2}c_b^2 + \frac{\text{w}}{3}c_b^3 = \frac{c_b}{N}\left(1 + \frac{v_N}{2} + \frac{2w_N}{3}\right)
\end{equation}
	    which coincides with Eq.(\ref{intr_flory_huggins_presure}).
%%%%%%%%%%%%%%%%%%%%%%%%%%%%%%%%%%%%%%%%%%%%%%%%%%%%%%%%%%%%%%%%%%%%%%%%%%%%%%%%%%%%%%%%%%%%%%%%%%%%%%%%%%%%%%%%%%%%%%%%%%%%%%%%%%%%%%%%%%%%%%%%%%%%%%%%%%%%%
%           Phase diagram
%%%%%%%%%%%%%%%%%%%%%%%%%%%%%%%%%%%%%%%%%%%%%%%%%%%%%%%%%%%%%%%%%%%%%%%%%%%%%%%%%%%%%%%%%%%%%%%%%%%%%%%%%%%%%%%%%%%%%%%%%%%%%%%%%%%%%%%%%%%%%%%%%%%%%%%%%%%%%            
\subsection{Phase diagram}	    
\begin{figure}[ht!]
\center{\includegraphics[width=0.7\linewidth]{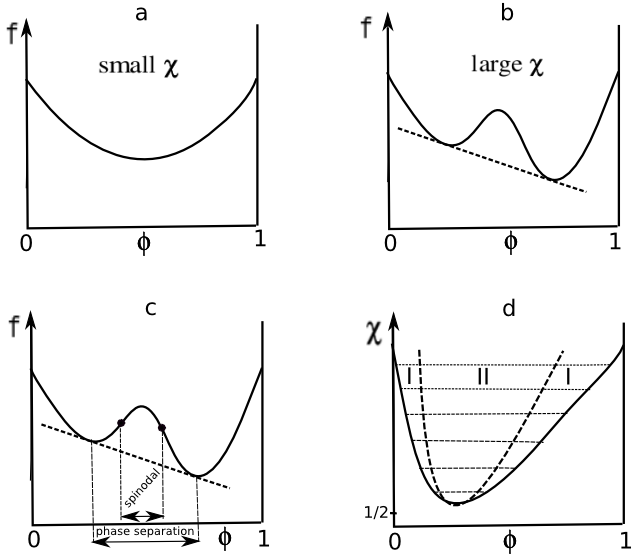}}
\caption{\small{Phase behavior of a polymer solution predicted by the Flory-Huggins theory. In figure \textbf{d}, I is the meta-stable region and II is the unstable region.}}
\label{phase_diagram_fig}
\end{figure}

Let us turn now to another aspect of Flory-Huggins theory, namely to the phase behavior of polymer solutions. The entropic term in Eq.(\ref{intr_polymer_flory_huggins_en}) always favors mixing while the energetic term favors demixing if $\chi > 0$.
In fact, the free energy develops a double minimum if $\chi$ becomes large enough, shown in Figs.\ref{phase_diagram_fig}a--\ref{phase_diagram_fig}b, and the Flory-Huggins theory predicts the liquid-liquid demixing, shown in Fig.\ref{phase_diagram_fig}c.             
%%%%%%%%%%%%%%%%%%%%%%%%%%%%%%%%%%%%%%%%%%%%%%%%%%%%%%%%%%%%%%%%%%%%%%%%%%%%%%%%%%%%%%%%%%%%%%%%%%%%%%%%%%%%%%%%%%%%%%%%%%%%%%%
%        phase diagram
%%%%%%%%%%%%%%%%%%%%%%%%%%%%%%%%%%%%%%%%%%%%%%%%%%%%%%%%%%%%%%%%%%%%%%%%%%%%%%%%%%%%%%%%%%%%%%%%%%%%%%%%%%%%%%%%%%%%%%%%%%%%%%%
The condition that the free energy has a double minimum is required a region where the second derivative $\mathrm{d}^2f/\mathrm{d}\phi^2 < 0$. The boundary of this "spinodal" region is the spinodal line $\mathrm{d}^2f/\mathrm{d}\phi^2= 0$ as indicated in Fig.\ref{phase_diagram_fig}c. The minimum value of $\chi$ on the spinodal curve is the point where additionally $\mathrm{d}^3f/\mathrm{d}\phi^3= 0$. This point is a fluid-fluid demixing critical point. Solving $\mathrm{d}^2f/\mathrm{d}\phi^2 = \mathrm{d}^3f/\mathrm{d}\phi^3= 0$ for the Flory-Huggins free energy shows that the demixing critical point occurs at $\phi_c = N^{-1/2}$ and $\chi_c=1/2 + N^{-1/2}$, where we also assumed $N\gg 1$. Thereby, when we increase $N$ it shifts the critical point to small volume fractions and closer to $\theta$-solvent conditions. The corresponding critical excluded volume parameter is $\text{v}_c = -b^3N^{-1/2}$ . 
Therefore, a small negative virial coefficient between segments will result in phase separation or, in other words, collapsed chains in a solution are dilute.
            
Due to the fact that the Flory-Huggins theory is a mean-field theory, there are obviously some things it does not get right, for example, the shape of the coexistence curve in the vicinity of the critical point. Less obviously, the prediction for the dilute-solution coexistence curve in bad solvent regime is poor. This is because in bad solvent conditions, the mean-field estimate of the energy is inappropriate for polymers, which are essentially dense collapsed coils \cite{Grosberg_1994, Rubinstein_book}.	    

%%%%%%%%%%%%%%%%%%%%%%%%%%%%%%%%%%%%%%%%%%%%%%%%%%%%%%%%%%%%%%%%%%%%%%%%%%%%%%%%%%%%%%%%%%%%%%%%%%%%%%%%%%%%%%%%%%%%%%%%%%%%%%%%%%%%%%%%%%%%%%%%%%%%%%%%%%%%%
%           Experimental determination of virial coefficients
%%%%%%%%%%%%%%%%%%%%%%%%%%%%%%%%%%%%%%%%%%%%%%%%%%%%%%%%%%%%%%%%%%%%%%%%%%%%%%%%%%%%%%%%%%%%%%%%%%%%%%%%%%%%%%%%%%%%%%%%%%%%%%%%%%%%%%%%%%%%%%%%%%%%%%%%%%%%%            
\subsection{Experimental determination of the virial coefficients} 
\label{sec:polymers_experiment}
Now we want to determine the numerical values of the second $\text{v}[\textup{\AA}^3]$ and third $\text{w}[\textup{\AA}^6]$ virial coefficients corresponding to real systems: polystyrene in toluene and polyethylene glycol in water. In real experiments it is more comfortable to work with molar concentration, $c_m[g/L]$ and measured virial coefficients are 
$A_2[\text{mol}\,\text{cm}^3\text{g}^{-2}]$, $A_3[\text{mol}\,\text{cm}^6\text{g}^{-3}]$. In this case the virial expansion for the osmotic pressure in terms of molar mass concentration is usually written as \cite{polymer_handbook} 
$$
\Pi = RTc_m\left(\frac{1}{M_n} + A_2c_m + A_3c_m^2 + \ldots\right) = \frac{k_BTN_Ac_m}{M_0} \left(\frac{1}{N} + A_2M_0c_m + A_3M_0c_m^2 + \ldots\right)
$$
where we used that $R=k_B N_A$ and $M_{\text{w}}=M_0N$ with $N$ as a polymerization index. Comparing the corresponding terms in the above equation with Eq.(\ref{intr_flory_huggins_presure}), we obtained relations between the quantities:
\begin{equation}
\label{intr_polymer_real_virial_relationship}
	    c_m = \frac{M_0}{N_A}c_b, \quad \text{v} = 2\frac{M_0^2}{N_A}A_2, \quad \text{w} = 3\frac{M_0^3}{N_A^2}A_3
\end{equation}	    
Another quantity that we need to know is polymer statistical segment, $a_s[\textup{\AA}]$. We can find it from the tabulated data of the gyration radius. In good solvent regime the empirical equation relates the gyration radius with the polymer statistical segment is \cite{stat_segment_1976}
\begin{equation}
\label{intr_polymer_real_stat_segment_rg}
	    R_g^2 = \frac{a_s^2N}{6}\psi(z)\theta(z)
\end{equation}	
with extrapolated functions
$$
	    \psi^5(z) = 1 + \frac{20}{3}z + 4\pi z^2, \quad \theta(z) = 0.933 + 0.067\exp\left(-(0.85 z + 1.39 z^2)\right)
$$
and the parameter $z$ defined as
$$
	    z = \left(\frac{3}{2\pi a_s^2}\right)^{3/2}\text{v}N^{1/2}
$$
Here, as before, $\text{v}$ is a second virial coefficient and $N$ is a polymerization index. 
Solving the nonlinear equation Eq.(\ref{intr_polymer_real_stat_segment_rg}) with respect to $a_s$, we will find the required value.
	
\textbf{Polystyrene(PS) in toluene}. Molar mass of one monomer is $M_0$(styrene)=104.15 g/mol. Theta temperature is $\theta \simeq 154 ^\circ$C. \\
1) \cite{handbook_vii_993} T=20 ($^\circ$C). Short polystyrene chains have been measured. The second virial coefficient is measured by small angle neutron scattering (SANS) and the radius of gyration by static light scattering (SCS). The resulting data are presented in Tab.\ref{tabular:osm_pres_polysterene}.

2) \cite{handbook_vii_814} T=25 ($^\circ$C). Long chains have been measured by low angle light scattering. The authors found extrapolated formulas for their results working for $3\times 10^4 < M_w < 24\times 10^6$
\begin{equation}
\label{intr_polymer_osm_pres_ps_a2a3rg}
\begin{array}{l}
	    A_2/(\text{mol}\,\text{cm}^3\text{g}^{-2}) = 5.24 \times 10^{-3}M_{\text{w}}^{-0.21}, \\
	    A_3/(\text{mol}\,\text{cm}^6\text{g}^{-3}) = 9.12 \times 10^{-6}M_{\text{w}}^{-0.58}, \\
	    R_g/(\text{cm}) = 1.1\times 10^{-9} M_{\text{w}}^{0.603}
\end{array}
\end{equation}
They used the Flory empirical relation $A_3=1/3A_2^2M_{\text{w}}$ ($\text{w} = \text{v}^2N/4$) for finding the third virial coefficient from the second one and claimed that it is in a good agreement with experimental results for such conditions.
We extended their relations, Eq.(\ref{intr_polymer_osm_pres_ps_a2a3rg}), for our purposes and drop the lower limit up to $M_{\text{w}} \simeq 10^4$. We chose the same set of mass as we have for T=20 ($^\circ$C) and calculated the corresponding quantities using Eq.(\ref{intr_polymer_osm_pres_ps_a2a3rg}). The data are present in Tab.\ref{tabular:osm_pres_polysterene}.
%%%%%%%%%%%%%%%%%%%%%%%%%%%%%%%%%%%%%%%%%%%%%%%%%%%%%%%%%%%%%%%%%%%%%%%%%%%%%%%%%%%%%%%%%%%%%%%%%%%%%%%%%%%%%%%%%%%%%%%%%%%%%%%
%        therm pot A = 200, A = 500
%%%%%%%%%%%%%%%%%%%%%%%%%%%%%%%%%%%%%%%%%%%%%%%%%%%%%%%%%%%%%%%%%%%%%%%%%%%%%%%%%%%%%%%%%%%%%%%%%%%%%%%%%%%%%%%%%%%%%%%%%%%%%%%
\begin{figure}[ht!]
\begin{minipage}[ht]{0.5\linewidth}
\center{\includegraphics[width=1\linewidth]{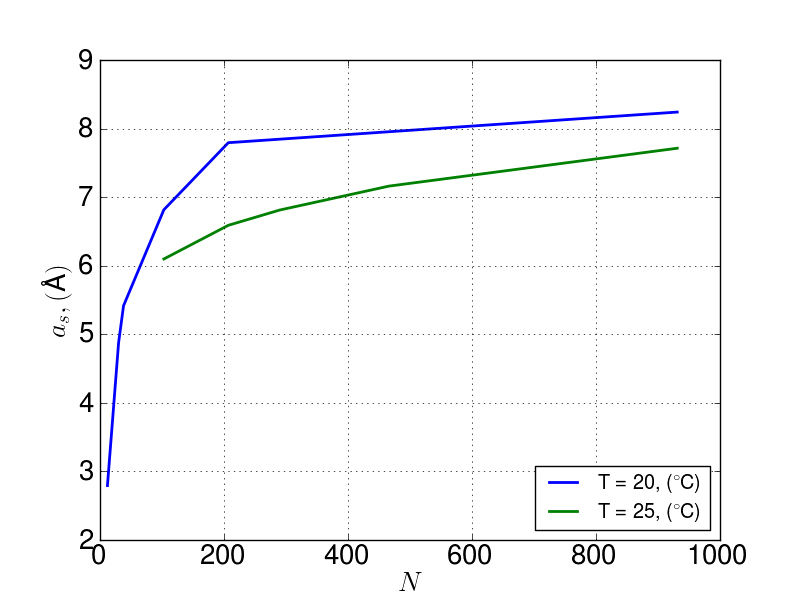}}
\caption{\small{The dependence of polymer statistical segment on the polymerization index for polystyrene in toluene solvent. 
		The data correspond to Tab.\ref{tabular:osm_pres_polysterene}.}}
\label{osmotic_pressure_segment_polysterene_fig}
\end{minipage}
\hfill
\begin{minipage}[ht]{0.5\linewidth}
\center{\includegraphics[width=1\linewidth]{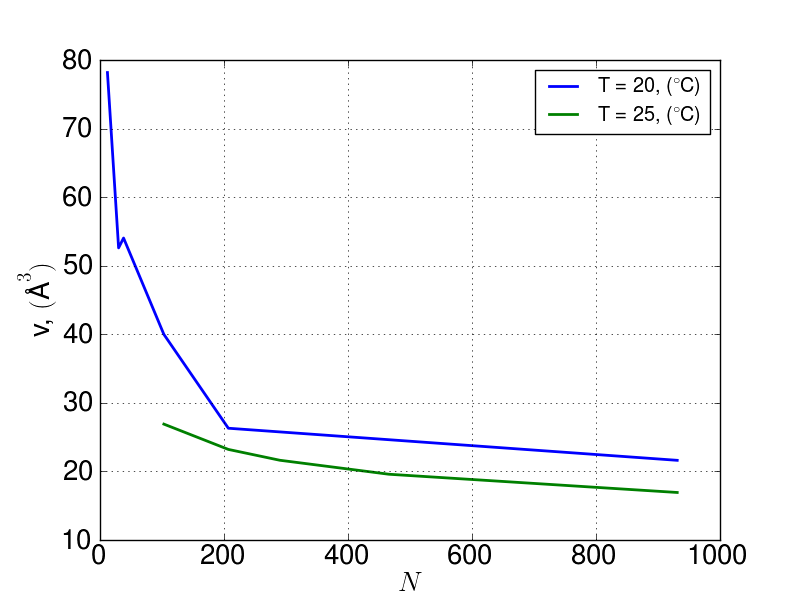}}
\caption{\small{The dependence of the second virial coefficient on the polymerization index for polystyrene in toluene.
		The data correspond to Tab.\ref{tabular:osm_pres_polysterene}}}
\label{osmotic_pressure_sec_virial_polysterene_fig}
\end{minipage}
\end{figure}
	
%%%%%%%%%%%%%%%%%%%%%%%%%%%%%%%%%%%%%%%%%%%%%%%%%%%%%%%%%%%%%%%%%%%%%%%%%%%%%%%%%%%%%%%%%%%%%%%%%%%%%%%%%%%%%%%%%%%%%%%%%%%%%%%
%        Table: comparisons
%%%%%%%%%%%%%%%%%%%%%%%%%%%%%%%%%%%%%%%%%%%%%%%%%%%%%%%%%%%%%%%%%%%%%%%%%%%%%%%%%%%%%%%%%%%%%%%%%%%%%%%%%%%%%%%%%%%%%%%%%%%%%%%	   
\begin{table}[ht!]
\caption{Polystyrene in toluene, experimental results.} 
\label{tabular:osm_pres_polysterene} 
\begin{center}
  \begin{tabular}{ | c | c | c | c | c | c | c | c | }
    \hline
                            \multicolumn{7}{|c|}{\cite{handbook_vii_993}, T=20 ($^\circ$C)} \\
    \hline
    $M_{\text{w}}$, g/mol   &   N   &   $R_g$, (nm)&  $a_s, (\textup{\AA})$& $A_2\times10^{-4}$, (cm$^3$mol/g$^2$)& $\text{v}, (\textup{\AA}^3)$ & $\text{w}\times 10^4, (\textup{\AA}^6)$ \\ 
    \hline
    1200                    &   12  &          0.66&                   2.79&                                  21.7&                         78.17& -\\
    \hline
    3100                    &   30  &          1.39&                   4.88&                                  14.6&                          52.6& -\\
    \hline
    4000                    &   38  &           1.7&                   5.41&                                  15.0&                         54.04& -\\ 
    \hline
    10700                   &  103  &          3.33&                   6.82&                                  11.1&                         39.99& -\\    
    \hline
    21600                   &  207  &          5.16&                   7.79&                                   7.3&                          26.3& -\\
    \hline
    30200                   &  290  &             -&                      -&                                   6.2&                         22.34& -\\ 
    \hline
    48500                   &  466  &             -&                      -&                                   5.8&                         20.89& -\\
    \hline
    97000                   &  931  &            12&                   8.24&                                   6.0&                         21.62& -\\
    \hline
  \end{tabular}
\end{center}
\begin{center}
  \begin{tabular}{ | c | c | c | c | c | c | c | c | }
    \hline
                            \multicolumn{7}{|c|}{\cite{handbook_vii_814}, Eq.(\ref{intr_polymer_osm_pres_ps_a2a3rg}), T=25 ($^\circ$C)} \\
    \hline
    $M_{\text{w}}$, g/mol   &   N   &   $R_g$, (nm)&  $a_s, (\textup{\AA})$& $A_2\times10^{-4}$, (cm$^3$mol/g$^2$)& $\text{v}, (\textup{\AA}^3)$ & $\text{w}\times 10^4, (\textup{\AA}^6)$ \\ 
    \hline
    10700                   &  103  &          2.96&                    6.1&                                  7.47&                          26.9& 1.86\\    
    \hline
    21600                   &  207  &          4.52&                   6.59&                                  6.44&                         23.21& 2.79\\
    \hline
    30200                   &  290  &          5.53&                   6.81&                                  6.01&                         21.63& 3.39\\ 
    \hline
    48500                   &  466  &          7.36&                   7.16&                                  5.44&                         19.59& 4.46\\
    \hline
    97000                   &  931  &         11.18&                   7.71&                                   4.7&                         16.93& 6.67\\
    \hline
  \end{tabular}
\end{center}
\end{table}
	  
	   \textbf{Polyethylene glycol (PEG) in water}. PEG is also known as polyethylene oxide (PEO) (for high molecular weight polymer) and generally it is called 
	   polyoxyethylene (POE). Molar mass of one monomer is $M_0$ (ethylene oxide)=$44.05 g/mol$. Theta temperature is $\theta \simeq 102 ^\circ$C. 
	   The tabulated data for the second virial coefficient are listed in \cite{polymer_handbook}. The data disagree with the original sources 
	   given in \cite{handbook_vii_965, handbook_vii_975}. Here, as before, we consider two cases with $T = 20$ and $T = 25$. \\
	   1) \cite{handbook_vii_965} T=20.15 ($^\circ$C). Light scattering in binary aqueous PEG solution. The authors found extrapolated formulas for the second and third virial
	   coefficients and measured the second virial coefficient and the data are present in Tab.\ref{tabular:osm_pres_polysterene}
\begin{equation}
\label{osm_pres_polyethylene_a2a3}
\begin{array}{l}
	   A_2/(\text{mol}\,\text{cm}^3\text{g}^{-2}) = a_2 + b_2M_{\text{w}}^{-c_2}, \\
	   A_3/(\text{mol}\,\text{cm}^6\text{g}^{-3}) = a_3 + b_3M_{\text{w}}^{-c_3}
\end{array}
\end{equation}
	   The coefficients $a_i, b_i, c_i$ depend on the temperature. For T=20.15 ($^\circ$C) the numerical values of the constants are present 
	   in Tab.\ref{tabular:osm_pres_polyethylene_analyt_param}. \\
%%%%%%%%%%%%%%%%%%%%%%%%%%%%%%%%%%%%%%%%%%%%%%%%%%%%%%%%%%%%%%%%%%%%%%%%%%%%%%%%%%%%%%%%%%%%%%%%%%%%%%%%%%%%%%%%%%%%%%%%%%%%%%%%%%%%%%%%%%%%
%  PEO in water table
%%%%%%%%%%%%%%%%%%%%%%%%%%%%%%%%%%%%%%%%%%%%%%%%%%%%%%%%%%%%%%%%%%%%%%%%%%%%%%%%%%%%%%%%%%%%%%%%%%%%%%%%%%%%%%%%%%%%%%%%%%%%%%%%%%%%%%%%%%%%
\begin{table}[ht!]
\caption{The parameters to Eq.(\ref{osm_pres_polyethylene_a2a3}) taken from \cite{handbook_vii_965} for T=20.15 ($^\circ$C).} 
\label{tabular:osm_pres_polyethylene_analyt_param} 
\begin{center}
  \begin{tabular}{ | c | c | c | c | c | c | }
    \hline
           $a_2$                &   $b_2$&        $c_2$&                 $a_3$&                 $b_3$&  $c_3$  \\ \hline
           $2.174\times 10^{-3}$&   0.374&        0.819&  4,09$\times 10^{-3}$&  3,26$\times 10^{-2}$&  0.134   \\ 
    \hline
  \end{tabular}
\end{center} 
\end{table}	 	
	 2) \cite{handbook_vii_975} T=25.2 ($^\circ$C). The virial coefficients were determined by simultaneously measured light scattering data (dilute solution) 
	 and isopiestic data (concentrated solution). The results for the second and third virial coefficient are given in Tab.\ref{tabular:osm_pres_polyethylene}.	 
%%%%%%%%%%%%%%%%%%%%%%%%%%%%%%%%%%%%%%%%%%%%%%%%%%%%%%%%%%%%%%%%%%%%%%%%%%%%%%%%%%%%%%%%%%%%%%%%%%%%%%%%%%%%%%%%%%%%%%%%%%%%%%%
%        Table: comparisons
%%%%%%%%%%%%%%%%%%%%%%%%%%%%%%%%%%%%%%%%%%%%%%%%%%%%%%%%%%%%%%%%%%%%%%%%%%%%%%%%%%%%%%%%%%%%%%%%%%%%%%%%%%%%%%%%%%%%%%%%%%%%%%%	   
\begin{table}[ht!]
\caption{Polyethylene glycol in water, experimental results.} 
\label{tabular:osm_pres_polyethylene} 
\begin{center}
  \begin{tabular}{ | c | c | c | c | c | c | c | }
    \hline
                            \multicolumn{6}{|c|}{\cite{handbook_vii_965}, T=20.15 ($^\circ$C)} \\
    \hline
    $M_{\text{w}}$, g/mol   &   N   & $A_2\times10^{-3}$, (cm$^3$mol/g$^2$)&  $A_3\times10^{-2}$, (cm$^6$mol/g$^3$)& $\text{v}, (\textup{\AA}^3)$ & $\text{w}\times 10^4, (\textup{\AA}^6)$ \\ 
    \hline
			     \multicolumn{6}{|c|}{Experimental} \\ 
    \hline
    1590                    &   27  &                                 2.963&                                     -&                         19.095& -\\
    \hline
    2880                    &   50  &                                 2.611&                                     -&                         16.826& -\\
    \hline
			     \multicolumn{6}{|c|}{Extrapolation, Eq.(\ref{osm_pres_polyethylene_a2a3}),} \\ 
    \hline
    1590                    &     27&                                 3.067&                                  1.623&                       19.766& 1.148\\ 
    \hline
    2880                    &     50&                                 2.723&                                  1.530&                       17.549& 1.082\\    
    \hline
    6902                    &    115&                                 2.442&                                  1.406&                        15.74& 0.994\\
    \hline
    20729                   &    345&                                 2.283&                                  1.269&                       14.713& 0.898\\ 
    \hline
    32483                   &    541&                                 2.249&                                  1.219&                       14.496& 0.862\\    
    \hline
  \end{tabular}
\end{center}
\begin{center}
  \begin{tabular}{ | c | c | c | c | c | c | c | }
    \hline
                            \multicolumn{6}{|c|}{\cite{handbook_vii_975}, T=25.2 ($^\circ$C)} \\
    \hline
    $M_{\text{w}}$, g/mol   &   N   & $A_2\times10^{-3}$, (cm$^3$mol/g$^2$)&  $A_3\times10^{-2}$, (cm$^6$mol/g$^3$)& $\text{v}, (\textup{\AA}^3)$ & $\text{w}\times 10^4, (\textup{\AA}^6)$ \\ 
    \hline
    6902                    &    115&                                   1.9&                                   2.18&                        12.244& 1.541\\
    \hline
    20729                   &    345&                                  1.67&                                   1.96&                        10.762& 1.386\\
    \hline
    32483                   &    541&                                   1.8&                                   2.24&                          11.6& 1.584\\     
    \hline
  \end{tabular}
\end{center}
\end{table}
	  
	  For our further purposes it is more convenient to assume that the virial coefficients are constant which do not depend on the molar mass of a polymer.
	  Thereby, we must fix them by choosing one from the tables.	  	  
	  For polystyrene in toluene we set $a_s = 7.6\textup{\AA}$, $\text{v} = 23\textup{\AA}^3$, $\text{w} = 2.79\times 10^4 \textup{\AA}^6$ 
	  and for polyethylene glycol in water we set\footnote{The value for polymer statistical segment, $a_s$ is taken from \cite{Vincent_peo_water}.} 
	  $a_s = 5.5\textup{\AA}$, $\text{v} = 12\textup{\AA}^3$, $\text{w} = 1.54\times 10^4 \textup{\AA}^6$.	  
%%%%%%%%%%%%%%%%%%%%%%%%%%%%%%%%%%%%%%%%%%%%%%%%%%%%%%%%%%%%%%%%%%%%%%%%%%%%%%%%%%%%%%%%%%%%%%%%%%%%%%%%%%%%%%%%%%%%%%%%%%%%%%%%%%%%%%%%%%%%%%%%%%%%%%%%%%%%%%%%%%%%%%%%%%%%%%%%%%%
%         SCFT for homopolymer solution.
%%%%%%%%%%%%%%%%%%%%%%%%%%%%%%%%%%%%%%%%%%%%%%%%%%%%%%%%%%%%%%%%%%%%%%%%%%%%%%%%%%%%%%%%%%%%%%%%%%%%%%%%%%%%%%%%%%%%%%%%%%%%%%%%%%%%%%%%%%%%%%%%%%%%%%%%%%%%%%%%%%%%%%%%%%%%%%%%%%%
\section{SCFT for homopolymer solution}
\label{sec:polymers_scft}
        Self-consistent field theory (SCFT) predicts the equilibrium structures formed in polymer solution and polymer melt. This is a mean-field theory, 
	neglecting concentration fluctuations. The description of the polymer is the coarse-grained Gaussian model for polymer chains and it treats their 
	interaction by mean-field theory (single chain subjected in external field).

	We will consider the following system of $n$ homopolymers, each of polymerization, $N$. As before, every polymer is 
	parametrized with a variable $s$ that increases continuously along its length. At the first-monomer end, $s = 0$,  and at the 
	other end, $s = N$. Using this parametrization, define functions, $\boldsymbol{r}_{i}(s)$, that specify the space curve occupied by 
	the $i$-th homopolymer. 

	In the grand canonical ensemble in which the number of chains is not fixed \cite{Landau_book_5} the configuration part of the partition function \cite{janert, muller}
	of our polymer solution  is 
\begin{equation}
\label{intr_polymer_grand_part_func_homo}
        \mathcal{Z} \propto \sum\limits_{n}^{\infty}\frac{z^{n}}{n!}\int\prod\limits_{i=1}^{n}\tilde{D}\boldsymbol{r}_{i}
        \exp\left\{-\int \mathrm{d}\boldsymbol{r}\,\left(\frac{\text{\text{v}}}{2}\hat{\rho}^2(\boldsymbol{r}) + 
	\frac{\text{w}}{6}\hat{\rho}^3(\boldsymbol{r}) + U(\boldsymbol{r})\hat{\rho}(\boldsymbol{r})\right)\right\}
\end{equation}
	where the external field $U(\boldsymbol{r})$ is considered along with self-constant field which was written with the help of interaction part	
	of free-energy density $f_{int}$ defined in Eq.(\ref{intr_polymer_flory_huggins_en_virial}). The tilde above the differential means that it involvs 
	the Wiener measure Eq.(\ref{intr_polymer_wiener_measure}).
	In the last expression $\text{v, w}$ are the second and third virial coefficients respectively.
        $z = \exp(N\mu_b)$ is the activity of a chain and $\mu_b$ is the chemical potential of a monomer unit in the bulk.
	
	Information about the configuration of the chain is contained in the local concentration
\begin{equation}
\label{intr_polymer_density_conf}
          \hat{\rho}(\boldsymbol{r}) = \sum\limits_{i=1}^{n}\int\limits_0^{N}\mathrm{d}s\delta(\boldsymbol{r}-\boldsymbol{r}_{i}(s)) 
\end{equation}
	  To make the expression Eq.(\ref{intr_polymer_grand_part_func_homo}) for the partition function more tractable, 
	  one inserts the following functional integrals
$$
          \int \mathcal{D} \rho\delta(\rho - \hat{\rho}) = 1
$$
	  and
$$
          \delta\left(\rho - \hat{\rho}\right) = \int\mathcal{D}W\, \exp\left\{i\int\mathrm{d}\boldsymbol{r}\, W\left(\rho-\hat{\rho}\right)\right\} 
					        \propto \int\mathcal{D}W\, \exp\left\{\int\mathrm{d}\boldsymbol{r}\, W\left(\rho-\hat{\rho}\right)\right\}    		  
$$  
          where we changed $iW\rightarrow W$ and the integration by the variable takes place in a complex region.
	  Those expressions permit us to replace the density $\hat{\rho}$, which depends on the polymer configuration, by the function $\rho$, which is independent of it. 
	  After that, our partition function converts into: 
\begin{equation}
\label{intr_polymer_partition_func_homo_1}
\begin{array}{l}
          \mathcal{Z} \propto \sum\limits_{n}^{\infty}\frac{z^{n}}{n!}\int\mathcal{D}\rho\mathcal{D}W\prod\limits_{i=1}^{n}\tilde{D}\boldsymbol{r}_{i}\times\\
          \times\exp\left\{-\int \mathrm{d}\boldsymbol{r}\,[\frac{\text{v}}{2}\rho^2(\boldsymbol{r}) + 
          \frac{\text{w}}{6}\rho^3(\boldsymbol{r}) + U(\boldsymbol{r})\rho(\boldsymbol{r})- W\rho] - \int\mathrm{d}\boldsymbol{r}\, W\hat{\rho}\right\}
\end{array}
\end{equation}
	  In order to further simplify it, consider separately the integral
$$
          \int\prod\limits_{i=1}^{n}\tilde{D}\boldsymbol{r}_{i} \exp\left\{-\int\mathrm{d}\boldsymbol{r}\, W\hat{\rho}\right\}
$$
          insert in the last expression the obvious expression for the density Eq.(\ref{intr_polymer_density_conf})
$$
\begin{array}{c}
          \int\prod\limits_{i=1}^{n}\tilde{D}\boldsymbol{r}_{i}\exp\left\{-\int\mathrm{d}\boldsymbol{r}\, W\hat{\rho}\right\} =
          \int\prod\limits_{i=1}^{n}\tilde{D}\boldsymbol{r}_{i}\exp\left\{-\int\mathrm{d}\boldsymbol{r}\, W\sum\limits_{i=1}^{n}
          \int\limits_0^{N}\mathrm{d}s\delta(\boldsymbol{r}-\boldsymbol{r}_{i}(s)) \right\} =\\\\
          \int\prod\limits_{i=1}^{n}\tilde{D}\boldsymbol{r}_{i}\exp\left\{-\sum\limits_{i=1}^n
          \int\limits_0^{N}\mathrm{d}sW(\boldsymbol{r}_{i}) \right\}  =
          \left(\int\tilde{D}\boldsymbol{r}\exp\left\{-\int\limits_0^{N}\mathrm{d}sW(\boldsymbol{r}) \right\}\right)^n = Q^n
\end{array}
$$
	  where we denoted 
\begin{equation}
\label{intr_polymer_partition_func_Q}
          Q = \int\tilde{D}\boldsymbol{r}\exp\left\{-\int\limits_0^{N}\mathrm{d}sW(\boldsymbol{r}) \right\}
\end{equation}
	  
          Then, after substitution this expression to Eq.(\ref{intr_polymer_partition_func_homo_1}) and carrying out the summations over $n$
          and using standard expansion for $\exp$ in the Taylor series, we obtained the following form of the partition function      
\begin{equation}
\label{intr_polymer_partition_func_homo_2}
          \mathcal{Z} \propto \int\mathcal{D}\rho\mathcal{D}W\,\exp\left\{-\beta \Omega[\rho, W]\right\}
\end{equation}
          where the free energy functional is
\begin{equation}
\label{intr_polymer_free_energy_homo}
          -\beta \Omega[\rho, W] = zQ[W] - \int\mathrm{d}\boldsymbol{r}\,
          \left\{\frac{\text{v}}{2}\rho^2(\boldsymbol{r}) + \frac{\text{w}}{6}\rho^3(\boldsymbol{r})+ 
          U(\boldsymbol{r})\rho(\boldsymbol{r}) - W(\boldsymbol{r})\rho(\boldsymbol{r})\right\}
\end{equation} 
	  where $Q[W]$ is the partition function of a single homopolymer in the external field $W$ \cite{Matsen_book}. 
	  This single polymer partition function can be obtained
	  from $Q = \int \mathrm{d}\boldsymbol{r}\,q_{\text{w}}(\boldsymbol{r}, N)$, where the end-segment distribution function is:
\begin{equation}
\label{intr_polymer_end_segment_distr}
          q_{\text{w}}(\boldsymbol{r}, N) = \int\tilde{\mathcal{D}}\boldsymbol{r}\,\delta(\boldsymbol{r} - \boldsymbol{r}(N))
          \exp\left\{-\int\mathrm{d}\boldsymbol{r}W\hat{\Phi}\right\}
\end{equation}
          Since the polymers are modeled as Gaussian chains, this distribution satisfies the diffusion equation Eq.(\ref{intr_polymer_edwards_eq_field})
          which in our case can be written as
$$
          \frac{\partial q}{\partial s} = a^2 \Delta q - W(\boldsymbol{r})q
$$
          where $a^2 = a_s^2/6$ and $W(\boldsymbol{r})$ is self-consistent field. The initial condition is $q(\boldsymbol{r}, 0) = 1$.	  
	  
In order to calculate the free energy Eq.(\ref{intr_polymer_free_energy_homo}) we use the self-consistent field theory ($SCFT$). This method helps us approximate the free energy functional $\Omega(\rho, W)$ by its extremum value. 
The concentration that extremizes the free energy,  we denote as $c$. It is satisfied the solution of the variational equations 
\begin{equation}
\label{intr_polymer_self_consistent_eq}
\begin{array}{ll}
          \text{a)} \quad \frac{\delta \Omega}{\delta \Phi} = 0  &\quad \Rightarrow W = \text{v}c + \frac{\text{w}}{2}c^2 + U(\boldsymbol{r})\\\\
          \text{b)} \quad \frac{\delta \Omega}{\delta W} = 0     &\quad \Rightarrow c = -z\frac{\delta Q[W]}{\delta W} 
\end{array} 
\end{equation}	  
Let us change the origin of the field $W(\boldsymbol{r})$, so that the new field in the bulk phase will be $W'(\boldsymbol{r}) = 0$. Thus
\begin{equation}
\label{intr_polymer_self_consistent_field_shift}
          W'(\boldsymbol{r}) = \text{v}(c-c_b) + \frac{\text{w}}{2}(c^2-c_b^2) + U(\boldsymbol{r}) = W - \mu_{int}(c_b)
\end{equation}
where $c_b$ is the concentration in the bulk phase. This transformation influences the partition function of a single homopolymer in external fields $W$, so that
$$
          Q[W'] = \int\tilde{D}\boldsymbol{r}\exp\left\{-\int\limits_0^{N}\mathrm{d}sW'(\boldsymbol{r}) \right\} =  \exp\left(N\mu_{int}(c_b)\right)Q[W] = \Lambda Q[W]
$$ 
          From Eq.(\ref{intr_polymer_self_consistent_eq}) and from definition of the single polymer partition functions Eq.(\ref{intr_polymer_partition_func_Q}), 
          it could be noticed \cite{Fredrickson_book, Matsen_book} that the concentration $c$ is an ensemble average of microscopic concentration $\hat{\rho}$. 
          Correspondingly, it can be written in terms of distribution function (like we did for the canonical ensemble in 
          Eq.(\ref{intr_polymer_part_fun_continuous_canonocal_conc}))
$$
	  c(\boldsymbol{r}) = z\int\limits_0^{N} \mathrm{d}s\, q(\boldsymbol{r}, s)q(\boldsymbol{r}, N - s)
$$
          where as before $z = \exp(N\mu_b)$ is the activity of the chain and $\mu$ is the chemical potential of a monomer. 
	  Using the definition for the end-segment distribution function Eq.(\ref{intr_polymer_end_segment_distr}) and changing the origin 
          of the field $W$, we can write
$$
          q_{\text{w}}(\boldsymbol{r}, s)q_{\text{w}}(\boldsymbol{r}, N - s) = 
          e^{-\mu_{int}s}q_{\text{w}'}(\boldsymbol{r}, s) e^{-\mu_{int}(N-s)}q_{\text{w}'}(\boldsymbol{r}, N - s) = 
          e^{-\mu_{int}N}q_{\text{w}'}(\boldsymbol{r}, s)q_{\text{w'}}(\boldsymbol{r}, N - s)
$$
          and expression for concentration profile becomes
\begin{equation}
\label{intr_polymer_density_homo}
          c(\boldsymbol{r}) = z_0\int\limits_0^{N} \mathrm{d}s\, q(\boldsymbol{r}, s)q(\boldsymbol{r}, N - s)
\end{equation}
	  where $z_0 = \Lambda z = e^{N\mu_b}e^{-N\mu_{int}(c_b)}$ and using Eq.(\ref{intr_polymer_flory_huggins_chem_pot_virial}) for the chemical potential in the third
	  virial approximation, we obtain $z_0 = \frac{c_b}{N}$.
	  	  
          Substituting the first condition of Eq.(\ref{intr_polymer_self_consistent_eq}) to Eq.(\ref{intr_polymer_free_energy_homo}), we get
\begin{equation}
\label{intr_polymer_free_energy_homo_SCFT}
\begin{array}{c}
          -\beta \Omega[c, W']  = z_0Q[W'] + \int\mathrm{d}\boldsymbol{r}\,\left\{\frac{\text{v}}{2}c^2(x) + \frac{\text{w}}{3}c^3(x)\right\} = \\
          = z_0Q[W'] - \mathrm{d}\boldsymbol{r}\, (\mu_{int}(c)c - f_{int}) = \frac{c_b}{N}Q[W'] + \int\mathrm{d}\boldsymbol{r}\,\Pi_{int}(c)
\end{array}
\end{equation}
where $\beta = 1/k_BT$ and $f_{int}, \mu_{int}, \Pi_{int}$ are contributions of the volume interactions to the free energy density, chemical potential and osmotic pressure 
which are defined in Eqs.(\ref{intr_polymer_flory_huggins_en_virial})--(\ref{intr_polymer_flory_huggins_pressure_virial}).           
The partition function is
\begin{equation}
\label{intr_polymer_partition_function_Q}
          Q[W'] = \int \mathrm{d}\boldsymbol{r} q(\boldsymbol{r}, N)
\end{equation}
          where $q(x, N)$ is the end-segment distribution function satisfying the Edwards equation in the self-consistent field $W'$.
%%%%%%%%%%%%%%%%%%%%%%%%%%%%%%%%%%%%%%%%%%%%%%%%%%%%%%%%%%%%%%%%%%%%%%%%%%%%%%%%%%%%%%%%%%%%%%%%%%%%%%%%%%%%%%%%%%%%%%%%%%%%%%%%%%%%%%%%%%%%%%%%%%%%%%%%%%%%%%%%%%%%%%%%%%%%%%%%%%%
%         The force between two parallel solid plates.
%%%%%%%%%%%%%%%%%%%%%%%%%%%%%%%%%%%%%%%%%%%%%%%%%%%%%%%%%%%%%%%%%%%%%%%%%%%%%%%%%%%%%%%%%%%%%%%%%%%%%%%%%%%%%%%%%%%%%%%%%%%%%%%%%%%%%%%%%%%%%%%%%%%%%%%%%%%%%%%%%%%%%%%%%%%%%%%%%%%
\section{The force between two parallel solid plates placed in polymer solution}          
\label{sec:polymers_force_2plates}
In order to calculate the force between two colloidal particles in polymer solution we consider two flat solid plates with polymer solution in between.
To calculate force per unit area between two flat plates, we employ the theorem of small increments \cite{Landau_book_5}
\begin{equation}
\label{intr_polymer_force_small_increments}
	\Pi  = - \langle\frac{\partial H}{\partial h}\rangle = - \frac{\partial \Omega}{\partial h}
\end{equation}                      
where $H$ is the Hamiltonian of the system, $\Omega$ is the grand thermodynamic potential of the system, $h$ is the separation between plates. 
In the previous section, based on the SCFT, we found the expression for the thermodynamic potential Eq.(\ref{intr_polymer_free_energy_homo_SCFT}). 	  
For the system between two identical plates, we can greatly simplify this expression and make it very convenient for the further numerical analysis.
First, note that the system consisting of two plates has a symmetry with respect to the midplane.
Correspondingly, we can consider only region $[0,h/2]$. Then it is convenient to consider all functions in the reduced variables: 
$s\leftarrow s/N, x\leftarrow x/R_g, h\leftarrow h/R_g$, where $R_g = (a_s/\sqrt{6})N^{1/2}$ is the unperturbed radius of gyration and $a_s$ is the polymer statistical segment. 	
In this case, the expression for the concentration profile Eq.(\ref{intr_polymer_density_homo}) takes the form
\begin{equation}
\label{intr_polymer_density_homo_dimless}
          c(x)/c_b = \int\limits_0^{1} \mathrm{d}s\, q(x, s)q(x, 1 - s)
\end{equation}
Correspondingly, we can rewrite the Edwards equation, Eq.(\ref{intr_polymer_edwards_eq_field}), in the self-consistent field, Eq.(\ref{intr_polymer_self_consistent_field_shift}), using the dimensionless variables 
\begin{equation}
 \label{intr_polymer_edwards_eq_field_dimless}
          \frac{\partial q}{\partial s} = \frac{\partial^2 q}{\partial x^2} - w(x)q
\end{equation}
	  where the self-consistent field is
\begin{equation}
\label{intr_polymer_self_consistent_field_dimless}
          w(x) = NW'(x) = v_N(c(x)/c_b - 1) + w_N((c(x)/c_b)^2 - 1) + u(x) 
\end{equation}
	  where $u(x) = NU(x)$ and we defined the dimensionless parameters: $v_N = \text{v}c_bN$, $w_N = \text{w}c_b^2N/2$, which we will call as 
	  the virial parameters throughout this thesis.	  	  
	  	  
Let us denote all dimensional thermodynamic functions using hats. 
Now, we can write the dimensionless form of Eq.(\ref{intr_polymer_free_energy_homo_SCFT}) for the system between plates as 
\begin{equation}
\label{intr_polymer_homo_vw_free_energy_in_reduced}
	-\Omega_{in}(c_b, h) = Q_{{in}}[w, h] + 2\int\limits_0^{h/2}\mathrm{d}x\, \left(\frac{v_N}{2}c^2 + \frac{2w_N}{3}c^3\right) = 
	 Q_{{in}}[w, h] + 2\int\limits_0^{h/2}\mathrm{d}x\, \Pi_{int}(c/c_b) 
\end{equation}
where $Q_{in}[w]$ is the partition function of single homopolymer in external fields, $w$ which can be found as
\begin{equation}
\label{intr_polymer_partition_function_Q_dimless} 
        Q_{in}[w, h] = \int\limits_0^h \mathrm{d}x\, q(x, 1) = 2\int\limits_0^{h/2} \mathrm{d}x\, q(x, 1) 
\end{equation} 
The relationships between the thermodynamic functions in real and dimensionless variables are	 
\begin{equation}
\begin{array}{l}
\label{intr_polymer_free_energy_homo_1d_link}
	\hat{\Omega}_{in}(c_b, \bar{h}) = \frac{\beta N}{c_b R_g}\Omega_{in}(c_b, h) \\
	\hat{Q}_{in}[w, \bar{h}] =  \frac{1}{R_g}Q_{in}[w, h]        \\
       	\hat{\Pi}_{int}(c) = \frac{N}{c_b}\Pi_{int}(c/c_b) 
\end{array}
\end{equation}	
where $\beta = 1/k_BT$ and the hats denote the dimensionless potentials.
In what follows we will omit hats where it does not leads to confusion\footnote{Similarly, we omit the bars for the reduced length variables: $\bar{x}=x/R_g, \bar{h}=h/R_g$.}.
	 
It is more convenient to distinguish two contribution in the polymer induced interaction between colloidal particles: the short-range depletion attraction 
created by bulk phase outside of the plates and the long-range depletion repulsion created by the polymer solution in the gap between plates. 
Based on that and using the definition given in Eq.(\ref{intr_polymer_force_small_increments}), the grand thermodynamic potential, Eq.(\ref{intr_polymer_homo_vw_free_energy_in_reduced}), can be rewritten as
$$
	 \Omega(c_b, h) = \Omega_{in}(c_b, h) + \Pi_{b}h
$$  
where $\Pi_{b}$ is the bulk osmotic pressure. Due to Eq.(\ref{intr_polymer_flory_huggins_pressure_virial}) for the bulk osmotic pressure we can write
$$
	 \Pi_b = \frac{c_b}{N} + \frac{\text{v}}{2}c_b^2 + \frac{\text{w}}{3}c_b^3 = \frac{c_b}{N}\left(1 + \frac{v_N}{2} + \frac{2w_N}{3}\right) = 
	         \frac{c_b}{N}\hat{\Pi}_{b}
$$  
Finally, the full thermodynamic potential in non-dimensional variables takes the following form
\begin{equation}
\label{intr_polymer_free_energy_homo_eq_0}
	\hat{\Omega}(c_b, \bar{h}) = - \left(\hat{Q}_{in}[w, \bar{h}] - \bar{h}\right) - \int\limits_0^{\bar{h}}\mathrm{d}\bar{x}\, \left(\frac{v_N}{2}\left(c/c_b)^2-1\right) +
       	 \frac{2w_N}{3}\left((c/c_b)^3-1\right)\right) 
\end{equation}	

For numerical analysis it is convenient to use the thermodynamic potential which at $h\rightarrow\infty$ is equal to 0, i.e
\begin{equation}
\label{intr_polymer_free_energy_homo_eq_inf}
	\Delta\Omega(c_b, h) = \Omega(c_b, h) - \Omega(c_b, \infty)
\end{equation}
	where $\Omega(c_b, \infty)$ corresponds to the double thermodynamic potential created by only one plate. 
	This value is usually found numerically.
%%%%%%%%%%%%%%%%%%%%%%%%%%%%%%%%%%%%%%%%%%%%%%%%%%%%%%%%%%%%%%%%%%%%%%%%%%%%%%%%%%%%%%%%%%%%%%%%%%%%%%%%%%%%%%%%%%%%%%%%%%%%%%%%%%%%%%%%%%%%%%%%%%%%%%%%%%%%%
%           GSD approximation
%%%%%%%%%%%%%%%%%%%%%%%%%%%%%%%%%%%%%%%%%%%%%%%%%%%%%%%%%%%%%%%%%%%%%%%%%%%%%%%%%%%%%%%%%%%%%%%%%%%%%%%%%%%%%%%%%%%%%%%%%%%%%%%%%%%%%%%%%%%%%%%%%%%%%%%%%%%%%	    
\section{GSD approximation}	  
	    One of the most important way to analyze the Edwards equations for the continuous chain models is to utilize eigenfunction expansions. 
	    Such expansions are familiar methods of analysis in quantum mechanics \cite{Landau_book_3}. To illustrate
	    this method, it is helpful to rewrite the diffusion equation in operator form 
\begin{equation}
\label{intr_polymer_edwards_liouville} 
	    \frac{\partial q(x, s)}{\partial s} = \hat{\mathcal{L}}q(x, s)
\end{equation}
	    where $\hat{\mathcal{L}}$ is a linear operator
$$
	    \hat{\mathcal{L}} = \frac{a_s^2}{6}\frac{\partial^2 }{\partial x^2} - W(x)
$$
	    For real $W(x)$ and suitable boundary conditions, the operator $\hat{\mathcal{L}}$ is Sturm-Liouville operator that has real eigenvalues $\Lambda_k$
	    and eigenfunctions $\psi_k(x)$. These eigenfunctions are orthogonal and complete. In addition, they satisfy to the following equation
\begin{equation}
\label{intr_polymer_storm_liouville}
	    \hat{\mathcal{L}}\psi_k(x) = - \Lambda_k\psi_k(x), \quad k=0, 1,2 \ldots
\end{equation}
	    where the index $k$ is chosen so that the eigenvalues are appropriately ordered to be less and less important, so $\Lambda_0$ is the smallest one. 
	    The solution of Eq.(\ref{intr_polymer_edwards_liouville}) can be represented as an expansion in the eigenfunctions 
	    of $\hat{\mathcal{L}}$ operator according to
\begin{equation}
\label{intr_polymer_edwards_eigenfunctions} 
	    q(x, s) = \sum\limits_{k=0}^{\infty}q_k\psi_k(x)\exp(-s\Lambda_k) 
\end{equation}
	    where the expansion coefficients are
$$
	    q_k = \int\mathrm{d}x\psi_k(x)q(x, 0) = \int\mathrm{d}x\psi_k(x)
$$
	    in the last expression we used that the initial condition is $q(x, 0) = 1$. 
	    
	    In practice, it is impossible to calculate the infinite sum in Eq.(\ref{intr_polymer_edwards_eigenfunctions}) in general case. 
	    An important approximation known as the \textit{ground state dominance} (GSD) approximation \cite{deGennes_book, Fredrickson_book}
	    is applied when an eigenfunction expansion can be truncated after the leading term $k=0$. Correspondingly
\begin{equation}
\label{intr_polymer_edwards_eigenfunctions_gsd} 
	    q(x, s) \simeq q_0\psi_0(x)\exp(-s\Lambda_0) 
\end{equation}	    
	    The partition function of the entire chain is
\begin{equation}
\label{intr_polymer_part_fun_gsd} 
	    Q(x) = \int\mathrm{d}x\,q(x, N) \simeq q_0^2\exp(-N\Lambda_0) 
\end{equation}	    
	    and the segment concentration is
\begin{equation}
\label{intr_polymer_conc_gsd} 
	    c(x) = \int\limits_0^N\mathrm{d}s\,q(x, s)q(x, N - s) \simeq q_0^2\exp(-N\Lambda_0) \psi_0^2 = \mathcal{N}\psi_0^2
\end{equation}	    
	    where the last equality is written taking into account the proper normalization, $\int\mathrm{d}x\,c(x) = \mathcal{N}$, 
	    where $\mathcal{N}$ is the total number of monomers in the system.
	    
	    It is obvious that the GSD approximation is the leading term in the asymptotic expansion for $N\gg1$ \cite{Grosberg_1994}. 
	    Two conditions are required to satisfy the GSD approximation: \\
	    1) The eigenvalues must be discrete. This implies that the polymer is localized in the final region of space.\\
	    2) The difference between the second and the first eigenvalues and chain length should be large enough. It 
	    allows us to neglect the rest terms in comparison to the first i.e
$$
	    \exp(-N(\Lambda_1 - \Lambda_0)) \ll 1
$$
	    The conditions are met in the situations where polymers are confined to a region of a characteristic size, which is smaller 
	    than the size of an unperturbed chain \cite{deGennes_book, Grosberg_1994}.
	    
	    From the computational standpoint, the GSD approximation greatly simplifies the calculations of the partition function and the segment concentration.
	    Instead of solving the Edwards partial differential equation in the GSD approximation, we eliminate the $s-$dependence and obtain an 
	    ordinary differential equation, which depends only on spatial variables, namely, Eq.(\ref{intr_polymer_storm_liouville}) for the ground state is
\begin{equation}
\label{intr_polymer_edwards_gsd} 
	    \frac{a_s^2}{6}\frac{\partial^2\psi(x)}{\partial x^2} - W(x)\psi(x) + \Lambda\psi(x) = 0
\end{equation}	    
	    where we denoted $\psi(x)\equiv \psi_0(x)$ and $\Lambda\equiv\Lambda_0$. The last equation has a variational basis because the functional
\begin{equation}
\label{intr_polymer_edwards_gsd_functional_psi}
	    F[\psi] = \mathcal{N}\int\mathrm{d}x\left\{\frac{a_s^2}{6}|\nabla\psi|^2 + W\psi^2\right\}
\end{equation}
	    has Eq.(\ref{intr_polymer_edwards_gsd}) as its Euler-Lagrange equation ( with the normalization condition $\int\mathrm{d}x \psi^2 = 1$). Moreover, we can rewrite the last functional
	    for the unit concentration $c(x)$ using Eq.(\ref{intr_polymer_conc_gsd})
\begin{equation}
\label{intr_polymer_edwards_gsd_functional_c}
	    F[c] = \int\mathrm{d}x\left\{\frac{a_s^2}{24c}|\nabla c|^2 + Wc\right\}
\end{equation}
The first term on the right hand side of Eq.(\ref{intr_polymer_edwards_gsd_functional_c}), taken with the opposite sign, is usually referred to as $\textit{Lifshitz entropy}$ \cite{Lifshitz_1978}.
	    It represents the appropriate expression for the conformational entropy of the continuous Gaussian chain with the prescribed segment concentration distribution.
%%%%%%%%%%%%%%%%%%%%%%%%%%%%%%%%%%%%%%%%%%%%%%%%%%%%%%%%%%%%%%%%%%%%%%%%%%%%%%%%%%%%%%%%%%%%%%%%%%%%%%%%%%%%%%%%%%%%%%%%%%%%%%%%%%%%%%%%%%%%%%%%%%%%%%%%%%%%
%           Correlation length
%%%%%%%%%%%%%%%%%%%%%%%%%%%%%%%%%%%%%%%%%%%%%%%%%%%%%%%%%%%%%%%%%%%%%%%%%%%%%%%%%%%%%%%%%%%%%%%%%%%%%%%%%%%%%%%%%%%%%%%%%%%%%%%%%%%%%%%%%%%%%%%%%%%%%%%%%%%%
\section{Correlation length}  
The correlation length $\xi$ is defined as a distance between two monomers so that at $r > \xi$ the effective interaction between the
monomers is expected to be very weak, and as a result concentration perturbations exponentially decrease with the characteristic length $\xi$. Let us find the expression for the correlation length in the semidilute solution 
for the marginal solvent regime. In the bulk phase we can rewrite Eq.(\ref{intr_polymer_edwards_gsd_functional_c}) 
replacing the interaction term $Wc$ by $f_{int}(c)$ (see Eq.(\ref{intr_polymer_flory_huggins_en_virial})):	    
\begin{equation}
\label{intr_polymer_free_energy_bulk_c}
            F[c]  = \int\mathrm{d}\boldsymbol{r}\left\{\frac{a^2}{4c}\left(\nabla c\right)^2 + 
	    \frac{\text{v}c^2}{2} + \frac{\text{w}c^3}{6} + \frac{c}{N}\ln\frac{c}{Ne}\right\} 
\end{equation} 
	    where $a=a_s/\sqrt{6}$ and we included the term describing the translational entropy of the chains in the bulk phase, $f^0_b(c_b)$. 
	    By virtue of the isotropy of the system we will consider $c(r)$ depending only on the absolute distance, $r$. 
	    Consider concentration fluctuations setting $c(r) = \bar{c} + \delta c$, where $\bar{c}$ is the mean (equilibrium) concentration, which is independent 
	    of $r$ and $\delta c(r)$ is the small deviation from $\bar{c}$, i.e. $c(r) \ll \delta c$ for any $r$ (see \cite{Grosberg_1994}). 
	    Let us separately write every integrand term. For the Lifshitz entropy, we have:
$$
	    \frac{1}{\bar{c}+\delta c}\left(\frac{\mathrm{d}(\bar{c} + \delta c)}{\mathrm{d}r}\right)^2 \simeq \frac{1}{\bar{c}}\left(\frac{\mathrm{d}\delta c}{\mathrm{d}r}\right)^2
$$
	    and for interaction term
$$
	    f_{int}(\bar{c}+\delta c) = \frac{\text{v}(\bar{c}+\delta c)^2}{2} + \frac{\text{w}(\bar{c}+\delta c)^3}{6} \simeq f_{int}(\bar{c}) + \frac{\mathrm{d}f_{int}(\bar{c})}{\mathrm{d}\bar{c}}\delta c + 
	    \frac{\delta c^2}{2}(\text{v}+\text{w}\bar{c}) 
$$
	    where we neglected the term proportional to $\delta c^3$. For the translational entropy, we can write
$$
	    \frac{(\bar{c}+\delta c)}{N}\ln\frac{(\bar{c}+\delta c)}{Ne} = \frac{(\bar{c}+\delta c)}{N}\left( \ln\frac{\bar{c}}{Ne} + \ln(1+\frac{\delta c}{\bar{c}})\right) = 
	    \frac{\bar{c}}{N}\ln\frac{\bar{c}}{Ne} + \frac{\delta c}{N}\left(\ln\frac{\bar{c}}{Ne}+1\right) + \frac{\delta c^2}{2\bar{c}N}
$$
	    We took into account only two terms in the logarithm expansion. After that, we can rewrite Eq.(\ref{intr_polymer_free_energy_bulk_c}) as 
	    the free energy expansion by $c_1$:
$$
	    F[c_b +c_1] = F[c_b] + F_1[c_b, c_1] +  F_2[c_b, c_1^2]
$$
	    where $F[c_b]$ is a constant that does not play any role, $F_1[c_b, c_1]$ vanishes in equilibrium. Correspondingly,
	    we will concern only the third term
\begin{equation}
\label{intr_polymer_free_energy_F2}
	    F_2[\bar{c}, \delta c^2] = \int\mathrm{d}r\left\{\frac{a^2}{4\bar{c}}\left(\frac{\mathrm{d}\delta c}{\mathrm{d}r}\right)^2 + 
	    \frac{\delta c^2}{2\bar{c}}\left(\bar{c}(\text{v} + \text{w}\bar{c}) + \frac{1}{N}\right) \right\} 
\end{equation}
	    using the above suggestion about equilibrium, we can write 
\begin{equation}
\label{intr_polymer_variational_rprinciple_c1}
	    \frac{\delta F}{\delta (\delta c)} \simeq \frac{\delta F_2}{\delta (\delta c)} = 0
\end{equation}	    
	    Let us consider separately the first gradient term
$$
	    \delta F^1_2(\bar{c}, \delta c^2) = \int\mathrm{d}r \frac{a^2}{2\bar{c}} \frac{\mathrm{d}\delta c}{\mathrm{d}r} \delta \frac{\mathrm{d}\delta c}{\mathrm{d}r} = 
	    \frac{a^2}{2\bar{c}} \frac{\mathrm{d}\delta c}{\mathrm{d}r} \delta (\delta c)\Big\vert_S - \int\mathrm{d}r \frac{a^2}{2\bar{c}} \frac{\mathrm{d}^2\delta c}{\mathrm{d}^2r}\delta (\delta c)
$$
	    where we integrated by parts and used the fact that variations at a border are equal to zero. 
	    Thus, Eq.(\ref{intr_polymer_variational_rprinciple_c1}) takes the following form:
\begin{equation}
\label{intr_polymer_diff_eq_xi}
	    \frac{\mathrm{d}^2\delta c}{\mathrm{d}r^2} - (1/\xi^2)\delta c = 0
\end{equation}
	    where we introduced the characteristic length, $\xi = a/\sqrt{2\bar{c}(\text{v}+\text{w}\bar{c})+2/N}$. The solution of the above equation is well known
	    and can be written as
$$
	    \delta c(r) = \frac{const}{r}e^{-r/\xi}
$$
	    thereby, we showed that $\xi$ is the correlation length. For our future aims it will be useful to rewrite the correlation length in the reduced variables
\begin{equation}
\label{intr_polymer_correlation_length}
            \xi = \frac{a}{\sqrt{2\bar{c}(\text{v}+\text{w}\bar{c})+2/N}} = \frac{R_g}{\sqrt{2(v_N+2w_N)+2}} 
\end{equation}
	    where we introduce the dimensionless virial parameters as: $v_N = \text{v}\bar{c}N$ and $w_N = \text{w}\bar{c}^2N/2$. The term "2" under the square root 
	    can be omitted in the $GSD$ approximation, since $\xi$ must be $\ll R_g$. Therefore $(\text{v} + \text{w})N\gg 1$, so 
\begin{equation}
\label{intr_polymer_correlation_length_gsd}
            \xi \simeq \frac{R_g}{\sqrt{2(v_N+2w_N})} 
\end{equation}
	    Due to reasons that will become apparent in the main text of this thesis, we will consider the virial parameters 
	    from zero up to $v_N = 500, w_N = 100$, where the corresponding dimensionless correlation length is $\xi/R_g = 0.027$.

%%%%%%%%%%%%%%%%%%%%%%%%%%%%%%%%%%%%%%%%%%%%%%%%%%%%%%%%%%%%%%%%%%%%%%%%%%%%%%%%%%%%%%%%%%%%%%%%%%%%%%%%%%%%%%%%%%%%%%%%%%%%%%%%%%%%%%%%%%%%%%%%%%%%%%%%%%%%%
%           Review on RPA, scattering functions of polymer solutions. Consistency of GSD and RPA.
%%%%%%%%%%%%%%%%%%%%%%%%%%%%%%%%%%%%%%%%%%%%%%%%%%%%%%%%%%%%%%%%%%%%%%%%%%%%%%%%%%%%%%%%%%%%%%%%%%%%%%%%%%%%%%%%%%%%%%%%%%%%%%%%%%%%%%%%%%%%%%%%%%%%%%%%%%%%%	    
\section{Review on RPA, scattering functions of polymer solutions. Consistency of GSD and RPA}
	    In the previous section we showed that the free energy in the concentrated regime $\bar{c} \gg \delta c$, Eq.(\ref{intr_polymer_free_energy_F2}),  
	    can be written as
\begin{equation}
\label{intr_polymer_free_energy_conc_tot}
	    F = F_0 + \frac{k_BT}{2}\int\mathrm{d}r\left\{\frac{a^2}{2\bar{c}}\left(\frac{\mathrm{d}\delta c}{\mathrm{d}r}\right)^2 + \nu^*(\delta c)^2
	     \right\} 
\end{equation}
	    where $F_0$ is the free energy for $\delta c = 0$ and 
$$
            \nu^* = \frac{1}{k_BT}\frac{\partial^2 f_{int}}{\partial c^2}\Big\vert_{c=\bar{c}} \simeq \text{v} + \text{w}\bar{c}
$$
	   The second term in the r.h.s of Eq.(\ref{intr_polymer_free_energy_conc_tot}) is the free energy due to $\delta c$ perturbation. We can easily write the term in the 
	   discrete Fourier space:
\begin{equation}
\label{intr_polymer_free_energy_F2_fourier}
	   F_2 = \frac{k_BT}{2V}\sum\limits_{q\ne0}A(q)c_qc_{-q} 
\end{equation}
	   where
$$
	   c_q = \int\limits_{V}\mathrm{d}^3 r c(r)e^{-iqr}
$$
	   is the Fourier transformation of the concentration profile, $c(r)$. The sum is taken over the spectrum of the wave vectors 
	   determined by the system size, $V$, and 
\begin{equation}
\label{intr_polymer_free_energy_kernal_fourier}	   
	   A(q) = \nu^* + \frac{a^2q^2}{2\bar{c}}
\end{equation}
	   This equation is valid for $aq\ll 1$, because the length scale of the concentration profile $1/q$ must be much larger than $a_s = \sqrt{6}a$.  
	   Applying the equipartition theorem \cite{Landau_book_5} to Eq.(\ref{intr_polymer_free_energy_kernal_fourier}), we can write
\begin{equation}
\label{intr_polymer_free_equip_theorem}
	   \langle c_qc_{-q}\rangle = \frac{V}{A(q)}
\end{equation}
	   Moreover, using the definition of the structure factor \cite{Teraoka_book}
\begin{equation}
\label{intr_polymer_structure_factor_def}
	   S(q) = \frac{1}{V}\langle c_qc_{-q}\rangle 
\end{equation}
	   immediately leads to the Ornstein-Zernike function:
\begin{equation}
\label{intr_polymer_structure_factor}
	   S(q) = \frac{1}{A(q)} = \frac{S(0)}{1 + (q\xi)^2}
\end{equation}	   
	   where $S(0) = 1/\nu^*$ and $\xi = a/\sqrt{2\bar{c}\nu^*} = a_s/\sqrt{12\bar{c}\nu^*}$. 
	   
	   The above results are also applicable to the ideal polymer system with no interactions between monomer units. In this case $\nu^*=0$ and 
	   the kernel Eq.(\ref{intr_polymer_free_energy_kernal_fourier}) is $A_{id} = (aq)^2/2\bar{c}$. Therefore, we can write 
	   Eq.(\ref{intr_polymer_free_energy_kernal_fourier}), in the form:
\begin{equation}
\label{intr_polymer_free_energy_kernal_fourier_improved}	   
	   A(q) = \nu^* + A_{id}
\end{equation}	   
	   The structure factor of the ideal system is $S_{id}(q) = 1/A_{id}(q)$. Correspondingly, we can write the following general relation:
\begin{equation}
\label{intr_polymer_structure_factor_improved}
	   \frac{1}{S(q)} = \nu^* + \frac{1}{S_{id}(q)}
\end{equation}	   
	   Therefore, we are able to calculate the structure factor, $S(q)$, for the concentrated system based on the known 
	   structure factor of the ideal system, $S_{id}(q)$. The last expression, Eq.(\ref{intr_polymer_structure_factor_improved}), is usually associated
	   with the \textit{random phase approximation} ($RPA$), which is equivalent to the mean field approximation \cite{deGennes_book, Semenov_review_2012}.
	   
	   In order to find the applicablity conditions for the above equations let us consider the correlation function of 
	   concentration fluctuations \cite{Teraoka_book, Semenov_review_2012}:
\begin{equation}
\label{intr_polymer_correlation_function}
	   G(\boldsymbol{r} - \boldsymbol{r'}) = \frac{ \langle c(\boldsymbol{r})c(\boldsymbol{r'})\rangle}{\bar{c}} - \bar{c}
\end{equation}	   	   
	   where $\bar{c}$ is the mean monomer concentration. 
	   The correlation function is related to the structure factor:
\begin{equation}
\label{intr_polymer_correlation_structure_functions}
	   S(q) = \bar{c}\int\limits_{V} \mathrm{d}r^3 G(r) e^{iqr} 
\end{equation}	   
	   and applying the inverse Fourier transformation, once the structure factor, $S(q)$ is known, the correlation function $G(r)$ can be obtained:
\begin{equation}
\label{intr_polymer_correlation_structure_functions_inverse}
	   G(r) = \frac{1}{(2\pi)^3\bar{c}}\int\limits_{q} \mathrm{d}q^3 S(q) e^{-iqr} 
\end{equation}	   
	   Suppose an arbitrary monomer unit is located at the origin. Then, the mean excess concentration $\langle c(r)\rangle - \bar{c}$ at the distance
	   $r$ from the origin coincides with the correlation function, $G(r)$. Therefore, substituting Eq.(\ref{intr_polymer_structure_factor}) 
	   in Eq.(\ref{intr_polymer_correlation_structure_functions_inverse}) we get 
\begin{equation}
\label{intr_polymer_correlation_function_last}
	   G(r) = \frac{1}{2\pi a^2r}e^{-r/\xi}
\end{equation}	   
	   For the mean excess concentration we have already obtained the similar expression in Eq.(\ref{intr_polymer_diff_eq_xi}). 
	   Based on the last equation, we deduce that the correlations in the system are screened at distances larger than the correlation length $\xi$. The typical
	   fluctuation $\delta c(r)$ averaged over the correlation volume $\xi^3$ must be small in comparison with $\bar{c}$. Thus, $G(\xi) \ll \bar{c}$, 
	   which leads to
\begin{equation}
\label{intr_polymer_correlation_function_condition_conc_bigger}
	   \bar{c} \gg c_{min} \sim \frac{\nu^*}{a_s^6}
\end{equation}	  	   
	   Note that we have already found $c_{min}$ in the r.h.s of the last equation when we considered the upper boundary of the semi-dilute solution regime, 
	   Eq.(\ref{intr_polymer_star_star}). Another condition can be obtained if we demand $\xi \gg a_s$, which leads to 
\begin{equation}
\label{intr_polymer_correlation_function_condition_conc_smaller}
	   \bar{c} \ll \frac{1}{\nu^*}
\end{equation}	   
%%%%%%%%%%%%%%%%%%%%%%%%%%%%%%%%%%%%%%%%%%%%%%%%%%%%%%%%%%%%%%%%%%%%%%%%%%%%%%%%%%%%%%%%%%%%%%%%%%%%%%%%%%%%%%%%%%%%%%%%%%%%%%%%%%%%%%%%%%%%%%%%%%%%%%%%%%%%%
%           GSDE theory, long-range effects. RPA limit of the GSDE. Basic results in this limit.
%%%%%%%%%%%%%%%%%%%%%%%%%%%%%%%%%%%%%%%%%%%%%%%%%%%%%%%%%%%%%%%%%%%%%%%%%%%%%%%%%%%%%%%%%%%%%%%%%%%%%%%%%%%%%%%%%%%%%%%%%%%%%%%%%%%%%%%%%%%%%%%%%%%%%%%%%%%%%	   
\section{Ground state dominance theory with end segments (GSDE), long-range effects. RPA limit of the GSDE. Basic results in this limit}
A localized external field applied to a polymer solution produces a response $\delta c(x)$ which can be considered as a sum of two terms: 
$\delta c(x) = \delta c^{sh}(x) + \delta c^{lr}(x)$, where $\delta c^{sh}(x)$ is a short-range contribution extending to the distances $x\lesssim\xi$ and 
$\delta c^{lr}(x)$ is a long-term contribution extending to the distances $x\sim R_g$.

In order to obtain the long-range concentration perturbations $\delta c^{lr}$, it is necessary to use a more accurate expression for the conformational free energy, Eq.(\ref{intr_polymer_edwards_gsd_functional_c})).
Such expression was firstly derived in \cite{semenov_1996}. Throughout the text we will call this approach as ground state dominance related with end segments (\textbf{GSDE}).
For a polymer solution layer between two opposite repulsive walls separated by the distance $h$, one can write
\begin{equation}
\label{intr_polymer_gsde_conf} 
F_{conf} = F_{conf}^{GSD} + \int\mathrm{d}x\,\frac{c}{N}\ln\left(\frac{c_b}{Ne}\right) - \frac{2c_b}{N}\eta_0 + \frac{2c_b}{hN}\sum\limits_{k}\,f(R^2_gk^2)|\eta_k|^2
\end{equation}
where the wave number $k$ adopts discrete values: $k_n=2\pi n/h$, $n=0, \pm 1, \pm 3, \ldots$, 
\begin{equation}
\label{intr_polymer_gsd_conf} 
F_{conf}^{GSD} = \frac{a}{4}\int\mathrm{d}x\,\frac{(\nabla c)^2}{c}
\end{equation}
is the conformational free energy in the GSD approximation,
\begin{equation}
\label{intr_polymer_gsde_etak} 
\eta_k = \int\limits_0^{\infty}\mathrm{d}x\,\{\sqrt{c(x)/c_b} - c(x)/c_b\}\cos(kx)
\end{equation}
and
\begin{equation}
\label{intr_polymer_gsde_fy} 
f(y) \equiv \frac{1-e^{-y}(1+y)}{y-1+e^{-y}}  
\end{equation}
The last two terms in Eq.(\ref{intr_polymer_gsde_conf}) represent the finite $N$ corrections which are proportional to $1/N$ and  $1/N^{3/2}$ correspondingly.
Moreover, the last term is non-local and it produces the long-range interaction (for $h\gg\xi$).

The grand thermodynamic potential can be written as
\begin{equation}
\label{intr_polymer_gsde_total_pre}
\Omega[c] = F_{conf} + F_{int} - \mu_b\int\mathrm{d}x\, c + \Pi_b\int\mathrm{d}x
\end{equation}
where $F_{int} = (\text{v}/2)\int\mathrm{d}x\, c^2$ is the interaction part of the free energy (written for simplicity in the second virial approximation),
$\mu_b = \partial f_b/\partial c_b$ is the chemical potential per repeat unit in the bulk, $\Pi_b = c_b\mu_b - f_b$ is the bulk osmotic pressure. Using 
the expression for the bulk free energy density, Eq.(\ref{intr_polymer_flory_huggins_en_virial}), we can write Eq.(\ref{intr_polymer_gsde_total_pre}) in the form:
\begin{equation}
\label{intr_polymer_gsde_total}
\Omega[c] = \tilde{F}_{conf} + \tilde{F}_{int} 
\end{equation}
where 
\begin{equation}
\label{intr_polymer_gsde_int_final} 
\tilde{F}_{int} = \frac{\text{v}}{2}\int\mathrm{d}x\, (c-c_b)^2,
\end{equation}
\begin{equation}
\label{intr_polymer_gsde_conf_final} 
\tilde{F}_{conf} = F_{conf}^{GSD} - \frac{1}{N}\int\mathrm{d}x\,(c-c_b) - \frac{2c_b}{N}\eta_0 + \frac{2c_b}{hN}\sum\limits_{k}\,f(R^2_gk^2)|\eta_k|^2
\end{equation}
Here, 
\begin{equation}
\label{intr_polymer_gsde_conc_profile} 
c = c(x) = c^{sh}(x) + \delta c^{lr}(x) 
\end{equation}
is the total concentration profile including the long-range perturbation which is always small, i.e. $\delta c^{lr}(x) \ll c_b$.
Hence, the integrand in Eq.(\ref{intr_polymer_gsde_etak}) can be approximated as
\begin{equation}
\label{intr_polymer_gsde_etax}
\eta(x) = \eta^{sh}(x) - \frac{\delta c^{lr}(x)}{2c_b}    
\end{equation}
where $\eta^{sh}(x) = \sqrt{c^{sh}(x)/c_b} - c^{sh}(x)/c_b$. The Fourier transformation of $\eta(x)$ can be split as
\begin{equation}
\label{intr_polymer_gsde_eta_k_appr}
\eta_k = \eta^{sh}_k + \eta^{lr}_k  
\end{equation}
where 
\begin{equation}
\label{intr_polymer_gsde_eta_k_lr}
\eta^{lr}_k \simeq -\frac{1}{2c_b}\int\limits_{0}^{\infty}\mathrm{d}x\, \delta c^{lr}(x)\cos(kx)      
\end{equation}
In particular, for $k=0$ we obtain:
\begin{equation}
\label{intr_polymer_gsde_eta_0}
\eta_0 = \Delta_e - \frac{1}{2c_b}\int\limits_{0}^{\infty}\mathrm{d}x\, \delta c^{lr}(x)         
\end{equation}
where 
\begin{equation}
\label{intr_polymer_gsde_delta_e_sh}
\Delta_e = \eta^{sh}_0 = \int\limits_0^{\infty}\mathrm{d}x\,\left\{\sqrt{c^{sh}(x)/c_b} - c^{sh}(x)/c_b\right\}
\end{equation}
Here, $c^{sh}(x)$ is the one-plate concentration profile obtained with the GSD, Eq.(\ref{intr_polymer_edwards_gsd}), with the boundary condition $c^{sh}(0)=0$.

Now, we are able to find the equilibrium $c^{lr}(x)$. Minimizing Eq.(\ref{intr_polymer_gsde_total}) with respect to $c^{lr}(x)$ and using 
Eqs.(\ref{intr_polymer_gsde_conc_profile}-\ref{intr_polymer_gsde_delta_e_sh}) we obtain
\begin{equation}
\label{intr_polymer_gsde_conc_profile_lr}
\delta c^{lr}(x)/c_b \simeq \frac{4\Delta_e\xi^2}{R_g^3}\chi(x/R_g, h/R_g)
\end{equation}
where we used the definition of the correlation length $\xi = R_g/\sqrt{2\text{v}c_bN}$ and introduced the function:
\begin{equation}
\label{intr_polymer_gsde_chi}
\chi(\bar{x}, \bar{h}) = \frac{1}{\bar{h}}\sum\limits_{k}\,f(k^2)\cos(k\bar{x}), \quad k=\frac{2\pi}{\bar{h}}n,\quad n=0, \pm 1, \pm 2, \ldots
\end{equation}
In Eq.(\ref{intr_polymer_gsde_conc_profile_lr}) we neglected the $k-$ dependence of $\eta_k^{sh}\simeq\eta_0^{sh}\simeq\Delta_e$ since the relevant wave numbers are small, $k\sim 1/h \ll 1/\xi$, where 
$\xi$ is the typical localization scale corresponding to the function $\eta(x)$. For further simplification, notice that $f(k^2) \simeq 1/(k^2-1)$ at $k^2\gg 1$ and
$$
\chi(\bar{x}, \bar{h}) \simeq \frac{1}{\bar{h}}\sum_k\frac{\cos(k\bar{x})}{k^2-1} = -\frac{\cos(\bar{h}/2-\bar{x})}{2\sin(\bar{h}/2)}
$$

Using that $c^{sh}(x) = c_b\tanh^2\left(\frac{x}{2\xi}\right)$ (see \cite{deGennes_book}), the total concentration profile between two plates is given by
\begin{equation}
\label{intr_polymer_gsde_conc_profile_tot}
c(x)/c_b \simeq \tanh^2\left(\frac{x}{2\xi}\right) + \frac{4\Delta_e\xi^2}{R_g^3}\chi(x/R_g, h/R_g)% \tanh^2\left(\frac{h-x}{2\xi}\right) +
\end{equation}

In order to get the long-range energy of interaction between the two plates we notice that the first three terms in Eq.(\ref{intr_polymer_gsde_conf_final}) (together with $\tilde{F}_{int}$, Eq.(\ref{intr_polymer_gsde_int_final})) give rise to an exponentially weak dependence on $h$ for $h\gg\xi$. 
The main long-range contribution to the interaction energy is due to the last term in Eq.(\ref{intr_polymer_gsde_conf_final}). So, we obtain
\begin{equation}
\label{intr_polymer_gsde_interaction} 
W_e = \frac{4c_b}{hN}\sum\limits_{k}\,f(R^2_gk^2)|\eta_k|^2 \simeq \frac{4\Delta_e^2 c_b}{hN}\sum\limits_{n}\,f(4\pi^2 R^2_g n^2/h^2)
\end{equation}
where $n=0, \pm 1, \pm 2, \ldots$. We can rewrite the last expression in the form 
\begin{equation}
\label{intr_polymer_gsde_interaction_new} 
W_{e} = \frac{4c_b\Delta_e^2}{NR_g}\left[u_{int}(h/R_g) + \kappa\right]
\end{equation}
where 
\begin{equation}
\label{intr_polymer_gsde_interaction_reduced} 
	    u_{int}(\bar{h}) = \frac{1}{\bar{h}}\sum\limits_{n=-\infty}^{\infty}f(4\pi^2n^2/\bar{h}^2) - \kappa
\end{equation}
and $\kappa\equiv\int\limits_{-\infty}^{\infty}\frac{\mathrm{d}k}{2\pi}f(k^2) \simeq 0.61617$ (see \cite{semenov_1996, semenov_obukhov}).
In the regime $\xi\ll h\ll R_g$ all terms in the sum can be neglected apart from $n=0$:
\begin{equation}
\label{intr_polymer_gsde_interaction_one_term} 
W_{e} = \frac{4c_b}{hN}\Delta_e^2 
\end{equation}
where $\Delta_e\simeq 2\xi(1-\ln 2)$ (see Sec.\ref{sec:repulsion_generalization_GSD}). This expression will be considered in Chs.\ref{chap:Chapter3}, \ref{chap:Chapter4}
along with the expression for the GSD free energy, $W_{gs}$, in order to compare the results of numerical calculation with the analytical results.

%%%%%%%%%%%%%%%%%%%%%%%%%%%%%%%%%%%%%%%%%%%%%%%%%%%%%%%%%%%%%%%%%%%%%%%%%%%%%%%%%%%%%%%%%%%%%%%%%%%%%%%%%%%%%%%%%%%%%%%%%%%%%%%%%%%%%%%%%%%%%%%%%%%%%%%%%%%%%
%           GSDE theory, long-range effects. RPA limit of the GSDE. Basic results in this limit.
%%%%%%%%%%%%%%%%%%%%%%%%%%%%%%%%%%%%%%%%%%%%%%%%%%%%%%%%%%%%%%%%%%%%%%%%%%%%%%%%%%%%%%%%%%%%%%%%%%%%%%%%%%%%%%%%%%%%%%%%%%%%%%%%%%%%%%%%%%%%%%%%%%%%%%%%%%%%%	   
\section{Polymer adsorption}
\label{sec:polymers_adsorption}
\begin{figure}[ht!]
\center{\includegraphics[width=0.6\linewidth]{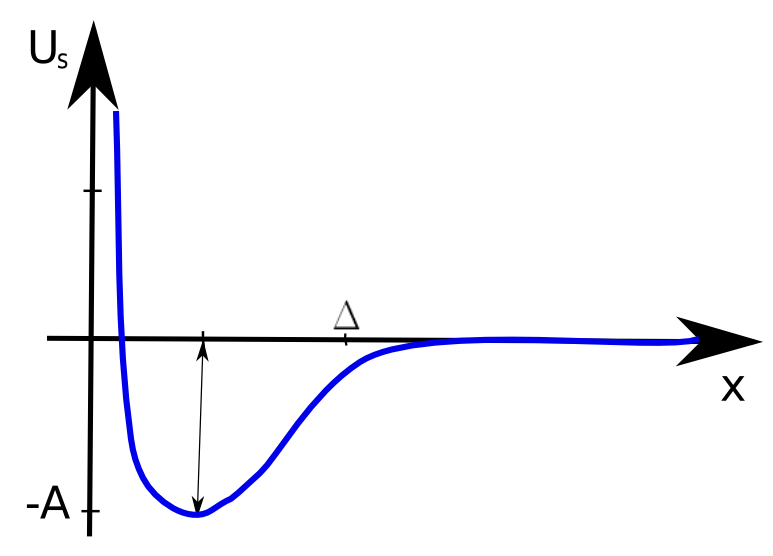}}
\caption{\small{The Potential well near an impenetrable adsorbing wall.}}
\label{surface_potential_fig}
\end{figure}

We now consider a polymer solution in contact with a solid surface plate, located at $x=0$. The solid surface is impenetrable to the polymer, so the concentration is defined only for $x>0$, and $c(0)=0$. Furthermore, a solid surface is usually characterized by a short-range potential $u_s$ (the monomer-surface interaction energy).
This potential can be included in Eq.(\ref{intr_polymer_edwards_gsd}) and for the ideal polymer solution we can write:
\begin{equation}
\label{intr_polymer_gsd_adsorption}
  \frac{a_s^2}{6}\frac{\partial^2 \psi}{\partial x^2} - U_s(x)\psi(x) + \Lambda\psi(x) = 0  
\end{equation}
with the boundary condition $\psi(0)=0$. The typical surface potential $U_s(x)=A\phi(x/\Delta)$ is shown in Fig.\ref{surface_potential_fig} ($A$ is the depth of the potential well). The potential is short-range: $\Delta\ll R_g$.
Integrating Eq.(\ref{intr_polymer_gsd_adsorption}) in $[0..\Delta]$ (in this range $\Lambda$ can be neglected), we obtain:
\begin{equation}
\label{intr_polymer_gsd_adsorption_bc_us}
\frac{\partial \psi}{\partial x}\Big\vert_{x=\Delta} = \kappa\psi(\Delta)
\end{equation}
The constant $\kappa$ depends on $A$: $\kappa>0$ for $A<A^*$,  $\kappa=0$ for $A=A^*$,  $\kappa<0$ for $A>A^*$. The critical adsorption strength $A^*$ is defined by the range of the potential: $A^*\sim a_s^2/\Delta^2$. 
\begin{figure}[ht!]
\center{\includegraphics[width=1\linewidth]{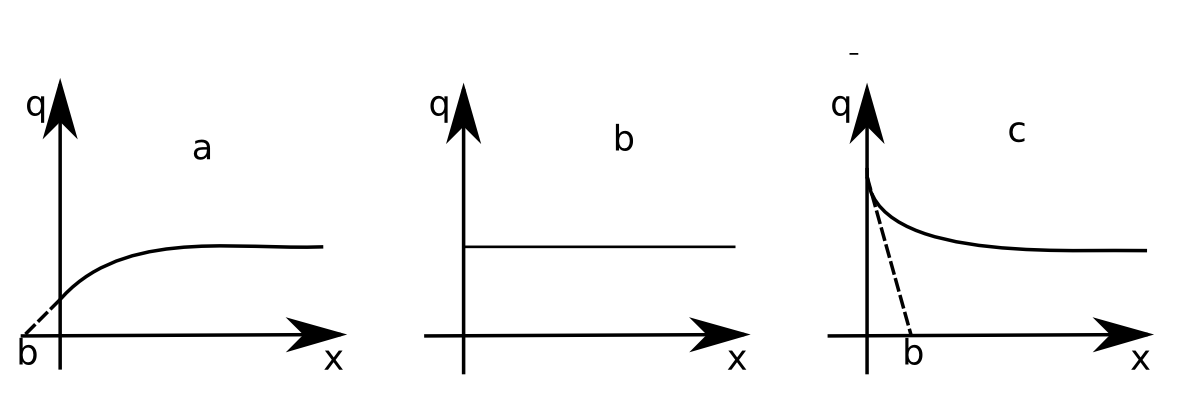}}
\caption{\small{Illustration of the extrapolation length parameter $b=-1/\kappa$ for (a) depletion, $b<0$, (b) neutral plates ($b=\infty$), and (c) adsorption, $b>0$.}}
\label{adsorption_3types_fig}
\end{figure}
For small $\tau\equiv A/A^*-1$ the parameter $\kappa$ is proportional to $\tau$: $\kappa \sim -\tau/\Delta$. The regime $A>A^*$ corresponds to polymer adsorption: monomer concentration near the wall (at $x\gtrsim\Delta$) is elevated in this case (see Fig.\ref{adsorption_3types_fig}c). For $x>\Delta$ the surface potential can be neglected, so Eq.(\ref{intr_polymer_gsd_adsorption}) with the effective boundary condition Eq.(\ref{intr_polymer_gsd_adsorption_bc}) can be easily solved yielding (for $A>A^*$) 
$$
\psi(x)\simeq const\,e^{-x/b}, \quad x>\Delta
$$                                                                     
with $b=-1/\kappa$, $\Lambda=-a_s^2/b^2$. The decay length $b$ (note that $b\ll 1/\tau$) is often called the extrapolation length (see Fig.\ref{adsorption_3types_fig}). The concentration profile is $c(x)\propto \psi^2(x)\propto e^{-2x/b}$, $x>\Delta$. 
Of course, these GSD results for $\psi(x)$, $c(x)$ are valid (for ideal polymers) only for $x\ll R_g$. However, they do show that $c(\Delta)$ can strongly exceed $c(R_g)\sim c_b$ if $b\ll R_g$.
The dependence of $b$ on the potential profile $U_s(x)$ is investigated further in Ch.\ref{chap:Chapter4}. Setting $\Delta\rightarrow 0$, the effective boundary condition can be rewritten (see also \cite{deGennes_1969, fleer_2010}) as
\begin{equation}
\label{intr_polymer_gsd_adsorption_bc}
\frac{\partial \psi}{\partial x}\Big\vert_{x=0} = -\frac{1}{b}\psi(0), \quad \frac{\partial c}{\partial x}\Big\vert_{x=0} = \frac{2}{b}c(0)
\end{equation}
These boundary conditions will be employed in Ch.\ref{chap:Chapter4}.

%% file: Chapters/repulsion.tex
% Chapter 3
\chapter{Purely repulsive surfaces} % Main chapter title
\label{chap:Chapter3} % For referencing the chapter elsewhere, use \ref{Chapter1} 
\lhead{Chapter 3. \emph{Purely repulsive walls}} % This is for the header on each page - perhaps a shortened title
%%%%%%%%%%%%%%%%%%%%%%%%%%%%%%%%%%%%%%%%%%%%%%%%%%%%%%%%%%%%%%%%%%%%%%%%%%%%%%%%%%%%%%%%%%%%%%%%%%%%%%%%%%%%%%%%%%%%%%%%%%%%%%%%%%%%%%%%%%%%%%%%%%%%%%%%%%%%%
%           Outline.
%%%%%%%%%%%%%%%%%%%%%%%%%%%%%%%%%%%%%%%%%%%%%%%%%%%%%%%%%%%%%%%%%%%%%%%%%%%%%%%%%%%%%%%%%%%%%%%%%%%%%%%%%%%%%%%%%%%%%%%%%%%%%%%%%%%%%%%%%%%%%%%%%%%%%%%%%%%%%
\section{Outline}
	In this section we consider the polymer-induced interaction between colloidal particles with purely repulsive surfaces.
	In order to simplify the problem in its original form, we use the Derjaguin approximation and consider locally interaction between two colloids 
	as interaction between two parallel solid plates, immersed in a polymer solution. 
	This approximation can be used when the polymer size is much smaller than the size of the colloidal particles.
	Thus, the problem is reduced to the one-dimensional case. 
	The polymer solution is characterized by the following parameters: $c_b$ is the bulk monomer concentration, 
	$a_s = a\sqrt{6}$ is the polymer statistical segment and $N$ is the number of monomer units per chain. 
%%%%%%%%%%%%%%%%%%%%%%%%%%%%%%%%%%%%%%%%%%%%%%%%%%%%%%%%%%%%%%%%%%%%%%%%%%%%%%%%%%%%%%%%%%%%%%%%%%%%%%%%%%%%%%%%%%%%%%%%%%%%%%%
%       colloidal particles in polymer solution and Derjagin approximation
%%%%%%%%%%%%%%%%%%%%%%%%%%%%%%%%%%%%%%%%%%%%%%%%%%%%%%%%%%%%%%%%%%%%%%%%%%%%%%%%%%%%%%%%%%%%%%%%%%%%%%%%%%%%%%%%%%%%%%%%%%%%%%%
\begin{figure}[ht!]
\begin{minipage}[ht]{0.43\linewidth}
\center{\includegraphics[width=1\linewidth]{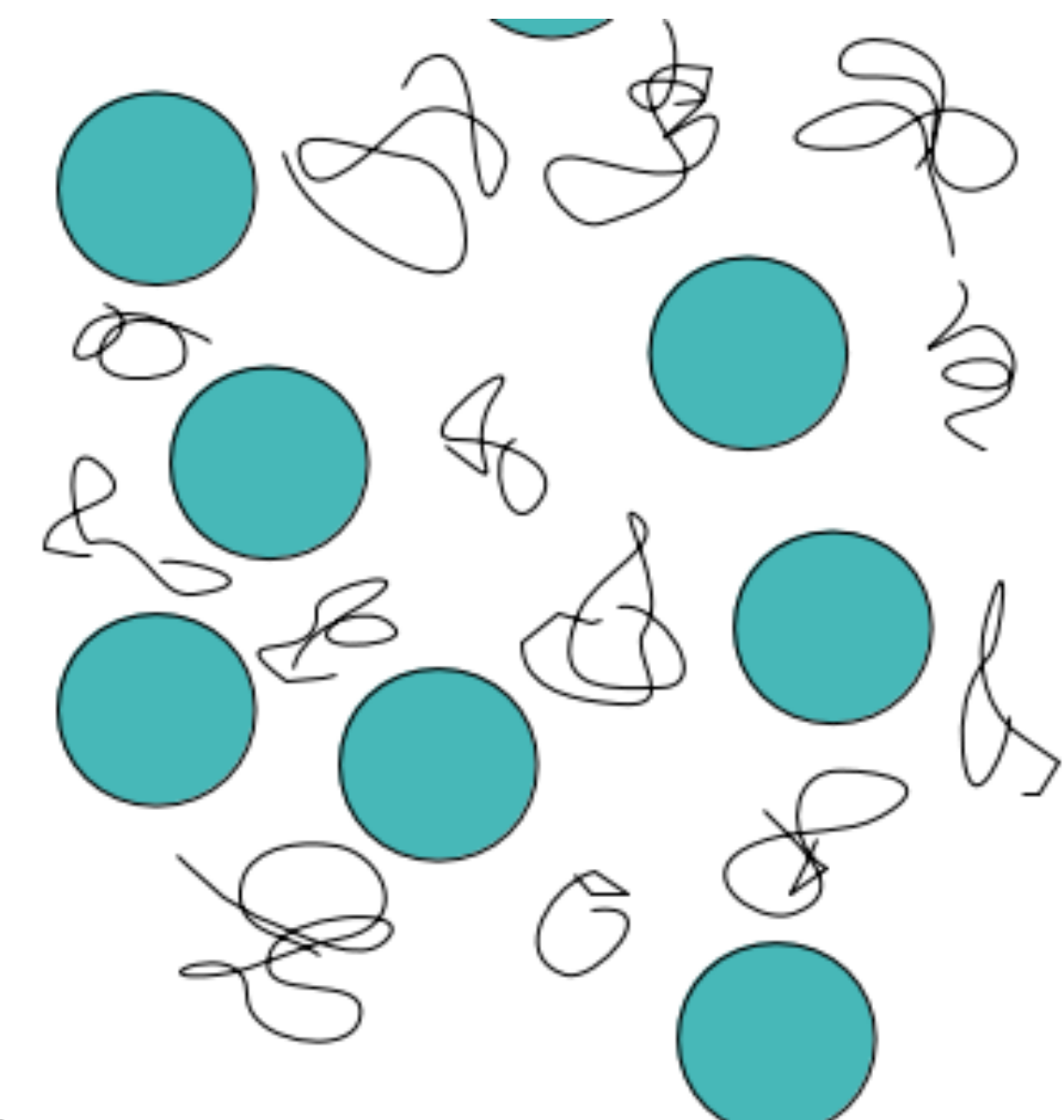}}
\caption{\small{Colloidal particles immersed in free polymer solution.}}
\label{colloids_fig}
\end{minipage}
\hfill
\begin{minipage}[ht]{0.43\linewidth}
\center{\includegraphics[width=1\linewidth]{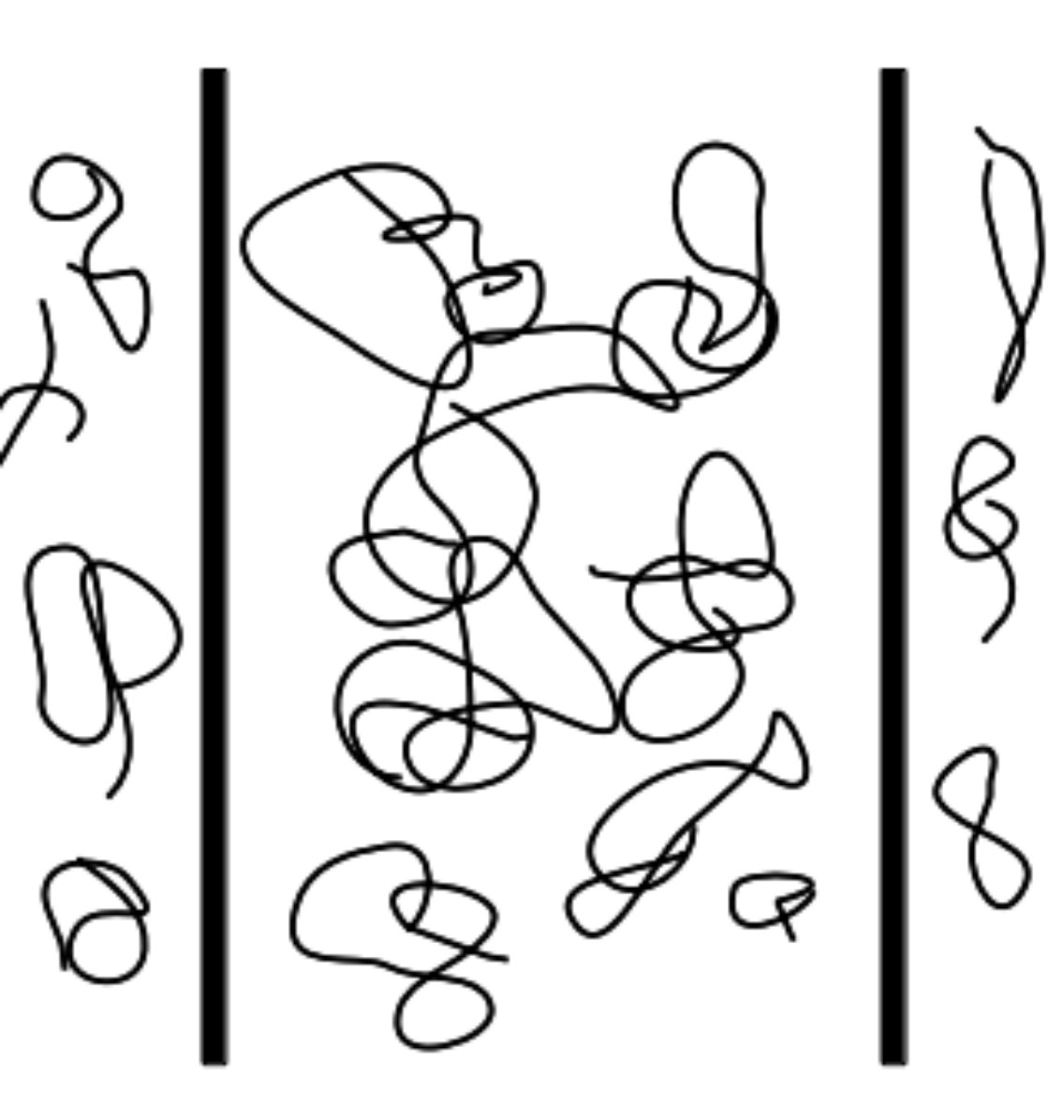}} 
\caption{\small{Representation of Derjaguin approximation.}}
\label{flat_plates_fig}
\end{minipage}
\end{figure}  
	For the excluded-volume interactions of monomers, the third virial approximation (with coefficients $\text{v}, \text{w}$) is used.
	We also assumed that the polymers in the gap between two plates are in thermodynamic equilibrium with the bulk solution
	(which is characterized by chemical potential $\mu_{int} = \mu_{int}(c_b)$), that allows us to use a grand canonical ensemble. 
	To analyze the system, we apply the  self-consistent field theory ($SCFT$), which is described in Sec.\ref{sec:polymers_scft}.
	First of all, we solve the SCFT equations for the ideal polymer solution ($\text{v} = \text{w} = 0$). In this case 
	we are able to find the analytical solutions of the SCFT equations and compare them with the numerical results. 
	Then, we widely investigate the non-ideal polymer solution.
	We develop the algorithm for the numerical solution of the SCFT equation and test the numerical solution for different virial coefficients and 
	different number of grid points.	
 	In addition, we improve the analytical GSDE theory and compare its results with the numerical SCFT solutions. 
	As a final step, we recalculate the thermodynamic potential in $k_BT$ units.
	Our main purpose here is to find the range of distances between plates, in which the polymer-induced interaction makes repulsive barrier, i.e. 
	stabilizes colloids. 
%%%%%%%%%%%%%%%%%%%%%%%%%%%%%%%%%%%%%%%%%%%%%%%%%%%%%%%%%%%%%%%%%%%%%%%%%%%%%%%%%%%%%%%%%%%%%%%%%%%%%%%%%%%%%%%%%%%%%%%%%%%%%%%%%%%%%%%%%%%%%%%%%%%%%%%%%%%%%
%           Exact solution.
%%%%%%%%%%%%%%%%%%%%%%%%%%%%%%%%%%%%%%%%%%%%%%%%%%%%%%%%%%%%%%%%%%%%%%%%%%%%%%%%%%%%%%%%%%%%%%%%%%%%%%%%%%%%%%%%%%%%%%%%%%%%%%%%%%%%%%%%%%%%%%%%%%%%%%%%%%%%%
\section{The boundary condition} 
	In this chapter, we consider a hard core repulsion between the colloidal surface and polymers. The surface potential contribution in the self-consistent 
	field takes the form: 
$$
U(x) = \left\{ 
\begin{array}{ll}
          0,           & \quad  0 < x < h  \\
                       &                   \\
          \infty,      & \quad  \text{otherwise}
\end{array} 
\right.
$$	
that leads to the boundary condition at the wall: $q(0, s) = 0$, because the polymer can not penetrate through the surface. \footnote{The Dirichlet boundary condition
is valid only for $N\gg 1$. In case of finite $N$, the boundary condition contains amendments that have been studied in details in \cite{Likhtman_2009}.} In case of attractive surfaces this boundary condition changes (See Ch.\ref{chap:Chapter4}, Ch.\ref{chap:Chapter5})
For the value of concentration at the surface, we can write:
$$
	c(0) = \int\limits_{0}^1\mathrm{d}s q(0, s) q(0, 1-s) = 0
$$
	Moreover, its derivative at $x=0$ is 
$$
	c_x(0) = \int\limits_{0}^1\mathrm{d}s q_x(0, s) q(0, 1-s) + \int\limits_{0}^1\mathrm{d}s q(0, s) q_x(0, 1-s)= 0
$$
	This means that the function, $c(x)$ is even with respect to $x=0$. Due to the definition of the self consistent field, $w(x)$ (see Eq.(\ref{intr_polymer_self_consistent_field_dimless})),
	the function, $w(x)$ is also an even function with respect to $x=0$. It will help us sufficiently improve the iterative algorithm of solving
	non homogeneous Edwards equation using the discrete cosine transform instead of the Fourier transform.
%%%%%%%%%%%%%%%%%%%%%%%%%%%%%%%%%%%%%%%%%%%%%%%%%%%%%%%%%%%%%%%%%%%%%%%%%%%%%%%%%%%%%%%%%%%%%%%%%%%%%%%%%%%%%%%%%%%%%%%%%%%%%%%%%%%%%%%%%%%%%%%%%%%%%%%%%%%%%
%           Exact solution.
%%%%%%%%%%%%%%%%%%%%%%%%%%%%%%%%%%%%%%%%%%%%%%%%%%%%%%%%%%%%%%%%%%%%%%%%%%%%%%%%%%%%%%%%%%%%%%%%%%%%%%%%%%%%%%%%%%%%%%%%%%%%%%%%%%%%%%%%%%%%%%%%%%%%%%%%%%%%%
\section{Ideal polymer solution: v = w = 0}
	It is interesting to study the ideal polymer solution separately from the non-ideal polymer solution, because, 
	in this case, we are able to solve the SCFT equations analytically. 
	The analytical solution can serve as a benchmark solution, which helps us to verify the accuracy of the appropriate numerical solution. 	
	In this case, the virial coefficients $\text{v} = \text{w} = 0$ and the resulting self-consistent field is $w(x) = 0$. 
	The grand thermodynamic potential, Eq.(\ref{intr_polymer_homo_vw_free_energy_in_reduced}), reduces to
\begin{equation}
\label{repulsion_partition_function_vw0}
	-\Omega(h) = Q(h) = \int\limits_0^h\mathrm{d}x\, q(x, 1)
\end{equation}
	 The expression for the concentration profile between two plates is (cf. Eq.(\ref{intr_polymer_density_homo_dimless}))
\begin{equation}
\label{repulsion_homo_conc_profile}
	c(x)/c_b = \int\limits_0^1\mathrm{d}s\, q(x, s)q(x, 1-s)
\end{equation}	
        The corresponding distribution function, $q(x, s)$ satisfies to the homogeneous diffusion equation:
\begin{equation}
\label{repulsion_homo_diff_eq} 
	\frac{\partial q(x, s)}{\partial s} = \frac{\partial^2 q(x, s)}{\partial x^2}
\end{equation}            
	with the boundary conditions $q(0, s) = q_x(h_m, s) = 0$ and the initial condition $q(x, 0) = 1$. 	 	
%%%%%%%%%%%%%%%%%%%%%%%%%%%%%%%%%%%%%%%%%%%%%%%%%%%%%%%%%%%%%%%%%%%%%%%%%%%%%%%%%%%%%%%%%%%%%%%%%%%%%%%%%%%%%%%%%%%%%%%%%%%%%%%%%%%%%%%%%%%%%%%%%%%%%%%%%%%%%
%           Analytical solution of the homogeneous diffusion equation.
%%%%%%%%%%%%%%%%%%%%%%%%%%%%%%%%%%%%%%%%%%%%%%%%%%%%%%%%%%%%%%%%%%%%%%%%%%%%%%%%%%%%%%%%%%%%%%%%%%%%%%%%%%%%%%%%%%%%%%%%%%%%%%%%%%%%%%%%%%%%%%%%%%%%%%%%%%%%%
\section{Analytical solution of the homogeneous diffusion equation}  
        In order to find the analytical solution for $q(x,s)$, we use the Fourier method of separation of variables.
        Assuming that the variables  $x$ and $s$ can be separated, we can write
$$
        q(x, s) = X(x)S(s) 
$$ 
        and Eq.(\ref{repulsion_homo_diff_eq}) takes the form:
$$
        X(x)S'(s) = S(s)X''(x) 
$$ 
        Dividing it by $X(x)S(s)$, we get
$$
        \frac{S'}{S} = \frac{X''}{X} = k
$$
        where $k$ is the separation constant. Here, the quantity $S'$ denotes the derivative of $S$ with respect to $s$ and $X'$ means the derivative
	of $X$ with respect to $x$. Let us to find all possible constants $k$ and the corresponding nonzero functions $S$ and $X$. For that, consider 
	the above equations:
$$
        S' =  kS ,\quad X'' - kX = 0
$$
        The solution of the first equation is
$$
        S(s) = S_0 e^{ks}
$$
        where $S_0$ is an arbitrary constant. Furthermore, the boundary conditions give
$$
	X(0)S(s) = 0, \quad X(h)S(s) = 0 \quad \text{for all } s
$$
	Since $S(s)$ is not identically zero, we obtain the eigenvalue problem:
$$
	X'' - kX = 0, \quad X(0) = X(h) = 0
$$
	There are, in general, three cases to solve it:\\
	a) \textbf{k}=0. Then, $X(x) = ax + b$, so applying the boundary conditions, we get
$$
	X(0) = b = 0, \quad X(h) = ah = 0 \Rightarrow a = 0 \quad \text{for any } h
$$
	Thus, the eigenvalue $k_0 = 0$, corresponds to $X(x) = 0$, which is impossible.\\
	b) \textbf{k}$>$0. Then, 
$$
	X(x) = ae^{kx} + be^{-kx}
$$
	Applying the boundary conditions, we have
$$
	X(0) = a + b = 0, \text{so} \quad a=-b
$$
	and
$$
	X(h) = a\left( e^{kh} - e^{-kh}\right) = 0, \text{so} \quad a = 0
$$
	Therefore, there are no positive eigenvalues.\\
	c) \textbf{k}$<$0. Let us denote $k = -p^2$, so:
$$
        X'' + p^2X = 0
$$
        This is the simple harmonic motion equation with trigonometric solutions. Thus, we can write
$$
        X(x) = a\sin(px) + b\cos(px)
$$
        Now, applying the boundary conditions, we find $X(0) = X(h) = 0$. Therefore, in the harmonic equation, $b = 0$, and 
        $\sin(ph)=0 \Rightarrow p_n = \pi n/h$. 
        The general solution can be constructed by adding all the possible solutions, satisfying the boundary conditions, together:
$$
        q(x, s) = \sum\limits_{n=0}^{\infty}A_n\sin\left(\frac{\pi nx}{h}\right)e^{-\frac{\pi^2n^2s}{h^2}}
$$  
        The final step is to apply the initial conditions, namely 
$$
        q(x, 0) = \sum\limits_{n=0}^{\infty}A_n\sin\left(\frac{\pi nx}{h}\right)=1
$$          
        The Fourier series must be multiplied  by $\sin m\pi x/h$ and integrated in the range $[0..h]$. 
        Therefore, for the Fourier coefficients we obtain:
$$
        A_{n} = \frac{2}{h}\int\limits_{0}^{h}\mathrm{d}x\, \sin\left(\frac{\pi n x}{h}\right) = \left\{ 
\begin{array}{ll}
        4/\pi n,      & \quad n = 1, 3 \ldots \\
                      &  \\
        0,            & \quad n = 0, 2 \ldots 
\end{array} 
\right.
$$
	Let us denote $n = 2m+1$ and $A_m = 4/\pi(2m+1)$. Thus,
\begin{equation}
\label{repulsion_analytical_homo_solution}
        q(x, s) = \sum\limits_{n=0}^{\infty}\frac{4}{\pi(2n+1)}\sin\left(\frac{\pi (2n+1)x}{h}\right)e^{-\frac{\pi^2(2n+1)^2s}{h^2}}
\end{equation}
	This expression will be used for comparisons with the numerical solutions. 
	Thereby, we should cut off the series. In order to find the cutting limit, we use the next conditions:\\
        1) $e^{-a_n}< 10^{-8}$ \quad $a_n > 8\ln10$, where $a_n = \pi^2(2n+1)^2s/h^2$. We cut the series when the terms become quite small. \\
        2) For small $h$, we will use at least $10$ first terms in the sum.

        $\textbf{Testing}$: 
        In order to exclude the numerical errors caused by the calculation of the infinite series, Eq.(\ref{repulsion_analytical_homo_solution}),
        it is usefull to consider the initial condition in Eq.(\ref{repulsion_homo_diff_eq}) taken in the form:
\begin{equation}
\label{repulsion_ic_sin}
        q(x, 0) = \sin\left(\frac{\pi x}{h}\right)
\end{equation}
        In this case, we have only one coefficient in the sum:
$$
        A_{n} = \frac{2}{h}\int\limits_{0}^{h}\mathrm{d}x\, \sin\left(\frac{\pi x}{h}\right) \sin\left(\frac{\pi n x}{h}\right) = \left\{ 
\begin{array}{ll}
        1,      & \quad n = 1 \\ 
                &  \\
        0,      & \quad n = 0, 2 \ldots 
\end{array} 
\right.
$$
        and the solution can be written as
\begin{equation}
\label{repulsion_solution_ic_sin}
        q(x, s) = \sin\left(\frac{\pi x}{h}\right)e^{-\frac{\pi^2s}{h^2}}
\end{equation} 
	This is the complete analytical solution with numerical errors comparable to machine epsilon.
		
        $\textbf{Partition \, function}$: Substituting Eq.(\ref{repulsion_analytical_homo_solution}) in Eq.(\ref{repulsion_partition_function_vw0}), we can find 
	the analytical expression for the partition function:
\begin{equation}
\label{repulsion_part_fun_analyt}
        Q(h) = \int\limits_0^h\mathrm{d}x\,q(x, 1) = \sum\limits_{n=0}^{\infty}\frac{8h}{\pi^2(2n+1)^2} e^{-\frac{\pi^2(2n+1)^2}{h^2}}
\end{equation}
        Lets us find asymptotics of this expression in two cases:\\
        a) $h \ll 1$ \quad in the sum of Eq.(\ref{repulsion_part_fun_analyt}), we leave only the first term, 
        because argument of the exponential factor is very large and negative,
        any other terms have to be very small. Thus,
$$
        Q(h) \simeq \frac{8h}{\pi^2} e^{-\frac{\pi^2}{h^2}} \rightarrow 0
$$
        b) $h \gg 1$ \quad in the sum of Eq.(\ref{repulsion_part_fun_analyt}), we can not leave only the $1-$st term, 
        because the power in $\exp$ is very small at least for several first terms. 
        We can estimate the first order correction. Let us consider the auxiliary function
$$
	I(t) = \sum\limits_{n=0}^{\infty}\frac{1}{(2n+1)^2} e^{-t(2n+1)^2}
$$
	After differentiation, we get 
$$
	\frac{\mathrm{d}I(t)}{\mathrm{d}t} = -\sum\limits_{n=0}^{\infty}e^{-t(2n+1)^2} \simeq -\int\limits_{0}^{\infty}\mathrm{d}xe^{-t(2x+1)^2} = 
	-\frac{1}{2\sqrt{t}}\int\limits_{\sqrt{t}}^{\infty}\mathrm{d}xe^{-x^2} \simeq -\frac{\sqrt{\pi}}{4\sqrt{t}}
$$
	This result is valid only for $t \ll 1$. After integration:
$$
	I(t) \simeq C - \frac{\sqrt{\pi t}}{2}
$$
	where $C$ is a constant which can be found as 
$$
	I(0) = C = \sum\limits_{n=0}^{\infty}\frac{1}{(2n+1)^2} = \frac{\pi^2}{8}
$$
	Thus, finally 
\begin{equation}
\label{repulsion_part_fun_asymptotic_hgg1}
        Q(h\rightarrow \infty) \simeq  h - \frac{4}{\sqrt{\pi}}
\end{equation}
        $\textbf{Force}$: Now we are able to calculate the force between the plates:
\begin{equation}
\label{repulsion_force_analyt_ideal}
        \Pi = \frac{\mathrm{d}Q(h)}{\mathrm{d}h}= 
	\sum\limits_{n=0}^{\infty}8\left(\frac{1}{\pi^2(2n+1)^2} + \frac{2}{h^2}\right) e^{-\frac{\pi^2(2n+1)^2}{h^2}}
\end{equation}
        Let us find asymptotics of this expression:\\
        a) $h \ll 1$: \quad in the sum of Eq.(\ref{repulsion_force_analyt_ideal}), we leave only the first term 
        (due to the same reason as we did for the partition function):
\begin{equation}
\label{repulsion_force_asymptotic_hll1}
	\Pi(h\ll1)\simeq 8\left(\frac{1}{\pi^2} + \frac{2}{h^2}\right)e^{-\frac{\pi^2}{h^2}}
\end{equation}                      
        b) $h \gg 1$. \quad In this case, $\Pi \simeq 1$ and it proved difficult to find corrections to this constant. \\
%\begin{equation}
%\label{repulsion_force_asymptotic_hgg1}
%	\Pi(h\rightarrow\infty)\simeq 1 - \frac{const}{h^2}e^{-\frac{h^2}{4}}
%\end{equation}	
%
        $\textbf{Concentration profile}$: Here, we can as well calculate the analytical expression for the concentration profile as:
\begin{equation}
\label{repulsion_concentration_analyt}
\begin{array}{l}
       	\frac{c(x, h)}{c_b} = \int\limits_{0}^{1}\mathrm{d}s\,q(x, s)q(x, 1-s) = 
        \sum\limits_{n=0}^{\infty}\frac{16}{\pi^2(2n+1)^2}\sin^2(\frac{\pi(2n+1)}{h}x)e^{-\frac{\pi^2(2n+1)^2}{h^2}} +\\\\
        +\sum\limits_{m<n}\frac{32h^2}{\pi^4(2n+1)(2m+1)}\sin(\frac{\pi(2n+1)}{h}x)\sin(\frac{\pi(2m+1)}{h}x)\times \\\\
	 \times\frac{1}{(2n+1)^2-(2m+1)^2}\left(e^{-\frac{\pi^2(2m+1)^2}{h^2}}-e^{-\frac{\pi^2(2n+1)^2}{h^2}}\right)
\end{array}
\end{equation}   
        $\textbf{Second derivative of the concentration profile}$: We also need to find the expression for a second derivative of 
	the concentration profile Eq.(\ref{repulsion_concentration_analyt}) at $x=0$:
\begin{equation}
\label{repulsion_second_der_con_analyt}
\begin{array}{l}
        \frac{1}{c_b}\frac{\partial^2c}{\partial x^2}\Big\vert_{x=0} = 
        \sum\limits_{n=0}^{\infty}\frac{32}{h^2}e^{-\frac{\pi^2(2n+1)^2}{h^2}} +\\\\
        +\sum\limits_{m<n}\frac{64}{\pi^2}\frac{1}{(2n+1)^2-(2m+1)^2}\left(e^{-\frac{\pi^2(2m+1)^2}{h^2}}-e^{-\frac{\pi^2(2n+1)^2}{h^2}}\right)
\end{array}
\end{equation}   
%%%%%%%%%%%%%%%%%%%%%%%%%%%%%%%%%%%%%%%%%%%%%%%%%%%%%%%%%%%%%%%%%%%%%%%%%%%%%%%%%%%%%%%%%%%%%%%%%%%%%%%%%%%%%%%%%%%%%%%%%%%%%%%%%%%%%%%%%%%%%%%%%%%%%%%%%%%%%
%       Numerical Solution.
%%%%%%%%%%%%%%%%%%%%%%%%%%%%%%%%%%%%%%%%%%%%%%%%%%%%%%%%%%%%%%%%%%%%%%%%%%%%%%%%%%%%%%%%%%%%%%%%%%%%%%%%%%%%%%%%%%%%%%%%%%%%%%%%%%%%%%%%%%%%%%%%%%%%%%%%%%%%%
\section{Numerical solution of the homogeneous diffusion equation}
	In this section, we develop the numerical scheme to solve one dimensional diffusion equation, Eq.(\ref{repulsion_homo_diff_eq}):
$$
        \frac{\partial q(x, s)}{\partial s} = \frac{\partial^2 q(x, s)}{\partial x^2}
$$
        with boundary conditions $q(0, s) = q_x(h_m, s) = 0$ and the initial condition $q(x, 0) = 1$. The numerical solution of the equation is then compared
        with the analytical one obtained in the previous section.
        
	Let us introduce the computational mesh. The interval $x\in [0, h_m]$, can be divided using $N_x$ equally spaced points
$$
        x_i = i\Delta x, \quad i = 0, 1, \ldots, N_x-1
$$
        where $\Delta x = h_m/(N_x-1)$ is the so-called grid spacing. A similar $N_s$-points discretization of $s\in[0, 1]$ leads to 
$$
        s_m = m\Delta s, \quad m = 0, 1, \ldots, N_s-1
$$
        where $\Delta s = 1/(N_s-1)$ is the contour step. The value of the propagator $q(x, s)$ at these discrete space-contour points will be 
        denoted by 
$$
        q_i^m = q(x_i, s_m) 
$$
	For the numerical solution of Eq.(\ref{repulsion_homo_diff_eq}) we chose the Crank-Nicolson method, 
	because it has the second order accuracy in $\Delta x$ and $\Delta s$.
	This method is discussed in details in Appendix.\ref{AppendixB} for various boundary conditions.         
%%%%%%%%%%%%%%%%%%%%%%%%%%%%%%%%%%%%%%%%%%%%%%%%%%%%%%%%%%%%%%%%%%%%%%%%%%%%%%%%%%%%%%%%%%%%%%%%%%%%%%%%%%%%%%%%%%%%%%%%%%%%%%%%%%%%%%%%%%%%%%%%%%%%%%%%%%%%%
%       Testing numerical scheme.
%%%%%%%%%%%%%%%%%%%%%%%%%%%%%%%%%%%%%%%%%%%%%%%%%%%%%%%%%%%%%%%%%%%%%%%%%%%%%%%%%%%%%%%%%%%%%%%%%%%%%%%%%%%%%%%%%%%%%%%%%%%%%%%%%%%%%%%%%%%%%%%%%%%%%%%%%%%%%
\section{Testing the numerical scheme} 
\begin{figure}[ht!]
\begin{minipage}[ht]{0.5\linewidth}
\center{\includegraphics[width=1\linewidth]{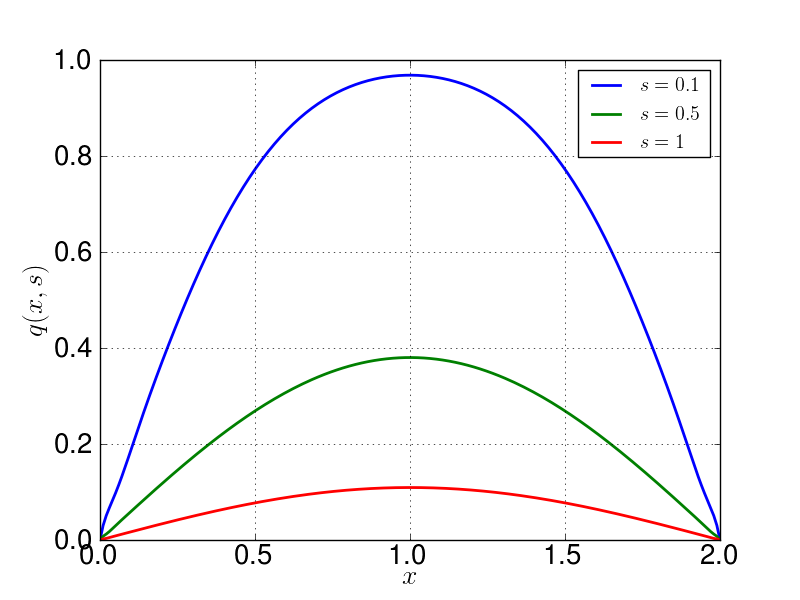}}
\caption{\small{The distribution function, $q(x, s)$, obtained from the numerical solution of the Edwards equation, Eq.(\ref{repulsion_homo_diff_eq}),
         for three different contour lengths. Fixed grid parameters: $N_{x}=200$, $N_{s}=100$.}}
\label{propagator_01_05_1_x200_s100_fig}
\end{minipage}
\hfill
\begin{minipage}[ht]{0.5\linewidth}
\center{\includegraphics[width=1\linewidth]{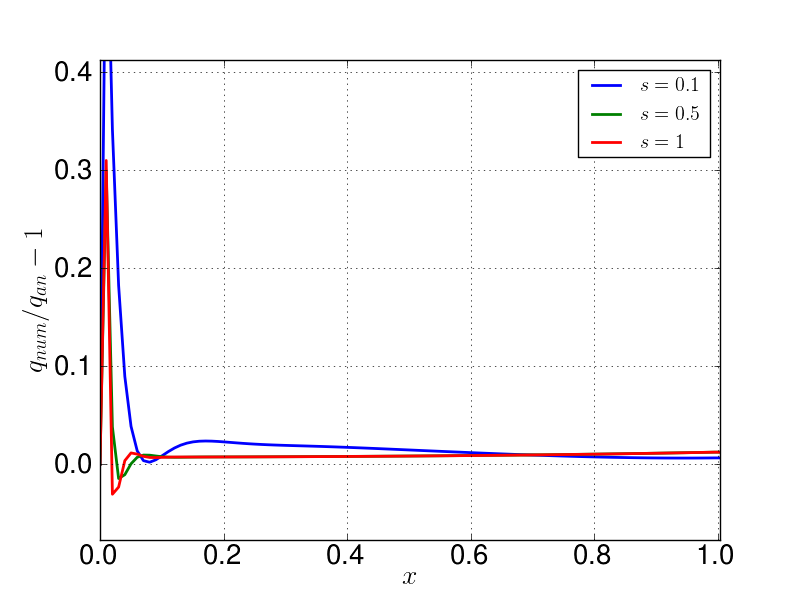}}
\caption{\small{The relative error, $q_{num}/q_{an} - 1$, between the numerical, Eq.(\ref{repulsion_homo_diff_eq}), and 
		the analytical, Eq.(\ref{repulsion_analytical_homo_solution}), solutions for three different contour lengths. Fixed grid parameters: $N_{x}=200$, $N_{s}=100$.}}
\label{propagator_diff_01_05_1_x200_s100_fig}
\end{minipage}
\end{figure}

\begin{figure}[ht!]
\begin{minipage}[ht]{0.5\linewidth}
\center{\includegraphics[width=1\linewidth]{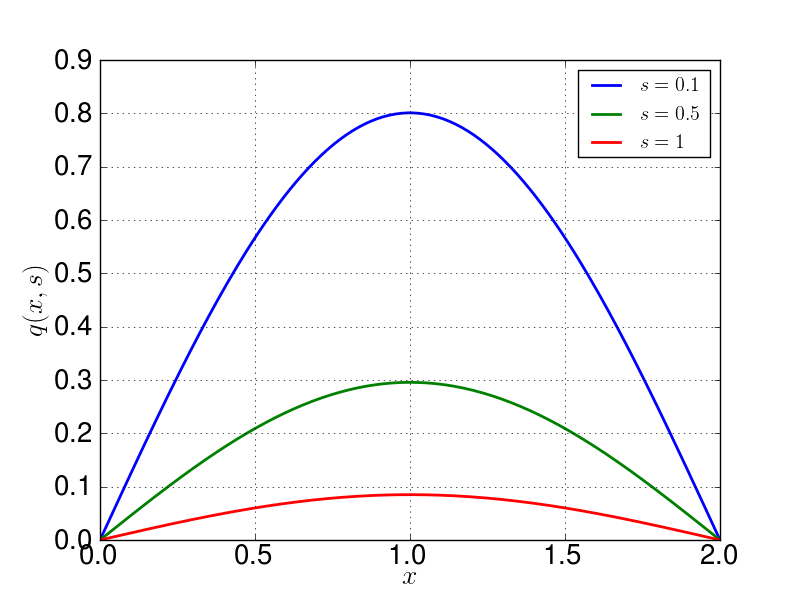}}
\caption{\small{The distribution function, $q(x, s)$, obtained from the numerical solution of the Edwards equation, Eq.(\ref{repulsion_homo_diff_eq}), 
	       with the initial condition, Eq.(\ref{repulsion_ic_sin}), for three different contour lengths. Fixed grid parameters: $N_{x}=200$, $N_{s}=100$.}}
\label{propagator_sin_01_05_1_x200_s100_fig}
\end{minipage}
\hfill
\begin{minipage}[ht]{0.5\linewidth}
\center{\includegraphics[width=1\linewidth]{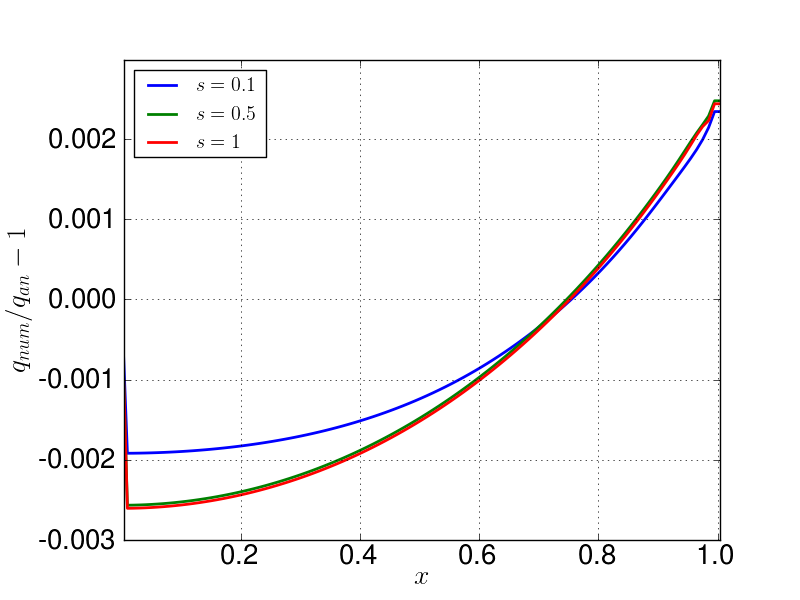}}
\caption{\small{The relative error, $q_{num}/q_{an} - 1$, between the numerical, Eq.(\ref{repulsion_homo_diff_eq}), and 
		the analytical, Eq.(\ref{repulsion_solution_ic_sin}), solutions calculated with the initial condition, Eq.(\ref{repulsion_ic_sin}), 
		for three different contour lengths. Fixed grid parameters: $N_{x}=200$, $N_{s}=100$.}}
\label{propagator_sin_diff_01_05_1_x200_s100_fig}
\end{minipage}
\end{figure}

	Below we give several examples of the numerical calculations and their comparison with the corresponding analytical solutions.        
	Adjusting the external parameters ($h, N_{x}, N_{s}$), we can control the accuracy of the numerical solution of 
	the diffusion equation, Eq.(\ref{repulsion_homo_diff_eq}).
	In Figs.\ref{propagator_01_05_1_x200_s100_fig}--\ref{propagator_diff_01_05_1_x200_s100_fig}, we present
	the distribution function, $q(x, s)$ and its relative error in comparison with the analytical solution, Eq.(\ref{repulsion_analytical_homo_solution}),
	calculated for different contour steps, $s$. 		
	The special case of the initial condition, Eq.(\ref{repulsion_ic_sin}), is considered in 
	Figs.\ref{propagator_sin_01_05_1_x200_s100_fig}--\ref{propagator_sin_diff_01_05_1_x200_s100_fig}.
%%%%%%%%%%%%%%%%%%%%%%%%%%%%%%%%%%%%%%%%%%%%%%%%%%%%%%%%%%%%%%%%%%%%%%%%%%%%%%%%%%%%%%%%%%%%%%%%%%%%%%%%%%%%%%%%%%%%%%%%%%%%%%%
%       Comparison of the analytical and numerical solutions for the distribution function 
%%%%%%%%%%%%%%%%%%%%%%%%%%%%%%%%%%%%%%%%%%%%%%%%%%%%%%%%%%%%%%%%%%%%%%%%%%%%%%%%%%%%%%%%%%%%%%%%%%%%%%%%%%%%%%%%%%%%%%%%%%%%%%%

%%%%%%%%%%%%%%%%%%%%%%%%%%%%%%%%%%%%%%%%%%%%%%%%%%%%%%%%%%%%%%%%%%%%%%%%%%%%%%%%%%%%%%%%%%%%%%%%%%%%%%%%%%%%%%%%%%%%%%%%%%%%%%%
%       Comparison of the analytical and numerical solutions for the distribution function with sin-like boundary conditions
%%%%%%%%%%%%%%%%%%%%%%%%%%%%%%%%%%%%%%%%%%%%%%%%%%%%%%%%%%%%%%%%%%%%%%%%%%%%%%%%%%%%%%%%%%%%%%%%%%%%%%%%%%%%%%%%%%%%%%%%%%%%%%%

%%%%%%%%%%%%%%%%%%%%%%%%%%%%%%%%%%%%%%%%%%%%%%%%%%%%%%%%%%%%%%%%%%%%%%%%%%%%%%%%%%%%%%%%%%%%%%%%%%%%%%%%%%%%%%%%%%%%%%%%%%%%%%%%%%%%%%%%%%%%%%%%%%%%%%%%%%%%%
%       Concentration profile.
%%%%%%%%%%%%%%%%%%%%%%%%%%%%%%%%%%%%%%%%%%%%%%%%%%%%%%%%%%%%%%%%%%%%%%%%%%%%%%%%%%%%%%%%%%%%%%%%%%%%%%%%%%%%%%%%%%%%%%%%%%%%%%%%%%%%%%%%%%%%%%%%%%%%%%%%%%%%%
\section{Concentration profile}
\begin{figure}[ht!]
\begin{minipage}[ht]{0.5\linewidth}
\center{\includegraphics[width=1\linewidth]{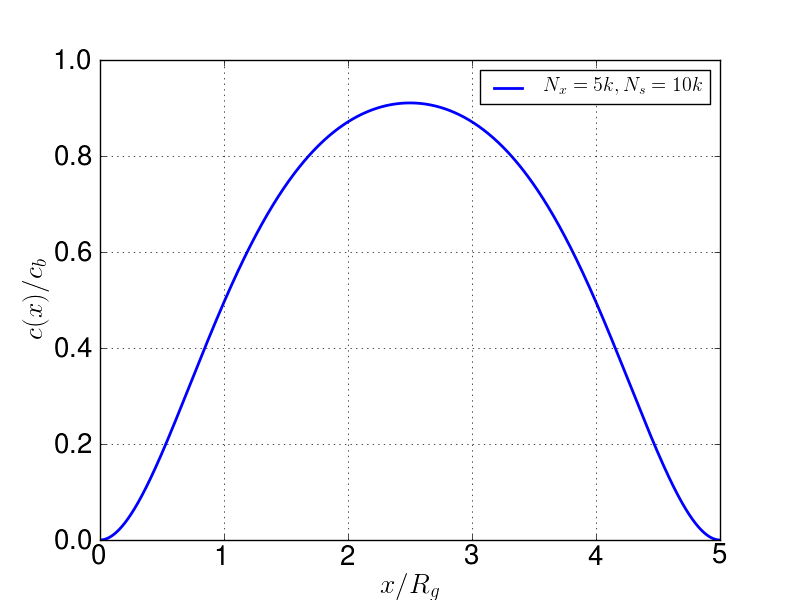}}
\caption{\small{The concentration profile, Eq.(\ref{repulsion_concentration_numerical}), obtained from the numerical solution of the Edwards equation,  
	       eq.(\ref{repulsion_homo_diff_eq}). Fixed grid parameters: $N_{x}=5000$, $N_{s}=10000$.}}
\label{conc_h5_x5k_s10k_fig}
\end{minipage}
\hfill
\begin{minipage}[ht]{0.5\linewidth}
\center{\includegraphics[width=1\linewidth]{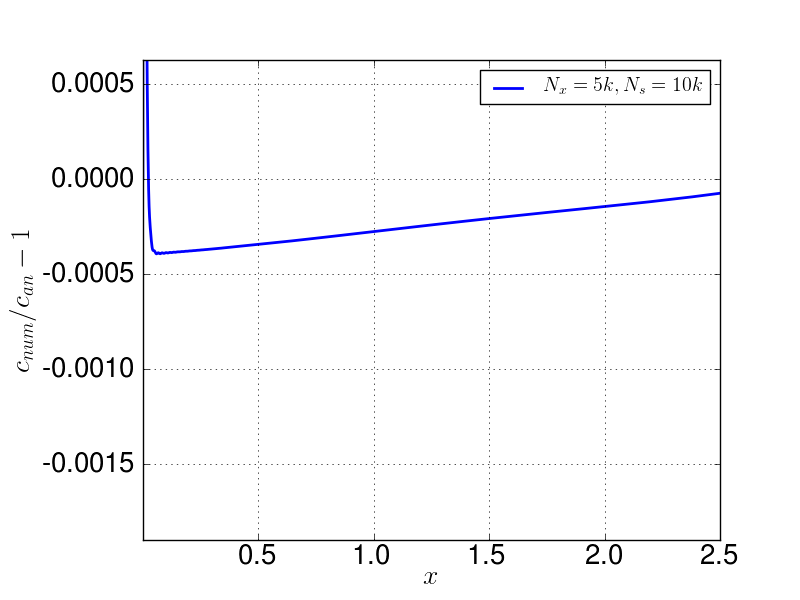}}
\caption{\small{The relative error, $c_{num}/c_{an} - 1$, between the numerical, Eq.(\ref{repulsion_concentration_numerical}), and the analytical, 
	 Eq.(\ref{repulsion_concentration_analyt}), concentration profiles.}}
\label{conc_diff_h5_x5k_s10k_fig}
\end{minipage}
\end{figure}

        In Eq.(\ref{repulsion_homo_conc_profile}), we wrote the concentration profile as an integral of the product of two distribution functions. 
	Since each of them is a smooth function, their product is also a smooth function. Based on that, we can use 
        a simple method for the integration like the Simpson's rule (third order accuracy) \cite{press_2007, hoffman_2001}. 
	In the expanded form, we can write
\begin{equation}
\label{repulsion_concentration_numerical}
\begin{array}{c}
        c(x)/c_b = \int\limits_0^1\mathrm{d}s\, q(x, s) q(x, 1 - s) \simeq 
        \frac{\Delta s}{3}\left(2q(x, 0)q(x, N_s-1) + 4q(x, 1)q(x, N_s-2) + \right. \\
	\left.\sum\limits_{m=2, 4, 6}^{N_s-2}(4q(x, m-1)q(x, N_s - m) + 2q(x, m)q(x, N_s-1-m))\right)
\end{array}
\end{equation}
        The numerical results for the concentration profile and its comparison with the analytical solution, Eq.(\ref{repulsion_concentration_analyt}), 
        can be found in Figs.\ref{conc_h5_x5k_s10k_fig}--\ref{conc_diff_h5_x5k_s10k_fig}.
%%%%%%%%%%%%%%%%%%%%%%%%%%%%%%%%%%%%%%%%%%%%%%%%%%%%%%%%%%%%%%%%%%%%%%%%%%%%%%%%%%%%%%%%%%%%%%%%%%%%%%%%%%%%%%%%%%%%%%%%%%%%%%%
%       concentration profile and its relative errors
%%%%%%%%%%%%%%%%%%%%%%%%%%%%%%%%%%%%%%%%%%%%%%%%%%%%%%%%%%%%%%%%%%%%%%%%%%%%%%%%%%%%%%%%%%%%%%%%%%%%%%%%%%%%%%%%%%%%%%%%%%%%%%%
%%%%%%%%%%%%%%%%%%%%%%%%%%%%%%%%%%%%%%%%%%%%%%%%%%%%%%%%%%%%%%%%%%%%%%%%%%%%%%%%%%%%%%%%%%%%%%%%%%%%%%%%%%%%%%%%%%%%%%%%%%%%%%%%%%%%%%%%%%%%%%%%%%%%%%%%%%%%%
%	Partition function.
%%%%%%%%%%%%%%%%%%%%%%%%%%%%%%%%%%%%%%%%%%%%%%%%%%%%%%%%%%%%%%%%%%%%%%%%%%%%%%%%%%%%%%%%%%%%%%%%%%%%%%%%%%%%%%%%%%%%%%%%%%%%%%%%%%%%%%%%%%%%%%%%%%%%%%%%%%%%%
\section{Partition function} 
\begin{figure}[ht!]
\begin{minipage}[ht]{0.5\linewidth}
\center{\includegraphics[width=1\linewidth]{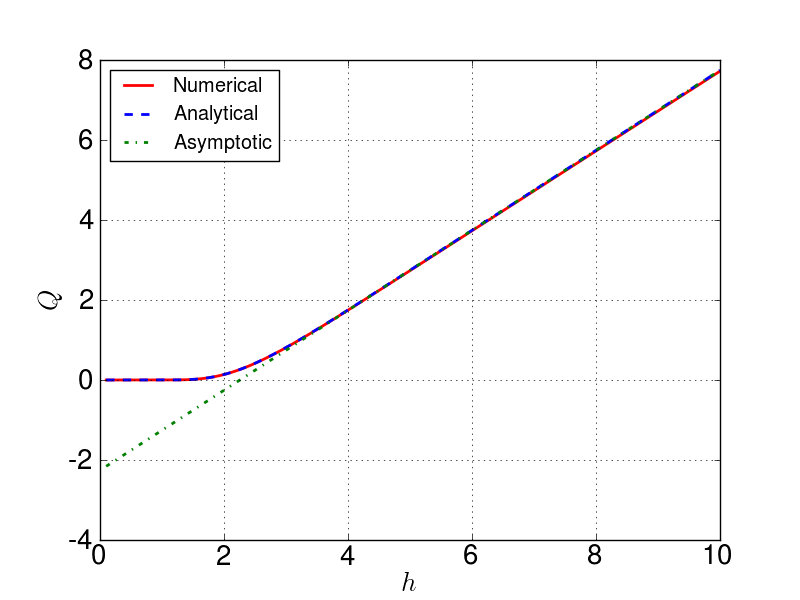}}
\caption{\small{The partition function calculated numerically, Eq.(\ref{repulsion_part_fun_numerical}), analytically, Eq.(\ref{repulsion_part_fun_analyt})
		and its asymptotics, Eq.(\ref{repulsion_part_fun_asymptotic_hgg1}).}}
\label{part_fun_num_an_asy_fig}
\end{minipage}
\hfill
\begin{minipage}[ht]{0.5\linewidth}
\center{\includegraphics[width=1\linewidth]{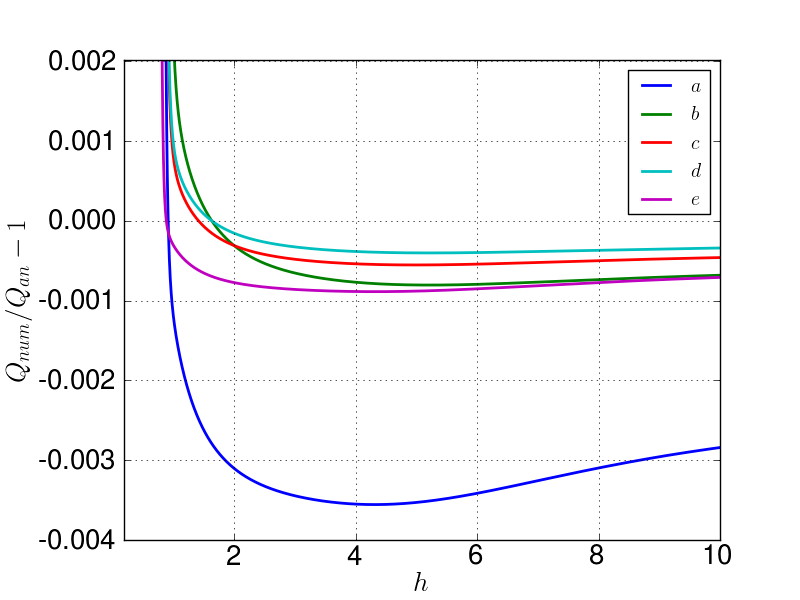}}
\caption{\small{The relative error between the numerical, Eq.(\ref{repulsion_part_fun_numerical}),  and the analytical, Eq.(\ref{repulsion_part_fun_analyt}), 
		solutions for the partition function. Different sets of grid parameters: 
	a) $N_x$ = 1k, $N_s$ = 2k, $N_h$ = 2k;  b) $N_x$ = 4k, $N_s$ = 2k, $N_h$ = 4k; c) $N_x$ = 6k, $N_s$ = 4k, $N_h$ = 4k;
	d) $N_x$ = 8k, $N_s$ = 4k, $N_h$ = 6k;  e) $N_x$ = 4k, $N_s$ = 8k, $N_h$ = 6k.}}
\label{part_fun_rel_err_ae_fig}
\end{minipage}
\end{figure}

        Repeating the same arguments as in the previous section, we use the same Simpson's integration rule for 
        the calculation of the partition function Eq.(\ref{repulsion_partition_function_vw0}). 
        In the expanded form it is
\begin{equation}
\label{repulsion_part_fun_numerical}
\begin{array}{l}
        Q(h) = \int\limits_0^h\mathrm{d}x\, q(x, 1) = 2 \int\limits_0^{h_m}\mathrm{d}x\, q(x, 1) \simeq \\ 
	\simeq \frac{2\Delta x}{3}\left(q(N_x-1, N_s-1) + 4q(N_x-2, N_s-1) +
	\sum\limits_{i=2, 4}^{N_x-2}(4q(i-1, N_s-1) + 2q(i, N_s-1))\right)
\end{array}
\end{equation}
        We used the symmetry of the integrand with respect to $h_m=h/2$ and the boundary conditions:  $q(0, s) = q_x(h_m, s) = 0$. 
        We depict the results for the partition functions in Fig.\ref{part_fun_num_an_asy_fig}. One can see that the numerical and 
	the analytical solutions coincide. Moreover, the asymptotics solution, Eq.(\ref{repulsion_part_fun_asymptotic_hgg1}), converges rather well. 
	In Fig.\ref{part_fun_rel_err_ae_fig}, we present the comparison between the numerical 
        and the analytical solutions for different values of the grid parameters. 
        The values of the functions for small $h$ seem to become unmanageably large. 
	This can be explained by two facts. First, the numerical values of the function in the region become less than the accuracy of the 
	numerical method. Second, if we look at the asymptotic behavior at $h\ll1$, we can see that the function has an exponential growth. 
	These reasons explain the sharp increase in the relative errors in this region. 
%%%%%%%%%%%%%%%%%%%%%%%%%%%%%%%%%%%%%%%%%%%%%%%%%%%%%%%%%%%%%%%%%%%%%%%%%%%%%%%%%%%%%%%%%%%%%%%%%%%%%%%%%%%%%%%%%%%%%%%%%%%%%%%
%        Partition function for a system between plates and its relative errors
%%%%%%%%%%%%%%%%%%%%%%%%%%%%%%%%%%%%%%%%%%%%%%%%%%%%%%%%%%%%%%%%%%%%%%%%%%%%%%%%%%%%%%%%%%%%%%%%%%%%%%%%%%%%%%%%%%%%%%%%%%%%%%%

%%%%%%%%%%%%%%%%%%%%%%%%%%%%%%%%%%%%%%%%%%%%%%%%%%%%%%%%%%%%%%%%%%%%%%%%%%%%%%%%%%%%%%%%%%%%%%%%%%%%%%%%%%%%%%%%%%%%%%%%%%%%%%%%%%%%%%%%%%%%%%%%%%%%%%%%%%%%%
%      	Force between plates and its comparison with analytical solution
%%%%%%%%%%%%%%%%%%%%%%%%%%%%%%%%%%%%%%%%%%%%%%%%%%%%%%%%%%%%%%%%%%%%%%%%%%%%%%%%%%%%%%%%%%%%%%%%%%%%%%%%%%%%%%%%%%%%%%%%%%%%%%%%%%%%%%%%%%%%%%%%%%%%%%%%%%%%%
\section{Force between plates} 
\begin{figure}[ht!]
\begin{minipage}[ht]{0.5\linewidth}
\center{\includegraphics[width=1\linewidth]{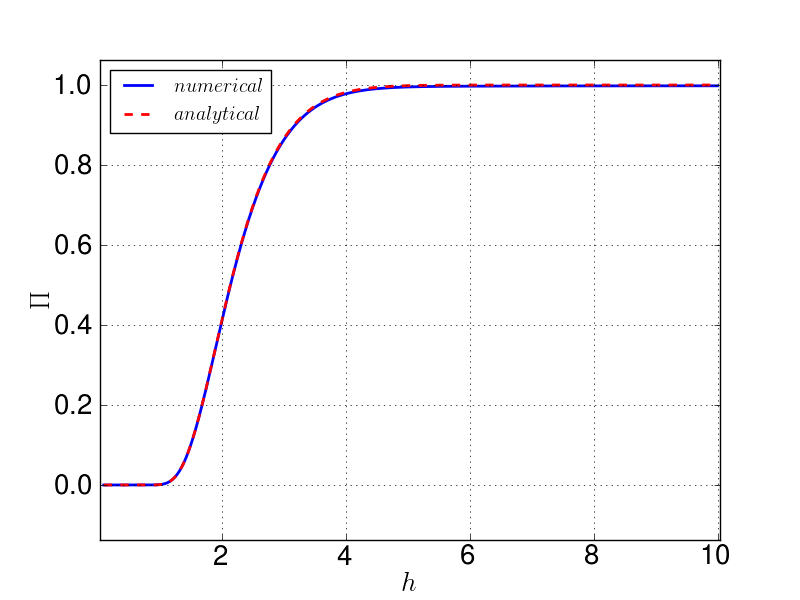}}
\caption{\small{The force calculated numerically, Eq.(\ref{repulsion_force_plates_numerical}) and analytically, Eq.(\ref{repulsion_force_analyt_ideal}).}}
\label{force_num_an_asy_fig}
\end{minipage}
\hfill
\begin{minipage}[ht]{0.5\linewidth}
\center{\includegraphics[width=1\linewidth]{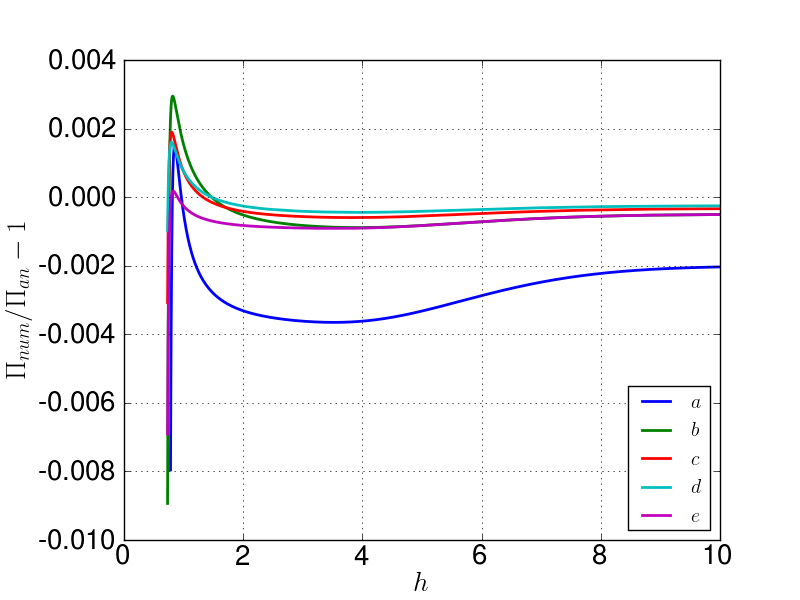}}
\caption{\small{The relative error between the numerical, Eq.(\ref{repulsion_force_plates_numerical}),  and the analytical, Eq.(\ref{repulsion_force_analyt_ideal}), 
		solutions for the force between plates. Different sets of grid parameters: 
	a) $N_x$ = 1k, $N_s$ = 2k, $N_h$ = 2k;  b) $N_x$ = 4k, $N_s$ = 2k, $N_h$ = 4k; c) $N_x$ = 6k, $N_s$ = 4k, $N_h$ = 4k;
	d) $N_x$ = 8k, $N_s$ = 4k, $N_h$ = 6k;  e) $N_x$ = 4k, $N_s$ = 8k, $N_h$ = 6k.}}
\label{force_rel_err_ae_fig}
\end{minipage}
\end{figure}

For the force between two plates, we can write:
$$
	\Pi(h) = -\frac{\mathrm{d}\Omega}{\mathrm{d}h} = \frac{\mathrm{d}Q(h)}{\mathrm{d}h}
$$
Numerically, the derivative with an accuracy of the second order, has the following form 
\begin{equation}
\label{repulsion_force_plates_numerical}
        \Pi_i = Q_h\Big\vert_i = \frac{Q_{i+1}-Q_{i-1}}{2\Delta h} - \frac{1}{6}Q_{hhh}(\xi)\Delta h^2
\end{equation} 
where $\xi\in [h_{i-1}, h_{i+1}]$. Many other approximations for numerical calculation of derivatives can be found in \cite{hoffman_2001}.
In Fig.\ref{force_num_an_asy_fig}, we represent the force for the numerical and the analytical solutions. These solutions coincide with each other. In Fig.\ref{force_rel_err_ae_fig}, we present the comparison between the numerical and the analytical solutions for different values of the grid parameters, $N_x, N_s$. 	The exponential growth of the errors in the leftmost region can be elucidated on the same basis as for the errors considered in the previous section.
%%%%%%%%%%%%%%%%%%%%%%%%%%%%%%%%%%%%%%%%%%%%%%%%%%%%%%%%%%%%%%%%%%%%%%%%%%%%%%%%%%%%%%%%%%%%%%%%%%%%%%%%%%%%%%%%%%%%%%%%%%%%%%%
%        Force between plates and its relative errors
%%%%%%%%%%%%%%%%%%%%%%%%%%%%%%%%%%%%%%%%%%%%%%%%%%%%%%%%%%%%%%%%%%%%%%%%%%%%%%%%%%%%%%%%%%%%%%%%%%%%%%%%%%%%%%%%%%%%%%%%%%%%%%%

%%%%%%%%%%%%%%%%%%%%%%%%%%%%%%%%%%%%%%%%%%%%%%%%%%%%%%%%%%%%%%%%%%%%%%%%%%%%%%%%%%%%%%%%%%%%%%%%%%%%%%%%%%%%%%%%%%%%%%%%%%%%%%%%%%%%%%%%%%%%%%%%%%%%%%%%%%%%%
%        Second derivative of the concentration at x = 0.
%%%%%%%%%%%%%%%%%%%%%%%%%%%%%%%%%%%%%%%%%%%%%%%%%%%%%%%%%%%%%%%%%%%%%%%%%%%%%%%%%%%%%%%%%%%%%%%%%%%%%%%%%%%%%%%%%%%%%%%%%%%%%%%%%%%%%%%%%%%%%%%%%%%%%%%%%%%%%
\section{Second derivative of the concentration profile at x = 0}
\begin{figure}[ht!]
\begin{minipage}[ht]{0.5\linewidth}
\center{\includegraphics[width=1\linewidth]{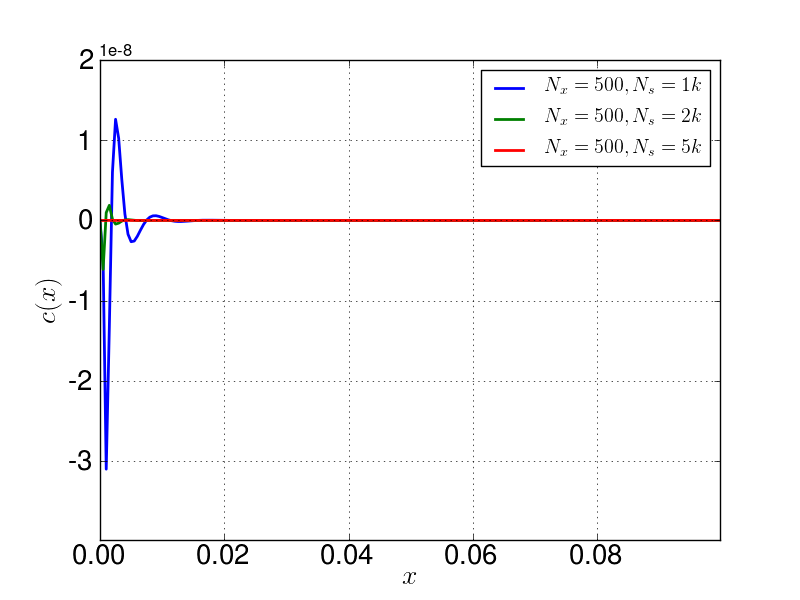}}
\caption{\small{The concentration profile at $h_m$ = 0.5, $N_x$ = 500 and different parameter $N_s$: a) $N_s$ = 1k;  
	b) $N_s$ = 2k; c) $N_s$ = 5k.}}
\label{concentration_125_fig}
\end{minipage}
\hfill
\begin{minipage}[ht]{0.5\linewidth}
\center{\includegraphics[width=1\linewidth]{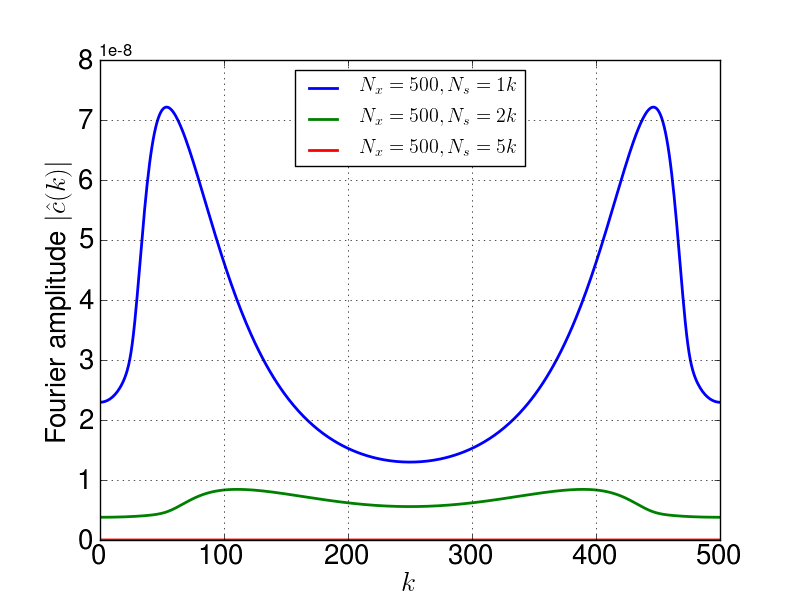}}
\caption{\small{The Fourier image of the concentration profiles, (Fig.\ref{concentration_125_fig}), as functions of the harmonic number.}}
\label{fourier_concentration_125_fig}
\end{minipage}
\end{figure}

Due to the reasons that will become apparent later in this section let us now numerically calculate the second derivative of the concentration profile at the surface.
For the numerical calculation of the second derivative of the concentration profile, we use the following second order approximation \cite{hoffman_2001}:
$$
         c_{xx}^{i} = \frac{2c_i-5c_{i+1}+4c_{i+2}-c_{i+3}}{\Delta x^2} + \frac{11}{12}c_{xxxx}(\xi)\Delta x^2
$$ 
where $\xi \in [x_{i}, x_{i+1}]$. At $x=0$ we can rewrite this expression as  
\begin{equation}
\label{repulsion_cxx0}
         c_{xx}\big\vert_{0} = \frac{2c_0-5c_{1}+4c_{2}-c_{3}}{\Delta x^2}
\end{equation}
	The numerical solution for the concentration profile has oscillations at small scales in the vicinity of the walls. 
	These oscillations might be even bigger than the maximum value of the concentration profile calculated for small 
	separations, $h$, thus, giving incorrect results for the second derivative. Moreover, it will strongly depend on the choice 
	of the spatial step $\Delta x$, for which, we calculated this derivative. The example of such oscillations can be found in 
	Fig.\ref{concentration_125_fig}. 
	Using the Fourier representation of the concentration profile, we can write
$$
	\hat{c}(k) = \sum\limits_{m=0}^{n-1}c(m)e^{-2\pi i\frac{mk}{N_x}} ,\,\,\, k = 0, \ldots, n - 1  
$$
	where $N_x$ corresponds to the input sequence of the concentration profile. We can analyze the Fourier image of the concentration profile as depicted
	in Fig.\ref{fourier_concentration_125_fig}. 
	One can see that these oscillations decrease upon increasing the parameter $N_s$. In principle, we can make these oscillations much less than square of 
	the spatial step, $(\Delta x)^2$ entering the expression for the numerical computation of the second derivative, but computationally it is very difficult.
%%%%%%%%%%%%%%%%%%%%%%%%%%%%%%%%%%%%%%%%%%%%%%%%%%%%%%%%%%%%%%%%%%%%%%%%%%%%%%%%%%%%%%%%%%%%%%%%%%%%%%%%%%%%%%%%%%%%%%%%%%%%%%%
%        concentration profiles and its fourtier images (1)
%%%%%%%%%%%%%%%%%%%%%%%%%%%%%%%%%%%%%%%%%%%%%%%%%%%%%%%%%%%%%%%%%%%%%%%%%%%%%%%%%%%%%%%%%%%%%%%%%%%%%%%%%%%%%%%%%%%%%%%%%%%%%%%
%%%%%%%%%%%%%%%%%%%%%%%%%%%%%%%%%%%%%%%%%%%%%%%%%%%%%%%%%%%%%%%%%%%%%%%%%%%%%%%%%%%%%%%%%%%%%%%%%%%%%%%%%%%%%%%%%%%%%%%%%%%%%%%
%        concentration profiles and its zoom in for h = 1
%%%%%%%%%%%%%%%%%%%%%%%%%%%%%%%%%%%%%%%%%%%%%%%%%%%%%%%%%%%%%%%%%%%%%%%%%%%%%%%%%%%%%%%%%%%%%%%%%%%%%%%%%%%%%%%%%%%%%%%%%%%%%%%
\begin{figure}[ht!]
\begin{minipage}[ht]{0.5\linewidth}
\center{\includegraphics[width=1\linewidth]{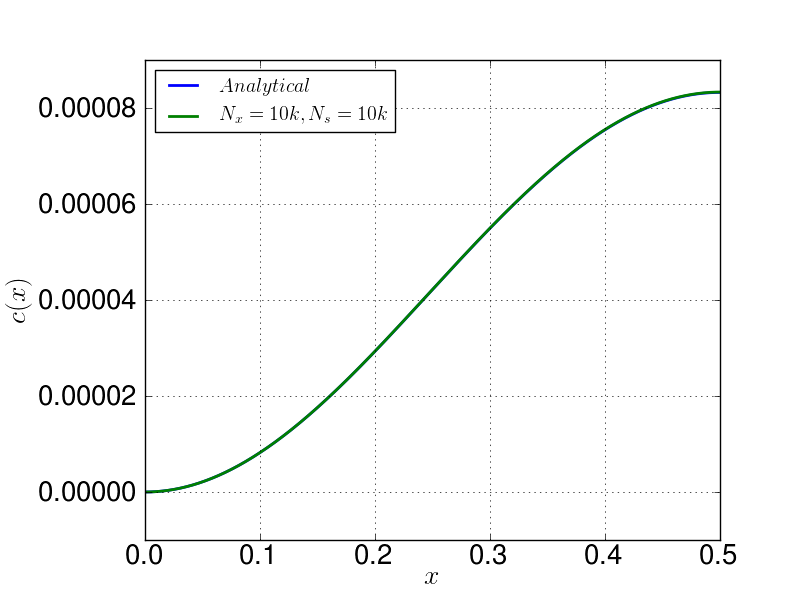}}
\caption{\small{The concentration profile calculated for $h_m=0.5$, $N_x=10k$, $N_s=10k$.}}
\label{concentration_h1_x1_s10_fig}
\end{minipage}
\hfill
\begin{minipage}[ht]{0.5\linewidth}
\center{\includegraphics[width=1\linewidth]{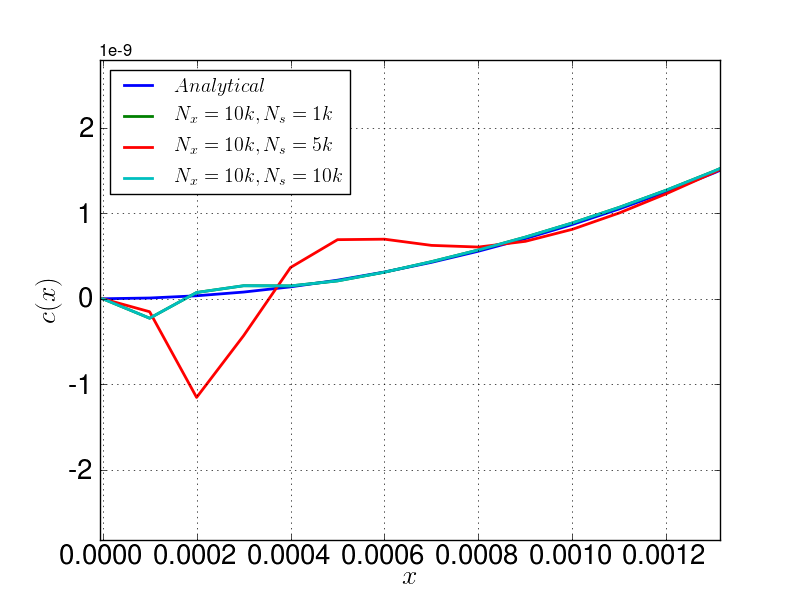}}
\caption{\small{Zoom in concentration profile (Fig.\ref{concentration_h1_x1_s10_fig}) calculated for $h_m$ = 0.5 and different values of grid dots, $N_x, N_s$.}}
\label{concentration_h1_x1_s1_10_zi_fig}
\end{minipage}
\end{figure} 
%%%%%%%%%%%%%%%%%%%%%%%%%%%%%%%%%%%%%%%%%%%%%%%%%%%%%%%%%%%%%%%%%%%%%%%%%%%%%%%%%%%%%%%%%%%%%%%%%%%%%%%%%%%%%%%%%%%%%%%%%%%%%%%
%        concentration profiles and its zoom in for h = 2
%%%%%%%%%%%%%%%%%%%%%%%%%%%%%%%%%%%%%%%%%%%%%%%%%%%%%%%%%%%%%%%%%%%%%%%%%%%%%%%%%%%%%%%%%%%%%%%%%%%%%%%%%%%%%%%%%%%%%%%%%%%%%%%
\begin{figure}[ht!]
\begin{minipage}[ht]{0.5\linewidth}
\center{\includegraphics[width=1\linewidth]{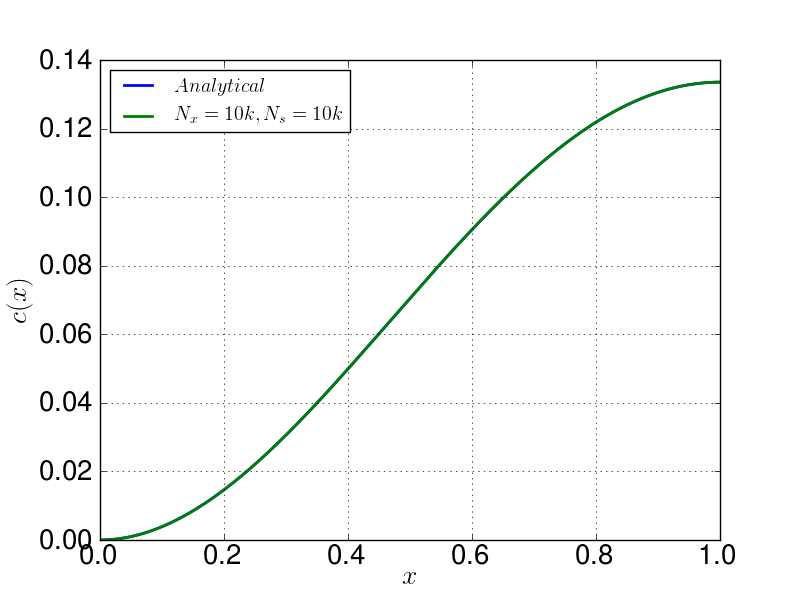}}
\caption{\small{The concentration profile calculated for $h_m=1$, $N_x=10k$, $N_s=10k$.}}
\label{concentration_h2_x10_s10_fig}
\end{minipage}
\hfill
\begin{minipage}[ht]{0.5\linewidth}
\center{\includegraphics[width=1\linewidth]{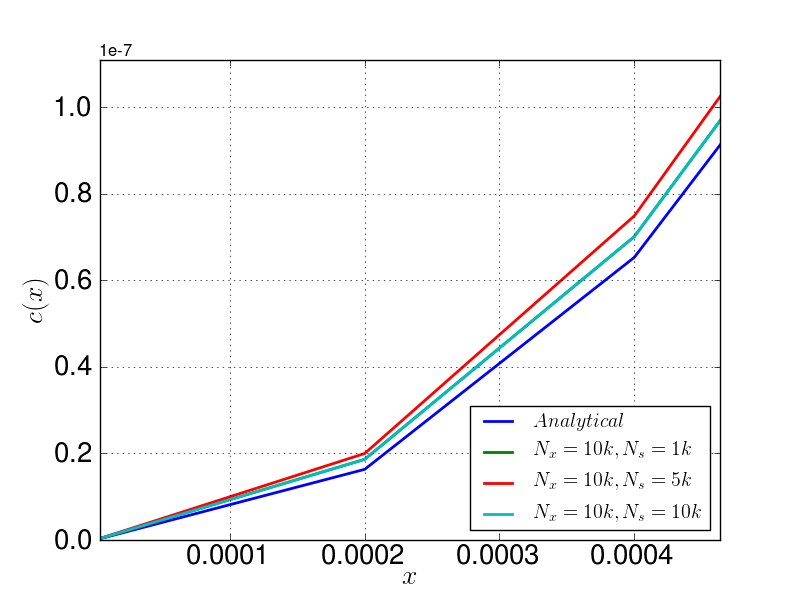}}
\caption{\small{Zoom in concentration profile (Fig.\ref{concentration_h2_x10_s10_fig}) calculated for $h_m=1$ and different values of grid dots, $N_x, N_s$.}}
\label{concentration_h2_x10_s1_10_zi_fig} 
\end{minipage}
\end{figure} 
	 
\begin{figure}[ht!]
\begin{minipage}[ht]{0.5\linewidth}
\center{\includegraphics[width=1\linewidth]{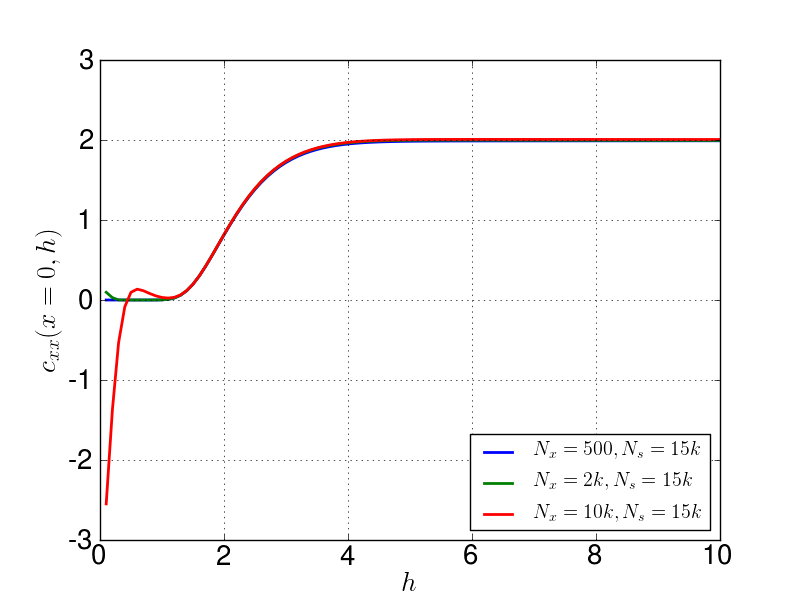}}
\caption{\small{The second derivative of the concentration profile.}}
\label{second_der_num3_fig}
\end{minipage}
\hfill
\begin{minipage}[ht]{0.5\linewidth}
\center{\includegraphics[width=1\linewidth]{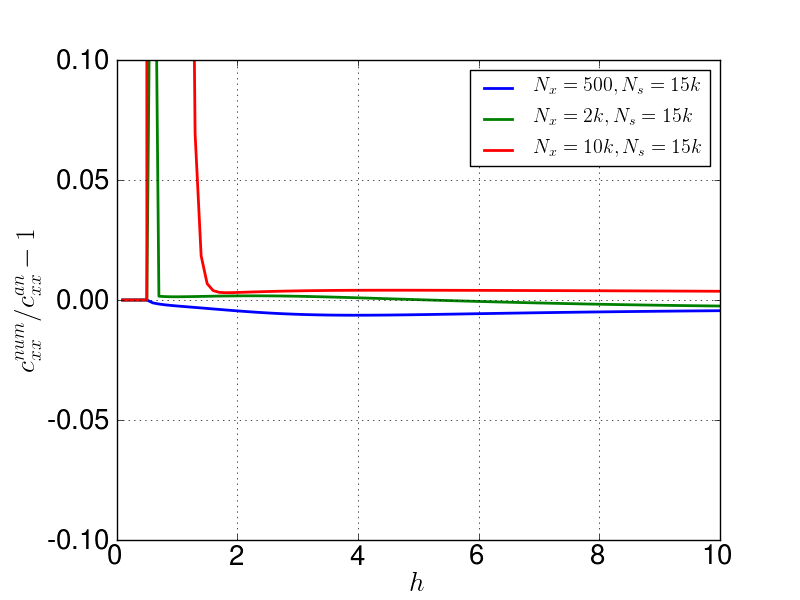}}
\caption{\small{The comparison of the second derivatives of the concentration profile with its analytical solution via the relative error.}}
\label{second_der_relerr_fig} 
\end{minipage}
\end{figure}

	 When the separation between the plates has been increased up to $h=2$, the oscillations have vanished for any values of the parameter $N_s$ (see 
	 Fig.\ref{concentration_h2_x10_s1_10_zi_fig}). In Fig.\ref{second_der_num3_fig}, one can see the second derivative of the concentration profile 
	 calculated for different values of the grid parameters, $N_x, N_s$. The comparison with its analytical solution via the relative error can be found in 
	 Fig.\ref{second_der_relerr_fig}. Based on this result, we can obtain the whole curve for the second derivative based on the following parts: 
	 a) $h=0.1..1.1$ with $N_x=500$, $N_s=15k$; b) $h = 1.1..1.8$ with $N_x=2k$, $N_s=15k$; c) $h = 1.8..10$ with $N_x=10k$, $N_s=15k$,
	 thus, providing better the accuracy along the whole curve.  
 
%%%%%%%%%%%%%%%%%%%%%%%%%%%%%%%%%%%%%%%%%%%%%%%%%%%%%%%%%%%%%%%%%%%%%%%%%%%%%%%%%%%%%%%%%%%%%%%%%%%%%%%%%%%%%%%%%%%%%%%%%%%%%%%
%        second derivative and its comparison with analytical solution
%%%%%%%%%%%%%%%%%%%%%%%%%%%%%%%%%%%%%%%%%%%%%%%%%%%%%%%%%%%%%%%%%%%%%%%%%%%%%%%%%%%%%%%%%%%%%%%%%%%%%%%%%%%%%%%%%%%%%%%%%%%%%%%

\begin{figure}[ht!]
\begin{minipage}[ht]{0.5\linewidth}
\center{\includegraphics[width=1\linewidth]{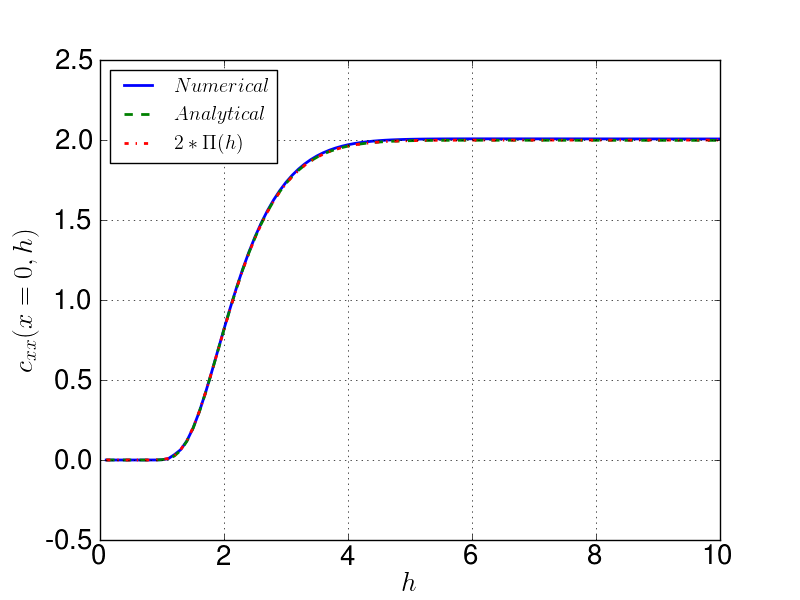}}
\caption{\small{The second derivative of the concentration profile. The resulting curves have been calculated for the following parameters: 
	1)\textbf{force} $N_x=8k$, $N_s=6k$, $N_h=100$;  2)\textbf{second derivative}. We divided $h$ by two intervals 
	a) $h_1\in[0.1..1.3]$, $N_x=500$, $N_s=15k$ and b) $h_2\in[1.3..10]$, $N_x=10k$, $N_s=15k$.}}
\label{second_der_num_an_pi_fig}
\end{minipage}
\hfill
\begin{minipage}[ht]{0.5\linewidth}
\center{\includegraphics[width=1\linewidth]{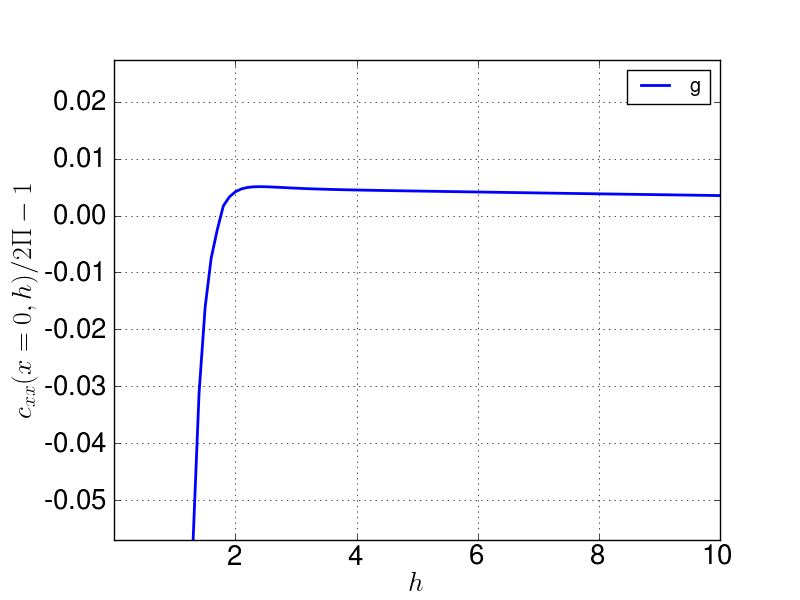}}
\caption{\small{Comparison of the second derivative with force between plates via Eq.(\ref{repulsion_cxx_2force}). 
	The resulting curves have been calculated for the following parameters: 
	1)\textbf{force} $N_x= 8k$, $N_s=6k$, $N_h=100$;  2)\textbf{second derivative}. We divided $h$ by two intervals: 
	a) $h_1\in[0.1..1.3]$, $N_x=500$, $N_s=15k$ and b) $h_2\in[1.3..10]$, $N_x=10k$, $N_s = 15k$.}}
\label{second_der_pi_num_fig}
\end{minipage}
\end{figure}

	The result for the second derivative of the concentration profile was required to verify the analytical relation: $\Pi(h) = c_{xx}(0)/2$. To this end
	we define the deviation function
\begin{equation}
\label{repulsion_cxx_2force}
        g(h) = \frac{c_{xx}(x = 0, h)}{2\Pi(h)} -1 
\end{equation}
	In Fig.\ref{second_der_num_an_pi_fig}, one can find the second derivative of the concentration profile and the force depicted together.
	The result of their comparison using Eq.(\ref{repulsion_cxx_2force}) is shown in Fig.\ref{second_der_pi_num_fig}.

%%%%%%%%%%%%%%%%%%%%%%%%%%%%%%%%%%%%%%%%%%%%%%%%%%%%%%%%%%%%%%%%%%%%%%%%%%%%%%%%%%%%%%%%%%%%%%%%%%%%%%%%%%%%%%%%%%%%%%%%%%%%%%%
%        second derivative and its comparison with force
%%%%%%%%%%%%%%%%%%%%%%%%%%%%%%%%%%%%%%%%%%%%%%%%%%%%%%%%%%%%%%%%%%%%%%%%%%%%%%%%%%%%%%%%%%%%%%%%%%%%%%%%%%%%%%%%%%%%%%%%%%%%%%%

%%%%%%%%%%%%%%%%%%%%%%%%%%%%%%%%%%%%%%%%%%%%%%%%%%%%%%%%%%%%%%%%%%%%%%%%%%%%%%%%%%%%%%%%%%%%%%%%%%%%%%%%%%%%%%%%%%%%%%%%%%%%%%%%%%%%%%%%%%%%%%%%%%%%%%%%%%%%%
%           General case.
%%%%%%%%%%%%%%%%%%%%%%%%%%%%%%%%%%%%%%%%%%%%%%%%%%%%%%%%%%%%%%%%%%%%%%%%%%%%%%%%%%%%%%%%%%%%%%%%%%%%%%%%%%%%%%%%%%%%%%%%%%%%%%%%%%%%%%%%%%%%%%%%%%%%%%%%%%%%%
\section{Non ideal polymer solution}
\label{sec:iterative_procedure}
	As we have already shown in Sec.\ref{sec:polymers_force_2plates}, the Edwards equation, written in the reduced variables, is	 
\begin{equation}
\label{repulsion_Edwards_diff_eq}
         \frac{\partial q(x, s)}{\partial s} = \frac{\partial^2 q(x, s)}{\partial x^2} - w(x)q(x, s)
\end{equation}
         where  
\begin{equation}
\label{repulsion_Edwards_field}
	 w(x) = v_N(c(x)/c_b-1)+w_N((c(x)/c_b)^2-1) 
\end{equation}
        and
\begin{equation}
\label{repulsion_Edwards_conc}
	 c(x)/c_b = \int\limits_{0}^{1}\mathrm{d}s\,q(x, s)q(x, 1-s)
\end{equation}        
        with $v_N=\text{v}c_bN$ and $w_N=\text{w}c_b^2N/2$. The propagator is defined in the range: $q(x, s)\in [0, h]\times [0, 1]$
        and the boundary conditions are: $q(0) = q_{x}(h_m) = 0$. 
        The SCFT equations are viewed as a set of simultaneous nonlinear equations. The available analytical tools for studying the SCFT equations 
        are rather crude techniques that rarely provide a complete picture of the solution. Therefore, it is important to have an effective numerical solution 
        of the equations. In this section, we will develop the strategy for the numerical solution of the SCFT equation in a computationally efficient manner
        following \cite{Fredrickson_book}.

	Usually, the solution of such self-consistent system of equations is sought iteratively, solving the modified diffusion equation 
	with the known field, $w(x)$, on each iterative step. As in the previous section, we use the Crank-Nicolson method for solving the modified
	diffusion equation with the known field, $w(x)$ (see AppendixB). 
	
	In order to develop the iterative algorithm, we apply the steepest descent strategy to find the thermodynamic potential, $\Omega$.
	Let us denote the effective Hamiltonian of the system as $\Phi[w]$. Then,	
$$
	-\Omega = \min\limits_{w}\Phi[w]
$$
	If one starts at some molecular field $w^{(0)}(x)$, which does not corresponds to the minimum of $\Phi[w]$, then the shift in the steepest descent 
	direction is required. The shift is defined by the gradient:
$$
	Y(x) \equiv \frac{\delta \Phi[w]}{\delta w(x)}
$$
	This gradient produces a new field:
$$
	w(x) = w^{(0)}(x) - \epsilon Y(x)
$$
	where the parameter $\epsilon>0$ is small and $\Phi[w] < \Phi[w^{(0)}]$. For a given $w(x)$, we can  write $Y(x) = \tilde{c}(x)-c(x)$, where
	$\tilde{c}(x)$ is defined by Eq.(\ref{repulsion_Edwards_field}) and $c(x)$ is defined by Eq.(\ref{repulsion_Edwards_conc}). 
	
	The new field $w(x)$ can be optimized treating $\delta w(x) = w(x)-w^{0}(x)$ as a perturbation and expanding the gradient, $Y(x)$, 
	at $w(x) = w^{(0)}(x) + \delta w(x)$ as $Y = Y^{(0)} + \hat{J}\delta w$, where $\hat{J}$ is the Jacobian operator of the field $Y = Y[w]$ with the kernel:
$$
	J(x, x')  = \frac{\delta^2\Phi[w]}{\delta w(x)\delta w(x')}
$$
	The minimization condition is $Y=0$, that leads to $\delta w =-JY^{(0)}$. Correspondingly, one can write $w(x) = w^{(0)}(x)-\hat{J}^{-1}Y^{(0)}$. 
	This result can be further simplified in the perturbative case of a nearly homogeneous system \cite{Fredrickson_book}. In this case, the Jacobian,
	written in the reciprocal space, has nearly diagonal highest wave vectors:
$$
	J_k \simeq \frac{1}{v^*} + cNg_D(k^2Na^2)
$$
	where $v^* = \partial \mu_{int}/\partial c$ and 
\begin{equation}
\label{repulsion_debye_fun}
	g_D(y) = \frac{2}{y}\left(y-1+e^{-y}\right)
\end{equation}
	is the Debye function. The field perturbation in the Fourier space can be written as
\begin{equation}
\label{repulsion_field_iterations}
	\delta w_k \simeq -\frac{Y^{(0)}_k}{1/v^* + cNg_D(k^2aN)} \simeq \frac{\tilde{w}_k-w^{(0)}_k}{1 + cv^*Ng_D(k^2aN)}
\end{equation}
	Here $\tilde{w}$ is defined by Eq.(\ref{repulsion_Edwards_field}) where $c(x)$ is defined in Eqs.(\ref{repulsion_Edwards_conc}, \ref{repulsion_Edwards_diff_eq})
	with $w(x) = w^{(0)}(x)$.
	
	The above analysis leads to the following iteration procedure, that is used to solve the SCFT equations. We start with some initial 
	guess $w^{(0)}$ for the molecular field. Then	
\begin{equation}
\label{repulsion_iteration_procedure}
	w^{new}_k = w^{(0)}_k + p(k)\left(\tilde{w}_k - w^{0}_k\right),
\end{equation}
	where $w^{(new)}_k$ is the cosine Fourier transformation of $w^{new}(x)$, and
\begin{equation}
\label{repulsion_mixing_p}
         p(k) = \frac{1}{\Lambda + Ag_D(k)}
\end{equation}
	where $A = (cv^*N)_{c=\bar{c}} = (v_N\bar{c} + w_N\bar{c}^2)N$ and 
$$
	\bar{c} = \frac{1}{h}\int\limits_{0}^{h}\frac{c(x)}{c_b}\mathrm{d}x
$$
	is the average concentration in the slit. 	
	The parameter $\Lambda$ is replacing $1$ in Eq.(\ref{repulsion_field_iterations}). In practice, it is necessary to increase $\Lambda$ beyond $1$
	to avoid instabilities which take place far from the equilibrium. Some other strategies of solving the modified diffusion equation for different 
	polymer systems are given in \cite{Fredrickson_book}.

	The iterative procedure for solving Eq.(\ref{repulsion_Edwards_diff_eq}) in the self consistent field Eq.(\ref{repulsion_Edwards_field}) is listed
	step-by-step below.\\
        \textbf{Iterative procedure} :

        0. Choose the initial field $w^{in}(x)$.

        1. Using the initial field $w^{in}(x)$, solve the diffusion equation Eq.(\ref{repulsion_Edwards_diff_eq}).

        2. Then, the new field $\tilde{w}(x)$ is generated using Eqs.(\ref{repulsion_Edwards_conc}, \ref{repulsion_Edwards_field}).

 	3. If, for some preselected convergence parameter, $\delta$, the condition
$$
        |\tilde{w}(x) - w^{in}(x)| < \delta    
$$
        is satisfied for all $x$, then we obtained a converged solution and can stop iterating here. In the opposite case, we go to $\textbf{Step.4}$. 
	Numerically, we will control it by a norm demanding
$$
        \text{norm}(w) = \frac{1}{N_x}\sum\limits_{i=0}^{N_x-1}\bigg\vert \tilde{w}_i - w_{i}^{in}\bigg\vert < \delta
$$
        where the index $i$ is the grid point number for the variable $x$. 

        4. Apply a discrete cosine Fourier transformation to the fields $w^{in}(x)$ and $\tilde{w}(x)$.

        5. Find the new field using simple mixing scheme in Fourier space
$$
        w_{k}^{new} = w_{k}^{in} + p(k)(\tilde{w}_{k} - w_{k}^{in})
$$
        where the index $k$ denotes the number of Fourier component of the field and the mixing parameter, $p(k)$, is defined in Eq.(\ref{repulsion_mixing_p}).

	6. Return to the real space, using the inverse discrete cosine Fourier transformation.

        7. $w^{in}(x) = w^{new}(x)$. Then return to $\textbf{Step.1}$. 

        8. The iterations are ceased when the solution is satisfied to the condition in $\textbf{Step.3}$, or else, when a
        preset maximum number of iterations, $N_{it}^{max}$ is reached.

        \textbf{Parameters}: For this iteration procedure we use the following numerical parameters: $\delta = 10^{-8}$, $N_{it}^{max} = 10000$.
        \begin{table}[ht!]
\caption{The dependence of $\Lambda$ on the virial parameters.} 
\label{tabular:table_Lambda} 
\begin{center}
  \begin{tabular}{ | c | c | c | c | c | }
    \hline
    $v_N$     & 10 & 30 & 100 & 100\\ \hline
    $w_N$     & 0  & 0  &   0 & 100\\ \hline
    $\Lambda$ & 1  & 5  &  10 &  30\\
    \hline
  \end{tabular}
\end{center} 
\end{table}
	In Table.$\ref{tabular:table_Lambda}$ one can see how the parameter, $\Lambda$, depends on the virial parameters.
	The iterative procedure has been tested for different values of parameter 
	$\Lambda$. The best convergence (fewest number of iterations) is reached at $\Lambda=1$. But at large virial parameters, to avoid the instabilities in the 
	iterative procedure, we have to increase it up to $\Lambda = 50..100$.        

	\textbf{Initial guess.}
	In  $\textbf{Step.0}$ of the iterative procedure, it is necessary to choose the initial guess, $w^{in}(x)$. There are several ways to make it.\\	
	\textbf{a. $w^{in}(x)=0$}:\\ 
	In this case, we solve the homogeneous diffusion equation, Eq.(\ref{repulsion_homo_diff_eq}), at the first iterative step. 
	Then, using the solution in Eqs.(\ref{repulsion_Edwards_conc}, \ref{repulsion_Edwards_field}), 
	we obtain the initial approximation, $w^{in}(x)\ne 0$ for the iterative procedure. 
	The iterative procedure with this initial guess works very well, but we should control the parameters $\Lambda$ and $h$, 
	because the convergence is rather sensitive to the parameters, especially for large virial parameters.			
%%%%%%%%%%%%%%%%%%%%%%%%%%%%%%%%%%%%%%%%%%%%%%%%%%%%%%%%%%%%%%%%%%%%%%%%%%%%%%%%%%%%%%%%%%%%%%%%%%%%%%%%%%%%%%%%%%%%%%%%%%%%%%%
%        Step-by-step  1) h = 0.31; v_N = 100; w = 0;   2) 1) h = 0.23; v_N = 100; w = 100;
%%%%%%%%%%%%%%%%%%%%%%%%%%%%%%%%%%%%%%%%%%%%%%%%%%%%%%%%%%%%%%%%%%%%%%%%%%%%%%%%%%%%%%%%%%%%%%%%%%%%%%%%%%%%%%%%%%%%%%%%%%%%%%%
\begin{figure}[ht!]
\begin{minipage}[ht]{0.5\linewidth}
\center{\includegraphics[width=1\linewidth]{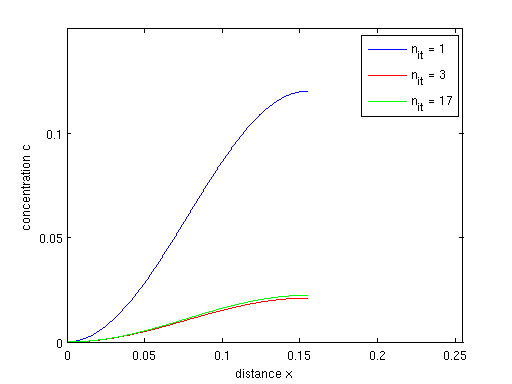}}
\caption{\small{The iterative procedure for the concentration profile. This procedure converges after $18$ iterations. 
	 Fixed parameters: $h = 0.31$, $v_N = 100$, $w_N = 0$, $\Lambda = 10$.}}
\label{iter_field_h031_v100_fig}
\end{minipage}
\hfill
\begin{minipage}[ht]{0.5\linewidth}
\center{\includegraphics[width=1\linewidth]{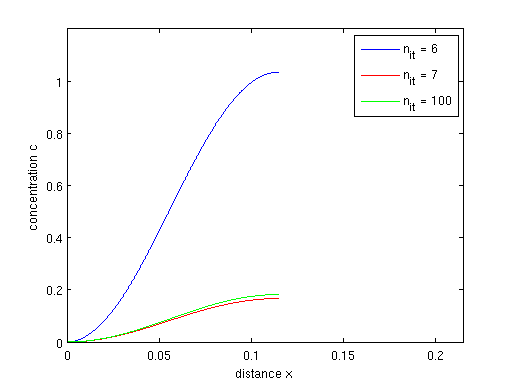}}
\caption{\small{The iterative procedure for the concentration profile. This procedure converges after $100$ iterations. 
	 Fixed parameters: $h = 0.23$, $v_N = 100$, $w_N = 100$, $\Lambda = 30$.}}
\label{iter_field_h023_v100_w100_fig}
\end{minipage}
\end{figure}
        In Figs.\ref{iter_field_h031_v100_fig}--\ref{iter_field_h023_v100_w100_fig}, we show the iterative procedure of finding the equilibrium concentration profile
	using the initial guess, $w^{in}(x) = 0$. \\
	\textbf{b. Add a small segment}:\\ 
	We start from a small initial half-separation, $h_{m0}=h_0/2$ (for example, $h_0$ = 0.1), for which the equilibrium solution is found using 
	the initial guess, $w^{in}(x)=0$. In order to find the initial guess for the half-separation, $h_m+\Delta h_m$, we 
	add a small segment, $\Delta h_m$, postulating that $c(x) = c(h_m)$, $x \in [h_m, h_m + \Delta h_m]$.		
	Therefore, as the initial guess for the half-separation $h_m+\Delta h_m$, we use $w^{in}$ obtained by Eq.(\ref{repulsion_Edwards_field}).	
%%%%%%%%%%%%%%%%%%%%%%%%%%%%%%%%%%%%%%%%%%%%%%%%%%%%%%%%%%%%%%%%%%%%%%%%%%%%%%%%%%%%%%%%%%%%%%%%%%%%%%%%%%%%%%%%%%%%%%%%%%%%%%%
%        Concentration profiles at diferent initial guess v_N = 10 and w_N = 0 for h = 1.5 and dh = 0.1
%%%%%%%%%%%%%%%%%%%%%%%%%%%%%%%%%%%%%%%%%%%%%%%%%%%%%%%%%%%%%%%%%%%%%%%%%%%%%%%%%%%%%%%%%%%%%%%%%%%%%%%%%%%%%%%%%%%%%%%%%%%%%%%
\begin{figure}[ht!]
\begin{minipage}[ht]{0.5\linewidth}
\center{\includegraphics[width=1\linewidth]{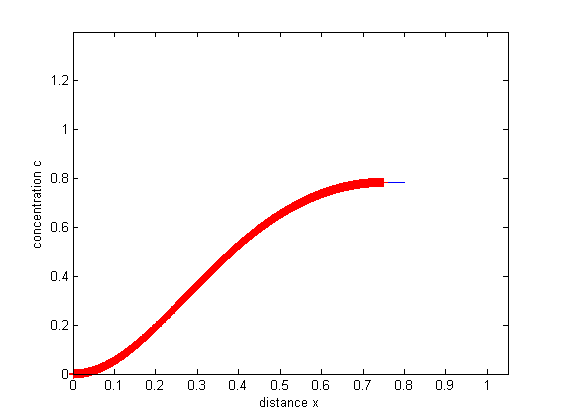}}
\caption{\small{We obtain new initial approximation using the solution obtained on the previous step, adding the segment, $\Delta h_m$ to the end of the curve. 
	        We rescale this approximation using the old number of space points, $N_x$.}}
\label{segment_fig}
\end{minipage}
\hfill
\begin{minipage}[ht]{0.5\linewidth}
\center{\includegraphics[width=1\linewidth]{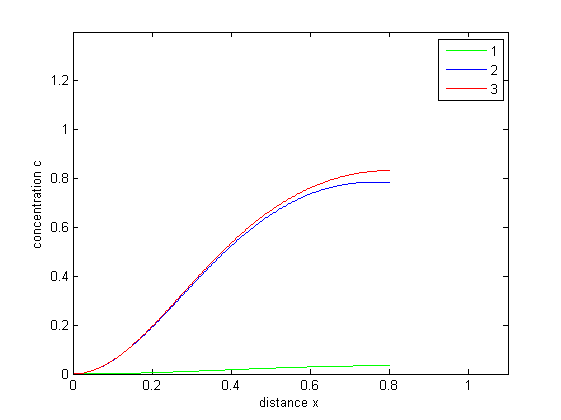}}
\caption{\small{The concentration profile obtained using: 1) the initial guess $w^{in}(x)=0$ ($n_{it} = 50$ iterations to find the equilibrium solution);
	         2) the initial guess obtained adding the segment ($n_{it} = 33$); 
		 3) the equilibrium concentration profile. 
	          Fixed parameters: $v_N = 10, w_N = 0$.}}
\label{initial_guess_fig}
\end{minipage}
\end{figure}	
	This procedure is shown in Figs.\ref{segment_fig}--\ref{initial_guess_fig}. 
	This is a suitable strategy, because our final aim is to calculate the thermodynamic potential for  
	different separations between plates, $h_m = h_{m0} + \sum\Delta h_m$. 
%%%%%%%%%%%%%%%%%%%%%%%%%%%%%%%%%%%%%%%%%%%%%%%%%%%%%%%%%%%%%%%%%%%%%%%%%%%%%%%%%%%%%%%%%%%%%%%%%%%%%%%%%%%%%%%%%%%%%%%%%%%%%%%
%        Thermodynamic potential and force between plates
%%%%%%%%%%%%%%%%%%%%%%%%%%%%%%%%%%%%%%%%%%%%%%%%%%%%%%%%%%%%%%%%%%%%%%%%%%%%%%%%%%%%%%%%%%%%%%%%%%%%%%%%%%%%%%%%%%%%%%%%%%%%%%%
\section{Numerical results} 
\label{sec:repulsion_num_res}
In the previous section, we described the numerical procedure of finding the equilibrium self-consistent field (or concentration).
	For the known field, we can calculate the thermodynamic potential of the system between two plates. As we have already shown 
	in Sec.\ref{sec:polymers_force_2plates} this potential is\footnote{For obvious reasons it is more convenient to use the thermodynamic potential, 
	which, at $h\rightarrow\infty$, is equal to $0$, i.e. $\Omega(c_b, h) \leftarrow \Omega(c_b, h) - \Omega(c_b, \infty)$,
	where $\Omega(c_b, \infty)$ physically corresponds to the thermodynamic potential created by two independent plates.}
\begin{equation}
\label{repulsion_therm_pot_general}
	\Omega(c_b, h) = - \left(Q_{in}[w, h] - h\right) - \int\limits_0^h\mathrm{d}x\, \left(\frac{v_N}{2}\left(c/c_b)^2-1\right) +
       	 \frac{2w_N}{3}\left((c/c_b)^3-1\right)\right) 
\end{equation}
	where 
\begin{equation}
\label{repulsion_part_fun}
	Q_{in}[w, h] = \int\limits_0^h\mathrm{d}x\, q(x, 1)
\end{equation}
	is the partition function of a chain placed in the field, $w(x)$.

\begin{figure}[ht!]
\begin{minipage}[ht]{0.5\linewidth}
\center{\includegraphics[width=1\linewidth]{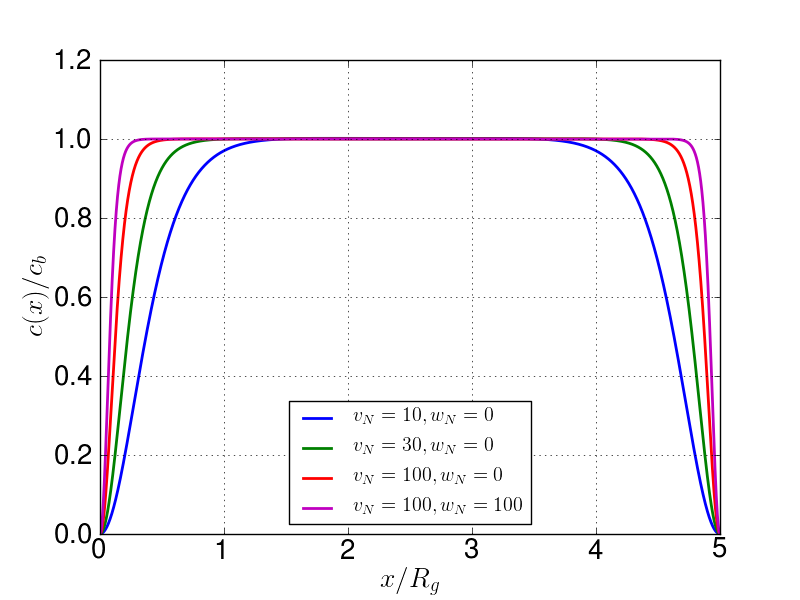}}
\caption{\small{Concentration profiles, Eq.(\ref{repulsion_Edwards_conc}), for different values of virial parameters. 
		Fixed parameters: $N_x = N_s = 5000$.}}
\label{concentrations_fig}
\end{minipage}
\hfill
\begin{minipage}[ht]{0.5\linewidth}
\center{\includegraphics[width=1\linewidth]{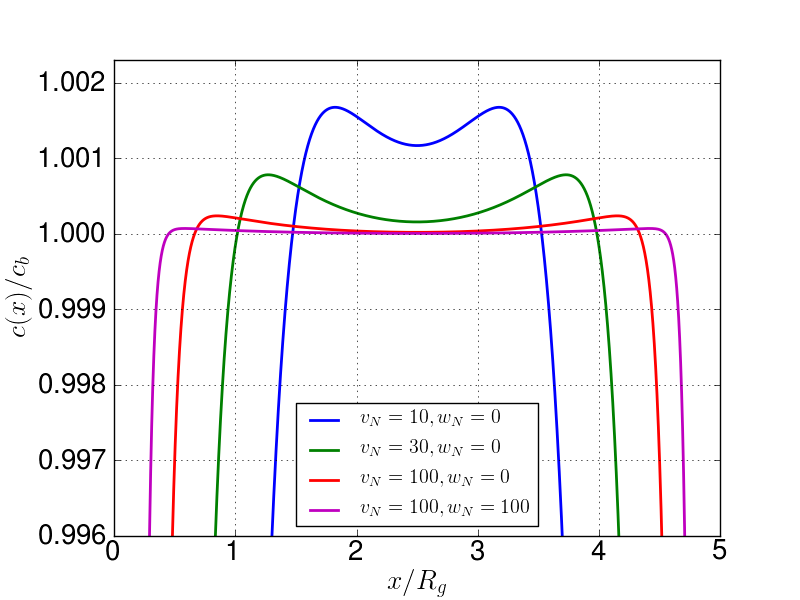}}
\caption{\small{Concentration profiles, Fig.\ref{concentrations_fig}, zoomed around $c(x)/c_b \simeq 1.0$.}}
\label{concentrations_scaled_fig}
\end{minipage}
\end{figure}

	In Figs.\ref{concentrations_fig}--\ref{concentrations_scaled_fig} we present the equilibrium concentration profiles calculated for different values 
	of the virial parameter.  The concentration profile in the slit becomes non monotonic for $0<x<h_m=h/2$ showing a maximum at $x<h_m$ and the local 
	minimum at $x=h_m$. The position of the maximum shifts towards the boundary as $v_N$ or $w_N$ increase.

\begin{figure}[ht!]
\begin{minipage}[ht]{0.5\linewidth}
\center{\includegraphics[width=1\linewidth]{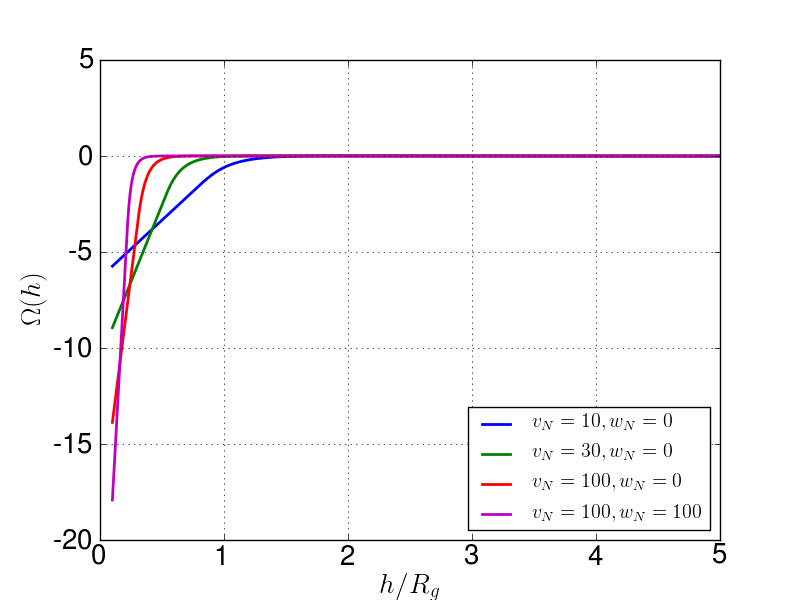}}
\caption{\small{Thermodynamic potential, Eq.(\ref{repulsion_therm_pot_general}), for different values of virial parameters. Fixed parameters: $N = N_x = N_s = 5000$.}}
\label{therm_pot_all_fig}
\end{minipage}
\hfill
\begin{minipage}[ht]{0.5\linewidth}
\center{\includegraphics[width=1\linewidth]{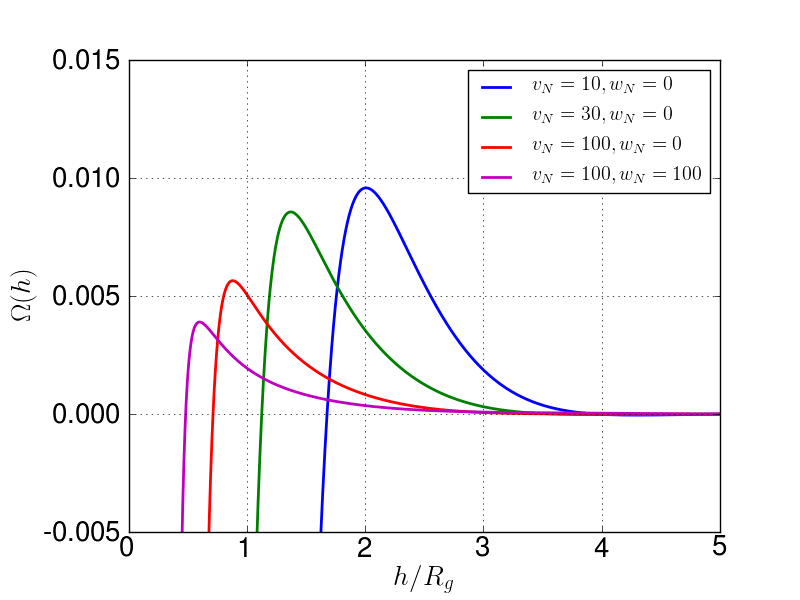}}
\caption{\small{Thermodynamic potential, Fig.\ref{therm_pot_all_fig}, zoomed in around $\Omega(x)\simeq 0$. }}
\label{therm_pot_all_scaled_fig}
\end{minipage}
\end{figure}

	The dependence $\Omega(h)$ is shown in Figs.\ref{therm_pot_all_fig}--\ref{therm_pot_all_scaled_fig} for different pairs $v_N, w_N$. 
	From the rescaled Fig.\ref{therm_pot_all_scaled_fig} one can observe a peak of the potential, $\Omega = \Omega^*$ attained at some $h=h^*$.
	Both quantities, $\Omega^*$ and $h^*$ decrease with the virial parameters, $v_N, w_N$. The critical separation, $h^*$ correlates very well with the 
	concentration maximum at large $h$.
%%%%%%%%%%%%%%%%%%%%%%%%%%%%%%%%%%%%%%%%%%%%%%%%%%%%%%%%%%%%%%%%%%%%%%%%%%%%%%%%%%%%%%%%%%%%%%%%%%%%%%%%%%%%%%%%%%%%%%%%%%%%%%%
%       concentration profile at v_N = 10, 30, 100, 100; w_N = 0, 0, 0, 100
%%%%%%%%%%%%%%%%%%%%%%%%%%%%%%%%%%%%%%%%%%%%%%%%%%%%%%%%%%%%%%%%%%%%%%%%%%%%%%%%%%%%%%%%%%%%%%%%%%%%%%%%%%%%%%%%%%%%%%%%%%%%%%%
%%%%%%%%%%%%%%%%%%%%%%%%%%%%%%%%%%%%%%%%%%%%%%%%%%%%%%%%%%%%%%%%%%%%%%%%%%%%%%%%%%%%%%%%%%%%%%%%%%%%%%%%%%%%%%%%%%%%%%%%%%%%%%%
%        Thermodynamic potential at v_N = 10, 30, 100, 100; w_N = 0, 0, 0, 100
%%%%%%%%%%%%%%%%%%%%%%%%%%%%%%%%%%%%%%%%%%%%%%%%%%%%%%%%%%%%%%%%%%%%%%%%%%%%%%%%%%%%%%%%%%%%%%%%%%%%%%%%%%%%%%%%%%%%%%%%%%%%%%%
	The interaction force becomes non monotonic as well; it is repulsive, for $h>h^*$, while attractive at $h<h^*$.

	Let us investigate the accuracy of the numerical calculations.	
	We should obtain the numerical solution with sufficiently good accuracy, while an elapsed time and computer resources should be acceptable. 
	Thereby, we want to find an optimal numerical solution. We consider many different combinations of grid dots, $N_x$, $N_s$, 
	for each set of the virial parameters, $\{v_N, w_N\}$ and 
	compare every solution with high resolution results ($N_x = $10k, $N_s = $10k).
	Quantitatively, as a measure of the accuracy, we take the ratio between the barrier hight, $\Omega^*$     
	calculated for any combination of grid dots and the barrier height of the high resolution calculation:
$$
	\epsilon = \Omega^*/\Omega^*_{hr} 
$$    
        In Tabs.\ref{tabular:rel_err_v10}--\ref{tabular:rel_err_v01k_w01k} one can find the results of computations 
	of the thermodynamic potential for different values of virial parameters, $v_N$, $w_N$ and different values of grid dots, $N_x$, $N_s$ 	
%%%%%%%%%%%%%%%%%%%%%%%%%%%%%%%%%%%%%%%%%%%%%%%%%%%%%%%%%%%%%%%%%%%%%%%%%%%%%%%%%%%%%%%%%%%%%%%%%%%%%%%%%%%%%%%%%%%%%%%%%%%%%%%%%%%%%%%%%%%%
% Table1 v = 10, w = 0
%%%%%%%%%%%%%%%%%%%%%%%%%%%%%%%%%%%%%%%%%%%%%%%%%%%%%%%%%%%%%%%%%%%%%%%%%%%%%%%%%%%%%%%%%%%%%%%%%%%%%%%%%%%%%%%%%%%%%%%%%%%%%%%%%%%%%%%%%%%%
\begin{table}[ht!]
\caption{The dependence of the thermodynamic potential barrier height on the grid parameters $N_x$ and $N_s$. Fixed parameters: $v_N$ = 10, $w_N$ = 0.} 
\label{tabular:rel_err_v10} 
\begin{flushleft}
  \begin{tabular}{ | c | c | c | c | c | c | c | c | c | c | c | }
    \hline
    $N_x$                           & 10000& 2000 & 3000 & 4000 & 6000 & 6000 & 6000 & 7000 & 7000 & 7000  \\ \hline
    $N_s$                           & 10000& 6000 & 6000 & 6000 & 2000 & 3000 & 4000 & 2000 & 3000 & 4000   \\ \hline
    $\Omega^*,\times 10^{-3}$ &10.022& 8.264& 8.983& 9.353& 9.768& 9.748& 9.738& 9.875& 9.856& 9.845      \\ \hline
    $\epsilon$                      &   1.0& 0.825& 0.896& 0.933& 0.975& 0.973& 0.972& 0.985& 0.983& 0.982      \\
    \hline
  \end{tabular}
\end{flushleft} 
%%%%%%%%%%%%%%%%%%%%%%%%%%%%%%%%%%%%%%%%%%%%%%%%%%%%%%%%%%%%%%%%%%%%%%%%%%%%%%%%%%%%%%%%%%%%%%%%%%%%%%%%%%%%%%%%%%%%%%%%%%%%%%%%%%%%%%%%%%%%
%v = 10, w = 0; extention of the table1 centered by left 
%%%%%%%%%%%%%%%%%%%%%%%%%%%%%%%%%%%%%%%%%%%%%%%%%%%%%%%%%%%%%%%%%%%%%%%%%%%%%%%%%%%%%%%%%%%%%%%%%%%%%%%%%%%%%%%%%%%%%%%%%%%%%%%%%%%%%%%%%%%%
\begin{flushleft}
  \begin{tabular}{ | c | c | c | c | c | }
    \hline
    $N_x$                           & 8000 & 8000 & 8000 & 5000\\ \hline
    $N_s$                           & 2000 & 3000 & 4000 & 5000\\ \hline
    $\Omega^*,\times 10^{-3}$ & 9.956& 9.936& 9.926& 9.582\\ \hline
    $\epsilon$                      & 0.993& 0.991& 0.990& 0.956\\
    \hline
  \end{tabular}
\end{flushleft}
\end{table}

%%%%%%%%%%%%%%%%%%%%%%%%%%%%%%%%%%%%%%%%%%%%%%%%%%%%%%%%%%%%%%%%%%%%%%%%%%%%%%%%%%%%%%%%%%%%%%%%%%%%%%%%%%%%%%%%%%%%%%%%%%%%%%%%%%%%%%%%%%%%
% Table2 v = 30, w = 0
%%%%%%%%%%%%%%%%%%%%%%%%%%%%%%%%%%%%%%%%%%%%%%%%%%%%%%%%%%%%%%%%%%%%%%%%%%%%%%%%%%%%%%%%%%%%%%%%%%%%%%%%%%%%%%%%%%%%%%%%%%%%%%%%%%%%%%%%%%%%
\begin{table}[ht!]
\caption{The dependence of the thermodynamic potential barrier height on the grid parameters $N_x$ and $N_s$. Fixed parameters: $v_N$ = 30, $w_N$ = 0.} 
\label{tabular:rel_err_v30} 
\begin{flushleft}
  \begin{tabular}{ | c | c | c | c | c | c | c | c | c | c | c | }
    \hline
    $N_x$                           & 10000& 2000 & 3000 & 4000 & 6000 & 6000 & 6000 & 7000 & 7000 & 7000   \\ \hline
    $N_s$                           & 10000& 6000 & 6000 & 6000 & 2000 & 3000 & 4000 & 2000 & 3000 & 4000   \\ \hline
    $\Omega^*,\times 10^{-3}$ & 8.989& 7.303& 7.979& 8.325& 8.794& 8.746& 8.725& 8.902& 8.856& 8.833  \\ \hline
    $\epsilon$                      &   1.0& 0.812& 0.888& 0.926& 0.978& 0.973& 0.971& 0.990& 0.985& 0.983  \\
    \hline
  \end{tabular}
\end{flushleft} 
%%%%%%%%%%%%%%%%%%%%%%%%%%%%%%%%%%%%%%%%%%%%%%%%%%%%%%%%%%%%%%%%%%%%%%%%%%%%%%%%%%%%%%%%%%%%%%%%%%%%%%%%%%%%%%%%%%%%%%%%%%%%%%%%%%%%%%%%%%%%
%v = 30, w = 0; extention of the table1 centered by left 
%%%%%%%%%%%%%%%%%%%%%%%%%%%%%%%%%%%%%%%%%%%%%%%%%%%%%%%%%%%%%%%%%%%%%%%%%%%%%%%%%%%%%%%%%%%%%%%%%%%%%%%%%%%%%%%%%%%%%%%%%%%%%%%%%%%%%%%%%%%%
\begin{flushleft}
  \begin{tabular}{ | c | c | c | c | c | c |}
    \hline
    $N_x$                           & 8000 & 8000 & 8000 & 5000\\ \hline
    $N_s$                           & 2000 & 3000 & 4000 & 5000 \\ \hline
    $\Omega^*,\times 10^{-3}$ & 8.983& 8.937& 8.914& 8.560\\ \hline
    $\epsilon$                      & 0.999& 0.994& 0.992& 0.952\\
    \hline
  \end{tabular}
\end{flushleft}
\end{table}

%%%%%%%%%%%%%%%%%%%%%%%%%%%%%%%%%%%%%%%%%%%%%%%%%%%%%%%%%%%%%%%%%%%%%%%%%%%%%%%%%%%%%%%%%%%%%%%%%%%%%%%%%%%%%%%%%%%%%%%%%%%%%%%%%%%%%%%%%%%%
% Table 3 v = 100, w = 0
%%%%%%%%%%%%%%%%%%%%%%%%%%%%%%%%%%%%%%%%%%%%%%%%%%%%%%%%%%%%%%%%%%%%%%%%%%%%%%%%%%%%%%%%%%%%%%%%%%%%%%%%%%%%%%%%%%%%%%%%%%%%%%%%%%%%%%%%%%%%
\begin{table}[ht!]
\caption{The dependence of the thermodynamic potential barrier height on the grid parameters $N_x$ and $N_s$. Fixed parameters $v_N$ = 100, $w_N$ = 0.} 
\label{tabular:rel_err_v01k} 
\begin{flushleft}
  \begin{tabular}{ | c | c | c | c | c | c | c | c | c | c | c |}
    \hline
    $N_x$                           & 10000 & 2000 & 3000 & 3000 & 3000 & 4000 & 4000 & 4000 & 5000 & 6000 \\ \hline
    $N_s$                           & 10000 & 6000 & 6000 & 7000 & 8000 & 6000 & 7000 & 8000 & 8000 & 2000 \\ \hline
    $\Omega^*,\times 10^{-3}$ & 5.995 & 4.755& 5.181& 5.169& 5.159& 5.453& 5.441& 5.431& 5.604& 5.946\\ \hline
    $\epsilon$                      &    1.0& 0.793& 0.864& 0.862& 0.860& 0.910& 0.906& 0.906& 0.935& 0.992 \\
    \hline
  \end{tabular}
\end{flushleft} 
%%%%%%%%%%%%%%%%%%%%%%%%%%%%%%%%%%%%%%%%%%%%%%%%%%%%%%%%%%%%%%%%%%%%%%%%%%%%%%%%%%%%%%%%%%%%%%%%%%%%%%%%%%%%%%%%%%%%%%%%%%%%%%%%%%%%%%%%%%%%
%v = 100, w = 0; extention of the table1 centered by left 
%%%%%%%%%%%%%%%%%%%%%%%%%%%%%%%%%%%%%%%%%%%%%%%%%%%%%%%%%%%%%%%%%%%%%%%%%%%%%%%%%%%%%%%%%%%%%%%%%%%%%%%%%%%%%%%%%%%%%%%%%%%%%%%%%%%%%%%%%%%%
\begin{flushleft}
  \begin{tabular}{ | c | c | c | c | c | c | c | c | c | c |}
    \hline
    $N_x$                           & 6000 & 6000 & 6000 & 7000 & 7000 & 8000 & 8000 & 8000 & 8000  \\ \hline
    $N_s$                           & 3000 & 4000 & 8000 & 3000 & 4000 & 3000 & 4000 & 5000 & 6000  \\ \hline
    $\Omega^*,\times 10^{-3}$ & 5.852& 5.805& 5.735& 5.949& 5.902& 6.022& 5.975& 5.947& 5.929 \\ \hline
    $\epsilon$                      &0.976 & 0.968& 0.957& 0.992& 0.985& 1.005& 0.997& 0.992& 0.989 \\
    \hline
  \end{tabular}
\end{flushleft}
\end{table}

%%%%%%%%%%%%%%%%%%%%%%%%%%%%%%%%%%%%%%%%%%%%%%%%%%%%%%%%%%%%%%%%%%%%%%%%%%%%%%%%%%%%%%%%%%%%%%%%%%%%%%%%%%%%%%%%%%%%%%%%%%%%%%%%%%%%%%%%%%%%
% Table4 v = 100, w = 100
%%%%%%%%%%%%%%%%%%%%%%%%%%%%%%%%%%%%%%%%%%%%%%%%%%%%%%%%%%%%%%%%%%%%%%%%%%%%%%%%%%%%%%%%%%%%%%%%%%%%%%%%%%%%%%%%%%%%%%%%%%%%%%%%%%%%%%%%%%%%
\begin{table}[ht!]
\caption{The dependence of the thermodynamic potential barrier height on the grid parameters $N_x$ and $N_s$. Fixed parameters $v_N$ = 100, $w_N$ = 100.} 
\label{tabular:rel_err_v01k_w01k} 
\begin{flushleft}
  \begin{tabular}{ | c | c | c | c | c | c | c | c | c | c | c | }
    \hline
    $N_x$                           & 10000& 3000 & 4000 & 4000 & 6000 & 6000 & 6000 & 7000 & 7000 & 8000  \\ \hline
    $N_s$                           & 10000& 6000 & 6000 & 7000 & 2000 & 3000 & 4000 & 2000 & 4000 & 2000   \\ \hline
    $\Omega^*,\times 10^{-3}$ & 4.106& 3.675& 3.772& 3.751& 4.259& 4.108& 4.032& 4.331& 4.104& 4.387 \\ \hline
    $\epsilon$                      &   1.0& 0.895& 0.918& 0.914& 1.037& 1.001& 0.982& 1.055& 0.999& 1.068 \\
    \hline
  \end{tabular}
\end{flushleft} 
%%%%%%%%%%%%%%%%%%%%%%%%%%%%%%%%%%%%%%%%%%%%%%%%%%%%%%%%%%%%%%%%%%%%%%%%%%%%%%%%%%%%%%%%%%%%%%%%%%%%%%%%%%%%%%%%%%%%%%%%%%%%%%%%%%%%%%%%%%%%
%v = 100, w = 100; extention of the table1 centered by left 
%%%%%%%%%%%%%%%%%%%%%%%%%%%%%%%%%%%%%%%%%%%%%%%%%%%%%%%%%%%%%%%%%%%%%%%%%%%%%%%%%%%%%%%%%%%%%%%%%%%%%%%%%%%%%%%%%%%%%%%%%%%%%%%%%%%%%%%%%%%%
\begin{flushleft}
  \begin{tabular}{ | c | c | c | c |}
    \hline
    $N_x$                           & 8000 & 8000 & 15000 \\ \hline
    $N_s$                           & 3000 & 4000 & 5000  \\ \hline
    $ \Omega^*,\times 10^{-3}$      & 4.235& 4.156& 4.339 \\ \hline
    $\epsilon$                      & 1.031& 1.012& 1.056 \\
    \hline
  \end{tabular}
\end{flushleft}
\end{table}
	Comparing the data for every set of virial parameters, we can see the general dependence. 
	One can notice that solution converges by the variable $N_s$ much faster than by the variable $N_x$.
	Moreover, the solution with the grid, $N_x=8k$, $N_s=3k$, have a sufficiently good accuracy for all values of the virial parameters.	
	Thus, we choose it for any further calculations.
%%%%%%%%%%%%%%%%%%%%%%%%%%%%%%%%%%%%%%%%%%%%%%%%%%%%%%%%%%%%%%%%%%%%%%%%%%%%%%%%%%%%%%%%%%%%%%%%%%%%%%%%%%%%%%%%%%%%%%%%%%%%%%%%%%%%%%%%%%%%
%        Large virial parameters
%%%%%%%%%%%%%%%%%%%%%%%%%%%%%%%%%%%%%%%%%%%%%%%%%%%%%%%%%%%%%%%%%%%%%%%%%%%%%%%%%%%%%%%%%%%%%%%%%%%%%%%%%%%%%%%%%%%%%%%%%%%%%%%%%%%%%%%%%%%%
\section{Large virial parameters} 
\begin{figure}[ht!]
\begin{minipage}[ht]{0.5\linewidth}
\center{\includegraphics[width=1\linewidth]{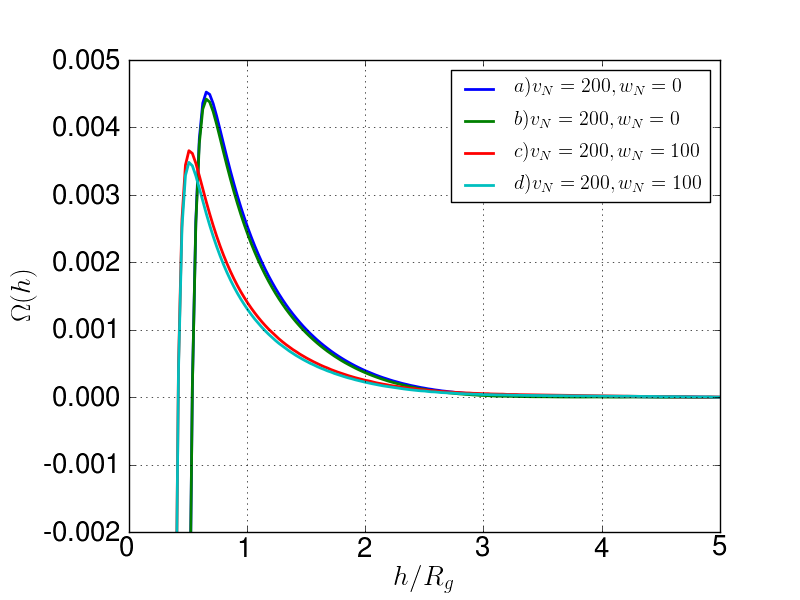}}
\caption{\small{Thermodynamic potentials for different virial parameters and: a) $N_x$ = 8k, $Ns$ = 3k, $\Omega^* = 4.524\times10^{-3}$; 
	 b) $N_x$ = 10k, $Ns$ = 10k, $\Omega^* = 4.421\times10^{-3}$;
	 c) $N_x$ = 8k, $Ns$ = 3k, $\Omega^* = 3.655\times10^{-3}$; 
	 d) $N_x$ = 10k, $Ns$ = 10k, $\Omega^* = 3.482\times10^{-3}$.}}
\label{omega_h5_v02k_w01k_fig}
\end{minipage}
\hfill
\begin{minipage}[ht]{0.5\linewidth}
\center{\includegraphics[width=1\linewidth]{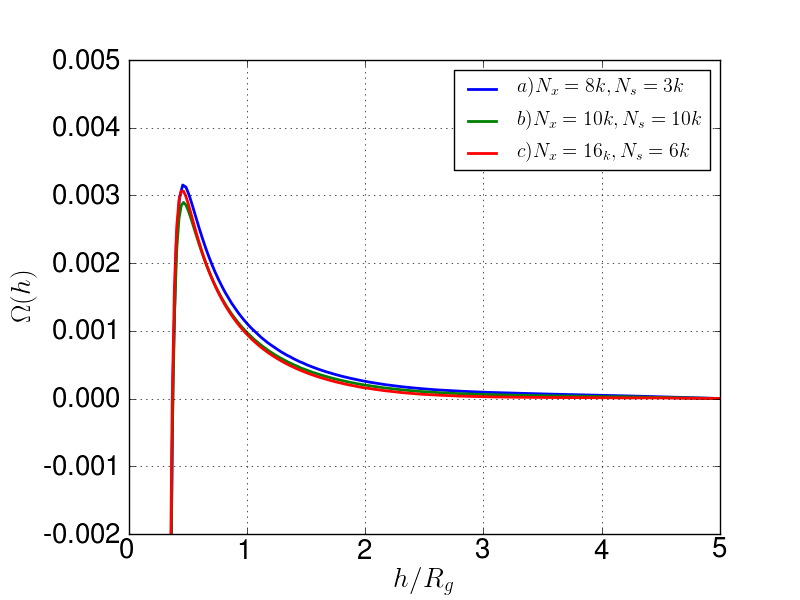}}
\caption{\small{Thermodynamic potentials for the virial parameters: $v_N$ = 500, $w_N$ = 0. The barrier heights for the corresponding curves are: 
         a)$\Omega^* = 3.154\times10^{-3}$; b) $\Omega^* = 2.896\times10^{-3}$; c) $\Omega^* = 3.063\times10^{-3}$.}}
\label{omega_h5_v05k_all_fig}
\end{minipage}
\end{figure}

\begin{figure}[ht!]
\begin{minipage}[ht]{0.5\linewidth}
\center{\includegraphics[width=1\linewidth]{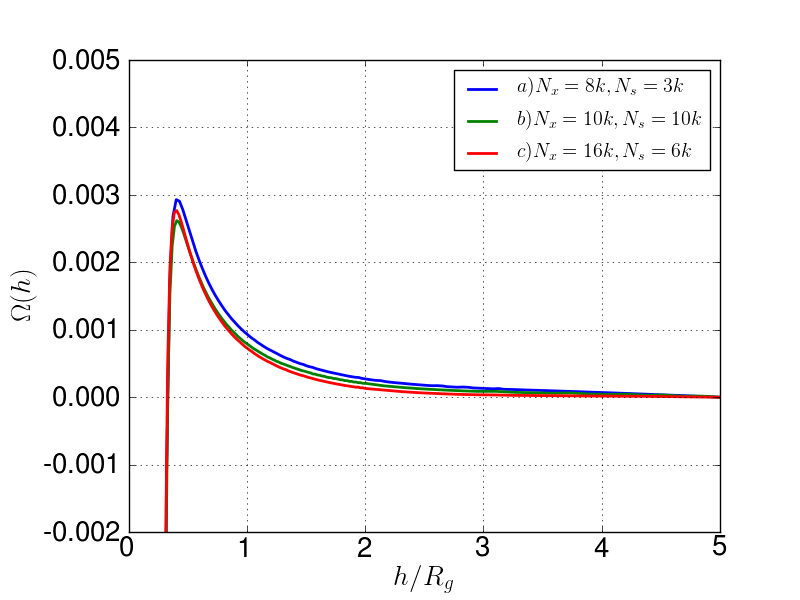}}
\caption{\small{Thermodynamic potentials for the virial parameters: $v_N$ = 500, $w_N$ = 100. The barrier heights for the corresponding curves are: 
         a)$\Omega^* = 2.93\times10^{-3}$; b) $\Omega^* = 2.618\times10^{-3}$; c) $\Omega^* = 2.767\times10^{-3}$.}}
\label{omega_h5_v05k_w01k_all_fig}
\end{minipage}
\hfill
\begin{minipage}[ht]{0.5\linewidth}
\center{\includegraphics[width=1\linewidth]{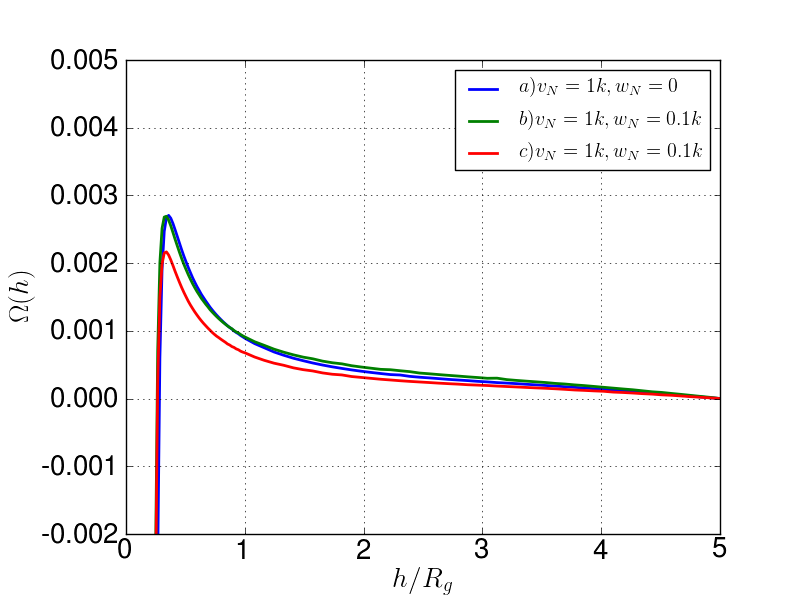}}
\caption{\small{Thermodynamic potentials for different virial parameters and: a) $N_x$ = 8k, $N_s$ = 3k, $\Omega^* = 2.705\times10^{-3}$; 
	 b) $N_x$ = 8k, $N_s$ = 3k, $\Omega^* = 2.692\times10^{-3}$;
	 c) $N_x$ = 10k, $N_s$ = 10k, $\Omega^* = 2.166\times10^{-3}$.}}
\label{omega_h5_v1k_w01k_fig}
\end{minipage}
\end{figure}

	In this section, we increase values of the virial parameters and consider the sequence which is given in Tab.\ref{tabular:large_virial_parameter}. 
	We use the method developed earlier to examine the thermodynamic potential for other virial parameters. 
	For example, we found there that grid dots,$N_x$ = 8k, $N_s$ = 3k correspond to a solution with a sufficiently good accuracy for all sets of the 
	virial parameters considered before. Based on that, we calculate the thermodynamic potentials for all sets of virial parameters
	(see Tab.\ref{tabular:large_virial_parameter}) and compare them with
	the high resolution calculations, performed for $N_x$ = 10k, $N_s$ = 10k.\footnote{The calculation of the 
	thermodynamic potential is computationally very expensive and for $v_N$ = 1000, $w_N$ = 100 and $N_x$ = 10k, $N_s$ = 10k  it took around $4$ days.} 	
	The thermodynamic potentials are shown in Figs.\ref{omega_h5_v02k_w01k_fig}--\ref{omega_h5_v1k_w01k_fig}. 
\begin{table}[ht!]
\caption{Large virial parameters.} 
\label{tabular:large_virial_parameter} 
\begin{center}
  \begin{tabular}{ | c | c | c | c | c | c | c | }
    \hline
    $v_N$   & 200& 200& 500& 500&1000&1000  \\ \hline
    $w_N$   &   0& 100&   0& 100&   0& 100  \\
    \hline
  \end{tabular}
\end{center}
\end{table}

%%%%%%%%%%%%%%%%%%%%%%%%%%%%%%%%%%%%%%%%%%%%%%%%%%%%%%%%%%%%%%%%%%%%%%%%%%%%%%%%%%%%%%%%%%%%%%%%%%%%%%%%%%%%%%%%%%%%%%%%%%%%%%%
%       Thermodynamic potential for a) v = 200, w = 0, 100; b) v = 500, w = 0
%%%%%%%%%%%%%%%%%%%%%%%%%%%%%%%%%%%%%%%%%%%%%%%%%%%%%%%%%%%%%%%%%%%%%%%%%%%%%%%%%%%%%%%%%%%%%%%%%%%%%%%%%%%%%%%%%%%%%%%%%%%%%%%

%%%%%%%%%%%%%%%%%%%%%%%%%%%%%%%%%%%%%%%%%%%%%%%%%%%%%%%%%%%%%%%%%%%%%%%%%%%%%%%%%%%%%%%%%%%%%%%%%%%%%%%%%%%%%%%%%%%%%%%%%%%%%%%
%       Thermodynamic potential for a) v = 500, w = 100; b) v = 1000, w = 0, 100 
%%%%%%%%%%%%%%%%%%%%%%%%%%%%%%%%%%%%%%%%%%%%%%%%%%%%%%%%%%%%%%%%%%%%%%%%%%%%%%%%%%%%%%%%%%%%%%%%%%%%%%%%%%%%%%%%%%%%%%%%%%%%%%%
	 One can notice that all the thermodynamic potentials except those shown in Fig.\ref{omega_h5_v1k_w01k_fig} converge rather well, 
	 while the thermodynamic potentials calculated for the virial parameters $\{v_N, w_N\}$ = $\{1000, 0\}$ and $\{v_N, w_N\}$ = $\{1000, 100\}$
	 do not converge even for the high resolution computation. 
	 Looking at the thermodynamic potentials for smaller virial parameters, one can see 
	 that tails of the curves reach a plateau, while for the larges virial parameters 
	 the conversion to the plateau does not seem to occur.
	 The similar behavior took place when we did calculations for virial parameters $v_N, v_N$ = 100 and grid dots ${N_x, N_s} = {1k, 1k}$. 
	 It was cured by decreasing the computational mesh size. 
	 In the case when the second virial parameter is $v_N=1000$, it is impossible to increase the number of grid dots significantly, because the 
	 computational time increases as well.
	 Thus, we restrict the SCFT calculations by the virial parameters, $v_N=500, w_N=100$. In order to overcome the restriction let us consider the 
	 analytical GSDE theory.

%%%%%%%%%%%%%%%%%%%%%%%%%%%%%%%%%%%%%%%%%%%%%%%%%%%%%%%%%%%%%%%%%%%%%%%%%%%%%%%%%%%%%%%%%%%%%%%%%%%%%%%%%%%%%%%%%%%%%%%%%%%%%%%%%%%%%%%%%%%%%%%%%%%%%%%%%%%%
%           deGennes ground state free energy 
%%%%%%%%%%%%%%%%%%%%%%%%%%%%%%%%%%%%%%%%%%%%%%%%%%%%%%%%%%%%%%%%%%%%%%%%%%%%%%%%%%%%%%%%%%%%%%%%%%%%%%%%%%%%%%%%%%%%%%%%%%%%%%%%%%%%%%%%%%%%%%%%%%%%%%%%%%%%
\section{Ground state free energy}
\label{sec:repulsion_gsd_free_en}
In this section, our aim is to find free energy in the gap between two flat plates in the $ground$ $state$ $dominance$ ($GSD$) approximation. For this reason, we use results, obtained by De Gennes. In particular, in \cite{deGennes_1981, deGennes_1982} it was shown that the corresponding force between two plates in the mean field approximation is 
\begin{equation}
\label{deGennes_force_gs_def}
	    f_{gs}(h) \simeq f^*_{int}(c_b) - f^*_{int}(c_m) 
\end{equation}
	    where $h$ is the distance between two plates, $c_m$ is the ground state concentration at the midpoint between plates, i.e. $c_m=c(h/2)$, $c_b$ is 
	    the bulk monomer concentration,
$$
	    f^*_{int}(c) = f_{int}(c) - \mu_{int}(c_b)c 
$$
	    here $f_{int}$ is the interaction part of free energy of the system between plates per unit volume and $\mu_{int}(c_b)$ is  
	    the chemical potential of the bulk phase defined as 
$$
	    f_{int}(c) = \frac{\text{v}}{2}c^2 + \frac{\text{w}}{6}c^3 , \quad \mu_{int}(c_b) = \frac{\partial f_{int}(c_b)}{\partial c_b} = \text{v}c_b + \frac{\text{w}}{2}c_b^2  
$$        
	    Combining all the above expressions together, we obtain
$$
	    f_{gs} \simeq \frac{\text{v}c^2_b}{2}\left(1-(c_m/c_b)^2\right) + \frac{\text{w}c_b^3}{6}\left(1-(c_m/c_b)^3\right) - \mu_{int}(c_b)c_b\left(1-(c_m/c_b)\right)
$$
	    Let us rewrite it in a more appropriate way. For that, consider separately the expression for $\text{v}=0$ and $\text{w}=0$.\\
	    \textbf{\text{w}=0}. In this case, we can write
$$		
	    f_{gs}(\text{w}=0) = \frac{\text{v}c^2_b}{2}\left(1-(c_m/c_b)^2\right) - \text{v}c_b^2\left(1-(c_m/c_b)\right) = -\frac{\text{v}c^2_b}{2}\left(1-(c_m/c_b)\right)^2
$$
	    \textbf{\text{v}=0}. Corresponingly, the force for $\text{v}=0$ is
$$
\begin{array}{c}
 	    f_{gs}(\text{v}=0) = \frac{\text{w}c_b^3}{6}\left(1-(c_m/c_b)^3\right) - \frac{\text{w}c_b^3}{2}\left(1-(c_m/c_b)\right) = 
				 -\frac{\text{w}c_b^3}{6}\left(1-(c_m/c_b)\right)^2(2+c_m/c_b) = \\
				 =-\frac{\text{w}c_b^3}{6}\left(1-(c_m/c_b)\right)^2(3 - (1 - c_m/c_b))
\end{array}
$$	    
	    Joining them together, we finally obtain:
\begin{equation}
\label{deGennes_force_gs}
            f_{gs}(h) = -\frac{c_b}{2}(1 - c_m/c_b)^2\left(\text{v}c_b + \text{w}c_b^2 - \frac{\text{w}c_b^2}{3}(1 - c_m/c_b)\right)
\end{equation}
In order to find the dependence $f_{gs}(h)$, we should firstly establish how the concentration at mid-plane, $c_m(h)$ depends on the distance between plates, $h$. Let us find now $c_m(h)$. Following the standard approach, \cite{Lifshitz_1978, deGennes_book}, the free energy functional including conformational and interaction part has the following form 
\begin{equation}
\label{deGennes_free_energy_functional}
            F = F_{conf} + F_{int} = \int\limits_{0}^{h}\left\{\frac{a^2}{4}\left(\frac{\mathrm{d}c}{\mathrm{d}x}\right)^2\frac{1}{c} + \frac{\text{v}c^2}{2} + 
            \frac{\text{w}c^3}{6}\right\}\mathrm{d}x
\end{equation} 
where $a^2=a_s^2/6$, $a_s$ is the polymer statistical segment, $\text{v}$, $\text{w}$ are second and third virial coefficients respectively
(For the interaction part of the free energy we used the virial expansion). For the concentration profile, we also have the normalization condition:
\begin{equation}
\label{deGennes_conc_restriction}
	    \int\limits_0^h c(x)\mathrm{d}x = \mathcal{N}
\end{equation}
	    where $\mathcal{N}$ is the total number of monomers in the system. Using the variational principle 
$$
            \frac{\delta}{\delta c}\left(F-\mu_{int} \int c(x)\mathrm{d}x\right) = 0,
$$ 
            we obtain the Edwards equation for the function $\psi$, where  $c(x) = \psi^2(x)$:
\begin{equation}
\label{deGennes_edwards_eq}
            a^2\frac{\mathrm{d^2}\psi}{\mathrm{d}x^2} - (\phi_{self} - \mu_{int})\psi = 0
\end{equation}
            where $\phi_{self} = \text{v}c + \text{w}c^2/2$ is the self-consistent mean field. Far away from any colloidal particle suspended in polymer solution 
	    the above equation must have a constant solution $\psi = \sqrt{c_b}$. For such solution, the gradient term is dropped down, hence
$$ 	
	    \mu_{int} = \phi_{self}(c_b) = \text{v}c_b + \frac{\text{w}}{2}c_b^2
$$
	    The boundary conditions for Eq.(\ref{deGennes_edwards_eq}): $\psi(0) = \psi(h) = 0$, mean that polymers 
	    can not penetrate through the plates, 
	    and by virtue of symmetry: $\mathrm{d}\psi/\mathrm{d}x = 0$ at $x = h_m = h/2$. Multiplying Eq.(\ref{deGennes_edwards_eq}) by $\frac{\mathrm{d}\psi}{\mathrm{d}x}$ 
	    and integrating it, we obtain the first integral:
\begin{equation}
\label{deGennes_first_integral}
            a^2\left(\frac{\mathrm{d}\psi}{\mathrm{d}x}\right)^2 = - \mu_{int}\psi^2 + \frac{\text{v}}{2}\psi^4 + \frac{\text{w}}{6}\psi^6 + A 
\end{equation}
	    where the integration constant $A$ can be found using the above boundary condition at mid-plane, i.e.
$$
	    A = \mu_{int}\psi_m^2 - \frac{\text{v}}{2}\psi_m^4 - \frac{\text{w}}{6}\psi_m^6
$$
	    Then,
\begin{equation}
\label{deGennes_edwards_eq_vw}
            a^2\left(\frac{\mathrm{d}\psi}{\mathrm{d}x}\right)^2 = - \mu_{int}(\psi^2-\psi_m^2) + \frac{\text{v}}{2}(\psi^4-\psi_m^4) + \frac{\text{w}}{6}(\psi^6-\psi_m^6)  
\end{equation}
	    where $\psi_m$ denotes the value of $\psi$ at mid-plane (where $\partial \psi/\partial x = 0$). Integrating this equation from $x = 0$ to $x=h/2$, we 
	    eventually obtain the solution expressed in quadratures. 
%%%%%%%%%%%%%%%%%%%%%%%%%%%%%%%%%%%%%%%%%%%%%%%%%%%%%%%%%%%%%%%%%%%%%%%%%%%%%%%%%%%%%%%%%%%%%%%%%%%%%%%%%%%%%%%%%%%%%%%%%%%%%%%%%%%%%%%%%%%%%%%%%%%%%%%%%%%%
%           Special case w=0
%%%%%%%%%%%%%%%%%%%%%%%%%%%%%%%%%%%%%%%%%%%%%%%%%%%%%%%%%%%%%%%%%%%%%%%%%%%%%%%%%%%%%%%%%%%%%%%%%%%%%%%%%%%%%%%%%%%%%%%%%%%%%%%%%%%%%%%%%%%%%%%%%%%%%%%%%%%%
\subsection{Special case (\text{w}=0)}	    
\begin{figure}[ht]
\begin{minipage}[ht]{0.47\linewidth}
\center{\includegraphics[width=1\linewidth]{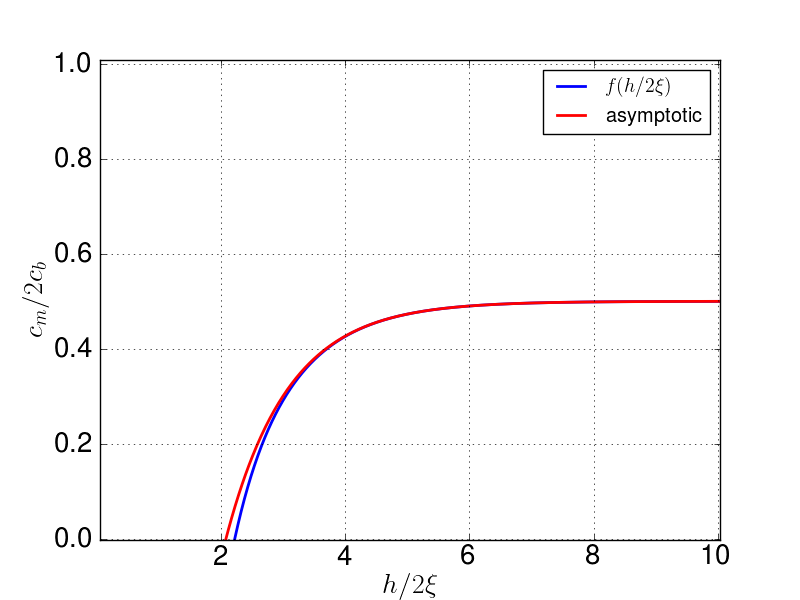}}
\caption{\small{Universal function $f$, Eq.(\ref{deGennes_universal_f}), and its asymptotics at $h\gg\xi$, Eq.(\ref{deGennes_elliptic_asy_gg}).}}
\label{deGennes_f_asy_fig}
\end{minipage}
\hfill
\begin{minipage}[ht]{0.47\linewidth}
\center{\includegraphics[width=1\linewidth]{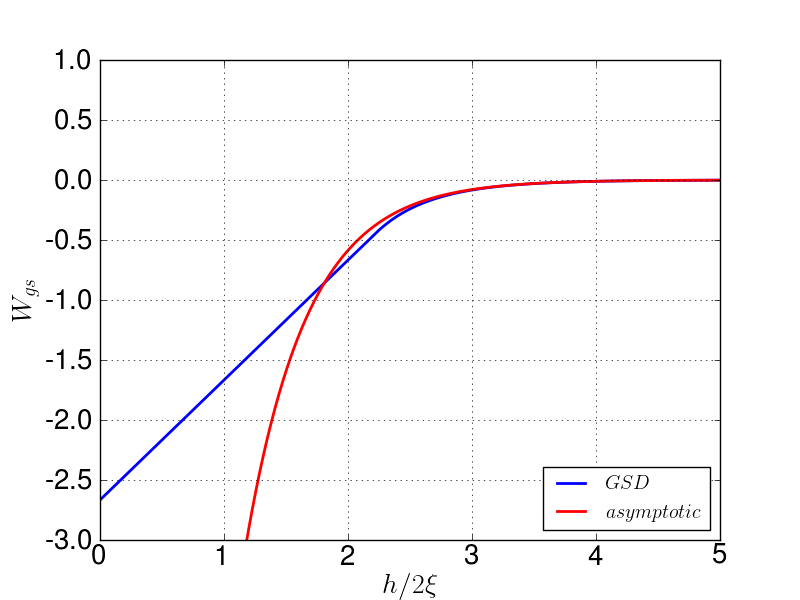}} 
\caption{\small{The universal function $w_{gs}(h/2\xi)$ for the ground state free energy Eq.(\ref{deGennes_w_univ_gs_v}) and its asymptotics,
	        Eq.(\ref{deGennes_w_univ_gs_v_gg}).}}
\label{deGennes_free_en_gs_v_fig}
\end{minipage}
\end{figure}

All of the results considered below have been originally obtained in \cite{Joanny_1979}. Consider separately the case: $\text{w}$ = 0. 
We can rewrite Eq.(\ref{deGennes_edwards_eq_vw}) as
$$
            a^2\left(\frac{\mathrm{d}\psi}{\mathrm{d}x}\right)^2 = \frac{\text{v}}{2}\left(\psi^4 - \psi^4_m\right) - \text{v}\psi^2_b\left(\psi^2 - \psi^2_m\right)
$$
	    or, after integration
\begin{equation}
\label{deGennes_pre_elliptic}
	    \frac{h}{2a} = \int\limits_0^{\psi_m}\frac{\mathrm{d}\psi}{\sqrt{\frac{\text{v}}{2}\left(\psi^4 - \psi^4_m\right) - \text{v}\psi^2_b\left(\psi^2 - \psi^2_m\right)}}   
\end{equation}
	    Let us use the following variables:
\begin{equation}
\label{deGennes_integrand_substitutions}
	    x^2 = \frac{\psi_m^2}{2\psi_b^2}, \quad \sin t = \frac{\psi}{\psi_m} %, \quad \mathrm{d}\psi = \psi_m\cos t \mathrm{d}t
\end{equation}
	    One can transform the denominator as
$$
\begin{array}{l}
	    \frac{\text{v}}{2}\left(\psi^4 - \psi^4_m\right) - \text{v}\psi^2_b\left(\psi^2 - \psi^2_m\right) = 
	    \text{v}\psi_b^2\psi_m^2\left(x^2(\sin^4 t-1) -(\sin^2 t-1) \right) = \\
	    =\text{v}\psi_b^2\psi_m^2\cos^2 t\left((1-x^2)-x^2\sin^2t\right) 
\end{array}
$$      
	    Finally, we obtain:
$$
	    \frac{h}{2\sqrt{2}\xi} = \frac{1}{\sqrt{1-x^2}}\int\limits_0^{\pi/2}\frac{\mathrm{d}t}{\sqrt{1-x^2/(1-x^2)\sin^2t}} = \sqrt{1+k^2}K(k)  
$$
	    or
\begin{equation}
\label{deGennes_elliptic}
	    \frac{h}{2\xi} = \frac{\sqrt{2}}{\sqrt{1-x^2}}\int\limits_0^{\pi/2}\frac{\mathrm{d}t}{\sqrt{1-x^2/(1-x^2)\sin^2t}} = \sqrt{2}\sqrt{1+k^2}K(k)  
\end{equation}
	    where we denoted $k^2 = x^2/(1-x^2)$, $\xi = a/\sqrt{2c_b\text{v}}$,  $K(k)$ is the complete elliptic integral of the first kind. 
	    Now, we can actually invert this expression as
\begin{equation}
\label{deGennes_universal_f}
	    \frac{c_m}{2c_b} = x^2 = f(h/2\xi)
\end{equation}
	    where $f$ is a universal function shown in Fig.\ref{deGennes_f_asy_fig}. Using this function, we can write Eq.(\ref{deGennes_force_gs}), for the case when
	    the third virial parameter is equal to zero, as
\begin{equation}
\label{deGennes_force_gs_v}
	    f_{gs} = -\frac{\text{v}c_b^2}{2}(1-(c_m/c_b))^2 = -\frac{\text{v}c_b^2}{2}(1-2f(h/2\xi))^2
\end{equation}
	    As shown in Fig.\ref{deGennes_f_asy_fig}, the universal function $f = c_m/2c_b$ is defined for $h$ exceeding some separation $h^{*}$ which can be found 
	    from Eq.(\ref{deGennes_elliptic}) setting $x=0$. Thereby,
\begin{equation}
\label{deGennes_h_star_gs_v}
	    \frac{h^{*}}{2\xi} = \frac{\pi}{\sqrt{2}}
\end{equation}
	    Hence, at $h\leqslant h^{*}$ the concentration $c_m = 0$, so $f(h/2\xi)=0$ and there is only the external osmotic pressure effected by the bulk phase. 
	    In Sec.\ref{sec:polymers_osmotic_pressure} we already found the expression for the bulk osmotic pressure and in the case $\text{w} = 0$ it is
\begin{equation}
\label{deGennes_osmoticP_v}
	    \Pi_b = \frac{c_b}{N} + \frac{\text{v}c_b^2}{2} = \frac{c_b}{N}\left(1 + \frac{v_N}{2}\right)
\end{equation}
	    where $v_N = \text{v}c_bN$.
	    In the $GSD$, the polymer length, $N$, is infinitely long. Therefore, we can neglect the first term. Accordingly, 
	    the two solutions for the force are matching at $h=h^*$.	    
  
	    The ground state free energy is obtained from Eq.(\ref{deGennes_force_gs_v}) by integrating it over $h$:
$$
	    W_{gs} = -\frac{\text{v}c_b^2}{2}\int\limits_h^{\infty}\mathrm{d}h(1-2f(h/2\xi))^2 = 
	    -\frac{\text{v}c_b^2}{2}\int\limits_0^{\infty}\mathrm{d}h(1-2f(h/2\xi))^2 + \frac{\text{v}c_b^2}{2}\int\limits_0^h\mathrm{d}h(1-2f(h/2\xi))^2
$$
	    or, in the reduced variables ($\bar{h} = h/R_g, \bar{\xi} = \xi/R_g$):
\begin{equation}
\label{deGennes_w_gs_v_num}
	    \hat{W}_{gs} = \frac{N}{R_gc_b}W_{gs} = -A + v_N\bar{\xi}\int\limits_0^{h/2\xi}\mathrm{d}h'(1-2f(h'))^2 
\end{equation}
	    where we introduced $v_N = \text{v}c_bN$. The constant $A$ is defined by the above integral with upper limit equal to infinity.
	    Numerically, the constant $A$ is equal to the integral in r.h.s. of the above expression at sufficiently large $h$,
	    which leads to zero value of the free energy, $W_{gs}$ at large separation. As we already defined in Eq.(\ref{deGennes_h_star_gs_v}) and 
	    Eq.(\ref{deGennes_osmoticP_v}) at distances less than $h^*$, there are no polymers inside the gap between plates and only external osmotic pressure
	    is exerted, so the integral in Eq.(\ref{deGennes_w_gs_v_num}) should be written as 
$$
 \int\limits_0^{h/2\xi}\mathrm{d}h'(1-2f(h'))^2 = \left\{
  \begin{array}{l l}
	    \frac{h}{2\xi} & \quad \text{if $h\leqslant h^*$}\\\\
            \frac{h^*}{2\xi} + \int\limits_{h^*/2\xi}^{h/2\xi}\mathrm{d}h'(1-2f(h'))^2 & \quad \text{if $h > h^*$}\\
  \end{array} \right.
$$
	    We can rewrite Eq.(\ref{deGennes_w_gs_v_num}) in a slightly different form:
\begin{equation}
\label{deGennes_w_univ_gs_v}
	    \hat{W}_{gs} = v_N\bar{\xi}w_{gs}(h/\xi)
\end{equation}
	    where $w_{gs}$ is a function that depends only on $h/\xi$. 
	    In terms of the function, we can rewrite the expression for the thermodynamic potential in real variables taking into account that the correlation
	    length, Eq.(\ref{intr_polymer_correlation_length_gsd}), in this case is $\bar{\xi}=1/\sqrt{2\text{v}c_bN}$. Thereby,
$$
	    W_{gs} = \frac{R_gc_b}{N}\hat{W}_{gs} = \frac{R_gc_b}{N}v_N\bar{\xi} w_{gs}(h/\xi) = \frac{a\sqrt{\text{v}}c_b^{3/2}}{\sqrt{2}}w_{gs}(h/\xi)
$$
	    where, as before, $a_s=a\sqrt{6}$ is the polymer statistical segment.
	    Therefore, we can explicitly see how the thermodynamic potential in real variables depends on $a_s, \text{v}, c_b$ for a given $h/\xi$ and 
	    also that it does not depend on $N$.  
%%%%%%%%%%%%%%%%%%%%%%%%%%%%%%%%%%%%%%%%%%%%%%%%%%%%%%%%%%%%%%%%%%%%%%%%%%%%%%%%%%%%%%%%%%%%%%%%%%%%%%%%%%%%%%%%%%%%%%%%%%%%%%%%%%%%%%%%%%%%
%          universal function f and its asymptotics at x \gg 1 for v
%%%%%%%%%%%%%%%%%%%%%%%%%%%%%%%%%%%%%%%%%%%%%%%%%%%%%%%%%%%%%%%%%%%%%%%%%%%%%%%%%%%%%%%%%%%%%%%%%%%%%%%%%%%%%%%%%%%%%%%%%%%%%%%%%%%%%%%%%%%%
	    
\textbf{The limit of large gap ($h\gg\xi$)}. In this case the parameter $x^2 = c_m/2c_b\rightarrow 1/2$ or $k\rightarrow 1$ in Eq.(\ref{deGennes_elliptic}). It is known that for such values of parameter $k$ the complete elliptic integral has the following asymptotics \cite{mathematica_K1}:
$$
	    K(k\rightarrow 1) \simeq -\frac{1}{2}\ln(1-k^2) + \ln 4 = -\frac{1}{2}\ln\left(\frac{1-2x^2}{16(1-x^2)}\right) \simeq -\frac{1}{2}\ln\left(\frac{1-2x^2}{8}\right)
$$
	    Substituting this expression in Eq.(\ref{deGennes_elliptic}) and then inverting it, we obtain 
\begin{equation}
\label{deGennes_elliptic_asy_gg}
	    c_m \simeq c_b\left(1-8e^{-\frac{h}{2\xi}}\right)
\end{equation}
	   or after substituting this expression into Eq.(\ref{deGennes_force_gs}), we eventually obtain
\begin{equation}
\label{deGennes_force_gs_v_asy}
	   f_{gs} \simeq -\frac{\text{v}c_b^2}{2}(1-(c_m/c_b))^2 \simeq  -32\text{v}c_b^2e^{-h/\xi}
\end{equation}

Semenov et all \cite{Grosberg_book} have shown that the same result for $h\gg\xi$ can be obtained using much easier way. They concluded that
for large separations $h\gg\xi$ the concentration perturbation for the system squeezed between two plates is twice larger than concentration 
perturbation caused by only one plate. The concentration perturbation caused by one plate was found in \cite{deGennes_book} and it equals
\begin{equation}
\label{deGennes_conc_single_wall_v}
	   c(x) = c_b\tanh^2\left(\frac{x}{2\xi}\right)
\end{equation} 
	   where $\xi = a/\sqrt{2\text{v}c_b}$. At $h\gg\xi$ the solution has the asymptotics:
$$
	   c(x) - c_b = c_b(\tanh^2(x/(2\xi))-1) = -\frac{c_b}{\cosh^2(x/(2\xi))}\simeq -4c_be^{-x/\xi}
$$
	   multiplying this result by $2$, replacing $x\rightarrow h/2$ and substituting it in Eq.(\ref{deGennes_force_gs_v_asy}), 
	   we immediately obtain the result, Eq.(\ref{deGennes_elliptic_asy_gg}).
	    
           As a final step we should integrate Eq.(\ref{deGennes_force_gs_v_asy}) in order to obtain the GSD free energy of the system between 
	   the plates at $h\gg\xi$:
$$
	   W_{gs} = \int\limits_h^{\infty}\mathrm{d}h'f_{gs}(h') = -32\text{v}c_b^2\xi e^{-h/\xi}
$$
	   This expression can be written in the reduced variables ($\bar{h}=h/R_g, \bar{\xi}=\xi/R_g$):
\begin{equation}
\label{deGennes_w_gs_v_gg}
	   \hat{W}_{gs} = \frac{N}{R_gc_b}W_{gs} = -32\text{v}c_b N (\xi/R_g) e^{-h/\xi} = -32 v_N \bar{\xi} e^{-h/\xi} 
\end{equation}	    
	   where, as before, we introduced $v_N = \text{v}c_bN$. In this case 
\begin{equation}
\label{deGennes_w_univ_gs_v_gg}
	   w_{gs}(h/\xi) = -32e^{-h/\xi} 
\end{equation}
	   The comparison between Eq.(\ref{deGennes_universal_f}) and its asymptotics, Eq.(\ref{deGennes_w_univ_gs_v_gg}) can be found in Fig.\ref{deGennes_free_en_gs_v_fig}. 
	   The last expression in the real variables can be rewritten as
$$
            W_{gs} \simeq -32\frac{a\sqrt{\text{v}}c_b^{3/2}}{\sqrt{2}}e^{-h/\xi}
$$	   
	   This expression is valid for $h\gg\xi$. 
%%%%%%%%%%%%%%%%%%%%%%%%%%%%%%%%%%%%%%%%%%%%%%%%%%%%%%%%%%%%%%%%%%%%%%%%%%%%%%%%%%%%%%%%%%%%%%%%%%%%%%%%%%%%%%%%%%%%%%%%%%%%%%%%%%%%%%%%%%%%%%%%%%%%%%%%%%%%
%           Generalization w \ne 0
%%%%%%%%%%%%%%%%%%%%%%%%%%%%%%%%%%%%%%%%%%%%%%%%%%%%%%%%%%%%%%%%%%%%%%%%%%%%%%%%%%%%%%%%%%%%%%%%%%%%%%%%%%%%%%%%%%%%%%%%%%%%%%%%%%%%%%%%%%%%%%%%%%%%%%%%%%%%  
\subsection{Generalization ($\text{w} \ne$ 0)}
\label{sec:repulsion_generalization_GSD}
\begin{figure}[ht]
\begin{minipage}[ht]{0.47\linewidth}
\center{\includegraphics[width=1\linewidth]{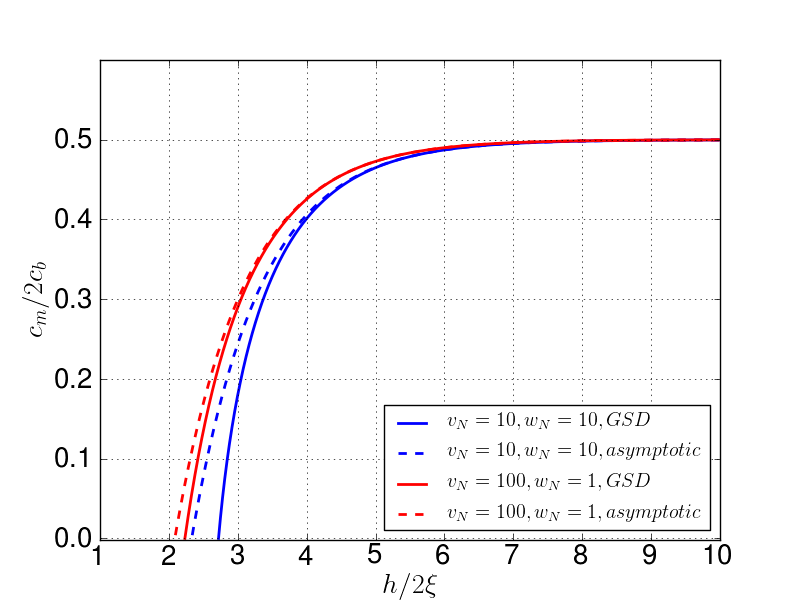}}
\caption{\small{Universal function $g$, Eq.(\ref{deGennes_universal_vw}), and its asymptotics, Eq.(\ref{deGennes_univ_fun_g_asy_gg_vw}),
		shown for different values of the virial parameters.}}
\label{deGennes_g_vw_asy_fig}
\end{minipage}
\hfill
\begin{minipage}[ht]{0.47\linewidth}
\center{\includegraphics[width=1\linewidth]{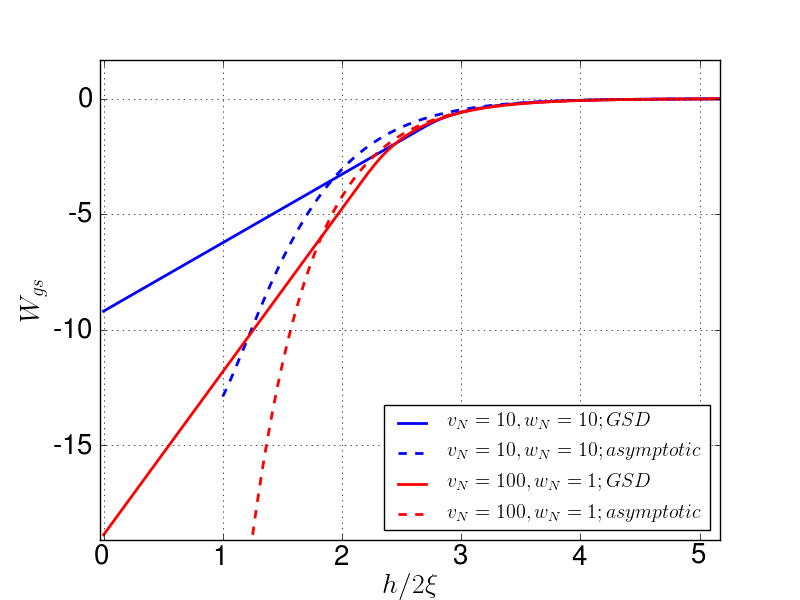}} 
\caption{\small{The ground state free energy, Eq.(\ref{deGennes_w_gs_vw_num}), and its asymptotics, Eq.(\ref{deGennes_w_gs_vw_gg}), obtained for 
		different values of the virial parameters $v_N, w_N$.}} 
\label{deGennes_free_en_gs_vw_fig}
\end{minipage}
\end{figure}

	   Let us generalize all of the above results for the case with the third virial coefficient, $\text{w}\ne 0$. 
	   We will follow the same integrand substitutions as in the previous section. 
	   After changing variables in Eq.(\ref{deGennes_integrand_substitutions}) and applying it to each term in Eq.(\ref{deGennes_edwards_eq_vw}): 
$$
\begin{array}{l}
	   \psi^6 - \psi_m^6 = \psi_m^6(\sin^6t-1) = \psi_m^6(\sin^6t-\sin^2t-\cos^2t) = \\ % 1 term
	   = \psi_m^6(\sin^2t(\sin^2t+1)(\sin^2t-1)-\cos^2t) = -\psi_m^6 \cos^2t (\sin^2t(\sin^2t+1)-1)\\\\ 
	   \psi^4 - \psi_m^4 = \psi_m^4(\sin^4t-1) = -\psi_m^4 \cos^2t (\sin^2t+1)\\\\ %2 term 
	   \psi^2 - \psi_m^2 = \psi_m^2(\sin^2t-1) = -\psi_m^2 \cos^2t %3 term 
\end{array}
$$
	   Now, we can rewrite Eq.(\ref{deGennes_edwards_eq_vw}) in quadratures:
$$
	   \frac{h}{2a} = \frac{1}{\sqrt{\text{w}c_b^2}}\int\limits_0^{\pi/2}\frac{\mathrm{d}t}{\sqrt{(\text{v}/\text{w}c_b)(1-x^2(\sin^2t+1)) + 
	   (1/2-(2/3)x^4(\sin^2t(\sin^2t+1)+1))}}  
$$
	   Multiplying both sides of the equation by $\sqrt{2c_b(\text{v}+\text{w}c_b)}$, we can write it as
\begin{equation}
\label{deGennes_quadratures_vw}
%\begin{array}{l}
	   \frac{h}{2\xi} = \sqrt{2\text{v}/(\text{w}c_b)+2}
	   \int\limits_0^{\pi/2}\frac{\mathrm{d}t}{\sqrt{(\text{v}/\text{w}c_b)(1-x^2(\sin^2t+1)) + (1/2-(2/3)x^4(\sin^2t(\sin^2t+1)+1))}}  
%\end{array}
\end{equation}
	    Now, as before, we can invert this expression as
\begin{equation}
\label{deGennes_universal_vw}
	    \frac{c_m}{2c_b} = x^2 = g(h/2\xi, \text{v}/\text{w}c_b)
\end{equation}
	    where $g$ is a universal function, shown in Fig.\ref{deGennes_g_vw_asy_fig} . As before, we can introduce the separation, 
	    $h^{*}$ setting $x = 0$ in Eq.(\ref{deGennes_quadratures_vw}), which leads to
\begin{equation}
\label{deGennes_h_star_gs_vw}
	    \frac{h^{*}}{2\xi} = \sqrt{\frac{\text{v}+\text{w}c_b}{2\text{v}+\text{w}c_b}}\pi = \sqrt{\frac{v_N/w_N+2}{v_N/w_N+1}}\frac{\pi}{\sqrt{2}}
\end{equation}
	    At $h\leqslant h^{*}$ the concentration $c_m = 0$, so $g(h/2\xi, \text{v}/\text{w}c_b)=0$ and there is only external osmotic pressure due to
	    the bulk phase. In Sec.\ref{sec:polymers_osmotic_pressure}, we already found the expression for the bulk osmotic pressure  
\begin{equation}
\label{deGennes_osmoticP_vw}
	    \Pi_b = \frac{c_b}{N} + \frac{\text{v}c_b^2}{2} + \frac{\text{w}c_b^3}{3} = \frac{c_b}{N}\left(1 + \frac{v_N}{2} + \frac{2w_N}{3}\right)
\end{equation}
	    where $v_N = \text{v}c_bN$ and $w_N = \text{w}c_b^2N/2$.
	    Since, in the $GSD$ the polymer length, $N$ is infinitely long, we neglect the first term for the time being. 	   

	    Now, using the universal function $g(h/2\xi, \text{v}/\text{w}c_b)$, we can write the expression for the force between the 
	    plates, Eq.(\ref{deGennes_force_gs}) in the general case, i.e. with $w_N\ne 0$:
$$
 	    f_{gs}(h/2\xi, \text{v}/\text{w}c_b) \simeq 
	    -\frac{c_b}{2}(1-2g(h/2\xi, \text{v}/\text{w}c_b))^2(\text{v}c_b + \text{w}c_b^2  - (\text{w}c_b^3/3)(1 - 2g(h/2\xi, \text{v}/\text{w}c_b)))
$$
	    For brevity, we will write the functions indicating the dependence on the distance, $h$, omitting $\text{v}/\text{w}c_b$. 
	    One can notice that $f_{gs}=-\Pi_b$ at $h\leqslant h^{*}$. 
	    Thereby, we extend Eq.(\ref{deGennes_force_gs_vw}) for $h\leqslant h^{*}$. 
	    In the dimensionless variables, we can rewrite it as
\begin{equation}
\label{deGennes_force_gs_vw}
\begin{array}{l}
 	    \hat{f}_{gs}(h/2\xi)= -\frac{N}{c_b}f_{gs} \simeq \frac{1}{2}\left(1-2g(h/2\xi, v_N/2w_N)\right)^2(v_N + 2w_N - (2w_N/3)(1 - 2g(h/2\xi, v_N/2w_N)))
\end{array}	    
\end{equation}	    	    	    
	    The GSD free energy is obtained from Eq.(\ref{deGennes_force_gs_vw}) by integrating it over $h$:
\begin{equation}
\label{deGennes_w_gs_vw_num}
	    \hat{W}_{gs} = \frac{N}{R_gc_b}W_{gs} = -\frac{1}{R_g}\int\limits_h^{\infty}\mathrm{d}h\hat{f}_{gs}(h/2\xi) = 
	    -B + 2\bar{\xi}\int\limits_0^{h/2\xi}\mathrm{d}h'\hat{f}_{gs}(h') 
\end{equation}	    
	    where the constant $B$ is defined to provide zero value of the free energy at very large separation. 
	    Using the same arguments as we did for the special case ($w_N=0$), we can write for the integral in r.h.s. of the above expression:	
$$
 \int\limits_0^{h/2\xi}\mathrm{d}h'\hat{f}_{gs}(h') = \left\{
  \begin{array}{l l}
    \left(\frac{v_N}{2}+\frac{2w_N}{3}\right)\frac{h}{2\xi} & \quad \text{if $h\leqslant h^*$}\\\\
    \left(\frac{v_N}{2}+\frac{2w_N}{3}\right)\frac{h^*}{2\xi} + \int\limits_{h^*/2\xi}^{h/2\xi}\mathrm{d}h'\hat{f}_{gs}(h') & \quad \text{if $h > h^*$}\\
  \end{array} \right.
$$	
	    Let us find now asymptotics expressions for the interaction energy and the universal function, $g(v_N/w_N, h/2\xi)$ at $h\gg \xi$.\\
	    \textbf{Asymptotics, $x\gg\xi$.} 
	    We will use the same approach as for the special case ($w_N=0$), to find the asymptotics, considering only the concentration perturbation 
	    created by only one plate at large enough separations. 
	    This approach dictates a new boundary condition for the distribution function:
$$
	    \psi(x\rightarrow\infty) = \sqrt {c_b}
$$ 
	    Substituting this condition into Eq.(\ref{deGennes_first_integral}) gives 
$$
	    A = \frac{\text{v}}{2}c_b^2 + \frac{\text{w}}{3}c_b^3
$$
	    Changing the variables in Eq.(\ref{deGennes_first_integral}) as $\psi(x) = \sqrt{c_b}f(x)$, where the corresponding boundary conditions for 
	    the function are: $f(0) = 0$ on the plate and $f(x\rightarrow\infty) = 1$ in the bulk phase. 
	    Substituting the constants, $\mu_{int}$ and $A$ in the expanded form, we can rewrite Eq.(\ref{deGennes_first_integral}) as
$$
	    a^2c_b\left(\frac{\mathrm{d}f}{\mathrm{d}x}\right)^2 = \frac{\text{v}c_b^2}{2}(f^4-2f^2+1) + \frac{\text{w}c_b^3}{6}(f^6-3f^2+2) =
	    \frac{\text{v}c_b^2}{2}(f^2-1)^2 + \frac{\text{w}c_b^3}{6}(f^2+2)(f^2-1)^2 
$$
	    Thus, 
$$
	    a^2\left(\frac{\mathrm{d}f}{\mathrm{d}x}\right)^2 = (f^2-1)^2\frac{c_b}{2}\left(\text{v} + \frac{\text{w}c_b}{3}(f^2+2)\right)
$$ 
	    In quadrature, we can rewrite this expression as
$$
	    \frac{x\sqrt{c_b}}{\sqrt{2}a}= \int\limits_0^f\frac{\mathrm{d}f}{(1-f^2)\sqrt{\text{v} + \frac{\text{w}c_b}{3}(f^2+2)}}
$$
	    where the minus sign before the integral is chosen for consistency between the left and the right hand sides of the equation (i.e. $f\le1$).  
	    In the reduced variables($\bar{x}=x/R_g, v_N=\text{v}c_bN, w_N=\text{w}c_b^2N/2$), the expression can be written as
\begin{equation}
\label{deGennes_quadrature_vw}
	    \frac{\bar{x}}{2} = \int\limits_0^f\frac{\mathrm{d}f}{(1-f^2)\sqrt{2(v_N + \frac{2w_N}{3}(f^2+2))}}
\end{equation}
	    One can see that one of the factors in the integrand coincides at $f = 1$ with the expression for the GSD correlation length, $\bar{\xi}$. 
	    Using this fact, we add and subtract under the integral the function
$$
	    y(f) = \frac{\bar{\xi}}{1-f^2} 
$$
	    Hence, we can rewrite Eq.(\ref{deGennes_quadrature_vw}) in the equivalent form
$$
	    \frac{\bar{x}}{2} = \bar{\xi}\int\limits_0^f\frac{\mathrm{d}f}{(1-f^2)}\left\{\frac{1}{\bar{\xi}\sqrt{2(v_N + \frac{2w_N}{3}(f^2+2))}}-1\right\}  + 
	                        \bar{\xi}\int\limits_0^f\frac{\mathrm{d}f}{(1-f^2)} = \bar{\xi}(I_1 + I_2)
$$
	    Let us analyze these integrals separately.\\
	    $I_1$) The integrand can be rewritten using the expression for the GSD correlation length:
$$
	    I_1 = \int\limits_0^f\frac{\mathrm{d}f}{(1-f^2)}\left\{\sqrt{\frac{2(v_N + 2w_N)}{2(v_N + \frac{2w_N}{3}(f^2+2))}}-1\right\} = 
		  \int\limits_0^f\frac{\mathrm{d}f}{(1-f^2)}\left\{\sqrt{\frac{1 + v_N/2w_N}{\frac{1}{3}(f^2+2) + v_N/2w_N}}-1\right\}
$$
	    It can be noted that the integrand is a smooth, slowly varying function, which depends on the ratio $v_N/2w_N$ as a parameter. Since we are 
	    interested in sufficiently big distances, $h\gg\xi$, where $f\rightarrow 1$, we can set the upper limit of the above integral $f_{up}=1$ 
	    without loss of accuracy. Therefore,
\begin{equation}
\label{deGennes_phase_i1_vw}
	    I_1(v_N/2w_N) \simeq \int\limits_0^1\frac{\mathrm{d}f}{(1-f^2)}\left\{\sqrt{\frac{1 + v_N/2w_N}{\frac{1}{3}(f^2+2) + v_N/2w_N}}-1\right\} 
\end{equation}
	    The numerical value for the integral, $I_1(v_N/2w_N)$, is changed from $I_1(\infty) = 0$ for $w_N = 0$ up to $I_1(0) = 0.203$ for $v_N=0$.\\
	    $I_2$) Let us consider the second integral:	    	    
$$
	    I_2 = \int\limits_0^f\frac{\mathrm{d}f}{(1-f^2)}
$$		
	    This integral can be easily calculated if we replace the integration variable as $f = \tanh(y)$ and it results in
$$
	    I_2 = {\rm arctanh}(f) 
$$
	    Joining the results from both integrals, we obtained
$$
	    f(x) = \tanh\left(\frac{x}{2\xi}-I_1(v_N/2w_N)\right), \quad \psi(x) = \sqrt{c_b}\tanh\left(\frac{x}{2\xi}-I_1(v_N/2w_N)\right)
$$
	    For $x\gg\xi$, the resulting concentration profile can be written as
\begin{equation}
\label{deGennes_conc_single_wall_vw}
	    c(x) = c_b\tanh^2\left(\frac{x}{2\xi}-I_1(v_N/2w_N)\right), \quad x\gg\xi
\end{equation}
	    or, asymptotically, the excess concentration at $x\gg\xi$ is
$$
	    c(x) - c_b = c_b\left(\tanh^2(\cdot)-1\right) = -\frac{c_b}{\cosh^2(\cdot)} \simeq -4c_bE_1(v_N/2w_N)e^{-x/\xi}
$$ 
	    where we introduced $E_1(v_N/2w_N)=\exp(2I_1(v_N/2w_N))$. 
	    As before, multiplying this result by $2$ (since the perturbation should be created by two plates) and replacing $x\rightarrow h/2$, we obtain 
\begin{equation}
\label{deGennes_conc_asy_gg_vw}
	    c_m - c_b \simeq -8c_bE_1(v_N/2w_N)e^{-h/2\xi}
\end{equation}
	    This expression is valid for $h\gg\xi$. Thus, we get the expression for universal function: 
\begin{equation}
\label{deGennes_univ_fun_g_asy_gg_vw}
	    g(h/2\xi,v_N/2w_N) = \frac{c_m}{2c_b} \simeq \frac{1}{2}\left(1 - 8E_1(v_N/2w_N)e^{-h/2\xi}\right) 
\end{equation}
	    Substituting it in Eq.(\ref{deGennes_force_gs_vw}), we obtain the result for the effective force between plates
$$
	    \hat{f}_{gs}(h/2\xi) \simeq 32E_1^2(v_N/2w_N)\left(\frac{1}{2\bar{\xi}^2}-\frac{16}{3}w_N E_1(v_N/2w_N) e^{-h/2\xi}\right) e^{-h/\xi}
$$
            Since we are interested in sufficiently big distances, we can neglect the second exponent. Therefore,
\begin{equation}
\label{deGennes_force_gs_vw_asy}
	    \hat{f}_{gs}(h/2\xi) \simeq \frac{16E_1^2(v_N/2w_N)}{\bar{\xi}^2} e^{-h/\xi}
\end{equation}
	    As a final step, we should integrate Eq.(\ref{deGennes_force_gs_vw_asy}) in order to obtain the GSD
	    expression for the interaction free energy at $h\gg\xi$:
\begin{equation}
\label{deGennes_w_gs_vw_gg}
	    \hat{W}_{gs} = -\frac{1}{R_g}\int\limits_h^{\infty}\mathrm{d}h\hat{f}_{gs}(h/2\xi) \simeq 
	    -\frac{16E_1^2(v_N/2w_N)}{\bar{\xi}}e^{-h/\xi} 
\end{equation}	    
	    One can notice that all of the above equations coincide with those obtained for the special case ($w_N=0$) due to the fact that $E_1(\infty)=1$. 
	    
	    In Fig.\ref{deGennes_g_vw_asy_fig} we draw the universal function $g=c_m/2c_b$ and its asymptotics obtained for different values of the virial parameters.
	    The corresponding curves for the free energy are depicted in Fig.\ref{deGennes_free_en_gs_vw_fig}. 	    
	    The difference between the curves for the different virial parameters and the convergence of the asymptotics become noticeable 
	    for roughly the same values of $h/2\xi$, as compared with those presented in Fig.\ref{deGennes_g_vw_asy_fig}. 
	    Based on the pictures, one can observe that the convergence of the asymptotics is reached earlier for bigger values of the ratio $v_N/w_N$.
%%%%%%%%%%%%%%%%%%%%%%%%%%%%%%%%%%%%%%%%%%%%%%%%%%%%%%%%%%%%%%%%%%%%%%%%%%%%%%%%%%%%%%%%%%%%%%%%%%%%%%%%%%%%%%%%%%%%%%%%%%%%%%%%%%%%%%%%%%%%
%          universal function g and its asymptotics at x \gg 1 for vw
%%%%%%%%%%%%%%%%%%%%%%%%%%%%%%%%%%%%%%%%%%%%%%%%%%%%%%%%%%%%%%%%%%%%%%%%%%%%%%%%%%%%%%%%%%%%%%%%%%%%%%%%%%%%%%%%%%%%%%%%%%%%%%%%%%%%%%%%%%%%
%%%%%%%%%%%%%%%%%%%%%%%%%%%%%%%%%%%%%%%%%%%%%%%%%%%%%%%%%%%%%%%%%%%%%%%%%%%%%%%%%%%%%%%%%%%%%%%%%%%%%%%%%%%%%%%%%%%%%%%%%%%%%%%%%%%%%%%%%%%%%%%%%%%%%%%%%%%%
%           Computational details
%%%%%%%%%%%%%%%%%%%%%%%%%%%%%%%%%%%%%%%%%%%%%%%%%%%%%%%%%%%%%%%%%%%%%%%%%%%%%%%%%%%%%%%%%%%%%%%%%%%%%%%%%%%%%%%%%%%%%%%%%%%%%%%%%%%%%%%%%%%%%%%%%%%%%%%%%%%%  
\subsection{Computational details}
	    In this section, we consider problems related to the calculation of the thermodynamic potential first in the case when the third virial coefficient
	    equals to $0$ and then in the general case. These problems are reduced to sufficiently accurate calculations of the universal functions $f$ and $g$.
	    At the beginning, we considered the calculation of the universal function $f$, Eq.(\ref{deGennes_universal_f}). 
	    A direct calculation of the elliptic integral, Eq.(\ref{deGennes_elliptic}) using, for example, the Simpon's rule, suffers from a divergency above 
	    certain value of the parameter $k$. One may overcome this problem using 
	    the appropriate numerical library \textbf{BOOST} \cite{Boost}, where all difficulties with the numerical integration
	    are remarkably resolved. As a result, we obtain a set of numerical values $\{h_i, c_{m,i}\}$ \footnote{The variable $h$ has different step sizes, while 
	    the corresponding concentration has equidistant values, $c_i$, which serve as the input values 
	    for calculation of the elliptic integral.} 
	    that is used to calculate the corresponding dependency for the thermodynamic potential.

	    In the general case ($w_N\ne 0$) of the universal function $g$, Eq.(\ref{deGennes_universal_vw}), we do not have an implemented numerical library for 
	    the corresponding integral, eq.(\ref{deGennes_quadratures_vw}). However, in this case the integrand is not so sensitive to the parameter $x$ 
	    as it occurred in the previous case.  
	    For calculation of the integral, we divided the range for $x\in[0..0.5]$ into three intervals: 
	    $[0..x_1], [x_1..x_2], [x_2..x_3]$ where $x_1 = 0.5-10^{-1}, x_2 = 0.5-10^{-5}, x_3 = 0.5-10^{-8}$.
	    We use in each of these intervals the same number of grid points.	    
	    
	    It is also noteworthy that the \textbf{BOOST} implementation is faster more than $10$ times in comparison with our numerical calculation 
	    for the general case using the same number of grid points.
%%%%%%%%%%%%%%%%%%%%%%%%%%%%%%%%%%%%%%%%%%%%%%%%%%%%%%%%%%%%%%%%%%%%%%%%%%%%%%%%%%%%%%%%%%%%%%%%%%%%%%%%%%%%%%%%%%%%%%%%%%%%%%%%%%%%%%%%%%%%%%%%%%%%%%%%%%%%
%           Comparison wiwh SCFT results
%%%%%%%%%%%%%%%%%%%%%%%%%%%%%%%%%%%%%%%%%%%%%%%%%%%%%%%%%%%%%%%%%%%%%%%%%%%%%%%%%%%%%%%%%%%%%%%%%%%%%%%%%%%%%%%%%%%%%%%%%%%%%%%%%%%%%%%%%%%%%%%%%%%%%%%%%%%%  
\section{Comparison with SCFT results}
\begin{figure}[ht]
\begin{minipage}[ht]{0.47\linewidth}
\center{\includegraphics[width=1\linewidth]{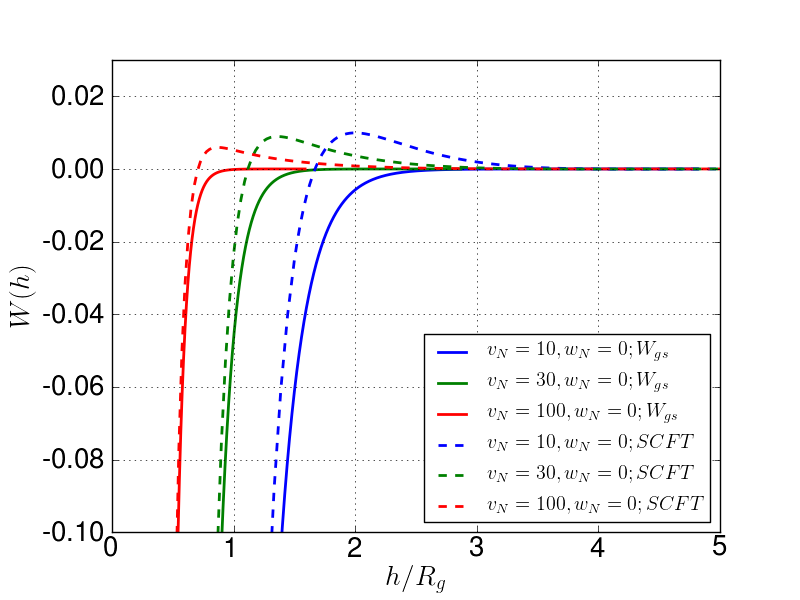}}
\caption{\small{The comparison between thermodynamic potentials obtained in the GSD approximation, Eq.(\ref{deGennes_w_gs_v_num}), (continuous lines) 
	        and in the SCFT (dashed lines) for different values of the virial parameter,$v_N$ at $w_N = 0$.}}
\label{deGennes_free_en_gs_scft_v1_fig}
\end{minipage}
\hfill
\begin{minipage}[ht]{0.47\linewidth}
\center{\includegraphics[width=1\linewidth]{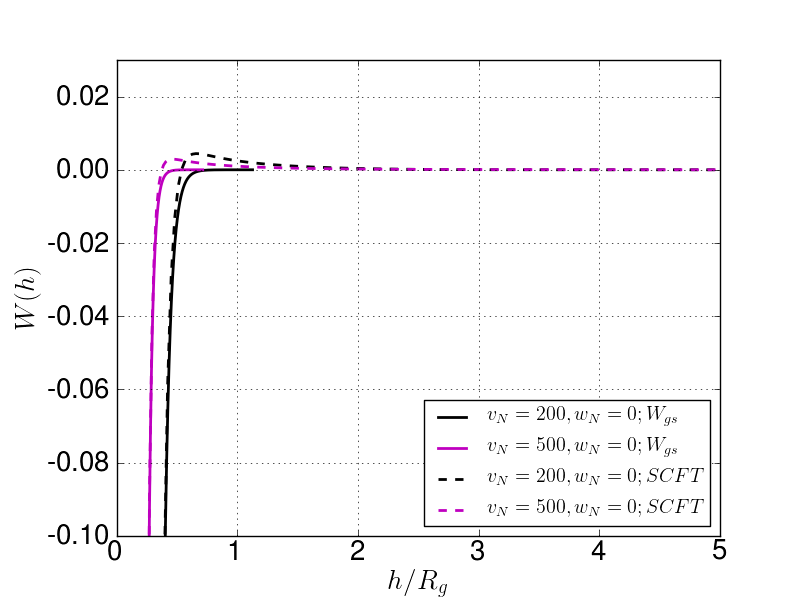}} 
\caption{\small{The comparison between thermodynamic potentials obtained in the GSD approximation, Eq.(\ref{deGennes_w_gs_v_num}), (continuous lines)  
		and in the SCFT (dashed lines) for different values of the virial parameter,$v_N$ at $w_N = 0$.}} 
\label{deGennes_free_en_gs_scft_v2_fig}
\end{minipage}
\end{figure}

\begin{figure}[ht]
\begin{minipage}[ht]{0.47\linewidth}
\center{\includegraphics[width=1\linewidth]{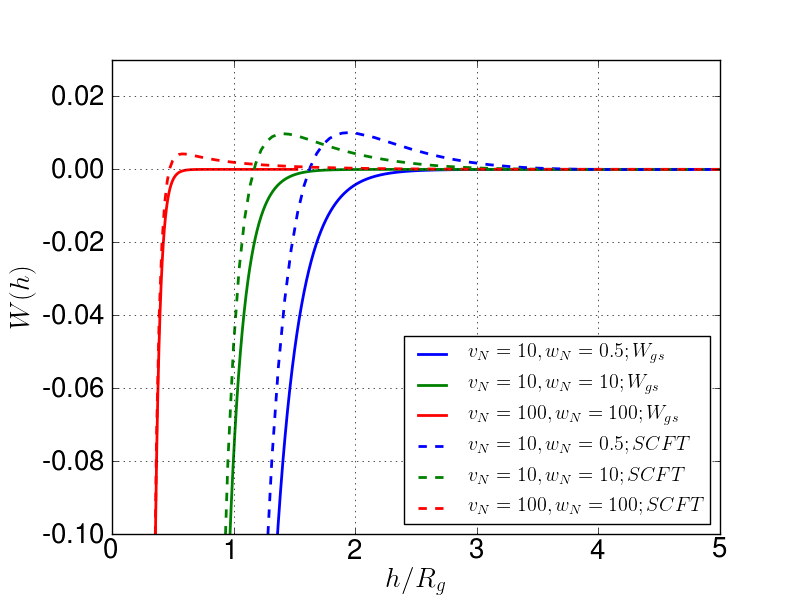}}
\caption{\small{The comparison between thermodynamic potentials obtained in the GSD approximation, Eq.(\ref{deGennes_w_gs_vw_num}), (continuous lines)
		and in the SCFT (dashed lines) for different values of the virial parameters,$v_N$, $w_N$.}}
\label{deGennes_free_en_gs_scft_vw1_fig}
\end{minipage}
\hfill
\begin{minipage}[ht]{0.47\linewidth}
\center{\includegraphics[width=1\linewidth]{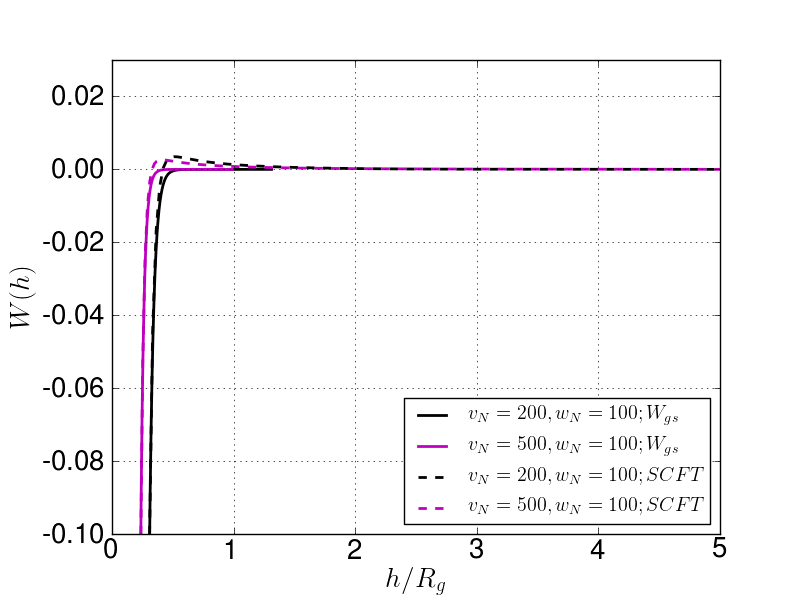}} 
\caption{\small{The comparison between thermodynamic potential obtained in the GSD approximation, Eq.(\ref{deGennes_w_gs_vw_num}), (continuous lines) 
		and in the SCFT (dashed lines) for different values of the virial parameters,$v_N$, $w_N$}}
\label{deGennes_free_en_gs_scft_vw2_fig}
\end{minipage}
\end{figure}

	    In this section, we compare the results for the ground state free energy $W_{gs}$, obtained in the limit of infinitely long chains, 
	    with the corresponding results obtained numerically by the SCFT (for finite chain lengths).  
	    In Figs.\ref{deGennes_free_en_gs_scft_v1_fig}--\ref{deGennes_free_en_gs_scft_vw2_fig} we present the comparison, 
	    showing at the same picture the results of both calculations for different values of the virial parameters. 
	    The continuous lines correspond the free energy obtained in the GSD approximation,
	    Eq.(\ref{deGennes_w_gs_v_num}), (\ref{deGennes_w_gs_vw_num}), 
	    and the dashed lines correspond to the SCFT calculations considered before. The monotonic behavior of the thermodynamic potential obtained in the 
	    GSD approximation corresponds to the attraction between plates confirming the well known-effect of the depletion attraction between colloidal 
	    particles in polymer solution. Due to the finite length of polymer chain in the SCFT, there are nonmonotonic behaviors 
	    of the thermodynamic potential with pronounced barrier corresponding to polymer-induced repulsion between the plates.
 	    
	    In the attractive range of the SCFT potential, we can describe the behavior of the thermodynamic potential by the ground state free energy, 
	    Eqs.(\ref{deGennes_w_gs_v_num}, \ref{deGennes_w_gs_vw_num}), (by their asymptotics) with exact RPA correlation length, Eq.(\ref{intr_polymer_correlation_length}). 
	    It ensures much better agreement between the attractive parts of the SCFT potential and the GSD free energy, 
	    especially, for small virial parameters in comparison with the calculations based on the GSD correlation length. 
 
	    We also demonstrated that the thermodynamic potentials corresponding to both of these approaches become closer and closer 
	    upon increasing of the virial parameters, $v_N = \text{v}c_bN$ and $w_N=\text{w}c_b^2N/2$ or for fixes
	    parameters: $c_b$, $\text{v}, \text{w}$ (defining blob size) and increasing $N$. 
	    The difference between the solutions decreases with increasing the polymer chain length, $N$, and in the limit, $N\rightarrow\infty$, they coincide.
	    
%%%%%%%%%%%%%%%%%%%%%%%%%%%%%%%%%%%%%%%%%%%%%%%%%%%%%%%%%%%%%%%%%%%%%%%%%%%%%%%%%%%%%%%%%%%%%%%%%%%%%%%%%%%%%%%%%%%%%%%%%%%%%%%%%%%%%%%%%%%%
%          Comparison of the deGennes results with SCFT for w = 0
%%%%%%%%%%%%%%%%%%%%%%%%%%%%%%%%%%%%%%%%%%%%%%%%%%%%%%%%%%%%%%%%%%%%%%%%%%%%%%%%%%%%%%%%%%%%%%%%%%%%%%%%%%%%%%%%%%%%%%%%%%%%%%%%%%%%%%%%%%%%
%%%%%%%%%%%%%%%%%%%%%%%%%%%%%%%%%%%%%%%%%%%%%%%%%%%%%%%%%%%%%%%%%%%%%%%%%%%%%%%%%%%%%%%%%%%%%%%%%%%%%%%%%%%%%%%%%%%%%%%%%%%%%%%%%%%%%%%%%%%%
%          Comparison of the deGennes results with SCFT for v, w \ne 0
%%%%%%%%%%%%%%%%%%%%%%%%%%%%%%%%%%%%%%%%%%%%%%%%%%%%%%%%%%%%%%%%%%%%%%%%%%%%%%%%%%%%%%%%%%%%%%%%%%%%%%%%%%%%%%%%%%%%%%%%%%%%%%%%%%%%%%%%%%%%
%%%%%%%%%%%%%%%%%%%%%%%%%%%%%%%%%%%%%%%%%%%%%%%%%%%%%%%%%%%%%%%%%%%%%%%%%%%%%%%%%%%%%%%%%%%%%%%%%%%%%%%%%%%%%%%%%%%%%%%%%%%%%%%%%%%%%%%%%%%%%%%%%%%%%%%%%%%%
%           Semenov's theory of colloidal stabilization and its comparison with SCFT
%%%%%%%%%%%%%%%%%%%%%%%%%%%%%%%%%%%%%%%%%%%%%%%%%%%%%%%%%%%%%%%%%%%%%%%%%%%%%%%%%%%%%%%%%%%%%%%%%%%%%%%%%%%%%%%%%%%%%%%%%%%%%%%%%%%%%%%%%%%%%%%%%%%%%%%%%%%%  
\section{The GSDE theory of colloidal stabilization and its comparison with the SCFT} 
	    In this section we consider the GSDE theory \cite{semenov_1996, semenov_2008}, which is more advanced compared to the classical GSD approach.
	    This theory accounts for the effect of the finite chain length, that provides a repulsion contribution in the thermodynamic potential.
	    
\begin{figure}[ht]
\begin{minipage}[ht]{0.47\linewidth}
\center{\includegraphics[width=1\linewidth]{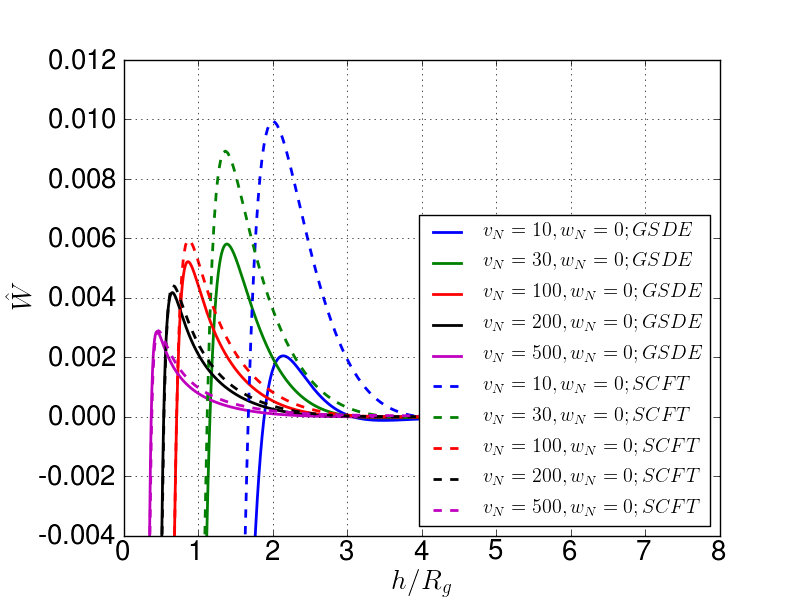}}
\caption{\small{The comparison between the thermodynamic potentials calculated by the GSDE theory (continuous lines), Eq.(\ref{ans_w_e_vw_general}), 
		with the ground state free energy, Eq.(\ref{deGennes_w_gs_v_gg}), and the SCFT (dashed lines) for different values of 
		the virial parameter, $v_N$, at $w_N=0$.}}
\label{ans_free_en_ans_scft_v_fig}
\end{minipage}
\hfill
\begin{minipage}[ht]{0.47\linewidth}
\center{\includegraphics[width=1\linewidth]{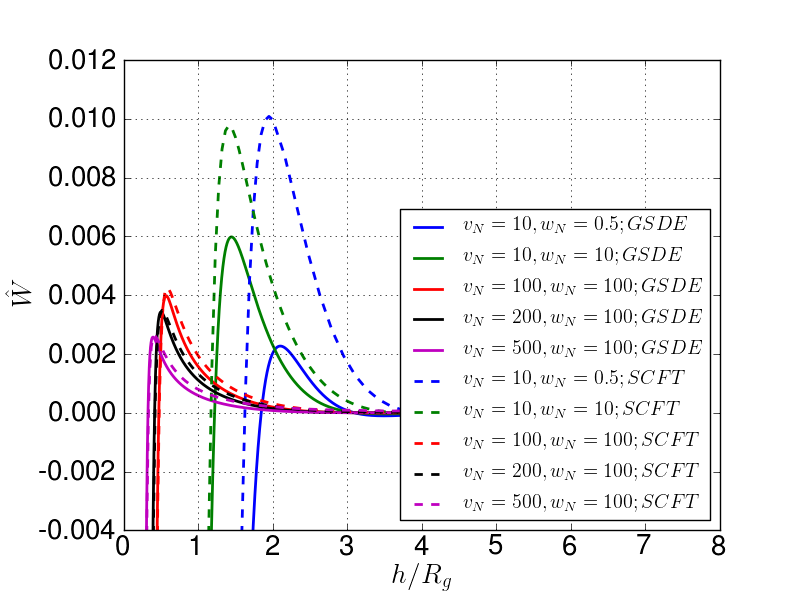}} 
\caption{\small{The comparison between the thermodynamic potentials calculated by the GSDE theory (continuous lines), Eq.(\ref{ans_w_e_vw_general}), with the ground state 
		free energy, Eq.(\ref{deGennes_w_gs_v_gg}), and SCFT (dashed lines) for different values of the virial parameter, $v_N$, $w_N$.}}
\label{ans_free_en_ans_scft_vw_fig}
\end{minipage}
\end{figure}

	    In the articles, \cite{semenov_1996, semenov_2008}, equivalent results were obtained for the long range interaction between 
	    the plates using completely different approaches. It was shown that the long range interaction corresponds to the repulsion caused by 
	    a redistribution of the chain ends. 
	    The repulsive part of the potential in the GSDE theory corresponds to the additional term complementing the GSD exponential function, i.e.
\begin{equation}
\label{ans_w_tot}
	   W = W_{gs} + W_{e}
\end{equation}
           where $W_{gs}$ is the ground-state free energy\footnote{The reduced GSD free energy is defined in Eq.(\ref{deGennes_w_gs_vw_gg}). 
           It is not necessarily to use more general Eqs.(\ref{deGennes_w_gs_v_num}-\ref{deGennes_w_gs_vw_num}), 
           because the analytical repulsive term $W_e$ is valid for $\xi \ll h$.}, 
	   Eq.(\ref{deGennes_w_gs_vw_gg}), and the long range interaction corresponds to
$$
	   W_{e} = \frac{4c_b}{R_gN}\Delta_e^2u_{int}(h/R_g) = \frac{c_bR_g}{N} \hat{W}_e
$$
	   Now, the reduced thermodynamic potential $\tilde{W}_e$ is 
\begin{equation}
\label{ans_w_e}
	   \hat{W}_e = 4\bar{\Delta}_e^2u_{int}(\bar{h})
\end{equation}
	   where $\bar{\Delta}_e = \Delta_e/R_g$, $\bar{h} = h/R_g$ and 
\begin{equation}
\label{ans_delta_e_def}
	\Delta_e = \int\limits_0^{\infty}\mathrm{d}x \left(\sqrt{c_1(x)/c_b} - c_1(x)/c_b\right)
\end{equation}
	is an effective length which defines the single wall excess of end points for the case of semi-infinite system with a single wall placed at $x=0$. 
	The concentration profile caused by the single wall is defined in Eq.(\ref{deGennes_conc_single_wall_vw}), only far from the wall, so it
	can not be used to calculate $\Delta_e$ in Eq.(\ref{ans_delta_e_def}). For the calculation we must use the numerical solution of 
	Eq.(\ref{deGennes_quadrature_vw}). In \cite{semenov_2008}, the analytical solution for $\Delta_e$, expressed via elementary 
	functions was found:\footnote{We verified that the analytical solution coincides with the numerical results based on Eq.(\ref{deGennes_quadrature_vw}).}   
\begin{equation}
\label{ans_delta_e}
	   \bar{\Delta}_e = \frac{\Delta_e}{R_g} = \frac{\sqrt{6}a}{\sqrt{\text{w}}c_bR_g}I\left(2+\frac{3\text{v}}{\text{w}c_b}\right) = 
						    \frac{\sqrt{3}}{\sqrt{w_N}}I\left(2+\frac{3v_N}{2w_N}\right)	
\end{equation}
	where we used that: $v_N=\text{v}c_bN, w_N=\text{v}c_b^2N/2$ and
\begin{equation}
\label{ans_i_alpha}
	I(\alpha) \equiv \left(1-\frac{1}{\sqrt{1+\alpha}}\right)\ln\frac{1+\sqrt{1+\alpha}}{\sqrt{\alpha}} - \frac{\ln 2}{\sqrt{1+\alpha}} + 
			      \frac{2}{\sqrt{1+\alpha}}\ln\left(\sqrt{1+\alpha}+1-\sqrt{\alpha}\right)
\end{equation}
	For $\text{w}c_b/\text{v}\ll1$, Eq.(\ref{ans_delta_e}), has the asymptotics:\footnote{It also can be found via concentration profile obtained in 
	Eq.(\ref{deGennes_conc_single_wall_v}) for the case when $w_N=0$.}
\begin{equation}
\label{ans_delta_e_asy}
	    \bar{\Delta}_e \simeq 2\bar{\xi}(1 - \ln 2) \simeq \sqrt{2}(1 - \ln 2)\frac{1}{\sqrt{v_N}}
\end{equation}
	 The last expression will be used for $\text{w}=0$. Next,
\begin{equation}
\label{ans_u_int}
	    u_{int}(r) = \frac{1}{r}\sum\limits_{n=-\infty}^{\infty}f(4\pi^2n^2/r^2) - \kappa
\end{equation}
	  where
$$
	    f(u) = \frac{1 - e^{-u}(u + 1)}{u - 1 + e^{-u}}
$$
	  and 
$$
	    \kappa \equiv \int\limits_{-\infty}^{\infty}\frac{\mathrm{d}k}{2\pi}f(k^2) \simeq 0.6161874
$$	    	 
	  Finally, we can write the expression for the thermodynamic potential for arbitrary virial parameters as
\begin{equation}
\label{ans_w_e_vw}
	   \hat{W}_e = \frac{12}{w_N}I^2\left(2+\frac{3v_N}{2w_N}\right)u_{int}(\bar{h}) 
\end{equation}
	    or for $w_N=0$
\begin{equation}
\label{ans_w_e_v}
	   \hat{W}_e = \frac{8}{v_N}(1-\ln 2)^2u_{int}(\bar{h})
\end{equation}
\textbf{Generalization.} For comparison with the SCFT we will use instead of Eqs.(\ref{ans_w_e_vw}, \ref{ans_w_e_v}), the more general equation 
proposed in \cite{semenov_2008}, namely, 
$$
	    W_e \simeq \frac{2c_b}{N}\left[2\Delta_e - h\ln\left(1+\frac{2\Delta_e}{h}\right)\right] H(h/R_g)
$$
Noting that $u_{int} \equiv (R_g/h) H(h/R_g)$, we can rewrite $W_e$ in the dimensionless variables:
\begin{equation}
\label{ans_w_e_vw_general}
	    \hat{W}_e = \frac{R_gc_b}{N}W_e \simeq 2\bar{h}\left[2\bar{\Delta}_e - \bar{h}_{b}\ln\left(1+\frac{2\bar{\Delta}_e}{\bar{h}_{b}}\right)\right] u_{int}(\bar{h})
\end{equation}
	    where we replaced in the square brackets the distance by the effective distance $h_b$:
$$
	    h_b = \frac{1}{c_b}\int\limits_0^h\mathrm{d}x c(x) = h + \frac{1}{c_b}\int\limits_0^h\mathrm{d}x \left(c(x) - 1\right) \simeq h + 2\Delta_c
$$
	    where 
$$
	   \Delta_c = \int\limits_0^{\infty} \mathrm{d}x \left(c_1(x)/c_b - 1\right)
$$
	   is the effective single-plate excess of polymer amount. Let us find analytical expressions for $\Delta_c$ in two cases:\\
	   \textbf{$\text{w}=0$.} Using the single-plate concentration profile, Eq.(\ref{deGennes_conc_single_wall_v}), we have obtained
$$
	   \Delta_c = \int\limits_0^{\infty}\mathrm{d}x \left(\tanh^2\left(\frac{x}{2\xi}\right) - 1\right) =
	              2\xi\int\limits_{0}^{\infty}\mathrm{d}y \left(\tanh^2(y) - 1\right) = -2\xi
$$
	   Therefore, we can write
$$
	   \bar{h}_b = \frac{h_b}{R_g} \simeq \bar{h} - 4\bar{\xi}
$$
	   \textbf{$\text{w}\ne0$.} In this more general case, we do not have the analytical expression for the concentration profile, but 
	   using Eq.(\ref{deGennes_quadrature_vw}), we can write
$$
	   \mathrm{d}\bar{x} = \frac{2\mathrm{d}f}{(1-f^2)\sqrt{2(v_N+2w_N(f^2+2)/3)}}
$$
	   Replacing the variable of integration in $\Delta_c$ by the above expression, we have 
$$
	   \bar{\Delta}_c = \int\limits_0^{\infty}\mathrm{d}\bar{x} (f^2-1) = -\sqrt{\frac{3}{w_N}}\int\limits_0^1\frac{\mathrm{d}f}{\sqrt{\frac{3v_N}{2w_N}+2 +f^2}}=
		             -\sqrt{\frac{3}{w_N}}\ln\left\{\sqrt{1+\frac{1}{\alpha}} + \frac{1}{\sqrt{\alpha}}\right\}
$$
	   where we introduced $\alpha\equiv 2 + 3v_N/2w_N$. Hence, 
$$
	   \bar{h}_b \simeq \bar{h} -  2\sqrt{\frac{3}{w_N}}\ln\left\{\sqrt{1+\frac{1}{\alpha}} + \frac{1}{\sqrt{\alpha}}\right\} 
$$ 
	   In Figs.\ref{ans_free_en_ans_scft_v_fig}--\ref{ans_free_en_ans_scft_vw_fig}, we present the comparison for the dimensionless 
	   thermodynamic potentials obtained by the GSDE theory, Eq.(\ref{ans_w_e_vw_general}), and by the $SCFT$.
	   The results obtained by numerical solution of the SCFT equations are in a reasonable agreement with the GSDE analytical theory and a 
	   full coincidence is nearly reached for big virial parameters. 
%%%%%%%%%%%%%%%%%%%%%%%%%%%%%%%%%%%%%%%%%%%%%%%%%%%%%%%%%%%%%%%%%%%%%%%%%%%%%%%%%%%%%%%%%%%%%%%%%%%%%%%%%%%%%%%%%%%%%%%%%%%%%%%%%%%%%%%%%%%%
%          Comparison of the ANS results with SCFT for v, w separately
%%%%%%%%%%%%%%%%%%%%%%%%%%%%%%%%%%%%%%%%%%%%%%%%%%%%%%%%%%%%%%%%%%%%%%%%%%%%%%%%%%%%%%%%%%%%%%%%%%%%%%%%%%%%%%%%%%%%%%%%%%%%%%%%%%%%%%%%%%%%
	  
Quantitatively, we compare the results for the barrier's heights:
$$
	 \hat{W}^* = \hat{W}_{max} - \hat{W}_{inf}
$$
and the corresponding dimensionless positions of the barrier's maximum: $\bar{h}^*=h^*/R_g$. The numerical values for those quantities are present in Tab.\ref{tabular:barrier_max_ans_scft_vw} for two different approaches (GSDE and SCFT).	
%%%%%%%%%%%%%%%%%%%%%%%%%%%%%%%%%%%%%%%%%%%%%%%%%%%%%%%%%%%%%%%%%%%%%%%%%%%%%%%%%%%%%%%%%%%%%%%%%%%%%%%%%%%%%%%%%%%%%%%%%%%%%%%%%%%%%%%%%%%%
% The height of the barrier for vw i.e comparison ANS and SCFT
%%%%%%%%%%%%%%%%%%%%%%%%%%%%%%%%%%%%%%%%%%%%%%%%%%%%%%%%%%%%%%%%%%%%%%%%%%%%%%%%%%%%%%%%%%%%%%%%%%%%%%%%%%%%%%%%%%%%%%%%%%%%%%%%%%%%%%%%%%%%
\begin{table}[ht!]
\caption{The thermodynamic potential barrier height $\hat{W}^*$ and its position $\bar{h}^*$ for the GSDE theory  
	 and SCFT obtained for different values of the virial parameters. 
	 The corresponding results are shown in Figs.\ref{ans_free_en_ans_scft_v_fig}--\ref{ans_free_en_ans_scft_vw_fig}.} 
\label{tabular:barrier_max_ans_scft_vw} 
\begin{flushleft}
  \begin{tabular}{ | c | c | c | c | c | c | c | c | c | c | c | }
    \hline
    $v_N$                           &    10&    30&   100&   200&   500&     10&   10&   100&   200&   500                          \\ \hline
    $w_N$                           &     0&     0&     0&     0&     0&    0.5&   10&   100&   100&   100                          \\ \hline
                                         \multicolumn{11}{|c|}{SCFT}                                                                \\ \hline
    $\hat{W}^*,\times 10^{-3}$      & 9.936& 8.937& 5.941& 4.421& 2.896& 10.083& 9.773& 4.235& 3.482& 2.618                         \\ \hline
    $\bar{h}^*$                     & 2.005& 1.364& 0.879& 0.656& 0.464&  1.947& 1.409& 0.576&  0.51& 0.407                         \\ \hline
                                             \multicolumn{11}{|c|}{GSDE, Eq.(\ref{ans_w_e_vw_general})}                            \\ \hline
    $\hat{W}^*,\times 10^{-3}$      & 2.049&  5.81& 5.216& 4.177& 2.864&  2.269& 5.431& 3.982& 3.432& 2.575                         \\ \hline
    $\bar{h}^*$                     & 2.139& 1.382& 0.861& 0.651& 0.446&    2.1&  1.57& 0.568& 0.499& 0.392                         \\ \hline    
					      \multicolumn{11}{|c|}{GSDE, Eq.(\ref{ans_w_e_vw}$, $\ref{ans_w_e_v})}                 \\ \hline
    $\hat{W}^*,\times 10^{-3}$      & 0.779& 3.657& 3.674& 3.051& 2.166&  1.086& 3.713& 2.903& 2.543& 1.955                         \\ \hline
    $\bar{h}^*$                     & 2.325& 1.466& 0.902& 0.678& 0.463&  2.236& 1.525& 0.591& 0.519& 0.406                         \\ \hline    
					      \multicolumn{11}{|c|}{GSDE asymptotics, Eq.(\ref{ans_barrier_height_vw})}              \\ \hline
    $\hat{W}^*,\times 10^{-2}$      &    - &    - & 2.007& 1.256& 0.689&     - &    - & 1.097& 0.878& 0.586                         \\ \hline
    $\bar{h}^*$                     &    - &    - & 0.373& 0.299& 0.218&     - &    - & 0.261& 0.236& 0.193     \\

    \hline
  \end{tabular}
\end{flushleft} 
\end{table}

%%%%%%%%%%%%%%%%%%%%%%%%%%%%%%%%%%%%%%%%%%%%%%%%%%%%%%%%%%%%%%%%%%%%%%%%%%%%%%%%%%%%%%%%%%%%%%%%%%%%%%%%%%%%%%%%%%%%%%%%%%%%%%%%%%%%%%%%%%%%%%%%%%%%%%%%%%%%
%           Scaling behavior at xi << 1
%%%%%%%%%%%%%%%%%%%%%%%%%%%%%%%%%%%%%%%%%%%%%%%%%%%%%%%%%%%%%%%%%%%%%%%%%%%%%%%%%%%%%%%%%%%%%%%%%%%%%%%%%%%%%%%%%%%%%%%%%%%%%%%%%%%%%%%%%%%%%%%%%%%%%%%%%%%%  	    
\section{Scaling behavior at $\bar{\xi} \ll$ 1} 
	As we have already seen in the previous section, the thermodynamic potential barrier height and its position calculated by the SCFT asymptotically coincide with 
	the corresponding quantities calculated by the GSDE theory at $\bar{\xi}\ll1$. Thus, based on the analytical formulae, we can find 
	the scaling behavior for the functions valid for $\bar{\xi}\ll1$. 
	At the beginning, we simplify the equations taking into account that the interesting range is $\bar{\xi}\ll \bar{h} \ll 1$. 
	Thereby, all terms in the sum of Eq.(\ref{ans_u_int}) can be neglected except 
	for $n=0$. Thus, we can write Eq.(\ref{ans_w_e}) as
$$
	\hat{W}_e \simeq \frac{4\bar{\Delta}_e^2}{\bar{h}}
$$
	The GSD free energy, Eq.(\ref{deGennes_w_gs_vw_gg}), has the following form
$$
	\hat{W}_{gs} \simeq -\frac{16E_1^2}{\bar{\xi}}e^{-h/\xi}
$$
	Combining these formulas, we can write the total free energy as	
\begin{equation}
\label{ans_w_e_simple}
	\hat{W} = \hat{W}_{gs} + \hat{W}_e \simeq -\frac{16E_1^2}{\bar{\xi}}e^{-h/\xi} + \frac{4}{\bar{h}}\bar{\Delta}_e^2 
\end{equation}
	Let us find the scaling dependence for the barrier position and the corresponding barrier height considering separately three different cases:\\
	a) $w_N=0$, $E_1=1$. Based on Eq.(\ref{ans_delta_e_asy}), we can write 
$$
	\bar{\Delta}_e \simeq \sqrt{2}(1 - \ln 2)\frac{1}{\sqrt{v_N}} = 2(1-\ln2)\bar{\xi}
$$
	Differentiating Eq.(\ref{ans_w_e_simple}) in order to find the maximum by the variable $\bar{h}$, we have
$$
	\hat{W}' \simeq \frac{16}{\bar{\xi}^2}e^{-h^*/\xi} - 16(1-\ln 2)^2\left(\frac{\xi}{h^*}\right)^2 = 0
$$
	or after simple transformations
$$
	-\frac{h^*}{2\xi} = \ln\xi + \ln\frac{\xi}{h^*} + \ln(1-\ln2)
$$
	Leaving only the leading term, we obtained
\begin{equation}
\label{ans_barrier_position_vw_scaling}
	h^* \simeq \xi \ln\frac{1}{\bar{\xi}^2}
\end{equation}
	The first term in Eq.(\ref{ans_w_e_simple}) by virtue of the exponential function tends to zero very quickly at $h\geqslant h^*$. 
	Hence, in this equation, we can neglect the negative part at large distances and take into account only the second term. 
	After substituting Eq.(\ref{ans_barrier_position_vw_scaling}) into Eq.(\ref{ans_w_e_simple}), 
	we obtain the expression for the barrier height in the case $w_N=0$:
$$
	\hat{W}^* \simeq \frac{16}{\bar{h}^*}(1-\ln 2)^2\bar{\xi}^2 = 16(1-\ln 2)^2 \frac{\bar{\xi}}{\ln\frac{1}{\bar{\xi}^2}} \sim \bar{\xi} \sim v_N^{-1/2}
$$
	b) $v_N=0$. In this case we can rewrite Eq.(\ref{ans_delta_e}) in terms of the GSD correlation length as  	
$$
	\bar{\Delta}_e = \frac{\sqrt{3}}{\sqrt{w_N}}I(2) = 2\sqrt{3}I(2)\bar{\xi}
$$
	One can notice that in Eq.(\ref{ans_w_e_simple}) the additional prefactor $E_1^2$ for the first term and the different 
	multiplier for the second term do not affect the leading term in the expression for $h^*$. Thereby, the answer for the 
	the barrier position is the same as in the previous case, Eq.(\ref{ans_barrier_position_vw_scaling}).
	Substituting Eq.(\ref{ans_w_e_simple}), we obtain
$$
	\hat{W}^* \simeq \frac{48}{\bar{h}^{*}}I^2(2)\bar{\xi}^2 = 48I^2(2)\frac{\bar{\xi}}{\ln\frac{1}{\bar{\xi}^2}} \sim \bar{\xi} \sim w_N^{-1/2}
$$
	c) $v_N\ne 0, w_N \ne 0$. Due to the fact that the logarithm is the slowly changing function, we can write 
$$
	I(\alpha) \sim \frac{1}{\sqrt{1+\alpha}} \quad \text{or} \quad I\left(2+\frac{3v_N}{2w_N}\right) \sim \frac{\sqrt{2w_N}}{\sqrt{3}}\bar{\xi}
$$
	Accordingly,
$$
	\bar{\Delta}_e \sim \sqrt{2}\bar{\xi}
$$  
	so, as before, $\bar{\Delta}_e$ is proportional to the GSD correlation length $\bar{\xi}$. 
	Thereby, the barrier position is the same function as we obtained in the previous cases, i.e. Eq.(\ref{ans_barrier_position_vw_scaling}). 
	The corresponding barrier height in this case is 
$$
	\hat{W}^* \sim \frac{8}{\bar{h}^*}\bar{\xi}^2 = 8\frac{\bar{\xi}}{\ln\frac{1}{\bar{\xi}^2}} \sim \bar{\xi} 
$$
	Therefore, it was shown in any of the previous cases that the scaling behavior for the barrier height and its position 
	are corresponingly $\hat{W}^*\sim \bar{\xi}$ and $\bar{h}^*\sim \bar{\xi}$.

	\textbf{Comparison}. Summarizing the previous analytical expressions for the thermodynamic potential barrier height and its position, 
	 we can write them in the more general form\footnote{Notice that the expression
	 for $\hat{W}^*(w_N = 0)$ can be derived from the more general expression when $w_N \ne 0$, but in order to simplify the numerical calculation using 
	 that formula, we show them separately.}	
\begin{equation}
\label{ans_barrier_height_vw}
\begin{array}{l}
	\bar{h}^* \simeq \bar{\xi}\ln\frac{1}{\bar{\xi}^2}, \\
	\hat{W}^*(w_N = 0) \simeq 16(1-\ln2)^2\frac{\bar{\xi}}{\ln\frac{1}{\bar{\xi}}}, \\
        \hat{W}^*(w_N\ne 0) \simeq \frac{24}{\sqrt{w_N}}\sqrt{1+\frac{v_N}{2w_N}}\frac{I^2\left(2+\frac{3v_N}{2w_N}\right)}{\ln\left(\frac{1}{\bar{\xi}^2}\right)}	
\end{array}
\end{equation}
	where $I(\alpha)$ is defined in Eq.(\ref{ans_i_alpha}).
	These expressions are valid for $\bar{\xi} \ll 1$ or in terms of the virial parameters for $(v_N+w_N)\gg1$. 
	Unfortunately, this condition is very crucial and those formulae are applicable only for extremely small $\bar{\xi}$(extremely large $(v_N+w_N)$).
        We add two additional rows in Tab.\ref{tabular:barrier_max_ans_scft_vw} using Eq.(\ref{ans_barrier_height_vw}). 
	We restricted the values of the virial parameters by the condition 
	$\bar{\xi} < 0.1$ (we started from $v_N=100$ that corresponds to $\bar{\xi}\simeq 0.07$). One can see that eq.(\ref{ans_barrier_height_vw}) 
	overestimates the barrier height by the factor of $\sim 2$ in the case with the largest considered virial parameters: $v_N=500, w_N=100$.
	Even for extremely big virial parameter: $v_N=20k, w_N=0$ the values for the barrier height and its position, 
	Eq.(\ref{deGennes_w_gs_v_num})-(\ref{ans_w_e_v}), are $\bar{h}^* \simeq 0.094, \hat{W}^* \simeq 3.559\times 10^{-4}$. 
	The corresponding values obtained with the analytical expressions, 
	Eq.(\ref{ans_barrier_height_vw}), are $\bar{h}^* \simeq 0.053, \hat{W}^* \simeq 7.109\times 10^{-4}$. 
	One can notice that even for such extremely large second virial parameter the difference is significant. 	

\section{The virial parameters providing the maximum barrier heights} 
\label{sec:repulsive_max_bar}
\begin{figure}[ht!]
\begin{minipage}[ht]{0.5\linewidth}
\center{\includegraphics[width=1\linewidth]{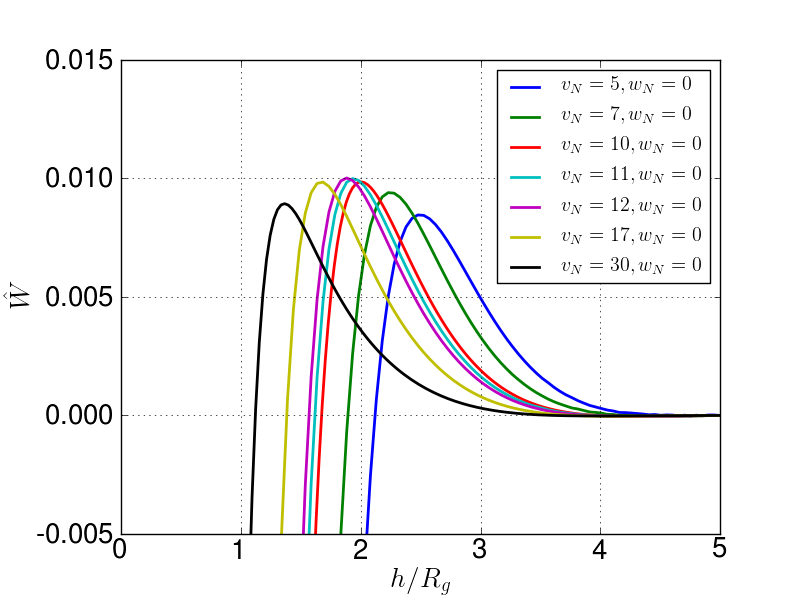}}
\caption{\small{The SCFT thermodynamic potential for different values of the virial parameter, $v_N$ ($w_N=0$).}}
\label{barrier_max_v_fig}
\end{minipage}
\hfill
\begin{minipage}[ht]{0.5\linewidth}
\center{\includegraphics[width=1\linewidth]{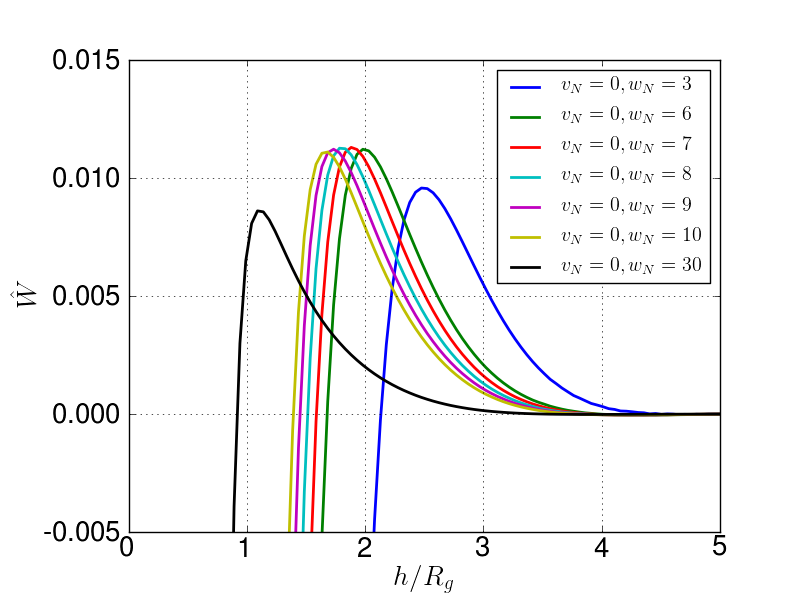}}
\caption{\small{The SCFT thermodynamic potential for the different values of the virial parameter, $w_N$ ($v_N=0$).}}
\label{barrier_max_w_fig}
\end{minipage}
\end{figure}

	The main purpose of the section is to find values of the virial parameters, $v_N$ and $w_N$ which correspond to the maximum height 
	of the barrier of the thermodynamic potential obtained in the SCFT for purely repulsive boundary conditions. 
	In order to do it, we considered separately the cases: $w_N=0$ and $v_N=0$ (theta solvent) and, after that,
	their various combinations. In Figs.\ref{barrier_max_v_fig}--\ref{barrier_max_w_fig} we show these thermodynamic potentials for some values of 
	the virial parameters which yield the maximum height of the barrier. For simplicity, all calculations are done for the 
	mesh size with: $N_x = 8k$, $N_s=3k$, $N_h=100$.
%%%%%%%%%%%%%%%%%%%%%%%%%%%%%%%%%%%%%%%%%%%%%%%%%%%%%%%%%%%%%%%%%%%%%%%%%%%%%%%%%%%%%%%%%%%%%%%%%%%%%%%%%%%%%%%%%%%%%%%%%%%%%%%
%        Thermodynamic potentials for w_N=0 and v_N=0
%%%%%%%%%%%%%%%%%%%%%%%%%%%%%%%%%%%%%%%%%%%%%%%%%%%%%%%%%%%%%%%%%%%%%%%%%%%%%%%%%%%%%%%%%%%%%%%%%%%%%%%%%%%%%%%%%%%%%%%%%%%%%%%
	
	As in the previous section, we compare different potentials by the dimensionless height of the thermodynamic potential barrier 
	that we denote as\footnote{Now, for consistency with the GSDE theory, we use the same notation for the potential, $W$ instead of $\Omega$.}    
$$
	\hat{W}^* = \hat{W}_{max} - \hat{W}_{inf}
$$
	and the corresponding dimensionless position of the maximum, $\bar{h}^*$. 
	We use the symbol $W_{scft}$ for the SCFT thermodynamic potential and $W_{gsde}$ for the GSDE potential, where it is necessary to distinguish them.  		
	
	In Tab.\ref{tabular:barrier_max_v} one can find numerical results for the height and the position
	of the thermodynamic potential for different values of the virial parameters. Firstly, we considered the case $w_N=0$ and found the value of the 
	second virial parameter that produces the maximum height of the barrier, $v_N=12$. Next, in the same table we present the results for $v_N = 5$ and $v_N=12$ 
	calculated for different values of $w_N$. 	
%%%%%%%%%%%%%%%%%%%%%%%%%%%%%%%%%%%%%%%%%%%%%%%%%%%%%%%%%%%%%%%%%%%%%%%%%%%%%%%%%%%%%%%%%%%%%%%%%%%%%%%%%%%%%%%%%%%%%%%%%%%%%%%%%%%%%%%%%%%%
% The height of the barrier for v \ne 0 and w = 0
%%%%%%%%%%%%%%%%%%%%%%%%%%%%%%%%%%%%%%%%%%%%%%%%%%%%%%%%%%%%%%%%%%%%%%%%%%%%%%%%%%%%%%%%%%%%%%%%%%%%%%%%%%%%%%%%%%%%%%%%%%%%%%%%%%%%%%%%%%%%
\begin{table}[ht!]
\caption{The SCFT thermodynamic potential barrier height $\hat{W}^*$ and its position $h^*$ for different values of the virial parameters. Based on $w_N=0$.} 
\label{tabular:barrier_max_v} 
\begin{flushleft}
  \begin{tabular}{ | c | c | c | c | c | c | c | c | c | c | c | }
    \hline
    $v_N$                                 &     5&    7 &    8 &     9&    10&    11&   12 &   13 &    14&    15     \\ \hline
    $w_N$                                 &     0&    0 &    0 &     0&    0 &    0 &    0 &    0 &     0&     0     \\ \hline
    $\hat{W}^*,\times 10^{-3}$            & 8.461& 9.402& 9.649& 9.834& 9.855& 9.996& 10.02&10.017& 9.996& 9.955     \\ \hline
    $h^*$                                 & 2.476& 2.228& 2.129&  2.08& 2.005& 1.931& 1.882& 1.832& 1.783& 1.733     \\
    \hline
  \end{tabular}
\end{flushleft} 
%%%%%%%%%%%%%%%%%%%%%%%%%%%%%%%%%%%%%%%%%%%%%%%%%%%%%%%%%%%%%%%%%%%%%%%%%%%%%%%%%%%%%%%%%%%%%%%%%%%%%%%%%%%%%%%%%%%%%%%%%%%%%%%%%%%%%%%%%%%%
% The height of the barrier for v \ne 0 and w - increasing
%%%%%%%%%%%%%%%%%%%%%%%%%%%%%%%%%%%%%%%%%%%%%%%%%%%%%%%%%%%%%%%%%%%%%%%%%%%%%%%%%%%%%%%%%%%%%%%%%%%%%%%%%%%%%%%%%%%%%%%%%%%%%%%%%%%%%%%%%%%%
\begin{flushleft}
  \begin{tabular}{ | c | c | c | c | c | c | c | c | c | c | c | }
    \hline 
    $v_N$                                 &     5&    5 &    5 &     5&     5&    12&    12&    12&    12&    12     \\ \hline
    $w_N$                                 &     3&    5 &    6 &     7&    10&     1&     2&     3&     4&     5     \\ \hline
    $\hat{W}^*,\times 10^{-3}$            &10.559&10.763&10.785& 10.72&10.421&10.143&10.204&10.185&10.142&10.081     \\ \hline
    $h^*$                                 & 1.981& 1.783& 1.733& 1.683& 1.535& 1.782& 1.733& 1.684& 1.585& 1.535     \\
    \hline
  \end{tabular}
\end{flushleft} 
\end{table}

	In Tab.\ref{tabular:barrier_max_w} we present the other set of the virial parameters. The first part of the table is based on the theta solvent ($v_N=0$). 
	We found that the maximum height of the barrier is reached at $w_N=7$. The barrier's height value is about $10$ percent higher than in the case of $w_N=0$. 
	One more thing that should be noticed is that for any other sets of the virial parameters with $w_N>7$ the highest altitude of the barrier 
	is reached at $v_N=0$. 
	In order to catch any functional relationship between the height and the virial parameters for $w_N < 7$ and $v_N < 12$, we need to 
	consider many different sets of the virial parameters. On the other hand, for real applications it is not necessary to improve the obtained accuracy 
	and the previous results for the optimal parameters seem to be suitable. 

\begin{figure}[ht!]
\begin{minipage}[ht]{0.5\linewidth}
\center{\includegraphics[width=1\linewidth]{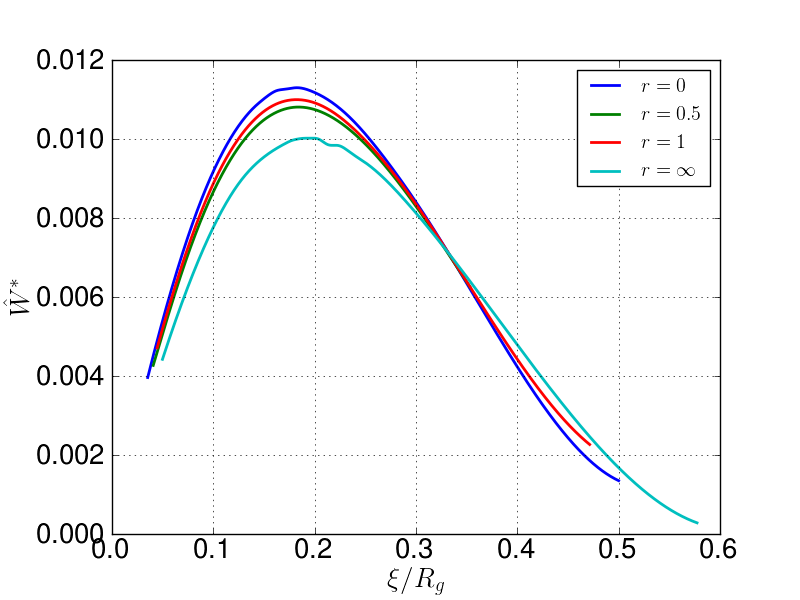}}
\caption{\small{The SCFT thermodynamic potential barrier height as a function of the correlation length, Eq.(\ref{intr_polymer_correlation_length}),
		for different values of fixed parameter, $r$ indicated in the figure. 
		The data are partially presented in Tabs.\ref{tabular:barrier_max_v}--\ref{tabular:barrier_max_w}. 
		The barrier height maximum always corresponds to $\xi/R_g\sim 0.2$.}}
\label{ans_scft_barrier_vs_xi_fig}
\end{minipage}
\hfill
\begin{minipage}[ht]{0.5\linewidth}
\center{\includegraphics[width=1\linewidth]{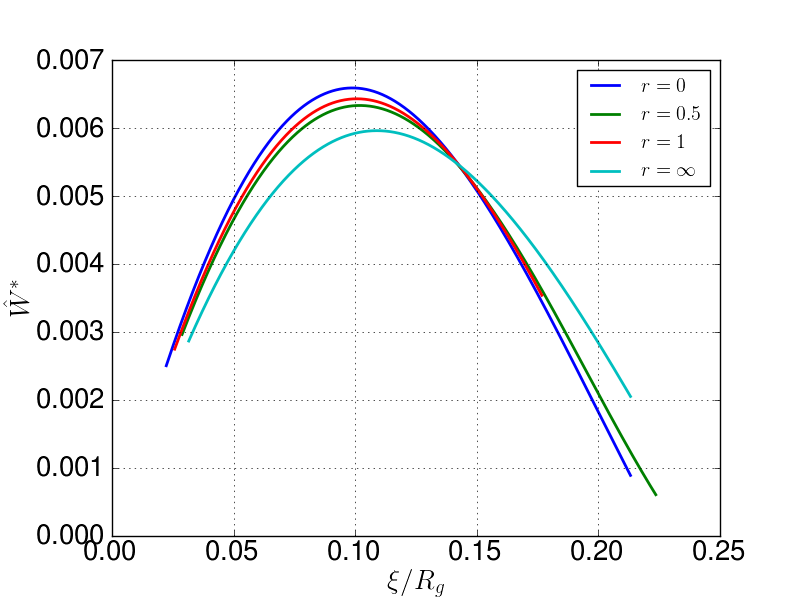}}
\caption{\small{The thermodynamic potential barrier height obtained by the GSDE theory, Eq.(\ref{ans_w_e_vw_general}), 
		as a function of the correlation length, Eq.(\ref{intr_polymer_correlation_length}),
		for different values of the parameter, $r$ indicated in the figure.}}
\label{ans_gsde_barrier_vs_xi_fig}
\end{minipage}
\end{figure}

%%%%%%%%%%%%%%%%%%%%%%%%%%%%%%%%%%%%%%%%%%%%%%%%%%%%%%%%%%%%%%%%%%%%%%%%%%%%%%%%%%%%%%%%%%%%%%%%%%%%%%%%%%%%%%%%%%%%%%%%%%%%%%%%%%%%%%%%%%%%
% The height of the barrier for w \ne 0 and v = 0
%%%%%%%%%%%%%%%%%%%%%%%%%%%%%%%%%%%%%%%%%%%%%%%%%%%%%%%%%%%%%%%%%%%%%%%%%%%%%%%%%%%%%%%%%%%%%%%%%%%%%%%%%%%%%%%%%%%%%%%%%%%%%%%%%%%%%%%%%%%%
\begin{table}[ht!]
\caption{The SCFT thermodynamic potential barrier height $\hat{W}^*$ and its position $h^*$ for different values of the virial parameters. Based on $v_N=0$.} 
\label{tabular:barrier_max_w} 
\begin{flushleft}
  \begin{tabular}{ | c | c | c | c | c | c | c | c | c | c | c | }
    \hline
    $v_N$                           &     0&    0 &    0 &     0&     0&     0&     0&     0&     0&     0     \\ \hline
    $w_N$                           &     5&    6 &    7 &     8&     9&    10&    11&    30&   100&   200     \\ \hline
    $\hat{W}^*,\times 10^{-3}$      &10.994&11.219&11.298&11.262&11.221&11.102&10.964& 8.608& 5.386&  3.96     \\ \hline
    $h^*$                           & 2.129& 1.981& 1.882& 1.782& 1.733& 1.683& 1.634&  1.09& 0.694& 0.545     \\
    \hline
  \end{tabular}
\end{flushleft} 
%%%%%%%%%%%%%%%%%%%%%%%%%%%%%%%%%%%%%%%%%%%%%%%%%%%%%%%%%%%%%%%%%%%%%%%%%%%%%%%%%%%%%%%%%%%%%%%%%%%%%%%%%%%%%%%%%%%%%%%%%%%%%%%%%%%%%%%%%%%%
%The height of the barrier for w \ne 0 and w increasing
%%%%%%%%%%%%%%%%%%%%%%%%%%%%%%%%%%%%%%%%%%%%%%%%%%%%%%%%%%%%%%%%%%%%%%%%%%%%%%%%%%%%%%%%%%%%%%%%%%%%%%%%%%%%%%%%%%%%%%%%%%%%%%%%%%%%%%%%%%%%
\begin{flushleft}
  \begin{tabular}{ | c | c | c | c | c | c | c | c | c | c | }
    \hline
    $v_N$                                 &   0.5&    1 &    5 &    12&     1&     2&     5&     1&     5     \\ \hline
    $w_N$                                 &     7&    7 &    7 &     7&    10&    10&    10&    30&    30     \\ \hline
    $\hat{W}^*,\times 10^{-3}$            &11.228&11.195& 10.72& 9.942&10.976&10.819&10.422& 8.556& 8.252     \\ \hline
    $h^*$                                 & 1.882& 1.832& 1.684& 1.486& 1.634& 1.584& 1.535&  1.09&  1.09     \\
    \hline
  \end{tabular}
\end{flushleft} 
\end{table}
	In Fig.\ref{ans_scft_barrier_vs_xi_fig} we present the dependence of the thermodynamic potential barrier height as a function  
	of the dimensionless correlation length, Eq.(\ref{intr_polymer_correlation_length}), obtained with the SCFT. 
	We fixed the parameter $r=v_N/2w_N$ and vary simultaneously the virial parameters, $v_N, w_N$. 
	For example, the cases with $r=0$ and $r=\infty$ correspond to $v_N=0$ and $w_N=0$. 
	For the purpose of comparison, we have also drawn in Fig.\ref{ans_gsde_barrier_vs_xi_fig} the same dependence 
	calculated using the GSDE theory, Eq.(\ref{ans_w_e_vw_general}). One can observe the same behavior of the curves for the corresponding virial parameters. 
	In the relevant range for the correlation length, when the interaction energy reaches its maximum values,
	all the curves are squeezed between marginal curves ($v_N=0$ for the upper and $w_N=0$ for the lower boundary curves). 	
	In contrast with the SCFT, whose maximum values for the marginal curves are listed above, the corresponding maximum values obtained by the GSDE 
	theory are $v_N=42$ for the lower boundary curve ($w_N=0$) and $w_N=28$ for the upper boundary curve ($v_N=0$).		  
%%%%%%%%%%%%%%%%%%%%%%%%%%%%%%%%%%%%%%%%%%%%%%%%%%%%%%%%%%%%%%%%%%%%%%%%%%%%%%%%%%%%%%%%%%%%%%%%%%%%%%%%%%%%%%%%%%%%%%%%%%%%%%%
%        Barrier heights Wm vs xi
%%%%%%%%%%%%%%%%%%%%%%%%%%%%%%%%%%%%%%%%%%%%%%%%%%%%%%%%%%%%%%%%%%%%%%%%%%%%%%%%%%%%%%%%%%%%%%%%%%%%%%%%%%%%%%%%%%%%%%%%%%%%%%%
	
\begin{figure}[ht!]
\begin{minipage}[ht]{0.5\linewidth}
\center{\includegraphics[width=1\linewidth]{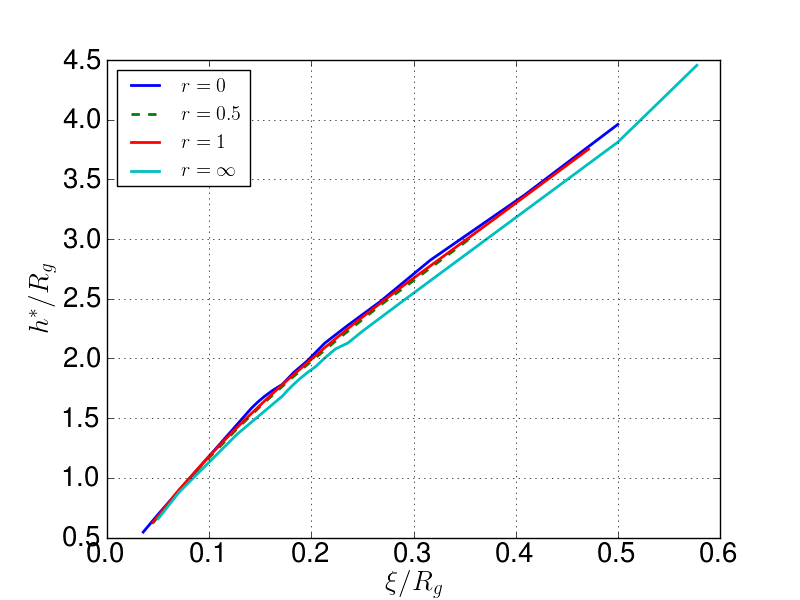}}
\caption{\small{The SCFT thermodynamic potential barrier position as a function of the correlation length, Eq.(\ref{intr_polymer_correlation_length}),
		for different values of fixed parameter $r$ indicated in the figure. 
		The data partially correspond to Tabs.\ref{tabular:barrier_max_v}, \ref{tabular:barrier_max_w}.}}
\label{ans_scft_hm_vs_xi_fig}
\end{minipage}
\hfill
\begin{minipage}[ht]{0.5\linewidth}
\center{\includegraphics[width=1\linewidth]{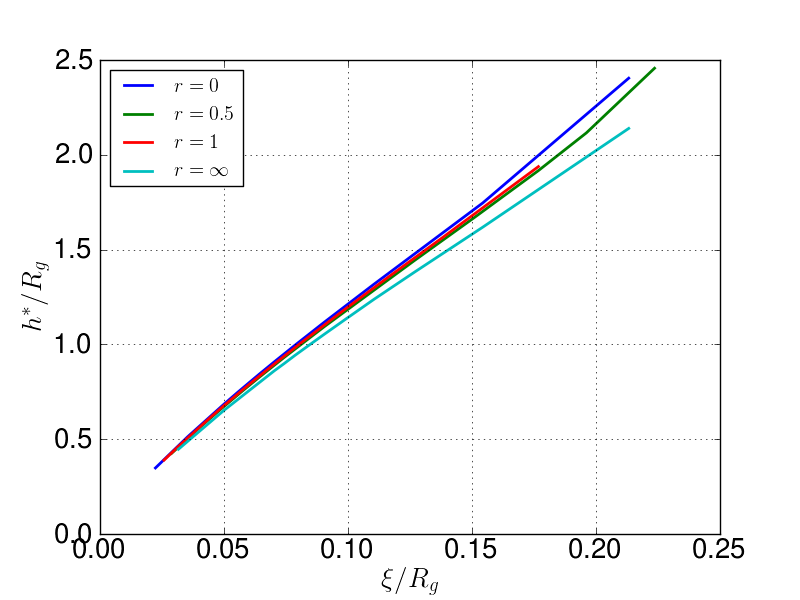}}
\caption{\small{The thermodynamic potential barrier position obtained by the GSDE theory, Eq.(\ref{ans_w_e_vw_general}),
		as a function of the correlation length, Eq.(\ref{intr_polymer_correlation_length}),
		for different values of fixed parameter $r$ indicated in the figure.}}
\label{ans_gsde_hm_vs_xi_fig}
\end{minipage}
\end{figure}

	Next, in Figs.\ref{ans_scft_hm_vs_xi_fig}-- \ref{ans_gsde_hm_vs_xi_fig} we present the thermodynamic potential barrier position 
	as a function of the dimensionless correlation length, Eq.(\ref{intr_polymer_correlation_length}), for the two different approaches. 
	These positions correspond to the barrier heights depicted in Figs.\ref{ans_scft_barrier_vs_xi_fig}--\ref{ans_gsde_barrier_vs_xi_fig}. 
	One can see in these pictures the linear or the linearithmic dependence. Similar behavior is predicted by the analytical expression,
	Eq.(\ref{ans_barrier_position_vw_scaling}), but again the quantitative comparison with the analytical expression is impossible, because 
	the expression is valid only for very low $\bar{\xi}\ll 1$.     	    
%%%%%%%%%%%%%%%%%%%%%%%%%%%%%%%%%%%%%%%%%%%%%%%%%%%%%%%%%%%%%%%%%%%%%%%%%%%%%%%%%%%%%%%%%%%%%%%%%%%%%%%%%%%%%%%%%%%%%%%%%%%%%%%
%        hm vs xi
%%%%%%%%%%%%%%%%%%%%%%%%%%%%%%%%%%%%%%%%%%%%%%%%%%%%%%%%%%%%%%%%%%%%%%%%%%%%%%%%%%%%%%%%%%%%%%%%%%%%%%%%%%%%%%%%%%%%%%%%%%%%%%%
	
\begin{figure}[ht!]
\begin{minipage}[ht]{0.5\linewidth}
\center{\includegraphics[width=1\linewidth]{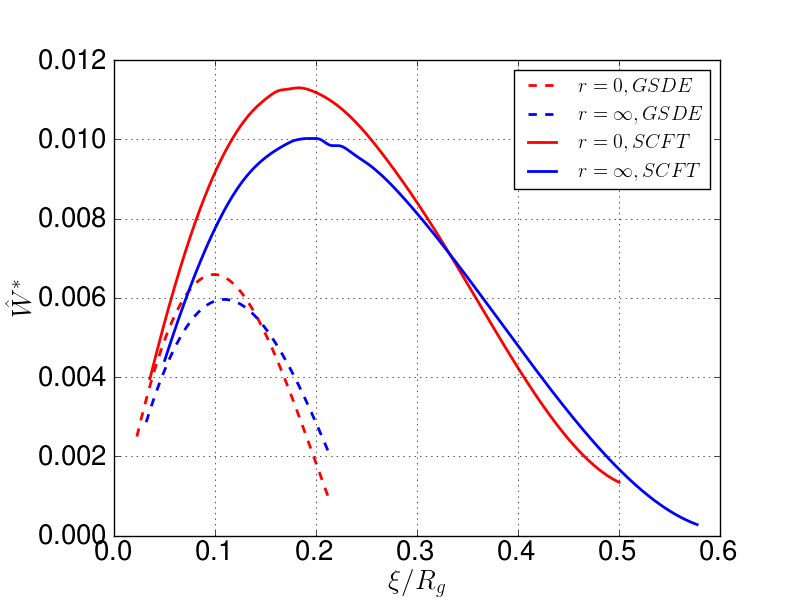}}
\caption{\small{The comparison between the SCFT results and the GSDE theory for the thermodynamic potential barrier height as a function of the 
	        correlation length $\xi$ for the boundary curves.}}
\label{ans_mix_barrier_vs_xi_fig}
\end{minipage}
\hfill
\begin{minipage}[ht]{0.5\linewidth}
\center{\includegraphics[width=1\linewidth]{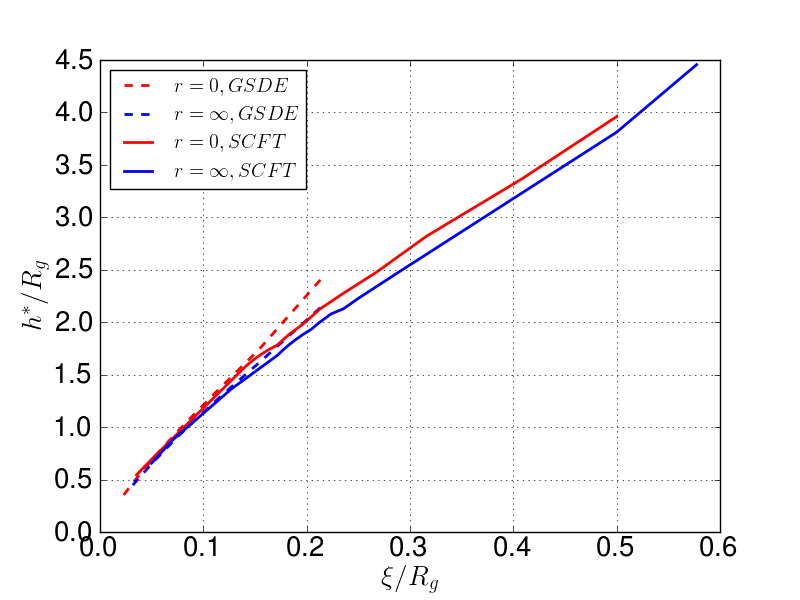}}
\caption{\small{The comparison between the SCFT results and the GSDE theory for the thermodynamic potential barrier position as a function of 
		the correlation length $\xi$  for the boundary curves.}}
\label{ans_mix_hm_vs_xi_fig}
\end{minipage}
\end{figure}

As one can notice from Figs.\ref{ans_scft_barrier_vs_xi_fig}--\ref{ans_gsde_hm_vs_xi_fig}, the interaction functions produced by both approaches have similar behavior, but at different scales. 
It is interesting to compare them in the same plots as we draw in Figs.\ref{ans_mix_barrier_vs_xi_fig}--\ref{ans_mix_hm_vs_xi_fig}.
The marginal curves with $w_N=0$ ($r=\infty$) are for the lower border and with $v_N=0$($r=0$), for the upper border. 
One can see that the difference between the peak values for the corresponding barrier heights is more than twice bigger for the SCFT results in comparison with the GSDE theory. 
The difference is even more significant for bigger values of the correlation length (smaller $(v_N+w_N)$), but it becomes less pronounced upon decreasing the correlation length (increasing $(v_N+w_N)$), asymptotically approaching the full agreement at $\bar{\xi}\ll 1$. 
The same effect characterizes the thermodynamic potential barrier position: one can see a convergence of the curves below certain value of the correlation length.
%%%%%%%%%%%%%%%%%%%%%%%%%%%%%%%%%%%%%%%%%%%%%%%%%%%%%%%%%%%%%%%%%%%%%%%%%%%%%%%%%%%%%%%%%%%%%%%%%%%%%%%%%%%%%%%%%%%%%%%%%%%%%%%
%       mix barrier and  hm vs xi comparison
%%%%%%%%%%%%%%%%%%%%%%%%%%%%%%%%%%%%%%%%%%%%%%%%%%%%%%%%%%%%%%%%%%%%%%%%%%%%%%%%%%%%%%%%%%%%%%%%%%%%%%%%%%%%%%%%%%%%%%%%%%%%%%%

	{\raggedright{\textbf{Concentration dependence.}}} Let us consider how the barrier height $W^{*}$ depends on the bulk polymer concentration, $c_b$. 
	Recall that the relationship between the real and the dimensionless thermodynamic potentials is
$$
	W = \frac{c_bR_g}{N}\hat{W}(v_N, w_N)
$$
	where $v_N=\text{v}c_bN$, $w_N=\text{w}c_b^2N/2$. Since, we are interested in the barrier height concentration dependence, 
	we adapted this expression in a more suitable form. 
	For that, the correlation length, Eq.(\ref{intr_polymer_correlation_length_gsd}), can be written as
$$
	\bar{\xi} = \frac{1}{\sqrt{2c_bN(\text{v}+c_b\text{w})}} = \frac{1}{\sqrt{2\text{w}N}c_b\sqrt{1+r}}
$$       
	where we introduced $r=\text{v}/c_b\text{w}=v_N/2w_N$, and inverting the last equation, we obtain
\begin{equation}
\label{ans_cb_vs_xi}
	c_b\sqrt{2\text{w}N} = \frac{1}{\sqrt{1+r}\bar{\xi}}
\end{equation}
	Substituting Eq.(\ref{ans_cb_vs_xi}) in the expression for the barrier height written in real variables, we have
$$
	W^{*} = \frac{\sqrt{2\text{w}N}}{\sqrt{2\text{w}N}}\frac{c_bR_g}{N}\hat{W}^{*} = \frac{a}{\sqrt{2\text{w}}N}\Tilde{W}^{*}
$$
	where 
\begin{equation}
\label{ans_wstar_real_vs_xi}
	\Tilde{W}^* = c_b\sqrt{2\text{w}N}\hat{W}^{*} = \frac{1}{\bar{\xi}\sqrt{1+r}}\hat{W}^{*}
\end{equation}
	is considered as a function of $c_b\sqrt{2\text{w}N}$.
%%%%%%%%%%%%%%%%%%%%%%%%%%%%%%%%%%%%%%%%%%%%%%%%%%%%%%%%%%%%%%%%%%%%%%%%%%%%%%%%%%%%%%%%%%%%%%%%%%%%%%%%%%%%%%%%%%%%%%%%%%%%%%%
%       mix barrier and  hm vs xi comparison
%%%%%%%%%%%%%%%%%%%%%%%%%%%%%%%%%%%%%%%%%%%%%%%%%%%%%%%%%%%%%%%%%%%%%%%%%%%%%%%%%%%%%%%%%%%%%%%%%%%%%%%%%%%%%%%%%%%%%%%%%%%%%%%
\begin{figure}[ht!]
\center{\includegraphics[width=0.6\linewidth]{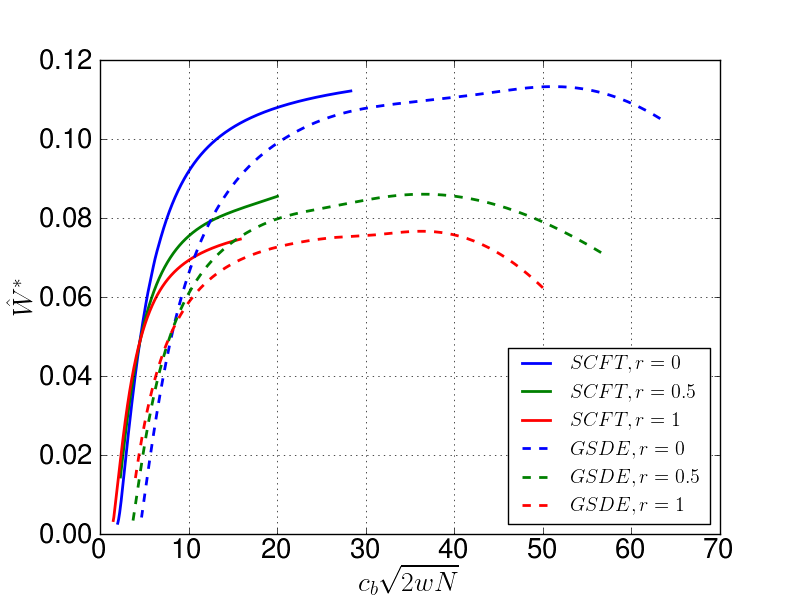}}
\caption{\small{The dependence of the barrier height, Eq.(\ref{ans_wstar_real_vs_xi}), on the polymer concentration, Eq.(\ref{ans_cb_vs_xi}). 
		The continuous lines correspond to the SCFT results and are taken from Fig.\ref{ans_scft_barrier_vs_xi_fig}. 
		The dashed lines correspond to the GSDE theory, Eq.(\ref{ans_w_e_vw_general}), and are taken from Fig.\ref{ans_gsde_barrier_vs_xi_fig}.}}
\label{ans_mix_barrier_vs_cb_fig}
\end{figure}

The barrier height increases with concentration for low $c_b$, reaches the maximum at the optimum concentration and then decreases at higher $c_b$. Thus, as $c_b$ increases, colloidal particles may change from instability to stability, and then to instability again (see Fig.\ref{ans_scft_barrier_vs_xi_fig}).\footnote{Note that $\xi/R_g$ monotonically decreases with $c_b$} From the physical point of view such behavior for the thermodynamic potential barrier height, as was noticed in \cite{semenov_2008}, corresponds to the effect of colloidal stabilization by adding a sufficient amount of free polymer to a colloidal dispersion. 
%%%%%%%%%%%%%%%%%%%%%%%%%%%%%%%%%%%%%%%%%%%%%%%%%%%%%%%%%%%%%%%%%%%%%%%%%%%%%%%%%%%%%%%%%%%%%%%%%%%%%%%%%%%%%%%%%%%%%%%%%%%%%%%%%%%%%%%%%%%%%%%%%%%%%%%%%%%%
%           Real variables
%%%%%%%%%%%%%%%%%%%%%%%%%%%%%%%%%%%%%%%%%%%%%%%%%%%%%%%%%%%%%%%%%%%%%%%%%%%%%%%%%%%%%%%%%%%%%%%%%%%%%%%%%%%%%%%%%%%%%%%%%%%%%%%%%%%%%%%%%%%%%%%%%%%%%%%%%%%%  	    
\section{Real variables} 
\label{sec:repulsive_real_var}
	In this section, we recalculate the thermodynamic potential in $k_BT$ units. 
	In order to stabilize colloidal particles, we should prevent the Van der Waals attraction between the particles using some agent, 
	in our case it is the free polymer. Thereby, the full thermodynamic potential is
$$
	U_{tot}(h) = U(h) + U_{VdW}(h)
$$
	where $U(h)$ corresponds to the thermodynamic potential produced by free polymers and $U_{VdW}(h)$ to the Van der Waals attraction potential. 
	Therefore, the condition imposed on the barrier height to provide the kinetic stabilization is $U_{tot}^* \gg k_BT$.
	Let us consider the two contributions in the total thermodynamic potential independently.
	
\begin{figure}[ht!]
\begin{minipage}[ht]{0.5\linewidth}
\center{\includegraphics[width=1\linewidth]{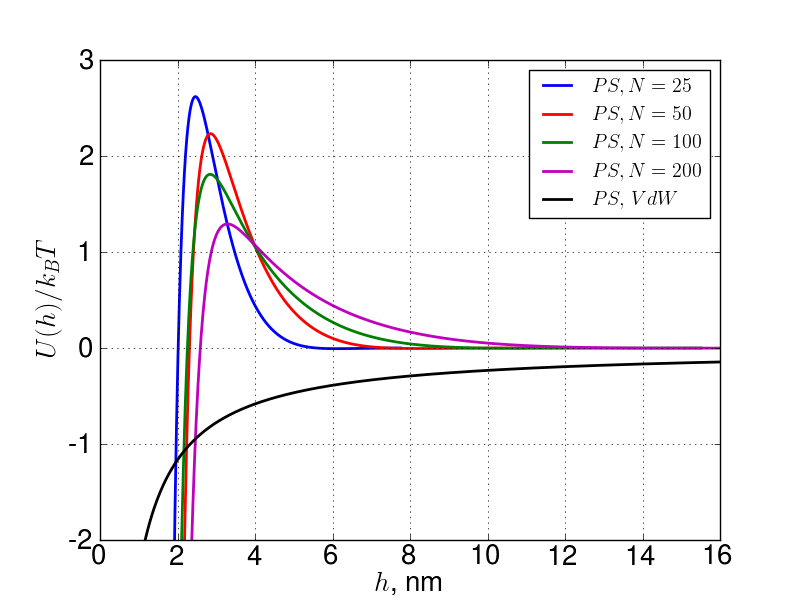}}
\caption{\small{The SCFT thermodynamic potential of interaction between two colloidal particles with radii $R_c = 200 \text{nm}$ 
		produced by free polystyrene in toluene calculated  for polymer volume fraction, $\phi\simeq 0.7$ and different chain lengths. 
		We also present Van der Waals potential between two PS particles with $R_c = 200 \text{nm}$ in toluene. }}
\label{ans_scft_therm_pot_ps_real_fig}
\end{minipage}
\hfill
\begin{minipage}[ht]{0.5\linewidth}
\center{\includegraphics[width=1\linewidth]{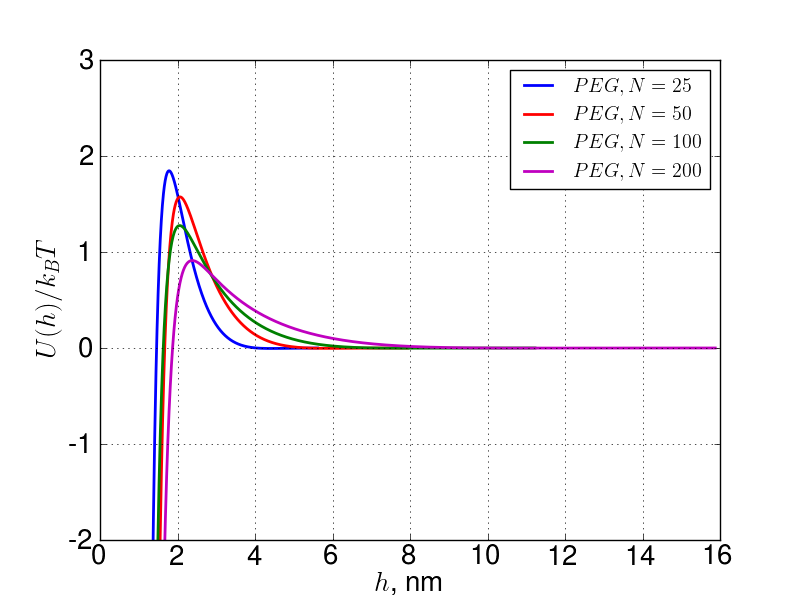}}
\caption{\small{The SCFT thermodynamic potential of interaction between two colloidal particles with radii $R_c = 200 \text{nm}$ 
		produced by free polystyrene in toluene calculated in real variables for polymer volume fraction, $\phi\simeq 0.4$ and different chain lengths.}}
\label{ans_scft_therm_pot_peg_real_fig}
\end{minipage}
\end{figure}

	\textbf{Van der Waals attraction}. As we already shown in the introduction chapter (see Sec.\ref{sec:colloids_VdW}),
	the Van der Waals interaction energy between spherical colloidal particles which have the same radius, $R_c$ is 
$$
	U_{VdW}(h) = -\frac{A_HR_c}{12h}
$$	
where $A_H$ is the Hamaker constant, $h$ is the distance between surfaces of the interacting particles. Let us denote the Hamaker constant of the colloidal material as $A_{11}$; the particles are interacting across a solvent with the Hamaker constant $A_{33}$. It is known \cite{Israelachvili_2011} that when the dielectric constants of the materials are quite small and the difference between the constants is also small, the following relation can be used:
$$
	A_{131} \simeq \left(\sqrt{A_{11}} - \sqrt{A_{33}}\right)^2
$$
        In one of our system we use the colloidal particles made up from crosslinked polystyrene with fraction $60\%$ of polymer and $40\%$ of solvent inside. 
	Such system allows us to prevent adsorption of the free polystyrene on the colloidal surfaces, and we can focus only on the depletion effect due to 
	free polymers.
	Let us cumpute the Hamaker constant for that system. Let us suppose (as quite crude approximation) that there is no solvent inside the colloid, 
	so the colloidal particles are composed solely of polystyrene. 
	Let $A_{11}$ be the Hamaker constant of polystyrene ($\varepsilon \simeq 2.6$) in vacuum. It is known that
	$A_{11} \simeq 6.57 \times 10^{-20}J$. Corresponingly, for toluene ($\varepsilon \simeq 2.38$) $A_{33} \simeq 5.4 \times 10^{-20}J$.
	Thus, $A_{131} \simeq 5.732\times 10^{-22} J$ or in $k_BT$ units at T=300 ($^\circ$K), we obtained $A_H/k_BT \simeq 0.138$. 
	The Van der Waals interaction profile is shown in Fig.\ref{ans_scft_therm_pot_ps_real_fig} \footnote{Where we compare it with the PI interaction.}.
	For micro-gel particles made up from crosslinked polystyrene with fraction $60\%$ of polymer and $40\%$ of toluene inside, it is obvious 
	that the interaction should be much weaker. Thus, we neglect the Van der Waals interaction for the present system. 
	We also suppose that the  colloidal particles are monodisperse in size: $R_c\simeq 200 \text{nm}$. 
	Moreover, we assume that the particles have an ideal spherical form and their surfaces are impenetrable to free polymers.

\begin{figure}[ht!]
\begin{minipage}[ht]{0.5\linewidth}
\center{\includegraphics[width=1\linewidth]{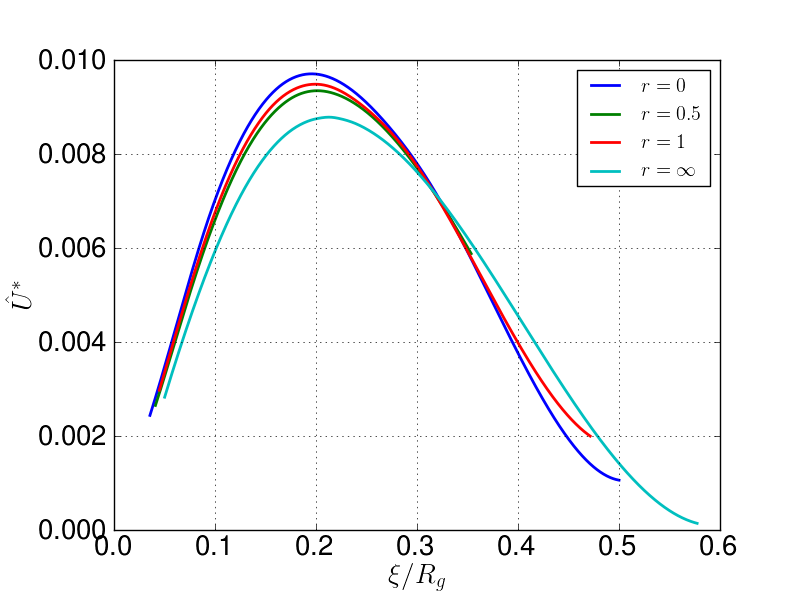}}
\caption{\small{The reduced SCFT barrier height calculated for spherical geometry via Eq.(\ref{ans_scft_therm_pot_real_reduced}) 
		as a function of the correlation length.
		The barrier height maximum always corresponds to $\xi/R_g\sim 0.2$.}}
\label{ans_scft_sph_barrier_vs_xi_fig}
\end{minipage}
\hfill
\begin{minipage}[ht]{0.5\linewidth}
\center{\includegraphics[width=1\linewidth]{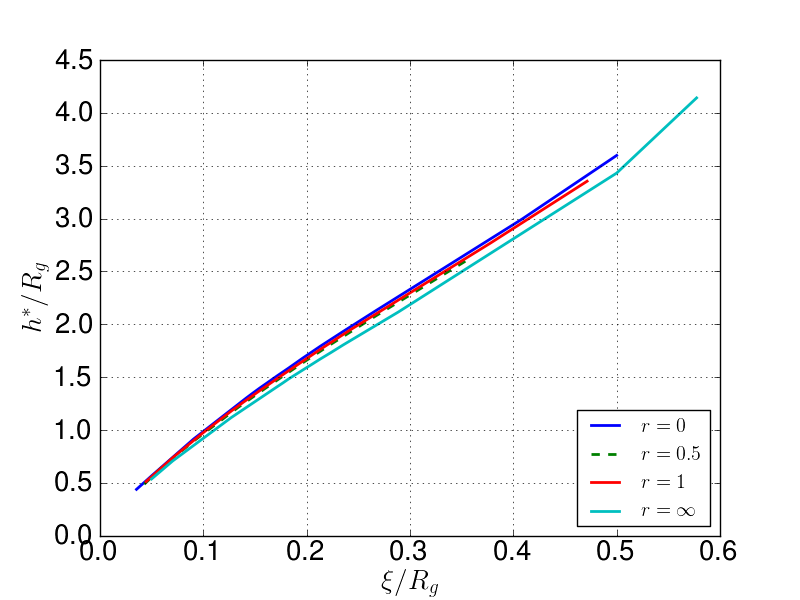}}
\caption{\small{The SCFT barrier position calculated for spherical geometry via Eq.(\ref{ans_scft_therm_pot_real_reduced}) as a function of the correlation length.}}
\label{ans_scft_sph_hm_vs_xi_fig}
\end{minipage}
\end{figure}

\begin{figure}[ht!]
\begin{minipage}[ht]{0.5\linewidth}
\center{\includegraphics[width=1\linewidth]{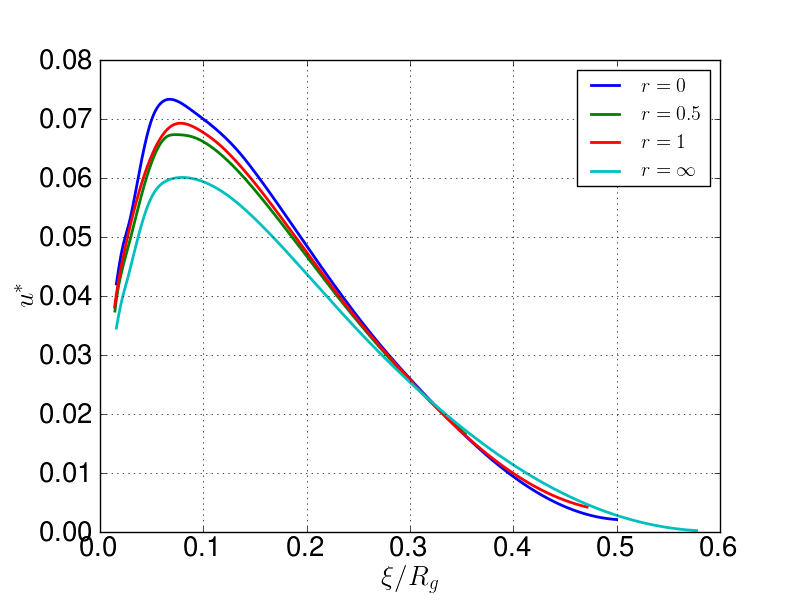}}
\caption{\small{The reduced SCFT barrier height calculated for spherical geometry via Eq.(\ref{ans_scft_therm_pot_real_reduced}) as a function of the 
		correlation length. The barrier height maximum always corresponds to $\xi/R_g\sim 0.075$.}}
\label{ans_scft_sph_ustar_vs_xi_fig}
\end{minipage}
\hfill
\begin{minipage}[ht]{0.5\linewidth}
\center{\includegraphics[width=1\linewidth]{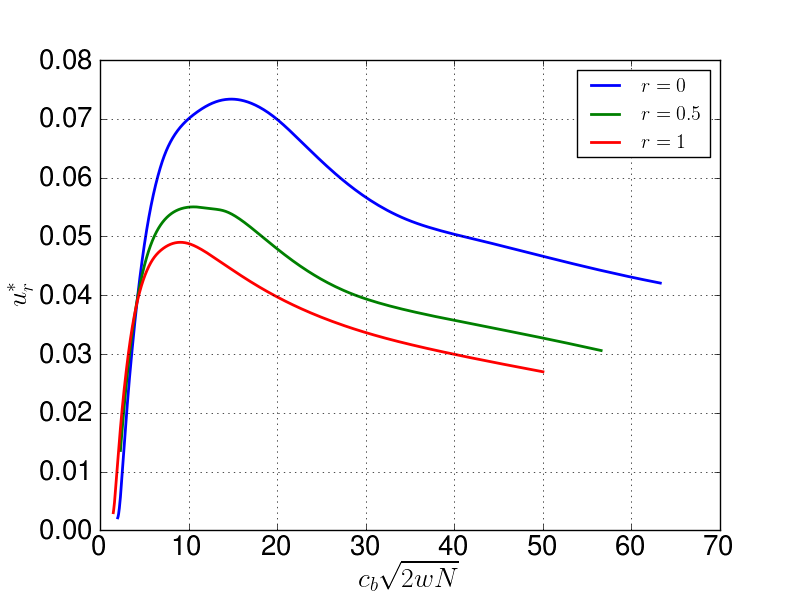}}
\caption{\small{The SCFT barrier height calculated for spherical geometry via Eq.(\ref{ans_real_ustar_r}) as a function of the reduced monomer concentration.}}
\label{ans_scft_sph_ustar_r_vs_cb_fig}
\end{minipage}
\end{figure}

	\textbf{Free polymer contribution.} Below, we consider real examples based on the following systems:\\
	\textbf{Polystyrene (PS) in toluene solvent}. We already found the parameters of polystyrene in toluene solvent in Sec.\ref{sec:polymers_experiment}. Thus,
$$
\begin{array}{c}
	M_0(\text{styrene})=104.15\,\text{g/mol}, \quad a_s = 7.6\textup{\AA}, \quad \rho = 960\,g/L, \\
	\text{v} = 23.21\textup{\AA}^3, \quad \text{w} = 2.79\times 10^4 \textup{\AA}^6. 
\end{array}															
$$	
	\textbf{Polyethylene glycol(PEG) in water solvent}. The corresponding parameters were found in the same section, and for T=25 ($^\circ$C) 
	they are equal:	
$$
\begin{array}{c}
	M_0(\text{ethylene oxide}) = 44.05\,\text{g/mol}, \quad a_s = 5.5\textup{\AA}, \quad \rho = 1128\,g/L, \\
	\text{v} = 12.24\textup{\AA}^3, \quad \text{w} = 1.54\times 10^4 \textup{\AA}^6.
\end{array}															    
$$
	\textbf{Thermodynamic potential}. The interaction energy of two spherical particles can be found using the Derjaguin approximation: 
$$
        U(h) = \pi R_c \int\limits_h^{\infty}\mathrm{d}h'W(h') = \pi R_c \frac{c_bR_g}{N} \int\limits_h^{\infty}\mathrm{d}h'\hat{W}(h') = 
                    \pi R_c \frac{c_bR_g^2}{N} \int\limits_{\bar{h}}^{\infty}\mathrm{d}\bar{h}\hat{W}(\bar{h}) = A_R \hat{U}(\bar{h})
$$
	Thereby,
\begin{equation}
\label{ans_scft_therm_pot_real}
	U(h) = A_R \hat{U}(h/R_g)
\end{equation}
	where we denoted
$$
	A_R = \pi R_c \frac{c_bR_g^2}{N} = \pi\left(\frac{R_c}{R_g}\right)\frac{c_bR_g^3}{N}
$$
	This prefactor defines the magnitude of the barrier and
\begin{equation}
\label{ans_scft_therm_pot_real_reduced}
	\hat{U}(\bar{h}) = \int\limits_{\bar{h}}^{\infty}\mathrm{d}\bar{h}\hat{W}(\bar{h})   
\end{equation}
	is the dimensionless thermodynamic potential for the spherical geometry expressed in terms of the thermodynamic potential obtained in the SCFT for the flat geometry. 
	We denote as $\hat{U}^*$ the corresponding barrier height. As we have already shown, Eq.(\ref{ans_cb_vs_xi}), the bulk polymer concentration can be written as
$$
	c_b = \frac{1}{\sqrt{2\text{w}N}\sqrt{1+r}\bar{\xi}}
$$ 
	where $\bar{\xi}$ is the dimensionless correlation length and $r=v_N/2w_N$ is the parameter that determines the solvent regime.
	In particular, in semidilute solution, the parameter $r$ slightly depends on the polymer concentration. 
	Hence, we can rewrite the prefactor as
$$
	A_R =  \frac{\pi a^3}{\sqrt{2\text{w}}}\left(\frac{R_c}{R_g}\right)\frac{1}{\sqrt{1+r}\bar{\xi}} = \frac{B}{\bar{\xi}}
$$
	where we introduced
$$
	B =  \frac{\pi a^3}{\sqrt{2\text{w}}}\left(\frac{R_c}{R_g}\right)\frac{1}{\sqrt{1+r}} 
$$
	This parameter does not depend on the virial parameters and we can write 
$$
	U(h, \text{v}, \text{w}, c_b, R_c, a_s, N) = B u(\bar{h}, \bar{\xi}, r)
$$
	where we introduced $u = \hat{U}/\bar{\xi}$. Therefore, we can find the stability condition for the dimensionless function $u$:
\begin{equation}
\label{ans_real_ustar_def}
\begin{array}{lr}
	\text{max}_{\bar{h}}\, u(\bar{h}, \bar{\xi}, r) = u^*(\bar{\xi}, r) & (a)\\
	\text{max}_{\bar{\xi}}\, u^*(\bar{\xi}, r) = u^{**}(r)              & (b)
\end{array}
\end{equation}
	The position of the maximum for the function $u$ by the variable $h$ is the same as for the function $\hat{U}$. We show in 
	Figs.\ref{ans_scft_sph_barrier_vs_xi_fig}--\ref{ans_scft_sph_hm_vs_xi_fig} the dependencies $\hat{U}^*(\bar{\xi})$ and $\bar{h}^*(\bar{\xi})$. 
%%%%%%%%%%%%%%%%%%%%%%%%%%%%%%%%%%%%%%%%%%%%%%%%%%%%%%%%%%%%%%%%%%%%%%%%%%%%%%%%%%%%%%%%%%%%%%%%%%%%%%%%%%%%%%%%%%%%%%%%%%%%%%%
%        spherical barrier max and hm vs xi
%%%%%%%%%%%%%%%%%%%%%%%%%%%%%%%%%%%%%%%%%%%%%%%%%%%%%%%%%%%%%%%%%%%%%%%%%%%%%%%%%%%%%%%%%%%%%%%%%%%%%%%%%%%%%%%%%%%%%%%%%%%%%%%
	As one can notice, the barrier height maximum corresponds $\xi/R_g\sim 0.2$ for different regimes. The same value for the 
	maximum barrier height had appeared in the flat case. Then, we should recalculate $\hat{U}^*$ for $u^*$. 
	We already demonstrated in Fig.\ref{ans_mix_barrier_vs_cb_fig} that the maximum barrier height, $W^*$, is reached for sufficiently big virial 
	parameters (small $\bar{\xi}$), where the numerical SCFT solution is difficult to obtain. 
	For the spherical case, we have the same problem.
	We already noticed, during the comparison of the SCFT results with those provided by the GSDE theory, that they are in good agreement. 
	Hence, we used the GSDE theory, Eq.(\ref{ans_w_e_vw_general}), in order to restore the low-$\bar{\xi}$ tail of the $u^*(\bar{\xi})$ dependence.
	We present it in Fig.\ref{ans_scft_sph_ustar_vs_xi_fig}. One can see that the maximum barrier height for the function $u^*(\bar{\xi}, r)$ 
	always corresponds to $\xi/R_g\sim 0.075.$
%%%%%%%%%%%%%%%%%%%%%%%%%%%%%%%%%%%%%%%%%%%%%%%%%%%%%%%%%%%%%%%%%%%%%%%%%%%%%%%%%%%%%%%%%%%%%%%%%%%%%%%%%%%%%%%%%%%%%%%%%%%%%%%
%        spherical u*  vs xi and u*  vs cb
%%%%%%%%%%%%%%%%%%%%%%%%%%%%%%%%%%%%%%%%%%%%%%%%%%%%%%%%%%%%%%%%%%%%%%%%%%%%%%%%%%%%%%%%%%%%%%%%%%%%%%%%%%%%%%%%%%%%%%%%%%%%%%%
	 
\begin{figure}[ht!]
\begin{minipage}[ht]{0.5\linewidth}
\center{\includegraphics[width=1\linewidth]{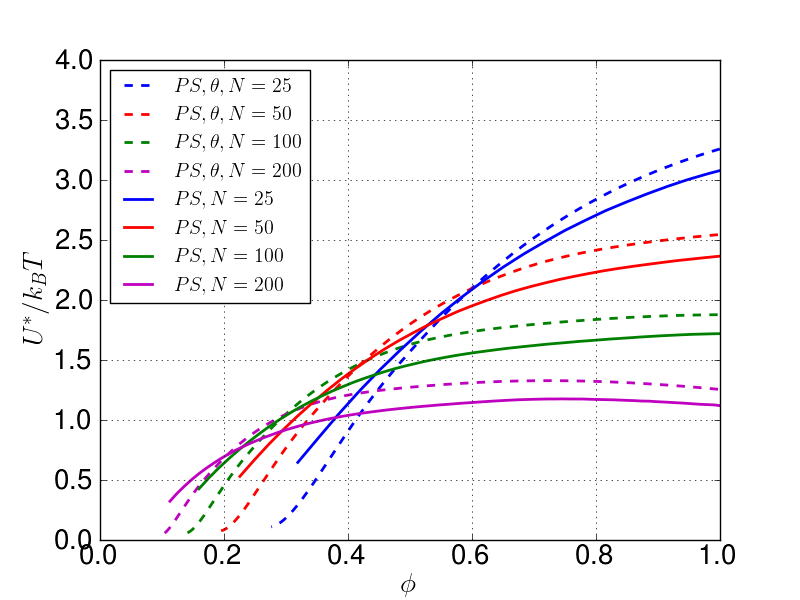}}
\caption{\small{The dependence of barrier height in $k_BT$ units on the volume fraction of polystyrene (PS) in toluene for different polymerization index, $N$.
		The radius of the colloid is $R_c=200\,nm$.}}
\label{ans_scft_therm_pot_vs_phi_ps_fig}
\end{minipage}
\hfill
\begin{minipage}[ht]{0.5\linewidth}
\center{\includegraphics[width=1\linewidth]{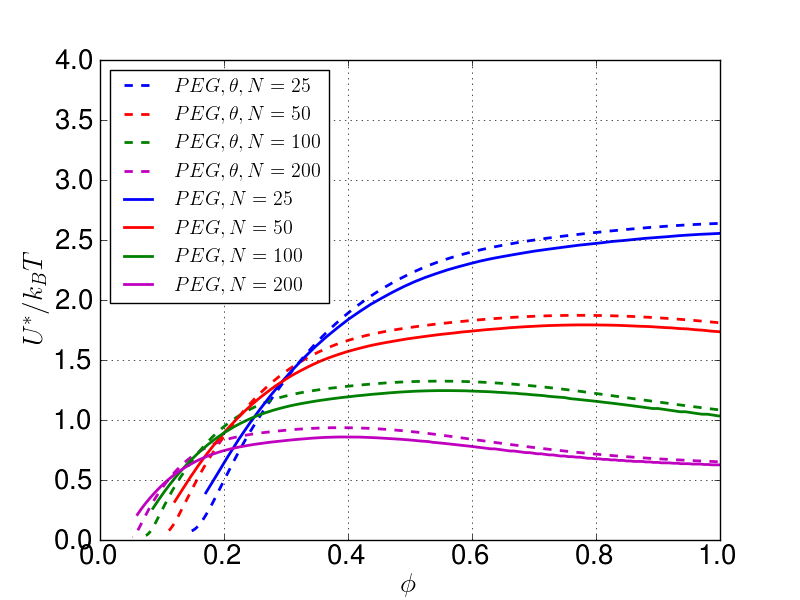}}
\caption{\small{The dependence of barrier height in $k_BT$ units on the volume fraction of polyethylene glycol (PEG) in water for different polymerization index, $N$.
		The radius of the colloid is $R_c=200\,nm$.}}
\label{ans_scft_therm_pot_vs_phi_peg_fig}
\end{minipage}
\end{figure}

	 Moreover, it is also very useful to represent 
\begin{equation}
\label{ans_real_ustar_r}
	  u^*_r = \frac{u^*}{\sqrt{1 + r}} \quad\text{as a function of}\quad c_b\sqrt{2\text{w}N} = \frac{1}{\sqrt{1+ r}\bar{\xi}} 
\end{equation}
	 that we show in Fig.\ref{ans_scft_sph_ustar_r_vs_cb_fig} for different values of the parameter $r$. 
	 One can compare it with the result obtained for the flat geometry in Fig.\ref{ans_mix_barrier_vs_cb_fig}.
	 In reality, $r=0$ corresponds to the theta solvent and, for most homopolymer systems, this regime is reached for sufficiently high temperatures. 
	 We will not restrict ourselves by such temperatures and consider the case with 
	 T=25 ($^\circ$C) for which both virial coefficients are positive. 
	 Below, we describe the algorithm of recovering the thermodynamic potential 
	 for real systems using the data for fixed $r$. Let us take a particular polymer, for example, polystyrene (PS) in toluene at certain polymerization index, $N$. 
	 For such system, we can recalculate the dependence $u_r^{*}(c_b)$ as shown in Fig.\ref{ans_scft_sph_ustar_r_vs_cb_fig}. 
	 The concentration, $c_b$ defines the parameter 
	 $r_c=\text{v}/\text{w}c_b$, where $\text{v}, \text{w}$ are the second and third virial coefficients of PS in toluene. 
	 At the same time, intersections of the line $c_b=const$ with $3$ curves corresponding to $r=\{0, 0.5, 1\}$, yield three value of the barrier 
	 heights, $\{u_{r1}^{*}, u_{r2}^{*}, u_{r3}^{*}\}$. Therefore,
	 we can approximate the function by interpolation based on the three above points. To this end, we choose the second-order polynomial 
	 $u_r^{*} = ar^2 + br + c$, with parameters
$$
\begin{array}{l}
	 a = \frac{u_{r3}^{*} - (r_3(u_{r2}^{*}- u_{r1}^{*}) + r_2u_{r1}^{*} - r_1u_{r2}^{*})/(r_2-r_1)}{r_3(r_3-r_1-r_2)+r_1r_2},\\\\
	 b = \frac{u_{r2}^{*}-u_{r1}^{*}}{r_2-r_1} - a(r_2+r_2),\quad  c = \frac{r_2u_{r1}^{*}-r_1u_{r2}^{*}}{r_2-r_1} + ar_1r_2 .
\end{array}
$$
	 Repeating the procedure for the entire range of $c_b$, we obtain the required dependency, $u_r^{*}(c_b)$. 
	 And, based on the above analysis for the prefactor, in real variables, we can write
$$
	 \frac{U^{*}(c_b)}{k_BT} = \frac{\pi a^3}{\sqrt{2\text{w}}}\left(\frac{R_c}{R_g}\right) u_r^{*}(c_b)
$$
	 Instead of the concentration it is more appropriate to use the polymer volume fraction, $\phi = c_m/\rho$, where $\rho$ is the density of the 
	 corresponding polymer and $c_m = M_0c_b/N_A$ is mass concentration (in $[g/L]$) of the polymer, $M_0$ is the molar mass of a monomer unit and $N_A$, the Avogadro constant.
	 The corresponding dependencies of the barrier height in the $k_BT$ units as a function of the volume fraction of the polymer are shown in 
	 Figs.\ref{ans_scft_therm_pot_vs_phi_ps_fig}--\ref{ans_scft_therm_pot_vs_phi_peg_fig} in the cases of PS in toluene and PEG in water for some values of the 
	 polymerization index, $N$. From Figs.\ref{ans_scft_therm_pot_vs_phi_ps_fig}--\ref{ans_scft_therm_pot_vs_phi_peg_fig} 
	 one can notice that not all polymer lengths produce barrier (confer the function presented in Fig.\ref{ans_scft_sph_ustar_r_vs_cb_fig}), 
	 because a certain range of the reduced concentration is forbidden for real systems. 
%%%%%%%%%%%%%%%%%%%%%%%%%%%%%%%%%%%%%%%%%%%%%%%%%%%%%%%%%%%%%%%%%%%%%%%%%%%%%%%%%%%%%%%%%%%%%%%%%%%%%%%%%%%%%%%%%%%%%%%%%%%%%%%
%        U vs phi real
%%%%%%%%%%%%%%%%%%%%%%%%%%%%%%%%%%%%%%%%%%%%%%%%%%%%%%%%%%%%%%%%%%%%%%%%%%%%%%%%%%%%%%%%%%%%%%%%%%%%%%%%%%%%%%%%%%%%%%%%%%%%%%%
	 
	 In Figs.\ref{ans_scft_therm_pot_vs_phi_ps_fig}--\ref{ans_scft_therm_pot_vs_phi_peg_fig} we have also presented results for the $\theta$-conditions ($r=0$). 	 	  
	 In this case, the third virial parameter, $w_N$, can be found from the expression
$$
	 \phi = \frac{c_m}{\rho} = \frac{M_0}{N_A\rho}c_b = \frac{M_0}{N_A\rho}\frac{1}{\sqrt{2\text{w}N}\bar{\xi}} = \frac{M_0}{N_A\rho}\frac{\sqrt{4w_N}}{\sqrt{2\text{w}N}}
$$
	 and the dimensionless thermodynamic potential can be then calculated using the obtained parameter, $w_N$.
	 The thermodynamic potentials for the $\theta$-conditions exceed those obtained by the interpolation procedure by only around 10$\%$.
	 
	 We present the thermodynamic potential calculated (in $k_BT$) unit as a function of the separation (in $nm$) 
	 only for the $\theta$-conditions\footnote{Because it is much easier to obtain it using the SCFT data.}	 	 
	 in Figs.\ref{ans_scft_therm_pot_ps_real_fig}--\ref{ans_scft_therm_pot_peg_real_fig} for different values of the chain length and colloidal size, $R_g=200nm$. 
	 For comparison, in Fig.\ref{ans_scft_therm_pot_ps_real_fig} we also show Van der Waals potential between two colloidal particles made up from 
	 polystyrene ($100\%$).
%%%%%%%%%%%%%%%%%%%%%%%%%%%%%%%%%%%%%%%%%%%%%%%%%%%%%%%%%%%%%%%%%%%%%%%%%%%%%%%%%%%%%%%%%%%%%%%%%%%%%%%%%%%%%%%%%%%%%%%%%%%%%%%
%        U(h) real, PS and PEG 
%%%%%%%%%%%%%%%%%%%%%%%%%%%%%%%%%%%%%%%%%%%%%%%%%%%%%%%%%%%%%%%%%%%%%%%%%%%%%%%%%%%%%%%%%%%%%%%%%%%%%%%%%%%%%%%%%%%%%%%%%%%%%%%
	One can see from these pictures that for short polymers the thermodynamic potential is sufficiently well localized and has quite big magnitude of the barrier.
	When we increase the length of the free polymers the barrier height decreases and spreads.
%%%%%%%%%%%%%%%%%%%%%%%%%%%%%%%%%%%%%%%%%%%%%%%%%%%%%%%%%%%%%%%%%%%%%%%%%%%%%%%%%%%%%%%%%%%%%%%%%%%%%%%%%%%%%%%%%%%%%%%%%%%%%%%%%%%%%%%%%%%%%%%%%%%%%%%%%%%%
%           Conclusion
%%%%%%%%%%%%%%%%%%%%%%%%%%%%%%%%%%%%%%%%%%%%%%%%%%%%%%%%%%%%%%%%%%%%%%%%%%%%%%%%%%%%%%%%%%%%%%%%%%%%%%%%%%%%%%%%%%%%%%%%%%%%%%%%%%%%%%%%%%%%%%%%%%%%%%%%%%%%  	    
\section{Summary and Conclusion} 
	  For systems of ideal homopolymers in solutions squeezed between flat plates, we found both numerical and conjugated analytical solutions 
	  for the following functions: the distribution function, $q(x, s)$, the concentration profile, $c(x)$, the partition function $Q(h)$, 
	  the force between plates $\Pi(h)$ and the second derivative of concentration profile, $c_{xx}(0)$. We compared the results between the analytical 
	  and corresponding numerical solutions. Based on the analytical solution, we found sufficient grid parameters which provide a good accuracy 
	  for the corresponding functions.  

	  In addition to the analysis of the ideal polymer systems, we developed the fast algorithm of solving the SCFT equations for non-ideal homopolymer 
	  solutions in a gap between two solid flat plates in a broad range of virial parameters $v_N\in[0..500]$ and $w_N\in[0..500]$. 
	  The numerical solution for thermodynamic potential obtained with the algorithm shows the existence of
	  a repulsive barrier due to the polymer-induced interaction between the plates. 
	  The calculated interaction curves are similar to the predictions of the analytical theory. 
	  
	  We have done calculations of the thermodynamic potentials in a broad range of virial parameters varying $v_N=0: 500$, $w_N=0: 100$.
	  Relying on the results given in Tabs.\ref{tabular:rel_err_v10}--\ref{tabular:rel_err_v01k_w01k}, we found as a general rule
	  that the numerical solution for the thermodynamic potential converges much faster as $N_s$ is increased than along the variable $N_x$. 
	  Comparing other results contained in those tables, we found the optimal parameters which provide the sufficient accuracy for solutions. 
	  We also analyzed the restrictions of the numerical scheme and found that 
	  the sufficient accuracy in computation of the thermodynamic potential can be achieved only for virial parameters that do not exceed: $v_N$ = 500, $w_N$ = 100.
	  We are able to use the numerical scheme for even larger virial parameters only for evaluating the range of the interaction and the magnitude of the barrier. 

	  Moreover, we performed the comparison between three different approaches related to the problem of colloidal stabilization in polymer solution. 
	  All these approaches are related to each other and are based on the Edwards equation, 
	  which is solved using different approximations. 
	    
	  Firstly, we examined the problem in the ground state dominance (GSD) approximation. 
	  All of the results are based on the theorem which was proposed by de Gennes \cite{deGennes_1981, deGennes_1982}
	  relating the force between flat plates with the mid-plane concentration. 
	  In the first instance, we explored the case when the third virial coefficient is equal to zero. 
	  We obtained the analytical expression for the concentration profile, and found its dependence on the separation, $h$, 
	  using the first kind of the elliptic integral. 
	  Then, we obtained the concentration at the mid-plane using the universal function depicted in Fig.\ref{deGennes_f_asy_fig}. 	  
	  In this picture one can see that the function becomes zero below a certain separation between the plates, which means that there are no polymer inside. 
	  At such separations we have only pressure exerted by the polymers outside. Based on that, we generalized the expression for the thermodynamic potential. 
	  We found the asymptotics of the thermodynamic potential for large separations between the plates in comparison with 
	  the correlation length in the bulk solution. In Fig.\ref{deGennes_free_en_gs_v_fig} it is shown that the asymptotics 
	  have very good accuracy for a fairly wide range of separations, $h>4\xi$.	  
	  After that, we generalized the approach for the case when the third virial parameter is not equal to zero. For the generalization we used the same
	  substitutions, as before, when $w_N=0$. We again introduced the new universal function, $g$ which is related to the concentration profile, separation $h$ 
	  and with the ratio $\text{v}/\text{w}c_b$. In Figs.\ref{deGennes_g_vw_asy_fig}--\ref{deGennes_free_en_gs_vw_fig} we
	  represented the universal function $g$ and the corresponding thermodynamic potential, with their asymptotics. 
	  In the general case the asymptotic expression for the thermodynamic potential shows also the exponential behavior with
	  the correlation length $\xi$ corresponding to the GSD correlation length, Eq.(\ref{intr_polymer_correlation_length_gsd}). 

	  Next, we compared the results for the thermodynamic potential obtained by the GSD aproximation and significantly improved
	  the analytical predictions by advancing the GSDE theory \cite{semenov_1996,semenov_2008}, where the finite chain length effect is taken into account. 
	  The GSDE potential includes repulsive as well as the attractive parts.
	  Using the asymptotics for the thermodynamic potential that we obtained in the GSD approximation, we can describe the attractive part of the 
	  thermodynamic potential calculated in SCFT by the asymptotic function using the exact RPA correlation length, $\xi$. 
	  Such improvement is significant, especially for small values of the virial parameters. 
	  It ensures the coincidence between the attractive parts of interaction profiles for small virial parameters.	  
	  As can be seen, the comparison shows a reasonable agreement which gets closer with increasing the reduced virial parameters 
	  (which are proportional to $N$).

	  In addition, we investigated the influence of the virial parameters on the thermodynamic potential barrier height and its position 
	  using the SCFT and the GSDE calculations. We found that both approaches reveal the same behavior in the dependence of the barrier height
	  on the exact RPA correlation length, considering it at a fixed value of one virial parameter, while changing the other one. 
	  In the relevant range for the correlation length, when the barrier has reached its maximum values,
	  all the curves are squeezed between the boundary curves: the lower one with $w_N=0$ and the upper one with $v_N=0$ (theta solvent).
	  The peak values for the curves are reached at $v_N=12$ and $w_N=7$ for the SCFT and at $v_N=42$ and $w_N=28$ for the GSDE theory. 
	  Due to the comparison between the SCFT and GSDE theory (valid for small $\xi/R_g$), we found that the applicability of the GSDE theory is 
	  restricted by the range $\xi<0.1R_g$.	  
	  It is also noteworthy that the difference between the peak values for the corresponding barrier heights is sufficiently big and 
	  for the SCFT barrier height it is more than twice bigger than the corresponding values for the GSDE theory. 
	  This difference is even more significant for the bigger values of the correlation length (smaller $(v_N+w_N)$) and becomes less upon
	  decreasing the correlation length (big $(v_N+w_N)$). The complete agreement is asymptotically approached in the latter regime. 	  
	  The same distinctions are valid for the thermodynamic potential barrier positions when the agreement between the curves is reached 
	  below a certain value of the correlation length corresponding to sufficiently big values of the virial parameters.	  
	  The results for the thermodynamic potential barrier height as a function of the correlation length can be used to describe the colloidal stability.
	  According to the most precise approach, the SCFT, the barrier height maximum always corresponds to $\xi/R_g \sim 0.2$. 
	  Also, note that $\xi/R_g$ is monotonically decreasing with $c_b$. 
	  Therefore, if we add free polymers to naked colloidal dispersion, 	   
	  the barrier height increases, up to a certain value, when it reaches its maximum, at the
	  optimal concentration. A further increase of the concentration results in a decrease of the barrier. 
	  Thus, as concentration increases, colloidal particles may change from instability to stability, and then to instability again. 

	  As a final step, we considered real polymer systems, where, as the stabilizing agent we used free, linear polystyrene dissolved in 
	  toluene and polyethylene glycol dissolved in water. 
	  For these polymer systems, we obtained the dependence of the barrier height on the volume 
	  fraction of the corresponding polymer. For short polymer chains, this is a monotonically increasing function reaching its maximum
	  in the concentrated solution regime, but for quite long polymers this dependence has a barrier-like behavior. 
	  The maximum height is reached at the optimum volume fraction and then it decreases at higher concentration. 
	  The value of the barrier height, $U^* \simeq 2-3 k_BT$, for the considered polymers is not enough to  
	  impart the kinetic stability. 
	  In order to improve the stabilization effect, we must consider either more rigid polymers characterized by smaller parameter $\text{v}/a_s^3$, where $\text{v}$ is 
	  the second virial coefficient of the polymer-solvent interaction, or take into account various effects associated with the polymer-surface adsorption.

%% file: Chapters/adsorption.tex
% Chapter 4

\chapter{Polymer-adsorbing surfaces} % Main chapter title
\label{chap:Chapter4} % For referencing the chapter elsewhere, use \ref{Chapter1} 
\lhead{Chapter 4. \emph{Polymer-adsorbing surfaces}} % This is for the header on each page - perhaps a shortened title
%%%%%%%%%%%%%%%%%%%%%%%%%%%%%%%%%%%%%%%%%%%%%%%%%%%%%%%%%%%%%%%%%%%%%%%%%%%%%%%%%%%%%%%%%%%%%%%%%%%%%%%%%%%%%%%%%%%%%%%%%%%%%%%%%%%%%%%%%%%%%%%%%%%%%%%%%%%%%
%           Outline.
%%%%%%%%%%%%%%%%%%%%%%%%%%%%%%%%%%%%%%%%%%%%%%%%%%%%%%%%%%%%%%%%%%%%%%%%%%%%%%%%%%%%%%%%%%%%%%%%%%%%%%%%%%%%%%%%%%%%%%%%%%%%%%%%%%%%%%%%%%%%%%%%%%%%%%%%%%%%%
\section{Outline}
In this chapter, we are going to consider a polymer solution placed in the gap between two plates, which are subject to adsorption effects on their surfaces. 
We will introduce the adsorption on the surface using two different approaches, which, in turn, are linked with each other. 
The first approach is related to the extrapolation length $b$, taking into account the boundary condition at the wall. 
The second one is related to the external surface potential, which is considered along with the self-consistent field $w(x)$ in the Edwards equation.   
We should compare these approaches with each other, find quantitative expressions linking them together and understand and find out which one is better for the accuracy and convergence problems. 
In addition, we will expand the analytical GSDE theory to the adsorption case and compare the results of the numerical SCFT with the analytical GSDE. 
As a final step, we recalculate the thermodynamic potential in $k_BT$ units and try to get information about the practical usage of the kinetic stability in this case.
%%%%%%%%%%%%%%%%%%%%%%%%%%%%%%%%%%%%%%%%%%%%%%%%%%%%%%%%%%%%%%%%%%%%%%%%%%%%%%%%%%%%%%%%%%%%%%%%%%%%%%%%%%%%%%%%%%%%%%%%%%%%%%%%%%%%%%%%%%%%%%%%%%%%%%%%%%%%%
%           The extrapolation length, b
%%%%%%%%%%%%%%%%%%%%%%%%%%%%%%%%%%%%%%%%%%%%%%%%%%%%%%%%%%%%%%%%%%%%%%%%%%%%%%%%%%%%%%%%%%%%%%%%%%%%%%%%%%%%%%%%%%%%%%%%%%%%%%%%%%%%%%%%%%%%%%%%%%%%%%%%%%%%%    
\section{The extrapolation length}	    
In the presence of an adsorption surface, the total potential felt by monomers is the sum of the self consistent field $W$ and the wall potential $U_s$.
We describe the statistics of a chain of $N$ monomers by the partition function $q(x, s)$. The partition function satisfied to the Edwards equation, 	    
$$
           \frac{\partial q(x, s)}{\partial s} = \frac{a_s}{6}\frac{\partial^2 q(x, s)}{\partial x^2} - (W(x)+U_{s}(x))q(x, s)
$$
where $a_s$ is the statistical segment. Exactly at the surface $q(0, s)=0$, since the polymer can not penetrate into the solid wall.	    
We are not interested in the details of physically observable quantity (such as concentration profile) 
in the vicinity of the wall, where it depends on the shape of the surface potential. 
When the surface potential has the short range $\Delta$, one may replace its effect by the effective boundary condition \cite{deGennes_book, deGennes_1969} 
for the partition function
\begin{equation}
\label{ads_bc_b}
	    q_x(0, s) = -\frac{1}{b}q(0,s)
\end{equation}
The inverse extrapolation length $1/b$ measures the strength of the adsorption and depends on the potential profile $U_s(x)$. 

At $x>\Delta$ the surface potential $U_s=0$, the partition function satisfies to the following equation:           
$$
           \frac{\partial q(x, s)}{\partial s} = \frac{a_s}{6}\frac{\partial^2 q(x, s)}{\partial x^2} - W(x)q(x, s)
$$	    
with the boundary condition Eq.(\ref{ads_bc_b}). Since we are interested in length scales $x\gg\Delta$ we can formally take the limit $\Delta\rightarrow 0$ \cite{Semenov_review_2012}.
%%%%%%%%%%%%%%%%%%%%%%%%%%%%%%%%%%%%%%%%%%%%%%%%%%%%%%%%%%%%%%%%%%%%%%%%%%%%%%%%%%%%%%%%%%%%%%%%%%%%%%%%%%%%%%%%%%%%%%%%%%%%%%%%%%%%%%%%%%%%%%%%%%%%%%%%%%%%%
%           Surface field
%%%%%%%%%%%%%%%%%%%%%%%%%%%%%%%%%%%%%%%%%%%%%%%%%%%%%%%%%%%%%%%%%%%%%%%%%%%%%%%%%%%%%%%%%%%%%%%%%%%%%%%%%%%%%%%%%%%%%%%%%%%%%%%%%%%%%%%%%%%%%%%%%%%%%%%%%%%%%    
\section{The Edwards equation with the surface potential}
The adsorption can be taken into account by considering the external surface field $u_s(x)=NU_s(x)$, which must be included in the Edwards equation 
along with self-consistent field. In the dimensionless variables we can write:    
\begin{equation}
\label{eqdwards_adsorption_surf}
\begin{array}{l}
           \frac{\partial q(x, s)}{\partial s} = \frac{\partial^2 q(x, s)}{\partial x^2} - (w(x)+u_{s}(x))q(x, s), \,\,\, x\in(0, h_m), \,\,\, s > 0 \\
           q(x, 0) = 1, \,\,\, x\in[0, h_m] \\
           q_x(0, s) = q_x(h_m, s) = 0, \,\,\, s \geq 0 	     
\end{array}
\end{equation}	     
where $h_m$ is the midplane separation. 
The boundary condition on the left plate $q_x(0, s) = 0$ is chosen in order to reduce computational errors because in such a way the function $q(x, s)$ 
is an even, smooth function with respect to $x=0$. This boundary condition corresponds to a neutral wall which can be formally derived from the 
change of the variable: $ u ^ {tot} _s (x) = u ^ {0} _s (x) + u_s (x) $, where the field $ u ^ {0 } _s (x) $ is a short ranged potential which provide the b.c. $ q (0, s) = 0 $ and 
$ u_s (x) = u ^ {tot} _s (x) - u ^ {0} _s (x)$ accounts for a deviation of the surface potential from $u_s^0(x)$. %we are interested in long-ranged effects  As we already noticed in Sec.\ref{sec:polymers_adsorption}

As for purely repulsive case, we use Crank-Nikolson scheme for the numerical solution of Eq.(\ref{eqdwards_adsorption_surf}) (see Appendix.\ref{AppendixB}).
Using the numerical scheme and the iterative algorithm, that we used in case of purely repulsive plates, 
we can calculate the concentration profile, $c(x)$ for different surface potentials and virial parameters. 	   

We also choose the surface potential $u_s(x)$ as an even, analytical function (having infinitely many derivatives) in the entire domain $-\infty<x<\infty$. 
In this case, the resulting solution should be also an even, analytical function in the whole space. 	   
Since an even function has smooth derivatives in the origin $x=0$, we thus avoid many problems in the numerical solution related to numerical calculations of derivatives. 
From this point, of view we propose three surface potentials satisfying to the aforementioned requirements:
\begin{equation}
\label{adsorption_surface_pot}         
\begin{array}{l}
           a)\quad u_s(x) = -A\exp(-(\alpha x)^2), \\
	   b)\quad u_s(x) = -A/\cosh((\alpha x)^2), \\ 
	   c)\quad u_s(x) = -A/\cosh^2((\alpha x)^2).
\end{array} 
\end{equation}
One may define the restrictions on the surface potential related to the constants $A$ and $\alpha$. 
First of all, the surface potential should spread on the microscopic distances. On the other hand, 
for numerical reasons, its range should be long enough compared to the grid spacing to reveal the smoothness of $u_s$.	   
	   
\begin{figure}[ht!]
\begin{minipage}[ht]{0.5\linewidth}
\center{\includegraphics[width=1\linewidth]{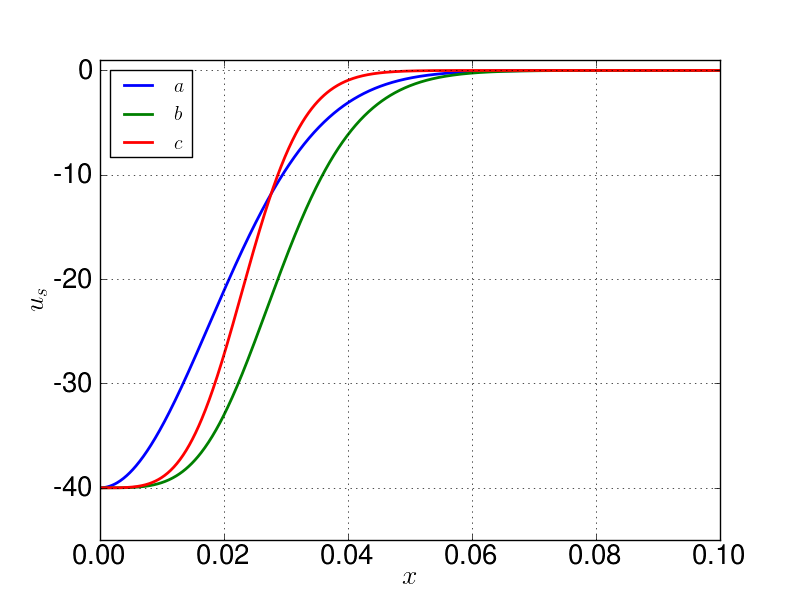}}
\caption{\small{The surface potentials, Eq.(\ref{adsorption_surface_pot}), for $A = 40$ and $\alpha = 40$.}}
\label{adsorption_surface_potentials_fig}
\end{minipage}
\hfill
\begin{minipage}[ht]{0.5\linewidth}
\center{\includegraphics[width=1\linewidth]{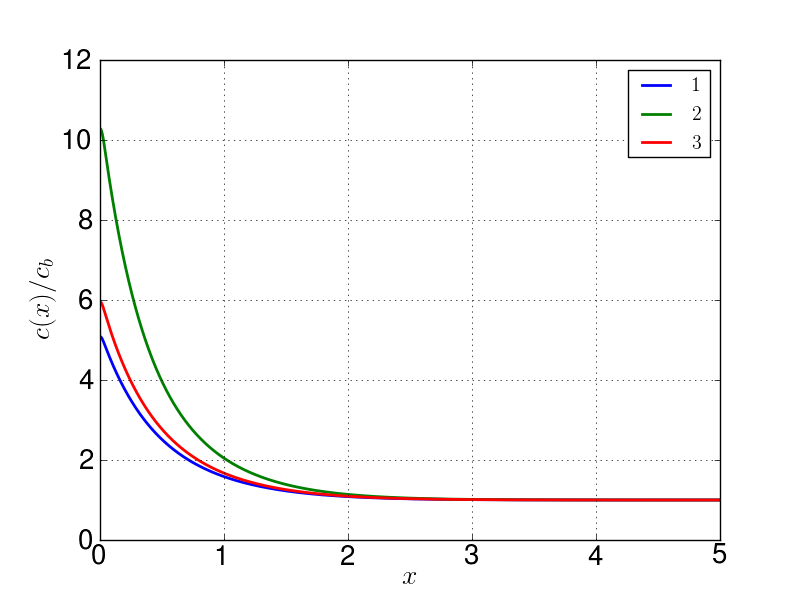}}
\caption{\small{The concentration profile, Eq.(\ref{ads_conc_profile}), for different surface potentials, 
		Eq.(\ref{adsorption_surface_pot}), with $A=40$, $\alpha=40$ and virial parameters $v_N = 0$, $w_N = 0$.}}
\label{adsorption_conc_ext_v10_w10_fig}
\end{minipage}
\end{figure}

For those reasons, we restrict our attention to $\alpha = 40$ defining the proper scope of the above potentials.
In this case, its range is $\Delta \thicksim 0.1$. 
In Fig.\ref{adsorption_surface_potentials_fig} one can find those potentials and in Fig.\ref{adsorption_surface_potentials_fig} the coresponding concentration profiles.
The parameter $A$ depends on the chain length, since $u_s(x) = U(x)N$ is actually the reduced surface potential. 
For sufficiently long chains it can be large: $|u_s^{max}| \gg 1$ or $A \gg 1$. 
%%%%%%%%%%%%%%%%%%%%%%%%%%%%%%%%%%%%%%%%%%%%%%%%%%%%%%%%%%%%%%%%%%%%%%%%%%%%%%%%%%%%%%%%%%%%%%%%%%%%%%%%%%%%%%%%%%%%%%%%%%%%%%%
%        concentration profiles for different external potentials
%%%%%%%%%%%%%%%%%%%%%%%%%%%%%%%%%%%%%%%%%%%%%%%%%%%%%%%%%%%%%%%%%%%%%%%%%%%%%%%%%%%%%%%%%%%%%%%%%%%%%%%%%%%%%%%%%%%%%%%%%%%%%%%
  
In what follows we perform calculations for just one surface potential profile, Eq.(\ref{adsorption_surface_pot}c). 
In addition we fix the parameter $\alpha = 40$. Correspondingly, in what follows we will vary just one parameter $A$ defining the adsorption strength.
	   
%%%%%%%%%%%%%%%%%%%%%%%%%%%%%%%%%%%%%%%%%%%%%%%%%%%%%%%%%%%%%%%%%%%%%%%%%%%%%%%%%%%%%%%%%%%%%%%%%%%%%%%%%%%%%%%%%%%%%%%%%%%%%%%%%%%%%%%%%%%%%%%%%%%%%%%%%%%%%
%           The extrapolation length, b
%%%%%%%%%%%%%%%%%%%%%%%%%%%%%%%%%%%%%%%%%%%%%%%%%%%%%%%%%%%%%%%%%%%%%%%%%%%%%%%%%%%%%%%%%%%%%%%%%%%%%%%%%%%%%%%%%%%%%%%%%%%%%%%%%%%%%%%%%%%%%%%%%%%%%%%%%%%%%    
\section{The Edwards equation with the extrapolation length}
In this section, we consider the numerical solution of the Edwards equation with Neumann boundary conditions, i.e.:
\begin{equation}
\label{eqdwards_adsorption_b}
\begin{array}{l}
	 \frac{\partial q(x, s)}{\partial s} = \frac{\partial^2 q(x, s)}{\partial x^2} - w(x)q(x, s), \,\,\, x\in(0, h_m), \,\,\, s > 0 \\
         q(x, 0) = 1, \,\,\, x\in[0, h_m] \\
         q_x(0, s) = -\frac{1}{b}q(0,s),\quad q_x(h_m, s) = 0, \,\,\, s \geq 0 	     
\end{array}
\end{equation}
where $h_m$ is the midplane separation, $b$ is the extrapolation length that measures the strength of the adsorption.
As before, we use Crank-Nikolson scheme for the numerical solution of Eq.(\ref{eqdwards_adsorption_b}) (see Appendix.\ref{AppendixB}).
Using the numerical scheme and the iterative algorithm, that we used in case of purely repulsive plates, 
we can obtain the concentration profile, $c(x)$ for different extrapolation lengths and virial parameters. 	   
	 
\begin{figure}[ht!]
\begin{minipage}[ht]{0.5\linewidth}
\center{\includegraphics[width=1\linewidth]{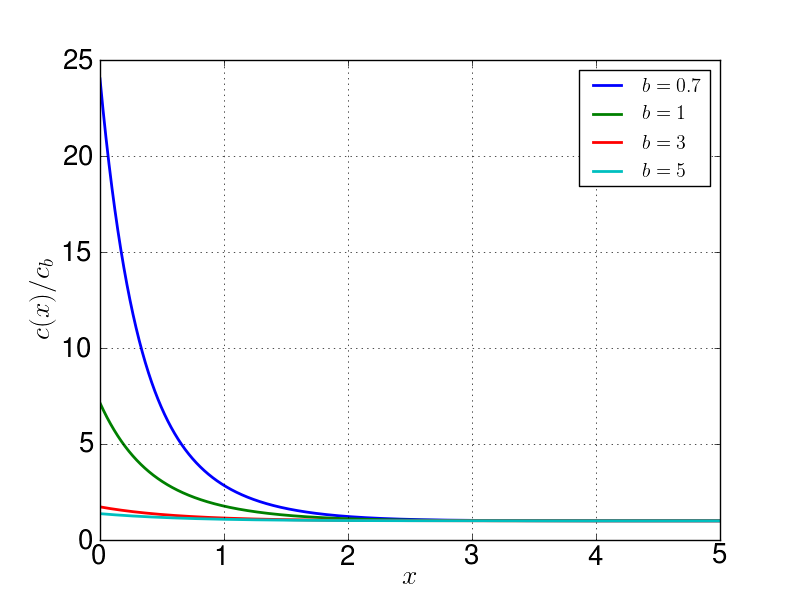}}
\caption{\small{The concentration profile, Eq.(\ref{ads_conc_profile}), for the following parameters: $v_N = 0 = w_N = 0$, $N_x=N_s = 10k$ 
		and different values of parameter $b$.}}
\label{conc_h5_xs10k_b_small_fig}
\end{minipage}
\hfill
\begin{minipage}[ht]{0.5\linewidth}
\center{\includegraphics[width=1\linewidth]{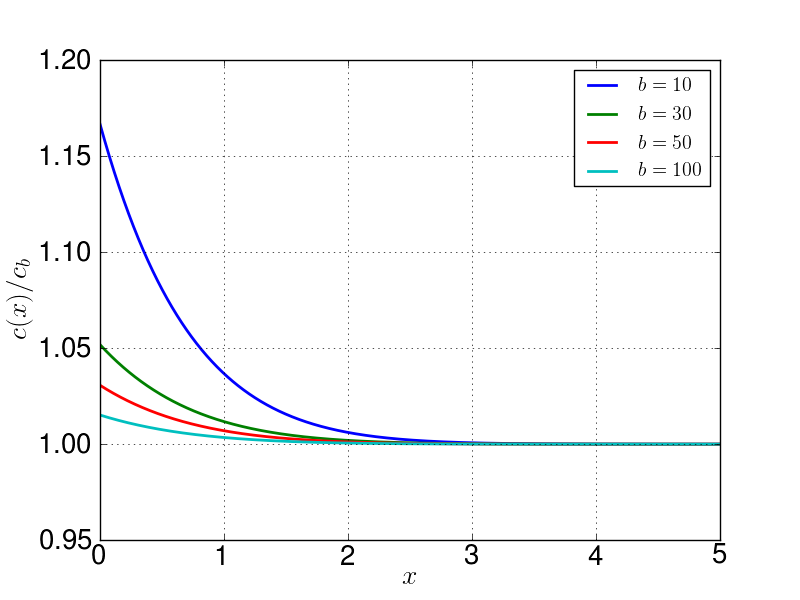}}
\caption{\small{The concentration profile, Eq.(\ref{ads_conc_profile}), for the following parameters: $v_N = 0 = w_N = 0$, $N_x=N_s = 10k$ 
		and different values of parameter $b$.}}
\label{conc_h5_xs10k_b_large_fig}
\end{minipage}
\end{figure}

Using the definition for the concentration profile, written via the distribution function:
\begin{equation}
\label{ads_conc_profile}
 	 c(x)/c_b = \int\limits_0^{1} \mathrm{d}s q(x, s) q(x, 1-s) 
\end{equation}
we can write the boundary condition in Eq.(\ref{eqdwards_adsorption_b}) in terms of the surface concentration:
$$
c_x(0)/c_b = \int\limits_0^{1} \mathrm{d}s q_x(0, s) q(0, 1-s) + \int\limits_0^{1} \mathrm{d}sq(0, s)q_x(0, 1-s) = - \frac{2}{b}c(0)/c_b 
$$
In Figs.\ref{conc_h5_xs10k_b_small_fig}--\ref{conc_h5_xs10k_b_large_fig} one can find the concentration profiles for different values of 
the adsorption parameter $b$ (ideal polymer solution, $v_N=0$, $w_N=0$). Non ideal polymer solution is considered later on.
%%%%%%%%%%%%%%%%%%%%%%%%%%%%%%%%%%%%%%%%%%%%%%%%%%%%%%%%%%%%%%%%%%%%%%%%%%%%%%%%%%%%%%%%%%%%%%%%%%%%%%%%%%%%%%%%%%%%%%%%%%%%%%%
%        concentration profiles for different parameter b
%%%%%%%%%%%%%%%%%%%%%%%%%%%%%%%%%%%%%%%%%%%%%%%%%%%%%%%%%%%%%%%%%%%%%%%%%%%%%%%%%%%%%%%%%%%%%%%%%%%%%%%%%%%%%%%%%%%%%%%%%%%%%%%
%%%%%%%%%%%%%%%%%%%%%%%%%%%%%%%%%%%%%%%%%%%%%%%%%%%%%%%%%%%%%%%%%%%%%%%%%%%%%%%%%%%%%%%%%%%%%%%%%%%%%%%%%%%%%%%%%%%%%%%%%%%%%%%%%%%%%%%%%%%%%%%%%%%%%%%%%%%%%
%          Correspondence between b and A
%%%%%%%%%%%%%%%%%%%%%%%%%%%%%%%%%%%%%%%%%%%%%%%%%%%%%%%%%%%%%%%%%%%%%%%%%%%%%%%%%%%%%%%%%%%%%%%%%%%%%%%%%%%%%%%%%%%%%%%%%%%%%%%%%%%%%%%%%%%%%%%%%%%%%%%%%%%%%    	   
\section{Correspondence between the extrapolation length and the amplitude of the surface field} 
\begin{figure}[ht!]
\center{\includegraphics[width=0.5\linewidth]{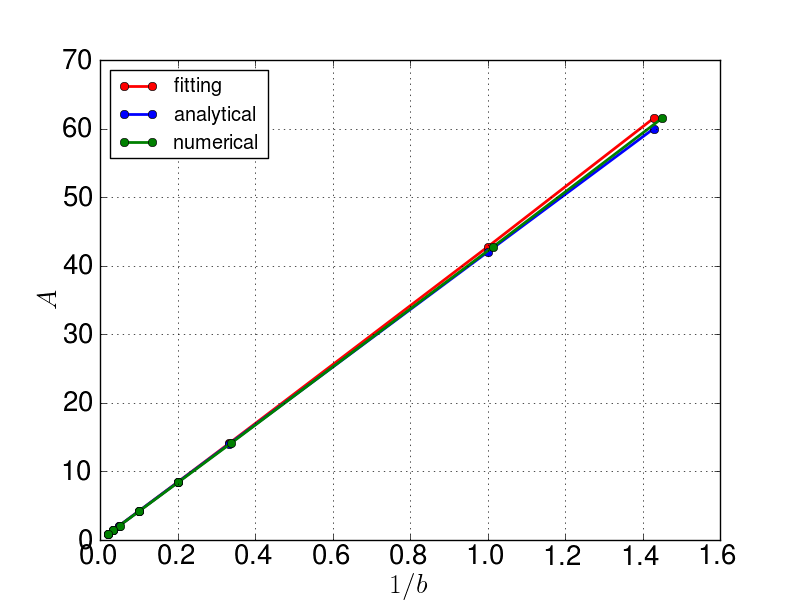}}
\caption{\small{The dependence of the surface potential parameter, $A$, on the inverse extrapolation length, $1/b$. The data correspond to the 
	  Tab.\ref{tabular:adsorption_comparison}.}}
\label{adsorption_correspondence_b_a_fig}
\end{figure}

	   In order to find the correspondence between $b$ and $A$ we use the following strategy.
	   First of all, we will consider the concentration profile at some value of the extrapolation length $b$ and pick up 
	   the corresponding surface potential, i.e. the parameter $A$, which produce the same concentration profile for the ideal polymer solution ($v_N=w_N =0$). 
	   After that, we will consider how $A=A(b)$ is changed when the virial parameters are varied. 
	   The global aim of this section is to find the relationship between $A$ and $b$. 
%%%%%%%%%%%%%%%%%%%%%%%%%%%%%%%%%%%%%%%%%%%%%%%%%%%%%%%%%%%%%%%%%%%%%%%%%%%%%%%%%%%%%%%%%%%%%%%%%%%%%%%%%%%%%%%%%%%%%%%%%%%%%%%
%        table: correspondence between b and A
%%%%%%%%%%%%%%%%%%%%%%%%%%%%%%%%%%%%%%%%%%%%%%%%%%%%%%%%%%%%%%%%%%%%%%%%%%%%%%%%%%%%%%%%%%%%%%%%%%%%%%%%%%%%%%%%%%%%%%%%%%%%%%%
\begin{table}[ht!]
\caption{The parameters producing the same concentration profiles in the three different approaches to the adsorption.} 
\label{tabular:correspondence_b_a} 
\begin{center}
  \begin{tabular}{ | c | c | c | c | c | c | c | c | c | }
    \hline
    $b$                                 &    0.7&      1&      3&     5&    10&    20&    30&    50 \\
    \hline
    $A_{fit}$                           &   61.5&  42.73&   14.1&  8.42&   4.2&   2.1&   1.4&  0.84   \\
    \hline
    $A_{an}$                            & 59.974& 41.982& 13.994& 8.396& 4.198& 2.099& 1.399&  0.84   \\
    \hline
    $A_{num}$                           & 60.559& 42.136& 13.915& 8.334& 4.161&  2.079& 1.385&  0.831   \\
    \hline
  \end{tabular}
\end{center}
\end{table}
	   In Tab.\ref{tabular:correspondence_b_a} we present the dependence obtained for different values of 
	   the extrapolation length, $b$ and the parameter of the external surface potential, $A$. 
	   We also depict the dependence for inverse extrapolation length 
	   and the corresponding parameter $A$, which can be found in Fig.\ref{adsorption_correspondence_b_a_fig}. 
	   One can see, that this is the straight line with a slope $\simeq 43$. 
	   Let us find the dependence $A(b)$ using different ways:\\	   
	   $\textbf{Analytical}$. The quantitative relationship can be found considering the Edwards equation written in the GSD approximation:
\begin{equation}
\label{adsorption_gsd_edwards}
	   q_{xx} - u_s(x)q = 0
\end{equation}
	   with the boundary condition on the plate, $q'(0) = 0$. The surface field, $u_s(x)$ is changed on the microscopic lengh-scale, 
	   where the function $q$ varies slightly. 
	   Thereby, we can neglect these variations and substitute it with the constant value $q(\delta)$, where $\delta$ defines the scope of the surface 
	   potential.\footnote{At distances $x>\delta$ we can neglect the surface potential and obviously $\delta$ must be larger than $1/\alpha$: $\delta\gg 1/\alpha$.}
	   We can rewrite the preceding equation as
$$
	    q_{xx} - u_s(x)q(\delta) = 0
$$
	   which can be easily integrated for $x\in[0..\delta]$ giving 
$$
	   q_{x}(\delta)-q_{x}(0) - q(\delta)\int\limits_0^{\delta}u_s(x) = 0
$$	   
	   or taking into account the boundary condition at $x=0$ we obtain
$$
	   q_{x}(\delta)/q(\delta) = \int\limits_0^{\delta}u_s(x) \simeq -1/b
$$
	   Using the appropriate expression for the surface potential Eq.(\ref{adsorption_surface_pot}c), we can easily write
$$
	   A \simeq \frac{1}{b\int\limits_0^{\delta}\mathrm{d}x/\cosh^2(\alpha x)^2}
$$	   
	   Since $\delta\gg 1/\alpha$, the upper limit in the integral can be replaced by infinity:\footnote{We used that 
	   $\int\limits_0^{\infty}\mathrm{d}x/\cosh^2(x^2)= 0.95278147 \simeq 1$} $A\simeq\alpha /b=40/b$. The corresponding results for 
	   $A=A_{an}$, defined by the above integral, may be found in Tab.\ref{tabular:correspondence_b_a}.  

	   $\textbf{Numerical}$. For finding the more accurate correspondence between adsorption parameters, $A$ and $b$, consider 
	   again Eq.(\ref{adsorption_gsd_edwards}) with "initial" conditions: $q_{x}(0) = 0$ and $q(0) = \sqrt{c(0)}$, where $c(0)$ is a value of 
	   the concentration profile at $x=0$ produced by solution of the Edwards equation, Eq.(\ref{eqdwards_adsorption_surf}).
	   Solving the equation using the Runge-Kutta method, we found the solution, which in the vicinity of the wall, produces the 
	   concentration profile $c(x) = q^2(x)$. This profile coincides with the concentration profile obtained by the Edwards equation. 
	   Then, after the bending becomes a straight line, the solution $q(x)$ has a certain slope.
	   In order to define the slope, we write the solution in the region as $q(x) = A_x + B_x x$, where the parameters $A_x$ and $B_x$ can be found through 
	   the standard equation of a straight line:
$$
	    \tan \phi = \frac{q_i - q_j}{x_i-x_j}; \quad A_x = q_j - x_j\tan \phi ; \quad B_x = \tan \phi
$$
	   where $i, j$ correspond to arbitrary points on the straight line. Using the adsorption boundary condition for the line, we can write
$$
	    q_x(0) = -\frac{1}{b}q(0) \rightarrow b = - \frac{A_x}{B_x}
$$
            The results of such calculation can be found in Tab.\ref{tabular:correspondence_b_a} as $A=A_{num}(b)$.\footnote{We found empirically 
            that the adsorption parameter, $b$ does not depend on the initial condition, $q(0)$. The same value, $b$ can be produced by different initial condition. 
            Using that we can solve the inverse problem: find adsorption parameter, $A$ via specified parameter, $b$. For this we should consider the function 
	    $F(A_0)$, which implements the above procedure and return a certain value $b_0$. Then, applying, for example, the bisection method for solving the equation
	    $F(A) - b = 0$, we can find the corresponding value of $A$.}
%%%%%%%%%%%%%%%%%%%%%%%%%%%%%%%%%%%%%%%%%%%%%%%%%%%%%%%%%%%%%%%%%%%%%%%%%%%%%%%%%%%%%%%%%%%%%%%%%%%%%%%%%%%%%%%%%%%%%%%%%%%%%%%
%        plot the depandence between 1/b and A
%%%%%%%%%%%%%%%%%%%%%%%%%%%%%%%%%%%%%%%%%%%%%%%%%%%%%%%%%%%%%%%%%%%%%%%%%%%%%%%%%%%%%%%%%%%%%%%%%%%%%%%%%%%%%%%%%%%%%%%%%%%%%%%
  	   
\begin{figure}[ht!]
\begin{minipage}[ht]{0.5\linewidth}
\center{\includegraphics[width=1\linewidth]{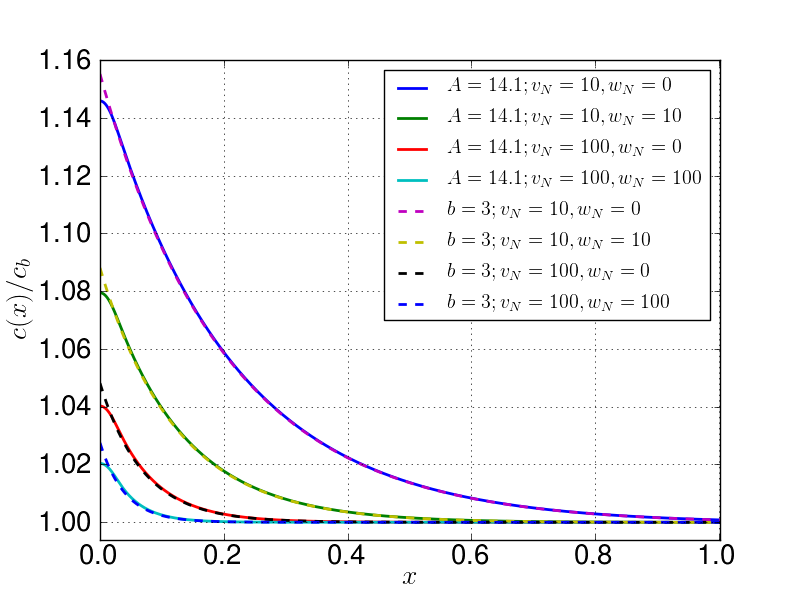}}
\caption{\small{The comparison between the concentration profiles, Eq.(\ref{ads_conc_profile}),
		calculated for adsorption, which is defined by the extrapolation lengh, $b$
		and by the surface potential, Eq.(\ref{adsorption_surface_pot}c), for different values of the virial parameters. The grid $N_x = 5k$, $N_s=5k$.}}
\label{adsorption_conc_comp_b3_a14_1_fig}
\end{minipage}
\hfill
\begin{minipage}[ht]{0.5\linewidth}
\center{\includegraphics[width=1\linewidth]{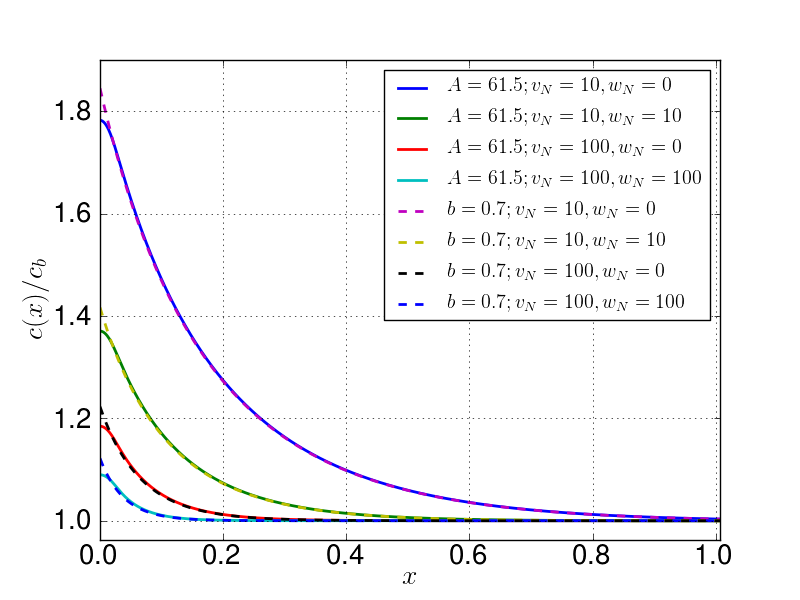}}
\caption{\small{The comparison between the concentration profiles, Eq.(\ref{ads_conc_profile}),
		calculated for adsorption, which is defined by the extrapolation lengh, $b$
		and by the surface potential, Eq.(\ref{adsorption_surface_pot}c), for different values of the virial parameters. The grid $N_x = 5k$, $N_s=5k$.}}
\label{adsorption_conc_comp_b07_a61_5_fig}
\end{minipage}
\end{figure}

  	   In Figs.\ref{adsorption_conc_comp_b3_a14_1_fig}--\ref{adsorption_conc_comp_b07_a61_5_fig} one can see the coincident concentration profiles 
  	   created for appropriate adsorption parameters and different values of the virial parameters. 
	   Despite small deviation, in the vicinity of the wall, the depicted concentration profiles are in good agreement with each other.	   
	   We summarize the results of the computation in Tab.\ref{tabular:adsorption_comparison}, where we present the comparison
	   between the two approaches and how many iterations, in the iterative procedure, are required. 
	   One can, obviously, see that the number of iterations is around two times smaller in the case when the adsorption is set by the external 
	   surface potential in comparison to the extrapolation length. 
	   Computationally, this is a very big advantage.
	   It is due to a rational way to introduce the adsorption effects via the analytical functions.

%%%%%%%%%%%%%%%%%%%%%%%%%%%%%%%%%%%%%%%%%%%%%%%%%%%%%%%%%%%%%%%%%%%%%%%%%%%%%%%%%%%%%%%%%%%%%%%%%%%%%%%%%%%%%%%%%%%%%%%%%%%%%%%
%        concentration profiles many and zoom in for a)  A = 14.1, b = 3; b) A = 61.5, b =0.7
%%%%%%%%%%%%%%%%%%%%%%%%%%%%%%%%%%%%%%%%%%%%%%%%%%%%%%%%%%%%%%%%%%%%%%%%%%%%%%%%%%%%%%%%%%%%%%%%%%%%%%%%%%%%%%%%%%%%%%%%%%%%%%%
%%%%%%%%%%%%%%%%%%%%%%%%%%%%%%%%%%%%%%%%%%%%%%%%%%%%%%%%%%%%%%%%%%%%%%%%%%%%%%%%%%%%%%%%%%%%%%%%%%%%%%%%%%%%%%%%%%%%%%%%%%%%%%%
%        Table: comparisons
%%%%%%%%%%%%%%%%%%%%%%%%%%%%%%%%%%%%%%%%%%%%%%%%%%%%%%%%%%%%%%%%%%%%%%%%%%%%%%%%%%%%%%%%%%%%%%%%%%%%%%%%%%%%%%%%%%%%%%%%%%%%%%%	   
\begin{table}[ht!]
\caption{The comparison of the two approaches.} 
\label{tabular:adsorption_comparison} 
\begin{center}
  \begin{tabular}{ | c | c | c | c | c | c | c | c | c | }
    \hline
            & \multicolumn{4}{|c|}{b=3} & \multicolumn{4}{|c|}{A=14.1}\\
    \hline
    $v_N$   &  10&  10& 100& 100&  10&  10& 100& 100 \\ 
    \hline
    $w_N$   &   0&  10&   0& 100&   0&  10&   0& 100 \\
    \hline
    $N_{it}$& 713& 752& 795& 887& 377& 390& 404& 524 \\
    \hline
            & \multicolumn{4}{|c|}{b=0.7} & \multicolumn{4}{|c|}{A=61.5}\\
    \hline
    $v_N$   &  10&  10& 100& 100&  10&  10& 100& 100 \\ 
    \hline
    $w_N$   &   0&  10&   0& 100&   0&  10&   0& 100 \\
    \hline
    $N_{it}$&1039&1318&1255&1767& 532& 682& 727& 1058 \\
    \hline
  \end{tabular}
\end{center}
\end{table}
%%%%%%%%%%%%%%%%%%%%%%%%%%%%%%%%%%%%%%%%%%%%%%%%%%%%%%%%%%%%%%%%%%%%%%%%%%%%%%%%%%%%%%%%%%%%%%%%%%%%%%%%%%%%%%%%%%%%%%%%%%%%%%%
%        Concentration profiles for small length, h; two approaches
%%%%%%%%%%%%%%%%%%%%%%%%%%%%%%%%%%%%%%%%%%%%%%%%%%%%%%%%%%%%%%%%%%%%%%%%%%%%%%%%%%%%%%%%%%%%%%%%%%%%%%%%%%%%%%%%%%%%%%%%%%%%%%%
\begin{figure}[ht!]
\begin{minipage}[ht]{0.5\linewidth}
\center{\includegraphics[width=1\linewidth]{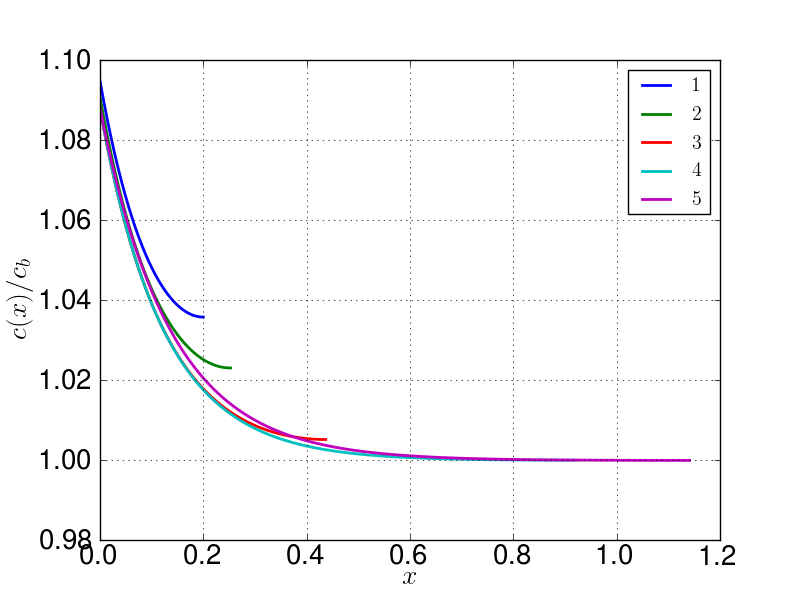}}
\caption{\small{The concentration profiles, Eq.(\ref{ads_conc_profile}), calculated for different separations, $h$. 
		The parameters: $v_N=10, w_N=10$, $b=3$ and $N_x = 7k$, $N_s=3k$.}}
\label{adsorption_conc_h_small_b3_v10_w10_73_fig}
\end{minipage}
\hfill
\begin{minipage}[ht]{0.5\linewidth}
\center{\includegraphics[width=1\linewidth]{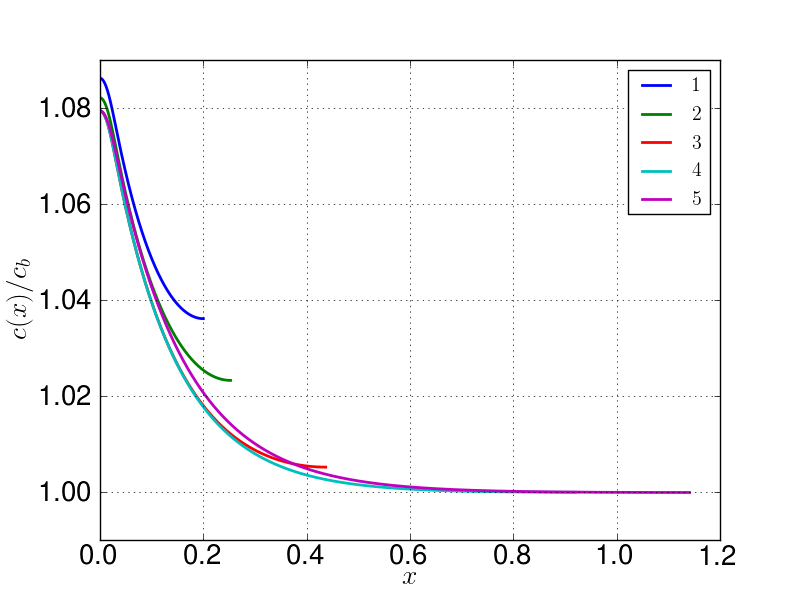}}
\caption{\small{The concentration profiles, Eq.(\ref{ads_conc_profile}), calculated for different separations, $h$. 
		The parameters: $v_N=10, w_N=10$, $A=14.1$ and $N_x = 7k$, $N_s=3k$.}}
\label{adsorption_conc_h_small_a14_1_v10_w10_73_fig}
\end{minipage}
\end{figure}
%%%%%%%%%%%%%%%%%%%%%%%%%%%%%%%%%%%%%%%%%%%%%%%%%%%%%%%%%%%%%%%%%%%%%%%%%%%%%%%%%%%%%%%%%%%%%%%%%%%%%%%%%%%%%%%%%%%%%%%%%%%%%%%
%        Concentration profiles for large length, h; two approaches
%%%%%%%%%%%%%%%%%%%%%%%%%%%%%%%%%%%%%%%%%%%%%%%%%%%%%%%%%%%%%%%%%%%%%%%%%%%%%%%%%%%%%%%%%%%%%%%%%%%%%%%%%%%%%%%%%%%%%%%%%%%%%%%
\begin{figure}[ht!]
\begin{minipage}[ht]{0.5\linewidth}
\center{\includegraphics[width=1\linewidth]{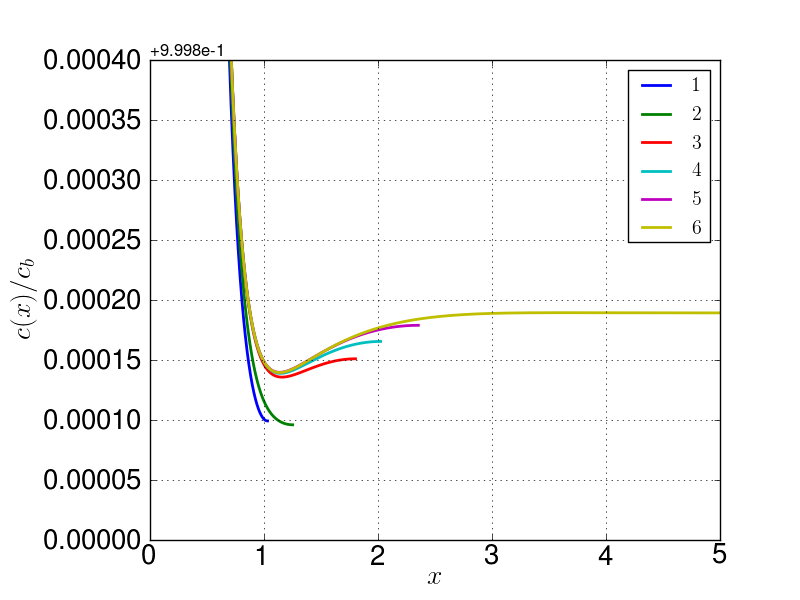}}
\caption{\small{The concentration profiles, Eq.(\ref{ads_conc_profile}), calculated for different separations, $h$. 
		The parameters: $v_N=10, w_N=10$, $b=3$ and $N_x = 7k$, $N_s=3k$.}}
\label{adsorption_conc_h_large_b3_v10_w10_73_fig}
\end{minipage}
\hfill
\begin{minipage}[ht]{0.5\linewidth}
\center{\includegraphics[width=1\linewidth]{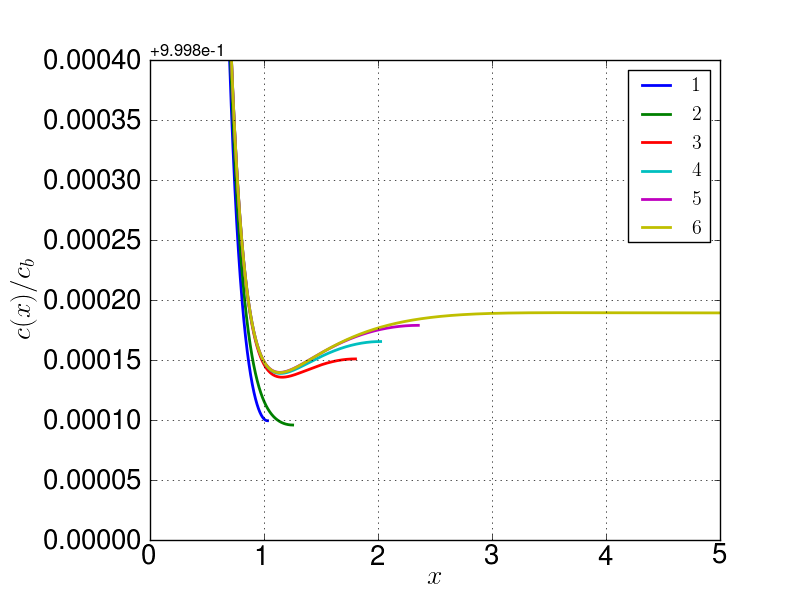}}
\caption{\small{The concentration profiles, Eq.(\ref{ads_conc_profile}), calculated for different separations, $h$. 
		The parameters: $v_N=10, w_N=10$, $A=14.1$ and $N_x = 7k$, $N_s=3k$.}}
\label{adsorption_conc_h_large_a14_1_v10_w10_73_fig}
\end{minipage}
\end{figure}
	  We also represent the dependence of the concentration profile as a function of the separation between plates for both approaches setting adsorption
	  via extrapolation length, Figs.\ref{adsorption_conc_h_small_b3_v10_w10_73_fig}--\ref{adsorption_conc_h_large_b3_v10_w10_73_fig}, and 
	  via surface potential, Fig.\ref{adsorption_conc_h_small_a14_1_v10_w10_73_fig}--\ref{adsorption_conc_h_large_a14_1_v10_w10_73_fig}. 
	  One can notice that the data are in good agreement with each other.	  
%%%%%%%%%%%%%%%%%%%%%%%%%%%%%%%%%%%%%%%%%%%%%%%%%%%%%%%%%%%%%%%%%%%%%%%%%%%%%%%%%%%%%%%%%%%%%%%%%%%%%%%%%%%%%%%%%%%%%%%%%%%%%%%%%%%%%%%%%%%%%%%%%%%%%%%%%%%%%
%          Thermodynamic potential
%%%%%%%%%%%%%%%%%%%%%%%%%%%%%%%%%%%%%%%%%%%%%%%%%%%%%%%%%%%%%%%%%%%%%%%%%%%%%%%%%%%%%%%%%%%%%%%%%%%%%%%%%%%%%%%%%%%%%%%%%%%%%%%%%%%%%%%%%%%%%%%%%%%%%%%%%%%%%    	   
\section{Thermodynamic potential} 
	   In Sec.\ref{sec:polymers_force_2plates} we obtained the expression for the thermodynamic potential (see Eq.(\ref{intr_polymer_free_energy_homo_eq_0})). 
	   In reduced, dimensionless variables, it has the following form
\begin{equation}
\label{adsorption_free_energy_homo}
	\Omega(c_b, h) = - \left(Q_{in}[w, h] - h\right) - \int\limits_0^h\mathrm{d}x\, \left(\frac{v_N}{2}\left(c(x)/c_b)^2-1\right) +
       	 \frac{2w_N}{3}\left((c(x)/c_b)^3-1\right)\right) 
\end{equation}
	  where $Q_{in}[w, h]$ is a partition function defined as
$$
	  Q_{in}[w, h] = \int\limits_0^h\mathrm{d}x q(x, 1), \quad x\in[0..h]
$$
	  $c(x)/c_b$ is defined from Eq.(\ref{ads_conc_profile}) and $q(x, s)$ is obtained from the numerical solution of the Edwards equation. 
	  Below we consider results obtained for the two different cases account for adsorption effects by extrapolation length, $b$, 
	  and external surface potential controlled by the parameter, $A$. 
	  
	  {\raggedright $\textbf{Initial guess}$.} 
	  The process of finding equilibrium thermodynamic potential Eq.(\ref{adsorption_free_energy_homo}) occurs via 
	  iterative procedure when we find the equilibrium self-consistent field in the Edwards equation Eq.(\ref{eqdwards_adsorption_surf}). 
	  Previously, when we examined purely repulsive walls, for the initial guess in the iterative procedure we used 
	  the results obtained for an ideal polymer solution ($v_N=w_N=0$). 	
	  The same strategy is appled for the adsorption case.
	  In the iterative procedure, we consider $c(x)=0$ as the initial guess for the lowest separation $h$. 
	  For all other separations the initial guesses are chosen in absolutely the same way as for purely repulsive case, 
	  i.e. we add a segment $\Delta h$ to the end of the concentration profile obtained with the previous $h$
	  (as described in Sec.\ref{sec:iterative_procedure}).	In addition,  
	  in Tab.\ref{tabular:adsorption_marginal_a} we present the values of the adsorption parameter $A$, at which the iterative 
	  procedure is still marginally converging.
%%%%%%%%%%%%%%%%%%%%%%%%%%%%%%%%%%%%%%%%%%%%%%%%%%%%%%%%%%%%%%%%%%%%%%%%%%%%%%%%%%%%%%%%%%%%%%%%%%%%%%%%%%%%%%%%%%%%%%%%%%%%%%%%%%%%%%%%%%%%
% The height of the barrier for vw: adsorption 
%%%%%%%%%%%%%%%%%%%%%%%%%%%%%%%%%%%%%%%%%%%%%%%%%%%%%%%%%%%%%%%%%%%%%%%%%%%%%%%%%%%%%%%%%%%%%%%%%%%%%%%%%%%%%%%%%%%%%%%%%%%%%%%%%%%%%%%%%%%%
\begin{table}[ht!]
\caption{The upper values of the adsorption parameter $A$, at which the iteration process is still converges. 
	  Fixed parameters: $N_x=7k, N_s=3k, \Lambda = 50$.} 
\label{tabular:adsorption_marginal_a} 
\begin{center}
  \begin{tabular}{ | c | c | c | c | c | c | c | }
    \hline
	 $v_N$                           &    10&    10&    10&    30&    30&    100   \\ \hline
         $w_N$                           &     0&   0.5&    10&     0&    10&      0   \\ \hline
           $A$                           &  1600&  1000&  1000&  1600&  1000&   1600   \\
    \hline
  \end{tabular}
\end{center} 
\end{table}

\begin{figure}[ht!]
\begin{minipage}[ht]{0.5\linewidth}
\center{\includegraphics[width=1\linewidth]{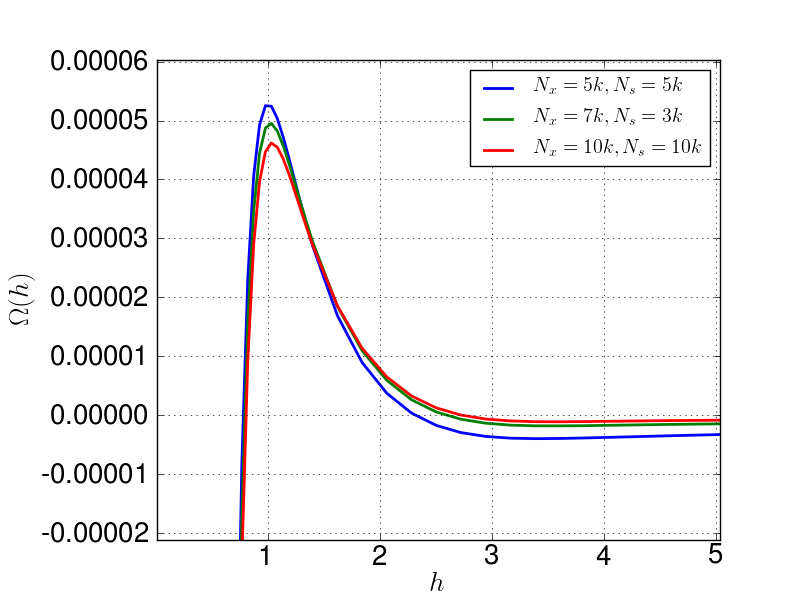}}
\caption{\small{The thermodynamic potential Eq.(\ref{adsorption_free_energy_homo}) for different 
		values of the grid parameters, $N_x$, $N_s$. Fixed parameters: $b = 3$ and $v_N = 10$, $w_N=10$.}}
\label{adsorption_Omega8_h5_b3_v10_w10_fig}
\end{minipage}
\hfill
\begin{minipage}[ht]{0.5\linewidth}
\center{\includegraphics[width=1\linewidth]{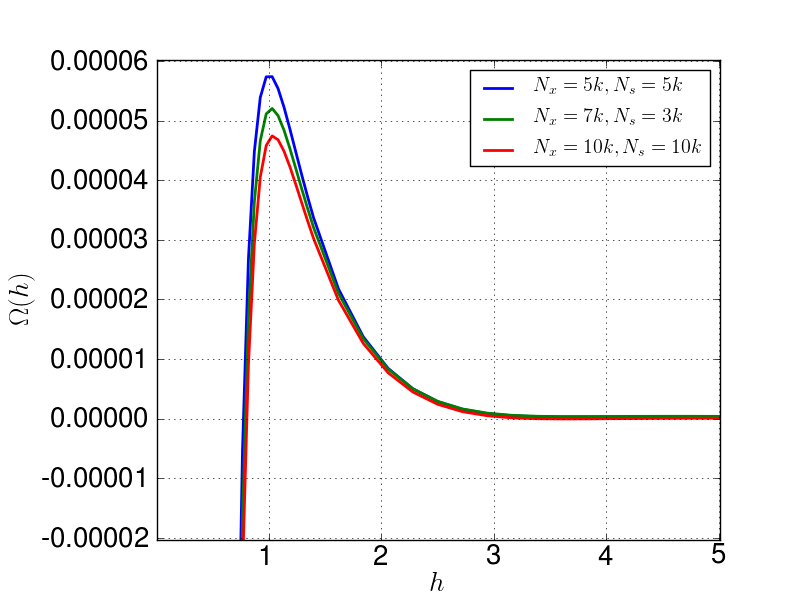}}
\caption{\small{The thermodynamic potentials Eq.(\ref{adsorption_free_energy_homo}) for different values of the grid parameters, $N_x$, $N_s$. 
		Fixed parameters: $A=14.1$ and $v_N = 10$, $w_N=10$.}}
\label{adsorption_Omega8_h5_a14_1_v10_w10_fig}
\end{minipage}
\end{figure}

	  {\raggedright $\textbf{Accuracy}$.} In order to find suitable mesh size and choose a better way of setting adsorption
	  we present in Figs.\ref{adsorption_Omega8_h5_b3_v10_w10_fig}--\ref{adsorption_Omega8_h5_a14_1_v10_w10_fig} 
	  thermodynamic potentials, Eq.(\ref{adsorption_free_energy_homo}), calculated for different number of the grid points $N_x, N_s$.	  
	  The long-$h$ deviation for the horizontal line in Fig.\ref{adsorption_Omega8_h5_b3_v10_w10_fig} is due to the fact that in our numerical scheme, 
	  we made the calculation with different spacial steps $\Delta x$ for different separations, $h$: $\Delta x = h/(N_x-1)$.
	  One can see that this step increases when we increase separation $h$ and as a consequence the accuracy of the calculation decreases. 	  
	  We implemented such storage of the array because it is convenient to keep the arrays for the concentration profiles with the same length, $N_x$ 
	  at different separations and write them in one file. Comparing the pictures with each other one can notice that 
	  the accuracy is much better when we set the adsorption by the external surface potential. 
	  Thereby, due to the fact that we need less iterations
	  to find the equilibrium solution and that the accuracy is better, for future purposes we consider the adsorption problem defined by
	  the external surface potential.	  
%%%%%%%%%%%%%%%%%%%%%%%%%%%%%%%%%%%%%%%%%%%%%%%%%%%%%%%%%%%%%%%%%%%%%%%%%%%%%%%%%%%%%%%%%%%%%%%%%%%%%%%%%%%%%%%%%%%%%%%%%%%%%%%
%        Thermodynamic potentials accuracy for different mesh size
%%%%%%%%%%%%%%%%%%%%%%%%%%%%%%%%%%%%%%%%%%%%%%%%%%%%%%%%%%%%%%%%%%%%%%%%%%%%%%%%%%%%%%%%%%%%%%%%%%%%%%%%%%%%%%%%%%%%%%%%%%%%%%%
	  
	  In the previous chapter when we compared the solutions with different mesh size, we  
	  found that the sufficient accuracy for the thermodynamic potential is reached at $N_x=7k, N_s=3k$ (see Sec.\ref{sec:repulsion_num_res}). 	  
	  As a quantitative measures of the accuracy, we introduced the relative error between barriers hights calculated for different mesh sizes:	  
$$
	  \epsilon = \Delta\Omega^*/\Delta\Omega^*_{hr} 
$$    
          where  $\Delta\Omega^*=\Omega_{max} - \Omega_{inf}$ is calculated for $N_x=7k, N_s=3k$ and $\Delta\Omega^*_{hr}$ corresponds to the height of the barrier 
          calculated for high resolution mesh, i.e. $N_x=10k, N_s=10k$. 
          Large virial parameters produce large amplitude of the self-consistent field. In turn, this leads to large numerical errors and worse accuracy for a fixed 
          adsorption parameter $A$.                    	
          In the considered case, the maximum amplitude of the self-consistent field is generated by the set of the virial parameters: $v_N=100, w_N=0$, and in 
          accordance with the above reason, any smaller virial parameters produce better accuracy. 
          Therefore, it is enough to consider only this set of virial parameters for different values of the adsorption strength $A$.          
          The corresponding results are shown in Tab.\ref{tabular:adsorption_comparison_max_scft_vw}. 
          
          In Tab.\ref{tabular:adsorption_barrier_max_scft_vw} we present all numerical values of the relative error for the barrier heights calculated 
          for different virial parameters and adsorption strength.
          One can notice that sufficient accuracy for the thermodynamic potential is reached when the  
          barriers height is $O(10^{-3})$ or bigger. Thus it is enough to restrict our consideration by $N_x=7k,N_s=3k$.                           
%%%%%%%%%%%%%%%%%%%%%%%%%%%%%%%%%%%%%%%%%%%%%%%%%%%%%%%%%%%%%%%%%%%%%%%%%%%%%%%%%%%%%%%%%%%%%%%%%%%%%%%%%%%%%%%%%%%%%%%%%%%%%%%%%%%%%%%%%%%%
% The height of the barrier for vw: adsorption 
%%%%%%%%%%%%%%%%%%%%%%%%%%%%%%%%%%%%%%%%%%%%%%%%%%%%%%%%%%%%%%%%%%%%%%%%%%%%%%%%%%%%%%%%%%%%%%%%%%%%%%%%%%%%%%%%%%%%%%%%%%%%%%%%%%%%%%%%%%%%
\begin{table}[ht!]
\caption{The comparison of the thermodynamic potential barrier heights expressed via the parameter $\epsilon$. 
	  $\Delta\Omega^*$ is calculated for $N_x=7k, N_s=3k$ for different adsorption parameters, $A$. Fixed virial parameters: $v_N=100, w_N=0$.} 
\label{tabular:adsorption_comparison_max_scft_vw} 
\begin{center}
  \begin{tabular}{ | c | c | c | c | c | c | c | }
    \hline
         $A$                           &  14.1& 42.73&  61.5&   100&   200&    500   \\ \hline
  $\epsilon$                           &  1.27&  1.08& 1.061& 1.045& 1.017&  1.009   \\
    \hline
  \end{tabular}
\end{center} 
\end{table}

\begin{figure}[ht!]
\begin{minipage}[ht]{0.5\linewidth}
\center{\includegraphics[width=1\linewidth]{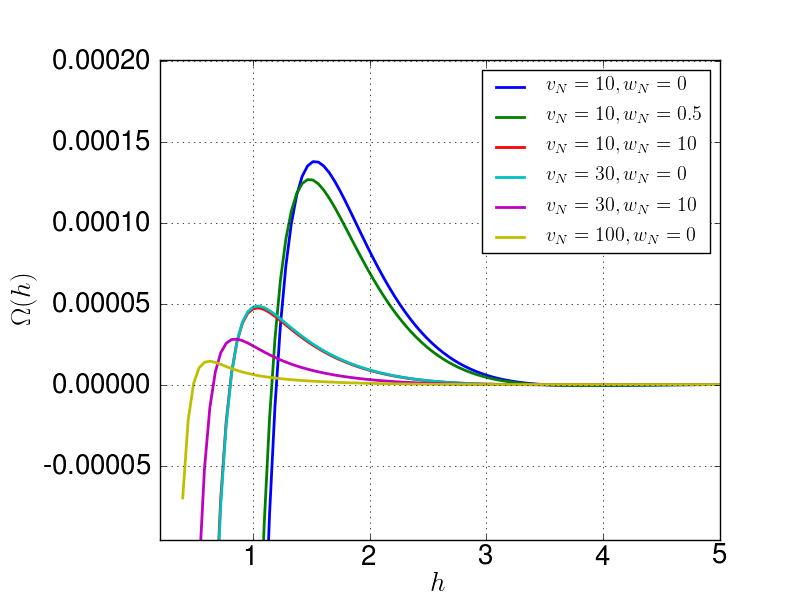}}
\caption{\small{The thermodynamic potential, Eq.(\ref{adsorption_free_energy_homo}), for different 
		values of the virial parameters, $v_N$, $w_N$. Fixed parameters: $A = 14.1$ and $N_x = 10k$, $N_s=10k$.}}
\label{adsorption_Omega8_h5_a14_1_vw_many_fig}
\end{minipage}
\hfill
\begin{minipage}[ht]{0.5\linewidth}
\center{\includegraphics[width=1\linewidth]{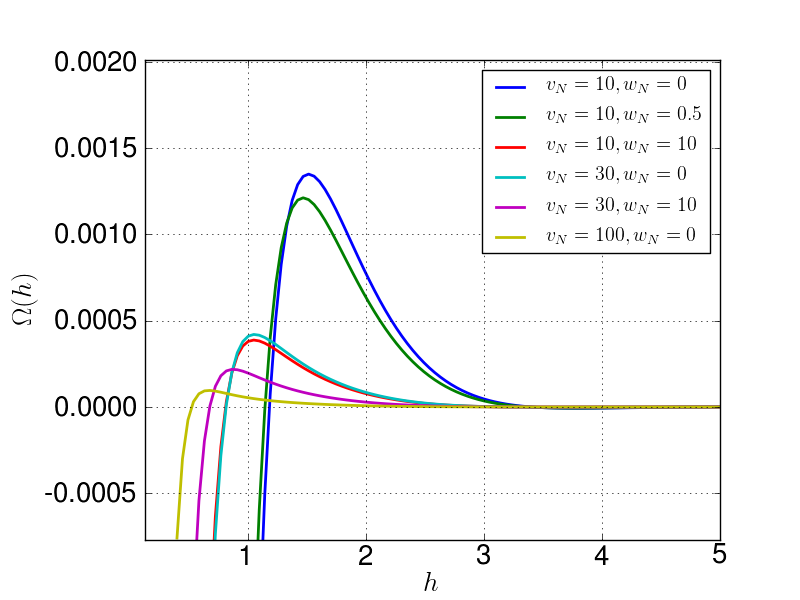}}
\caption{\small{The thermodynamic potential, Eq.(\ref{adsorption_free_energy_homo}), for different 
		values of the virial parameters, $v_N$, $w_N$. Fixed parameters: $A = 42.73$ and $N_x = 10k$, $N_s=10k$.}}
\label{adsorption_Omega8_h5_a42_7_vw_many_fig}
\end{minipage}
\end{figure}

\begin{figure}[ht!]
\begin{minipage}[ht]{0.5\linewidth}
\center{\includegraphics[width=1\linewidth]{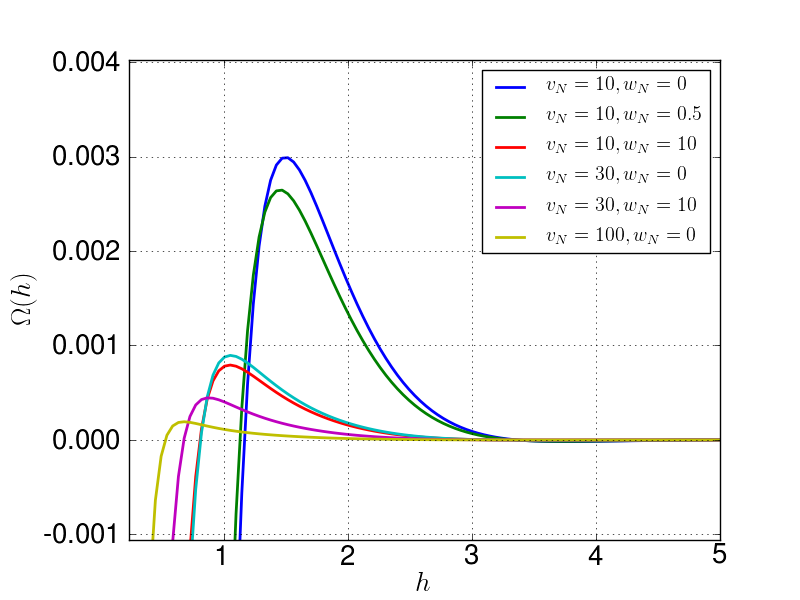}}
\caption{\small{The thermodynamic potential, Eq.(\ref{adsorption_free_energy_homo}), for different 
		values of the virial parameters, $v_N$, $w_N$. Fixed parameters: $A = 61.5$ and $N_x = 10k$, $N_s=10k$.}}
\label{adsorption_Omega8_h5_a61_5_vw_many_fig}
\end{minipage}
\hfill
\begin{minipage}[ht]{0.5\linewidth}
\center{\includegraphics[width=1\linewidth]{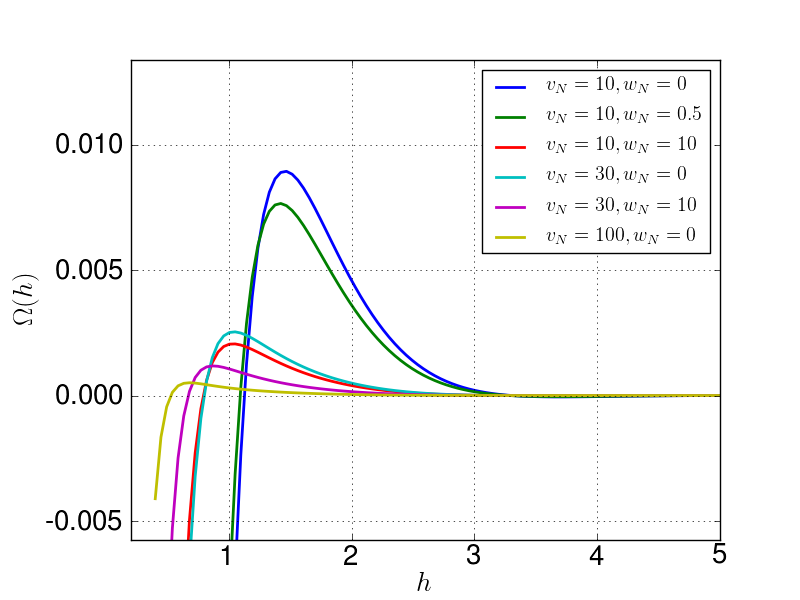}}
\caption{\small{The thermodynamic potential, Eq.(\ref{adsorption_free_energy_homo}), for different 
		values of the virial parameters, $v_N$, $w_N$. Fixed parameters: $A = 100$ and $N_x = 10k$, $N_s=10k$.}}
\label{adsorption_Omega8_h5_a100_vw_many_fig}
\end{minipage}
\end{figure}

\begin{figure}[ht!]
\begin{minipage}[ht]{0.5\linewidth}
\center{\includegraphics[width=1\linewidth]{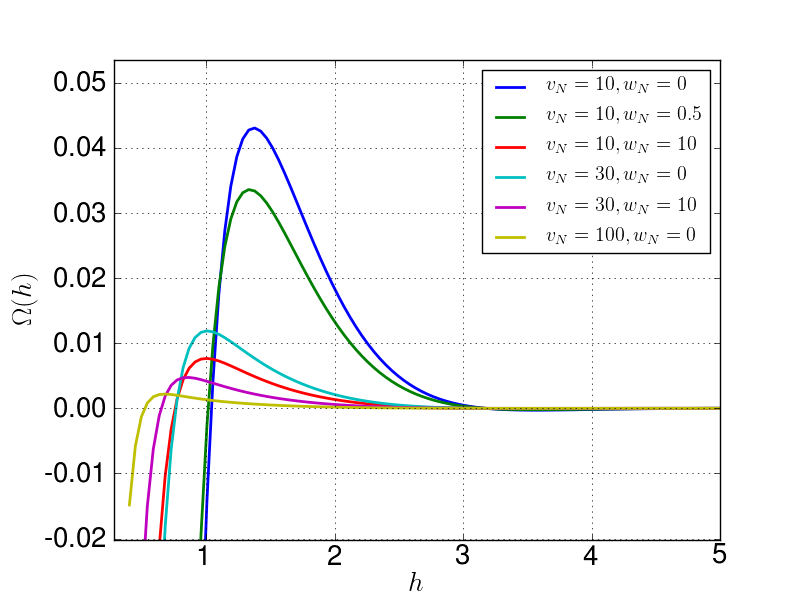}}
\caption{\small{The thermodynamic potential, Eq.(\ref{adsorption_free_energy_homo}), for different 
		values of the virial parameters parameters, $v_N$, $w_N$. Fixed parameters: $A = 200$ and $N_x = 10k$, $N_s=10k$.}}
\label{adsorption_Omega8_h5_a200_vw_many_fig}
\end{minipage}
\hfill
\begin{minipage}[ht]{0.5\linewidth}
\center{\includegraphics[width=1\linewidth]{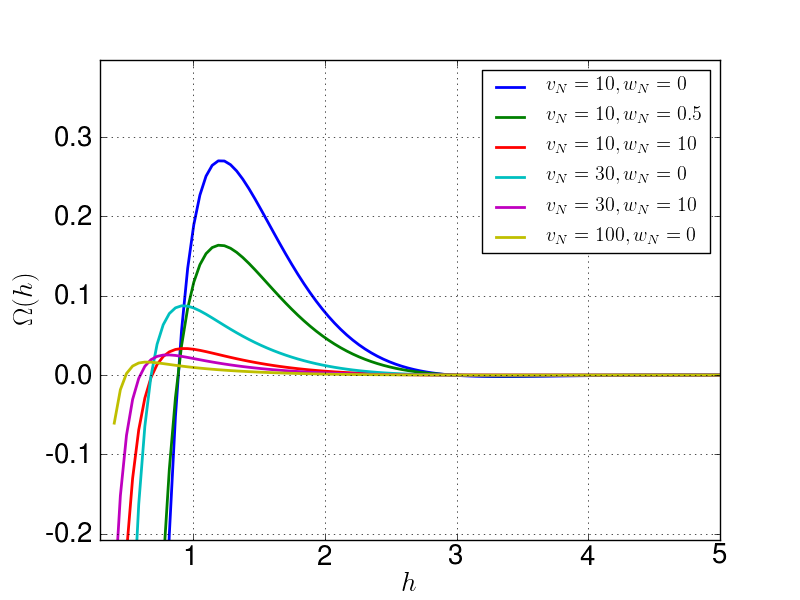}}
\caption{\small{The thermodynamic potential, Eq.(\ref{adsorption_free_energy_homo}), for different 
		values of the virial parameters parameters, $v_N$, $w_N$. Fixed parameters: $A = 500$ and $N_x = 10k$, $N_s=10k$.}}
\label{adsorption_Omega8_h5_a500_vw_many_fig}
\end{minipage}
\end{figure}

	  {\raggedright $\textbf{Different A}$.} We also compare the thermodynamic potentials for different virial parameters and adsorption strength. 	  
	  One can find the results of the calculation in Figs.\ref{adsorption_Omega8_h5_a14_1_vw_many_fig}--\ref{adsorption_Omega8_h5_a500_vw_many_fig}.	  
	  Comparing the thermodynamic potentials for the same virial parameters, one can see that the position of a barrier's maximum and 
	  the scope of the barrier does not change or changes very slightly when we vary the adsorption parameter. Despite that the magnitude 
	  of the barrier increases $1000$ times when we increase the adsorption strength from $A = 14.1$ up to $A = 500$. 	  
	  The numerical valus of the barrier's height $ \Delta\Omega^* = \Omega_{max} - \Omega_{inf}$ and its position, $h_m$ corresponding to all of the 
	  figures are indicated in Tab.\ref{tabular:adsorption_barrier_max_scft_vw}. 	  	  
%%%%%%%%%%%%%%%%%%%%%%%%%%%%%%%%%%%%%%%%%%%%%%%%%%%%%%%%%%%%%%%%%%%%%%%%%%%%%%%%%%%%%%%%%%%%%%%%%%%%%%%%%%%%%%%%%%%%%%%%%%%%%%%
%        Thermodynamic potentials: different A(b) and virial parameters: A = 14.1, A = 42.73
%%%%%%%%%%%%%%%%%%%%%%%%%%%%%%%%%%%%%%%%%%%%%%%%%%%%%%%%%%%%%%%%%%%%%%%%%%%%%%%%%%%%%%%%%%%%%%%%%%%%%%%%%%%%%%%%%%%%%%%%%%%%%%%
%%%%%%%%%%%%%%%%%%%%%%%%%%%%%%%%%%%%%%%%%%%%%%%%%%%%%%%%%%%%%%%%%%%%%%%%%%%%%%%%%%%%%%%%%%%%%%%%%%%%%%%%%%%%%%%%%%%%%%%%%%%%%%%
%        Thermodynamic potentials: different A(b) and virial parameters: A = 61.5, 100
%%%%%%%%%%%%%%%%%%%%%%%%%%%%%%%%%%%%%%%%%%%%%%%%%%%%%%%%%%%%%%%%%%%%%%%%%%%%%%%%%%%%%%%%%%%%%%%%%%%%%%%%%%%%%%%%%%%%%%%%%%%%%%%
%%%%%%%%%%%%%%%%%%%%%%%%%%%%%%%%%%%%%%%%%%%%%%%%%%%%%%%%%%%%%%%%%%%%%%%%%%%%%%%%%%%%%%%%%%%%%%%%%%%%%%%%%%%%%%%%%%%%%%%%%%%%%%%
%        Thermodynamic potentials: different A(b) and virial parameters: A = 200, 500
%%%%%%%%%%%%%%%%%%%%%%%%%%%%%%%%%%%%%%%%%%%%%%%%%%%%%%%%%%%%%%%%%%%%%%%%%%%%%%%%%%%%%%%%%%%%%%%%%%%%%%%%%%%%%%%%%%%%%%%%%%%%%%%

%%%%%%%%%%%%%%%%%%%%%%%%%%%%%%%%%%%%%%%%%%%%%%%%%%%%%%%%%%%%%%%%%%%%%%%%%%%%%%%%%%%%%%%%%%%%%%%%%%%%%%%%%%%%%%%%%%%%%%%%%%%%%%%%%%%%%%%%%%%%
% The height of the barrier for vw: adsorption 
%%%%%%%%%%%%%%%%%%%%%%%%%%%%%%%%%%%%%%%%%%%%%%%%%%%%%%%%%%%%%%%%%%%%%%%%%%%%%%%%%%%%%%%%%%%%%%%%%%%%%%%%%%%%%%%%%%%%%%%%%%%%%%%%%%%%%%%%%%%%
\begin{table}[ht!]
\caption{The value of the thermodynamic potential barrier height $\Delta\Omega^*$ and its position $h_m$ corresponding to 
	 Figs.\ref{adsorption_Omega8_h5_a14_1_vw_many_fig}--\ref{adsorption_Omega8_h5_a500_vw_many_fig}.} 
\label{tabular:adsorption_barrier_max_scft_vw} 
\begin{center}
  \begin{tabular}{ | c | c | c | c | c | c | c | }
    \hline
    $v_N$                           &    10&    10&    10&    30&    30&    100   \\ \hline
    $w_N$                           &     0&   0.5&    10&     0&    10&      0   \\ \hline
                                             \multicolumn{7}{|c|}{A=14.1}         \\ \hline
    $\Delta\Omega^*,\times 10^{-4}$ & 1.376& 1.265& 0.473& 0.485& 0.280&  0.144   \\ \hline
    $h_m$                           & 1.515& 1.469& 1.051& 1.051& 0.865&  0.632   \\ \hline
                                             \multicolumn{7}{|c|}{A=42.73}        \\ \hline
    $\Delta\Omega^*,\times 10^{-3}$ & 1.350& 1.213& 0.388&  0.42& 0.218&  0.096   \\ \hline
    $h_m$                           & 1.515& 1.469& 1.051& 1.051& 0.865&  0.679   \\ \hline
				              \multicolumn{7}{|c|}{A=61.5}         \\ \hline
    $\Delta\Omega^*,\times 10^{-3}$ & 2.988& 2.644& 0.793& 0.895& 0.446&  0.193   \\ \hline
    $h_m$                           & 1.515& 1.469& 1.051& 1.051& 0.865&  0.679   \\ \hline
                                             \multicolumn{7}{|c|}{A=100}          \\ \hline                                             
    $\Delta\Omega^*,\times 10^{-3}$ &  8.94& 7.658& 2.057& 2.538& 1.181&  0.509   \\ \hline
    $h_m$                           & 1.469& 1.422& 1.051& 1.051& 0.865&  0.678   \\ \hline
                                             \multicolumn{7}{|c|}{A=200}          \\ \hline                                             
    $\Delta\Omega^*,\times 10^{-2}$ & 4.304& 3.358& 0.766& 1.189& 0.474&  0.219   \\ \hline
    $h_m$                           & 1.376& 1.329& 1.004& 1.004& 0.864&  0.679   \\ \hline
                                             \multicolumn{7}{|c|}{A=500}          \\ \hline                                                                                          
    $\Delta\Omega^*,\times 10^{-1}$ & 2.701& 1.637& 0.334& 0.873& 0.254& 0.164    \\ \hline
    $h_m$                           &  1.19&  1.19& 0.911& 0.911& 0.818& 0.632    \\

    \hline
  \end{tabular}
\end{center} 
\end{table}
	    
	    Are there any restriction for the allowed adsorption strength? To clarify this point, let us list the concentration values at the surface.	  
	    For $A=500$ and $v_N=10, w_N=0$ the value of concentration at surfaces is $c(0)/c_b=18.3c_b$. It decreases when we 
	    increase the virial parameters: it becomes $c(0)/c_b=3.1c_b$ at $v_N=100, w_N=0$. Since we are interested 
	    in weak or moderate adsorption it is not useful to consider the adsorption strength $A>500$.
%%%%%%%%%%%%%%%%%%%%%%%%%%%%%%%%%%%%%%%%%%%%%%%%%%%%%%%%%%%%%%%%%%%%%%%%%%%%%%%%%%%%%%%%%%%%%%%%%%%%%%%%%%%%%%%%%%%%%%%%%%%%%%%%%%%%%%%%%%%%%%%%%%%%%%%%%%%%%
%          Repulsive surface potential i.e $A<0$.
%%%%%%%%%%%%%%%%%%%%%%%%%%%%%%%%%%%%%%%%%%%%%%%%%%%%%%%%%%%%%%%%%%%%%%%%%%%%%%%%%%%%%%%%%%%%%%%%%%%%%%%%%%%%%%%%%%%%%%%%%%%%%%%%%%%%%%%%%%%%%%%%%%%%%%%%%%%%%    	   
\section{The repulsive surface potential} 
\begin{figure}[ht!]
\begin{minipage}[ht]{0.5\linewidth}
\center{\includegraphics[width=1\linewidth]{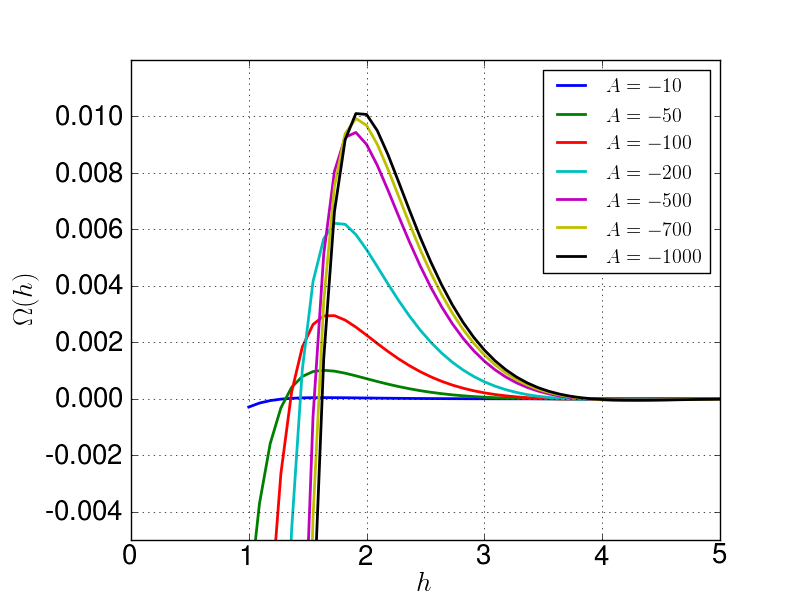}}
\caption{\small{The thermodynamic potential, Eq.(\ref{adsorption_free_energy_homo}), for different adsorption strength $A$ in order to find the saturation value. 
		Fixed parameters: $N_x=7k$, $N_s=3k$ and $v_N=10$, $w_N=10$.}}
\label{adsorption_Omega8_h5_an_v10_fig}
\end{minipage}
\hfill
\begin{minipage}[ht]{0.5\linewidth}
\center{\includegraphics[width=1\linewidth]{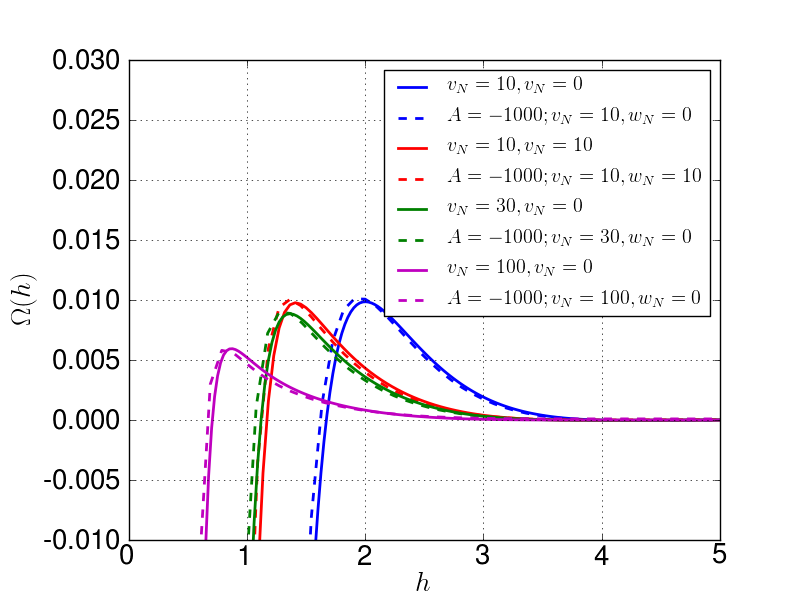}}
\caption{\small{The comparison between thermodynamic potential calculated for purely repulsive walls (continuous curves) and the thermodynamic potential calculated 
		for the external surface potential (dashed curves), for different values of the virial parameters $v_N$, $w_N$. 
		Fixed parameters: $A=-1000.1$ and $N_x=7k, N_s=3k$.}}
\label{adsorption_Omega8_h5_an_v_comparison_fig}
\end{minipage}
\end{figure}

	    In this section we compare thermodynamic potentials obtained, before, for a purely repulsive walls with the thermodynamic potential calculated  
	    with repulsive adsorption potential, i.e. $A<0$.
	    At the beginning we should find the value of the adsorption strength at which the thermodynamic potential reaches the saturation and 
	    then does not change. 
	    In Fig.\ref{adsorption_Omega8_h5_an_v10_fig} we show the searching process and found out that the saturation value $A=-1000$ is sufficient. 
	    Next, using the value of the adsorption strength $A=-1000$ we calculate the thermodynamic potentials for different values of the virial parameters.	    
            In Fig.\ref{adsorption_Omega8_h5_an_v_comparison_fig} we present the comparison with the hard wall case.
            One can see that the results obtained by different approaches for the same value of the virial parameters are similar.
%%%%%%%%%%%%%%%%%%%%%%%%%%%%%%%%%%%%%%%%%%%%%%%%%%%%%%%%%%%%%%%%%%%%%%%%%%%%%%%%%%%%%%%%%%%%%%%%%%%%%%%%%%%%%%%%%%%%%%%%%%%%%%%
%        Thermodynamic potentials negative A
%%%%%%%%%%%%%%%%%%%%%%%%%%%%%%%%%%%%%%%%%%%%%%%%%%%%%%%%%%%%%%%%%%%%%%%%%%%%%%%%%%%%%%%%%%%%%%%%%%%%%%%%%%%%%%%%%%%%%%%%%%%%%%%

%%%%%%%%%%%%%%%%%%%%%%%%%%%%%%%%%%%%%%%%%%%%%%%%%%%%%%%%%%%%%%%%%%%%%%%%%%%%%%%%%%%%%%%%%%%%%%%%%%%%%%%%%%%%%%%%%%%%%%%%%%%%%%%%%%%%%%%%%%%%%%%%%%%%%%%%%%%%%
\section{GSDE theory. Reversible adsorption} 
In this section, we extend the GSDE theory for adsorption and then compare it with corresponding numerical SCFT results.
Below, as before, we consider polymer solution in a gap between two flat plates, with reversibly adsorbed chains, i.e. 
when the polymers in the gap are in equilibrium with the bulk polymers.		
Following the GSDE theory \cite{semenov_1996, semenov_2008}, we can write the free energy in the form
\begin{equation}
\label{ans_adsorption_free_en_tot}
	W = W_{gs} + W_e
\end{equation}
	where $W_{gs}$ is the GSD free energy corresponding to attraction. We can write it as 
\begin{equation} 		
\label{ans_adsorption_free_en_gs}
	W_{gs} = \int\limits_h^{\infty}\mathrm{d}h f_{gs}(h) 
\end{equation}
where $f_{gs}$ is the GSD force. We have already shown using the deGennes assumption \cite{deGennes_1981, deGennes_1982}, that the GSD force is 
$$
	f_{gs}(h) = \frac{\text{v}c_b^2}{2}\left(1-(c_m/c_b)^2\right) + \frac{\text{w}c_b^3}{6}\left(1-(c_m/c_b)^3\right)-\mu(c_b)c_b(1-c_m/c_b)
$$
	or in more appropriate form
\begin{equation}
\label{ans_adsorption_force_gs}
	f_{gs}(h) = -\frac{c_b}{2}(c_m/c_b - 1)^2\left(\text{v}c_b + \text{w}c_b^2 + \frac{\text{w}c_b^2}{3}(c_m/c_b - 1)\right)
\end{equation}
	where $\text{v}, \text{w}$ are second and third virial coefficients, $c_b$ is bulk monomer concentration, $c_m = c(h/2)$ is monomer concentration at midplane.
	Another term in Eq.(\ref{ans_adsorption_free_en_tot}) is the contribution to the free energy due to finite molecular weight of polymers (end-segment effect) 
	which corresponds to repulsion. Following \cite{semenov_1996, semenov_2008}  we can write it as
\begin{equation}
\label{ans_adsorption_free_en_end}
	W_e = \frac{4c_b}{NR_g}\Delta_e^2 u_{int}(h/R_g)
\end{equation}
where 
\begin{equation}
\label{ans_adsorption_delta_def}
	\Delta_e = \int\limits_0^{\infty}\mathrm{d}x \left(\psi_1((x))/\psi_b - (\psi_1(x)/\psi_b)^2\right) = 
		    \int\limits_0^{\infty}\mathrm{d}x \left(\sqrt{c_1(x)/c_b} - c_1(x)/c_b \right)
\end{equation}
	$|\Delta_e|$ is an effective length\footnote{In adsorption case, $c(x)\geqslant c_b$ near the wall, so that $\Delta_e < 0$.} 
	related to the single wall excess of end points for the case of semi-infinite system with single wall at $x=0$, 
$$
	u_{int}(r) = \frac{1}{r}\sum\limits_{n=-\infty}^{\infty}f(4\pi^2n^2/r^2) - \kappa
$$
	with
$$
	f(u) = \frac{1 - e^{-u}(u + 1)}{u - 1 + e^{-u}}
$$
	and 
$$
	\kappa \equiv \int\limits_{-\infty}^{\infty}\frac{\mathrm{d}k}{2\pi}f(k^2) \simeq 0.6161874
$$	    	 
	In Eq.(\ref{ans_adsorption_delta_def}) we used the fact that the GSD distribution function $\psi$ is related to the concentration profile as $c(x) = \psi^2(x)$. 
	This function can be found via solution of Edwards equation written in the GSD approximation:
\begin{equation}
\label{ans_adsorption_edwards_eq}
            a^2\frac{\mathrm{d^2}\psi}{\mathrm{d}x^2} - (\phi_{self} - \mu_{int})\psi = 0
\end{equation}
        where $a_s = \sqrt{6}a$ is a polymer statistical segment, $\phi_{self} = \text{v}c + \text{w}c^2/2$ is a self-consistent mean field, 
        $\mu_{int} = \text{v}c_b + \text{w}c_b^2/2$ is the interaction part of the chemical potential. 
        The boundary condition for Eq.(\ref{ans_adsorption_edwards_eq}) on the plate is $(\psi'/\psi)_{x=0} = -1/b$,  
        where $b$ is the extrapolation length that defines adsorption strength. 
	Another boundary condition by virtue of symmetry is $\mathrm{d}\psi/\mathrm{d}x = 0$ at $x = h/2$. 
	Multiplying Eq.(\ref{ans_adsorption_edwards_eq}) by $\frac{\mathrm{d}\psi}{\mathrm{d}x}$ and integrating it we obtain the following first integral
\begin{equation}
\label{ans_adsorption_first_integral_general}
        a^2\left(\frac{\mathrm{d}\psi}{\mathrm{d}x}\right)^2 = - \mu\psi^2 + \frac{\text{v}}{2}\psi^4 + \frac{\text{w}}{6}\psi^6 + A 
\end{equation}
	In this section we restrict ourselves only by the reversible adsorption. We start with the case of a weak adsorption strength 
	resulting in a weak concentration perturbation.	
	In addition, the main interest for us is related to the free energy barrier that occurs at large enough separations.
	Therefore, we can assume that the concentration perturbation at midplane, for such separations, is a sum of independent contributions from two plates, i.e.
	$(c_m-c_b) \simeq 2\delta c_1(h/2)$, where $\delta c_1(x) = c_1(x)-c_b$ is an excess concentration caused by only one plate. 
	Thus, we can find only one solution of the Eq.(\ref{ans_adsorption_first_integral_general}) corresponding to concentration profile caused by a single
	weakly adsorbing plate. We will use the result in both terms of Eq.(\ref{ans_adsorption_free_en_tot}).
	To define the constant $A$, one may notice that $c_1(x\rightarrow\infty) = c_b$. Thus
$$
	A = \mu c_b - \frac{\text{v}}{2}c_b^2 - \frac{\text{w}}{6}c_b^3 = \frac{\text{v}}{2}c_b^2 + \frac{2\text{w}}{3}c_b^3
$$	    
	For simplicity, consider separately two different cases: firstly $w_N = 0$ and then the case with $w_N\ne0$.

        \textbf{Special case: $w_N=0$}. In this part our purpose is to obtain expressions for one plate concentration profile and GSD free energy for a case when
        the third virial parameter $w_N=0$.
        As we have already seen the first integral to the Edwards equation in GSD approximation is
$$
	a^2\left(\frac{\mathrm{d}\psi_1}{\mathrm{d}x}\right)^2 = - \mu_{int}\psi_1^2 + \frac{\text{v}}{2}\psi_1^4 + \frac{\text{w}}{6}\psi_1^6 + A 
$$
	where 
$$
	\mu_{int} = \text{v}c_b \quad \text{and}\quad A = \text{v}c_b^2/2
$$
	thereby, after substitution $\psi_1=\sqrt{c_b}f$ we can rewrite it as      
$$
	a^2\left(\frac{\mathrm{d}f}{\mathrm{d}x}\right)^2 = \frac{\text{v}c_b}{2}(f^4-2f^2+1) = \frac{\text{v}c_b}{2}(f^2-1)^2
$$
	Taking the square root of the both parts of the expression we get
$$
	\frac{\mathrm{d}f}{\mathrm{d}x} = -\sqrt{\frac{\text{v}c_b}{2a^2}}(f^2-1)
$$
	Suppose we know the value of the function on the wall, i.e. at $x=0$, and denote it as $f_0$. Note that the value is a maximum value for the function $f$. 
	Thus, after integration we have 
$$
	-\int\limits_{f_0}^f\frac{\mathrm{d}f}{f^2-1} = \sqrt{\frac{\text{v}c_b}{2a^2}}\int\limits_{0}^x\mathrm{d}x = \frac{1}{2}\sqrt{\frac{2\text{v}c_b}{a^2}}x = \frac{x}{2\xi}
$$
	where as before the bulk correlation length is $\xi = a/\sqrt{2\text{v}c_b}$. 
	The integral in the r.h.s of the above equation can be easily calculated if we replace the integration variable as $f = \coth(y)$. The result is
$$
	{\rm arccoth}(f) = \frac{x}{2\xi} + {\rm arccoth}(f_0) = \frac{x+x_0}{2\xi}
$$
	Inverting the expression, we obtain
$$
	f(x) = \coth\left(\frac{x+x_0}{2\xi}\right), \quad f(0) \equiv f_0 = \coth\left(\frac{x_0}{2\xi}\right)
$$
	where we introduced $x_0 = 2\xi{\rm arccoth}(f_0)$. Now try to satisfy the boundary condition and find relationship between extrapolation length $b$ and $x_0$:
$$
	\frac{f'}{f}\Bigg\vert_{x=0} = -\frac{1}{2\xi\sinh^2(\frac{x_0}{2\xi})\coth(\frac{x_0}{2\xi})} = -\frac{\coth^2(\frac{x_0}{2\xi}) - 1}{2\xi\coth(\frac{x_0}{2\xi})} = 
					-\frac{1}{b}
$$
	or in terms of $f_0$, we can write the relation as
$$
	\frac{f_0^2 - 1}{f_0} = \frac{2\xi}{b}
$$
	Let us consider the case $f_0\gtrsim 1$ when $c_1(0)$ just slightly differ from $c_b$.
	This case correspods to weak adsorption and it requires $\xi/b\ll1$. The last condition is the condition for weak adsorption regime ($WAR$). 
	Solving the above equation we get two solutions where a physical meaning has only the solution with sign $"+"$, which 
	corresponds to an attractive plate ($f_0\geqslant 1$): 
\begin{equation}
\label{ans_adsorption_gs_f0}
	f_0 = \frac{\xi}{b}+\sqrt{1+\left(\frac{\xi}{b}\right)^2}
\end{equation}	
	Therefore, we can write expression for $x_0$ with the explicit dependence from the adsorption parameter $b$:
\begin{equation}
\label{ans_adsorption_gs_x0}
	x_0 = 2\xi{\rm arccoth}\left(\frac{\xi}{b}+\sqrt{1+\left(\frac{\xi}{b}\right)^2}\right)
\end{equation}
	Correspondingly, the concentration perturbation caused by one adsorbed plate is
\begin{equation}
\label{ans_adsorption_gs_concentration_v}
	c_1(x) = c_bf^2(x) = c_b\coth^2\left(\frac{x+x_0}{2\xi}\right)
\end{equation}
	At $x\gg\xi$ it has the asymptotics 
$$
	c_1(x) - c_b = c_b\left(\coth^2\left(\frac{x+x_0}{2\xi}\right) - 1\right) = \frac{c_b}{\sinh^2\left(\frac{x+x_0}{2\xi}\right)} \simeq 4c_b e^{-(x+x_0)/\xi}    
$$
	In order to get the concentration profile created by two plates we should multiply the last result by $2$ and change $x\rightarrow h/2$. 
	Thus the concentration perturbation for midplane between two plates separated by $h$ is equal to
$$
	c_m - c_b \simeq 8c_b e^{-x_0/\xi}e^{-h/2\xi}     
$$
        This expression is valid for $h\gg\xi$. Based on that, the force between two plates, Eq.(\ref{ans_adsorption_force_gs}), is
$$
	f_{gs}(h) \simeq -\frac{\text{v}c_b^2}{2}\left(c_m/c_b - 1\right)^2 \simeq -32\text{v}c_b^2e^{-2x_0/\xi}e^{-h/\xi}
$$
	Integrating it over $h$ we obtain the ground state expression for the free energy of the system between plates
$$
	W_{gs}(h) \simeq \int\limits_h^{\infty}\mathrm{d}h f_{gs}(h/\xi) = -32\text{v}c^2_b \xi e^{-2x_0/\xi}e^{-h/\xi} 
$$
	or in dimensionless variables($\bar{h}=h/R_g, \bar{\xi}=\xi/R_g, v_N=\text{v}c_bN$)
\begin{equation}
\label{ans_adsorption_free_en_gs_v}
	\hat{W}_{gs}(h) = \frac{N}{R_gc_b}W_{gs} \simeq -32v_N\bar{\xi}e^{-2x_0/\xi}e^{-h/\xi}
\end{equation}
	where $\bar{\xi}=1/\sqrt{2v_N}$ is the dimensionless correlation length. Consider now the general case. 

	\textbf{General case: $w_N\ne0$}. We now find the expression for the concentration profile and free energy caused by one adsorbed plate placed at $x=0$. 
	After changing the variable $\psi_1(x) = \sqrt{c_b}f(x)$ we can write Eq.(\ref{ans_adsorption_first_integral_general}) as
$$
	a^2c_b\left(\frac{\mathrm{d}f}{\mathrm{d}x}\right)^2 = \frac{\text{v}c_b^2}{2}(f^4-2f^2+1) + \frac{\text{w}c_b^3}{6}(f^6-3f^2+2) =
        \frac{\text{v}c_b^2}{2}(f^2-1)^2 + \frac{\text{w}c_b^3}{6}(f^2+2)(f^2-1)^2 
$$
	or after simple transformations 
\begin{equation}
\label{ans_adsorption_first_integral_vw}
	a^2\left(\frac{\mathrm{d}f}{\mathrm{d}x}\right)^2 = \frac{c_b}{2}(f^2-1)^2\left(\text{v} + \frac{\text{w}c_b}{3}(f^2+2)\right)
\end{equation}
	Extracting the square root from both sides of the last expression we get:
$$
	\frac{\mathrm{d}f}{\mathrm{d}x} = -\sqrt{\frac{c_b}{2a^2}}(f^2-1)\sqrt{\left(\text{v} + \frac{\text{w}c_b}{3}(f^2+2)\right)}
$$	
	Suppose again that we know the value of the function on the wall, i.e. $f_0 = f(x=0)$, which serves as a 
	boundary condition at the wall.
	Thereby we can integrate the above expression:
$$
	\frac{x\sqrt{c_b}}{\sqrt{2}a}= -\int\limits_{f_0}^f\frac{\mathrm{d}f}{(f^2-1)\sqrt{\text{v} + \frac{\text{w}c_b}{3}(f^2+2)}} \quad \Bigg\vert 
	\times \frac{a}{\sqrt{2c_b}\xi}
$$
	where $\xi$ is the bulk correlation length. For convenience we use the reduced virial parameters $v_N=\text{v}c_bN, w_N=\text{w}c_b^2N/2$ 
	and dimensionless correlation length $\bar{\xi} = \xi/R_g = 1/\sqrt{2(v_N+2w_N)}$ where $R_g=aN^{1/2}$ is the radius of gyration of the ideal Gaussian chain.
	So, in dimensionless variables we can rewrite the last expression as
\begin{equation}
\label{ans_adsorption_differentials}
	\frac{x}{2\xi}= \frac{1}{\bar{\xi}}\int\limits_{f}^{f_0}\frac{\mathrm{d}f}{(f^2-1)\sqrt{2\left(v_N + \frac{2w_N}{3}(f^2+2)\right)}} 	
\end{equation}
	One can notice that far away from the plate, when $f\rightarrow 1$, it leads to singularity in the integrand. We should treat the point carefully.
	Moreover, since we consider weak adsorption we have the following condition $f_0-f\ll1$. As we did it for the purely repulsive case, notice that
	the second multiplier in the integrand at $f=1$ coincides with the bulk correlation length.
	Due to that we make a simple transformation, namely, add and subtract under the integral the following function
$$
	y(f) = \frac{1}{f^2-1}
$$
        So that
$$
	\frac{x}{2\xi}= \int\limits_f^{f_0} \frac{\mathrm{d}f}{(f^2-1)}\left\{\frac{1}{\bar{\xi}\sqrt{2\left(v_N + \frac{2w_N}{3}(f^2+2)\right)}} - 1\right\}+ 
	\int\limits_f^{f_0}\frac{\mathrm{d}f}{(f^2-1)} = I_1 + I_2
$$	
	Consider these integrals separately.\\
	$I_1$) Rewrite this integral using explicit value of the dimensionless correlation length:
$$
	I_1 = \int\limits_f^{f_0}\frac{\mathrm{d}f}{(f^2-1)}\left\{\sqrt{\frac{2(v_N + 2w_N)}{2(v_N + \frac{2w_N}{3}(f^2+2))}}-1\right\} = 
	      \int\limits_f^{f_0}\frac{\mathrm{d}f}{(f^2-1)}\left\{\sqrt{\frac{1 + v_N/2w_N}{\frac{1}{3}(f^2+2) + v_N/2w_N}}-1\right\}
$$
       It can be noticed that the integrand is smooth, slowly varying function. Let us denote the integrand as $y(f, v_N/w_N)$. The function  $y(f, v_N/w_N) < 0$  
       and it shows its mininum  at $|y(1, 0)|=0.166$. Using the mean value theorem we can write $|I_1|=|y(f^*)|(f_0-f) \leqslant |y(1, 0)|(f_0-f)$ 
       where $f^*\in[f..f_0]$. 
       Since we consider weak adsorption $f_0-f\ll1$,  $|I_1|\ll1$, so in the WAR we neglect the $I_1$-term.
       Outside the WAR we must keep this contribution and set $f=1$ because we are interested in the concentration profile for $x\gg 2\xi$. Therefore
\begin{equation}
\label{ans_adsorption_i1_general}
 	I_1 =  \int\limits_1^{f_0}\frac{\mathrm{d}f}{(f^2-1)}\left\{\sqrt{\frac{1 + v_N/2w_N}{\frac{1}{3}(f^2+2) + v_N/2w_N}}-1\right\}	
\end{equation}
       For further analysis, the integral can be treated numerically.\\
       $I_2$) The second integral:
$$
	I_2 = \int\limits_f^{f_0}\frac{\mathrm{d}f}{(f^2-1)}
$$	
	was already treated before:
$$
	I_2 = {\rm arccoth}(f) - {\rm arccoth}(f_0)
$$
       Joining together those expressions, we obtain:
$$
	{\rm arccoth}(f) = \frac{x}{2\xi} + {\rm arccoth}(f_0) - I_1 = \frac{x+x_0}{2\xi} - I_1 \simeq \frac{x+x_1}{2\xi}
$$
       where, as before, we introduced $x_0 = 2\xi{\rm arccoth}(f_0)$ and $x_1 = x_0 - 2\xi I_1$. 
       As we have already seen $I_1$ is very small quantity, so in the WAR we will neglect it and write $x_1 \simeq x_0$. 
       Inverting the above expression we get
\begin{equation}
\label{ans_adsorption_conc_profile_general_x_gg_xi}
       f(x) = \coth\left(\frac{x+x_1}{2\xi}\right) 
\end{equation}
       Note that the expression is valid for $x\gg\xi$.
       Thereby we obtained the same expression as in the special case, but with more general expression for the correlation length, $\xi$ and additional 
       factor $x_1$. Since the above expression is valid for $x\gg\xi$ we can not use it for $x=0$. In order to obtain $f_0$  
       we should consider again the first integral of the Edwards equation, Eq.(\ref{ans_adsorption_first_integral_vw}):
$$
       a^2\left(\frac{\mathrm{d}f}{\mathrm{d}x}\right)^2 = \frac{c_b}{2}(f^2-1)^2\left(\text{v} + \frac{\text{w}c_b}{3}(f^2+2)\right)
$$
       For the contact point, $x=0$,using the  boundary condition, $(f'/f)_{x=0} = -1/b$, we then rewrite it as
$$
       \frac{a^2}{b^2}f^2_0 = \frac{c_b}{2}(f^2_0-1)^2\left(\text{v} + \frac{\text{w}c_b}{3}(f^2_0+2)\right)
$$	
       where as before $f_0^2 = c_0/c_b$. Introducing new variable, $g_0=f^2_0$, and then multiplying both sides of the above equation 
       by $N$ we can rewrite it in our reduced variables:
$$
       \frac{\bar{b}^2}{2}(g_0-1)^2\left(v_N + \frac{2w_N}{3}(g_0+2)\right) - g_0 = \frac{\bar{b}^2}{2}(g_0-1)^2\left(v_N + 2w_N + \frac{2w_N}{3}(g_0-1)\right) - g_0 = 0
$$	
       or in more suitable form: 
\begin{equation}
\label{ans_adsorption_boundary_cubic_eq}
         \frac{\bar{b}^2}{4\bar{\xi}^2}(g_0-1)^2\left(1 + \frac{4\bar{\xi}^2w_N}{3}(g_0-1)\right) - g_0 = 0
\end{equation}
	Solving this cubic equation, for the certain input parameters, $v_N, w_N$ and $\bar{b}$ we obtain the corresponding contact concentration:
\begin{equation}
\label{ans_adsorption_boundary_eq_solution}
       g_0 = G(v_N, w_N, \bar{b}) \quad \text{and}\quad f_0 = \sqrt{g_0}
\end{equation}	
	For example, we can solve Eq.(\ref{ans_adsorption_boundary_cubic_eq}) by the bisection method seeking the corresponding 
	root for $g_0$ in the range $g_0\in[1..500]$. After that, we can reproduce $x_0 = 2\xi{\rm arccoth}(f_0)$ and $x_1 = x_0 - 2\xi I_1$ 
	using the profile Eq.(\ref{ans_adsorption_conc_profile_general_x_gg_xi}).
        Therefore, the concentration perturbation created by one plate is
\begin{equation}
\label{ans_adsorption_gs_concentration_vw}
       c_1(x) = c_bf^2(x) = c_b\coth^2\left(\frac{x+x_1}{2\xi}\right)
\end{equation}
       At $x\gg\xi$ we can simplify showing the asymptotic behavior:
$$
       c_1(x) - c_b = c_b\left(\coth^2\left(\frac{x+x_1}{2\xi}\right) - 1\right) = \frac{c_b}{\sinh^2\left(\frac{x+x_1}{2\xi}\right)} \simeq 4c_b e^{-(x+x_1)/\xi}    
$$
       Thus, the concentration perturbation for midplane between two plates separated by $h$ is equal to
$$
       c_m - c_b \simeq 8c_b e^{-x_1/\xi}e^{-h/2\xi}     
$$
       The last expression is valid for $h\gg\xi$. 
       Since we are interested in the behavior of the system at $h\gg\xi$, where $c_m \to c_b$, we can neglect the term propotional to $(c_m/c_b-1)^3$ 
       in Eq.(\ref{ans_adsorption_force_gs}). Therefore, substituting the expression for the excess concentration in Eq.(\ref{ans_adsorption_force_gs}), 
       we obtain the GSD expression for the force between plates in polymer solution: 
$$
       f_{gs}(h) \simeq -32c_b e^{-2x_1/\xi} e^{-h/\xi}\left(\text{v}c_b + \text{w}c_b^2\right) = -\frac{16c_b}{N\bar{\xi}^2}e^{-2x_1/\xi} e^{-h/\xi}
$$
       Integrating it over $h$, we obtain the ground state expression for the free energy of the system between plates:
$$
       W_{gs}(h) \simeq \int\limits_h^{\infty}\mathrm{d}h f_{gs}(h/\xi) \simeq -\frac{16c_bR_g}{N\bar{\xi}}e^{-2x_1/\xi} e^{-h/\xi}
$$
       or, in dimensionless variables:
\begin{equation}
\label{ans_adsorption_free_en_gs_vw}  
       \hat{W}_{gs}(h) = \frac{N}{R_gc_b}W_{gs} \simeq -\frac{16}{\bar{\xi}}e^{-2x_1/\xi} e^{-h/\xi}
\end{equation}
       where, as before, $\bar{\xi} = 1/\sqrt{2(v_N+2w_N)}$ is the dimensionless correlation length. 
       Notice that the above equation in the WAR coincides with Eq.(\ref{ans_adsorption_free_en_gs_v}) that was valid for $w_N=0$.

	\textbf{Repulsive term}. We have just shown that the concentration profile caused by one adsorbed plate is given in Eq.(\ref{ans_adsorption_gs_concentration_vw}). 
	In the general case the expression is valid only for $h\gg\xi$, but in the WAR (or in case when $w_N=0$), we can neglect the term related to $I_1$, i.e.
	use Eq.(\ref{ans_adsorption_gs_concentration_v}). Let us consider the repulsive term in this case and the general case independently.\\
	1) $w_N=0$ or the WAR. Due to the defenition we can rewrite Eq.(\ref{ans_adsorption_delta_def}) as
$$
	\bar{\Delta}_e = \int\limits_0^{\infty}\mathrm{d}\bar{x} \left(\sqrt{c(\bar{x})/c_b} - c(\bar{x})/c_b\right) = 
		          \int\limits_0^{\infty}\mathrm{d}\bar{x} \left(f - f^2\right) = 
			  \int\limits_0^{\infty}\mathrm{d}\bar{x} \left(f - 1\right) - \int\limits_0^{\infty}\mathrm{d}\bar{x} \left(f^2 - 1\right) = J_1 - J_2
$$
	Consider those integrals separately. Using the expression for the concentration profile, Eq.(\ref{ans_adsorption_gs_concentration_v}),
	in the first integral, we get: 	
$$
\begin{array}{c}
	J_1 = \int\limits_0^{\infty}\mathrm{d}\bar{x} \left(f - 1\right) = \int\limits_0^{\infty}\mathrm{d}x \left(\coth\left(\frac{x+x_0}{2\xi}\right) - 1\right) = 
	      2\xi\int\limits_{x_0/2\xi}^{\infty}\mathrm{d}y \left(\coth(y) - 1\right) = \\
	    = 2\bar{\xi}\left\{\frac{x_0}{2\xi} - \log\left(2\sinh\left(\frac{x_0}{2\xi}\right)\right)\right\}
\end{array}
$$	     
	Correspondingly, for the second integral we have
$$
\begin{array}{c}
 	J_2 = \int\limits_0^{\infty}\mathrm{d}\bar{x} \left(f^2 - 1\right) = \int\limits_0^{\infty}\mathrm{d}x \left(\coth^2\left(\frac{x+x_0}{2\xi}\right) - 1\right) = 
	      2\xi\int\limits_{x_0/2\xi}^{\infty}\mathrm{d}y \left(\coth^2(y) - 1\right) = \\
	    =  2\bar{\xi}\left\{\coth\left(\frac{x_0}{2\xi}\right) - 1\right\} 
\end{array}
$$	     	
        Thus  
\begin{equation}
\label{ans_adsorption_delta_calc_war}
       \bar{\Delta}_e = J_1 - J_2 = 2\xi\left(1 + \frac{x_0}{2\xi} - \coth\left(\frac{x_0}{2\xi}\right) - \ln\left(2\sinh\left(\frac{x_0}{2\xi}\right)\right)\right) 
\end{equation}
        Note that the expression is valid either for $w_N=0$ or in the WAR.\\
        2) The general case. Now we do not have any restrictions for the virial parameters and adsorption strength. 
        For convenience we will work with dimensionless parameter $\bar{\Delta}_e = \Delta_e/R_g$. 
        We split the expression for $\Delta_e$ in two independent integrals:
$$
	\bar{\Delta}_e = \int\limits_0^{\infty}\mathrm{d}\bar{x} (f - f^2) = \int\limits_0^{\infty}\mathrm{d}\bar{x} (f - 1) - \int\limits_0^{\infty}\mathrm{d}\bar{x} (f^2 - 1) = J_1 - J_2
$$
        Let us change in both integrals the integration variable. Based on Eq.(\ref{ans_adsorption_differentials}), we can write
$$
	\mathrm{d}\bar{x} = -\frac{2\mathrm{d}f}{(f^2 - 1)\sqrt{2(v_N+2w_N(f^2+2)/3)}}
$$
        where $\bar{x} = x/R_g$. The minus sign is important here, since for adsorption case when we increase $x$, the function $f$ always decreases, 
        i.e. $\mathrm{d}f<0$. As before, let us consider those integrals independently. \\
        $J_1$) After simple transformations, we get
$$
	J_1 = \int\limits_0^{\infty}\mathrm{d}\bar{x} (f - 1) = -\int\limits_{f_0}^{1}\frac{2\mathrm{d}f}{(1+f)\sqrt{2(v_N+2w_N(f^2+2)/3)}} = 
	      \sqrt{\frac{3}{w_N}}\int\limits_{1}^{f_0}\frac{\mathrm{d}f}{(1+f)\sqrt{\frac{3v_N}{2w_N} + 2 + f^2}}
$$
	 where we have changed the sign of the integral by the inversion of the integration limits.
	 It is convenient to introduce the variable $\alpha = 2 + 3v_N/2w_N$. Therefore, we get the tabulated integral:
$$
	 \int\limits\frac{\mathrm{d}f}{(1+f)\sqrt{\alpha + f^2}} = \frac{1}{\sqrt{1+\alpha}}\left\{\ln(1+f) - \ln(\sqrt{(\alpha + 1)(\alpha + f^2)} + \alpha - f)\right\} + const
$$
	 Finally, we can write the expression for the first integral:
$$
	 J_1 =  \sqrt{\frac{3}{w_N(1 + \alpha)}}\left\{\ln\alpha + \ln(1+f_0) - \ln(\sqrt{(\alpha + 1)(\alpha + f_0^2)} + \alpha - f_0)\right\}
$$
	We not treat the second integral.\\
        $J_2$)  After the same transformations, we get:
$$
	J_2 = \int\limits_0^{\infty}\mathrm{d}\bar{x} (f^2 - 1) = -\int\limits_{f_0}^{1}\frac{2\mathrm{d}f}{\sqrt{2(v_N+2w_N(f^2+2)/3)}} = 
	      \sqrt{\frac{3}{w_N}}\int\limits_{1}^{f_0} \frac{\mathrm{d}f}{\sqrt{\frac{3v_N}{2w_N} + 2 + f^2}}
$$
	As before, this integral is also tabulated:
$$
        \int\limits\frac{\mathrm{d}f}{\sqrt{\alpha + f^2}} = \ln(\sqrt{\alpha + f^2} + f) + const
$$
	and finally for the second integral we write the expression:
$$
	J_2 = \sqrt{\frac{3}{w_N}}\left\{\ln(f_0 + \sqrt{\alpha + f_0^2 }) - \ln(1 + \sqrt{\alpha + 1})\right\}
$$
	Joining together the expression for the integrals, we can write the analytical expression for the repulsive part of the thermodynamic potential 
	in the general case: 
\begin{equation}
\label{ans_adsorption_delta_calc_general}
	\bar{\Delta}_e = J_1 - J_2 = \sqrt{\frac{3}{w_N}}J(f_0, \alpha)
\end{equation}
	where we introduced the function:
\begin{equation}
\label{ans_adsorption_delta_fun}
\begin{array}{rr}	
	J(f_0, \alpha) \equiv \frac{1}{\sqrt{1 + \alpha}}\left\{\ln\alpha + \ln(1+f_0) - \ln(\sqrt{(\alpha + 1)(\alpha + f_0^2)} + \alpha - f_0)\right\} + & \\
	+ \ln(1 + \sqrt{\alpha + 1}) - \ln(f_0 + \sqrt{\alpha + f_0^2 })  & 
\end{array}
\end{equation}
	where $\alpha = 2 + 3v_N/2w_N$ and $f_0$ is defined in Eq.(\ref{ans_adsorption_boundary_eq_solution}).
        Now we are able to calculate the total GSDE free energy, Eq.(\ref{ans_adsorption_free_en_tot}).
               
        \textbf{Comparison and errors.} Now we are going to make a comparison between the general solution and solution valid in the WAR. 
        For this, we subscribe to each quantity the index, $"g"$ related to the general solution and correspondingly $"w"$ related to the WAR. 
        Below, we present the quantities, which correspond to repulsive part of the potential. \\
        1) $v_N=500$, $w_N=100$ and $b=2.96082$. Thus, $\xi/b = 0.009 \ll 1$ and the WAR can be applied. Then, we have:
$$
\begin{array}{ll}
	 f_0^g = 1.00905       & f_0^w = 1.00906\\ 
	 J_1^g = 0.000241      & J_1^w = 0.000241 \\ 
	 J_2^g = 0.0004837     & J_2^w = 0.000484 \\
	 \bar{\Delta}_e^g = -0.000242  & \bar{\Delta}_e^w = -0.000242
\end{array}
$$
	 2) $v_N=500$, $w_N=100$ and $b=0.429342$. Thus, $\xi/b = 0.0622 \ll 1$ and the WAR can be applied as well. Then we have:
$$
\begin{array}{ll}
	 f_0^g = 1.0637       & f_0^w = 1.0641\\ 
	 J_1^g = 0.00167      & J_1^w = 0.00169 \\ 
	 J_2^g = 0.00339      & J_2^w = 0.00343 \\
	 \bar{\Delta}_e^g = -0.00172  & \bar{\Delta}_e^w = -0.00174
\end{array}
$$
	 3) $v_N=500$, $w_N=100$ and $b=0.0965361$. Thus, $\xi/b = 0.277 \lesssim 1$; this case correspods to the boundary of the WAR. Then
$$
\begin{array}{ll}
	 f_0^g = 1.303        & f_0^w = 1.314\\ 
	 J_1^g = 0.00743      & J_1^w = 0.00779 \\ 
	 J_2^g = 0.01595      & J_2^w = 0.01678 \\
	 \bar{\Delta}_e^g = -0.00852  & \bar{\Delta}_e^w = -0.00898
\end{array}
$$
	One can see that these "general" and "WAR" values are in good agreement with each other.	
	There are some issues related to numerical calculation of the integral in Eq.(\ref{ans_adsorption_i1_general}). 
	The value of $f_0$ could be small, especially in the WAR. We should be careful when
	we numerically calculate it, because the numerical error could be bigger than the value of the corresponding integral.
       
\begin{figure}[ht!]
\begin{minipage}[ht]{0.5\linewidth}
\center{\includegraphics[width=1\linewidth]{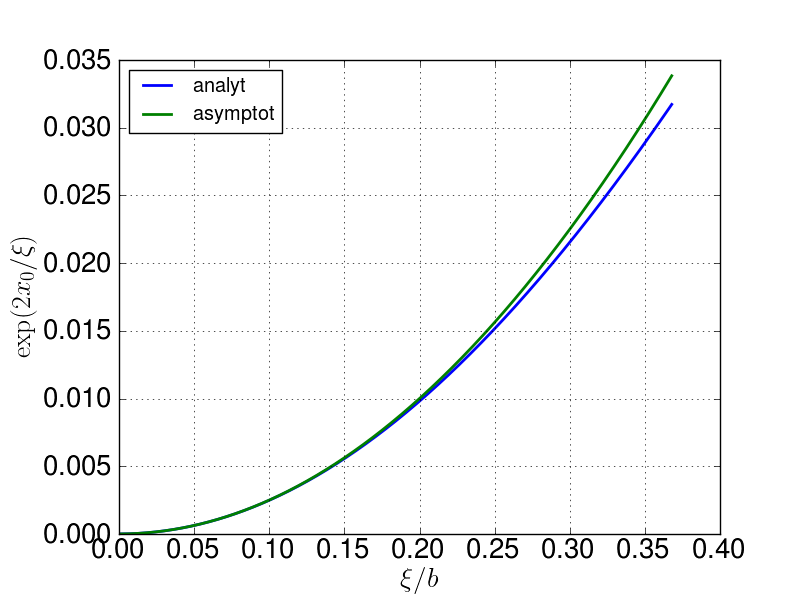}}
\caption{\small{The prefactor $\exp(2x_0/\xi)$ in Eq.(\ref{ans_adsorption_free_en_gs_v}) (the same in Eq.(\ref{ans_adsorption_free_en_gs_vw}) for WAR) 
		as a function of adsorption strength $\xi/b$ and its asymptotics Eq.(\ref{ans_adsorption_exp_x0_asy}).}}
\label{ans_adsorption_prefactor_vs_b_fig}
\end{minipage}
\hfill
\begin{minipage}[ht]{0.5\linewidth}
\center{\includegraphics[width=1\linewidth]{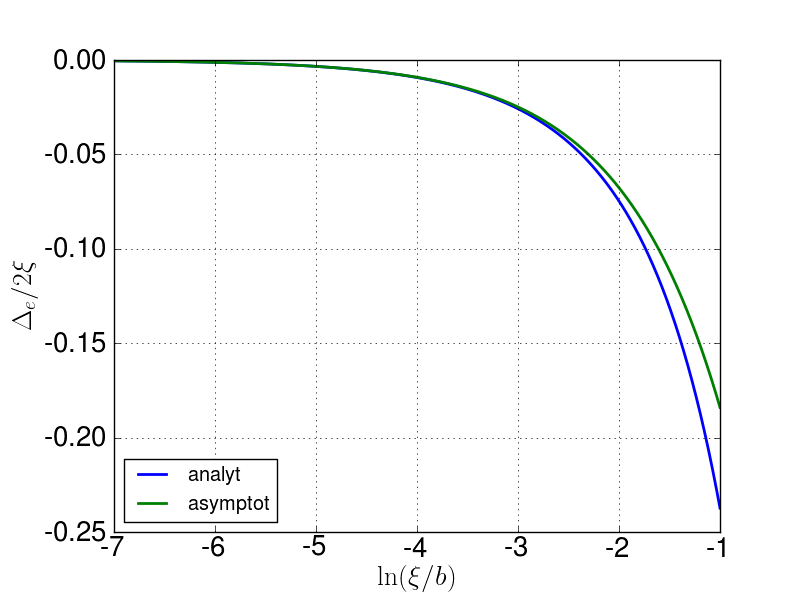}}
\caption{\small{The effective length, Eq.(\ref{ans_adsorption_delta_calc_war}), as a function of the adsorption strength, $\xi/b$, and its asymptotics, 
		Eq.(\ref{ans_adsorption_delta_war_asy}).}}
\label{ans_adsorption_delta_vs_b_fig}
\end{minipage}
\end{figure}

        \textbf{Weak adsorption regime (WAR).} 
        We can simplify the expression for the total GSDE free energy in the weak adsorption regime, i.e. when $\xi/b\ll1$. Joining together
        attractive part, Eq.(\ref{ans_adsorption_free_en_gs_vw}), and repulsive part, Eq.(\ref{ans_adsorption_free_en_end}), we obtain:        
$$
	\hat{W}_{tot}(h) = -\frac{16}{\bar{\xi}}e^{-2x_0/\xi} e^{-h/\xi} + 4\Delta_e^2u_{int}(\bar{h})   
$$
        Let us consider separately each terms. The prefactor of the first term is
$$
	e^{-2x_0/\xi} = e^{-4{\rm arccoth}(f_0)} = \frac{1}{(f_0^2-1)^2}\left\{f_0^4+6f_0^2+1 - 4f_0(f_0^2+1)\right\}
$$
        Substitute explicit expression for $f_0$ via Eq.(\ref{ans_adsorption_gs_f0}) and leaving only the leading term we 
        obtain:
\begin{equation}
\label{ans_adsorption_exp_x0_asy}
	e^{-2x_0/\xi} \simeq \frac{1}{4}\left(\frac{\xi}{b}\right)^2
\end{equation}
        Let us move now to the second term in the expression for the total free energy. We should find the asymptotics for $\Delta_e$.
        We consider separately each term in Eq.(\ref{ans_adsorption_delta_calc_war}). At the beginning the term:
$$
	\coth\left(\frac{x_0}{2\xi}\right) = \coth({\rm arccoth}(f_0)) = f_0 \simeq 1 + \frac{\xi}{b}
$$
	Before treating the second term, one may note that $\sinh({\rm arccoth}(y)) = 1/\sqrt{y^2-1} = 1\sqrt{(y-1)(y+1)}$, so
$$
	\ln\left(2\sinh\left(\frac{x_0}{2\xi}\right)\right) \simeq \ln(2) - \frac{1}{2}\left\{\ln\left(2+\frac{\xi}{b}\right) + \ln\left(\frac{\xi}{b}\right)\right\}
$$
	and
$$
	\frac{x_0}{2\xi} \simeq {\rm arccoth}\left(1 + \frac{\xi}{b}\right) = \frac{1}{2}\left\{\ln\left(2+\frac{\xi}{b}\right) - \ln\left(\frac{\xi}{b}\right)\right\}  
$$
	Substituting all the above expressions in Eq.(\ref{ans_adsorption_delta_calc_war}), we obtain:
$$
	\Delta_e \simeq 2\xi\left(\ln\left(2+\frac{\xi}{b}\right) - \frac{\xi}{b} - \ln(2)\right) \simeq -2\xi\left(\frac{\xi}{2b}\right)   
$$
	Thereby
\begin{equation}
\label{ans_adsorption_delta_war_asy}
       \Delta_e \simeq -\frac{\xi^2}{b}
\end{equation}
	Joining together those expressions, we can write the free energy as 
\begin{equation}
\label{ans_adsorption_free_en_war}
       \hat{W}_{tot}(h) \simeq -\frac{4\bar{\xi}}{\bar{b}^2}e^{-h/\xi} + \frac{4\bar{\xi}^4}{\bar{b}^2}u_{int}(\bar{h}) = 
	                  -\frac{4\bar{\xi}}{\bar{b}^2}\left(e^{-h/\xi} - \bar{\xi}^3u_{int}(\bar{h})\right)
\end{equation}
	In Figs.\ref{ans_adsorption_prefactor_vs_b_fig}--\ref{ans_adsorption_delta_vs_b_fig} we present
	the dependence of the prefactors $\exp(2x_0/\xi)$ and $\Delta_e/2\xi$, and of their asymptotics on the parameter $\xi/b$. 
	So, one can explicitly see, on the pictures, that differences between the analytical dependences and their asymptotics are
	non-negligible for $\xi/b\gtrsim 0.3$.
%%%%%%%%%%%%%%%%%%%%%%%%%%%%%%%%%%%%%%%%%%%%%%%%%%%%%%%%%%%%%%%%%%%%%%%%%%%%%%%%%%%%%%%%%%%%%%%%%%%%%%%%%%%%%%%%%%%%%%%%%%%%%%%
%        adsorption delta and exp_x0
%%%%%%%%%%%%%%%%%%%%%%%%%%%%%%%%%%%%%%%%%%%%%%%%%%%%%%%%%%%%%%%%%%%%%%%%%%%%%%%%%%%%%%%%%%%%%%%%%%%%%%%%%%%%%%%%%%%%%%%%%%%%%%%
	Since $\Delta_e/2\xi$ linearly depends on $\xi/b$, we considered logarithmic scale of $\xi/b$ to distinguish the curves.

	\textbf{Strong adsorption regime (SAR), $\xi\gg b$.} We just obtained the asymptotic expression for $\Delta_e$ valid in the WAR, Eq.(\ref{ans_adsorption_delta_war_asy}). 
	Let us derive a similar expression valid in the SAR. For that, consider again Eq.(\ref{ans_adsorption_delta_calc_war}), that (in the case $w_N=0$) can be used 
	to find the SAR asymptotic.
	Following absolutely the same strategy as we did in the WAR, consider separately each term in the brackets of Eq.(\ref{ans_adsorption_delta_calc_war}).
	For the first term we can write:
$$
	\coth\left(\frac{x_0}{2\xi}\right) = \coth({\rm arccoth}(f_0)) = f_0 = \frac{\xi}{b} + \sqrt{1 + \left(\frac{\xi}{b}\right)^2} \simeq \frac{2\xi}{b}
$$
	Next, using, as before, that $\sinh({\rm arccoth}(f_0)) = 1/\sqrt{f_0^2-1} \simeq 1/f_0 \simeq \frac{b}{2\xi}$,  we obtain
$$
	\ln\left(2\sinh\left(\frac{x_0}{2\xi}\right)\right) \simeq \ln(2) - \ln\frac{2\xi}{b} = -\ln\frac{\xi}{b} 
$$
	and 
$$
	\frac{x_0}{2\xi} \simeq {\rm arccoth}\left(\frac{2\xi}{b}\right) = \frac{1}{2}\left\{\ln\left(\frac{2\xi}{b} + 1\right) - \ln\left(\frac{2\xi}{b}-1\right)\right\} \simeq 0  
$$		
	Thus, joining together all of the terms and leaving only the leading term, we get:
\begin{equation}
\label{ans_adsorption_delta_sar_asy}
       \Delta_e \simeq -\frac{4\xi^2}{b}
\end{equation}
	Note again that the result is valid only when $w_N=0$, so $\bar{\xi}^2 = 1/(2v_N)$.
	
	\textbf{One-parameter model.} When we considered the GSDE theory we found that in the WAR the free energy in the $2$-parametric model 
	depends only on one single parameter, which is expressed through the combination of these parameters. 
	Let us clarify this result on more general ground.
	For this, we note that in the WAR $\delta c(x) = c(x) - c_b \ll c_b$. Correspondingly, the interaction part of the free energy is 
$$
	\mathcal{F}_{int}[c(x)] = \int\mathrm{d}x\left\{\frac{\text{v}}{2}c^2 + \frac{\text{w}}{6}c^3\right\} - \int\mathrm{d}x \mu_{int}(c_b) c 
$$
	where $\mu_{int} = \text{v}c_b + \text{w}c_b^2/2$. 
	Then, expanding it as a series of $\delta c$, neglecting constant and linear terms  
	(at equilibrium the linear term is equal to $0$ anyway) and neglecting any terms with the order higher than $2$, we obtain:
$$
	\mathcal{F}_{int}[\delta c(x)] = \frac{1}{2}\int\mathrm{d}x\left\{\text{v} + \text{w}c_b\right\}\delta c^2 \sim \frac{1}{\xi^2}\int\mathrm{d}x\delta c^2
$$
	Thereby, we found that $\mathcal{F}_{int}$ to the second order in $\delta c$ involves exactly the same combination of parameters which is included in the correlation length. 
	
	\textbf{Generalization.}  For the strong adsorption regime (small $b$) instead of Eq.(\ref{ans_adsorption_free_en_end}) 
	for the repulsive part of the free energy we should use more general expression proposed in \cite{semenov_2008}, namely:
$$
	W_e \simeq \frac{2c_b}{N}\left[2\Delta_e - h\ln\left(1+\frac{2\Delta_e}{h}\right)\right] H(h/R_g)
$$
	Recalling that $H(\bar{h}) \equiv \bar{h}\,u_{int}(\bar{h})$, we can rewrite it in our dimensionless variables:
\begin{equation}
\label{ans_adsorption_free_en_end_general}
	\hat{W}_e = \frac{R_gc_b}{N}W_e \simeq 2\left[2\bar{\Delta}_e - \bar{h}_{b}\ln\left(1+\frac{2\bar{\Delta}_e}{\bar{h}_{b}}\right)\right] H(\bar{h})
\end{equation}
	where in the square brackets we replaced the distance $h$ by the effective distance $h_b$ that is defined as: 
$$
	h_b = \frac{1}{c_b}\int\limits_0^h\mathrm{d}x c_1(x) = h + \int\limits_0^h\mathrm{d}x \left(c_1(x)/c_b - 1\right) \simeq h + 2\Delta_c
$$
	where 
$$
	\Delta_c = \int\limits_0^{\infty} \mathrm{d}x \left(c_1(x)/c_b - 1\right)
$$
	is the effective single-plate excess of polymer amount. As before, consider the quantity in two regimes:\\
	1) $w_N=0$ or WAR. Using the expression for the single-plate concentration profile, Eq.(\ref{ans_adsorption_gs_concentration_vw}), we can
	see that $\bar{\Delta_c}$ can be identified with the integral $J_2$ included in Eq.(\ref{ans_adsorption_delta_calc_war}):
$$
	\bar{\Delta}_c = J_2 = 2\xi\left(\coth\left(\frac{x_0}{2\xi}\right) - 1\right)
$$
	Therefore, using Eq.(\ref{ans_adsorption_gs_x0}), we can write
\begin{equation}
\label{ans_adsorption_hb_war}
	\bar{h}_b = \frac{h_b}{R_g} \simeq \bar{h} + 4\bar{\xi}\left(\frac{\xi}{b} + \sqrt{1 + \left(\frac{\xi}{b}\right)^2} - 1 \right)
\end{equation}
	2) The general case. In the general case the integral $J_2$ also coincides with $\bar{\Delta}_c$:
$$
	\bar{\Delta}_c = J_2 = \sqrt{\frac{3}{w_N}}\left\{\ln\left(f_0 + \sqrt{\alpha + f_0^2 }\right) - \ln\left(1 + \sqrt{\alpha + 1}\right)\right\}
$$
	and 
\begin{equation}
\label{ans_adsorption_hb_general}
	\bar{h}_b \simeq \bar{h} + 2\sqrt{\frac{3}{w_N}}\left\{\ln\left(f_0 + \sqrt{\alpha + f_0^2 }\right) - \ln\left(1 + \sqrt{\alpha + 1}\right)\right\}
\end{equation}
	where $\alpha = 2 + 3v_N/2w_N$ and $f_0$ is defined in Eq.(\ref{ans_adsorption_boundary_eq_solution}).
	The main improvement of Eq.(\ref{ans_adsorption_free_en_end_general}), in comparison with Eq.(\ref{ans_adsorption_free_en_end}) is that it is 
	valid for any $h\gg\xi$ (if $\bar{\Delta}_e\ll1$), i.e. $h$ may be shorter than $\Delta_e$. 
	We will use Eq.(\ref{ans_adsorption_free_en_end_general}) for the numerical comparison of the GSDE theory with the SCFT results.
		
	\textbf{Scaling behavior.}
	The expression for the free energy, Eq.(\ref{ans_adsorption_free_en_war}), is valid in the WAR for $\bar{h}\gg\bar{\xi}$. 
	The next step: we are going to find the scaling behavior for the free energy barrier height and its position in the region $\bar{\xi}\ll\bar{h}\ll 1$.
	Such distances allow us to neglect all terms in the sum for $u_{int}(\bar{h})$ except the one with $n=0$.
	Thus, we can write $u_{int}(\bar{h}) \simeq 1/\bar{h}$, and the total free energy, Eq.(\ref{ans_adsorption_free_en_war}), is
$$
	\hat{W}_{tot}(h) \simeq -\frac{4\bar{\xi}}{\bar{b}^2}\left(e^{-\bar{h}/\bar{\xi}} - \frac{\bar{\xi}^3}{\bar{h}}\right),\quad \xi\ll h\ll R_g
$$	
	Differentiating the equation, we find the condition for the maximum of $W_{tot}$:
$$
	\hat{W}'_{tot} \simeq -\frac{4\bar{\xi}}{\bar{b}^2}\left(-\frac{1}{\bar{\xi}}e^{-\bar{h}/\bar{\xi}} + \frac{\bar{\xi}^3}{\bar{h}^2}\right) = 0
$$
	or, after simple transformations:
$$
	-\frac{\bar{h}^*}{\bar{\xi}} =\ln\left(\frac{\bar{\xi}^2}{\bar{h}^*}\right)^2 = 2\ln\bar{\xi} + 2\ln\frac{\xi}{h^*}
$$
	Since $\bar{\xi}/\bar{h}\ll 1$ and leaving only the leading term, we obtain:
\begin{equation}
\label{ans_adsorption_barrier_position}
	\bar{h}^* \simeq \bar{\xi}\ln\frac{1}{\bar{\xi}^2} \sim \bar{\xi}
\end{equation}
	The first term in the expression for the free energy, by virtue of the exponential behavior at $\bar{\xi}\ll 1$, is a rapidly changing function,
        which at $h\geqslant h^*$ tends to zero very quickly. Thereby, we can neglect the negative part at such distances 
	and take into account only the second term. After using Eq.(\ref{ans_adsorption_barrier_position}), we obtain the expression for the barrier height:
\begin{equation}
\label{ans_adsorption_barrier_height}
	\hat{W}^*_{tot} \simeq \frac{4\bar{\xi}^4}{\bar{b}^2\bar{h}^*} \simeq \frac{2\bar{\xi}^3}{\bar{b}^2\ln(1/\bar{\xi})} \sim \frac{\bar{\xi}^3}{\bar{b}^2}
\end{equation}
	 Therefore, the crude scaling behaviors for the barrier height and its position are given, correspondingly, by $\hat{W}^*_{tot} \sim \bar{\xi}^3/\bar{b}^2$ and 
	 $\bar{h}^*\sim \bar{\xi}$. One can notice that the barrier position does not depend on the adsorption parameter $\bar{b}$. 
	 We already saw that for the thermodynamic potential calculated by the SCFT,
	 the barrier position is independent on the adsorption parameter in a wide range of the virial parameters and the adsorption strength.

%%%%%%%%%%%%%%%%%%%%%%%%%%%%%%%%%%%%%%%%%%%%%%%%%%%%%%%%%%%%%%%%%%%%%%%%%%%%%%%%%%%%%%%%%%%%%%%%%%%%%%%%%%%%%%%%%%%%%%%%%%%%%%%%%%%%%%%%%%%%%%%%%%%%%%%%%%%%%
%           Comparison with ANS theory
%%%%%%%%%%%%%%%%%%%%%%%%%%%%%%%%%%%%%%%%%%%%%%%%%%%%%%%%%%%%%%%%%%%%%%%%%%%%%%%%%%%%%%%%%%%%%%%%%%%%%%%%%%%%%%%%%%%%%%%%%%%%%%%%%%%%%%%%%%%%%%%%%%%%%%%%%%%%%
\section{Perturbation SCFT theory for weak adsorption} 
	In the GSDE theory of colloidal stabilization, the free energy is derived as a correction to the GSD free energy,
	based on the difference in distributions for a middle unit and the ends of a chain.	
	Another approach, the perturbation SCFT theory for weak adsorption proposed in \cite{avalos_2003},
	considers small variations of concentration caused by the weak adsorption field that allowed to linearize the Edwards equation.	
	Bellow, as before, we consider polymer solution in a gap between two flat plates, in a case of reversible adsorbed chains, i.e.
	when the polymers are in equilibrium with the bulk solution.		
	Following the authors in \cite{avalos_2003}, we can write the expression for the free energy in the following form:
$$
	\frac{U(h)}{k_BT} = -\frac{4a^2_sc_b\xi}{3\pi b^2}\sum\limits_{n=1}^{\infty}\int\limits_{0}^{\infty}\mathrm{d}k\cos\left(\frac{hnk}{\xi}\right)f(k, \bar{\xi})
$$
	or, in the dimensionless variables: 
\begin{equation}
\label{avalos_adsorption_free_en_f}
	\hat{U}(h) = \frac{N}{c_bR_g} \frac{U(h)}{k_BT} = 
		     -\frac{8\bar{\xi}}{\pi\bar{b}^2}\sum\limits_{n=1}^{\infty}\int\limits_{0}^{\infty}\mathrm{d}k\cos\left(\frac{hnk}{\xi}\right)f(k, \bar{\xi})
\end{equation}
       where $R_g = a_sN^{1/2}/6$ is the radius of gyration of an ideal polymer chain, $a_s$ is the polymer statistical segment, 
       $N$ is the polymerization index of the chain, $\bar{b} = b/R_g, \bar{h}=h/R_g$ are the dimensionless extrapolation length and
       the distance between plates, $\bar{\xi} = \xi/R_g \simeq 1/\sqrt{2v_N}$ is the dimensionless GSD correlation length in the bulk 
       phase\footnote{Note that, the correlation length $\xi_b$, which is defined in the article \cite{avalos_2003}, is different from our $\xi$ 
       and they are related as $\xi_b=2\xi$. We also excluded the variable $\varepsilon \equiv 4\xi^2/R_g^2=4\bar{\xi}^2$ from the expression for $f(\varepsilon, k)$.}, 
       $v_N = \text{v}c_bN$ is the reduced second virial coefficient. 
       The theory does not take into account the third virial coefficient, i.e. $\text{w}=0$.        
       The integrand function is 
\begin{equation}
\label{avalos_adsorption_integrand}
	f(k, \bar{\xi}) = \frac{k^2-\bar{\xi}^2 + \bar{\xi}^2\exp(-k^2/\bar{\xi}^2)}{k^4 + k^2 - \bar{\xi}^2 + \bar{\xi}^2\exp(-4k^2/\bar{\xi}^2)}
\end{equation}		
	The expression for the free energy, Eq.(\ref{avalos_adsorption_free_en_f}), is arranged in such a way that the free energy at infinite separation 
	is vanishing, thus $\hat{U}(h)\rightarrow 0$ at large separations.
	In order to complete the comparison of the theory with the numerical SCFT and the analytical GSDE, we should calculate  
	the free energy, Eq.(\ref{avalos_adsorption_free_en_f}), numerically. 
	For that, we must truncate the summation after certain upper value and cut the upper limit of the integral as well. 
	
	Let us estimate the truncated error and find the value for the upper summation limit that provide the sufficient accuracy. 	
	The terms in the above sum decrease with their number. 
	Therefore, for simplicity, we consider the integrand $hnk/\xi\ll 1$ and the cosine factor close to $1$.	 
        For large enough values of $k$, one can notice that the function $f(k, \bar{\xi})$,
        which is defined in Eq.(\ref{avalos_adsorption_integrand}), has the asymptotics $f(k, \bar{\xi})\simeq 1/k^2$, thus 
$$
	\int\limits_{0}^{\infty}\mathrm{d}kf(k, \bar{\xi}) = \int\limits_{0}^{k_{up}}\mathrm{d}kf(k, \bar{\xi}) + \int\limits_{k_{up}}^{\infty}\mathrm{d}kf(k, \bar{\xi}) = 
								I_k(\bar{\xi}) + \delta_f(k_{up}, \bar{\xi})    
$$
	Using the asymptotics, we can write:
$$
	\delta_f(k_{up}, \bar{\xi}) \simeq \int\limits_{k_{up}}^{\infty}\frac{\mathrm{d}k}{k^2} = \frac{1}{k_{up}}      
$$
	Therefore, to provide the acceptable truncated accuracy, say, $10^{-6}$, we should take the upper summation limit $k_{up}\simeq 10^{6}$.	
	In this way, to take such upper limit and provide sufficient accuracy is numerically impossible, because there are also errors related to
	the integration that depend on the integration step.	
	In order to treat it, we should use another trick and make the integral appropriate for numerical calculations. 	
	We consider the simple transformation of the function $f(k, \bar{\xi})$, namely:
$$
	f(k, \bar{\xi}) = \left(f(k, \bar{\xi}) - h(k)\right) + h(k) = g(k, \bar{\xi}) + h(k)
$$
	where the function $h(k)$ coincides with $f(k, \bar{\xi})$ at $k\gg 1$ and does not have a singularity at $k=0$. This allows us to perform the 
	integration from $k=0$. In addition, 	
	the function $g(k, \bar{\xi})$ converges to $0$ faster than $1/k^2$ at large $k$. 
		
	\textbf{Asymptotics}. Since $\bar{\xi}^2 = 1/2v_N$ has limited value, for big $k$, using Eq.(\ref{avalos_adsorption_integrand}), we can write that 
$$
	h(k) = \frac{1}{k^2+1} \simeq f(k\gg1, \bar{\xi}) 
$$
	After integration we obtain: 
$$
	\int\limits_0^{\infty}\mathrm{d}k\frac{\cos(hnk/\xi)}{k^2+1} = \frac{\pi}{2}e^{-hn/\xi}
$$
	Correspondingly, after summation
$$
	\sum\limits_{n=1}^{\infty}\frac{\pi}{2}e^{-hn/\xi} = \frac{\pi}{2(e^{h/\xi} - 1)}
$$
	Therefore, we can rewrite Eq.(\ref{avalos_adsorption_free_en_f}) as
\begin{equation}
\label{avalos_adsorption_free_en_g}
	\hat{U}(h) = -\frac{8\bar{\xi}}{\pi\bar{b}^2}\left(\frac{\pi}{2(e^{h/\xi} - 1)} + 
					      \sum\limits_{n=1}^{\infty}\int\limits_{0}^{\infty}\mathrm{d}k\cos\left(\frac{hnk}{\xi}\right)g(\bar{\xi}, k)\right)	
\end{equation}
	\textbf{Upper limit of integration}. Let us find now the appropriate integration range for the function $g(\bar{\xi}, k)$ that satisfy a certain accuracy. 
	Using the cut-off value $k_{up}$ for the upper limit of the function $g(k, \bar{\xi})$, we can write: 
$$
	\int\limits_0^{\infty}\mathrm{d}k g(k, \bar{\xi}) = \int\limits_0^{k_{up}}\mathrm{d}k g(k, \bar{\xi}) + \delta_g(k_{up}, \bar{\xi})
$$
	Numerically, one can be sure that $g(10^3, \bar{\xi}) \simeq 10^{-4}g(10^2, \bar{\xi}) \simeq 10^{-8}g(10, \bar{\xi})$.
	Thus, at big value of $k$ we have $g(k, \bar{\xi}) \simeq C_{\xi}/k^{4}$ and 
$$
	\delta_g(k_{up}, \bar{\xi}) \equiv \int\limits_{k_{up}}^{\infty}\mathrm{d}kg(k,\bar{\xi}) \simeq 
	C_{\xi}\int\limits_{k_{up}}^{\infty}\frac{\mathrm{d}k}{k^4} = \frac{C_{\xi}}{4k_{up}^3}	
$$
	The constant $C_{\xi}$ is varied in the range $ 10^{-1} \div  10^{-4}$ when the second virial parameter is changed in the range $v_N \in [10^{-1} .. 10^3]$.
	Therefore, if we want to make the cutting error $\delta_g \simeq 10^{-6}$, we should set the upper limit in the integral at $k_{up} \simeq 30$ 
	(we chose the maximum value for the upper limit to avoid different $k_{up}$ for different $\xi$. It corresponds to the highest value of $C_{\xi}(v_N)$).

	\textbf{Zero point feature}. If we put the value $k=0$ in Eq.(\ref{avalos_adsorption_integrand}), we get  $0/0$. In order to evaluate it, 
	we use the Taylor series up to the second order for the numerator and denominator in Eq.(\ref{avalos_adsorption_integrand}). 
	Correspondingly, we have:
$$
	f(0, \bar{\xi}) = \frac{1}{2\bar{\xi}^2+1}
$$
	and
$$
	g(0, \bar{\xi}) = f(0, \bar{\xi}) - 1 =  -\frac{2\bar{\xi}^2}{2\bar{\xi}^2 + 1}
$$

	We extend the value from $k=0$ up to $k_{min} \simeq 10^{-3}$ for the function $g(k, \bar{\xi})$, 
	because both the numerator and denominator of the function $f(k, \xi)$ are very small $k<k_{min}$.
	
	\textbf{Other errors}. To evaluate the integral in Eq.(\ref{avalos_adsorption_free_en_g}) we use the Simpson integration rule with upper integral 
	limit $k_{up}=30$. 
	Since the integrand drastically drops when we increase $k$ we sum up it in the Simpson rule in the reverse order, starting from $k_{up}$.
	It helps us to keep the accuracy related to computer representation of the floating point numbers. Next, we should sum up every integral 
	in Eq.(\ref{avalos_adsorption_free_en_g}). As we already noticed, a certain number of terms in the sum defines the accuracy of the free energy
	by virtue of the periodic behavior of the integrand. 
	When the period becomes small enough, the function $g(k, \xi)$ changes insignificantly within a period. 
	Thus, the integral becomes very small and we can neglect it for large $n$. 
	We found empirically the parameters providing sufficiently good accuracy
	for the number\footnote{We tested the sum with $n=500$ terms and the number of integration points $n_{int}=10^5$. 
	The results coincide with $n_{sum}=50$ for small distances, capturing the barrier.
	For big distances the free energy with $n_{sum}=500$ in the sum generates artificially big oscillations 
	caused by the fact that there are not enough integration points per a period.} 
	of terms in the sum $n_{up} = 50$.
%%%%%%%%%%%%%%%%%%%%%%%%%%%%%%%%%%%%%%%%%%%%%%%%%%%%%%%%%%%%%%%%%%%%%%%%%%%%%%%%%%%%%%%%%%%%%%%%%%%%%%%%%%%%%%%%%%%%%%%%%%%%%%%%%%%%%%%%%%%%%%%%%%%%%%%%%%%%%
%       Comparison with SCFT results
%%%%%%%%%%%%%%%%%%%%%%%%%%%%%%%%%%%%%%%%%%%%%%%%%%%%%%%%%%%%%%%%%%%%%%%%%%%%%%%%%%%%%%%%%%%%%%%%%%%%%%%%%%%%%%%%%%%%%%%%%%%%%%%%%%%%%%%%%%%%%%%%%%%%%%%%%%%%%
\section{Comparison with SCFT results} 
	  We have already shown in subsection "$\textbf{one-parameter model}$" that in the WAR the free energy depends on the second and 
	  third virial parameters not independently, but via the correlation length $\bar{\xi}\simeq 1/\sqrt{2(v_N+2w_N)}$. 
	  Thereby, only one parameter, $v_N+2w_N$ is relevant in the WAR.
	  Based on that, we can also generalize the perturbation SCFT in weak adsorption regime, Eq.(\ref{avalos_adsorption_free_en_f}),
	  including the case with $w_N\ne 0$ via $\bar{\xi}\simeq 1/\sqrt{2(v_N+2w_N)}$.
	  
\begin{figure}[ht!]
\begin{minipage}[ht]{0.5\linewidth}
\center{\includegraphics[width=1\linewidth]{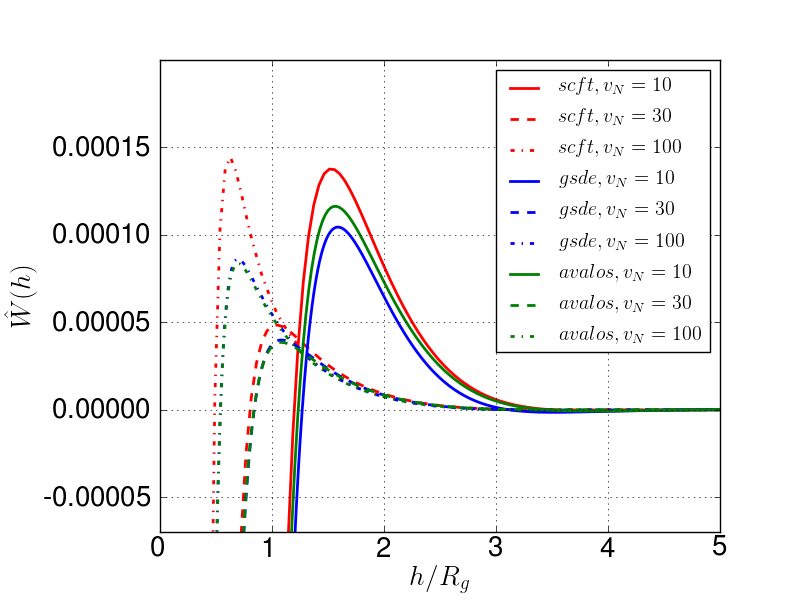}}
\caption{\small{$A = 14.1 (\bar{b}\simeq 2.961)$; $v_N=10: \bar{\Delta}_e = 0.018$; $v_N=30: \bar{\Delta}_e = 0.006$; $v_N=100: \bar{\Delta}_e = 0.002$.}}
\label{ans_adsorption_omega_h5_a14_comparison_fig}
\end{minipage}
\hfill
\begin{minipage}[ht]{0.5\linewidth}
\center{\includegraphics[width=1\linewidth]{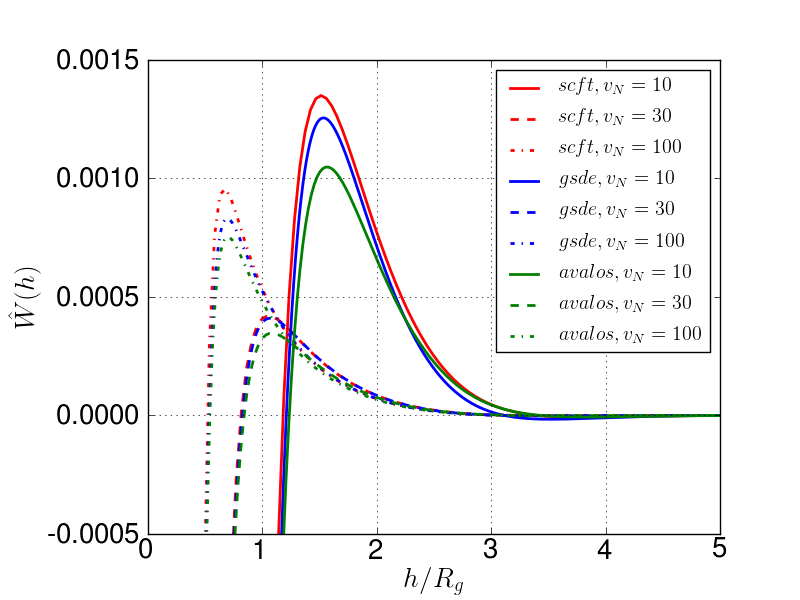}}
\caption{\small{$A = 42.73 (\bar{b}\simeq 0.986)$; $v_N=10:\bar{\Delta}_e = 0.06$; $v_N=30:\bar{\Delta}_e = 0.019$; $v_N=100:\bar{\Delta}_e = 0.005$.}}
\label{ans_adsorption_omega_h5_a42_comparison_fig}
\end{minipage}
\end{figure}

\begin{figure}[ht!]
\begin{minipage}[ht]{0.5\linewidth}
\center{\includegraphics[width=1\linewidth]{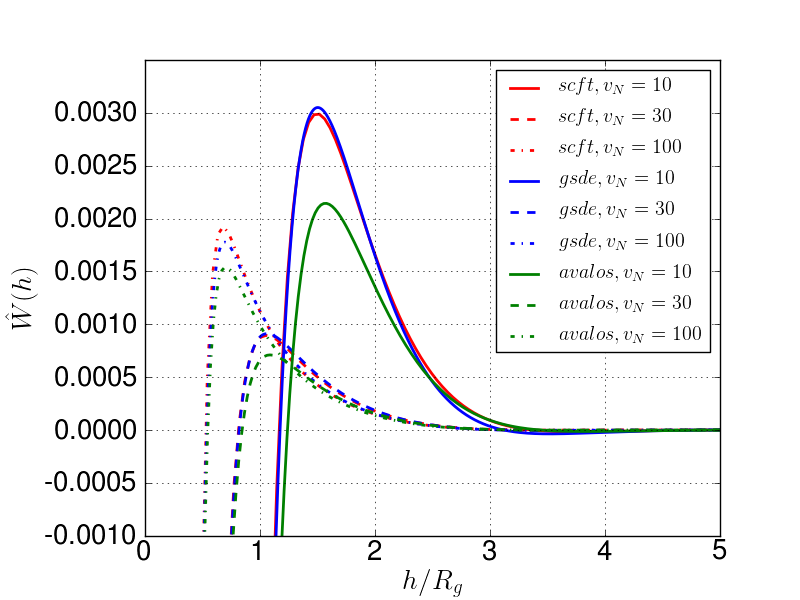}}
\caption{\small{$A = 61.5 (\bar{b}\simeq 0.689)$; $v_N=10:\bar{\Delta}_e = 0.091$; $v_N=30:\bar{\Delta}_e = 0.028$; $v_N=100:\bar{\Delta}_e = 0.008$.}}
\label{ans_adsorption_omega_h5_a61_comparison_fig}
\end{minipage}
\hfill
\begin{minipage}[ht]{0.5\linewidth}
\center{\includegraphics[width=1\linewidth]{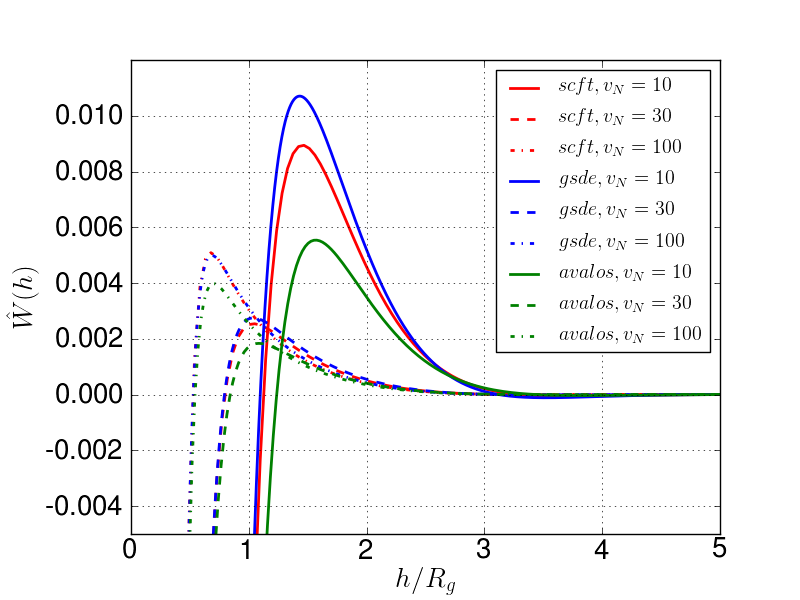}}
\caption{\small{$A = 100 (b\simeq 0.429)$; $v_N=10:\bar{\Delta}_e = 0.164$; $v_N=30:\bar{\Delta}_e = 0.048$; $v_N=100:\bar{\Delta}_e = 0.013$.}}
\label{ans_adsorption_omega_h5_a100_comparison_fig}
\end{minipage}
\end{figure}

\begin{figure}[ht!]
\begin{minipage}[ht]{0.5\linewidth}
\center{\includegraphics[width=1\linewidth]{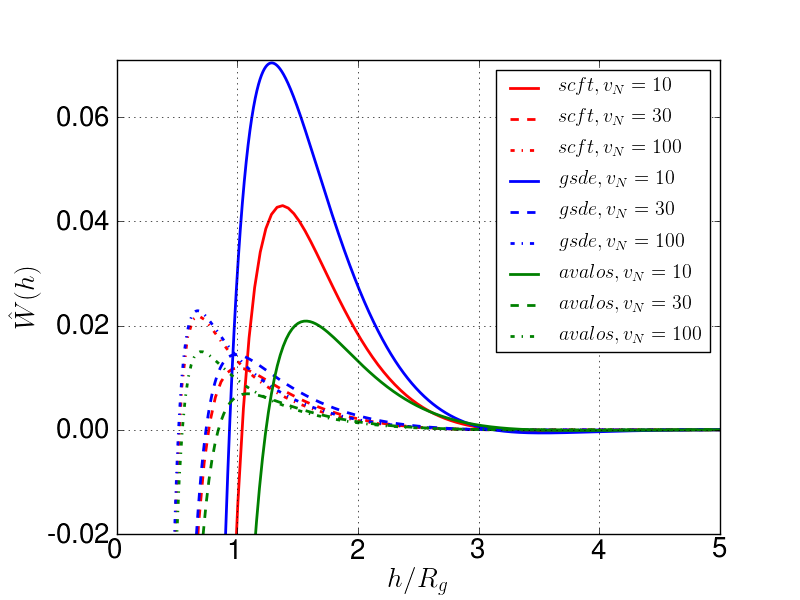}}
\caption{\small{$A = 200 (\bar{b}\simeq 0.221)$; $v_N=10:\bar{\Delta}_e = 0.398$; $v_N=30:\bar{\Delta}_e = 0.11$; $v_N=100:\bar{\Delta}_e = 0.028$.}}
\label{ans_adsorption_omega_h5_a200_comparison_fig}
\end{minipage}
\hfill
\begin{minipage}[ht]{0.5\linewidth}
\center{\includegraphics[width=1\linewidth]{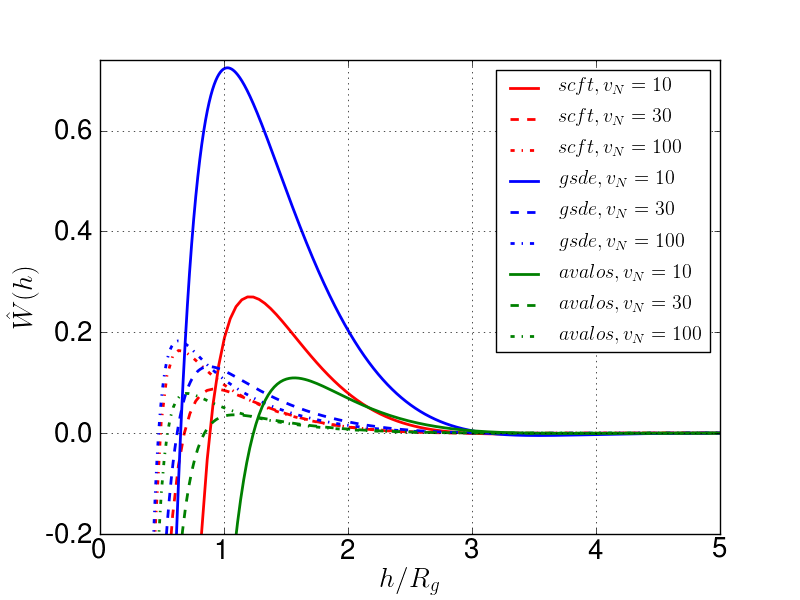}}
\caption{\small{$A = 500 (\bar{b}\simeq 0.097)$; $v_N=10:\bar{\Delta}_e = 1.238$; $v_N=30:\bar{\Delta}_e = 0.339$; $v_N=100:\bar{\Delta}_e = 0.081$.}}
\label{ans_adsorption_omega_h5_a500_comparison_fig}
\end{minipage}
\end{figure}

	  The free energy potentials obtained by the GSDE theory, Eqs.(\ref{ans_adsorption_free_en_gs_vw}, \ref{ans_adsorption_free_en_end_general}),
	  are valid either for $w_N=0$ or for the WAR. Thus, to include a greater range of the comparison
	  between the different approaches we restrict ourselves to the case with $w_N=0$.
	  In Figs.\ref{ans_adsorption_omega_h5_a14_comparison_fig}--\ref{ans_adsorption_omega_h5_a500_comparison_fig} we represent the comparison between
	  the thermodynamic potentials calculated by different ways. We denoted them correspondingly as $scft$ for the SCFT thermodynamic potential\footnote{All of the 
	  SCFT thermodynamic potentials for the comparison are calculated with high resolution grid with $N_x=10k, N_s=10k$.}, 
	  $gsde$ for the free energy potential obtained by the GSDE theory that corresponds to dimensionless Eq.(\ref{ans_adsorption_free_en_tot}) 
	  with end-segment contribution defined in 
	  Eq.(\ref{ans_adsorption_free_en_end_general}) and $\Delta_e$ in Eqs.(\ref{ans_adsorption_delta_calc_war}, \ref{ans_adsorption_delta_calc_general}). 
	  The perturbation SCFT calculations in weak adsorption case for the free energy, Eq.(\ref{avalos_adsorption_free_en_f}), we are denoted as 
	  $avalos$.
	  We indicated under each picture the value of the adsorption strength $A$ (it is expressed terms of $\bar{b} = b/R_g$) 
	  and corresponding $\bar{\Delta}_e$ calculated with Eq.(\ref{ans_adsorption_delta_calc_war}) for different values of the second virial parameter.
	  To distinguish the curves for $v_N=100$ we multiplied the thermodynamic potentials corresponding to $v_N=100$ by $10$ in each picture.
%%%%%%%%%%%%%%%%%%%%%%%%%%%%%%%%%%%%%%%%%%%%%%%%%%%%%%%%%%%%%%%%%%%%%%%%%%%%%%%%%%%%%%%%%%%%%%%%%%%%%%%%%%%%%%%%%%%%%%%%%%%%%%%
%        therm pot A = 14, A = 42
%%%%%%%%%%%%%%%%%%%%%%%%%%%%%%%%%%%%%%%%%%%%%%%%%%%%%%%%%%%%%%%%%%%%%%%%%%%%%%%%%%%%%%%%%%%%%%%%%%%%%%%%%%%%%%%%%%%%%%%%%%%%%%%

%%%%%%%%%%%%%%%%%%%%%%%%%%%%%%%%%%%%%%%%%%%%%%%%%%%%%%%%%%%%%%%%%%%%%%%%%%%%%%%%%%%%%%%%%%%%%%%%%%%%%%%%%%%%%%%%%%%%%%%%%%%%%%%
%        therm pot A = 61, A = 100
%%%%%%%%%%%%%%%%%%%%%%%%%%%%%%%%%%%%%%%%%%%%%%%%%%%%%%%%%%%%%%%%%%%%%%%%%%%%%%%%%%%%%%%%%%%%%%%%%%%%%%%%%%%%%%%%%%%%%%%%%%%%%%%
%%%%%%%%%%%%%%%%%%%%%%%%%%%%%%%%%%%%%%%%%%%%%%%%%%%%%%%%%%%%%%%%%%%%%%%%%%%%%%%%%%%%%%%%%%%%%%%%%%%%%%%%%%%%%%%%%%%%%%%%%%%%%%%
%        therm pot A = 200, A = 500
%%%%%%%%%%%%%%%%%%%%%%%%%%%%%%%%%%%%%%%%%%%%%%%%%%%%%%%%%%%%%%%%%%%%%%%%%%%%%%%%%%%%%%%%%%%%%%%%%%%%%%%%%%%%%%%%%%%%%%%%%%%%%%%
	    For the weak adsorption strength and small virial parameters the curves are in a reasonable agreement with each other.
	    The SCFT curves exceed the analytical curves by around 20$\%$, but the difference is more significant for big second virial coefficient,
	    despite that the analytical results coincide between each other.     
%%%%%%%%%%%%%%%%%%%%%%%%%%%%%%%%%%%%%%%%%%%%%%%%%%%%%%%%%%%%%%%%%%%%%%%%%%%%%%%%%%%%%%%%%%%%%%%%%%%%%%%%%%%%%%%%%%%%%%%%%%%%%%%%%%%%%%%%%%%%
% The height of the barrier for vw: adsorption 
%%%%%%%%%%%%%%%%%%%%%%%%%%%%%%%%%%%%%%%%%%%%%%%%%%%%%%%%%%%%%%%%%%%%%%%%%%%%%%%%%%%%%%%%%%%%%%%%%%%%%%%%%%%%%%%%%%%%%%%%%%%%%%%%%%%%%%%%%%%%
\begin{table}[ht!]
\caption{The relative error of the SCFT thermodynamic potential barrier height defined as $\epsilon = \Delta\Omega^*_{77}/\Delta\Omega^{*}_{hr} -1$
	  calculated for grid with $N_x=7k, N_s=3k$ and for high resolution (hr) grid corresponding to $N_x=10k, N_s=10k$.} 
\label{tabular:ans_adsorption_comparison_max_scft_vw} 
\begin{center}
    \begin{tabular}{ | c | c | c | c | c | c | c | }
    \hline
         $A$                                    &  14.1&  42.73&  61.5&   100&   200&    500   \\ \hline
  $\epsilon, v_N=10$                            &  0.043& 0.006&  0.01& 0.008& 0.003&  0.002   \\ \hline
  $\epsilon, v_N=30$                            &   0.08& 0.029& 0.019& 0.007& 0.006&  0.003   \\ \hline
  $\epsilon, v_N=100$                           &  0.267&  0.08& 0.061& 0.045& 0.017&  0.009   \\
    \hline
  \end{tabular}

\end{center} 
\end{table}
	    We present in Tab.\ref{tabular:ans_adsorption_comparison_max_scft_vw} the relative error for the SCFT thermodynamic potential 
	    barrier height. 
	    One can observe that in the weakest adsorption case with the highest second virial parameter the numerical error contributes around $30\%$ and
	    the discrepancy in the data is shown in Fig.\ref{ans_adsorption_omega_h5_a14_comparison_fig}. The discrepancy, for big virial parameters, 
	    can be explained as a numerical error emerging from the lack of precision in the calculations of the SCFT thermodynamic potential.
	    
	    Then, when we enhance the adsorption strength, the GSDE potential in a wide range of adsorption strength is in very good agreement with 
	    the numerical SCFT results.
	    This is especially manifested for the region of big second virial parameter, $v_N$, where the potentials are similar throughout the whole considered range of the 
	    adsorption strength.
	    
	    For strong adsorption and small virial parameters (when $\xi/b>1$), the GSDE potential gives an overestimated result for the barrier height
	    in comparison to the SCFT potential, whereas the analytical perturbative SCFT potential is significantly less than the numerical SCFT result. 	    
	    
	    The difference between the SCFT and the GSDE analytical theory can be explained if we 
	    recall an applicability condition to the GSDE, namely  $\bar{\Delta}_e\ll 1$. According to
	    Figs.\ref{ans_adsorption_omega_h5_a14_comparison_fig}--\ref{ans_adsorption_omega_h5_a500_comparison_fig} the GSDE theory works 
	    well for $\bar{\Delta}_e\lesssim 0.2$.
	    
	    It is also noticeable that the position of the energy barrier for the numerical SCFT potential and analytical GSDE potential are shifted to smaller 
	    separations while the barrier position for the analytical perturbation SCFT potential ($avalos$)
	    does not move with increasing of the adsorption strength, so it becomes invalid for strong adsorption.
%%%%%%%%%%%%%%%%%%%%%%%%%%%%%%%%%%%%%%%%%%%%%%%%%%%%%%%%%%%%%%%%%%%%%%%%%%%%%%%%%%%%%%%%%%%%%%%%%%%%%%%%%%%%%%%%%%%%%%%%%%%%%%%%%%%%%%%%%%%%%%%%%%%%%%%%%%%%%
%           The virial parameters providing the maximum barrier height.
%%%%%%%%%%%%%%%%%%%%%%%%%%%%%%%%%%%%%%%%%%%%%%%%%%%%%%%%%%%%%%%%%%%%%%%%%%%%%%%%%%%%%%%%%%%%%%%%%%%%%%%%%%%%%%%%%%%%%%%%%%%%%%%%%%%%%%%%%%%%%%%%%%%%%%%%%%%%%
\section{The virial parameters providing the maximum barrier height. Flat case} 
\begin{figure}[ht!]
\begin{minipage}[ht]{0.5\linewidth}
\center{\includegraphics[width=1\linewidth]{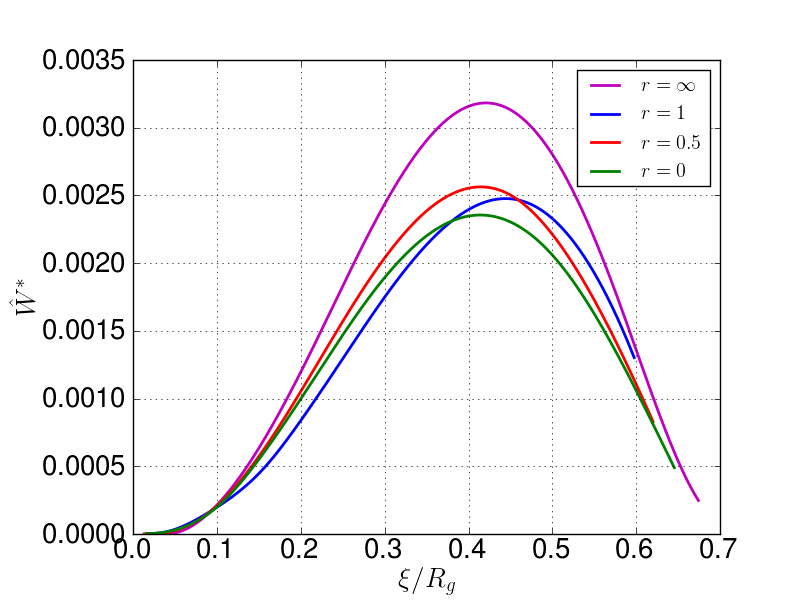}}
\caption{\small{The SCFT thermodynamic potential barrier height as a function of the correlation length, $\xi$,
		for different values of fixed virial parameters indicated in the figure. The adsorption strength is $A = 42.73(\bar{b}\simeq 0.986)$. }}
\label{therm_barrier_vs_xi_a42_fig}
\end{minipage}
\hfill
\begin{minipage}[ht]{0.5\linewidth}
\center{\includegraphics[width=1\linewidth]{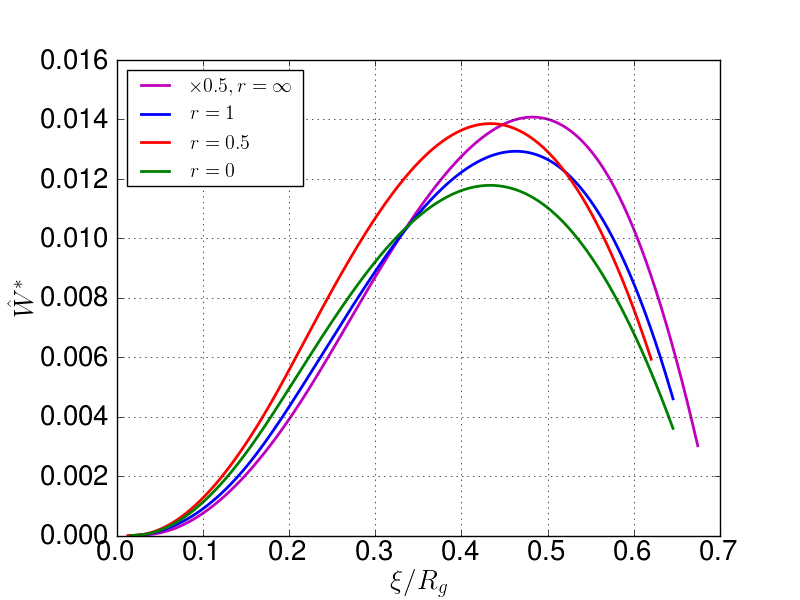}}
\caption{\small{The SCFT thermodynamic potential barrier height as a function of the correlation length, $\xi$, 
		for different values of fixed virial parameters indicated in the figure. The adsorption strength is $A = 100(\bar{b}\simeq 0.429)$. 
		We multiplied the curve for $w_N=0$ by 0.5.}}
\label{therm_barrier_vs_xi_a100_fig}
\end{minipage}
\end{figure}

\begin{figure}[ht!]
\begin{minipage}[ht]{0.5\linewidth}
\center{\includegraphics[width=1\linewidth]{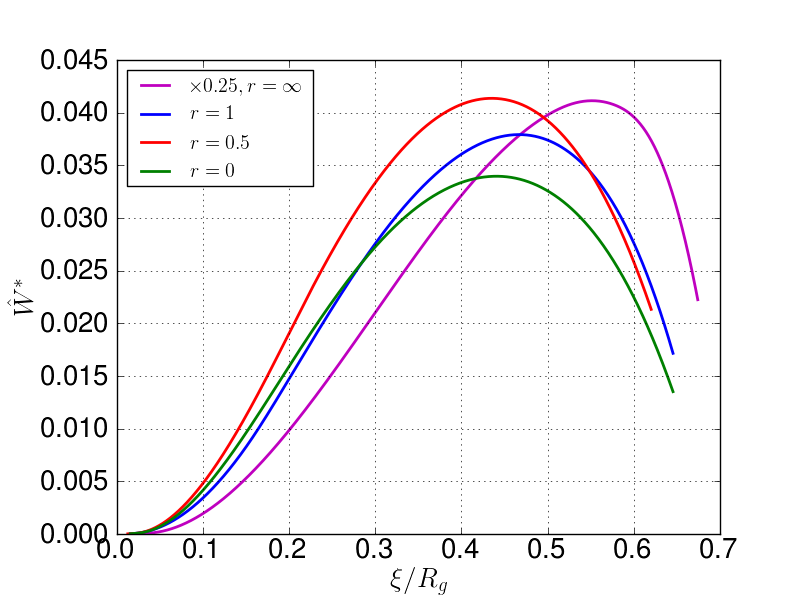}}
\caption{\small{The SCFT thermodynamic potential barrier height as a function of the correlation length, $\xi$, 
		for different values of fixed virial parameters indicated in the figure. 
		The adsorption strength is $A = 200(\bar{b}\simeq 0.221)$. We multiplied the curve for $w_N=0$ by 0.25.}}
\label{therm_barrier_vs_xi_a200_fig}
\end{minipage}
\hfill
\begin{minipage}[ht]{0.5\linewidth}
\center{\includegraphics[width=1\linewidth]{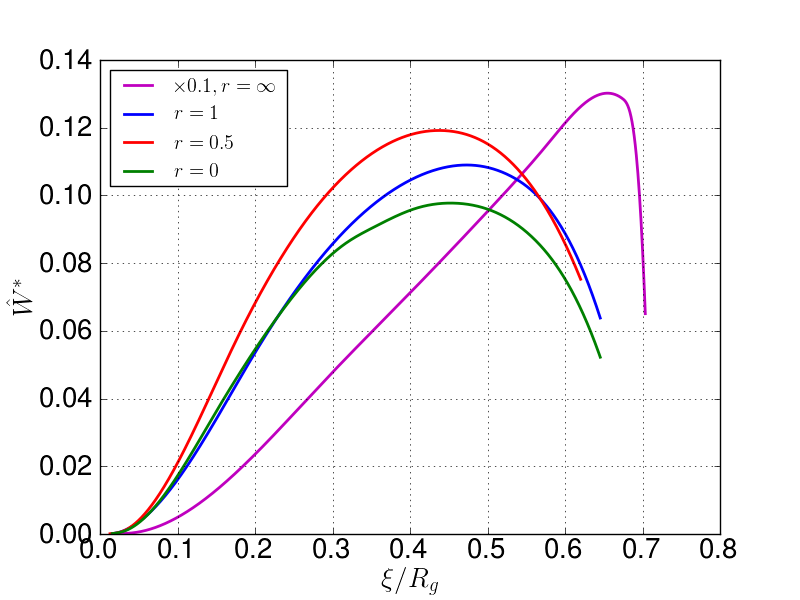}}
\caption{\small{The SCFT thermodynamic potential barrier height as a function of the correlation length, $\xi$, 
		for different values of fixed virial parameters indicated in the figure. 
		The adsorption strength, $A = 500(\bar{b}\simeq 0.097)$. We multiplied the curve for $w_N=0$ by 0.1.}}
\label{therm_barrier_vs_xi_a500_fig}
\end{minipage}
\end{figure}

	 In this section we will find the values of the virial parameters, which provide the highest potential barriers.
	 Similarly to the case of purely repulsive walls, we denote the SCFT thermodynamic potential barrier height as
$$
       	 \hat{W}^* = \hat{W}_{max} - \hat{W}_{inf}
$$
	and the corresponding dimensionless position of the maximum, $\bar{h}^*$. Then, varying the virial parameters we find the maximum repulsion between 
	plates for different values of the adsorption strength, $A$. 
	In Figs.\ref{therm_barrier_vs_xi_a42_fig}--\ref{therm_barrier_vs_xi_a500_fig} 
	we present the dependence of the SCFT barrier height on the bulk correlation length for different values of the fixed parameter,
	$r=v_N/2w_N=\text{v}/\text{w}c_b$. The bulk correlation length is related to the virial parameters as
	$\bar{\xi}=1/\sqrt{2(v_N+2w_N)+2}$, where $v_N=\text{v}c_bN$ and $w_N=\text{w}c_b^2N/2$ are reduced virial coefficients. 	
	The virial parameters that produce the corresponding peak values are shown separately in Tab.\ref{tabular:therm_barrier_vs_xi}. 
	Since the position of the peak corresponds to quite big correlation length, $\bar{\xi}\sim 0.45$($v_N, w_N \sim 1$), we used the SCFT calculations
	to establish the position of the peak and its surroundings. We already demonstrated that the SCFT calculations coincide with the GSDE results for small 
	correlation length. We used the fact to reproduce the low-$\bar{\xi}$ tails for the curves obtained with the GSDE. 
	The tails correspond to high concentration of polymer solution and their importance will become clear later when we 
	consider real examples of polymer solutions.
	The intermediate values for the curves are obtained using an interpolation procedure.
%%%%%%%%%%%%%%%%%%%%%%%%%%%%%%%%%%%%%%%%%%%%%%%%%%%%%%%%%%%%%%%%%%%%%%%%%%%%%%%%%%%%%%%%%%%%%%%%%%%%%%%%%%%%%%%%%%%%%%%%%%%%%%%
%        barrier height vs xi for the different A = 42.73, 100
%%%%%%%%%%%%%%%%%%%%%%%%%%%%%%%%%%%%%%%%%%%%%%%%%%%%%%%%%%%%%%%%%%%%%%%%%%%%%%%%%%%%%%%%%%%%%%%%%%%%%%%%%%%%%%%%%%%%%%%%%%%%%%%
%%%%%%%%%%%%%%%%%%%%%%%%%%%%%%%%%%%%%%%%%%%%%%%%%%%%%%%%%%%%%%%%%%%%%%%%%%%%%%%%%%%%%%%%%%%%%%%%%%%%%%%%%%%%%%%%%%%%%%%%%%%%%%%
%        barrier height vs xi for the different A = 200, 500
%%%%%%%%%%%%%%%%%%%%%%%%%%%%%%%%%%%%%%%%%%%%%%%%%%%%%%%%%%%%%%%%%%%%%%%%%%%%%%%%%%%%%%%%%%%%%%%%%%%%%%%%%%%%%%%%%%%%%%%%%%%%%%%

	One can notice in the figures that all curves are squeezed between those for $w_N=0$ and for $v_N=0$.
	In comparison with the case of purely repulsive walls, the boundary curves are reversed. 
	Next, the barrier heights for $w_N=0$ grow much faster than for $w_N\ne0$ when we increase the adsorption strength. 	
	Therefore, the system described by only the second virial coefficient differs significantly from the systems described by 
	the second and third virial coefficients.		
	
%%%%%%%%%%%%%%%%%%%%%%%%%%%%%%%%%%%%%%%%%%%%%%%%%%%%%%%%%%%%%%%%%%%%%%%%%%%%%%%%%%%%%%%%%%%%%%%%%%%%%%%%%%%%%%%%%%%%%%%%%%%%%%%%%%%%%%%%%%%%
% The height of the barrier for vw: adsorption 
%%%%%%%%%%%%%%%%%%%%%%%%%%%%%%%%%%%%%%%%%%%%%%%%%%%%%%%%%%%%%%%%%%%%%%%%%%%%%%%%%%%%%%%%%%%%%%%%%%%%%%%%%%%%%%%%%%%%%%%%%%%%%%%%%%%%%%%%%%%%
\begin{table}[ht!]
\caption{The value of the thermodynamic potential barrier height, $\hat{W}^*$, and its position $\bar{h}^*$ corresponding to the peak values in
	 Figs.\ref{therm_barrier_vs_xi_a42_fig}--\ref{therm_barrier_vs_xi_a500_fig}.} 
\label{tabular:therm_barrier_vs_xi} 
\begin{center}
  \begin{tabular}{ | c | c | c | c | c |}
    \hline
                                             \multicolumn{5}{|c|}{A=42.73}        \\ \hline
      $r$                      &$\infty$&     0&   0.5 &    1    \\ \hline                                             
    $v_N$                      &       2&     0&   0.8 &  0.4    \\ \hline
    $w_N$                      &       0&     1&   0.8 &  0.8    \\ \hline                                             
    $\hat{W}^*,\times 10^{-3}$ &   3.173& 2.354&  2.52 &  2.475  \\ \hline
    $\bar{h}^*$                      &   2.351& 2.312& 2.218 &  2.314  \\ \hline	
                                             \multicolumn{5}{|c|}{A=100}          \\ \hline                           
      $r$                      &$\infty$&     0&   0.5 &    1    \\ \hline                                                                                          
    $v_N$                      &       1&     0&   0.5 &  0.3    \\ \hline
    $w_N$                      &       0&   0.8&   0.5 &  0.6    \\ \hline                                             
    $\hat{W}^*,\times 10^{-2}$ &   2.802& 1.178& 1.382 &  1.289  \\ \hline
    $h_m$                      &   2.494& 2.226& 2.271 &  2.266  \\ \hline
                                             \multicolumn{5}{|c|}{A=200}          \\ \hline
      $r$                      &$\infty$&     0&   0.5 &  1      \\ \hline                                                                                          
    $v_N$                      &     0.8&     0&   0.5 &  0.3    \\ \hline
    $w_N$                      &       0&   0.8&   0.5 &  0.6    \\ \hline                                             
    $\hat{W}^*,\times 10^{-2}$ &  16.323& 3.396& 4.129 &  3.787  \\ \hline
    $h_m$                      &   2.264& 2.024& 2.069 &  2.059  \\ \hline
                                             \multicolumn{5}{|c|}{A=500}          \\ \hline                                                                                          
      $r$                      &$\infty$&     0&   0.5 &    1    \\ \hline                                                                                          
    $v_N$                      &     0.1&     0&    0.5&   0.3   \\ \hline
    $w_N$                      &       0&   0.7&    0.5&   0.6   \\ \hline                                             
    $\hat{W}^*,\times 10^{-2}$ & 128.591& 9.772&  11.91&   10.896\\ \hline
    $h_m$                      & 2.333  & 1.829&  1.819&   1.814 \\
    \hline
  \end{tabular}
\end{center} 
\end{table}

%%%%%%%%%%%%%%%%%%%%%%%%%%%%%%%%%%%%%%%%%%%%%%%%%%%%%%%%%%%%%%%%%%%%%%%%%%%%%%%%%%%%%%%%%%%%%%%%%%%%%%%%%%%%%%%%%%%%%%%%%%%%%%%%%%%%%%%%%%%%%%%%%%%%%%%%%%%%%
%           The virial parameters providing the maximum barrier height.
%%%%%%%%%%%%%%%%%%%%%%%%%%%%%%%%%%%%%%%%%%%%%%%%%%%%%%%%%%%%%%%%%%%%%%%%%%%%%%%%%%%%%%%%%%%%%%%%%%%%%%%%%%%%%%%%%%%%%%%%%%%%%%%%%%%%%%%%%%%%%%%%%%%%%%%%%%%%%
\section{The virial parameters providing the maximum barrier height. Spherical case} 
\begin{figure}[ht!]
\begin{minipage}[ht]{0.5\linewidth}
\center{\includegraphics[width=1\linewidth]{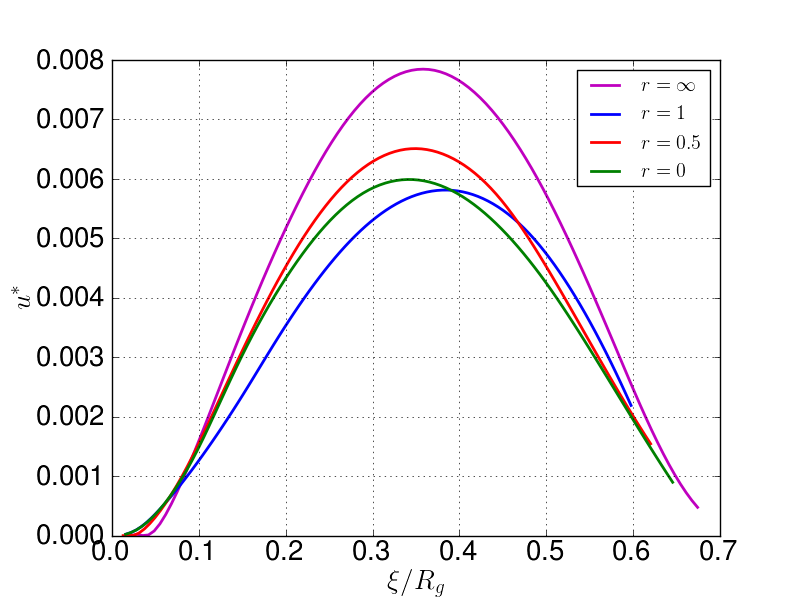}}
\caption{\small{The reduced SCFT barrier height, Eq.(\ref{ads_ans_real_ustar_def}), calculated for spherical geometry as a function of the correlation length.
		The barrier height maximum for $w_N\ne 0$ corresponds to $\xi/R_g\sim 0.35$. The adsorption strength is $A = 42.73 (\bar{b}\simeq 0.986)$.}}
\label{sph_u_star_vs_xi_a42_fig}
\end{minipage}
\hfill
\begin{minipage}[ht]{0.5\linewidth}
\center{\includegraphics[width=1\linewidth]{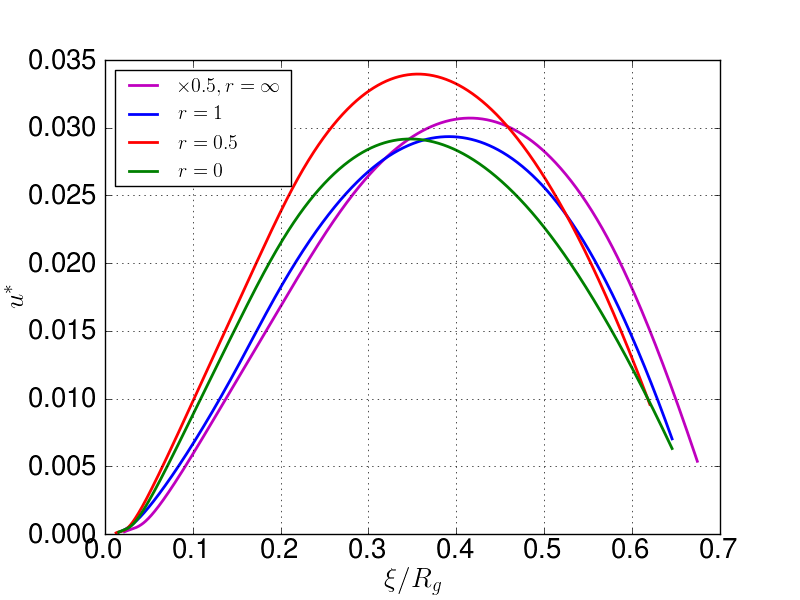}}
\caption{\small{The reduced SCFT barrier height, Eq.(\ref{ads_ans_real_ustar_def}), calculated for spherical geometry as a function of the correlation length.
		The barrier height maximum for $w_N\ne 0$ corresponds to $\xi/R_g\sim 0.35$. The adsorption strength is $A = 100 (\bar{b}\simeq 0.429)$. 
		We multiplied the curve for $w_N=0$ by 0.5}}
\label{sph_u_star_vs_xi_a100_fig}
\end{minipage}
\end{figure}

\begin{figure}[ht!]
\begin{minipage}[ht]{0.5\linewidth}
\center{\includegraphics[width=1\linewidth]{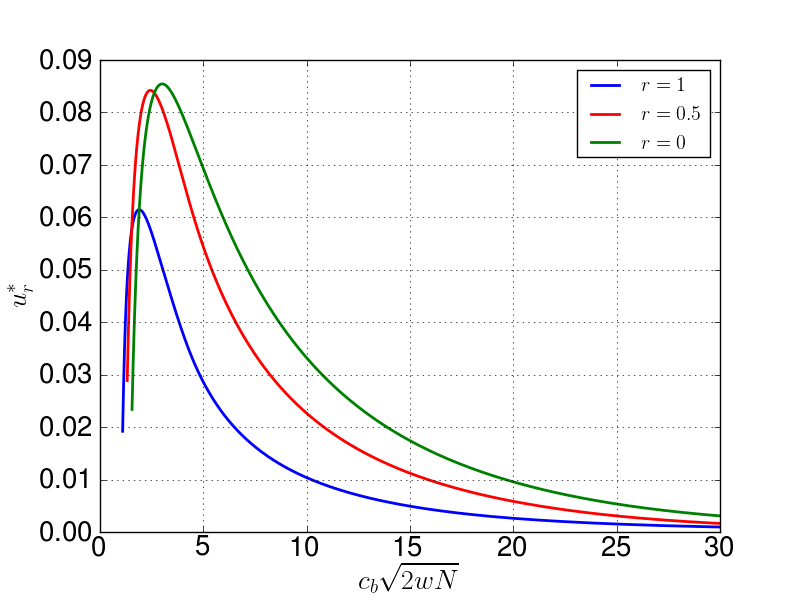}}
\caption{\small{The reduced SCFT barrier height, Eq.(\ref{ads_ans_real_ustar_def}), calculated for spherical geometry as a function of the correlation length.
		The barrier height maximum for $w_N\ne 0$ corresponds to $\xi/R_g\sim 0.35$. 
		The adsorption strength is $A = 200 (\bar{b}\simeq 0.221)$. We multiplied the curve for $w_N=0$ by 0.25.}}
\label{sph_u_star_vs_xi_a200_fig}
\end{minipage}
\hfill
\begin{minipage}[ht]{0.5\linewidth}
\center{\includegraphics[width=1\linewidth]{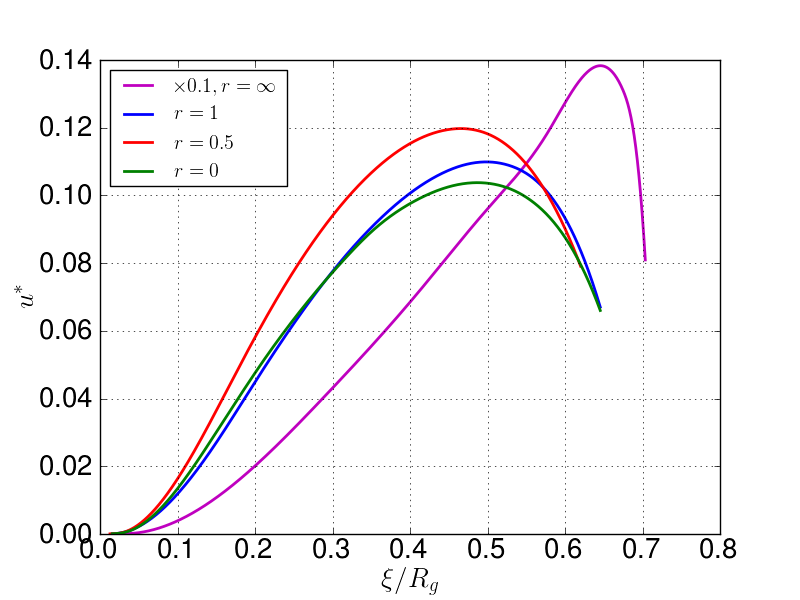}}
\caption{\small{The reduced SCFT barrier height, Eq.(\ref{ads_ans_real_ustar_def}), calculated for spherical geometry as a function of the correlation length.
		The barrier height maximum for $w_N\ne 0$ corresponds to $\xi/R_g\sim 0.45$. 
		The adsorption strength is $A = 500 (\bar{b}\simeq 0.097)$. We multiplied the curve for $w_N=0$ by 0.1.}}
\label{sph_u_star_vs_xi_a500_fig}
\end{minipage}
\end{figure}

	Using the Derjaguin approximation, in a spherical geometry we can write the corresponding expression for the thermodynamic potential:
$$
        U(h) = \pi R_c \int\limits_h^{\infty}\mathrm{d}h'W(h') = \pi R_c \frac{c_bR_g}{N} \int\limits_h^{\infty}\mathrm{d}h'\hat{W}(h') = 
                    \pi R_c \frac{c_bR_g^2}{N} \int\limits_{\bar{h}}^{\infty}\mathrm{d}\bar{h}\hat{W}(\bar{h}) = A_R \hat{U}(\bar{h})
$$
	Thereby:
\begin{equation}
\label{ads_ans_scft_therm_pot_real}
	U(h) = A_R \hat{U}(h/R_g)
\end{equation}
	where we denoted:
$$
	A_R = \pi R_c \frac{c_bR_g^2}{N} = \pi\left(\frac{R_c}{R_g}\right)\frac{c_bR_g^3}{N}
$$
	is a prefactor that defines the magnitude of the barrier and
\begin{equation}
\label{ads_ans_scft_therm_pot_real_reduced}
	\hat{U}(\bar{h}) = \int\limits_{\bar{h}}^{\infty}\mathrm{d}\bar{h}\hat{W}(\bar{h})   
\end{equation}
	is the dimensionless thermodynamic potential for the spherical geometry, which is expressed via the thermodynamic potential obtained in the SCFT 
	for the flat geometry. We denote the corresponding barrier height as $\hat{U}^*$. 
	As we already did it for the purely repulsive case, the bulk polymer concentration can be written in the form:
$$
	c_b = \frac{1}{\sqrt{2\text{w}N}\sqrt{1+r}\bar{\xi}}
$$ 
	where $\bar{\xi}$ is the dimensionless correlation length and $r=v_N/2w_N$ is the parameter that determines the solvent regime.  
	For semidilute solution, this parameter slightly depends on polymer concentration. 
	So, we can rewrite the prefactor as
$$
	A_R =  \frac{\pi a^3}{\sqrt{2\text{w}}}\left(\frac{R_c}{R_g}\right)\frac{1}{\sqrt{1+r}\bar{\xi}} = \frac{B}{\bar{\xi}}
$$
	where we introduced
$$
	B =  \frac{\pi a^3}{\sqrt{2\text{w}}}\left(\frac{R_c}{R_g}\right)\frac{1}{\sqrt{1+r}} 
$$
	Therefore, we can write 
$$
	U(h, \text{v}, \text{w}, c_b, R_c, a_s, N, b) = B u(\bar{h}, \bar{\xi}, \bar{b}, r)
$$
        where we introduced $u = \hat{U}/\bar{\xi}$. In such a way, we can find the stability condition for the dimensionless function $u$ as
\begin{equation}
\label{ads_ans_real_ustar_def}
\begin{array}{lr}
	\text{max}_{\bar{h}}\, u(\bar{h}, \bar{\xi}, \bar{b}, r) = u^*(\bar{\xi}, \bar{b}, r) & (a)\\
	\text{max}_{\bar{\xi}}\, u^*(\bar{\xi}, \bar{b}, r) = u^{**}(\bar{b}, r)              & (b)
\end{array}
\end{equation}
	The position of the maximum for the function $u$ by the variable $h$ is the same as for the function $\hat{U}$. 
	In Figs.\ref{sph_u_star_vs_xi_a42_fig}--\ref{sph_u_star_vs_xi_a500_fig} we present the dependences $u^*(\bar{\xi})$ for different adsorption strengths. 

%%%%%%%%%%%%%%%%%%%%%%%%%%%%%%%%%%%%%%%%%%%%%%%%%%%%%%%%%%%%%%%%%%%%%%%%%%%%%%%%%%%%%%%%%%%%%%%%%%%%%%%%%%%%%%%%%%%%%%%%%%%%%%%
%        barrier height vs xi for the different A = 42.73, 100
%%%%%%%%%%%%%%%%%%%%%%%%%%%%%%%%%%%%%%%%%%%%%%%%%%%%%%%%%%%%%%%%%%%%%%%%%%%%%%%%%%%%%%%%%%%%%%%%%%%%%%%%%%%%%%%%%%%%%%%%%%%%%%%
%%%%%%%%%%%%%%%%%%%%%%%%%%%%%%%%%%%%%%%%%%%%%%%%%%%%%%%%%%%%%%%%%%%%%%%%%%%%%%%%%%%%%%%%%%%%%%%%%%%%%%%%%%%%%%%%%%%%%%%%%%%%%%%
%        barrier height vs xi for the different A = 200, 500
%%%%%%%%%%%%%%%%%%%%%%%%%%%%%%%%%%%%%%%%%%%%%%%%%%%%%%%%%%%%%%%%%%%%%%%%%%%%%%%%%%%%%%%%%%%%%%%%%%%%%%%%%%%%%%%%%%%%%%%%%%%%%%%
%	  \\
          It is also very useful to represent 
\begin{equation}
\label{ans_scft_ur_vs_cb}
	  u^*_r = \frac{u^*}{\sqrt{1 + r}} \quad\text{as a function of}\quad c_b\sqrt{2\text{w}N} = \frac{1}{\sqrt{1+ r}\bar{\xi}} 
\end{equation}

%%%%%%%%%%%%%%%%%%%%%%%%%%%%%%%%%%%%%%%%%%%%%%%%%%%%%%%%%%%%%%%%%%%%%%%%%%%%%%%%%%%%%%%%%%%%%%%%%%%%%%%%%%%%%%%%%%%%%%%%%%%%%%%
%        barrier height vs xi for the different A = 42.73, 100
%%%%%%%%%%%%%%%%%%%%%%%%%%%%%%%%%%%%%%%%%%%%%%%%%%%%%%%%%%%%%%%%%%%%%%%%%%%%%%%%%%%%%%%%%%%%%%%%%%%%%%%%%%%%%%%%%%%%%%%%%%%%%%%
\begin{figure}[ht!]
\begin{minipage}[ht]{0.5\linewidth}
\center{\includegraphics[width=1\linewidth]{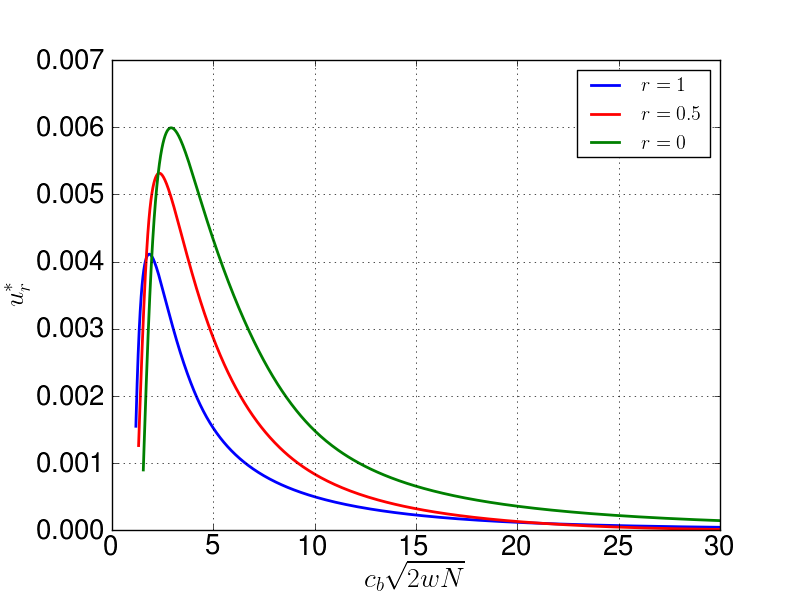}}
\caption{\small{The reduced SCFT barrier height, Eq.(\ref{ans_scft_ur_vs_cb}), calculated for spherical geometry as a function of the reduced monomer concentration.
		The adsorption strength is $A = 42.73 (\bar{b}\simeq 0.986)$.}}
\label{sph_u_star_vs_cb_a42_fig}
\end{minipage}
\hfill
\begin{minipage}[ht]{0.5\linewidth}
\center{\includegraphics[width=1\linewidth]{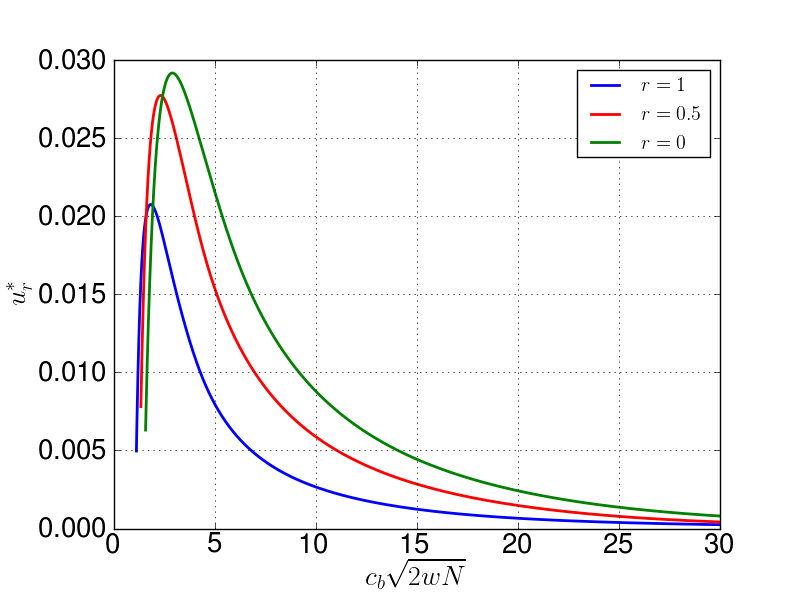}}
\caption{\small{The reduced SCFT barrier height, Eq.(\ref{ans_scft_ur_vs_cb}), calculated for spherical geometry as a function of the reduced monomer concentration.
		The adsorption strength is $A = 100(\bar{b}\simeq 0.429)$.}}
\label{sph_u_star_vs_cb_a100_fig}
\end{minipage}
\end{figure} 
%%%%%%%%%%%%%%%%%%%%%%%%%%%%%%%%%%%%%%%%%%%%%%%%%%%%%%%%%%%%%%%%%%%%%%%%%%%%%%%%%%%%%%%%%%%%%%%%%%%%%%%%%%%%%%%%%%%%%%%%%%%%%%%
%        barrier height vs xi for the different A = 200, 500
%%%%%%%%%%%%%%%%%%%%%%%%%%%%%%%%%%%%%%%%%%%%%%%%%%%%%%%%%%%%%%%%%%%%%%%%%%%%%%%%%%%%%%%%%%%%%%%%%%%%%%%%%%%%%%%%%%%%%%%%%%%%%%%
\begin{figure}[ht!]
\begin{minipage}[ht]{0.5\linewidth}
\center{\includegraphics[width=1\linewidth]{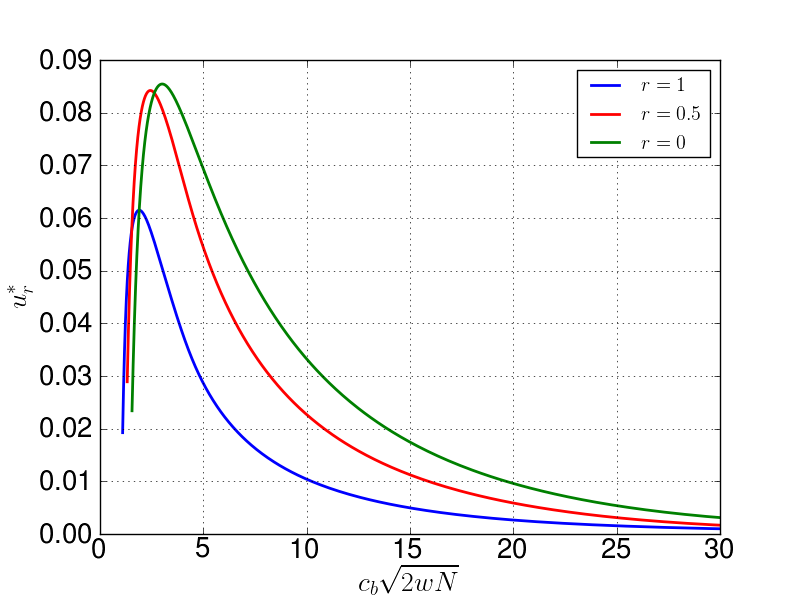}}
\caption{\small{The reduced SCFT barrier height, Eq.(\ref{ans_scft_ur_vs_cb}), calculated for spherical geometry as a function of the reduced monomer concentration.
		The adsorption strength is $A = 200 (\bar{b}\simeq 0.221)$.}}
\label{sph_u_star_vs_cb_a200_fig}
\end{minipage}
\hfill
\begin{minipage}[ht]{0.5\linewidth}
\center{\includegraphics[width=1\linewidth]{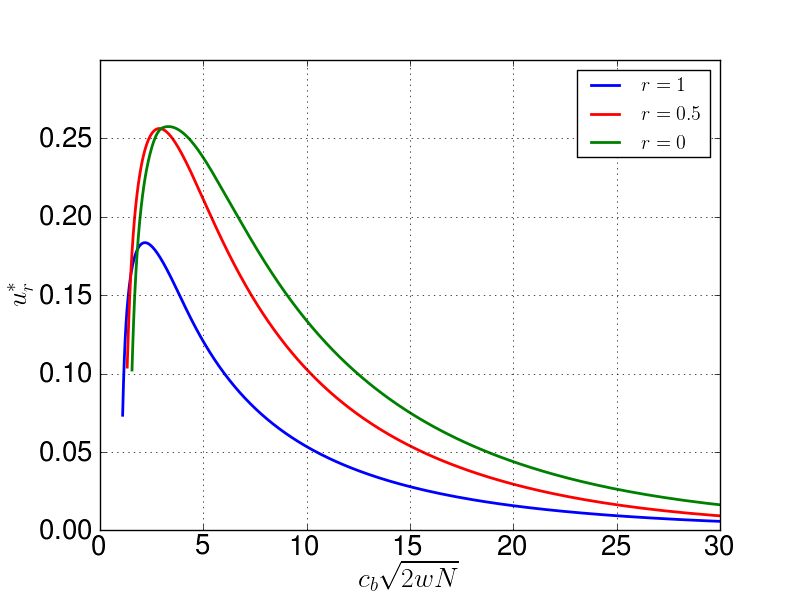}}
\caption{\small{The reduced SCFT barrier height, Eq.(\ref{ans_scft_ur_vs_cb}), calculated for spherical geometry as a function of the reduced monomer concentration.
		The adsorption strength is $A = 500 (\bar{b}\simeq 0.097)$.}}
\label{sph_u_star_vs_cb_a500_fig}
\end{minipage}
\end{figure} 
	  
	  \textbf{Upper limits for the virial parameters}. In order to analyse the accessible range of the correlation length, we can rewrite the expression for the polymer 
	  volume fraction as
$$
	  \phi_b = \frac{M_0}{N_A\rho}c_b = \frac{M_0}{N_A\rho\sqrt{2\text{w}}}\frac{c_b\sqrt{2\text{w}N}}{\sqrt{N}} = \frac{\phi_1}{\sqrt{N}} c_b\sqrt{2\text{w}N} =
			                    \frac{\phi_1}{\sqrt{N}} \frac{1}{\sqrt{1+r}\bar{\xi}}
$$
	  where we introduced $\phi_1 = M_0/(N_A\rho\sqrt{2\text{w}})$. This constant does not depend on the polymer length. 
	  For the model polymers
	  $\phi_1(\text{PS}) = 0.747$ and $\phi_1(\text{PEG}) = 0.37$, so that now we can easily find the dimensionless correlation length, $\bar{\xi}$, for any $\phi_b$. 
	  In particular, to get the virial parameters, $v_N$ and $w_N$, we can use the RPA definition of the correlation length, 
	  $\bar{\xi}=1/\sqrt{2(v_N+2w_N)+2}$ written via the parameters, $v_N$ and $r=v_N/(2w_N)$. Correspondingly, we have
\begin{equation}
\label{ads_volume_fraction_vN_threshold}
	  v_N = \frac{r}{2(1+r)}\left(N(1+r)\left(\frac{\phi_b}{\phi_1}\right)^2 - 2\right) \simeq \frac{rN}{2}\left(\frac{\phi_b}{\phi_1}\right)^2, \quad w_N = \frac{v_N}{2r}
\end{equation}
	  Alternatively, in the case $r=0$ starting from the definition of the correlation length, we get
\begin{equation}
\label{ads_volume_fraction_wN_threshold}
	  w_N = \frac{1}{4}\left(N\left(\frac{\phi_b}{\phi_1}\right)^2-2\right) \simeq \frac{N}{4}\left(\frac{\phi_b}{\phi_1}\right)^2
\end{equation}
	  Therefore, to set the upper bound for the virial parameters, we should find the accessible bulk volume fraction for the 
	  corresponding polymer.
%%%%%%%%%%%%%%%%%%%%%%%%%%%%%%%%%%%%%%%%%%%%%%%%%%%%%%%%%%%%%%%%%%%%%%%%%%%%%%%%%%%%%%%%%%%%%%%%%%%%%%%%%%%%%%%%%%%%%%%%%%%%%%%%%%%%%%%%%%%%%%%%%%%%%%%%%%%%%
%           Bulk concentration restrictions.
%%%%%%%%%%%%%%%%%%%%%%%%%%%%%%%%%%%%%%%%%%%%%%%%%%%%%%%%%%%%%%%%%%%%%%%%%%%%%%%%%%%%%%%%%%%%%%%%%%%%%%%%%%%%%%%%%%%%%%%%%%%%%%%%%%%%%%%%%%%%%%%%%%%%%%%%%%%%%
\section{Bulk concentration restrictions}
	 For adsorption case, the contact wall concentration is always bigger than the concentration 
	 in the bulk phase, i.e. $c_0>c_b$. In addition, the contact volume fraction should be $\phi_0 < 1$. 
	 Therefore, we have the restriction for the bulk monomer concentration that depends on the adsorption strength and the degree of polymerization. 
	 The easiest way to assess the contact polymer volume fraction, $\phi_0$ is to use a complete analytical formula. 
	 The expression, Eq.(\ref{ans_adsorption_gs_f0}), for the contact concentration on the adsorption wall depends on the correlation 
	 length and adsorption strength (in the case when the third virial coefficient is not zero it is valid in the WAR):	 
$$
       	 \sqrt{c_0/c_b} = f_0 = \frac{\bar{\xi}}{\bar{b}} + \sqrt{1 + \left(\frac{\bar{\xi}}{\bar{b}}\right)^2}
$$
	 In terms of volume fraction we can rewrite it as
$$
	 \phi_0 = \left(\frac{\bar{\xi}}{\bar{b}} + \sqrt{1 + \left(\frac{\bar{\xi}}{\bar{b}}\right)^2}\right)^2\phi_b 
$$
	 with the dimensionless correlation length
$$
	 \bar{\xi} = \frac{1}{\sqrt{2(v_N + 2w_N) + 2}}
$$
	 where $v_N = \text{v}c_bN$, $w_N = \text{w}c_b^2N/2$, as before, are the second and third virial parameters. 
	 The relation between monomer concentration and polymer volume fraction is 
$$
	\phi_b = \frac{M_0}{N_a\rho}c_b ,\quad  \quad c_b = \phi_b \frac{N_a\rho}{M_0}
$$
	Therefore, finally we can write  
\begin{equation}
\label{volume_fraction_phi0_phib_war}
	\phi_0 = F(\text{v}, \text{w}, b, N, \phi_b)
\end{equation}
	A more general relationship can be found from the expression for the first integral of the Edwards equation in the GSD approximation 
$$
        a^2\left(\frac{\mathrm{d}f}{\mathrm{d}x}\right)^2 = \frac{c_b}{2}(f^2-1)^2\left(\text{v} + \frac{\text{w}c_b}{3}(f^2+2)\right)
$$
	At the wall, $x=0$ we can use the  boundary condition, $(f'/f)_{x=0} = -1/b$, that leads to
$$
        \frac{a^2}{b^2}f^2_0 = \frac{c_b}{2}(f^2_0-1)^2\left(\text{v} + \frac{\text{w}c_b}{3}(f^2_0+2)\right)
$$	
	where, as before, $f_0^2 = c_0/c_b = \phi_0/\phi_b$. Introducing a new variable, $g_0=f^2_0$ and then multiplying both sides of the above equation 
	by $N$ we can rewrite it in our reduced variables 
$$
         \frac{\bar{b}^2}{2}(g_0-1)^2\left(v_N + \frac{2w_N}{3}(g_0+2)\right) - g_0 = \frac{\bar{b}^2}{2}(g_0-1)^2\left(v_N + 2w_N + \frac{2w_N}{3}(g_0-1)\right) - g_0 = 0
$$	
	This is equivalent to 
\begin{equation}
\label{volume_fraction_cubic_eq}
         \frac{\bar{b}^2}{4\bar{\xi}^2}(g_0-1)^2\left(1 + \frac{4\bar{\xi}^2w_N}{3}(g_0-1)\right) - g_0 = 0
\end{equation}
	So, solving this cubic equation for the certain input parameters ($N, \text{v}, \text{w}, c_b, \bar{b}$) 
	we obtain the corresponding contact volume fraction
\begin{equation}
\label{volume_fraction_phi0_phib_general}
       \phi_0 = G(\text{v}, \text{w}, b, N, \phi_b)
\end{equation}	
	For simplicity, we can solve Eq.(\ref{volume_fraction_cubic_eq}), for example, by the bisection method seeking the corresponding 
	root for $g_0$ in the range $g_0\in[1..500]$.  The difference between Eq.(\ref{volume_fraction_phi0_phib_war}) obtained in the WAR and 
	the more general Eq.(\ref{volume_fraction_phi0_phib_general}) is sufficiently small for the weak adsorption strength (big $b$).
	Then, when we increase the adsorption strength (decrease $b$) the results obtained with Eq.(\ref{volume_fraction_phi0_phib_war}) (valid in the WAR)	
        significantly overstates the results obtained with more general Eq.(\ref{volume_fraction_phi0_phib_general}). 
	We are interested in the following real polymer systems:\\	
	\textbf{Polystyrene (PS) in toluene solvent}. In Sec.\ref{sec:polymers_experiment} we found the parameters of polystyrene in toluene solvent at T=25 ($^\circ$C): 	
$$
\begin{array}{c}
	M_0(\text{styrene})=104.15\,\text{g/mol}, \quad a_s = 7.6\textup{\AA}, \quad \text{v} = 23.21\textup{\AA}^3, \quad \text{w} = 2.79\times 10^4 \textup{\AA}^6, \\ 
	\quad \rho = 960\,g/L
\end{array}	
$$
	\\
	\textbf{Polyethylene glycol(PEG) in water solvent}. The corresponding parameters are found in the same section for T=25 ($^\circ$C):
$$	  
\begin{array}{c}
	M_0(\text{ethylene oxide}) = 44.05\,\text{g/mol}, \quad a_s = 5.5\textup{\AA}, \quad \text{v} = 12.24\textup{\AA}^3, \quad \text{w} = 1.54\times 10^4 \textup{\AA}^6,\\
        \quad \rho = 1128\,g/L    
\end{array}      
$$
	
%%%%%%%%%%%%%%%%%%%%%%%%%%%%%%%%%%%%%%%%%%%%%%%%%%%%%%%%%%%%%%%%%%%%%%%%%%%%%%%%%%%%%%%%%%%%%%%%%%%%%%%%%%%%%%%%%%%%%%%%%%%%%%%
%        barrier height vs xi for the different A = 42.73, 100
%%%%%%%%%%%%%%%%%%%%%%%%%%%%%%%%%%%%%%%%%%%%%%%%%%%%%%%%%%%%%%%%%%%%%%%%%%%%%%%%%%%%%%%%%%%%%%%%%%%%%%%%%%%%%%%%%%%%%%%%%%%%%%%
\begin{figure}[ht!]
\begin{minipage}[ht]{0.5\linewidth}
\center{\includegraphics[width=1\linewidth]{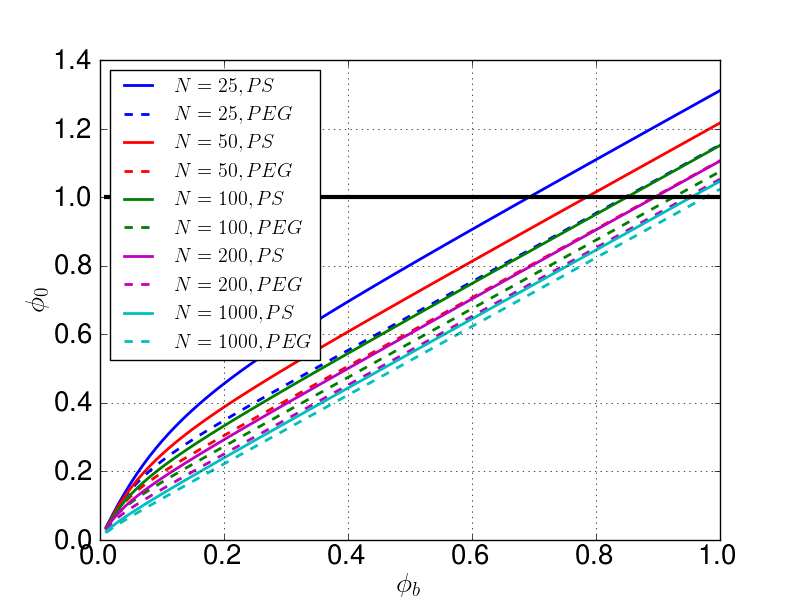}}
\caption{\small{The dependence of the volume fraction at the wall as a function of the bulk volume fraction for different chain length. 
		The adsorption strength is $A = 42.73 (\bar{b}\simeq 0.986)$.}}
\label{phi0_phib_a42_fig}
\end{minipage}
\hfill
\begin{minipage}[ht]{0.5\linewidth}
\center{\includegraphics[width=1\linewidth]{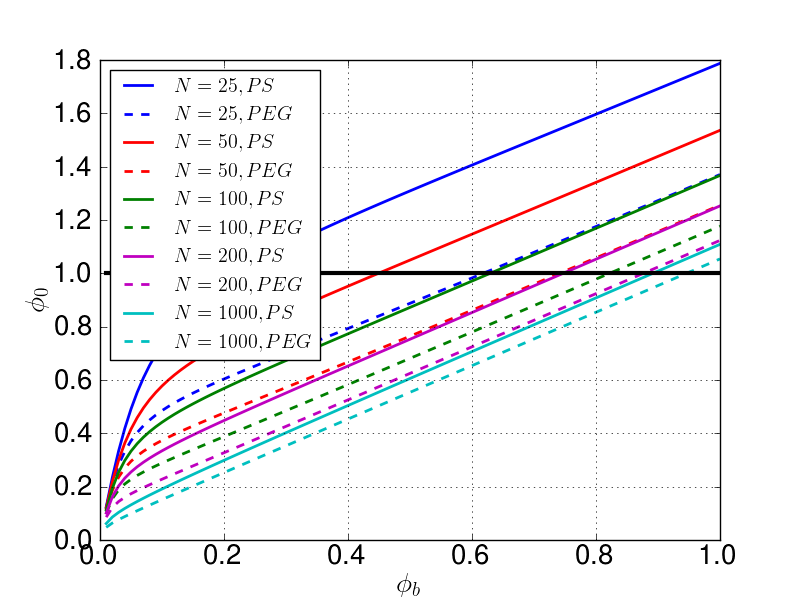}}
\caption{\small{The dependence of the volume fraction at the wall as a function of the bulk volume fraction for different chain length. 
		The adsorption strength is $A = 100 (\bar{b}\simeq 0.429)$. }}
\label{phi0_phib_a100_fig}
\end{minipage}
\end{figure} 
%%%%%%%%%%%%%%%%%%%%%%%%%%%%%%%%%%%%%%%%%%%%%%%%%%%%%%%%%%%%%%%%%%%%%%%%%%%%%%%%%%%%%%%%%%%%%%%%%%%%%%%%%%%%%%%%%%%%%%%%%%%%%%%
%        barrier height vs xi for the different A = 200, 500
%%%%%%%%%%%%%%%%%%%%%%%%%%%%%%%%%%%%%%%%%%%%%%%%%%%%%%%%%%%%%%%%%%%%%%%%%%%%%%%%%%%%%%%%%%%%%%%%%%%%%%%%%%%%%%%%%%%%%%%%%%%%%%%
\begin{figure}[ht!]
\begin{minipage}[ht]{0.5\linewidth}
\center{\includegraphics[width=1\linewidth]{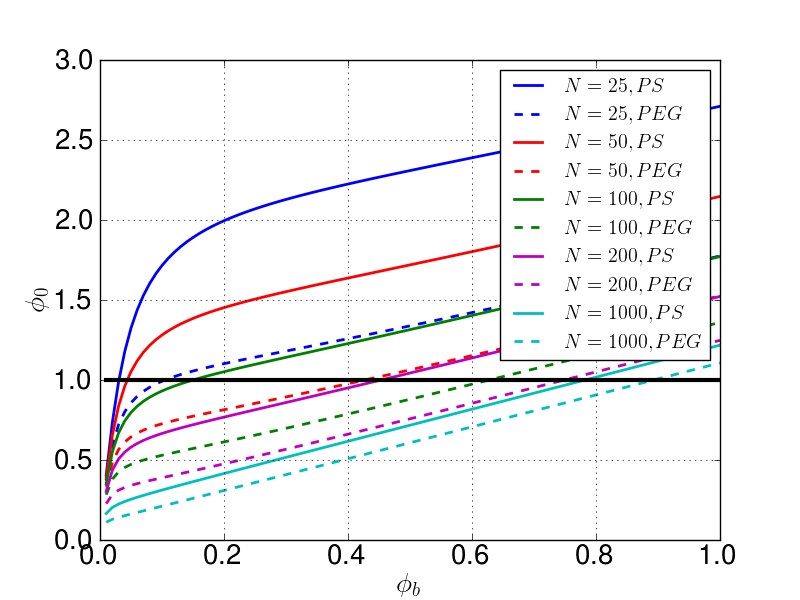}}
\caption{\small{The dependence of the volume fraction at the wall as a function of the bulk volume fraction for different chain length. 
		The adsorption strength is $A = 200 (\bar{b}\simeq 0.221)$.}}
\label{phi0_phib_a200_fig}
\end{minipage}
\hfill
\begin{minipage}[ht]{0.5\linewidth}
\center{\includegraphics[width=1\linewidth]{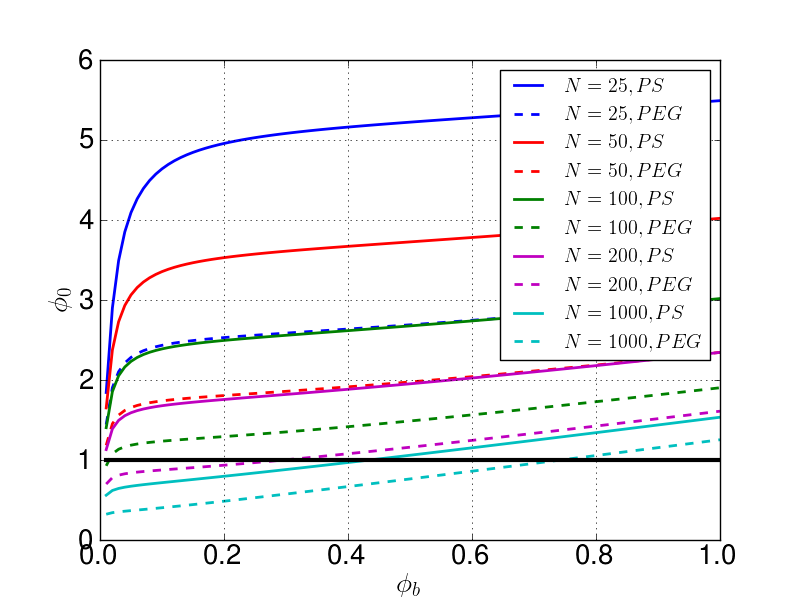}}
\caption{\small{The dependence of the volume fraction at the wall as a function of the bulk volume fraction for different chain length. 
		The adsorption strength is $A = 500 (\bar{b}\simeq 0.097)$.}}
\label{phi0_phib_a500_fig}
\end{minipage}
\end{figure} 
	
	  One can notice in Figs.\ref{phi0_phib_a42_fig}--\ref{phi0_phib_a500_fig} that for certain bulk polymer volume fractions 
	  the contact volume fraction, $\phi_0$, has no physical meaning since $\phi_0$ exceeds $1$. 
	  All the curves are increasing functions of the bulk concentration, so setting $\phi_0=1$ we can find the boundary
	  of the applicability of our calculations.
	  We can not use our model for larger values of the volume fraction. 
	  The range of allowed bulk volume fractions at fixed polymer length is decreasing for higher adsorption strength. 	  
	  For example, at the strongest considered adsorption strength, Fig.\ref{phi0_phib_a500_fig}, the range of bulk volume fractions 
	  does not vanish for long chains. 	 
	  Using the obtained values for the marginal bulk volume fraction along with Eqs.(\ref{ads_volume_fraction_vN_threshold}, \ref{ads_volume_fraction_wN_threshold}),
	  we can find the range for the virial parameters that cover all the accessible values for the bulk volume fraction. 
	  For simplicity, we used in Eq.(\ref{ads_volume_fraction_vN_threshold}) only one value, $r=1$, since the dependence is linear and 
	  to establish the results with any other $r$ is simple.
	  The results for the marginal value of the virial parameters are present in Tab.\ref{tabular:therm_barrier_restriction_phib} for different adsorption strength. 	  	  
	  
	  Now, we should find out if the maximum is included or not in the accessible range of the virial parameters. 
	  We already noticed that the maximum corresponds to $\bar{\xi} \simeq 0.35$ in Figs.\ref{sph_u_star_vs_xi_a42_fig}--\ref{sph_u_star_vs_xi_a200_fig}. 
	  On the other hand, in the GSD we can write
$$
	  v_N \simeq \frac{r}{2(1+r)\bar{\xi}^2}\quad w_N = \frac{v_N}{2r} 
$$
	  for $r\neq 0$ and $w_N = 1/(4\bar{\xi}^2)$ for $r=0$. 
	  Therefore, we can assess the values of virial parameters that produce maximum. For example,
	  for $r=0$ the maximum corresponds to $w_N \simeq 2$ and $v_N \simeq 2$ (for $r=1$), so for any bigger values of the virial parameters
	  the maximum is included. In this way, using the data listed in Tab.\ref{tabular:therm_barrier_restriction_phib}, we can find
	  that for the adsorption strength, $A = 43.73$ the maximum is included for all the considered values of the polymer length.
	  For $A=100$, the maximum is included for the polymer length starting from $N=50$, and for $A=200$ starting from $N=100$. 
	  	  
	  We can extract from Tab.\ref{tabular:therm_barrier_restriction_phib} other useful informations.
	  For different chain length there are different accessible ranges of the virial parameters. 
	  Technically, it would be much easier to have only one function valid for different considered chain lengths. 	  	  
	  Thereby, we can expand our calculations for the thermodynamic potential,
	  Figs.\ref{sph_u_star_vs_xi_a42_fig}--\ref{sph_u_star_vs_xi_a500_fig}, up to the following virial parameters:
	  $A=42.73, v_N=700, w_N=350$; $A=100, v_N=600, w_N=300$; $A=200, v_N=500, w_N=250$; $A=500, v_N=200, w_N=100$. 
	  It turns out that the potential is sufficiently difficult to calculate at such high values of the virial parameters. 
	  As a result, we can use the coincidence between the SCFT calculations and the GSDE theory that we already established for sufficiently big values of the virial parameters. 
	  We used the same approach to recover the tails of the corresponding curves when we considered depletion stabilization for the purely repulsive walls 
	  (see Sec.\ref{sec:repulsive_max_bar}).
	  
%%%%%%%%%%%%%%%%%%%%%%%%%%%%%%%%%%%%%%%%%%%%%%%%%%%%%%%%%%%%%%%%%%%%%%%%%%%%%%%%%%%%%%%%%%%%%%%%%%%%%%%%%%%%%%%%%%%%%%%%%%%%%%%%%%%%%%%%%%%%
% The height of the barrier for vw: adsorption 
%%%%%%%%%%%%%%%%%%%%%%%%%%%%%%%%%%%%%%%%%%%%%%%%%%%%%%%%%%%%%%%%%%%%%%%%%%%%%%%%%%%%%%%%%%%%%%%%%%%%%%%%%%%%%%%%%%%%%%%%%%%%%%%%%%%%%%%%%%%%
\begin{table}[ht!]
\caption{Evaluation for marginal values of the virial parameters for different chain length and adsorption strength.} 
\label{tabular:therm_barrier_restriction_phib} 
\begin{center}
  \begin{tabular}{ | c | c | c | c | c |c | c | c | c | c | c | c |}
    \hline
	          \multicolumn{6}{|c|}{PS}        & \multicolumn{5}{|c|}{PEG}        \\ \hline
	                                       \multicolumn{11}{|c|}{A=42.73}        \\ \hline
      $N$                      &      25&    50&    100&    200&    1000&     25&    50&    100&    200&    1000  \\ \hline                                             
 $\phi_b$                      &   0.692& 0.785&  0.850&  0.895&   0.954&  0.847& 0.893&  0.925&  0.947&    0.976 \\ \hline                                             
 	                                       \multicolumn{11}{|c|}{r=0}        \\ \hline    
    $w_N$                      &       5&    13&     31&     69&     391&     33&    73&    157&    328&    1744  \\ \hline                                             
 	                                       \multicolumn{11}{|c|}{r=1}        \\ \hline    
    $v_N$                      &      10&    26&     62&    138&     782&     66&   146&    314&    656&    3488 \\ \hline                                                 
                                               \multicolumn{11}{|c|}{A=100}          \\ \hline                           
      $N$                      &      25&    50&    100&    200&    1000&     25&    50&    100&    200&    1000 \\ \hline                                             
 $\phi_b$                      &   0.223& 0.449&  0.629&  0.747&   0.892&  0.618& 0.741&  0.822&  0.876&    0.946 \\ \hline                                                 
       	                                       \multicolumn{11}{|c|}{r=0}        \\ \hline    
    $w_N$                      &     0.5&     5&     17&     48&     342&     17&    51&    124&    281&    1639 \\ \hline                                             
 	                                       \multicolumn{11}{|c|}{r=1}        \\ \hline    
    $v_N$                      &       1&     9&     34&     96&     684&     35&   101&    248&    562&    3278 \\ \hline                                                 
                                              \multicolumn{11}{|c|}{A=200}          \\ \hline                           
      $N$                      &      25&    50&    100&    200&    1000&     25&    50&    100&    200&    1000 \\ \hline                                             
 $\phi_b$                      &    0.03& 0.044&   0.15&  0.453&   0.782&  0.105& 0.427&  0.628&  0.748&    0.893 \\ \hline                                                 
       	                                       \multicolumn{11}{|c|}{r=0}        \\ \hline    
    $w_N$                      &    0.01&  0.04&      1&     18&     263&    0.5&    17&     72&    205&    1460 \\ \hline                                             
 	                                       \multicolumn{11}{|c|}{r=1}        \\ \hline    
    $v_N$                      &    0.02&  0.08&      2&     36&     526&      1&    34&    144&    410&    2920 \\ \hline                                                 
                                              \multicolumn{11}{|c|}{A=500}          \\ \hline                           
      $N$                      &      25&    50&    100&    200&    1000&     25&    50&    100&    200&    1000 \\ \hline                                             
 $\phi_b$                      &       -&     -&      -&      -&   0.435&      -&     -&      -&  0.296&    0.743 \\ \hline                                                 
       	                                       \multicolumn{11}{|c|}{r=0}        \\ \hline    
    $w_N$                      &       -&     -&      -&      -&      82&      -&     -&      -&     32&    1011 \\ \hline                                             
 	                                       \multicolumn{11}{|c|}{r=1}        \\ \hline    
    $v_N$                      &       -&     -&      -&      -&     164&      -&     -&      -&     64&    2022 \\                                              
    \hline
  \end{tabular}
\end{center} 
\end{table}

\begin{figure}[ht!]
\begin{minipage}[ht]{0.5\linewidth}
\center{\includegraphics[width=1\linewidth]{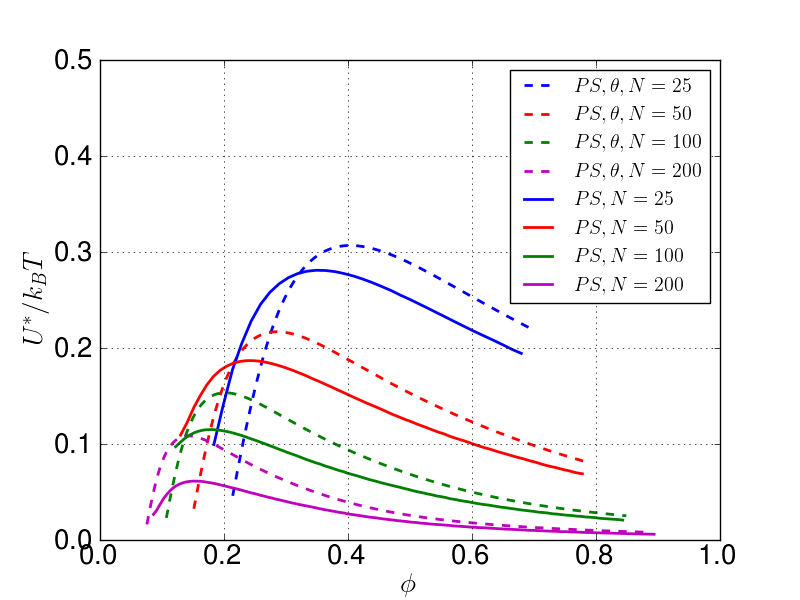}}
\caption{\small{The dependence of barrier height in $k_BT$ units on the volume fraction of polystyrene(PS) in toluene for different polymerization index, $N$.
		The radius of the colloid is $R_c=200\,nm$. The adsorption strength, $A = 42.73(\bar{b}\simeq 0.986)$.}}
\label{barrier_ps_vs_phi_a42_mix_fig}
\end{minipage}
\hfill
\begin{minipage}[ht]{0.5\linewidth}
\center{\includegraphics[width=1\linewidth]{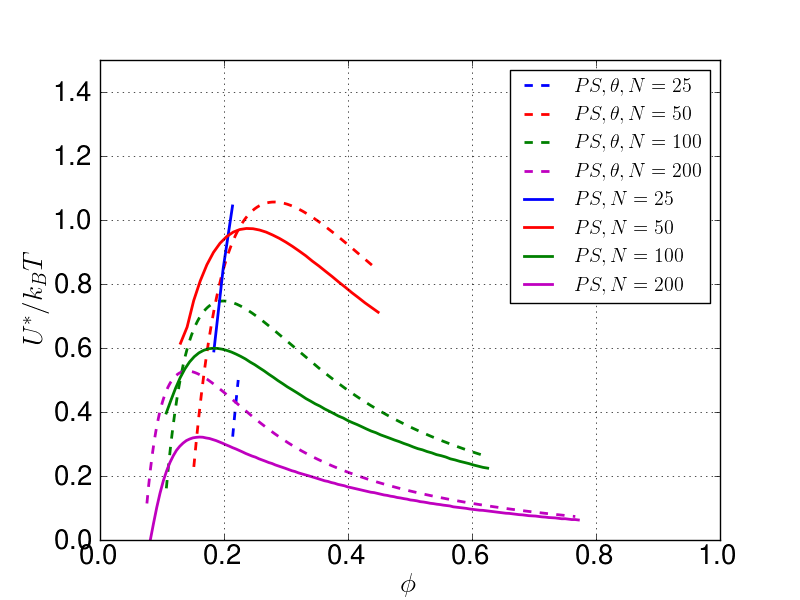}}
\caption{\small{The dependence of barrier height in $k_BT$ units on the volume fraction of polystyrene(PS) in toluene for different polymerization index, $N$.
		The radius of the colloid is $R_c=200\,nm$. The adsorption strength, $A = 100(\bar{b}\simeq 0.429)$.}}
\label{barrier_ps_vs_phi_a100_mix_fig}
\end{minipage}
\end{figure}

\begin{figure}[ht!]
\center{\includegraphics[width=0.5\linewidth]{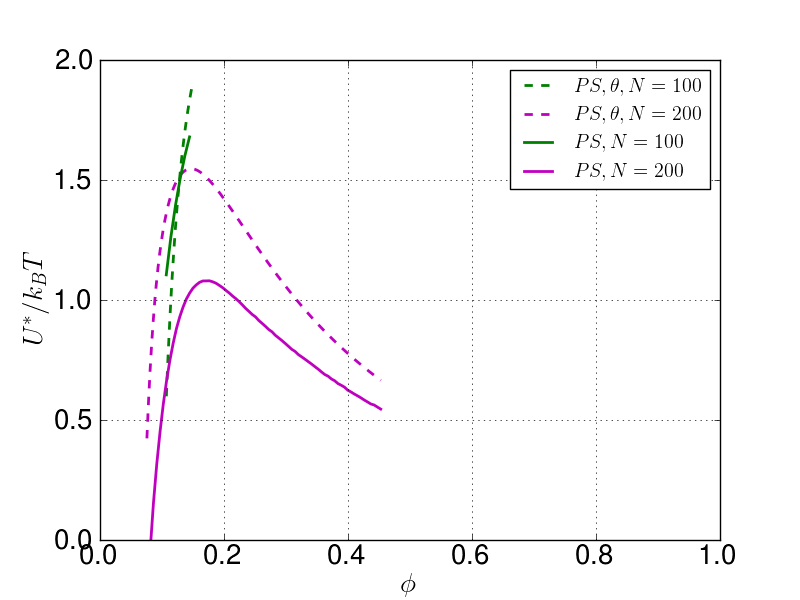}}
\caption{\small{The dependence of the barrier height, in $k_BT$ units, on the volume fraction of polystyrene(PS) in toluene for different polymerization index, $N$.
		The radius of the colloid is $R_c=200\,nm$. The adsorption strength is $A = 200 (\bar{b}\simeq 0.221)$.}}
\label{barrier_ps_vs_phi_a200_mix_fig}
\end{figure}

\begin{figure}[ht!]
\begin{minipage}[ht]{0.5\linewidth}
\center{\includegraphics[width=1\linewidth]{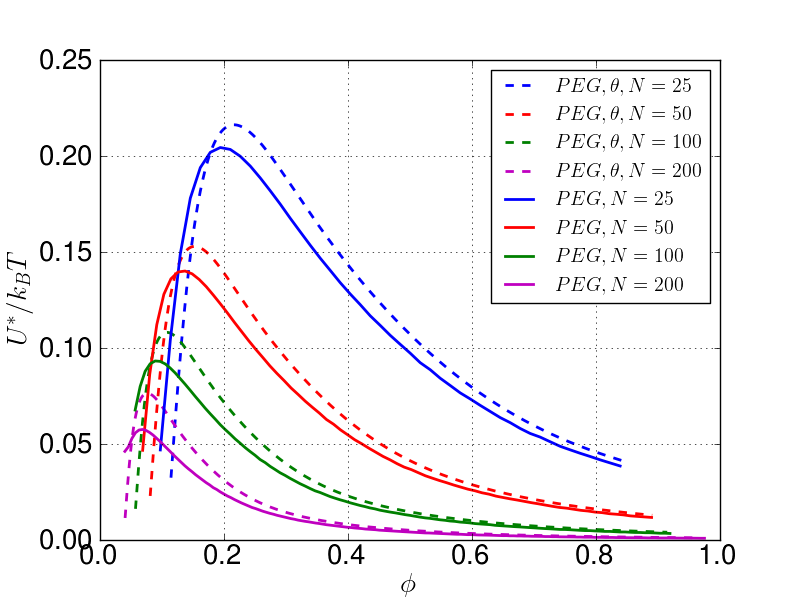}}
\caption{\small{The dependence of the barrier height, in $k_BT$ units, on the volume fraction of polyethylene glycol(PEG) in water for different polymerization index, $N$.
		The radius of the colloid is $R_c=200\,nm$. The adsorption strength is $A = 42.73 (\bar{b}\simeq 0.986)$.}}
\label{barrier_peg_vs_phi_a42_mix_fig}
\end{minipage}
\hfill
\begin{minipage}[ht]{0.5\linewidth}
\center{\includegraphics[width=1\linewidth]{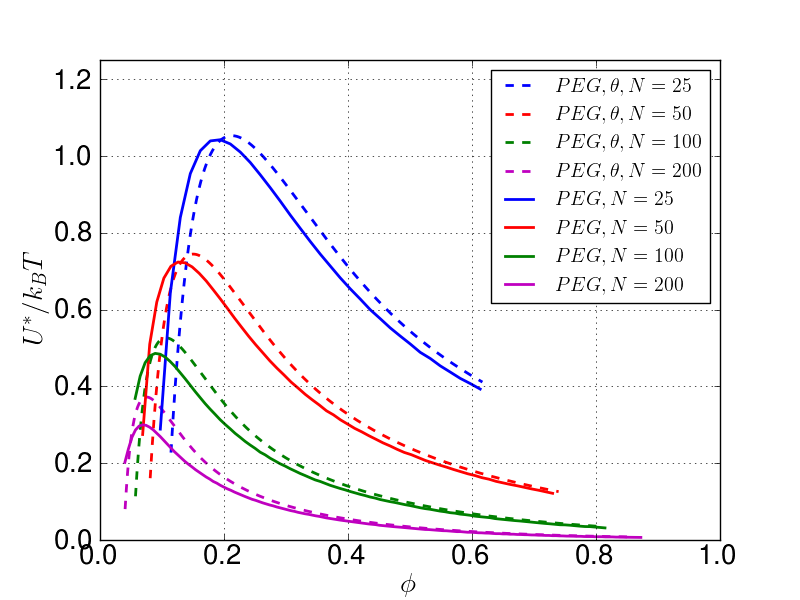}}
\caption{\small{The dependence of the barrier height, in $k_BT$ units, on the volume fraction of polyethylene glycol(PEG) in water for different polymerization index, $N$.
		The radius of the colloid is $R_c=200\,nm$. The adsorption strength is $A = 100 (\bar{b}\simeq 0.429)$.}}
\label{barrier_peg_vs_phi_a100_mix_fig}
\end{minipage}
\end{figure}

\begin{figure}[ht!]
\begin{minipage}[ht]{0.5\linewidth}
\center{\includegraphics[width=1\linewidth]{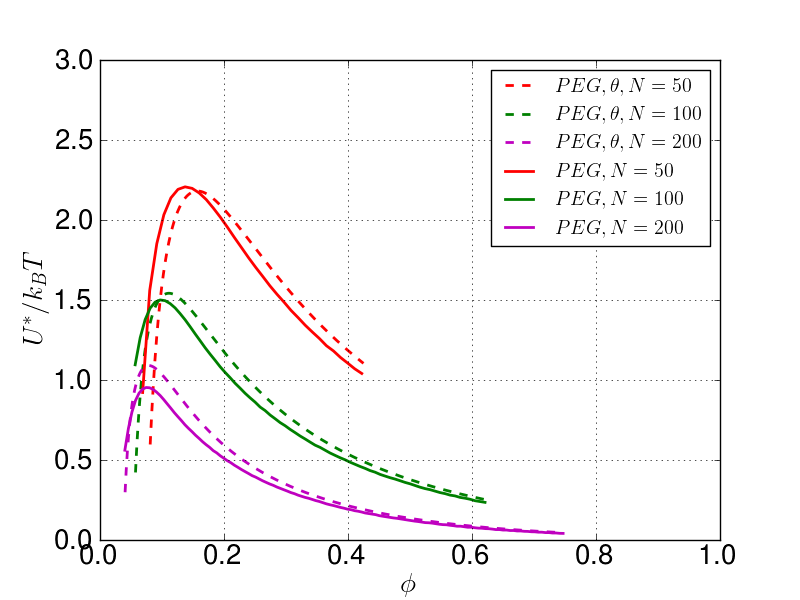}}
\caption{\small{The dependence of barrier, height in $k_BT$ units, on the volume fraction of polyethylene glycol(PEG) in water for different polymerization index, $N$.
		The radius of the colloid is $R_c=200\,nm$. The adsorption strength is $A = 200 (\bar{b}\simeq 0.221)$.}}
\label{barrier_peg_vs_phi_a200_mix_fig}
\end{minipage}
\hfill
\begin{minipage}[ht]{0.5\linewidth}
\center{\includegraphics[width=1\linewidth]{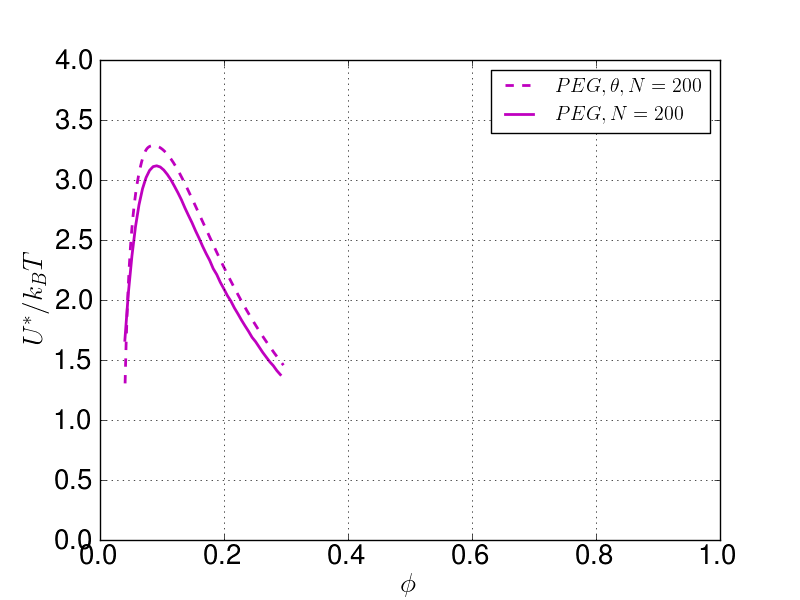}}
\caption{\small{The dependence of the barrier height, in $k_BT$ units, on the volume fraction of polyethylene glycol(PEG) in water for different polymerization index, $N$.
		The radius of the colloid is $R_c=200\,nm$. The adsorption strength is $A = 500 (\bar{b}\simeq 0.097)$.}}
\label{barrier_peg_vs_phi_a500_mix_fig}
\end{minipage}
\end{figure}

	Finally, taking into account all the above, we can use the same interpolation procedure by the variable $r$ that we did for the purely repulsive case. 
	As a result of the procedure we find the dependence of the barrier height in the $k_BT$ units on the volume fraction of the corresponding polymer.
	The corresponding curves are presented in Figs.\ref{barrier_ps_vs_phi_a42_mix_fig}--\ref{barrier_peg_vs_phi_a500_mix_fig} for different 
	polymerization index, $N$ and the adsorption strength, $A(\text{or}\,\bar{b})$ defined by $A$ (or $\bar{b}$). 
	All of the calculations are implemented for the radius of the colloidal particle, $R_c=200\,nm$. 
	Looking at the pictures one can notice that, in contrast to the purely repulsive case,
	the peak values of the barriers are well localized for all considered chain length rather than spread over the accessible range as was previously 
	occured for purely repulsive case.
	Another feature is related to the bulk concentration restriction: we can not make the effect as strong as we want. On the one hand, when  	
        the chain length is increased, the accessible range of the bulk volume fraction becomes wider, but simultaneously the barrier height decreases for larger $N$.        
	
%%%%%%%%%%%%%%%%%%%%%%%%%%%%%%%%%%%%%%%%%%%%%%%%%%%%%%%%%%%%%%%%%%%%%%%%%%%%%%%%%%%%%%%%%%%%%%%%%%%%%%%%%%%%%%%%%%%%%%%%%%%%%%%
%        barrier height(kBT) vs \phi_ps for the different A = 42.73, 100
%%%%%%%%%%%%%%%%%%%%%%%%%%%%%%%%%%%%%%%%%%%%%%%%%%%%%%%%%%%%%%%%%%%%%%%%%%%%%%%%%%%%%%%%%%%%%%%%%%%%%%%%%%%%%%%%%%%%%%%%%%%%%%%
%%%%%%%%%%%%%%%%%%%%%%%%%%%%%%%%%%%%%%%%%%%%%%%%%%%%%%%%%%%%%%%%%%%%%%%%%%%%%%%%%%%%%%%%%%%%%%%%%%%%%%%%%%%%%%%%%%%%%%%%%%%%%%%
%        barrier height(kBT) vs \phi_ps for the different A = 200
%%%%%%%%%%%%%%%%%%%%%%%%%%%%%%%%%%%%%%%%%%%%%%%%%%%%%%%%%%%%%%%%%%%%%%%%%%%%%%%%%%%%%%%%%%%%%%%%%%%%%%%%%%%%%%%%%%%%%%%%%%%%%%%

%%%%%%%%%%%%%%%%%%%%%%%%%%%%%%%%%%%%%%%%%%%%%%%%%%%%%%%%%%%%%%%%%%%%%%%%%%%%%%%%%%%%%%%%%%%%%%%%%%%%%%%%%%%%%%%%%%%%%%%%%%%%%%%
%        barrier height(kBT) vs \phi_peg for the different A = 42.73, 100
%%%%%%%%%%%%%%%%%%%%%%%%%%%%%%%%%%%%%%%%%%%%%%%%%%%%%%%%%%%%%%%%%%%%%%%%%%%%%%%%%%%%%%%%%%%%%%%%%%%%%%%%%%%%%%%%%%%%%%%%%%%%%%%

%%%%%%%%%%%%%%%%%%%%%%%%%%%%%%%%%%%%%%%%%%%%%%%%%%%%%%%%%%%%%%%%%%%%%%%%%%%%%%%%%%%%%%%%%%%%%%%%%%%%%%%%%%%%%%%%%%%%%%%%%%%%%%%
%        barrier height(kBT) vs \phi_ps for the different A = 42.73, 100
%%%%%%%%%%%%%%%%%%%%%%%%%%%%%%%%%%%%%%%%%%%%%%%%%%%%%%%%%%%%%%%%%%%%%%%%%%%%%%%%%%%%%%%%%%%%%%%%%%%%%%%%%%%%%%%%%%%%%%%%%%%%%%%
	    
%%%%%%%%%%%%%%%%%%%%%%%%%%%%%%%%%%%%%%%%%%%%%%%%%%%%%%%%%%%%%%%%%%%%%%%%%%%%%%%%%%%%%%%%%%%%%%%%%%%%%%%%%%%%%%%%%%%%%%%%%%%%%%%%%%%%%%%%%%%%%%%%%%%%%%%%%%%%%
%           Conclusion
%%%%%%%%%%%%%%%%%%%%%%%%%%%%%%%%%%%%%%%%%%%%%%%%%%%%%%%%%%%%%%%%%%%%%%%%%%%%%%%%%%%%%%%%%%%%%%%%%%%%%%%%%%%%%%%%%%%%%%%%%%%%%%%%%%%%%%%%%%%%%%%%%%%%%%%%%%%%%    	   
\section{Summary and Conclusion}
	    In this chapter we implemented the algorithm for numerical solution of SCFT equations taking into account adsorption effect 	    
	    using two different approaches. The adsorption was taken into consideration via the extrapolation length, $b$, as well as 
	    via external surface potential, $u_s(x)$. 
	    We found the quantitative expressions linking together the extrapolation length, $b$, with the parameters of the surface potential, $u_s$,
	    and thus we showed interchangeability of the approaches.
	    
	    We compared the numerical results for the concentration profile and the thermodynamic potential setting adsorption with both of these approaches.
	    We found that, in the case when adsorption is taken into account by the surface potential, the iterative procedure 
	    took almost two times less iterations in comparison to the case associated with the extrapolation length. 
	    In Tab.\ref{tabular:adsorption_comparison} it is obviously shown that the approach related to the surface potential is computationally 
	    more beneficial. Thus, for further consideration, we chose the adsorption defined by the external potential. In this case we found 
	    the optimal values for the grid points.				      
	    	    
	    We  examined how the thermodynamic potentials depend on the adsorption strength, $A$, and calculated them for different virial parameters.
	    For the thermodynamic potential at fixed virial parameters, we found that the position of the barrier's maximum and its
	    scope slightly change when the adsorption strength is varied.
	    It occurred, even when the barrier height changes by a factor of $1000$ (when the adsorption strength is varied in the range $A = 14..500$).

	    The thermodynamic potential for the repulsive surface potential ($A<0$) was also considered. 
	    We found that the saturation value of the adsorption strength is reached at $A = -1000$; with a further decrease of the adsorption strength the potential 
	    behaves like for a purely repulsive walls solid wall.  	    
	    We compared the thermodynamic potentials calculated in this manner with the thermodynamic potentials calculated before for purely repulsive walls.
	    For the same virial parameters the potentials coincide with a small deviation.
	     	    	    
	    In addition we extended the analytical GSDE theory for adsorption case valid in both weak adsorption regime and beyond it. 
	    We compared the results for the numerical SCFT thermodynamic potential with the analytical GSDE potential and 
	    analytical perturbation SCFT potential. The results are in reasonable agreement with each other in the field of applicability of the corresponding theories.
	    
	    As a final step of the chapter we recalculate the thermodynamic potential in $k_BT$ units. 
	    As for purely repulsive walls, combining GSDE and SCFT approaches, 
	    we obtained the colloidal interaction potential as a function of separation and external parameters like bulk monomer concentration and chain length. 
	    Our general results for the interaction potential turn out to be nicely consistent with calculations \cite{avalos_2003} 
	    done in the weak adsorption regime. We found that the barrier height is an increasing function 
	    of adsorption strength, but due to the elevated monomer concentration at the surface this increase is limited for strong adsorption. 
	    In accordance with this restriction, the value of the barrier is $U_m = 1-3 k_BT$ for the same colloid/polymer systems (PS and PEG solutions).

%% file: Chapters/brushes.tex
% Chapter 5
\chapter{Surfaces covered by a soft shell layer} % Main chapter title
\label{chap:Chapter5} % For referencing the chapter elsewhere, use \ref{Chapter1} 
\lhead{Chapter 5. \emph{Surfaces covered by a soft shell layer}} % This is for the header on each page - perhaps a shortened title
%%%%%%%%%%%%%%%%%%%%%%%%%%%%%%%%%%%%%%%%%%%%%%%%%%%%%%%%%%%%%%%%%%%%%%%%%%%%%%%%%%%%%%%%%%%%%%%%%%%%%%%%%%%%%%%%%%%%%%%%%%%%%%%%%%%%%%%%%%%%%%%%%%%%%%%%%%%%%
%        Outline.
%%%%%%%%%%%%%%%%%%%%%%%%%%%%%%%%%%%%%%%%%%%%%%%%%%%%%%%%%%%%%%%%%%%%%%%%%%%%%%%%%%%%%%%%%%%%%%%%%%%%%%%%%%%%%%%%%%%%%%%%%%%%%%%%%%%%%%%%%%%%%%%%%%%%%%%%%%%%%
\section{Outline}
Let us consider the following physical situation. Suppose we have a surface with attractive centers immersed in a polymer solution. When the centers are far away from each other they can reversibly adsorb monomers of a chain. But, when the concentration of attractive centers is not too low, they can form large enough cluster capable to irreversibly adsorb the chain segments. Based on this notion, we consider two surfaces in a polymer solution and assume that irreversibly adsorbed layers are built on each surface independently (i.e. before they come close to each other). In this chapter we develop a theory of polymer-induced interaction between such surfaces, taking into account both irreversibly adsorbed and free chains. 
%%%%%%%%%%%%%%%%%%%%%%%%%%%%%%%%%%%%%%%%%%%%%%%%%%%%%%%%%%%%%%%%%%%%%%%%%%%%%%%%%%%%%%%%%%%%%%%%%%%%%%%%%%%%%%%%%%%%%%%%%%%%%%%%%%%%%%%%%%%%%%%%%%%%%%%%%%%%%
%           Concentrated polymer solution. 
%%%%%%%%%%%%%%%%%%%%%%%%%%%%%%%%%%%%%%%%%%%%%%%%%%%%%%%%%%%%%%%%%%%%%%%%%%%%%%%%%%%%%%%%%%%%%%%%%%%%%%%%%%%%%%%%%%%%%%%%%%%%%%%%%%%%%%%%%%%%%%%%%%%%%%%%%%%%%
\section{Concentrated polymer solution}		
Let us consider firstly a concentrated polymer solution ($\bar{\xi}\ll 1$). As in the previous chapters, we put into the solution two infinite plates, separated from each other by the distance, $h$.
Therefore, the inhomogeneity of the solution involves only one spatial variable, $x\in[0..h]$. 
Due to the symmetry of the system with respect to the mid-plane, we consider only the left half of the space, restricted by the left plate and the mid-plane, i.e. $x\in[0..h_m]$, where $h_m=h/2$. 
Suppose that there is a neutral interaction between polymers and the flat surface (no steric and no depletion interactions). Thus, we can write the expression for the interaction part of the free energy density\footnote{Actually, this is the Taylor expansion, where we 
restricted ourselves by the second-order term in the expansion and dropped the first two terms by virtue of their constant contribution to the energy and the equilibrium state of the system. The prefactor $\text{v}^*$ corresponds to the second derivative of the free energy by concentration at $c_b$.} as
\begin{equation}
\label{brushes_free_en_interaction}
f_{int} = \frac{\text{v}^*}{2}\left(c - c_b\right)^2
\end{equation}
where $c$ is the concentration of monomers in the gap, $c_b$ is the concentration of monomers in the bulk phase and the constant $\text{v}^*$ plays the role of the second virial coefficient (but differ from the actual second virial coefficient).
Correspondingly, the chemical potential is 
\begin{equation}
\label{brushes_cham_pot}
       	\mu_{int} = \frac{\partial f_{int}}{\partial c} = \text{v}^*\left(c - c_b\right) 
\end{equation}
It plays the role of the self-consistent mean field. Therefore, the Edwards equation with the corresponding boundary conditions can be wrritten as
\begin{equation}
\label{brushes_edwards_eqdwards}
\begin{array}{l}
\begin{cases}
       \frac{\partial q(x, s)}{\partial s} = \frac{\partial^2 q(x, s)}{\partial x^2} - w(x)q(x, s), \quad x\in(0, h_m), \quad s > 0 \\       
       q(x, 0) = 1, \quad x\in[0, h_m] \\       
       \partial q(0, s) /\partial x= 0,\quad \partial q(h_m, s)/\partial x = 0, \quad s \in[0, 1] 
\end{cases}
\end{array}
\end{equation}	
Recall that the variables are normalized according to $s \leftarrow s/N$ and  $x \leftarrow x/R_g$, 
where $N$ is the polymer length and $R_g$ is its radius of gyration. Taking it into account, we introduced the reduced self-consistent field $w(x) = \mu(x)N$. The variable $h_m=h/2$ corresponds to the distance from the left wall to the mid-plane and for $h\rightarrow \infty$ we get a system with just one wall. Thus, the one-plate system can be considered as a reference state.
The boundary condition on the wall is chosen due to the neutral character of the interaction between the polymer solution and the surface.
It is true for a concentrated polymer solution or melt on characteristic scales $\gg \xi$. The molecular field is automatically adjusted in order to satisfy this boundary condition. It is related with the fact that the system is incompressibile on the scale $> \xi$ and the concentration profile does not change which leads to $c_x(0)=0$.
Another boundary condition is taken due to the symmetry of the solution with respect to the mid-plane \footnote{It could be also obtained for large separations using the fact that far away from the wall the chains are distributed uniformly $q(h_m, s) = const$)}.

\textbf{Melt}. For a polymer melt we have $\phi \simeq \phi_b = 1$, where $\phi=c/c_b$ is the volume fraction of monomers. 
In the expression for the self-consistent field $w=v_N^*(\phi/\phi_b - 1)$, the constant $v_N^*=\text{v}^*N\phi_b$ is unknown and can be quite big. 
Therefore, we can not always set $w = 0$ in the melt. However, $w\simeq 0$ in a nearly homogeneous melt of free chains.
In fact the self-consistent field, $w(x)$ is determined by the concentration or, vice versa the concentration is determined by the field (see Eq.(\ref{brushes_cham_pot})). In a melt, a neutral-interacting wall does not produce any perturbations in the polymer concentration. Therefore, 
given a constant concentration profile, we have to choose a field that produces such concentration distribution. It does not matter what kind of 
the field it is, an external one or the one created by the surrounding chains. Such homogeneous concentration can be produced only by a constant external field and we can set it 
as $w \simeq 0$. Hence, we can write Eq.(\ref{brushes_edwards_eqdwards}) as\footnote{Strictly viewed, the boundary condition at $x=0$ implies that there is some attraction of free polymer segments to the surface. Physically, this attraction can be provided by those surface centers that are not clustered and are therefore responsible for reversible polymer/surface interaction.\label{fnt:neumann_bc}}
\begin{equation}
\label{brushes_edwards_eqdwards_melt}
\begin{array}{l}
\begin{cases}
       \frac{\partial q(x, s)}{\partial s} = \frac{\partial^2 q(x, s)}{\partial x^2}, \quad x\in(0, h_m), \quad s > 0 \\       
       q(x, 0) = 1, \quad x\in[0, h_m] \\       
       \partial q(0, s) /\partial x= 0,\quad \partial q(h_m, s)/\partial x = 0, \quad s \in[0, 1] 
\end{cases}
\end{array}
\end{equation}	
Such kind of the PDE can be solved successfully by means of Fourier method of separation variables. 
We already used the method when we considered an ideal polymer solution squeezed between purely repulsive surfaces (see Ch.\ref{chap:Chapter3}). 
Therefore, let us focus only on the main points. The solution for the above equation is sought in the form $q(x, s) = X(x)S(s)$, where every function, $X(x)$ and $S(s)$ depends on only one variable. If we put the solution in Eq.(\ref{brushes_edwards_eqdwards}), we get two ODE\footnote{Those ODEs are absolutely the same as for the case with purely repulsive walls.}. 
The first one, for the function $S(s)$, is the first order ODE which has the solution: $S(s) = \exp(-p^2s)$, where $p$ is a free parameter.
The second equation for the function $X(x)$ is the second order linear ODE, for which the solution is sought in the form        
$$
        X(x) = ae^{ipx} + be^{-ipx}
$$	
where $a$ and $b$ are free parameters, $i$ is an imaginary unit. Let us find the parameters using the boundary conditions:\\
1) $X'(0) = 0$, leads to $a=b$ and after simple transformations, we can rewrite the solution as $X(x) = A\cos(px)$, where we have introduced the new constant $A = 2ai$.\\
2) $X'(h_m) = 0$, leads to $\sin(ph_m)=0$ and as a consequence $p_n=\pi n/h_m,\, n\in \mathbb{Z}$. 
Therefore, the general solution of Eq.(\ref{brushes_edwards_eqdwards_melt}) can be written in the form:
$$
	q(x,s) = A_0 + \sum\limits_{n=1}^{\infty}A_n\cos\left(\frac{\pi n x}{h_m}\right)e^{-\pi^2n^2s/h_m^2}
$$
Using the initial condition, we can define the constants, $A_n$, i.e.,
$$
	q(x,0) = 1 = A_0 + \sum\limits_{n=0}^{\infty}A_n\cos\left(\frac{\pi n x}{h_m}\right)
$$	
In order to find $A_n$ we can invert the last expression using orthogonal properties of the trigonometric functions. 
For that, multiplying the expression by $\cos\left(\pi n x/h_m\right)$ and integrating it in the range $[0, h_m]$, we obtain for $n=0$:
$$
	A_0 = \frac{1}{h}\int\limits_0^{h_m} \mathrm{d}x = 1
$$
Correspondingly, for $n>1$:
$$
	A_n = \frac{2}{h}\int\limits_{0}^{h_m}\mathrm{d}x\,\cos\left(\frac{\pi n x}{h_m}\right) = 0 
$$
Therefore, the general solution for Eq.(\ref{brushes_edwards_eqdwards_melt}) is
\begin{equation}
\label{brushes_edwards_solution_q}
\begin{array}{l}       
       q(x, s) = 1, \quad x\in [0..h_m], \quad s\in [0..1]
\end{array}
\end{equation}	
Using this result, it is quite easy to find the concentration profile, 
\begin{equation}
\label{brushes_edwards_solution_conc}
\begin{array}{l}       
\phi(x)/\phi_b = \int\limits_0^1\mathrm{d}s\, q(x, s)q(x, 1-s)  = 1, \quad x\in [0..h_m]
\end{array}
\end{equation}	
i.e. the volume fraction of monomers $\phi(x) = \phi_b$ is a constant in space. Therefore, we found the solution of Eq.(\ref{brushes_edwards_eqdwards})
for the special case of the polymer melt analytically. For the concentrated solution, the boundary condition at the wall should be changed.
%%%%%%%%%%%%%%%%%%%%%%%%%%%%%%%%%%%%%%%%%%%%%%%%%%%%%%%%%%%%%%%%%%%%%%%%%%%%%%%%%%%%%%%%%%%%%%%%%%%%%%%%%%%%%%%%%%%%%%%%%%%%%%%%%%%%%%%%%%%%%%%%%%%%%%%%%%%%%
%           Adsorbed and free chains
%%%%%%%%%%%%%%%%%%%%%%%%%%%%%%%%%%%%%%%%%%%%%%%%%%%%%%%%%%%%%%%%%%%%%%%%%%%%%%%%%%%%%%%%%%%%%%%%%%%%%%%%%%%%%%%%%%%%%%%%%%%%%%%%%%%%%%%%%%%%%%%%%%%%%%%%%%%%%
\section{Adsorbed and free chains}	
Further, we will distinguish between the chains that have a contact with the surface from the free chains. Suppose, that all chains having a monomer located at distance $\lambda \sim \xi$ are adsorbed chains and the rest chains are free.
We will subscribe the variables related to free polymers by index $f$ and adsorbed polymers by the index $a$.
Since the polymer solution is concentrated, we have $\phi = \phi_a + \phi_f \simeq \phi_b$, where $\phi_a$ corresponds to the monomer volume fraction of adsorbed polymers and correspondingly, $\phi_f$ to free (unattached) polymers. The distribution function can be decomposed as a sum of two terms corresponding to adsorbed and free chains:   
$$
	q(x, s) = q_f(x, s) + q_a(x, s)
$$
Using the expression for the chemical potential $\mu_{int} = \text{v}^*(\phi_a + \phi_f - \phi_b)$ (keep in mind that $w=\mu N$), we can write the Edwards equations for adsorbed and free chains as
\begin{equation}
\label{brushes_edwards_eqdwards_af}
\begin{array}{l}
\begin{cases}
       \frac{\partial q_a(x, s)}{\partial s} = \frac{\partial^2 q_a(x, s)}{\partial x^2} - w(x)q_a(x, s), \quad x\in(0, h_m), \quad s > 0 \\
       q_a(x, 0) = 0, \quad x\in(0, h_m] \\
       q_a(0, s) = \text{unknown},\quad \partial q_a(h_m, s) /\partial x = 0, \quad s\in [0, 1] 
\end{cases}       
        \\
\begin{cases}
       \frac{\partial q_f(x, s)}{\partial s} = \frac{\partial^2 q_f(x, s)}{\partial x^2} - w(x)q_f(x, s), \quad x\in(\lambda, h_m), \quad s > 0 \\
       q_f(x, 0) = 1 \quad x\in[\lambda, h_m]\\       
       q_f(\lambda, s) = 0,\quad \partial q_f(h_m, s)/\partial x = 0, \quad s\in [0, 1] 	     
\end{cases}              
\end{array}
\end{equation}
For the adsorbed chains the boundary condition on the surface is unknown and depends on the free polymers, i.e., the condition can not be considered separately from free polymers. For another boundary condition, we use the symmetry of the system with respect to the 
mid-plane\footnote{Far away from the plate there are no adsorbed chains, so we could also write $q_a(h\rightarrow\infty, s) = 0$.}. 
In order to justify the initial condition for the adsorbed chains, we suppose that a polymer chain is fixed by one end at $x>0$. 
For the initial condition the chain length should be: $s\to 0$. 
Thereby, for the adsorbed chain to have any contact with the wall, we should stretch the chain and as a result to expand the energy. The probability of the process is proportional to $\exp(-E_{stretched})\to 0$, where $E_{stretched}$ is the energy of the stretched chain. This energy is large with respect to the energy of a free chain. Therefore, we get the initial condition for the adsorbed chains.
For the free chains, the boundary conditions are defined due to fact that there are no free chains for $x<\lambda$ (where $\lambda$ is a microscopic contact length). 
On the other hand, at the mid-plane we have an even function with respect to the mid-plane. The initial condition for the free chains is the same as before, since $q_f(x, 0) + q_a(x, 0) = 1$. Again, for the melt, $\phi_a + \phi_f \simeq \phi_b = 1$ and we can find analytical expressions for the functions, $q_a(x, s)$ and $q_f(x, s)$. In order to simplify the task, we skip the process of finding the solution for the adsorbed chains. For that we use the solution for the function $q(x, s)$ and obtain: $q_a(x, s) = q(x, s) - q_f(x, s)$. Correspondingly, for the volume fraction, we have: $\phi_a = \phi - \phi_f$. In case of the melt, these expressions undergo further simplifications, namely, due to Eqs.(\ref{brushes_edwards_solution_q}, \ref{brushes_edwards_solution_conc}), we can write $q_a(x, s) = 1 - q_f(x, s)$ and $\phi_a = 1 - \phi_f$. 
	
\textbf{Melt of free chains.} For the melt we have $\phi_a + \phi_f = \phi_b = 1$. The self-consistent field coincides with the case considered above for $w = 0$. Correspondingly, Eq.(\ref{brushes_edwards_eqdwards_af}) takes the form
\begin{equation}
\label{brushes_edwards_eqdwards_free_melt}
\begin{array}{l}       
\begin{cases}
       \frac{\partial q_f(x, s)}{\partial s} = \frac{\partial^2 q_f(x, s)}{\partial x^2}, \quad x\in(\lambda, h_m), \quad s\in (0, 1) \\       
       q_f(x, 0) = 1 \quad x\in[\lambda, h_m]\\       
       q_f(\lambda, s) = 0,\quad \partial q_f(h_m, s)/\partial x = 0, \quad s\in [0, 1] 	     
\end{cases}
\end{array}
\end{equation}
       One can notice that except one boundary condition this equation coincides with Eq.(\ref{brushes_edwards_eqdwards_melt}). So, as before,
       $q_f(x, s) = X(x)S(s)$ and for one of the function we can write: $S(s) = \exp(-p^2s)$. The solution for another function we find in the following form:
$$
       X(x) = ae^{ipx} + be^{-ipx}
$$
       or, in the view of the boundary condition at $x=\lambda$, we can renormalize constants correspondingly, $a\leftarrow a\exp(ip\lambda)$ and 
       $b\leftarrow b\exp(-ip\lambda)$. Instead of the above expression, we can write 
$$  
       X(x) = ae^{ip(x-\lambda)} + be^{-ip(x-\lambda)}
$$	
Using the boundary conditions, we obtain:\\
1) $X(\lambda) = 0$, leads to $a=-b$ and $X(x) = A\sin\left(p(x-\lambda)\right)$, where we introduced the constant $A=2ai$. \\
2) $X'(h_m) = 0$, leads to
$$
	\left.\frac{\partial X}{\partial x}\right\vert_{h_m} = Ap\cos\left(p(h_m-\lambda)\right) = 0 \quad \rightarrow \quad p_n = \frac{\pi(1+2n)}{2(h_m-\lambda)}, \quad n\in \mathbb{Z} 
$$
Therefore, the general solution can be written as
$$
	q_f(x, s) = \sum\limits_{n=0}^{\infty} A_n\sin\left(p_n(x-\lambda)\right)e^{-p_n^2s}
$$
Let us find now the coefficients $A_n$. Using the initial condition, we have
$$
       q_f(x, 0) = 1 = \sum_{n=0}^{\infty} A_n\sin\left(p_n(x-\lambda)\right)
$$
Due to the orthogonality of trigonometric functions, we invert the above equation multiplying it by $\sin (p_m(x-\lambda)$ and integrating it in the range $[\lambda, h_m]$:
$$
       A_n = \frac{2}{(h_m-\lambda)}\int\limits_{\lambda}^{h_m}\mathrm{d}x\,\sin\left(p_n(x-\lambda)\right) = \frac{2}{p_n(h_m-\lambda)}\int\limits_{0}^{p_n(h_m-\lambda)}\mathrm{d}x\,\sin(x) = 
       \frac{2}{p_n(h_m-\lambda)} = \frac{4}{\pi(1+2n)}
$$
Finally, the solution is
\begin{equation}
\label{brushes_edwards_solution_q_free}
\begin{array}{l}       
       q_f(x, s) = \sum\limits_{n=0}^{\infty} \frac{4}{\pi(1+2n)}\sin\left(p_n(x-\lambda)\right)e^{-p_n^2s}, \quad p_n = \frac{\pi(1+2n)}{2(h_m-\lambda)}, \quad x\in [0..h_m], \quad s\in [0..1]
\end{array}
\end{equation}
Thus, we are able to find the concentration profile of the free chains.

\textbf{Asymptotic, $h_m\to\infty$}. At such values of $h$ the variable, $p_n$ is becoming continuous. Thus we can rewrite the above equation as
$$
       q_f(x, s) = \frac{2}{(h_m-\lambda)}\sum\limits_{n=0}^{\infty} \frac{1}{p_n}\sin\left(p_n(x-\lambda)\right)e^{-p_n^2s} \xrightarrow[h_m\to\infty]{} 
       \frac{2}{h_m}\int\limits_{p_0}^{\infty} \frac{\mathrm{d}p}{p}\sin\left(p(x-\lambda)\right)e^{-p^2s}
$$
where $p_0=\pi/2h_m$. In the above equation we neglected $\lambda$ in comparison to $h_m$.

\begin{figure}[ht!]
\begin{minipage}[h]{0.5\linewidth}
\center{\includegraphics[width=1\linewidth]{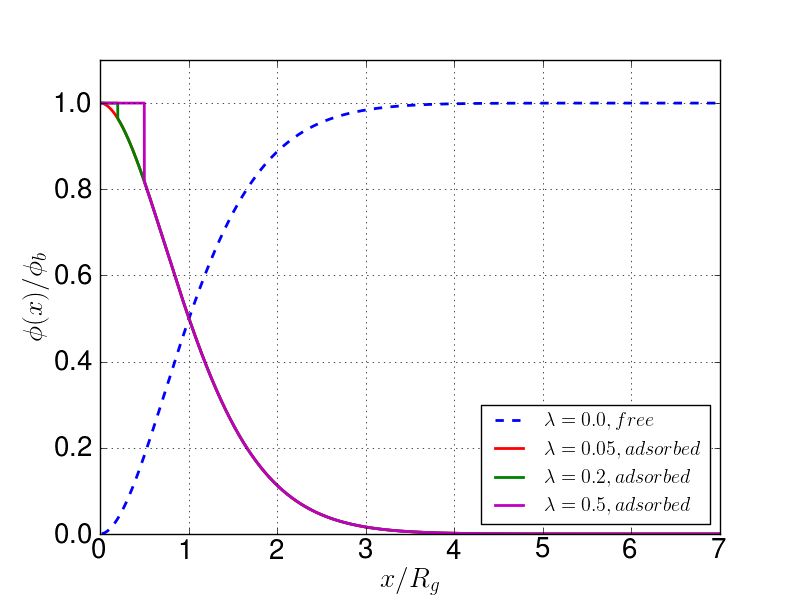}}
\caption{\small{The concentration profiles of adsorbed, Eq.(\ref{brushes_edwards_solution_conc_adsorbed}), and free, Eq.(\ref{brushes_edwards_solution_conc_free}), chains
	       for different values of the parameter $\lambda$.}}
\label{brush_conc_melt_af_fig}
\end{minipage}
\hfill
\begin{minipage}[h]{0.5\linewidth}
\center{\includegraphics[width=1\linewidth]{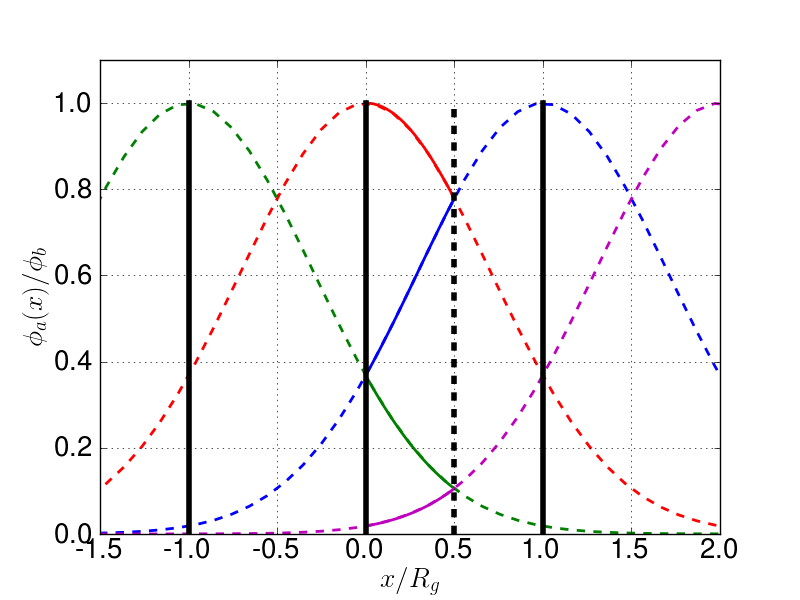}}
\caption{\small{Designing of the soft adsorbed layer based on Eqs.(\ref{brushes_two_plates_profile_ads}, \ref{brushes_x_plates_profile_ads}), for a fixed 
		separation between the plates.}}
\label{brush_total_ads_field_fig}
\end{minipage}
\end{figure}

\textbf{Concentration profile}. In order to find the concentration profile, we use the general expression (see Eq.\ref{intr_polymer_density_homo_dimless})
$$
       \phi_f(x)/\phi_b = \int\limits_0^1\mathrm{d}s\, q_f(x,s)q_f(x,1-s)
$$
where 
$$       
       q_f(x, s) = \sum\limits_{n=0}^{\infty} \frac{4}{\pi(1+2n)}\sin\left(p_n(x-\lambda)\right)e^{-p_n^2s}, \quad p_n = \frac{\pi(1+2n)}{2(h_m-\lambda)}
$$     
and\footnote{Here, the variable $h_m$ does not include any dependency on $m$. The subscription index, $m$ refers to the mid-plane. } 
$$       
       q_f(x, 1 - s) = \sum\limits_{m=0}^{\infty} \frac{4}{\pi(1+2m)}\sin\left(p_m(x-\lambda)\right)e^{-p_m^2(1-s)}, \quad p_m = \frac{\pi(1+2m)}{2(h_m-\lambda)}
$$     
Since only parts of the above expressions depend on the polymer contour parameter, $s$, we perform the integration independently for those parts. For simplicity, we separate the diagonal terms with $m=n$:
$$
       \int\limits_0^1\mathrm{d}s\, e^{-p_n^2s}e^{-p_n^2(1-s)} = e^{-p_n^2}\int\limits_0^1\mathrm{d}s = e^{-p_n^2}
$$
For the rest terms with $m\ne n$, we get:
$$
       \int\limits_0^1\mathrm{d}s\, e^{-p_n^2s}e^{-p_m^2(1-s)} = \frac{1}{p_n^2-p_m^2}\left\{e^{-p_m^2} - e^{-p_n^2}\right\}
$$
One can notice that this is a symmetric expression with respect to $n \leftrightarrow m$. Therefore, we reduce twice the summation multiplying by $2$ the appropriate summation term (with $n\ne m$). Further, since our aim is to obtain the soft surface layer caused by one plate,
we should set $h_m\rightarrow\infty$, but for computational purposes we use the finite $h_m=h_{max}$, at which the profile does not change if we further increase $h_m$\footnote{Usually $h_{max}=10$ is enough.}.
Hence, we obtain
\begin{equation}
\label{brushes_edwards_solution_conc_free}
\begin{array}{l}       
       \phi_f(x)/\phi_b = \sum\limits_{n=0}^{\infty} \frac{16}{\pi^2(1+2n)^2}\sin^2\left(p_n(x-\lambda)\right)e^{-p_n^2} + \\
       +\sum\limits_{m<n}^{\infty} \frac{32}{\pi^2(1+2n)(1+2m)(p_n^2-p_m^2)}\sin\left(p_n(x-\lambda)\right)\sin\left(p_m(x-\lambda)\right)\left\{e^{-p_m^2} - e^{-p_n^2}\right\}
\end{array}
\end{equation}
Correspondingly, for the adsorbed chains, we have
\begin{equation}
\label{brushes_edwards_solution_conc_adsorbed}
\begin{array}{l}       
       \phi_a(x)/\phi_b = 1 - \phi_f(x)/\phi_b
\end{array}
\end{equation}       
This result is presented in Fig.\ref{brush_conc_melt_af_fig} for some values of the parameter, $\lambda$. 
According to the definition, this parameter should be microscopic size or, more precisely, $\lambda \sim a_s$. In the reduced 
variables it can be written as $\bar{\lambda} = \lambda/R_g \sim\sqrt{6/N}$ and for $N = 100$ it leads to $\bar{\lambda}\sim 0.2$. 
From Fig.\ref{brush_conc_melt_af_fig} one can notice that there is a weak dependence on the microscopic length $\lambda$. Let us set $\lambda = 0$. 
%%%%%%%%%%%%%%%%%%%%%%%%%%%%%%%%%%%%%%%%%%%%%%%%%%%%%%%%%%%%%%%%%%%%%%%%%%%%%%%%%%%%%%%%%%%%%%%%%%%%%%%%%%%%%%%%%%%%%%%%%%%%%%%
%        analytical concentration profile and N_ads polymers 2+many
%%%%%%%%%%%%%%%%%%%%%%%%%%%%%%%%%%%%%%%%%%%%%%%%%%%%%%%%%%%%%%%%%%%%%%%%%%%%%%%%%%%%%%%%%%%%%%%%%%%%%%%%%%%%%%%%%%%%%%%%%%%%%%%
%%%%%%%%%%%%%%%%%%%%%%%%%%%%%%%%%%%%%%%%%%%%%%%%%%%%%%%%%%%%%%%%%%%%%%%%%%%%%%%%%%%%%%%%%%%%%%%%%%%%%%%%%%%%%%%%%%%%%%%%%%%%%%%%%%%%%%%%%%%%%%%%%%%%%%%%%%%%%
%           Numerical solution
%%%%%%%%%%%%%%%%%%%%%%%%%%%%%%%%%%%%%%%%%%%%%%%%%%%%%%%%%%%%%%%%%%%%%%%%%%%%%%%%%%%%%%%%%%%%%%%%%%%%%%%%%%%%%%%%%%%%%%%%%%%%%%%%%%%%%%%%%%%%%%%%%%%%%%%%%%%%%
\section{Soft adsorbed layer}
\begin{figure}[ht!]
\begin{minipage}[h]{0.5\linewidth}
\center{\includegraphics[width=1\linewidth]{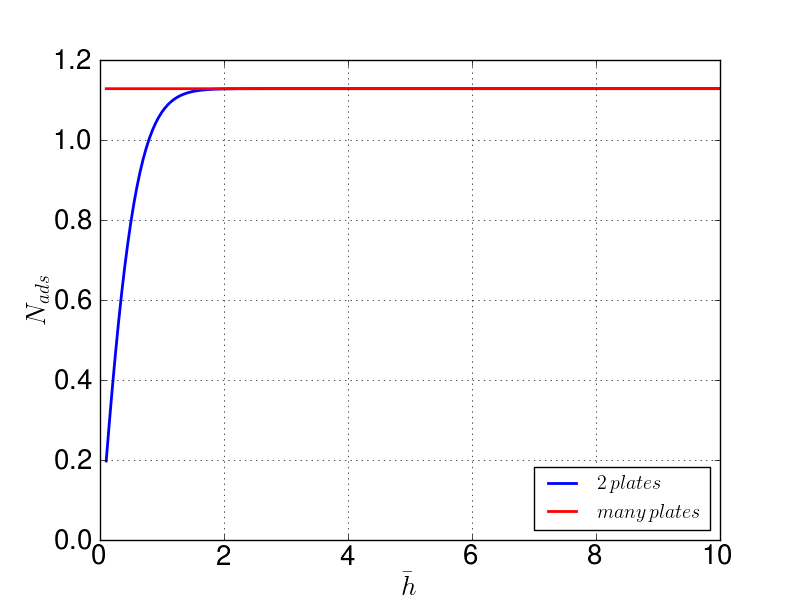}}
\caption{\small{The dependence of the number of adsorbed monomers, Eq.(\ref{brush_num_ads_monomers}), on the separation between the plates 
		calculated for the adsorbed field $\phi_a(x)/\phi_b$ defined in Eq.(\ref{brushes_two_plates_profile_ads}) (blue line) 
		and in Eq.(\ref{brushes_x_plates_profile_ads_calc}) (red line).}}
\label{brush_nads_2p_fig}
\end{minipage}
\hfill
\begin{minipage}[h]{0.5\linewidth}
\center{\includegraphics[width=1\linewidth]{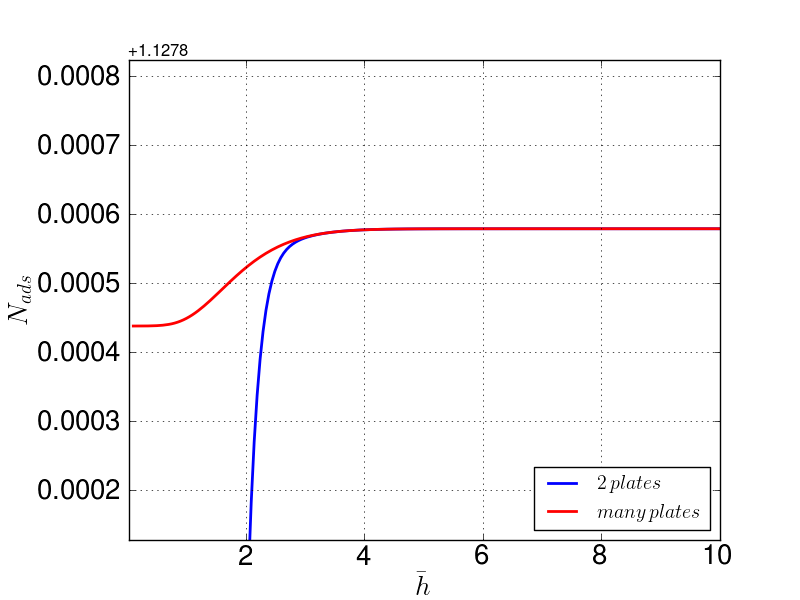}}
\caption{\small{Zoomed in Fig.\ref{brush_nads_2p_fig}.}}
\label{brush_nads_2p_zoomed_fig}
\end{minipage}
\end{figure}

As a next step we construct the surface field that corresponds to one plate soft surface layer. For that, in 
Eq.(\ref{brushes_edwards_solution_conc_free}) we set $h_m\rightarrow\infty$. For computational purposes, it makes sence to consider the finite $h_m=h_{max}$, at which the profile does not change upon the further increase of $h_m$.
Thus, we can rewrite Eq.(\ref{brushes_edwards_solution_conc_free}) as
\begin{equation}
\label{brushes_one_plate_profile_free}
\begin{array}{l}       
       \phi_{1|f}(x)/\phi_b = f_1(x) \equiv \sum\limits_{n=0}^{\infty} \frac{16}{\pi^2(1+2n)^2}\sin^2\left(p_nx\right)e^{-p_n^2} + \\
       +\sum\limits_{m<n}^{\infty} \frac{32}{\pi^2(1+2n)(1+2m)(p_n^2-p_m^2)}\sin\left(p_nx\right)\sin\left(p_mx\right)\left\{e^{-p_m^2} - e^{-p_n^2}\right\}       
\end{array}
\end{equation}
where $p_k = \pi(1+2k)/(2h_{max})$, $k=n,m$ and the index $1$ which corresponds to one plate profile. Therefore, one plate profile of adsorbed chains has the following
form
\begin{equation}
\label{brushes_one_plate_profile_ads}
\begin{array}{l}       
       \phi_{1|a}(x)/\phi_b = a_1(x) \equiv 1 - f_1(x)
\end{array}
\end{equation}	
The function $a_1(x)$ must be tabulated for $x\in[0..h_{max}]$ with, for example, $N_x=20k$ and its calculation is reasonably limited in the number 
of $25k$ terms\footnote{This number of terms becomes clear if we make a Fourier transformation of the function $a_1(x)$. 
At smaller number of the terms in the sums there are artifacts in the Fourier image of the function.}
in the sums of Eq.(\ref{brushes_one_plate_profile_free}). To recover the function for any intervals $[h_1..h_2]\in[0..h_{max}]$ (with any computational mesh), we can use 
the linear interpolation. Now, we should construct the field existing between the two plates. The easiest way to do it is to take the linear superposition of one plate profiles
\begin{equation}
\label{brushes_two_plates_profile_ads}
\begin{array}{l}       
	 \phi_{2|a}(x) = \phi_{1|a}(x) + \phi_{1|a}(h-x)
\end{array}
\end{equation}	
where the first term in r.h.s is a contribution from the left plate and the second one is a contribution from the right plate ($h=2h_m$). The concentration profile constructed in such a way has at least two drawbacks. First of all, at small and moderate separations between the plates, the two plates profile loses its symmetry with respect to $x=0$ due to the contribution of the second term. As a result the self consistent field loses its symmetry too. This problem influences the iterative procedure (see Ch.\ref{chap:Chapter3}), because the symmetry of the corresponding functions allowed us to use the cosine Fourier transformation (see Appendix.C). Another drawback is related to the conservation of the total number of adsorbed monomers. 
%%%%%%%%%%%%%%%%%%%%%%%%%%%%%%%%%%%%%%%%%%%%%%%%%%%%%%%%%%%%%%%%%%%%%%%%%%%%%%%%%%%%%%%%%%%%%%%%%%%%%%%%%%%%%%%%%%%%%%%%%%%%%%%
%        the field of adsorbed polymers 2+many
%%%%%%%%%%%%%%%%%%%%%%%%%%%%%%%%%%%%%%%%%%%%%%%%%%%%%%%%%%%%%%%%%%%%%%%%%%%%%%%%%%%%%%%%%%%%%%%%%%%%%%%%%%%%%%%%%%%%%%%%%%%%%%%
\begin{figure}[ht!]
\begin{minipage}[h]{0.5\linewidth}
\center{\includegraphics[width=1\linewidth]{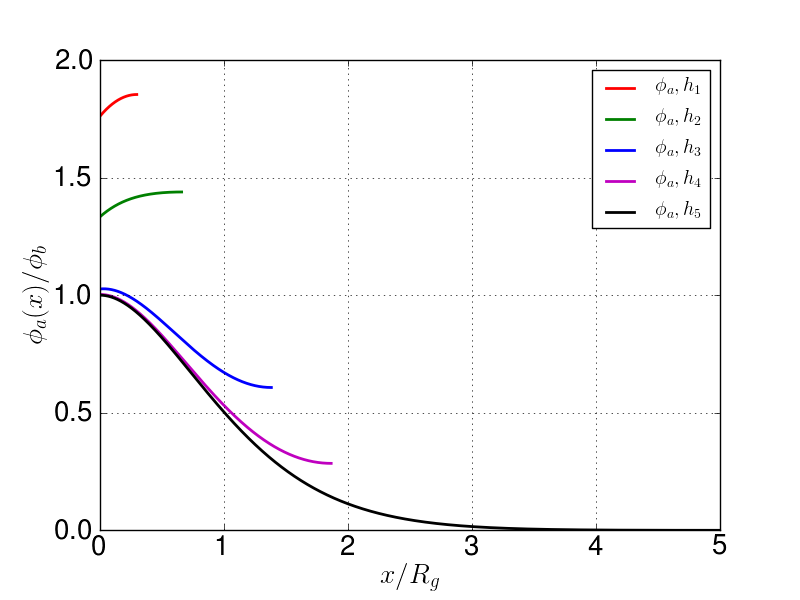}}
\caption{\small{The left-half of soft adsorbed layer, Eq.(\ref{brushes_two_plates_profile_ads}), for some separations between the plates.}}
\label{brush_ads_field_2p_fig}
\end{minipage}
\hfill
\begin{minipage}[h]{0.5\linewidth}
\center{\includegraphics[width=1\linewidth]{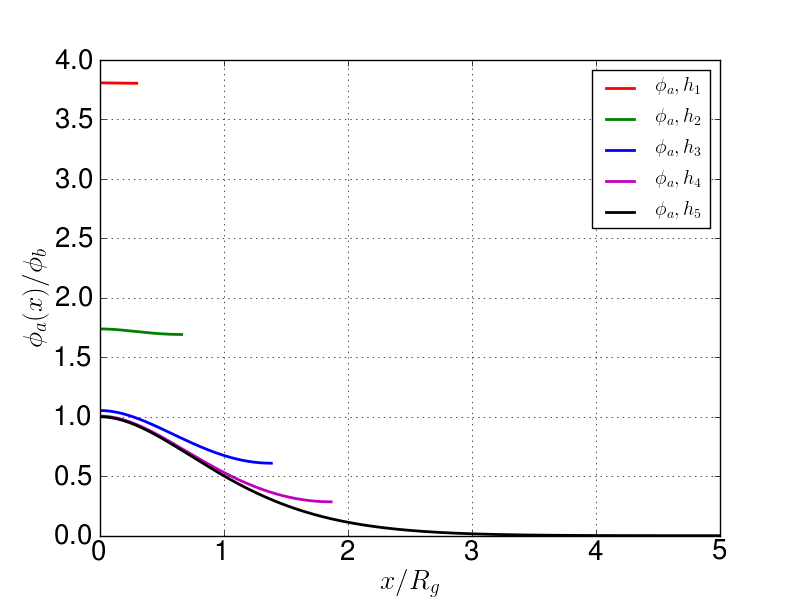}}
\caption{\small{The left-half of soft adsorbed layer, Eq.(\ref{brushes_x_plates_profile_ads_calc}), for some separations between the plates.}}
\label{brush_ads_field_xp_fig}
\end{minipage}
\end{figure}         
Recall that we consider the irreversibly adsorbed layer of polymers. Hence, upon compressing the system, the number of adsorbed monomers should stay unchanged. However, the number of adsorbed monomers is proportional to the following quantity\footnote{Here, our main suggestion is that the profile of adsorbed chains is slightly depends on changes in the solution (concentration, distance between plates). In particular, the property $c_a'(0)=0$ is retained and using that $c_{tot}'(0)$ leads again to $c_f'(0)=0$.}	 	 	
\begin{equation}
\label{brush_num_ads_monomers}
	  N_{ads} = \int\limits_0^{\bar{h}_m}\mathrm{d}\bar{x}\, \frac{\phi_{2|a}(\bar{x})}{\phi_b} = 
	            \int\limits_0^{\bar{h}_m}\mathrm{d}\bar{x}\, (a_1(\bar{x}) + a_1(2\bar{h}_m-\bar{x}))
\end{equation}
Its dependence on the separation between the plates, obtained with Eq.(\ref{brushes_two_plates_profile_ads}), is presented in 
Figs.\ref{brush_nads_2p_fig}--\ref{brush_nads_2p_zoomed_fig}. One can notice that at small separations ($\bar{h} < 1$) there is a linear behavior followed by a smooth transition to a constant value. This behavior at small separations contradicts the demand $N_{ads} = const$.  
%%%%%%%%%%%%%%%%%%%%%%%%%%%%%%%%%%%%%%%%%%%%%%%%%%%%%%%%%%%%%%%%%%%%%%%%%%%%%%%%%%%%%%%%%%%%%%%%%%%%%%%%%%%%%%%%%%%%%%%%%%%%%%%
%        N_ads polymers 2+many
%%%%%%%%%%%%%%%%%%%%%%%%%%%%%%%%%%%%%%%%%%%%%%%%%%%%%%%%%%%%%%%%%%%%%%%%%%%%%%%%%%%%%%%%%%%%%%%%%%%%%%%%%%%%%%%%%%%%%%%%%%%%%%%
In order to correct both of these drawbacks, we consider the following improvement. Let us take into account the truncated parts of the above potential, Eq.(\ref{brushes_two_plates_profile_ads}), and reflect the curves from the opposite plates instead of truncating them. This procedure is schematically shown in Fig.\ref{brush_total_ads_field_fig} for two plates. Exactly the same values of the potential are obtained if we fill the whole space by the bell functions, placing them equidistantly at $h=2h_m$ from each other. The result of this transformation can be analytically written as
\begin{equation}
\label{brushes_x_plates_profile_ads}
\begin{array}{l}       
	 \phi_{a}(x) = \sum\limits_{n=-\infty}^{\infty}\left\{\phi_{1|a}(nh+x) + \phi_{1|a}(nh-x)\right\}
\end{array}
\end{equation}	
This is the periodic function (with the period $h=2h_m$) defined in the whole space i.e., $x\in[-\infty..\infty]$. One can also notice that derivatives of the function at $x=nh_m$, where $n\in\mathbb{Z}$, are all equal to zero, i.e. $\phi_{a}'(nh_m)=0$. We present the function in Fig.\ref{brush_total_ads_field_fig}. For computational purposes it is more convenient to consider only values of the potential in $x\in[0..h_m]$. For that, we consider the range $x\in[0..h_m]$ and write the dimensionless function $\phi_{a}(x)/\phi_b = a(x)$ as
\begin{equation}
\label{brushes_x_plates_profile_ads_calc}
\begin{array}{ll}       
	a(x) \equiv &  a_{1}(x) + a_{1}(2h_m-x) + a_{1}(2h_m+x) + a_{1}(4h_m-x) + a_{1}(4h_m+x) + \ldots =  \\
	                                  &=  a_{1}(x) + \sum\limits_{n=1}^{\infty}\left\{a_{1}(2nh_m-x) + a_{1}(2nh_m+x)\right\}, \\
	     f(x) = & 1 - a(x)                          
\end{array}
\end{equation}	
The corresponding procedure of the function derivation is presented in Fig.\ref{brush_total_ads_field_fig}. The resuling function is presented in Fig.\ref{brush_ads_field_xp_fig} and one can compare it with the previous function, Eq.(\ref{brushes_two_plates_profile_ads}),
that is shown in Fig.\ref{brush_ads_field_2p_fig}. Concerning the computational problems we should consider one more question, namely, how many terms in the sum of
Eq.(\ref{brushes_x_plates_profile_ads_calc}) should be taken into account. We remember that the function $a_1(x)$ is tabulated in the interval $x\in[0..h_{max}]$, so we should use all terms in the sum with $arg(a_1(x)) < h_{max}$. Therefore, the smallest argument should be 
$2nh_m \ge h_{max}$ or $n=ceil(h_{max}/(2h_m))$. Thus, in order to feel this number, we have at $h_m=0.1$ the sum with $n_{max}=50$ terms and at $h_m=10$ the sum with $n_{max}=1$ term.
Moreover, we can also calculate the function $a(x)$ using a different approach. For that, recall that the function, $a_1(x)$ has the derivative $a_1'(x=0)=0$. Therefore, we 
can use the discrete cosine transformation for the function to obtain its Fourier image. Using the Fourier image and the inverse discrete cosine transformation, we 
can recover the periodic function with the period equal to $h=2h_m$. This function corresponds to $a(x)$ and is defined in Eq.(\ref{brushes_x_plates_profile_ads_calc}).
	
The last point of the section is to consider how the number of adsorbed monomers, $N_{ads}$ (see Eq.(\ref{brush_num_ads_monomers})) which corresponds to the function, Eq.(\ref{brushes_x_plates_profile_ads_calc}), depends on the distance between the plates in accordance with the compression and stretching. 
The dependence is shown in Figs.\ref{brush_nads_2p_fig}--\ref{brush_nads_2p_zoomed_fig} in different scales. One can notice that the conservation of the adsorbed
monomers for the new function is performed with a sufficiently qood accuracy $\sim 10^{-4}$.  
These errors caused by many numerical inefficiencies including those related with the implementation of the calculation for the function $a(x)$ as discussed above.
	
%%%%%%%%%%%%%%%%%%%%%%%%%%%%%%%%%%%%%%%%%%%%%%%%%%%%%%%%%%%%%%%%%%%%%%%%%%%%%%%%%%%%%%%%%%%%%%%%%%%%%%%%%%%%%%%%%%%%%%%%%%%%%%%%%%%%%%%%%%%%%%%%%%%%%%%%%%%%%
%           Numerical solution
%%%%%%%%%%%%%%%%%%%%%%%%%%%%%%%%%%%%%%%%%%%%%%%%%%%%%%%%%%%%%%%%%%%%%%%%%%%%%%%%%%%%%%%%%%%%%%%%%%%%%%%%%%%%%%%%%%%%%%%%%%%%%%%%%%%%%%%%%%%%%%%%%%%%%%%%%%%%%
\section{Numerical solution}	 
\label{sec:brush_num_solutions}
Let us append a soft shell layer to a colloidal surface. For that, we use the concentration profile produced by adsorbed polymers, Eq.(\ref{brushes_x_plates_profile_ads_calc}) which is formed at certain bulk volume fraction, $\phi_{b0}$.
Then, we add free chains increasing the bulk concentration to $\phi_b\ge\phi_{b0}$, and consider interaction between two flat plates caused by the free polymers at this higher concentration. In this case, the self-consistent field can be written as
\begin{equation}
\label{brushes_scft}
	 w(x) = \text{v}^*N(\phi_a + \phi_f - \phi_b) =  v_N\left(r_ba(x) + (\phi_f/\phi_b) - 1\right)
\end{equation}
Here, we introduced $r_b=\phi_{b0}/\phi_b$ and $v_N=\text{v}^*N\phi_b$ that plays the role of the second virial parameter, 
but has a different meaning. Using the above approximations, we can write the Edwards equation for the free polymers:
\begin{equation}
\label{brushes_edwards_eqdwards_free_scft}
\begin{array}{l}       
\begin{cases}
         \frac{\partial q_f(x, s)}{\partial s} = \frac{\partial^2 q_f(x, s)}{\partial x^2} - w(x)q(x, s), \quad x\in(0, h_m), \quad s\in (0, 1) \\       
         q_f(x, 0) = 1 \quad x\in[0, h_m]\\       
         q_f(0, s) = 0,\quad \partial q_f(h_m, s)/\partial x = 0, \quad s\in [0, 1] 	     
\end{cases}
\end{array}
\end{equation}
In order to find the numerical solution of Eq.(\ref{brushes_edwards_eqdwards_free_scft}) let us introduce the computational mesh for the space variable $x\in [0..h_m]$:
$$
x_i = i\Delta x, \quad i = 0, 1, \ldots, N_x-1, \quad \Delta x = \frac{h_m}{(N_x-1)}
$$
and for the contour length variable, $s\in[0, 1]$:
$$
         s_m = m\Delta s, \quad m = 0, 1, \ldots, N_s-1, \quad \Delta s = \frac{1}{(N_s-1)}
$$	 
In order to solve Eq.(\ref{brushes_edwards_eqdwards_free_scft}), we use again the Crank-Nicolson procedure (see Appendix.B).

This task is similar to the adsorption task, except the boundary condition on the surface. Now, the role of the microscopic surface field, $u_s(x)$, plays the adsorbed polymers: $\phi_a(x) = r_ba(x)$. Let us set $r_b=1$ to begin with. In the iterative procedure of finding the equilibrium solution, we use the initial guess: $\phi_f(x)/\phi_b=0$. After several calculations for different, $v_N = 10..100$, we get the following observations:

1) The number of iterations required in order to get the equilibrium solution is $n_{it}(v_N=10, \Lambda=50) \sim 800$, 
$n_{it}(v_N=30, \Lambda=50) \sim 1500$, $n_{it}(v_N=50, \Lambda=500) \sim 4500$, where $\Lambda$ is the convergence parameter defined in Ch.\ref{chap:Chapter3}. For larger values of the virial parameters the iterative procedure diverges.

\begin{figure}[ht!]
\begin{minipage}[h]{0.5\linewidth}
\center{\includegraphics[width=1\linewidth]{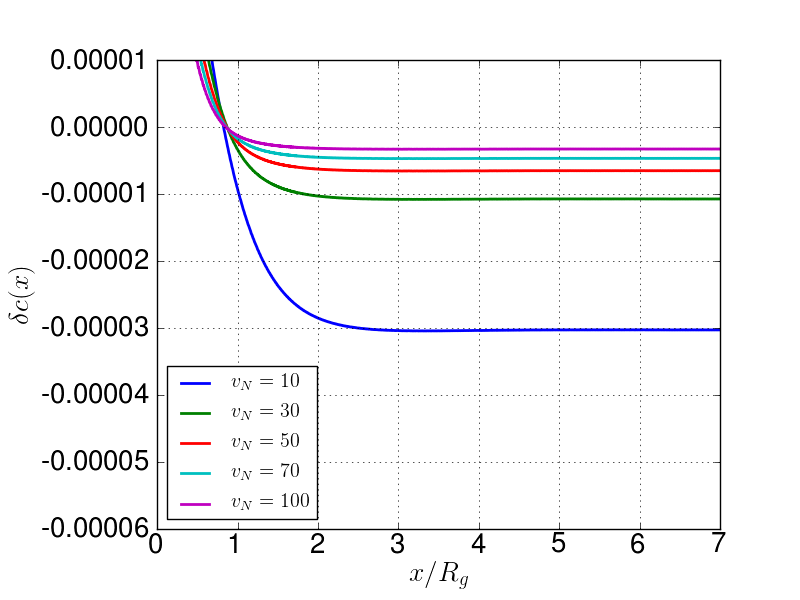}}
\caption{\small{The relative difference between the SCFT and the analytical concentration profiles,  
		$\delta c(x) = (c(x)-f(x))/f(x)$, where $f(x)$ is the concentration profile obtained analytically, Eq.(\ref{brushes_x_plates_profile_ads_calc}), and $c(x)=\phi_f(x)/\phi_b$ is the solution obtained numerically, for different $v_N$.
	        Fixed parameters:  $r_b=1$, $N_x$=8k, $N_s$=3k.}}
\label{brush_conc_diff_v_83_fig}
\end{minipage}
\hfill
\begin{minipage}[h]{0.5\linewidth}
\center{\includegraphics[width=1\linewidth]{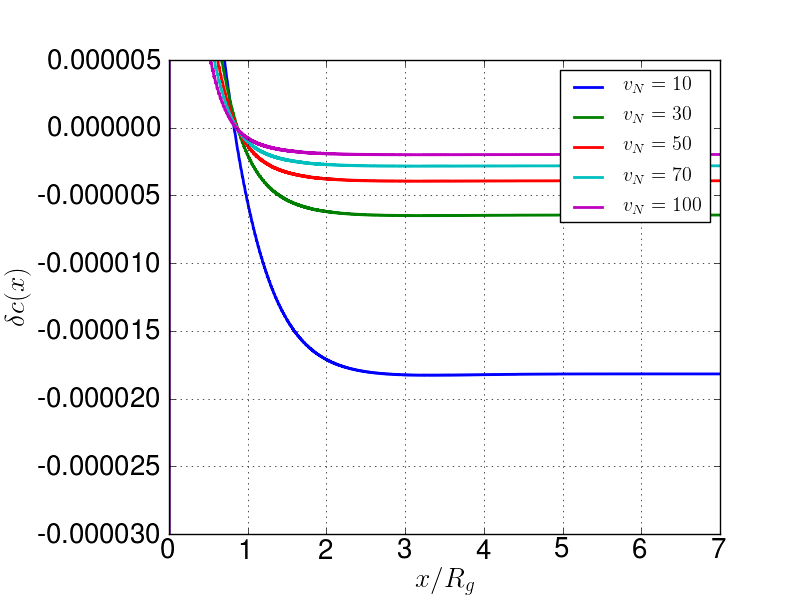}}
\caption{\small{The relative difference between the SCFT and the analytical concentration profiles,
		$\delta c(x) = (c(x)-f(x))/f(x)$, where $f(x)$ is the concentration profile obtained analytically, Eq.(\ref{brushes_x_plates_profile_ads_calc}), and $c(x)=\phi_f(x)/\phi_b$ is the solution obtained numerically, for different $v_N$.
	        Fixed parameters:  $r_b=1$, $N_x$=15k, $N_s$=5k.}}
\label{brush_conc_diff_v_155_fig}
\end{minipage}
\end{figure}

2) The concentration profiles for different virial parameters slightly differ from each other and from the analytical solution 
$f(x)=1-a(x)$, Eq.(\ref{brushes_x_plates_profile_ads_calc}), as one can see in 
Figs.\ref{brush_conc_diff_v_83_fig}--\ref{brush_conc_diff_v_155_fig}. 
It seems very reasonable to change the initial guess for the iterative procedure and start to use the analytical solution $f(x)=1-a(x)$  
defined in Eq.(\ref{brushes_x_plates_profile_ads_calc}) (instead of $\phi_f(x)/\phi_b=0$). Therefore, the initial guess
$\phi(x)/\phi_b= 1 - r_ba(x)$ leads to $w(x) = 0$ in Eq.(\ref{brushes_edwards_eqdwards_free_scft}). The initial guess taken in such a way lets us drastically decrease the number of iterations in the iterative procedure (see Ch.\ref{chap:Chapter3}) and also extends the accessible range for the virial parameters (see Tab.\ref{tabular:brush_iterations_rb1}). Another point concerns Figs.\ref{brush_conc_diff_v_83_fig}--\ref{brush_conc_diff_v_155_fig}. We made the pictures for different computational mesh sizes which are subscribed under the plots. One can notice that on both pictures at large $x$, the relative error $\delta c(x) = (c(x)-f(x))/f(x)$ does not tend to $0$, where $f(x)$ is the concentration profile obtained analytically, Eq.(\ref{brushes_x_plates_profile_ads_calc}), and $c(x)=\phi_f(x)/\phi_b$ is the solution obtained numerically. Moreover, it tends to a negative constant. The constant is caused by numerical errors and diminishes upon more precise calculations.
%%%%%%%%%%%%%%%%%%%%%%%%%%%%%%%%%%%%%%%%%%%%%%%%%%%%%%%%%%%%%%%%%%%%%%%%%%%%%%%%%%%%%%%%%%%%%%%%%%%%%%%%%%%%%%%%%%%%%%%%%%%%%%%%%%%%%%%%%%%%
% The height of the barrier for vw: adsorption 
%%%%%%%%%%%%%%%%%%%%%%%%%%%%%%%%%%%%%%%%%%%%%%%%%%%%%%%%%%%%%%%%%%%%%%%%%%%%%%%%%%%%%%%%%%%%%%%%%%%%%%%%%%%%%%%%%%%%%%%%%%%%%%%%%%%%%%%%%%%%
\begin{table}[h!]
\caption{The number of iterations required for convergence of the iterative procedure with the initial guess $\phi_f(x)/\phi_b= 1 - r_ba(x)$. 
	  Fixed parameters: $v_N=10$, $N_x=8k, N_s=3k, \Lambda = 50$.} 
\label{tabular:brush_iterations_rb1} 
\begin{center}
  \begin{tabular}{ | c | c | c | c | c | c |}
    \hline
	 $v_N$                           &    10&    30&    50&    70&   100   \\ \hline
         $n_{it}$                        &    94&    86&   116&   163&   202  \\
    \hline
  \end{tabular}
\end{center} 
\end{table}

%%%%%%%%%%%%%%%%%%%%%%%%%%%%%%%%%%%%%%%%%%%%%%%%%%%%%%%%%%%%%%%%%%%%%%%%%%%%%%%%%%%%%%%%%%%%%%%%%%%%%%%%%%%%%%%%%%%%%%%%%%%%%%%
%        concentration profile
%%%%%%%%%%%%%%%%%%%%%%%%%%%%%%%%%%%%%%%%%%%%%%%%%%%%%%%%%%%%%%%%%%%%%%%%%%%%%%%%%%%%%%%%%%%%%%%%%%%%%%%%%%%%%%%%%%%%%%%%%%%%%%%

%%%%%%%%%%%%%%%%%%%%%%%%%%%%%%%%%%%%%%%%%%%%%%%%%%%%%%%%%%%%%%%%%%%%%%%%%%%%%%%%%%%%%%%%%%%%%%%%%%%%%%%%%%%%%%%%%%%%%%%%%%%%%%%%%%%%%%%%%%%%%%%%%%%%%%%%%%%%%
%           Thermodynamic potential.
%%%%%%%%%%%%%%%%%%%%%%%%%%%%%%%%%%%%%%%%%%%%%%%%%%%%%%%%%%%%%%%%%%%%%%%%%%%%%%%%%%%%%%%%%%%%%%%%%%%%%%%%%%%%%%%%%%%%%%%%%%%%%%%%%%%%%%%%%%%%%%%%%%%%%%%%%%%%%
\section{Thermodynamic potential}	
\label{sec:brush_therm_pot}
\begin{figure}[ht!]
\begin{minipage}[h]{0.5\linewidth}
\center{\includegraphics[width=1\linewidth]{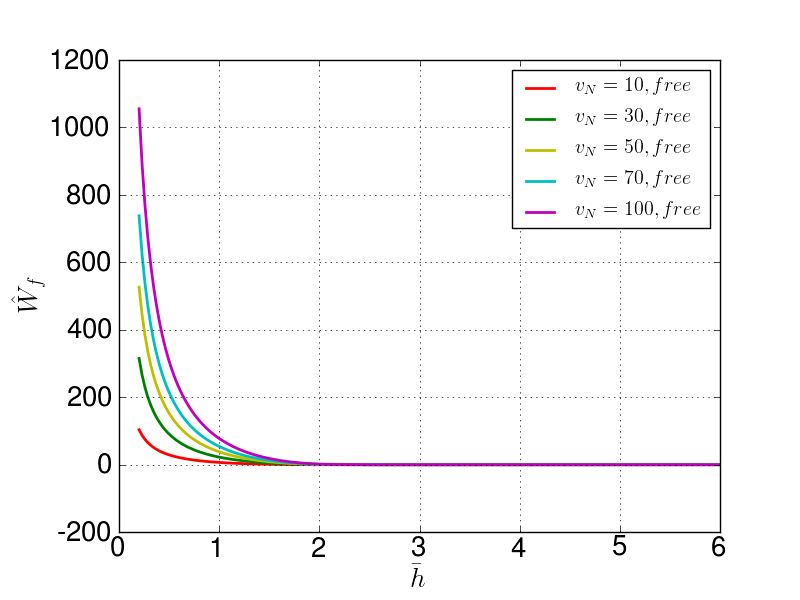}}
\caption{\small{The dimensionless thermodynamic potential, Eq.(\ref{brush_therm_pot_dimless}), as a function of the reduce separation between plates calculated 
		in the adsorbed field, Eq.(\ref{brushes_x_plates_profile_ads_calc}), for some values of the virial parameter, $v_N$. 
		Fixed parameters: $r_b$=1, $N_x$ = 15k, $N_s$=5k.}}
\label{brush_therm_pot_rb1_fig}
\end{minipage}
\hfill
\begin{minipage}[h]{0.5\linewidth}
\center{\includegraphics[width=1\linewidth]{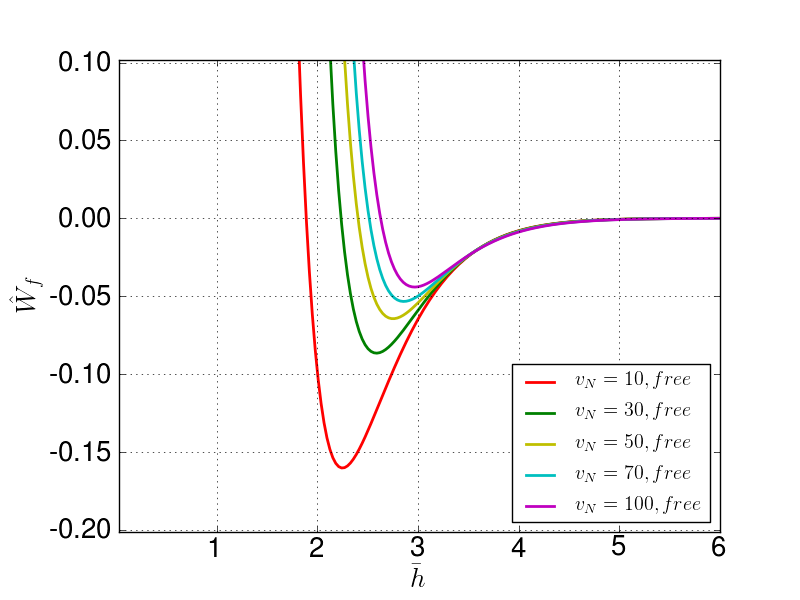}}
\caption{\small{Thermodynamic potentials depicted in Fig.\ref{brush_therm_pot_rb1_fig}, zoomed in around $\hat{W} = 0$.}}
\label{brush_therm_pot_rb1_zoomed_fig}
\end{minipage}
\end{figure}

In this section we derive the expression for the thermodynamic potential. First of all, we consider the general expression for the thermodynamic potential of the system between flat plates that we already derived in Ch.\ref{chap:Chapter2}
\begin{equation}
\label{brush_therm_pot_dim}
W_{f|in}(c_b, h_m) = -\frac{2c_b}{N}Q_{f|in}[w, h_m] - 2\int\limits_0^{h_m}\mathrm{d}x\, p_{f|int}(c)
\end{equation}
Here, we multiplied the expression by $2$, because there are contributions from two surfaces. 
As before, we denoted the mid-plane separation as $h_m = h/2$. The partition function of the free chains is defined as
$$
Q_{f|in}[w, h_m] = \int\limits_0^{h_m}\mathrm{d}x q_f(x, 1), \quad x \in[0..h_m]
$$ 
where $q_f(x, 1)$ is the propagator of a free chain.The contribution to the pressure due to interaction of free chains is 
$p_{f|int} = \mu_{int}c_f - f_{int}$, where $f_{int} = \text{v}^*(c_a + c_f - c_b)^2/2$ (see Eq.(\ref{brushes_free_en_interaction})). 
Therefore, we can write the pressure in the form: 
$$
p_{f|int} = \frac{\partial f_{int}}{\partial c_f} c_f - f_{int} = 
-\frac{\text{v}^*c_b^2}{2}\left\{\left(1 - \frac{c_a}{c_b}\right)^2 - \left(\frac{c_f}{c_b}\right)^2\right\}
$$
or, using the expression for the soft shell layer, $c_a(\bar{x})/c_b \equiv \phi_a(\bar{x}) = r_b a(\bar{x})$, we can rewrite it as
$$
p_{f|int}(\bar{x}) = -\frac{\text{v}^*c_b^2}{2}\left\{(1 - r_b a(\bar{x}))^2 - \left(\frac{c_f}{c_b}\right)^2\right\}
$$
Based on the results for concentration profiles given in Figs.\ref{brush_conc_diff_v_83_fig}--\ref{brush_conc_diff_v_155_fig}, the expression in the curly brackets is small, because the free polymer concentration profile slightly differs from the melt phase. In cases of purely repulsive surfaces and adsorbing surfaces, we extracted the bulk ideal-gas pressure contribution from the thermodynamic potential:
$$
\Pi_b = \frac{c_b}{N} \quad \text{or in dimensionless units} \quad \hat{\Pi}_b = 1
$$
So, we can write the total reduced thermodynamic potential in the form: $\hat{W}_{f}(\phi_b, \bar{h}) = \hat{W}_{f|in}(\phi_b, \bar{h}) + \hat{\Pi}_b\bar{h}$ or, finally, in the expanded form:
\begin{equation}
\label{brush_therm_pot_dimless}
\hat{W}_f(\phi_b, \bar{h}_m) = - 2\left(\hat{Q}_{f|in}[w, \bar{h}_m] - \bar{h}_m\right) +
v_N^*\int\limits_0^{\bar{h}_m}\mathrm{d}\bar{x}	\, \left\{(1 - r_b a(\bar{x}))^2 - \left(\frac{\phi_f}{\phi_b}\right)^2\right\}
\end{equation}
The above dimensionless thermodynamic potential is related with the potential in real ($k_BT$) units as 
$$
W_f(c_b, h) = \frac{R_g c_b}{N}\hat{W}(c_b, \bar{h}_m)
$$
In Figs.\ref{brush_therm_pot_rb1_fig}--\ref{brush_therm_pot_rb1_zoomed_fig} we have presented the result of the numerical calculation for the dimensionless thermodynamic potential as a function of the separation, $\bar{h}=2\bar{h}_m$ between the plates.
The calculations are done using the adsorbed field, Eq.(\ref{brushes_x_plates_profile_ads_calc}), for different virial parameter, $v_N$ 
and the results are presented on different scales. 
The thermodynamic potential is shown in Fig.\ref{brush_therm_pot_rb1_zoomed_fig} for small separations. It contains the repulsive part due to the hindrance between polymers attached to different plates. Then, for larger distances between the plates,
the attraction part is emerged due to the depletion attraction. However, for larger distances, the repulsive part of the potential is not manifested.               
%%%%%%%%%%%%%%%%%%%%%%%%%%%%%%%%%%%%%%%%%%%%%%%%%%%%%%%%%%%%%%%%%%%%%%%%%%%%%%%%%%%%%%%%%%%%%%%%%%%%%%%%%%%%%%%%%%%%%%%%%%%%%%%
%        thermodynamic potential for different vN=10..100
%%%%%%%%%%%%%%%%%%%%%%%%%%%%%%%%%%%%%%%%%%%%%%%%%%%%%%%%%%%%%%%%%%%%%%%%%%%%%%%%%%%%%%%%%%%%%%%%%%%%%%%%%%%%%%%%%%%%%%%%%%%%%%%
This issue is simply signalling us that we should take more care about the boundary conditions. Now, let us turn to a detailed investigation of the system with "neutral" surfaces of solid plates (colloidal particles). Formally, this means that we should consistently impose the Neumann boundary conditions at all solid walls, i.e., in particular, $q_x(0,s)=0$.\textsuperscript{\ref{fnt:neumann_bc},}\footnote{Note that the procedure used to construct the adsorbed concentration profile (see Fig.\ref{brush_ads_field_xp_fig}	) implies that the chains irreversibly adsorbed on plate $1$ satisfy the Neumann boundary condition on plate $2$. However, as these chains are not physically bound to plate $2$, their boundary condition must coincide with that for free chains (on the same plate), hence we again arrive at the Neumann boundary condition for free polymers.}

%%%%%%%%%%%%%%%%%%%%%%%%%%%%%%%%%%%%%%%%%%%%%%%%%%%%%%%%%%%%%%%%%%%%%%%%%%%%%%%%%%%%%%%%%%%%%%%%%%%%%%%%%%%%%%%%%%%%%%%%%%%%%%%%%%%%%%%%%%%%%%%%%%%%%%%%%%%%%
%           Comparison with bc $q_x(0, s)=0$.
%%%%%%%%%%%%%%%%%%%%%%%%%%%%%%%%%%%%%%%%%%%%%%%%%%%%%%%%%%%%%%%%%%%%%%%%%%%%%%%%%%%%%%%%%%%%%%%%%%%%%%%%%%%%%%%%%%%%%%%%%%%%%%%%%%%%%%%%%%%%%%%%%%%%%%%%%%%%% 
\section{Comparison with the Neumann b.c.}
\begin{figure}[ht!]
\begin{minipage}[h]{0.5\linewidth}
\center{\includegraphics[width=1\linewidth]{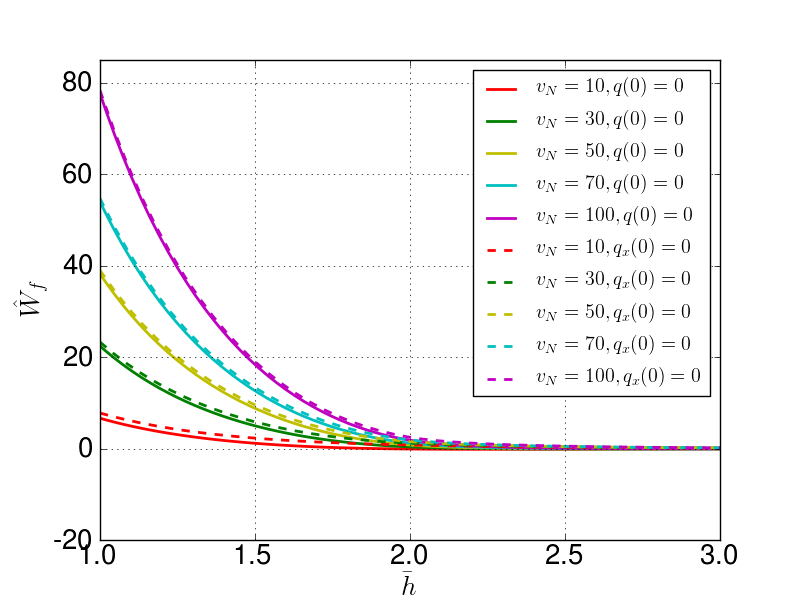}}
\caption{\small{Comparison between the thermodynamic potentials calculated in the adsorbed field, Eq.(\ref{brushes_x_plates_profile_ads_calc}),
                for some values of the virial parameter, $v_N$. The continuous lines correspond to the boundary condition $q(0, s)=0$ and the dashed lines to $q_x(0, s)=0$.
		Fixed parameters: $r_b$=1, $N_x$ = 15k, $N_s$=5k.}}
\label{brush_therm_pot_rb1_q0qx0_fig}
\end{minipage}
\hfill
\begin{minipage}[h]{0.5\linewidth}
\center{\includegraphics[width=1\linewidth]{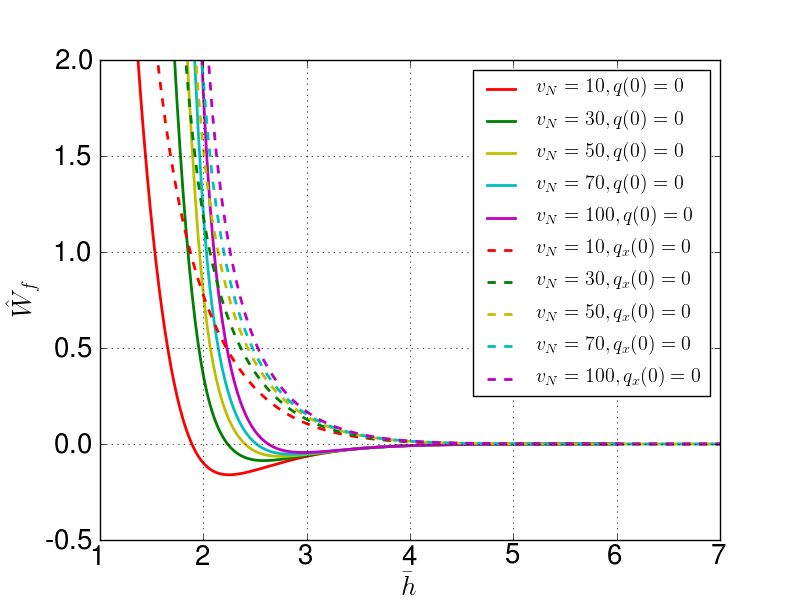}}
\caption{\small{Thermodynamic potentials depicted in Fig.\ref{brush_therm_pot_rb1_q0qx0_fig}, zoomed around $\hat{W} = 0$.}}
\label{brush_therm_pot_rb1_q0qx0_zoomed_fig}
\end{minipage}
\end{figure}

In the previous section we considered the numerical solution of Eq.(\ref{brushes_edwards_eqdwards_free_scft}) with the boundary condition, $q(0,s )=0$ at the wall. This condition leads to $c'(0)=0$ at the wall. The same is true with the boundary condition $q'(0,s )=0$.
From this point of view it is interesting to compare the results for the thermodynamic potential obtained using 
Eq.(\ref{brushes_edwards_eqdwards_free_scft}) with the thermodynamic potential obtained using the solution of the following equation (here $x=\bar{x}$):
\begin{equation}
\label{brushes_edwards_eqdwards_free_scft_qx0}
\begin{array}{l}       
\begin{cases}
         \frac{\partial q_f(x, s)}{\partial s} = \frac{\partial^2 q_f(x, s)}{\partial x^2} - w(x)q(x, s), \quad x\in(0, h_m), \quad s\in (0, 1) \\       
         q_f(x, 0) = 1 \quad x\in[0, h_m]\\       
         \partial q_f(0, s)/\partial x = 0,\quad \partial q_f(h_m, s)/\partial x = 0, \quad s\in [0, 1] 	     
\end{cases}
\end{array}
\end{equation}
The results of the comparison are presented in Figs.\ref{brush_therm_pot_rb1_q0qx0_fig}--\ref{brush_therm_pot_rb1_q0qx0_zoomed_fig}.
%%%%%%%%%%%%%%%%%%%%%%%%%%%%%%%%%%%%%%%%%%%%%%%%%%%%%%%%%%%%%%%%%%%%%%%%%%%%%%%%%%%%%%%%%%%%%%%%%%%%%%%%%%%%%%%%%%%%%%%%%%%%%%%
%        comparison of thermodynamic potentials with diff bc for different vN=10..100
%%%%%%%%%%%%%%%%%%%%%%%%%%%%%%%%%%%%%%%%%%%%%%%%%%%%%%%%%%%%%%%%%%%%%%%%%%%%%%%%%%%%%%%%%%%%%%%%%%%%%%%%%%%%%%%%%%%%%%%%%%%%%%%
One can notice that on the normal scale, Fig.\ref{brush_therm_pot_rb1_q0qx0_fig}, the results predicted using both boundary conditions are similar with minor deviations. In contrast with that, at the zoomed scale, Fig.\ref{brush_therm_pot_rb1_q0qx0_zoomed_fig}, one can observe the fundamental difference between the thermodynamic potentials. The previous calculations that were based on the boundary condition 
$q(0, s)=0$ produced an attractive part of the potential, while potentials calculated with the boundary condition $q_x(0, s)=0$ show basically the repulsive properties. This mismatch follows from the inconsistency of the boundary condition $q(0, s) = 0$ (see the end of Sec.\ref{sec:brush_therm_pot}). We can also use the following argument for a formal justification of the new boundary condition, 
$q_x(0, s)=0$. In the case of melt with $r_b = 0.5$, we can write the volume fraction of the adsorbed monomers at the wall as $\phi_a(0) = 1/2$ and its derivative as $\phi_a'(0) = 0$. Meanwhile, the similar expressions for free polymers are $\phi_f(0) = 1/2$ and $\phi_f'(0) = 0$ which formally correspond to the boundary condition $q_f'(0, s) = 0$\footnote{Or we can use incompressibility condition and that $c_a'(0)=0$ which leads to $c_f'(0)=0$ and $q_f'(0, s)=0$.} 
\footnote{Here and below we use the effective boundary conditions of Neumann type which are valid for length-scales $\gg\xi$.
There is no contradiction with Ch.\ref{chap:Chapter3}, because we consider scales of inhomogeneity $\gg\xi$, while in Ch.\ref{chap:Chapter3} we considered scales $\sim\xi$ and could use the Dirichlet boundary condition.}.
%%%%%%%%%%%%%%%%%%%%%%%%%%%%%%%%%%%%%%%%%%%%%%%%%%%%%%%%%%%%%%%%%%%%%%%%%%%%%%%%%%%%%%%%%%%%%%%%%%%%%%%%%%%%%%%%%%%%%%%%%%%%%%%%%%%%%%%%%%%%%%%%%%%%%%%%%%%%%
%           surface adsorbed field.
%%%%%%%%%%%%%%%%%%%%%%%%%%%%%%%%%%%%%%%%%%%%%%%%%%%%%%%%%%%%%%%%%%%%%%%%%%%%%%%%%%%%%%%%%%%%%%%%%%%%%%%%%%%%%%%%%%%%%%%%%%%%%%%%%%%%%%%%%%%%%%%%%%%%%%%%%%%%%	 
\section{Surface adsorbed field}
\begin{figure}[ht!]
\begin{minipage}[h]{0.5\linewidth}
\center{\includegraphics[width=1\linewidth]{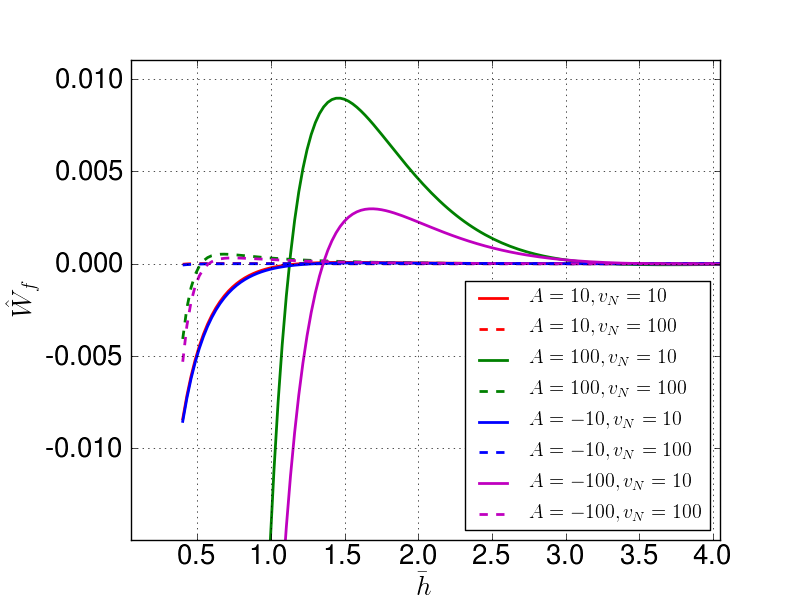}}
\caption{\small{The thermodynamic potential calculated in microscopic surface field, $u_s(x)$ with boundary condition at the wall $q_x(0, s)=0$.
		Fixed parameters: $N_x$ = 15k, $N_s$=5k.}}
\label{brush_therm_pot_ads_micro_qx0_fig}
\end{minipage}
\hfill
\begin{minipage}[h]{0.5\linewidth}
\center{\includegraphics[width=1\linewidth]{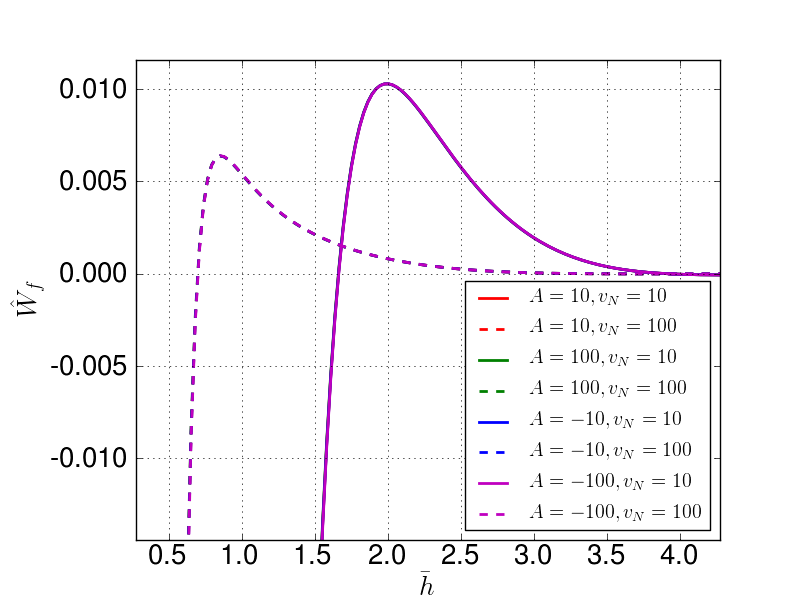}}
\caption{\small{The thermodynamic potential calculated in microscopic surface field, $u_s(x)$ with boundary condition at the wall $q(0, s)=0$.
	        Fixed parameters: $N_x$ = 15k, $N_s$=5k.}}
\label{brush_therm_pot_ads_micro_q0_fig}
\end{minipage}
\end{figure}

Since we do not have analytical benchmarks for comparison with the new results, it is important to compare at least the numerical procedure with something that we have already obtained. If we change the left boundary condition in Eq.(\ref{brushes_edwards_eqdwards_free_scft}) to $q_x(0, s)=0$ and replace the adsorbed field $a(x)$ by the surface field $u_s(x)/v_N$, we obtain the problem which we have already considered, namely, free polymers in the surface attraction field. We define the surface attraction field, $u_s(x)$, as before,
$$
u_s(\bar{x}) = -A/\cosh^2((\alpha \bar{x})^2)
$$
where the parameter $A$ defines the adsorption strength and $\alpha$ defines the range of the surface field. Since we consider a microscopic field, it should be localized in the vicinity of the wall on a microscopic scale. Thus, we fix $\alpha = 40$. In addition, we can relate the parameters of the adsorbed field with the parameter of the surface field. Both of the fields have similar features, namely, both functions tend to zero at infinity and have zero derivative at the origin. Therefore, equating their values at $x=0$, we obtain
$$
A = - r_bv_N^*
$$
In order to obtain the expression for the thermodynamic potential, we continue to use the analogy between the surface field and the adsorbed field. In the case of surface field, the second term in Eq.(\ref{brush_therm_pot_dimless}) should involve excess osmotic pressure of free chains corresponding to $f_{int} = u_sc_f/c_b + v_N^*(c_f/c_b -1)^2/2$: $p_{f|int}=v_N^*((c_f/c_b)^2 -1)/2$. Therefore, the expression for the thermodynamic potential for the case with surface field is:
$$	
\hat{W}_f(\phi_b, \bar{h}_m) = - 2\left(\hat{Q}_{f|in}[w, \bar{h}_m] - \bar{h}_m\right) +
v_N^*\int\limits_0^{\bar{h}_m}\mathrm{d}\bar{x}\, \left\{1 - \left(\frac{c_f}{c_b}\right)^2\right\}	
$$	                                   
The rest of the numerical procedure is still the same. Let us consider calculations of the thermodynamic potential for different boundary condition on the surface: $q(0, s)=0$ and $q_x(0, s)=0$, for different values of the virial parameter, $v_N$, and different values of the amplitude of the surface adsorbed field, $A$. The results of the calculations are shown in Figs.\ref{brush_therm_pot_ads_micro_qx0_fig}--\ref{brush_therm_pot_ads_micro_q0_fig}. The thermodynamic potential with the boundary condition $q_x(0, s)=0$ has absolutely the same behavior as in the case of the adsorbed surfaces (see in Ch.\ref{chap:Chapter4}). The dependence of the barrier height on the amplitude of the surface adsorbed field is shown in Fig.\ref{brush_therm_pot_ads_micro_qx0_fig}. For large and negative values of the amplitude, the surface field produces the results which are similar with the results obtained for the purely repulsive surface. Correspondingly, the constant value of the barrier in this region coincides with the value of the barrier in the purely repulsive case. Then, when the amplitude is around zero, the interaction between the wall and monomers become neutral. For large and positive values of the amplitude the barrier increases with the amplitude.
%%%%%%%%%%%%%%%%%%%%%%%%%%%%%%%%%%%%%%%%%%%%%%%%%%%%%%%%%%%%%%%%%%%%%%%%%%%%%%%%%%%%%%%%%%%%%%%%%%%%%%%%%%%%%%%%%%%%%%%%%%%%%%%
%        comparison of thermodynamic potentials with diff bc for different vN=10..100
%%%%%%%%%%%%%%%%%%%%%%%%%%%%%%%%%%%%%%%%%%%%%%%%%%%%%%%%%%%%%%%%%%%%%%%%%%%%%%%%%%%%%%%%%%%%%%%%%%%%%%%%%%%%%%%%%%%%%%%%%%%%%%%
By contrast, the results shown in Fig.\ref{brush_therm_pot_ads_micro_q0_fig}, corresponding to the boundary condition, $q(0, s)=0$, demonstrate that the thermodynamic potential does not depend on the amplitude of the adsorbed surface field: it is defined only by the virial parameter, $v_N$. As explained above, for the further analysis of the thermodynamic potential, Eq.(\ref{brush_therm_pot_dimless}), we will use the Edwards equation, Eq.(\ref{brushes_edwards_eqdwards_free_scft_qx0}), with the boundary condition at the wall $q_x(0, s)=0$.
	
%%%%%%%%%%%%%%%%%%%%%%%%%%%%%%%%%%%%%%%%%%%%%%%%%%%%%%%%%%%%%%%%%%%%%%%%%%%%%%%%%%%%%%%%%%%%%%%%%%%%%%%%%%%%%%%%%%%%%%%%%%%%%%%%%%%%%%%%%%%%%%%%%%%%%%%%%%%%%
%           Adsorbed layer with different density.Universal behavior.
%%%%%%%%%%%%%%%%%%%%%%%%%%%%%%%%%%%%%%%%%%%%%%%%%%%%%%%%%%%%%%%%%%%%%%%%%%%%%%%%%%%%%%%%%%%%%%%%%%%%%%%%%%%%%%%%%%%%%%%%%%%%%%%%%%%%%%%%%%%%%%%%%%%%%%%%%%%%%	 
\section{Adsorbed layer with different density. Universal behavior}
\label{sec:brush_univ_behavior}
\begin{figure}[ht!]
\begin{minipage}[h]{0.5\linewidth}
\center{\includegraphics[width=1\linewidth]{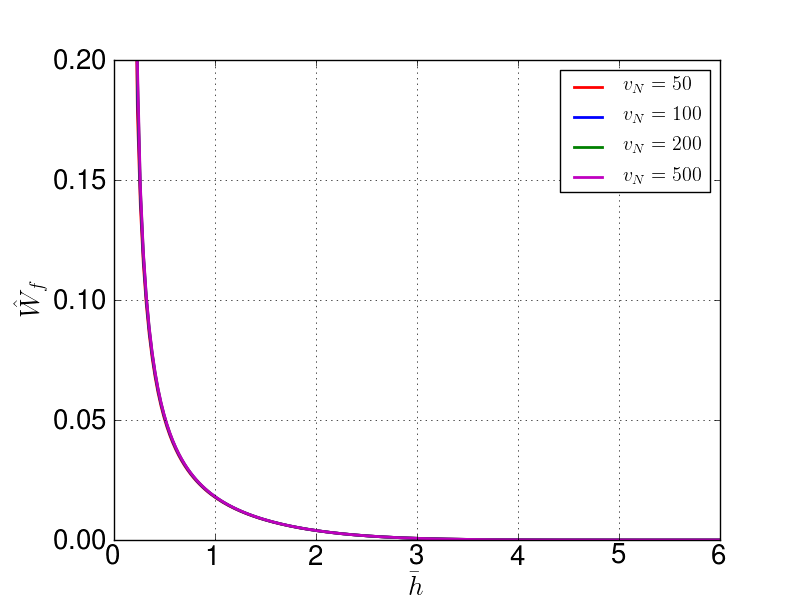}}
\caption{\small{The thermodynamic potential as a function of the separation between the plates calculated for soft adsorbed layer with $r_b=0.1$. Fixed numerical parameters: $N_x = 8k$, $N_s=3k$.}}
\label{brush_therm_pot_rb01_diff_vN_fig}
\end{minipage}
\hfill
\begin{minipage}[h]{0.5\linewidth}
\center{\includegraphics[width=1\linewidth]{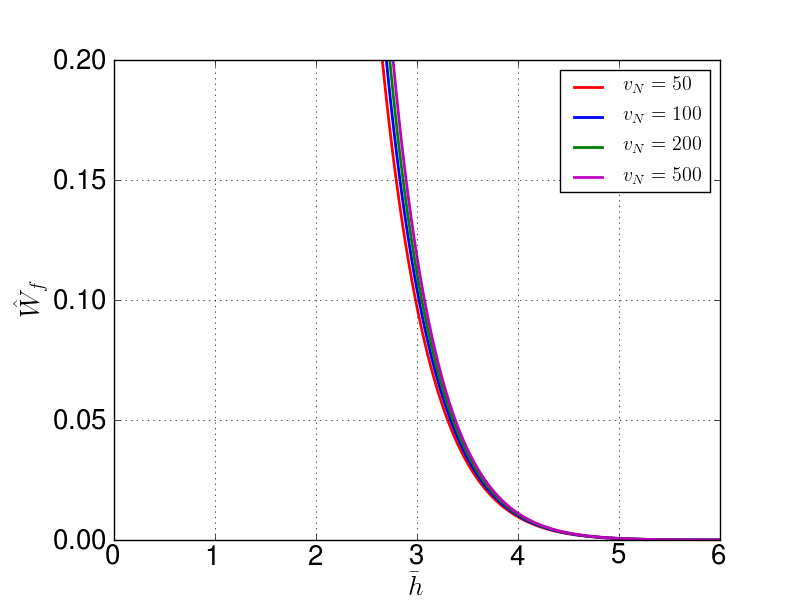}}
\caption{\small{The thermodynamic potential as a function of the separation between the plates calculated for soft adsorbed layer with $r_b=0.9$. Fixed numerical parameters: $N_x = 8k$, $N_s=3k$.}}
\label{brush_therm_pot_rb09_diff_vN_fig}
\end{minipage}
\end{figure}

In this section, we will vary the density of the soft shell layer changing the parameter $r_b$. This parameter is included in the expression for the self-consistent field, Eq.(\ref{brushes_scft}), as well as in the expression for the thermodynamic potential, 
Eq.(\ref{brush_therm_pot_dimless}). First of all, in Figs.\ref{brush_therm_pot_rb01_diff_vN_fig}--\ref{brush_therm_pot_rb09_diff_vN_fig}, we represented the thermodynamic potential as a function of the separation between the plates, calculated for different virial parameter $v_N$ in the soft adsorbed layer with different density $r_b$. The scale of the represnetation for the function is chosen based on the analysis implemented for the purely repulsive case when we observed the barrier height at $\hat{W}^*\simeq 0.02$, for the similar potential. For the pictures we chose even a broader scale in order to show that the functions are very close even for such scale.	 One can notice that the thermodynamic potential slightly depends on the virial parameter when $r_b \lesssim 1$ and it becomes independent from the virial parameters for smaller $r_b$. For further analysis, we fix the value of the virial parameter at $v_N=500$. It corresponds to the concentrated polymer solutions (see Ch.\ref{chap:Chapter3}). Recall that for the purely repulsive walls the thermodynamic potential barrier height reaches its maximum at sufficiently high polymer concentration.
%%%%%%%%%%%%%%%%%%%%%%%%%%%%%%%%%%%%%%%%%%%%%%%%%%%%%%%%%%%%%%%%%%%%%%%%%%%%%%%%%%%%%%%%%%%%%%%%%%%%%%%%%%%%%%%%%%%%%%%%%%%%%%%
%        Thermodynamic potential for different vN at fixed rb
%%%%%%%%%%%%%%%%%%%%%%%%%%%%%%%%%%%%%%%%%%%%%%%%%%%%%%%%%%%%%%%%%%%%%%%%%%%%%%%%%%%%%%%%%%%%%%%%%%%%%%%%%%%%%%%%%%%%%%%%%%%%%%%
	  
\begin{figure}[ht!]
\begin{minipage}[h]{0.5\linewidth}
\center{\includegraphics[width=1\linewidth]{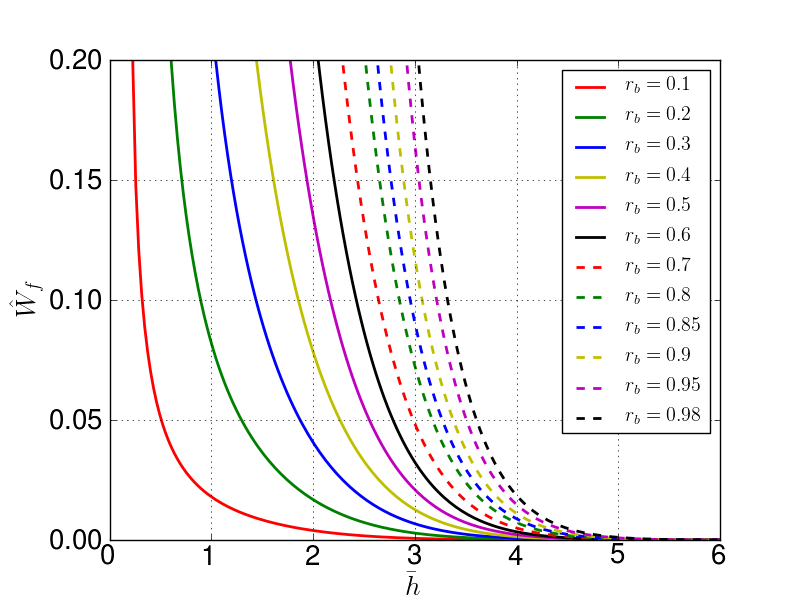}}
\caption{\small{The thermodynamic potential as a function of the separation between the plates calculated for different $r_b$. 
		Fixed numerical parameters: $v_N=500$, $N_x = 8k$, $N_s=3k$.}}
\label{brush_therm_pot_vN500_diff_rb_fig}
\end{minipage}
\hfill
\begin{minipage}[h]{0.5\linewidth}
\center{\includegraphics[width=1\linewidth]{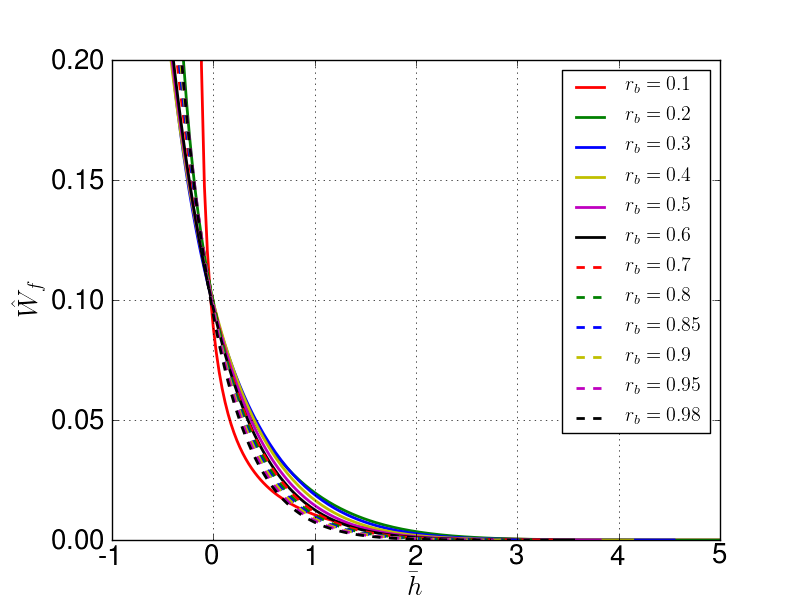}}
\caption{\small{The imposed thermodynamic potentials correspond to Fig.\ref{brush_therm_pot_vN500_diff_rb_fig}.}}
\label{brush_therm_pot_vN500_diff_rb_imposed_fig}
\end{minipage}
\end{figure}

\begin{figure}[ht!]
\begin{minipage}[h]{0.5\linewidth}
\center{\includegraphics[width=1\linewidth]{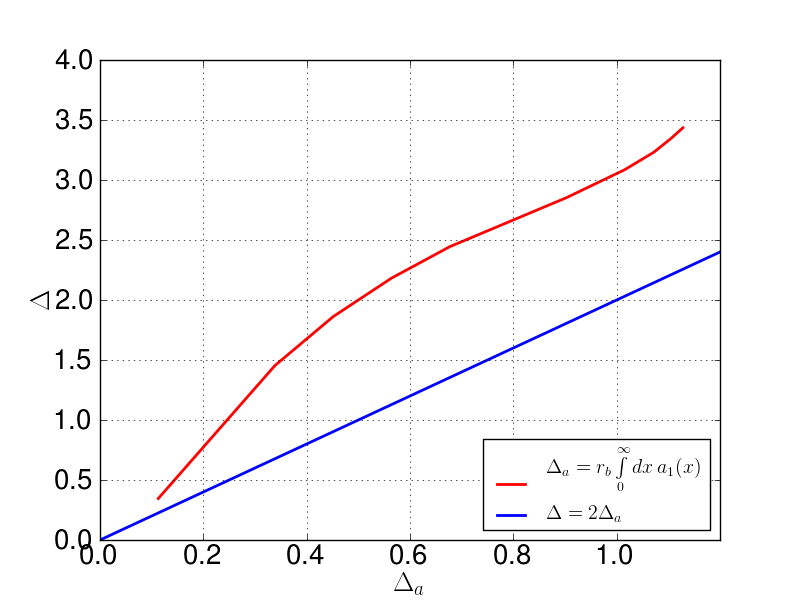}}
\caption{\small{The shift $\Delta$ as a function of $\Delta_a$ obtained from Fig.\ref{brush_therm_pot_vN500_diff_rb_fig} and the theoretical prediction $\Delta \simeq 2\Delta_a$.}}
\label{brush_delta_vs_deltaa_fig}
\end{minipage}
\hfill
\begin{minipage}[h]{0.5\linewidth}
\center{\includegraphics[width=1\linewidth]{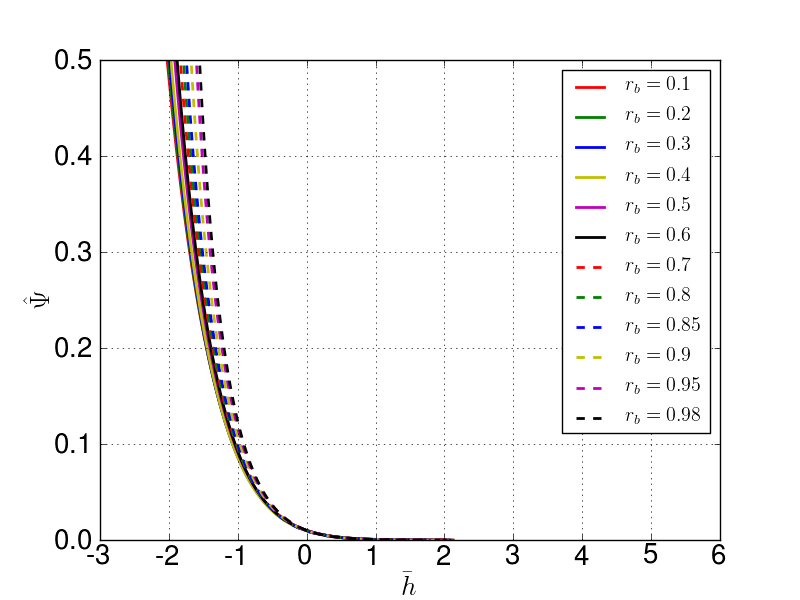}}
\caption{\small{The universal function $\hat{\Psi}(h) = \hat{W}_f(h+\Delta)/r_b^2$ calculated for several $r_b$.}}
\label{brush_therm_pot_univ_psi_fig}
\end{minipage}
\end{figure}

It is also interesting to investigate the dependence of the thermodynamic potential on the parameter $r_b$. In 
Fig.\ref{brush_therm_pot_vN500_diff_rb_fig} we have presented the thermodynamic potential for different values of the parameter $r_b$. As one can notice from the picture, the curves are similar, but are shifted upon changing the parameter $r_b$. We can try to superimpose them on each other shifting every curve to the origin. Every shift is measured by the parameter $\Delta(r_b)$. The shifted thermodynamic potentials are shown in Fig.\ref{brush_therm_pot_vN500_diff_rb_imposed_fig}. The curves with close $r_b$ are similar and we can easily combine them, whilst curves with significantly different $r_b$ are much harder to combine with a good accuracy. 
%%%%%%%%%%%%%%%%%%%%%%%%%%%%%%%%%%%%%%%%%%%%%%%%%%%%%%%%%%%%%%%%%%%%%%%%%%%%%%%%%%%%%%%%%%%%%%%%%%%%%%%%%%%%%%%%%%%%%%%%%%%%%%%
%        Thermodynamic potential for different rb at fixed vN=500
%%%%%%%%%%%%%%%%%%%%%%%%%%%%%%%%%%%%%%%%%%%%%%%%%%%%%%%%%%%%%%%%%%%%%%%%%%%%%%%%%%%%%%%%%%%%%%%%%%%%%%%%%%%%%%%%%%%%%%%%%%%%%%%
In order to understand the origin of the shift $\Delta(r_b)$ let us consider the following quantity 
$$
	  \Delta_a = r_b\int\limits_{0}^{h/2}\mathrm{d}x\,a(x)= r_b\int\limits_{0}^{\infty}\mathrm{d}x\,a_1(x)
$$
The similar quantities are already considered in the parts devoted to purely repulsive (see \textbf{Chapter.3}) and reversibly adsorbed surfaces (see Ch.\ref{chap:Chapter4}). These quantities played the role of the depletion layer thickness. 
In Fig.\ref{brush_delta_vs_deltaa_fig} we present how $\Delta_a$ depends on the shift $\Delta$ along with the theoretical dependence $\Delta \simeq 2\Delta_a$, where the factor $2$ takes into account the contributions from both plates. 
These curves have a similar behavior, but have slightly different slopes. 
%%%%%%%%%%%%%%%%%%%%%%%%%%%%%%%%%%%%%%%%%%%%%%%%%%%%%%%%%%%%%%%%%%%%%%%%%%%%%%%%%%%%%%%%%%%%%%%%%%%%%%%%%%%%%%%%%%%%%%%%%%%%%%%
%        Delta_a vs Delta and universal function \Psi 
%%%%%%%%%%%%%%%%%%%%%%%%%%%%%%%%%%%%%%%%%%%%%%%%%%%%%%%%%%%%%%%%%%%%%%%%%%%%%%%%%%%%%%%%%%%%%%%%%%%%%%%%%%%%%%%%%%%%%%%%%%%%%%%

In order to complete this section and obtain the results for the additional generalization, let us do the following. 
Based on the equation for the thermodynamic potential, Eq.(\ref{brush_therm_pot_dimless}), at least for small values of $r_b$, we can write\footnote{In this case, the shift is selected again for a match between the functions $\hat{W}_f(h+\Delta)/r_b^2$.} $\hat{W}_f(h_{eff}+\Delta) \simeq r_b^2\hat{\Psi}(h_{eff})$, where $\hat{\Psi}(h_{eff})$ is an universal function that does not depend on $r_b$. 
Without loss of generality, we can try to find the universal function for any $r_b \in [0..1)$. The results for the function $\hat{\Psi}$ are shown in Fig.\ref{brush_therm_pot_univ_psi_fig}. Again, the universality is achieved only in a certain range of $h$ depending on $r_b$. 
Due to the definition of the universal function $\hat{\Psi}(h) = \hat{W}_f(h+\Delta)/r_b^2$, the coincidence between the functions with large values of $r_b$ is reached for small values of the function $\hat{\Psi}$. Conversely, for small values of $r_b$ the coincidence occurs for bigger values of $\hat{\Psi}$.
%%%%%%%%%%%%%%%%%%%%%%%%%%%%%%%%%%%%%%%%%%%%%%%%%%%%%%%%%%%%%%%%%%%%%%%%%%%%%%%%%%%%%%%%%%%%%%%%%%%%%%%%%%%%%%%%%%%%%%%%%%%%%%%
%        Delta va rb and  Delta vs Delta_a for universal function \Psi 
%%%%%%%%%%%%%%%%%%%%%%%%%%%%%%%%%%%%%%%%%%%%%%%%%%%%%%%%%%%%%%%%%%%%%%%%%%%%%%%%%%%%%%%%%%%%%%%%%%%%%%%%%%%%%%%%%%%%%%%%%%%%%%%
\begin{figure}[ht!]
\begin{minipage}[h]{0.5\linewidth}
\center{\includegraphics[width=1\linewidth]{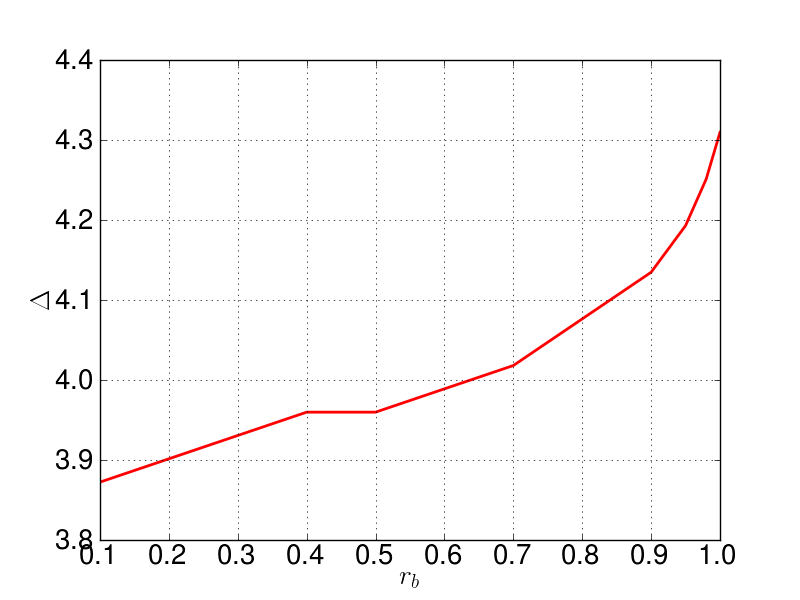}}
\caption{\small{The shift $\Delta$ as a function of $r_b$, obtained from Fig.\ref{brush_therm_pot_univ_psi_fig} for the universal function $\hat{\Psi}(h_{eff})$.}}
\label{brush_delta_vs_rb_for_psi}
\end{minipage}
\hfill
\begin{minipage}[h]{0.5\linewidth}
\center{\includegraphics[width=1\linewidth]{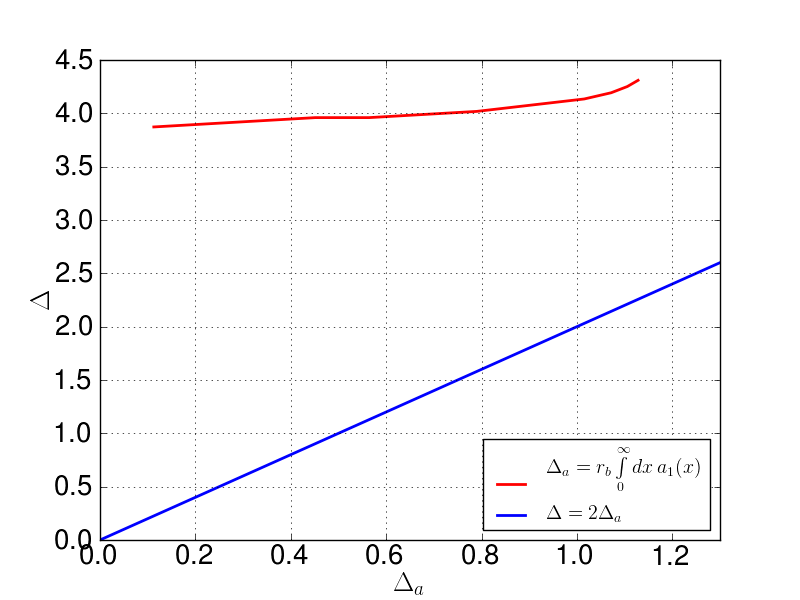}}
\caption{\small{The shift $\Delta$ as a function of $\Delta_a$, obtained from Fig.\ref{brush_therm_pot_univ_psi_fig}, for the universal function $\hat{\Psi}(h_{eff})$ and the theoretical prediction $\Delta \simeq 2\Delta_a$.}}
\label{brush_delta_vs_deltaa_for_psi_fig}
\end{minipage}
\end{figure}	  
For comparison, in Fig.\ref{brush_delta_vs_deltaa_for_psi_fig}, we also present how the shift $\Delta$ for the universal function $\hat{\Psi}(h_{eff})$ depends on $\Delta_a$. In this case, as one can notice, the shift $\Delta$ is almost independent on $\Delta_a$.
%%%%%%%%%%%%%%%%%%%%%%%%%%%%%%%%%%%%%%%%%%%%%%%%%%%%%%%%%%%%%%%%%%%%%%%%%%%%%%%%%%%%%%%%%%%%%%%%%%%%%%%%%%%%%%%%%%%%%%%%%%%%%%%%%%%%%%%%%%%%%%%%%%%%%%%%%%%%%
%           Adsorbed layer with different layer thickness
%%%%%%%%%%%%%%%%%%%%%%%%%%%%%%%%%%%%%%%%%%%%%%%%%%%%%%%%%%%%%%%%%%%%%%%%%%%%%%%%%%%%%%%%%%%%%%%%%%%%%%%%%%%%%%%%%%%%%%%%%%%%%%%%%%%%%%%%%%%%%%%%%%%%%%%%%%%%%	 
\section{Adsorbed layer with different layer thickness}
In this section, we generalize the soft shell layer problem allowing that the chain length of the adsorbed polymers, $N_a$ is different from the length $N$ of the free polymers in solution. In Eq.(\ref{brushes_scft}) for the self-consistent field, the function $a(x)$, that is responsible for the soft shell layer, depends on the variable $x \leftarrow x/R_g$. If we fix the chain length $N_a$ and put the expression for the soft shell layer in Eq.(\ref{brushes_scft}), that is suitable for description of free chains with length $N$, we should rescale the spacial variable by the following way
$$
	   x_{old} \leftarrow \frac{x}{R_g^{old}} = \frac{R_g^{new}}{R_g^{old}}\frac{x}{R_g^{new}} = \frac{x_{new}}{n_a}
$$
where, based on the definition of radius of gyration, we introduced the parameter $n_a = \sqrt{N_a/N}$ which is the reduced length of the adsorbed polymers. Therefore, the equation for self-consistent field  can be rewritten more generally:
\begin{equation}
\label{brushes_scft_improved}
	   w(x) = v_N\left(r_ba(n_a, x) + (\phi_f/\phi_b) - 1\right)
\end{equation}	   
which now contains two parameters. Thus, we can control the density and the thickness of the soft shell layer. As it was shown earlier, in 
Eq.(\ref{brushes_x_plates_profile_ads_calc}), the function $a(x)$ is composed of many one-plate concentration profiles of adsorbed chains. 
Thus, explicitly, we can write the generalized form of one-plate profile of the adsorbed chains as
\begin{equation}
\label{brushes_one_plate_profile_free_diff}
\begin{array}{l}       
       a_1(n_a, x) \equiv 1 - f_1(n_a, x), \\
       f_1(n_a, x) \equiv a_1\left(\frac{x}{n_a}\right)
\end{array}
\end{equation}
The function $a(n_a, x)$ is calculated\footnote{One more point should be mentioned here. For the numerical calculations, in the case $n_a =1$, we used $h_{max}= 10$ replacing infinity. When $n_a \gg 1$, the condition for $h_{max}$ should be written more carefully, because the function $a_1(x/n_a)$ could become periodic. Thereby, we will use  $h_{max}/n_a = 10$.} 
similarly with Eq.(\ref{brushes_x_plates_profile_ads}). 

%%%%%%%%%%%%%%%%%%%%%%%%%%%%%%%%%%%%%%%%%%%%%%%%%%%%%%%%%%%%%%%%%%%%%%%%%%%%%%%%%%%%%%%%%%%%%%%%%%%%%%%%%%%%%%%%%%%%%%%%%%%%%%%%%%%%%%%%%%%%%%%%%%%%%%%%%%%%%
%           Concentration restriction.
%%%%%%%%%%%%%%%%%%%%%%%%%%%%%%%%%%%%%%%%%%%%%%%%%%%%%%%%%%%%%%%%%%%%%%%%%%%%%%%%%%%%%%%%%%%%%%%%%%%%%%%%%%%%%%%%%%%%%%%%%%%%%%%%%%%%%%%%%%%%%%%%%%%%%%%%%%%%% 
%\section{Concentration restriction.}

%%%%%%%%%%%%%%%%%%%%%%%%%%%%%%%%%%%%%%%%%%%%%%%%%%%%%%%%%%%%%%%%%%%%%%%%%%%%%%%%%%%%%%%%%%%%%%%%%%%%%%%%%%%%%%%%%%%%%%%%%%%%%%%%%%%%%%%%%%%%%%%%%%%%%%%%%%%%%
%           The virial parameters providing the maximum barrier height.
%%%%%%%%%%%%%%%%%%%%%%%%%%%%%%%%%%%%%%%%%%%%%%%%%%%%%%%%%%%%%%%%%%%%%%%%%%%%%%%%%%%%%%%%%%%%%%%%%%%%%%%%%%%%%%%%%%%%%%%%%%%%%%%%%%%%%%%%%%%%%%%%%%%%%%%%%%%%%
\section{The virial parameters providing the maximum barrier height}
\begin{figure}[ht!]
\begin{minipage}[h]{0.5\linewidth}
\center{\includegraphics[width=1\linewidth]{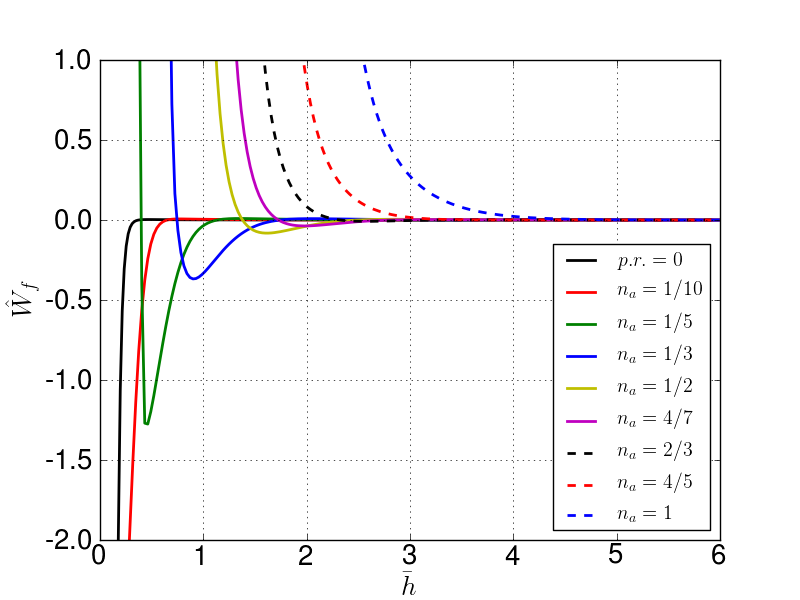}}
\caption{\small{The thermodynamic potential calculated in the adsorbed field, Eq.(\ref{brushes_one_plate_profile_free_diff}), for different values of the parameter $n_a$. 
                Fixed parameters: $r_b=1$, $v_N=500$, $N_x=8k$, $N_s=3k$.}}
\label{brush_therm_pot_diff_na_fig}
\end{minipage}
\hfill
\begin{minipage}[h]{0.5\linewidth}
\center{\includegraphics[width=1\linewidth]{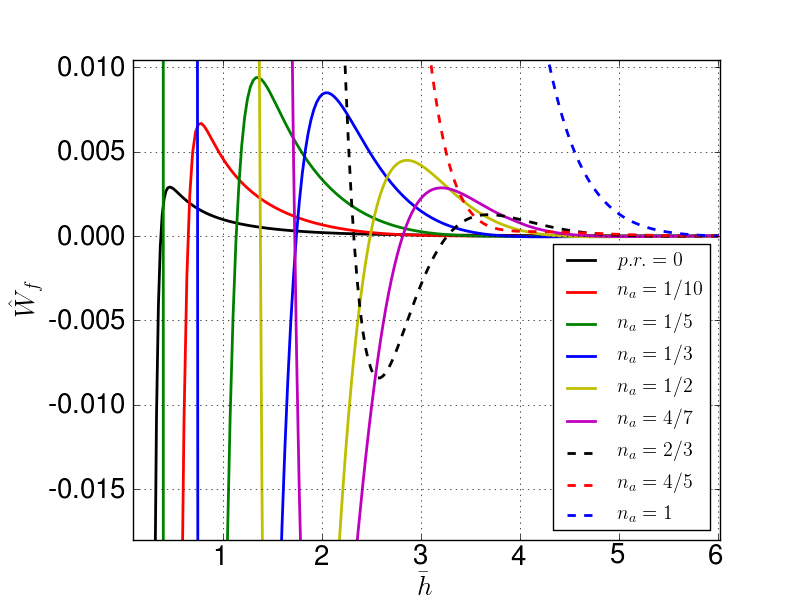}}
\caption{\small{Zoomed in thermodynamic potentials shown in Fig.\ref{brush_therm_pot_diff_na_fig}.}}
\label{brush_therm_pot_diff_na_zoomed_fig}
\end{minipage}
\end{figure}

The main purpose of the section is to find values of the parameters $v_N$, $r_b$ and $n_a$ which correspond to the maximum barrier height 
of the thermodynamic potential, Eqs.(\ref{brush_therm_pot_dimless}, \ref{brushes_edwards_eqdwards_free_scft_qx0}), obtained in the SCFT for the soft shell layer, Eq.(\ref{brushes_x_plates_profile_ads_calc})
using as the one-plate profile of the adsorbed chains Eq.(\ref{brushes_one_plate_profile_free_diff}).
For simplicity, all calculations are done for the mesh size with: $N_x = 8k$, $N_s=3k$, $N_h=100$.
As in the previous chapters, we denote the dimensionless barrier height as
$$
	\hat{W}^* = \hat{W}_{max} - \hat{W}_{inf}
$$
and the corresponding dimensionless position of the barrier as $\bar{h}^*$. 
The results for the thermodynamic potential calculated in the adsorbed field, Eq.(\ref{brushes_one_plate_profile_free_diff}), for different parameter $n_a$ and fixed parameters: $r_b=1$, $v_N=500$ are presented in Figs.\ref{brush_therm_pot_diff_na_fig}--\ref{brush_therm_pot_diff_na_zoomed_fig}. 
%%%%%%%%%%%%%%%%%%%%%%%%%%%%%%%%%%%%%%%%%%%%%%%%%%%%%%%%%%%%%%%%%%%%%%%%%%%%%%%%%%%%%%%%%%%%%%%%%%%%%%%%%%%%%%%%%%%%%%%%%%%%%%%
%        Thermodynamic potential for different n_a
%%%%%%%%%%%%%%%%%%%%%%%%%%%%%%%%%%%%%%%%%%%%%%%%%%%%%%%%%%%%%%%%%%%%%%%%%%%%%%%%%%%%%%%%%%%%%%%%%%%%%%%%%%%%%%%%%%%%%%%%%%%%%%%
We represented the pictures in the two different scales. In Fig.\ref{brush_therm_pot_diff_na_fig} one can see the strong repulsive part of the potential 
produced due to interaction of adsorbed polymers from the opposite surfaces. However, when we decrease the length of the adsorbed polymers, in comparison with the free polymers, the attractive well appears and becomes deeper upon decreasing the parameter $n_a$. 
From physical point of view, when the soft layer becomes thinner it can not override the depletion attraction.
In both pictures, we put the thermodynamic potential obtained for the purely repulsive surface with the same value of the virial parameter, i.e. $v_N = 500$. We denoted the potential in these pictures as $p.r$. Unfortunately, we are not prepared within our computational abilities to consider the thermodynamic potentials with $n_a < 0.1$. Nevertheless, one can extrapolate that a further decrease of $n_a$ leads to the curve for $p.r$.

%%%%%%%%%%%%%%%%%%%%%%%%%%%%%%%%%%%%%%%%%%%%%%%%%%%%%%%%%%%%%%%%%%%%%%%%%%%%%%%%%%%%%%%%%%%%%%%%%%%%%%%%%%%%%%%%%%%%%%%%%%%%%%%
%        barrier(h) vs n_a
%%%%%%%%%%%%%%%%%%%%%%%%%%%%%%%%%%%%%%%%%%%%%%%%%%%%%%%%%%%%%%%%%%%%%%%%%%%%%%%%%%%%%%%%%%%%%%%%%%%%%%%%%%%%%%%%%%%%%%%%%%%%%%%
\begin{figure}[ht!]
\begin{minipage}[h]{0.5\linewidth}
\center{\includegraphics[width=1\linewidth]{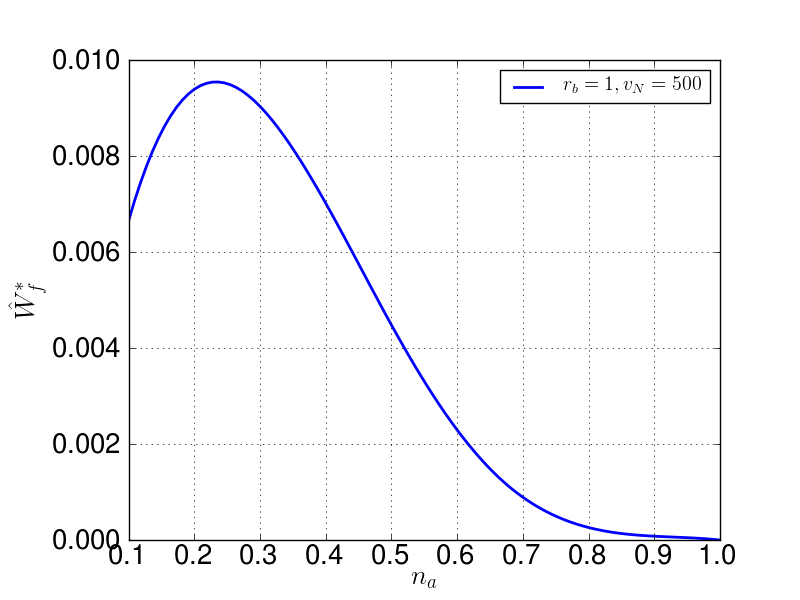}}
\caption{\small{The barrier height as a function of $n_a = \sqrt{N_a/N}$. 
                The data correspond to Fig.\ref{brush_therm_pot_diff_na_zoomed_fig}.}}
\label{brush_barrier_w_vs_na_fig}
\end{minipage}
\hfill
\begin{minipage}[h]{0.5\linewidth}
\center{\includegraphics[width=1\linewidth]{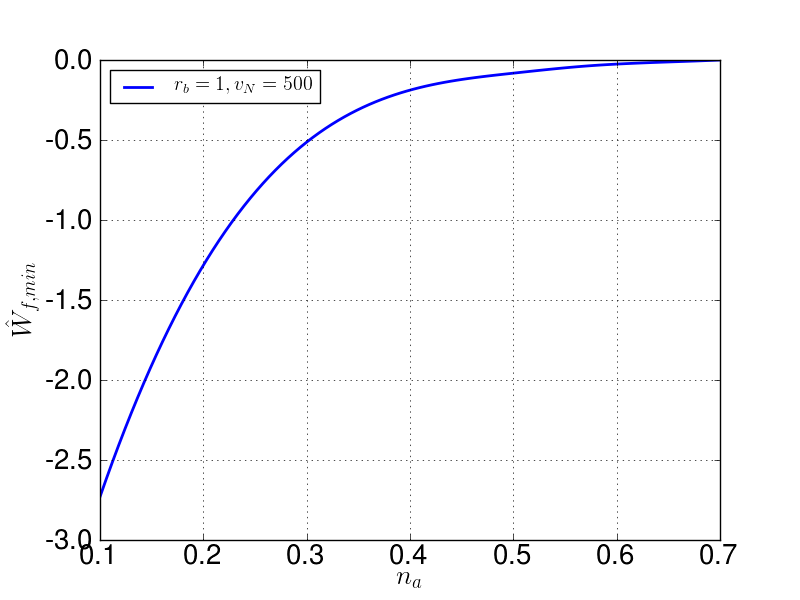}}
\caption{\small{The depth of the well as a function of $n_a = \sqrt{N_a/N}$. 
                The data correspond to Fig.\ref{brush_therm_pot_diff_na_zoomed_fig}.}}
\label{brush_min_w_vs_na_fig}
\end{minipage}
\end{figure}
\begin{figure}[ht!]
\begin{minipage}[h]{0.5\linewidth}
\center{\includegraphics[width=1\linewidth]{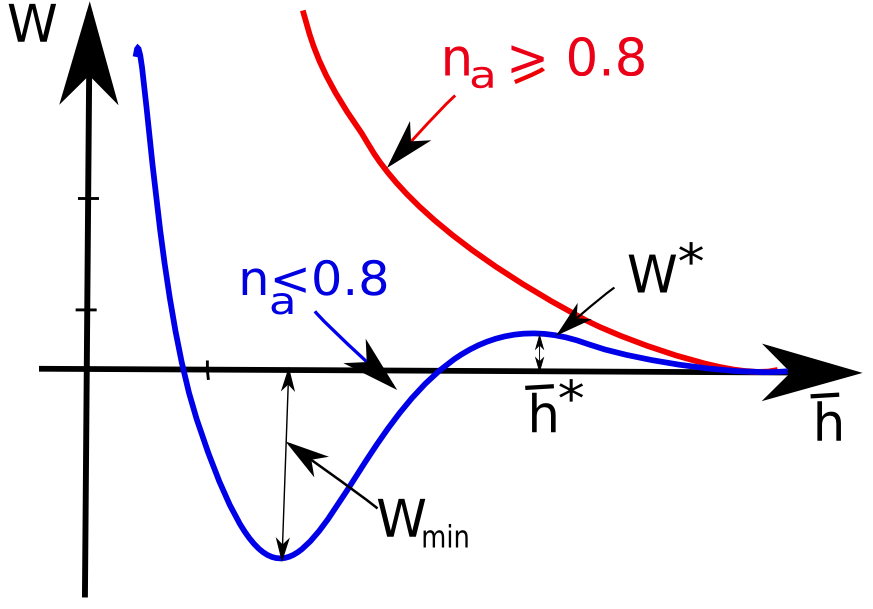}}
\caption{\small{Schematic representation of the thermodynamic potential in two regimes.}}
\label{brush_therm_pot_regimes_fig}
\end{minipage}
\hfill
\begin{minipage}[h]{0.5\linewidth}
\center{\includegraphics[width=1\linewidth]{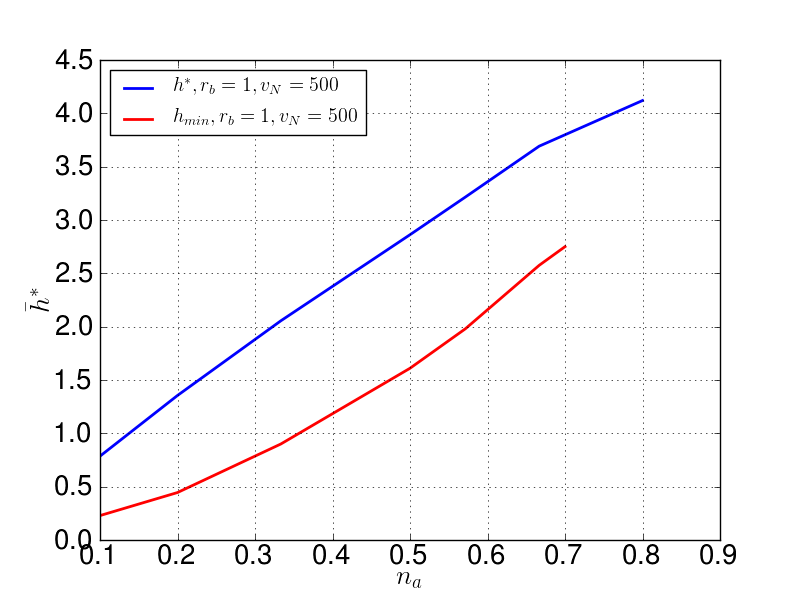}}
\caption{\small{The barrier position and the position of $W_{f, min}$ as a function of $n_a$. 
                The data correspond to Fig.\ref{brush_therm_pot_diff_na_zoomed_fig}.}}
\label{brush_barrier_h_vs_na_fig}
\end{minipage}
\end{figure}

 
In Figs.\ref{brush_barrier_w_vs_na_fig}--\ref{brush_min_w_vs_na_fig} we show how the barrier height and $W_{f, min}$ depend on the parameter $n_a$. $W_{f, min}$ is monotonically grow with the parameter $n_a$. The barrier height increases with the parameter for low $n_a$, reaches 
the maximum at optimum $n_a$ and then decreases at higher $n_a$. Thus, as $n_a$ increases, colloidal particles may change from instability to stability, and then to instability again. As one can notice from Fig.\ref{brush_barrier_w_vs_na_fig}, the maximum repulsion, due to free polymers, is achieved for $n_a \simeq 1/4$ or, in terms of the polymer lengths: $N\simeq 16N_a$. Based on the results shown in Fig.\ref{brush_therm_pot_diff_na_fig}--\ref{brush_therm_pot_diff_na_zoomed_fig}, we schematically represented the thermodynamic potential in two regimes: $n_a<0.8$ and $n_a\ge 0.8$, see Fig.\ref{brush_therm_pot_regimes_fig}. One can notice that in case $n_a\ge 0.8$ the thermodynamic potential is monotonically decreasing function, while in the case $n_a<0.8$ the thermodynamic potential is not-monotonic function having a local maximum and a global minimum.
%%%%%%%%%%%%%%%%%%%%%%%%%%%%%%%%%%%%%%%%%%%%%%%%%%%%%%%%%%%%%%%%%%%%%%%%%%%%%%%%%%%%%%%%%%%%%%%%%%%%%%%%%%%%%%%%%%%%%%%%%%%%%%%
%        barrier(h) vs n_a
%%%%%%%%%%%%%%%%%%%%%%%%%%%%%%%%%%%%%%%%%%%%%%%%%%%%%%%%%%%%%%%%%%%%%%%%%%%%%%%%%%%%%%%%%%%%%%%%%%%%%%%%%%%%%%%%%%%%%%%%%%%%%%%
In the case $n_a\leq 0.8$, the thermodynamic potential has only the repulsive part and for the colloidal stabilization it will compete only with the van der Waals interaction. On the other hand, in the case $n_a<0.8$, the thermodynamic potential has repulsive as well as attractive parts. The magnitudes of these parts are defined by the parameters $n_a$ and $r_b$. 

In addition, we show in Fig.\ref{brush_barrier_h_vs_na_fig} the dependence of barrier height position and $W_{f, min}$ position on $n_a$. One can notice that in both cases these are linear functions.
%%%%%%%%%%%%%%%%%%%%%%%%%%%%%%%%%%%%%%%%%%%%%%%%%%%%%%%%%%%%%%%%%%%%%%%%%%%%%%%%%%%%%%%%%%%%%%%%%%%%%%%%%%%%%%%%%%%%%%%%%%%%%%%
%        barrier(h) vs n_a
%%%%%%%%%%%%%%%%%%%%%%%%%%%%%%%%%%%%%%%%%%%%%%%%%%%%%%%%%%%%%%%%%%%%%%%%%%%%%%%%%%%%%%%%%%%%%%%%%%%%%%%%%%%%%%%%%%%%%%%%%%%%%%%
\begin{figure}[ht!]
\begin{minipage}[h]{0.5\linewidth}
\center{\includegraphics[width=1\linewidth]{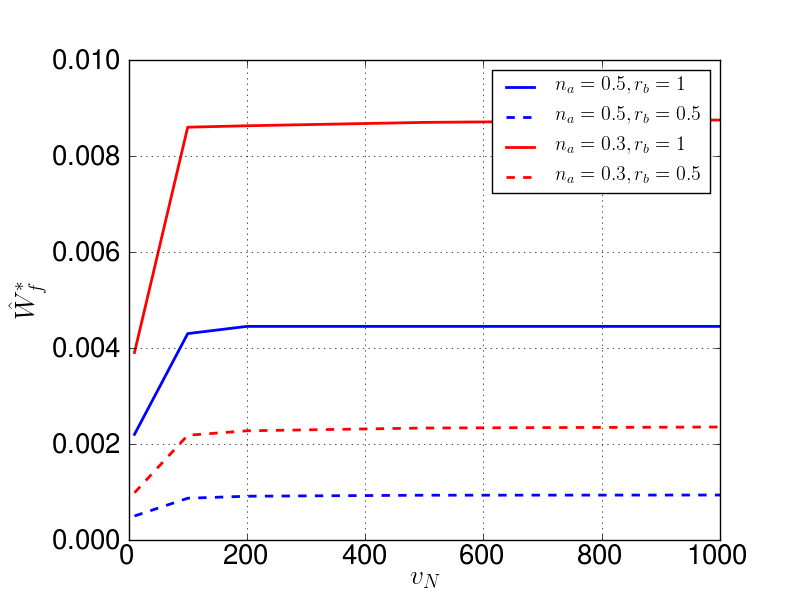}}
\caption{\small{The barrier height as a function of $v_N$ for different values of $n_a$ and $r_b$. 
                The data correspond to Fig.\ref{brush_therm_pot_diff_na_zoomed_fig}.}}
\label{brush_barrier_w_vs_vN_fig}
\end{minipage}
\hfill
\begin{minipage}[h]{0.5\linewidth}
\center{\includegraphics[width=1\linewidth]{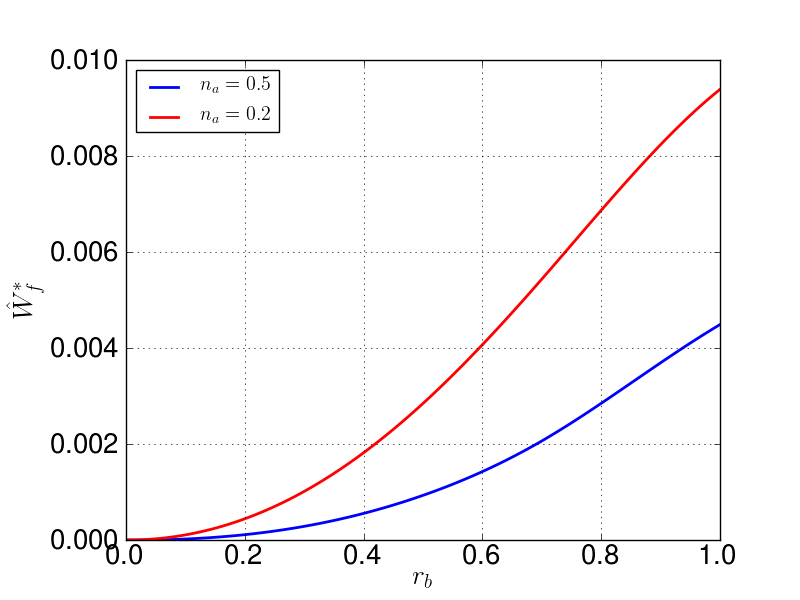}}
\caption{\small{The barrier position as a function of $r_b$ for different values of $n_a$. 
                The data correspond to Fig.\ref{brush_therm_pot_diff_na_zoomed_fig}.}}
\label{brush_barrier_w_vs_rb_fig}
\end{minipage}
\end{figure}          	 	 	 

In Sec.\ref{sec:brush_univ_behavior} we have already seen that the thermodynamic potential slightly depends on the parameter $v_N$ for $n_a=1$. In Fig.\ref{brush_barrier_w_vs_vN_fig} we present the barrier height as a function of $v_N$ for some values of $r_b$ and $n_a$. One can notice that starting from a certain value of $v_N^*$ ($v_N^* > 50$) the barrier height is independent of $v_N$. This allows us to use the results obtained for $v_N=500$ for any other $v_N$ when we will reconstruct the thermodynamic potential in the real variables.

As a final step of this section we investigate the dependence of the barrier height on the parameter $r_b$ (see 
Sec.\ref{sec:brush_num_solutions}). This dependence is shown in Fig.\ref{brush_barrier_w_vs_rb_fig} for several parameters $n_a$. One can notice that $W^*(r_b)$ is a monotonically increasing function which reaches its maximum value at $r_b=1$. Correspondingly, the best stabilization is achieved at $r_b=1$.
%%%%%%%%%%%%%%%%%%%%%%%%%%%%%%%%%%%%%%%%%%%%%%%%%%%%%%%%%%%%%%%%%%%%%%%%%%%%%%%%%%%%%%%%%%%%%%%%%%%%%%%%%%%%%%%%%%%%%%%%%%%%%%%%%%%%%%%%%%%%%%%%%%%%%%%%%%%%%
%        Real variables.
%%%%%%%%%%%%%%%%%%%%%%%%%%%%%%%%%%%%%%%%%%%%%%%%%%%%%%%%%%%%%%%%%%%%%%%%%%%%%%%%%%%%%%%%%%%%%%%%%%%%%%%%%%%%%%%%%%%%%%%%%%%%%%%%%%%%%%%%%%%%%%%%%%%%%%%%%%%%%
\section{Real variables}
In order to find the condition for the colloidal stabilization, we recalculate the SCFT thermodynamic potential in $k_BT$ units for the spherical geometry and compare it with the Van der Waals attraction. Thereby, the full thermodynamic potential is
$$
	U_{tot}(h) = U(h) + U_{VdW}(h)
$$
where $U(h)$ corresponds to the thermodynamic potential produced by free polymers and $U_{VdW}(h)$ to the Van der Waals attraction potential. In Sec.\ref{sec:colloids_VdW} we found the expression for the van der Waals attraction
$$
	U_{VdW}(h) = -\frac{A_HR_c}{12h}
$$	
where $A_H$ is the Hamaker constant, $h$ is the distance between surfaces of the interacting particles. In Sec.\ref{sec:repulsive_real_var} the Hamaker constant was found for polystyrene latex-particles: $A_H\simeq 0.138 k_BT$. 	
	
Using the Derjaguin approximation (see Sec.\ref{sec:colloids_derjaguin}), in a spherical geometry we can write the corresponding expression for the SCFT thermodynamic potential:
$$
U(h) = \pi R_c \int\limits_h^{\infty}\mathrm{d}h'W(h') = \pi R_c \frac{c_bR_g}{N} \int\limits_h^{\infty}\mathrm{d}h'\hat{W}(h') = 
\pi R_c \frac{c_bR_g^2}{N} \int\limits_{\bar{h}}^{\infty}\mathrm{d}\bar{h}\hat{W}(\bar{h}) = A_R \hat{U}(\bar{h})
$$
Thereby,
\begin{equation}
\label{brush_scft_therm_pot_real}
U(h) = A_R \hat{U}(h/R_g)
\end{equation}
where we denoted:
$$
A_R = \pi R_c \frac{c_bR_g^2}{N} = \pi\left(\frac{R_c}{R_g}\right)\frac{c_bR_g^3}{N}
$$
is a prefactor that defines the magnitude of the barrier and
\begin{equation}
\label{brush_scft_therm_pot_real_reduced}
\hat{U}(\bar{h}) = \int\limits_{\bar{h}}^{\infty}\mathrm{d}\bar{h}\hat{W}(\bar{h})   
\end{equation}
is the dimensionless thermodynamic potential for the spherical geometry, which is expressed via the thermodynamic potential obtained in the SCFT for the flat geometry. We denote the corresponding barrier height as $\hat{U}^*$. As we already did it for the purely repulsive case, the bulk polymer concentration can be written in the form:
$$
c_b = \frac{N_A\rho}{M_0}\phi_b
$$ 
where $N_A$ is  the Avogadro constant, $\rho$ is the density of the corresponding polymer and $M_0$ is the molar mass of a monomer unit.
Correspondingly, we can rewrite the prefactor $A_R$ as	
$$
	A_R = \pi N_A\left(\frac{R_c}{R_g}\right)\left(\frac{R_g^3\rho}{NM_0}\right)\phi_b
$$
One can notice that the quantity is independent of $N$. We already found the parameters of polystyrene in toluene solvent (see Sec.\ref{sec:polymers_experiment}): $M_0(\text{styrene})=104.15\,\text{g/mol}, \quad a_s = 7.6\textup{\AA}, \quad \rho = 960\,g/L$. 
%%%%%%%%%%%%%%%%%%%%%%%%%%%%%%%%%%%%%%%%%%%%%%%%%%%%%%%%%%%%%%%%%%%%%%%%%%%%%%%%%%%%%%%%%%%%%%%%%%%%%%%%%%%%%%%%%%%%%%%%%%%%%%%
%        therm pot total vs h in nm
%%%%%%%%%%%%%%%%%%%%%%%%%%%%%%%%%%%%%%%%%%%%%%%%%%%%%%%%%%%%%%%%%%%%%%%%%%%%%%%%%%%%%%%%%%%%%%%%%%%%%%%%%%%%%%%%%%%%%%%%%%%%%%%
\begin{figure}[ht!]
\begin{minipage}[h]{0.5\linewidth}
\center{\includegraphics[width=1\linewidth]{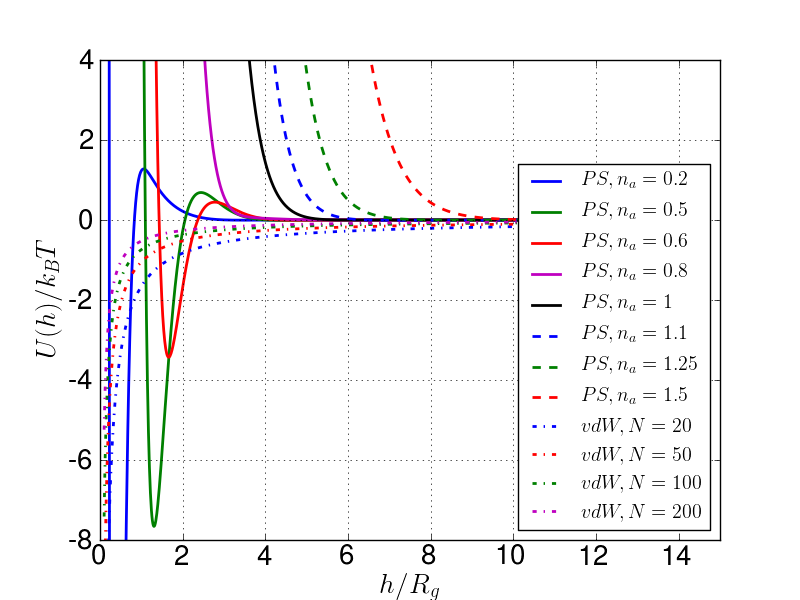}}
\caption{\small{The thermodynamic potential and the van der Waals potential as a function of dimensionless separation $\bar{h}$. 
                Fixed parameters: $R_c = 200 \text{nm}$, $\phi\simeq 0.5$, $A_H\simeq 0.138k_BT$.}}
\label{brush_therm_pot_ps_all_hRg_fig}
\end{minipage}
\hfill
\begin{minipage}[h]{0.5\linewidth}
\center{\includegraphics[width=1\linewidth]{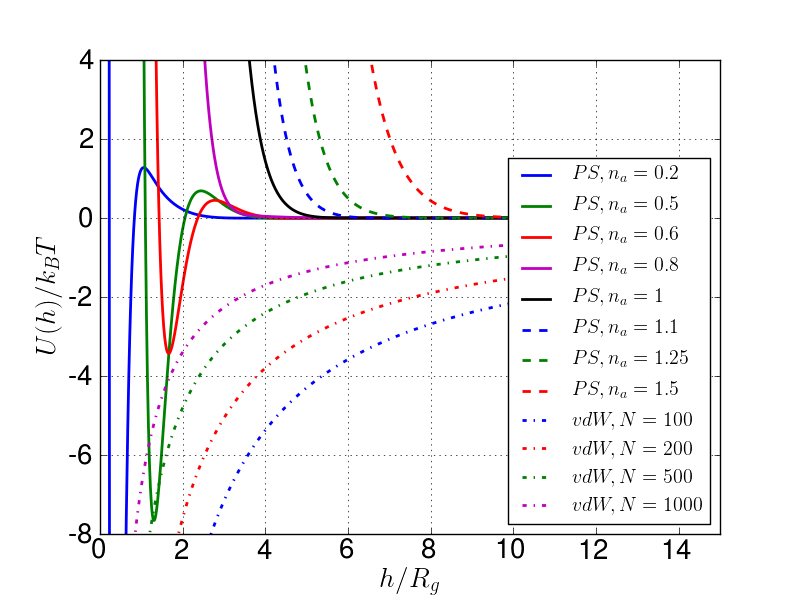}}
\caption{\small{The thermodynamic potential and the van der Waals potential as a function of dimensionless separation $\bar{h}$. 
                Fixed parameters: $R_c = 200 \text{nm}$, $\phi\simeq 0.5$, $A_H\simeq 4k_BT$.}}
\label{brush_therm_pot_ps_all_hRg_largeAh_fig}
\end{minipage}
\end{figure}          	 	 	 
  	 	 	 
\begin{figure}[ht!]
\begin{minipage}[h]{0.5\linewidth}
\center{\includegraphics[width=1\linewidth]{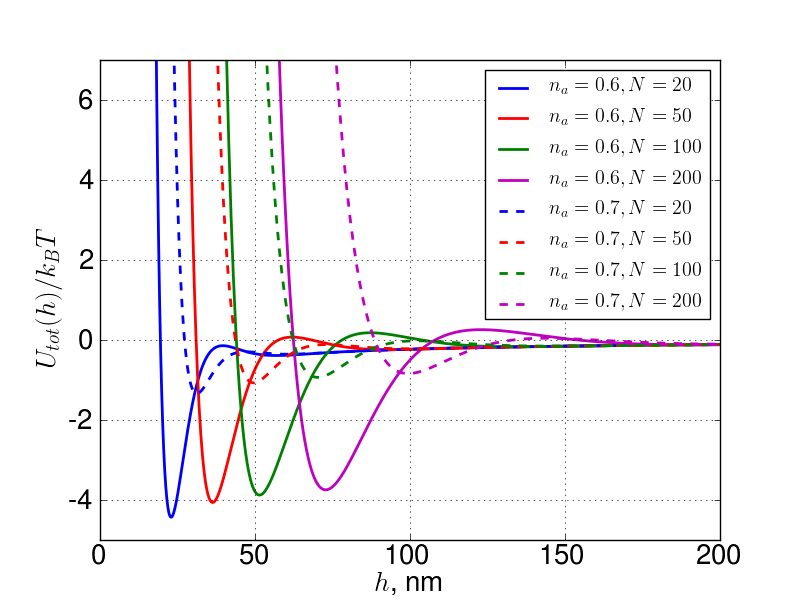}}
\caption{\small{The total thermodynamic potential of interaction between two colloidal particles with radii $R_c = 200 \text{nm}$ 
		produced by free polystyrene in toluene calculated  for polymer volume fraction, $\phi\simeq 0.5$ and different chain lengths.}}
\label{brush_therm_ps_tot_na67_fig}
\end{minipage}
\hfill
\begin{minipage}[h]{0.5\linewidth}
\center{\includegraphics[width=1\linewidth]{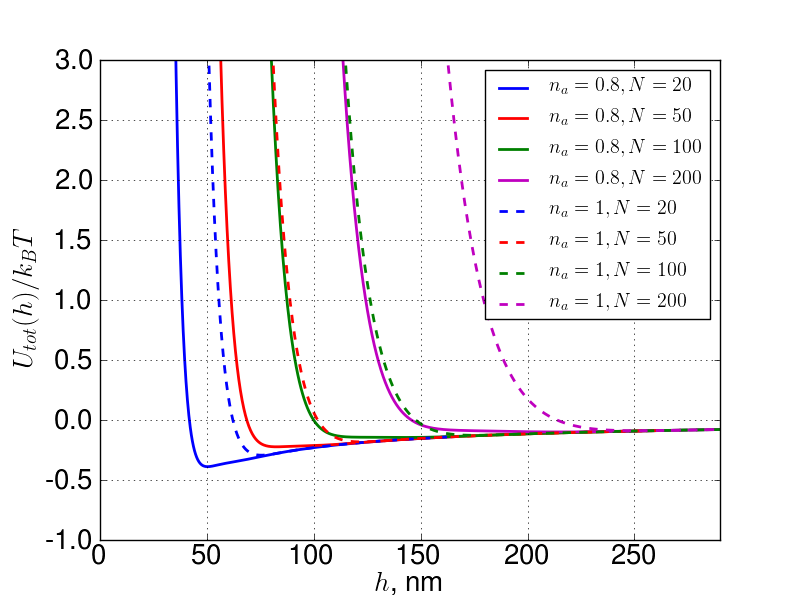}}
\caption{\small{The total thermodynamic potential of interaction between two colloidal particles with radii $R_c = 200 \text{nm}$ 
		produced by free polystyrene in toluene calculated in real variables for polymer volume fraction, $\phi\simeq 0.5$ and different chain lengths.}}
\label{brush_therm_ps_tot_na81_fig}
\end{minipage}
\end{figure}

In order to find the region of absolute (not just kinetic) stability in terms of $N$, $n_a$ we use the condition that the depth of the minimum must be $\lesssim 1k_BT$. Since the SCFT potential in the real variables is independent of $N$ it is convenient to represent this potential in real variables as a function of $h/R_g$ for different values of $n_a$ as shown in Fig.\ref{brush_therm_pot_ps_all_hRg_fig}. In this graph we also added the van der Waals energy $U_{VdW} = -A_HR_c/(12R_g\bar{h})$ calculated for several $N$. One can notice that the van der Waals interaction is weak for the relevant separations. Thus, for the system, the stability is defined almost totally by the 
parameter $n_a$. The resulting total thermodynamic potential calculated for the real systems is presented in 
Figs.\ref{brush_therm_ps_tot_na67_fig}--\ref{brush_therm_ps_tot_na81_fig} for several values of $N$ and $n_a$. One can notice that
the criterion of the absolute stability is satisfied for $n_a\geq 0.7$. The situation changes for large values of the Hamaker constant.
%%%%%%%%%%%%%%%%%%%%%%%%%%%%%%%%%%%%%%%%%%%%%%%%%%%%%%%%%%%%%%%%%%%%%%%%%%%%%%%%%%%%%%%%%%%%%%%%%%%%%%%%%%%%%%%%%%%%%%%%%%%%%%%
%        therm pot total vs h in nm
%%%%%%%%%%%%%%%%%%%%%%%%%%%%%%%%%%%%%%%%%%%%%%%%%%%%%%%%%%%%%%%%%%%%%%%%%%%%%%%%%%%%%%%%%%%%%%%%%%%%%%%%%%%%%%%%%%%%%%%%%%%%%%%
For example, in Fig.\ref{brush_therm_pot_ps_all_hRg_largeAh_fig} we represent the thermodynamic potential as a function of the dimensionless separation for $A_H = 4k_BT$. 
In this case the criterion of the absolute stability would be $n_a>1$ and $N>500$ or upon further increase of $N$ we can decrease $n_a$. 
%%%%%%%%%%%%%%%%%%%%%%%%%%%%%%%%%%%%%%%%%%%%%%%%%%%%%%%%%%%%%%%%%%%%%%%%%%%%%%%%%%%%%%%%%%%%%%%%%%%%%%%%%%%%%%%%%%%%%%%%%%%%%%%%%%%%%%%%%%%%%%%%%%%%%%%%%%%%%
%           Conclusion
%%%%%%%%%%%%%%%%%%%%%%%%%%%%%%%%%%%%%%%%%%%%%%%%%%%%%%%%%%%%%%%%%%%%%%%%%%%%%%%%%%%%%%%%%%%%%%%%%%%%%%%%%%%%%%%%%%%%%%%%%%%%%%%%%%%%%%%%%%%%%%%%%%%%%%%%%%%% 
\section{Summary and Conclusion}
In this chapter we developed the theory of polymer-induced interaction between surfaces with irreversibly adsorbed layers, taking into account both irreversibly adsorbed and free chains. 

We constructed the soft shell layer using the analytical solution of the Edwards equation written for the concentrated polymer solution. 
When the colloidal particles collide with each other their layers overlap. In order to keep the number of monomers belonging to these layers constant, 
we reflect one-plate concentration profile from the opposite wall when it intersects with the wall. 
Then, we investigated the numerical solution of the Edwards equation for free polymers which are squeezed between these layers.

We found that the thermodynamic potential is nearly independent of the parameter $v_N^*=\text{v}^*c_bN$ and it is defined only by the reduced thickness $n_a=\sqrt{N_a/N}$ and the 
density $r_b$ of the adsorbed layer. While varying the density of the layer, we found that the thermodynamic potential can be described using the universal function
$\hat{\Psi}(h) = \hat{W}_f(h+\Delta)/r_b^2$, where $\hat{W}_f$ is the dimensionless thermodynamic potential of free polymers and $\Delta = \Delta(r_b)$ is the shift along $h$ direction.
Moreover, we found that the thermodynamic potential is an increasing function of $r_b$. Varying the thickness of the layer, we found that the thermodynamic potential in the case $n_a\ge 0.8$ is a positive monotonically decreasing function of $h$, while in the case $n_a<0.8$ the thermodynamic potential 
is not-monotonic function having a global minimum and a local maximum whose peak value is reached for $n_a \simeq 0.25$.

The thickness and the density of the layers were varied in the effort to find the best conditions for the stabilization. We found that the interaction potential profiles are significantly
affected by the presence of soft layers. Most importantly, we revealed a strongly enhanced cumulative stabilization effect of free polymers in tandem with the soft adsorbed layers for $n_a\gtrsim 0.8$.

We also considered real colloidal system consistent of micro-gel PS particles ($R_c \simeq 200 nm$) and established that for $n_a>0.7$ the resultant potential energy barrier is high enough to prevent colloid coagulation/flocculation. 
This effect can be traced back to a favorable modulation (by the adsorbed chains) of the distribution of the unattached polymers increasing the free energy and the pressure of their free ends.
%\cite{Israelachvili_2011}

%% file: Chapters/conclusion.tex
% Chapter 6
\chapter{Summary and outlook} % Main chapter title
\label{chap:Chapter6} % For referencing the chapter elsewhere, use \ref{Chapter1} 
\lhead{Chapter 6. \emph{Summary and outlook}} % This is for the header on each page - perhaps a shortened title

Stable colloidal dispersions with evenly distributed particles are important for many technological applications \cite{Napper_book}. 
Due to Brownian motion colloidal particles have constant collisions with each other which often lead to their aggregation driven by the long range van der Waals 
attraction. As a result the colloidal systems often tend to precipitate. A number of methods have been devised to minimize the effect of long-range van der Waals 
attraction between colloidal particles or to override the influence of the attraction in order to provide the colloidal stability \cite{Israelachvili_2011}.

In this PhD thesis we investigated the colloidal stabilization in solutions of free polymers which is commonly referred to as depletion stabilization \cite{Napper_book}. 
Previous theoretical studies of free-polymer induced (FPI) stabilization were based on oversimplified models involving uncontrolled approximations \cite{Napper_book, scheijtens_1982}. 
Even the most basic features of the depletion stabilization phenomenon were unknown. It was unclear how the PI repulsion depends on the solution parameters, 
polymer structure and monomer/surface interactions. Thus, the following problems arose naturally as the main goals of this thesis: \\
\textbf{a)} To elucidate the nature of the free polymer induced interaction and to quantitatively assess their range and magnitude as well as their potential for colloidal stabilization. \\
\textbf{b)} To establish how the stabilization effect of FPI interaction depends on the main macroscopic, mesoscopic and molecular parameters of the system. \\
\textbf{c)} To generalize both the numerical and analytical SCFT approaches to the depletion stabilization phenomenon to account for wider concentration and molecular weight regimes and for surface effects (surface repulsion vs. attraction, reversible vs. irreversible adsorption). \\
\textbf{d)} To identify the optimum regimes for exploiting the FPI interaction stabilization effect.\\
\textbf{e)} To choose or optimize the model parameters so that the theoretical calculation corresponds as closely as possible to the parameters of the experimental model systems in order to allow a decent comparison of experimental and theoretical results.

To reach these goals we had to analyse the intricate interplay of colloid-solvent, colloid-polymer, polymer-solvent interactions and  polymer chains conformations. To this end
we developped the self-consistent field theory (SCFT) for the colloid/polymer systems. In this approach polymer chains are treated as random walks propagating in the molecular field created by all the polymer segments. The SCFT then served as a basis for our numerical calculations of the chain conformational distributions, and of  various thermodynamic parameters and interaction energy profiles. 

Noteworthy, we also treated the problem analytically. The classical analytical theory of FPI interactions was developped by De Gennes et al \cite{deGennes_1981, deGennes_1982, Joanny_1979}, based on the ground state dominance (GSD) approximation which is strictly valid for very long (infinite) chains. This approach was improved by taking into account the finite chain length effects \cite{semenov_1996, semenov_2008}. These effects qualitatively change the interaction potential between colloidal particles. This advanced theory which we called the GSDE theory \cite{shvets_2013, semenov2015theory} (where "E" stands for the end-effects) is shown to be in good agreement with computational SCFT results.   
Moreover, we find that the two approaches (analytical GSDE and numerical SCFT) consistently complement each other allowing to fully quantify the FPI interactions in a wide range of conditions in the most important regimes.

\textbf{1) Purely repulsive colloidal surfaces.} \\
In this case the potential of the FPI interaction between two colloidal particles in a semidilute polymer solution generically shows a repulsion energy peak associated with the polymer chain end effects. The barrier height $U_m$ increases with bulk polymer concentration $\phi$ in the semidilute regime; $U_m$ saturates or slightly decreases in  the concentrated regime (as revealed for real systems of polystyrene (PS) in toluene and polyethylene glycol (PEG) in water).  In all the cases considered the repulsion energy due to free polymers in the concentrated solution regime is about $U_m=1-3k_BT$ for colloidal particle size $R_c=100nm$. For the typical range of free chain polymerization degree, $N=25-200$, the potential barrier is developed at separations $h\sim 2nm$ between the solid surfaces. These results published in \cite{shvets_2013} show that  colloidal stability can be significantly improved  in a concentrated solution of free chains. However, in most cases a pre-stabilization by other means would be required. Therefore, it is important to consider FPI interaction in combination with other means enhancing the colloid stabilization effect (reversible or irreversible polymer surface adsorption, grafted polymer layers, soft penetrable rather than solid interacting surfaces). 

\textbf{2) Reversibly adsorbed colloidal surfaces.}\\
Further, we considered colloidal solid surfaces with weak or moderate affinity for segments of dissolved homopolymers leading to formation of reversible polymer adsorption layers. As before, combining the GSDE and the SCFT approaches, which have been generalized for the adsorption case, we obtained  the colloidal interaction potential as a function of separation and external parameters like bulk monomer concentration and chain length.  Our general results for the interaction potential turn out to be nicely consistent with calculations \cite{avalos_2003} done in the weak adsorption regime. We found that the barrier height is an increasing function of adsorption strength, but due to the elevated monomer concentration at the surface this increase is limited for strong adsorption. In accordance with this restriction, the value of the barrier is $U_m=1-3 k_BT$ for the same colloid/polymer systems (PS and PEG solutions). 

\textbf{3) Colloidal surfaces with soft shell layer.}\\
Seeking to enhance the free polymer mediated repulsion, we advanced our studies to consider the possibility of irreversible adsorption of some free polymers on the colloidal surfaces. In these model systems colloidal particles covered by the adsorbed soft layers interact with each other and with free polymers in solution. The thickness and density of the layers were varied in the effort to find the best conditions for the stabilization. We found that the interaction potential profiles are significantly affected by the presence of soft layers both in the case when adsorbed and free chains are identical and when adsorbed chains are shorter.  Most importantly, we revealed a strongly enhanced cumulative stabilization effect of free polymers in tandem with soft adsorbed layers. For typical sizes of  colloidal particles, the resultant potential energy barrier is high enough to prevent colloid coagulation/flocculation.  This effect can be traced back to a favorable modulation (by the adsorbed chains) of the distribution of unattached polymers increasing the free energy and the pressure of their free ends.  

In order to develop further the model and to find new regimes of enhanced colloid stabilization we can consider end-segment adsorption effects, as well as hetero-polymeric systems such as amphiphilic block-copolymers of $HP$, $HPH$ or $HPHP\ldots PH$ types. 
The most important further step would be related to the experimental verification and confirmation of the above theoretical predictions.

%% file: Appendices/AppendixA.tex
% Appendix A
%%%%%%%%%%%%%%%%%%%%%%%%%%%%%%%%%%%%%%%%%%%%%%%%%%%%%%%%%%%%%%%%%%%%%%%%%%%%%%%%%%%%%%%%%%%%%%%%%%%%%%%%%%%%%%%%%%%%%%%%%%%%%%%%%%%%%%%%%%%%%%%%%%%%%%%%%%%%%
%        Tridiagonal matrix algorithm
%%%%%%%%%%%%%%%%%%%%%%%%%%%%%%%%%%%%%%%%%%%%%%%%%%%%%%%%%%%%%%%%%%%%%%%%%%%%%%%%%%%%%%%%%%%%%%%%%%%%%%%%%%%%%%%%%%%%%%%%%%%%%%%%%%%%%%%%%%%%%%%%%%%%%%%%%%%%%
\chapter{Tridiagonal matrix algorithm} % Main appendix title
\label{AppendixA} % For referencing this appendix elsewhere, use \ref{AppendixA}

\lhead{Appendix A. \emph{Tridiagonal matrix algorithm}} % This is for the header on each page - perhaps a shortened title
	The system of equations for the Crank-Nicolson scheme can be represented in the matrix form as $Aq = d$, 
	or if we expand it:
\begin{equation}
\label{full_tridiag_matrix}
\left(
\begin{array}{cccccc}
    b_0  &  c_0  &   0  &     0   &  \ldots&   0\\
    a_1&  b_1&   c_1&     0   &  \ldots&   0\\
     0 &  a_2&   b_2&   c_2   &  \ldots&   0\\
    \ldots&  &\vdots&     & \ddots&  \vdots \\
    \ldots& & 0&      a_{n-2}&    b_{n-2} &  c_{n-2}    \\
    \ldots& &  &               &    a_{n-1} &  b_{n-1}   
\end{array}
\right) \left(
\begin{array}{c}
    q_{0}\\
    q_{1}\\
    q_{2}\\
    \vdots\\
    q_{n-2}\\ 
    q_{n-1}
\end{array}
\right) = 
\left(
\begin{array}{c}
    d_{0}\\
    d_{1}\\
    d_{2}\\
    \vdots\\
    d_{n-2}\\
    d_{n-1}
\end{array}
\right)
\end{equation}
	This system of linear equations can be efficiently solved using $LU$ factorization with the backward substitution. 
	Since the matrix coefficients are known and positive, we can use the $LU$ factorization.
	For the matrix $A$, we want to find matrices $L$ and $U$ such that, $A = LU$. Moreover, the matrices $L$ and $U$ have the form:
\begin{equation}
\label{LU_tridiag_matrix}
L = \left(
\begin{array}{cccccc}
    e_0 &     &      &          &           &    \\
    a_1 &  e_1&      &          &           &    \\
        &  a_2&   e_2&          &           &    \\
        &     &\ddots&    \ddots&           &    \\
        &     &      &   a_{n-2}&    e_{n-2}&     \\
        &     &      &          &    a_{n-1}& e_{n-1}   
\end{array}
\right); \quad 
U = \left(
\begin{array}{cccccc}
      1 &  f_0&      &          &           &    \\
        &    1&   f_1&          &           &    \\
        &     &     1&       f_2&           &    \\
        &     &      &    \ddots&     \ddots&    \\
        &     &      &          &          1&  f_{n-2}   \\
        &     &      &          &           &  1   
\end{array}
\right)
\end{equation}
	Evaluating each non-zero element, in the product $LU$, and setting it equal to the corresponding element in the matrix $A$, leads to 
$$
\begin{array}{rcl}
            e_{0} & = &  b_{0}      \\
       e_{0}f_{0} & = &  c_{0}      \\\\
          a_{1}   & = &  a_{1}      \\
a_{1}f_{0} + e_{1}& = &  b_{1}      \\
           e_1f_1 & = &  c_1      \\\\
          a_{i}   & = &  a_{i}      \\
a_{i}f_{i-1} + e_i& = &  b_{i}      \\
       e_{i}f_{i} & = &  c_{i}      \\\\
          a_{n-1}   & = &  a_{n-1}      \\
a_{n-1}f_{n-2} + e_{n-1}& = &  b_{n-1}      
\end{array}
$$
	Solving the system of equations for the unknown elements, $e_{i}$ and $f_{i}$, gives:
$$
\begin{array}{rcl}
            f_{0} & = &  c_{0}/e_{0} = c_{0}/b_{0}   \\\\
            e_{1} & = &  b_{1} - a_{1}f_{0}          \\
            f_{1} & = &  c_{1}/e_{1}                 \\\\
            e_{i} & = &  b_{i} - a_{i}f_{i-1}         \\
            f_{i} & = &  c_{i}/e_{i}                 \\\\
          e_{n-1} & = &  b_{n-1} - a_{n-1}f_{n-2}      
\end{array}
$$
	The equivalent \verb!C++! code is 
\begin{lstlisting}
e[0] = b[0];
f[0] = c[0]/b[0];
for (int i = 1; i <= n-1; i++){
  e[i] = b[i] - a[i]*f[i-1];
  f[i] = c[i]/e[i];
} 
\end{lstlisting}
	Using that, $A = LU$, the system of equations $Aq = d$, can be rewritten as: $(LU )q = d$ or $L(U q) = d$. Introducing the vector $w$, wich is defined as $w = Uq$, the system of equations becomes $Lw = d$. The matrix $L$ is a lower triangular matrix, correspondingly, the solution for the vector $w$, can  be easily obtained solving $Lw = d$. 
	
When the vector $w$ is known, for the upper triangular matrix $U$, we can easily obtain $q$, solving $Uq = w$. 
Thus, once the matrices $L$ and $U$ are found, the vector $q$, can be computed in two step process:
$$
\begin{array}{ccc}
        \text{solve} & &  Lw = d   \\\\
        \text{solve} & &  Uq = w                 
\end{array}
$$
where for the lower triangular matrix, $L$, the solution is the forward substitution and 
for the upper triangular matrix, $U$, the solution is the backward substitution.

When the vectors $a$, $e$, and $f$, that define the matrices $L$ and $U$, are given, the solution code can be embodied in two loops:
\begin{lstlisting}
// Forward substitution to solve L*w = d
w[0] = d[0]/e[1];
for (int i = 1; i <= n-1; i++)
  w[i] = (d[i] - a[i]*w[i-1])/e[i];
 
// Backward substitution to solve U*v = w
q[n-1] = w[n-1];
for (int i = n-2; i >= 0; i--)
  q[i] = w[i] - f[i]*w[i+1];
\end{lstlisting}
In the algorithm we can optimize the memory usage. 
	The two preceding loops shows that, the vector $q$, is used only when the vector $w$ is already created. Furthermore, the element $q[i]$ is obtained only after the element $q[i+1]$ is found. These facts enable us to eliminate the vector $w$, entirely and use the vector $q$, in its place. This trick saves us from the task of creating the temporary vector $w$.

%% file: Appendices/AppendixB.tex
% Appendix B
%%%%%%%%%%%%%%%%%%%%%%%%%%%%%%%%%%%%%%%%%%%%%%%%%%%%%%%%%%%%%%%%%%%%%%%%%%%%%%%%%%%%%%%%%%%%%%%%%%%%%%%%%%%%%%%%%%%%%%%%%%%%%%%%%%%%%%%%%%%%%%%%%%%%%%%%%%%%%
%        The Crank-Nicolson scheme for the Edwards equation.
%%%%%%%%%%%%%%%%%%%%%%%%%%%%%%%%%%%%%%%%%%%%%%%%%%%%%%%%%%%%%%%%%%%%%%%%%%%%%%%%%%%%%%%%%%%%%%%%%%%%%%%%%%%%%%%%%%%%%%%%%%%%%%%%%%%%%%%%%%%%%%%%%%%%%%%%%%%%%
\chapter{The Crank-Nicolson scheme for the Edwards equation} % Main appendix title
\label{AppendixB} % For referencing this appendix elsewhere, use \ref{AppendixA}

\lhead{Appendix B. \emph{The Crank-Nicolson scheme for the Edwards equation}} % This is for the header on each page - perhaps a shortened title

In the main text we obtained the modified diffusion equation, which define the distribution function, $q(x)$ in the self consistent field, $w$ and in one dimensional case can be written as 
\begin{equation}
\label{app_homo_diff_eq_field}
         \frac{\partial q(x, s)}{\partial s} = \frac{\partial^2 q(x, s)}{\partial x^2} - w(x)q(x, s)
\end{equation}
        where 
\begin{equation}
\label{homo_field}
	 w(x) = v_N(c(x)/c_b-1)+w_N((c(x)/c_b)^2-1) \quad \text{and} \quad c(x)/c_b = \int\limits_{0}^{1}\mathrm{d}s\,q(x, s)q(x, 1-s)
\end{equation}
        with $v_N=\text{v}c_bN$ and $w_N=\text{w}c_b^2N/2$, and the range of definition $q(x, s)\in [0, h_m]\times [0, 1]$. Usually, the solution of such self-consistent system of equations is sought iteratively, solving the modified diffusion equation with known field $w(x)$, on each iterative step. Let us consider the last procedure more deeply.

In order to find the numerical solution of Eq.(\ref{app_homo_diff_eq_field}) with known field $w(x)$
we apply the Crank–Nicolson method, which is usually used for numerical solution of the heat (diffusion) equation. This is a finite difference method, which have a second-order accuracy in time and space. In addition, the Crank-Nikolson method is implicit in time and numerically stable. 	

Over the interval $x\in [0, h]$, we could use $N_x$ equally spaced points:
$$
        x_i = i\Delta x, \quad i = 0, 1, \ldots, N_x-1
$$
where $\Delta x = h_m/(N_x-1)$ is so-called grid spacing. Similarly, for the variable $s\in[0, 1]$, we use $N_s$  discretization points:
$$
        s_m = m\Delta s, \quad m = 0, 1, \ldots, N_s-1
$$
where $\Delta s = 1/(N_s-1)$ is the contour step. The value of the propagator $q(x, s)$ at these discrete space-contour points we denote as
$$
        q_i^m = q(x_i, s_m) 
$$
The next step is to introduce approximations for the relevant derivatives. The left hand side of the diffusion equation Eq.(\ref{app_homo_diff_eq_field}) is approximated by the backward contour difference. The right hand side of the diffusion equation, Eq.(\ref{app_homo_diff_eq_field}), is approximated by the average of the 
central difference scheme evaluated at the current and the previous contour steps. 
Thus, Eq.(\ref{app_homo_diff_eq_field}) is approximated by       
\begin{equation}
\label{app_cranc_nic_field}
       \frac{q_{i}^{m}-q_{i}^{m-1}}{\Delta s} = \frac{1}{2}\left[\frac{q_{i-1}^{m}-2q_{i}^{m}+q_{i+1}^{m}}{\Delta x^2}+
       \frac{q_{i-1}^{m-1}-2q_{i}^{m-1}+q_{i+1}^{m-1}}{\Delta x^2}\right] - \frac{1}{2}w_i\left(q_{i}^{m}+q_{i}^{m-1}\right)
\end{equation}   
This equation is used to predict the value of $q$ at the contour step $m$, so all the value of $q$ at contour step $m-1$ are assumed to be known. Let us rearrange Eq.(\ref{app_cranc_nic_field}) putting values of $q$ at the contour step $m$ to the left, and values of $q$ at the contour step $m-1$ to the right:
\begin{equation}
\label{app_cranc_nic_scheme_field}
         -\gamma q_{i-1}^{m} + (1+ \beta w_i + 2\gamma)q_{i}^{m} - \gamma q_{i+1}^{m} = 
         \gamma q_{i-1}^{m-1} + (1-\beta w_i-2\gamma)q_{i}^{m-1} + \gamma q_{i+1}^{m-1}
\end{equation}
where $\gamma = \Delta s/2\Delta x^2$ and $\beta = \Delta s/2$. The system of equations 
Eq.(\ref{app_cranc_nic_scheme_field}) can be rewritten in the following matrix form:
\begin{equation}
\label{app_homo_vw_tridiag_matrix}
\left(
\begin{array}{cccccc}
    b_0&  c_0&   0  &     0   &  \ldots&   0\\
    a_1&  b_1&   c_1&     0   &  \ldots&   0\\
     0 &  a_2&   b_2&   c_2   &  \ldots&   0\\
    \ldots&  &\vdots&     & \ddots&  \vdots \\
    \ldots& & 0&      a_{N_x-2}&    b_{N_x-2} &  c_{N_x-2}    \\
    \ldots& &  &              &      a_{N_x-1}&  b_{N_x-1}   
\end{array}
\right) \left(
\begin{array}{c}
    q_{0}^m\\
    q_{1}^m\\
    q_{2}^m\\
    \vdots\\
    q_{N_x-2}^m\\ 
    q_{N_x-1}^m
\end{array}
\right) = 
\left(
\begin{array}{c}
    d_0\\
    d_{1}\\
    d_{2}\\
    \vdots\\
    d_{N_x-2}\\
    d_{N_x-1}
\end{array}
\right)
\end{equation}
where the matrix coefficients of internal nodes are
\begin{equation}
\label{app_homo_vw_tridiag_coefficients}
\begin{array}{ll}
         a_i = -\gamma,        & \quad i = 1, 2, \ldots, N_x-2 \\
         b_i = 1+\beta w_i+2\gamma,      & \\
         c_i = -\gamma         & \\
         d_i = \gamma q_{i-1}^{m-1} + (1-\beta w_i-2\gamma)q_{i}^{m-1} + \gamma q_{i+1}^{m-1} &  
\end{array} 
\end{equation}     
Due to the symmetry of the task, the boundary condition at midplate is $q_x(h_m, s) = 0$. In order to maintain the second order accuracy of the algorithm, 
we have to introduce a so called $ghost$ point, $x_{N_x} = h_m + \Delta x$ outside the region, and approximate the boundary condition by
$$
	 \frac{q_{N_x}^{m} - q_{N_x-2}^{m}}{2\Delta x} = 0
$$	
which, in turn, means that $q^{m}_{N_x} = q^{m}_{N_x-2}$. Using the last result and 
Eq.(\ref{app_cranc_nic_scheme_field}), written at $i = N_x-1$, we obtain the expression:
$$
        (1 + \beta w_i + 2\gamma)q_{N_x-1}^{m} - 2\gamma q_{N_x-2}^{m} = (1 -\beta w_i - 2\gamma)q_{N_x-1}^{m-1} + 2\gamma q_{N_x-2}^{m-1}
$$
which immediately leads to
\begin{equation}
\label{app_homo_vw_tridiag_neuman_bc}
\begin{array}{ll}    
        a_{N_x-1} = -2\gamma, & b_{N_x-1} = (1+\beta w_i+2\gamma), \\
	    d_{N_x-1} = (1- \beta w_i-2\gamma)q_{N_x-1}^{m-1} + 2\gamma q_{N_x-2}^{m-1}. & 
\end{array} 
\end{equation}
Then, depending on the boundary conditions on the wall, the boundary nodes will be different. Consider consequently different boundary conditions.\\
1) Purely repulsive surface, $q(0, s)=0$ (Dirichlet b.c.).
For the the boundary condition ($q_0^m = 0$). In order to obtain the matrix coefficients, $b_0, c_0, d_0$, we should 
continuously extend the solution, $q(x, s)$ for negative values of $x$. Ensuring the continuity of the derivative at $x=0$, we should demand from the function $q(x, s)$, to be an odd function with respect to $x=0$. Thereby, we can write $q_{1}^m = -q_{-1}^m$, that is valid for any contour step, $m$. Then, putting the expression
to Eq.(\ref{app_cranc_nic_scheme_field}), for $i=0$, we obtain
$$
\begin{array}{l}    
-\gamma (q_{-1}^{m} + q_{1}^{m}) + (1+\beta w_0+2\gamma)q_{0}^{m} = c_0 (q_{-1}^{m} + q_{1}^{m}) + b_0q_{0}^{m} = \\ =d_0 = \gamma (q_{-1}^{m-1} + q_{1}^{m-1}) + (1-2\gamma)q_{0}^{m-1} = 0
\end{array}
$$
Due to the boundary condition: $q_{0}^{m} = 0$ the coefficient $b_0$, can be any finite number, and for simplicity we fix it at $b_0=1$. Thus:
\begin{equation}
\label{tridiag_dirichlet_bc}
\begin{array}{lll}
        b_0 = 1,     & \quad c_0 = 0, & \quad d_0 = 0 
\end{array} 
\end{equation} 	
2) Adsorbed surface, $q_x(0, s)=-q(0, s)/b$ (Robin b.c.). In order to provide the second order approximation to the Robin boundary condition, we apply the center difference at the boundary, which goes beyond the region. 
As before, let us introduce, so called $ghost$ point, $x_{-1} = - \Delta x$, which spread outside the region, 
Correspondingly, we approximate the boundary conditions by the following expressions:
$$
	\frac{q_{1}^{m} - q_{-1}^{m}}{2\Delta x} = -\frac{1}{b}q_{0}^{m} 
$$
which, in turn, means that, $q^{m}_{-1} = q^{m}_{1}+2\Delta x/b q_{0}^{m}$. 
Using the result and Eq.(\ref{app_cranc_nic_scheme_field}), written for $i = 0$, we obtain the following expressions:
\begin{equation}
\label{app_ads_eqdwards_bc}
(1 + \beta w_0 + 2\gamma(1-\Delta x/b))q_{0}^{m} - 2\gamma q_{1}^{m} = (1 - \beta w_0 - 2\gamma(1-\Delta x/b))q_{0}^{m-1} + 2\gamma q_{1}^{m-1} 
\end{equation}
Correspondingly, for the boundary nodes, we can write:
\begin{equation}
\label{app_ads_adsorption_coeff_b}         
\begin{array}{l}
b_0 = 1+\beta w_0+2\gamma(1-\Delta x/b), \quad c_0 = -2\gamma, \\
d_0 = (1 - \beta w_0 - 2\gamma(1-\Delta x/b))q_{0}^{m-1} + 2\gamma q_{1}^{m-1}	
\end{array} 
\end{equation}

3) External surface potential, $u_s(x)$, with $q_x(0, s)=0$ .	  	  
In case when the adsorption is set by the surface potential $u_s(x)$, the "surface" boundary condition is changed to $q_x(0, s) = 0$. 
In order to write new numerical scheme we should add the the surface field, $u_s(x)$, to the self-consistent field, $w(x)$, for the interior nodes, Eq.(\ref{app_homo_vw_tridiag_coefficients}), and use new boundary conditions:   
\begin{equation}
\label{adsorption_coeff_surf}         
\begin{array}{l}
b_0 = 1+\beta (w_0+u_s^0)+2\gamma, \quad c_0 = -2\gamma, \quad a_{N_x-1} = -2\gamma, \\
d_0 = (1 - \beta (w_0+u_s) - 2\gamma q_{0}^{m-1} + 2\gamma q_{1}^{m-1}, \\ 
\end{array} 
\end{equation}
which, in turn, can be obtained from Eq.(\ref{app_ads_adsorption_coeff_b}) setting $b=\infty$.   

The system of equations, Eq.(\ref{app_homo_vw_tridiag_matrix}), can be efficiently solved using the $LU$ factorization with backward substitution. This method use only $O(N_x)$ operations for solving the three diagonal system of equations on each contour step. The description of $LU$ factorization, in the context of the three diagonal matrix, can be found in Appendix.\ref{AppendixA}. For more details see \cite{thomas_1995, press_2007}.

%% file: Appendices/AppendixC.tex
% Appendix C
%%%%%%%%%%%%%%%%%%%%%%%%%%%%%%%%%%%%%%%%%%%%%%%%%%%%%%%%%%%%%%%%%%%%%%%%%%%%%%%%%%%%%%%%%%%%%%%%%%%%%%%%%%%%%%%%%%%%%%%%%%%%%%%%%%%%%%%%%%%%%%%%%%%%%%%%%%%%%
%        The Crank-Nicolson scheme for the Edwards equation.
%%%%%%%%%%%%%%%%%%%%%%%%%%%%%%%%%%%%%%%%%%%%%%%%%%%%%%%%%%%%%%%%%%%%%%%%%%%%%%%%%%%%%%%%%%%%%%%%%%%%%%%%%%%%%%%%%%%%%%%%%%%%%%%%%%%%%%%%%%%%%%%%%%%%%%%%%%%%%
\chapter{Fourier transformation} % Main appendix title
\label{AppendixC} % For referencing this appendix elsewhere, use \ref{AppendixA}

\lhead{Appendix C. \emph{Fourier transformation}} % This is for the header on each page - perhaps a shortened title

\textbf{Even and odd functions.}\\
Let us consider the Fourier transformation of an even ($f(-x) = f(x)$) and an odd ($f(-x) = f(x)$) functions. Any function we can represent in the form
$$
	f(x) = \frac{1}{2}(f(x) + f(-x)) + \frac{1}{2}(f(x) - f(-x)) \equiv f_{ev}(x) + f_{od}(x)
$$		
where $f_{ev}(x)$ is an even function and $f_{od}(x)$ is an odd function.
If we apply the Fourier transformation to the function $f(x)$, we obtain:
$$
\hat{F}(k) = \int\limits_{-\infty}^{\infty} f(x)e^{-2\pi ixk} \mathrm{d}x= 
\int\limits_{-\infty}^{\infty} f_{ev}(x)\cos(2\pi xk)\mathrm{d}x - 
i\int\limits_{-\infty}^{\infty} f_{od}(x)\sin(2\pi xk)\mathrm{d}x
$$	
other terms in last expression equal to $0$ because the integral from an odd function in symmetric limits is always  $0$. When $f(x)$ is an even function ($ f(x) = f_{ev}$)
$$
\hat{F}(k) = 2\int\limits_{0}^{\infty} f(x)\cos(2\pi xk)\mathrm{d}x 
$$	
or when $f(x)$ is an odd function ($f(x) = f_{od}$) 
$$
\hat{F}(k) = 2i\int\limits_{0}^{\infty} f(x)\sin(2\pi xk)\mathrm{d}x 
$$		
All these comments were attributed to the self-consistent field $w(x)$, because this function has the symmetry with respect to $h_m=h/2$, where $h$, is the distance between plates. Using that we can reduce the dimension of the vector $w$ twice. This allows to reduce the computation costs. Using the fact that the first derivative of self-consistent field at $x = 0$ is $w'(x) = 0$, one can  consider the reduced self-consistent field as even function. This gives us the reason to use the cosine Fourier transformation. \\\\
%%%%%%%%%%%%%%%%%%%%%%%%%%%%%%%%%%%%%%%%%%%%%%%%%%%%%%%%%%%%%%%%%%%%%%%%%%%%%%%%%%%%%%%%%%%%%%%%%%%%%%%%%%%%%%%%%%%%%%%%%%%%%%%
%        Discrete Fourier cosine transformation.
%%%%%%%%%%%%%%%%%%%%%%%%%%%%%%%%%%%%%%%%%%%%%%%%%%%%%%%%%%%%%%%%%%%%%%%%%%%%%%%%%%%%%%%%%%%%%%%%%%%%%%%%%%%%%%%%%%%%%%%%%%%%%%%
\textbf{Discrete Fourier cosine transformation.}\\
Formally, the discrete cosine transform is a linear, invertible function $F: RN \rightarrow RN$ (where $R$ denotes the set of real numbers), or equivalently an invertible $N \times N$ square matrix. There are several variants of the the discrete Fourier transformations (DCT) with slightly modified definitions (we will use only one of them calls as DCT-I). $N$ real numbers $w_0, \ldots, w_{N-1}$ are transformed into $N$ real numbers $\hat{w}_{0}, \ldots, \hat{w}_{N-1}$ according to the formula:
$$
	\hat{w}_{k} = \frac{1}{2}(w_0 + (-1)^{k}w_{N-1}) +\sum\limits_{i=1}^{N-2}w_{i}\cos(\pi ik/(N-1))
$$
The DCT-I corresponds to the boundary conditions: $x_{i}$ is the even function around $i=0$ and around $i=N-1$; similarly for $X_{k}$. The inverse of DCT-I is DCT-I, but the output array should be multiplied by $2/(N-1)$.\\
In the calculations in the main part of the thesis we used \textbf{FFTW}, \cite{FFTW3}, library for getting the fast Fourier cosine transformation. This is a standard fast Fourier transformation library, which is used as a main instrument for getting Fourier transformation in many computational packages and also serves as the etalon of implementation of fast Fourier transformation on different computer languages.